\documentclass[12pt]{memoir}

\usepackage[utf8]{inputenc}
\usepackage[T1]{fontenc}

\usepackage{palatino}
\usepackage{mathpazo}
\usepackage{microtype}
\usepackage{mathtools}
\usepackage{amssymb}
\usepackage{tikz}
\usepackage{tikz-3dplot}
\usepackage{circuitikz}
\usepackage{blochsphere}
\usepackage{pgfplots} \pgfplotsset{compat=1.18}
\usepackage{booktabs}
\usepackage{enumitem}
\usepackage{geometry}
\usepackage{scalerel}
\usepackage{titletoc}
\usepackage{enumitem}
\usepackage{hyperref}

\pgfmathsetmacro{\samples}{400}

\definecolor{Blue40}{HTML}{78A9FF}
\definecolor{Blue60}{HTML}{0F62FE}
\definecolor{Purple60}{HTML}{8A3FFC}
\definecolor{Purple50}{HTML}{A56EFF}
\definecolor{Purple40}{HTML}{BE95FF}
\definecolor{Magenta50}{HTML}{EE5396}

\colorlet{Gray}{black!20}
\colorlet{LightGray}{black!10}

\colorlet{Highlight}{Purple60}
\colorlet{CircuitBlue}{Blue40}
\colorlet{DataColor0}{Blue60}
\colorlet{DataColor1}{Purple60}
\definecolor{DataColor2}{HTML}{005D5D}
\definecolor{DataColor3}{HTML}{9F1853}

\usetikzlibrary{
  shapes,
  shapes.misc,
  patterns,
  positioning,
  calc,
  arrows.meta,
  angles,
  quotes,
  3d
}

\title{%
  {\LARGE\normalfont Understanding}\\[-1mm]
  Quantum Information\\[-1mm]
  and Computation\\
  {\large\normalfont A Course on the Theory of Quantum Computing}}
\author{John Watrous}
\date{}

\sloppybottom
\setlist[enumerate,itemize]{itemsep=3pt, parsep=1pt, topsep=6pt}

\makeatletter
\renewcommand{\@endpart}{}
\makeatother

\newcommand{\parttoc}{%
  \begingroup
  \vspace{2\baselineskip}
  \startcontents[part]
  \printcontents[part]{}{0}{\setcounter{tocdepth}{1}}
  \vspace{2\baselineskip}
  \endgroup
}

\newcommand{\unit}[2][]{%
  \ifx\relax#1\relax
    \part{#2}%
  \else
    \part[#1]{#2}%
  \fi
  \parttoc
  \clearpage
}

\newcommand{\lesson}[1]{\chapter{#1}}

\hypersetup{
    colorlinks=true,
    linkcolor=blue,
    urlcolor=cyan,
    pdftitle={Understanding Quantum Information and Computation},
    pdfauthor={John Watrous}
}

\usepackage[skins,breakable]{tcolorbox}
\usepackage{amsthm}

\newtcolorbox{callout}[1][]{
  colback=black!10,
  colframe=Gray,
  coltitle=black,
  boxrule=0pt,
  arc=0pt,
  left=10pt,
  right=8pt,
  top=8pt,
  bottom=8pt,
  fonttitle=\bfseries,
  before title={\vspace{3pt}},
  after title={\vspace{3pt}},
  #1
}

\newcommand{\readout}[1]{%
  \draw[thick, color=white] (#1.center) ++ (0.33,-0.12) arc (0:180:3.3mm);
  \draw[thick, color=white] (#1.center) ++ (0,-0.12) -- +(0.33,0.33);
  \fill[color=white] (#1.center) ++ (0,-0.12) circle (0.4mm);
}

\newcommand{\smallreadout}[1]{%
  \draw[thick, color=white] (#1.center) ++ (0.15,-0.07) arc (0:180:1.5mm);
  \draw[thick, color=white] (#1.center) ++ (0,-0.05) -- +(0.16,0.16);
  \fill[color=white] (#1.center) ++ (0,-0.05) circle (0.225mm);
}

\tikzset{
  ground/.pic={
    \draw (-2mm,0) to (2mm,0);
    \draw (-1.3mm,-2pt) to (1.3mm,-2pt);
    \draw (-0.6mm,-4pt) to (0.6mm,-4pt);    
  }
}

\DeclarePairedDelimiter\bra{\langle}{\rvert}
\DeclarePairedDelimiter\ket{\lvert}{\rangle}

\newenvironment{problem} {%
  \begin{description}[
      leftmargin=5.5em,
      labelwidth=5em,
      labelsep=0.5em,
      align=left,
      topsep=0pt,
      partopsep=0pt,
      itemsep=2pt,
      parsep=0pt
    ]
    \abovedisplayskip=6pt
    \belowdisplayskip=6pt
    \abovedisplayshortskip=3pt
    \belowdisplayshortskip=3pt
}{%
    \end{description}
}

\newcommand{\Input}[1]{\item[\textbf{Input:}] #1}
\newcommand{\Output}[1]{\item[\textbf{Output:}] #1}
\newcommand{\Promise}[1]{\item[\textbf{Promise:}] #1}

\usepackage{listings}
\usepackage{seqsplit}

\definecolor{codebg}{RGB}{247,247,247}
\definecolor{outputbg}{RGB}{255,255,255}
\definecolor{bordercolor}{RGB}{207,207,207}
\definecolor{promptcolor}{RGB}{0,0,255}

\newtcolorbox{code}[1][]{
  colback=LightGray,
  colframe=Gray,
  boxrule=1pt,
  arc=0pt,
  boxsep=0pt,
  left=10pt,
  right=10pt,
  top=2pt,
  bottom=2pt,
  breakable,
  before skip=\medskipamount,
  after skip=\medskipamount,
  #1
}

\newenvironment{invbox}[2]{%
  \begin{minipage}[c][#2][t]{#1}
    \raggedright
}{
  \end{minipage}
}

\makeatletter
\renewcommand\partnumberlinehook[1]{%
  \settowidth{\@tempdima}{\bfseries #1 }%
  \addtolength{\@tempdima}{0.25em}%
}

\renewcommand{\l@part}[2]{%
  \ifnum \c@tocdepth >-2\relax
  \addpenalty{-\@highpenalty}%
  \addvspace{1em \@plus\p@}%
  \begingroup
  \parindent \z@ \rightskip \@pnumwidth
  \parfillskip -\@pnumwidth {%
    \leavevmode \large \bfseries
    Unit #1\hfill #2}\par
  \addvspace{0.5em}
  \nobreak
  \endgroup
  \fi}
\makeatother

\begin{document}

\hypersetup{
    linktocpage=true,
    linkcolor=Highlight,
    urlcolor=Highlight
}

\pretitle{\begin{center}\huge\bfseries}
\posttitle{\end{center}}
\setlength{\droptitle}{-2cm}

\frontmatter
\pagestyle{empty}
\hypersetup{pageanchor=false}
\renewcommand{\thepage}{}  

\maketitle

\vspace*{-100pt}

\setlength{\cftbeforechapterskip}{0pt}
\setlength{\cftbeforepartskip}{0pt}
\renewcommand{\cftpartaftersnumb}{\par\vskip 6pt}
\renewcommand{\cftchapterfont}{\normalfont}
\maxtocdepth{chapter}
\renewcommand{\contentsname}{}
\tableofcontents*
\setlength{\cftbeforechapterskip}{10pt}

\chapter*{Preface}

Welcome to \emph{Understanding Quantum Information and Computation}, a course
on the theory of quantum computing that I created while working for IBM as
Technical Director for Quantum Education from 2022 to 2025.
It covers subject matter roughly corresponding to a one-semester university
course at the advanced undergraduate or introductory graduate level.
It has 16 lessons divided into four units, with each lesson including a video
and a written component.
The videos are available through the
\href{https://www.youtube.com/@qiskit}{Qiskit YouTube channel}
and the written material is available through
\href{https://quantum.cloud.ibm.com/learning}{IBM Quantum Learning},
where it is split into four courses named for the four units.

This document essentially represents a ``Director's Cut''
(literally perhaps, given my former title) of the written material.
It is mostly unchanged from that which is available from IBM Quantum Learning
aside from its typesetting, a few minor tweaks here and there, the addition of
a bibliography, and a reversion to the structure of four units rather than four
separate courses, as it was originally envisioned.

I wish to thank IBM, and especially Jay Gambetta, for supporting the creation
of this course with an understanding from day one that it would be made freely
available to the community.
Those knowledgeable about the history of quantum information and computation,
from its early days to the present, will recognize the massive role that IBM
has played in the development of the field, and it is my honor to add this
course to the body of work done under its banner.

As a professor, I taught university students about quantum computing in
classrooms for over 20 years.
But eventually I had this thought that I couldn't let go:
\emph{not everyone who wants to learn this stuff can be here.}
Not every university offers courses on quantum computing, and those that do
turn many people away.
For some the days of being a student have passed, and for too many others the
opportunity to attend a university has never existed.
This was the motivation behind the creation of this course:
to explain quantum computing to anyone interested in learning, wherever they
are and whenever they choose to get started.
(And make no mistake: this course is just a start.)

I could not have done this alone and thank all those who contributed and
offered input, guidance, and support.
This includes the incredible video team at IBM Quantum
(Clinton Herrick, Joshua Luna, David Rodriguez, and Paul Searle), as well as
Frank Harkins,
Jacob Watkins,
Leron Gil,
Russell Huffman,
Sanket Panda,
Beatriz Carramolino Arranz,
Olivia Lanes,
Chris Porter,
Katie McCormick,
Nathaniel DePue,
Abby Cross,
Becky Dimock,
Grace Lindsell,
Pedro Rivero,
Manfred Oevers,
Sergey Bravyi,
Ted Yoder,
Borja Peropadre,
Scott Crowder,
and
Katie Pizzolato,
not to mention the magnificent designers and developers at IBM Quantum who
built a web platform from the ground up in no small part to support this
course.
I am grateful to have worked with such talented people and appreciate you all!

This course is licensed under the Creative Commons Attribution-ShareAlike 4.0
International (CC BY-SA 4.0) license.
This means that it may be copied, redistributed, remixed, transformed, and
built upon provided that the
\href{https://creativecommons.org/licenses/by-sa/4.0/legalcode.en}{%
  terms of the license} are respected.
In particular, educators can freely use this content in courses, handouts, and
online materials, and adapt it to their needs.
I sincerely hope that it will find value in the hands of learners and educators
alike.
\vspace{5mm}

\noindent
Waterloo, Canada, July 2025

\chapter*{Dedication}

This course is dedicated to the memory of David Poulin (1976--2020), an
outstanding scientist and the very best of colleagues, from whom I first
learned about the toric code.


\mainmatter
\hypersetup{pageanchor=true}
\setcounter{page}{1}
\pagenumbering{arabic}
\pagestyle{headings}


\unit[Basics of Quantum Information]{Basics of\\[-1mm] Quantum Information}
\label{unit:basics-of-quantum-information}

This unit introduces the mathematics of quantum information, including a
description of quantum information for both single and multiple systems;
quantum circuits, which provide a standard way to describe quantum
computations; and three fundamentally important examples connected with the
phenomenon of quantum entanglement.

\begin{trivlist}
  \setlength{\parindent}{0mm}
  \setlength{\parskip}{2mm}
  \setlength{\itemsep}{1mm}
  
\item
  \textbf{Lesson 1: Single Systems}
  
  This lesson introduces the basics of quantum information for single systems,
  including the description of quantum states as vectors with complex number
  entries, measurements that allow classical information to be extracted from
  quantum states, and operations on quantum states that are described by
  unitary matrices.

  Lesson video URL: \url{https://youtu.be/3-c4xJa7Flk}

\item
  \textbf{Lesson 2: Multiple Systems}
  
  This lesson extends the description of quantum information presented in the
  previous lesson to multiple systems, such as collections of qubits.
  
  Lesson video URL: \url{https://youtu.be/DfZZS8Spe7U}
  
\item
  \textbf{Lesson 3: Quantum Circuits}
  
  This lesson introduces the quantum circuit model, as well as some
  mathematical concepts that are important to quantum information including
  inner products, orthogonality, and projections. Fundamental limitations of
  quantum information, including the no-cloning theorem, are also discussed.

  Lesson video URL: \url{https://youtu.be/30U2DTfIrOU}
  
\item
  \textbf{Lesson 4: Entanglement in Action}
  
  This lesson covers three fundamentally important examples in quantum
  information: the teleportation and superdense coding protocols and an
  abstract game known as the CHSH game. The interesting and important
  phenomenon of entanglement plays a key role in all three examples.

  Lesson video URL:
  \url{https://youtu.be/GSsElSQgMbU}
  
\end{trivlist}

\newpage


\lesson{Single Systems}
\label{lesson:single-systems}

This lesson introduces the basic framework of quantum information, including
the description of quantum states as vectors with complex number entries,
measurements that allow classical information to be extracted from quantum
states, and operations on quantum states that are described by unitary
matrices.

We will restrict our attention in this lesson to the comparatively simple
setting in which a \emph{single system} is considered in isolation.
In the next lesson, we'll expand our view to \emph{multiple systems}, which can
interact with one another and be correlated.

\section{Classical information}

To describe quantum information and how it works, we'll begin with an
overview of classical information.
It is natural to wonder why so much attention is paid to classical information
in a course on quantum information, but there are good reasons.

For one, although quantum and classical information are different in some
spectacular ways, their mathematical descriptions are actually quite similar.
Classical information also serves as a familiar point of reference when
studying quantum information, as well as a source of analogy that goes a
surprisingly long way.
It is common that people ask questions about quantum information that have
natural classical analogs, and often those questions have simple answers that
can provide both clarity and insight into the original questions about quantum
information.
Indeed, it is not at all unreasonable to claim that one cannot truly understand
quantum information without understanding classical information.

Some readers may already be familiar with the material to be discussed in this
section, while others may not --- but the discussion is meant for both
audiences.
In addition to highlighting the aspects of classical information that are most
relevant to an introduction to quantum information, this section introduces the
\emph{Dirac notation}, which is often used to describe vectors and matrices in
quantum information and computation.
As it turns out, the Dirac notation is not specific to quantum information; it
can equally well be used in the context of classical information, as well as
for many other settings in which vectors and matrices arise.

\subsection{Classical states and probability vectors}

Suppose that we have a system that stores information.
More specifically, we shall assume that this system can be in one of a finite
number of \emph{classical states} at each instant.
Here, the term classical state should be understood in intuitive terms, as a
configuration that can be recognized and described unambiguously.

The archetypal example, which we will come back to repeatedly, is that of a
\emph{bit}, which is a system whose classical states are $0$ and $1$.
Other examples include a standard six-sided die, whose classical states are
$1$, $2$, $3$, $4$, $5$, and $6$ (represented by the corresponding number of
dots on whatever face is on top); a nucleobase in a strand of DNA, whose
classical states are \emph{A}, \emph{C}, \emph{G}, and \emph{T};
and a switch on an electric fan, whose classical states are (commonly)
\emph{high}, \emph{medium}, \emph{low}, and \emph{off}.
In mathematical terms, the specification of the classical states of a system
are, in fact, the starting point: we \emph{define} a bit to be a system that has
classical states $0$ and $1$, and likewise for systems having different
classical state sets.

For the sake of this discussion, let us give the name $\mathsf{X}$ to the
system being considered, and let us use the symbol $\Sigma$ to refer to the set
of classical states of $\mathsf{X}$.
In addition to the assumption that $\Sigma$ is finite, which was already
mentioned, we naturally assume that $\Sigma$ is \emph{nonempty} --- for it is
nonsensical for a physical system to have no states at all.
And while it does make sense to consider physical systems having
\emph{infinitely} many classical states, we will disregard this possibility,
which is certainly interesting but is not relevant to this course.
For these reasons, and for the sake of convenience and brevity, we will
hereafter use the term \emph{classical state set} to mean any finite and
nonempty set.

Here are a few examples:

\begin{enumerate}
\item
  If $\mathsf{X}$ is a bit, then $\Sigma = \{0,1\}$. In words, we refer to this
  set as the \emph{binary alphabet}.
\item
  If $\mathsf{X}$ is a six-sided die, then $\Sigma = \{1,2,3,4,5,6\}$.
\item
  If $\mathsf{X}$ is an electric fan switch, then
  $\Sigma = \{\mathrm{high}, \mathrm{medium}, \mathrm{low}, \mathrm{off}\}$.
\end{enumerate}

When thinking about $\mathsf{X}$ as a carrier of information, the different
classical states of $\mathsf{X}$ could be assigned certain meanings, leading to
different outcomes or consequences.
In such cases, it may be sufficient to describe $\mathsf{X}$ as simply being in
one of its possible classical states.
For instance, if $\mathsf{X}$ is a fan switch, we might happen to know with
certainty that it is set to \emph{high}, which might then lead us to switch it
to \emph{medium}.

Often in information processing, however, our knowledge is uncertain.
One way to represent our knowledge of the classical state of a system
$\mathsf{X}$ is to associate \emph{probabilities} with its different possible
classical states, resulting in what we shall call a \emph{probabilistic state}.

For example, suppose $\mathsf{X}$ is a bit.
Based on what we know or expect about what has happened to $\mathsf{X}$ in the
past, we might perhaps believe that $\mathsf{X}$ is in the classical state $0$
with probability $3/4$ and in the state $1$ with probability $1/4$.
We may represent these beliefs by writing this:
\[
\operatorname{Pr}(\mathsf{X}=0) = \frac{3}{4}
\quad\text{and}\quad
\operatorname{Pr}(\mathsf{X}=1) = \frac{1}{4}.
\]
A more succinct way to represent this probabilistic state is by a column vector.
\[
\begin{pmatrix}
  \frac{3}{4}\\[2mm]
  \frac{1}{4}
\end{pmatrix}
\]
The probability of the bit being $0$ is placed at the top of the vector and the
probability of the bit being $1$ is placed at the bottom, because this is the
conventional way to order the set $\{0,1\}$.

In general, we can represent a probabilistic state of a system having any
classical state set in the same way, as a vector of probabilities.
The probabilities can be ordered in any way we choose, but it is typical that
there is a natural or default way to do this.
To be precise, we can represent any probabilistic state through a column vector
satisfying two properties:

\begin{enumerate}
\item
  All entries of the vector are nonnegative real numbers.
\item
  The sum of the entries is equal to $1$.
\end{enumerate}

\noindent
Conversely, any column vector that satisfies these two properties can be taken
as a representation of a probabilistic state.
Hereafter, we will refer to vectors of this form as \emph{probability vectors}.

Alongside the succinctness of this notation, identifying probabilistic states
as column vectors has the advantage that operations on probabilistic states are
represented through matrix-vector multiplication, as will be discussed shortly.

\subsection{Measuring probabilistic states}

Next let us consider what happens if we \emph{measure} a system when it is in a
probabilistic state.
In this context, by measuring a system we simply mean that we look at the
system and recognize whatever classical state it is in without ambiguity.
Intuitively speaking, we can't ``see'' a probabilistic state of a
system; when we look at it, we just see one of the possible classical states.

By measuring a system, we may also change our knowledge of it, and therefore
the probabilistic state we associate with it can change.
That is, if we recognize that $\mathsf{X}$ is in the classical state
$a\in\Sigma$, then the new probability vector representing our knowledge of the
state of $\mathsf{X}$ becomes the vector having a $1$ in the entry
corresponding to $a$ and $0$ for all other entries.
This vector indicates that $\mathsf{X}$ is in the classical state $a$ with
certainty --- which we know having just recognized it --- and we denote this
vector by $\vert a\rangle$, which is read as ``ket $a$'' for a reason that will
be explained shortly.
Vectors of this sort are also called \emph{standard basis} vectors.

For example, assuming that the system we have in mind is a bit, the standard
basis vectors are given by
\[
\vert 0\rangle =
\begin{pmatrix}
  1\\[1mm]
  0
\end{pmatrix}
\quad\text{and}\quad
\vert 1\rangle = \begin{pmatrix}0\\[1mm] 1\end{pmatrix}.
\]
Notice that any two-dimensional column vector can be expressed as a linear
combination of these two vectors.
For example,
\[
\begin{pmatrix}
  \frac{3}{4}\\[2mm]
  \frac{1}{4}
\end{pmatrix}
= \frac{3}{4}\,\vert 0\rangle + \frac{1}{4}\,\vert 1\rangle.
\]
This fact naturally generalizes to any classical state set: any column vector
can be written as a linear combination of standard basis states.
Quite often we express vectors in precisely this way.

Returning to the change of a probabilistic state upon being measured, we may
note the following connection to our everyday experiences.
Suppose we flip a fair coin, but cover up the coin before looking at it.
We would then say that its probabilistic state is
\[
\begin{pmatrix}
  \frac{1}{2}\\[2mm]
  \frac{1}{2}
\end{pmatrix}
= \frac{1}{2}\,\vert\text{heads}\rangle + \frac{1}{2}\,\vert\text{tails}\rangle.
\]
Here, the classical state set of our coin is $\{\text{heads},\text{tails}\}$.
We'll choose to order these states as heads first, tails second. 
\[
\vert\text{heads}\rangle = \begin{pmatrix}1\\[1mm] 0\end{pmatrix}
\qquad
\vert\text{tails}\rangle = \begin{pmatrix}0\\[1mm] 1\end{pmatrix}
\]

If we were to uncover the coin and look at it, we would see one of the two
classical states: heads or tails.
Supposing that the result were tails, we would naturally update our description
of the probabilistic state of the coin so that it becomes
$|\text{tails}\rangle$.
Of course, if we were then to cover up the coin, and then uncover it and look
at it again, the classical state would still be tails, which is consistent with
the probabilistic state being described by the vector $|\text{tails}\rangle$.

This may seem trivial, and in some sense it is.
However, while quantum systems behave in an entirely analogous way, their
measurement properties are frequently considered strange or unusual.
By establishing the analogous properties of classical systems, the way quantum
information works might seem less unusual.

One final remark concerning measurements of probabilistic states is this:
probabilistic states describe knowledge or belief, not necessarily something
actual, and measuring merely changes our knowledge and not the system itself.
For instance, the state of a coin after we flip it, but before we look, is
either heads or tails --- we just don't know which until we look.
Upon seeing that the classical state is tails, say, we would naturally update
the vector describing our knowledge to $|\text{tails}\rangle$, but to someone
else who didn't see the coin when it was uncovered, the probabilistic state
would remain unchanged.
This is not a cause for concern; different individuals may have different
knowledge or beliefs about a particular system, and hence describe that system
by different probability vectors.

\subsection{Classical operations}

In the last part of this brief summary of classical information, we will
consider the sorts of operations that can be performed on a classical system.

\subsubsection{Deterministic operations}

First, there are deterministic operations, where each classical state
$a\in\Sigma$ is transformed into $f(a)$ for some function $f$ of the form
$f:\Sigma\rightarrow\Sigma$.

For example, if $\Sigma = \{0,1\}$, there are four functions of this form,
$f_1$, $f_2$, $f_3$, and $f_4$, which can be represented by tables of values as
follows: 
\[
\begin{array}{c|c}
  a & f_1(a)\\
  \hline
  0 & 0\\
  1 & 0
\end{array}
\qquad
\begin{array}{c|c}
  a & f_2(a)\\
  \hline
  0 & 0\\
  1 & 1
\end{array}
\qquad
\begin{array}{c|c}
  a & f_3(a)\\
  \hline
  0 & 1\\
  1 & 0
\end{array}
\qquad
\begin{array}{c|c}
  a & f_4(a)\\
  \hline
  0 & 1\\
  1 & 1
\end{array}
\]
The first and last of these functions are \emph{constant:}
$f_1(a) = 0$ and $f_4(a) = 1$ for each $a\in\Sigma$.
The middle two are not constant, they are \emph{balanced:}
each of the two output values occurs the same number of times
(once, in this case) as we range over the possible inputs.
The function $f_2$ is the identity function: $f_2(a) = a$ for each
$a\in\Sigma$.
And $f_3$ is the function $f_3(0) = 1$ and $f_3(1) = 0$, which is better-known
as the NOT function.

The actions of deterministic operations on probabilistic states can be
represented by matrix-vector multiplication.
Specifically, the matrix $M$ that represents a given function
$f:\Sigma\rightarrow\Sigma$ is the one that satisfies
\[
M \vert a \rangle = \vert f(a)\rangle
\]
for every $a\in\Sigma$.
Such a matrix always exists and is uniquely determined by this requirement.
Matrices that represent deterministic operations always have exactly one $1$ in
each column, and $0$ for all other entries.

For instance, the matrices $M_1,\ldots,M_4$ corresponding to the functions
$f_1,\ldots,f_4$ above are as follows:
\[
M_1 =
.
\]
Here's a quick verification showing that the first matrix is correct.
The other three can be checked similarly.
\[
\begin{aligned}
M_1 \vert 0\rangle
& =
%
= \vert 0\rangle = \vert f_1(1)\rangle
\end{aligned}
\]

A convenient way to represent matrices of these and other forms makes use of an
analogous notation for row vectors to the one for column vectors discussed
previously: we denote by $\langle a \vert$ the \emph{row} vector having a $1$
in the entry corresponding to $a$ and zero for all other entries, for each
$a\in\Sigma$.
This vector is read as ``bra $a$.''

For example, if $\Sigma = \{0,1\}$, then
\[
\langle 0 \vert = %
.
\]

For any classical state set $\Sigma$, we can view row vectors and column
vectors as matrices, and perform the matrix multiplication $\vert b\rangle
\langle a\vert$.
We obtain a square matrix having a $1$ in the entry corresponding to the pair
$(b,a)$, meaning that the row of the entry corresponds to the classical state
$b$ and the column corresponds to the classical state $a$, with $0$ for all
other entries.
For example,
\[
\vert 0 \rangle \langle 1 \vert =
%
.
\]

Using this notation, we may express the matrix $M$ that corresponds to any
given function $f:\Sigma\rightarrow\Sigma$ as
\[
M = \sum_{a\in\Sigma} \vert f(a) \rangle \langle a \vert.
\]
For example, consider the function $f_4$ above, for which $\Sigma = \{0,1\}$.
We obtain the matrix
\[
M_4 =
\vert f_4(0) \rangle \langle 0 \vert + \vert f_4(1) \rangle \langle 1 \vert
= \vert 1\rangle \langle 0\vert + \vert 1\rangle \langle 1\vert
= %
.
\]

The reason why this works is as follows.
If we again think about vectors as matrices, and this time consider the
multiplication $\langle a \vert \vert b \rangle$, we obtain a $1\times 1$
matrix, which we can think about as a scalar (i.e., a number).
For the sake of tidiness, we write this product as $\langle a \vert b\rangle$
rather than $\langle a \vert \vert b \rangle$.
This product satisfies the following simple formula.
\[
\langle a \vert b \rangle
= \begin{cases}
  1 & a = b\\[1mm]
  0 & a \neq b
\end{cases}
\]
Using this observation, together with the fact that matrix multiplication is
associative and linear, we obtain
\[
M \vert b \rangle =
\Biggl(
\sum_{a\in\Sigma} \vert f(a) \rangle \langle a \vert
\Biggr)
\vert b\rangle
= \sum_{a\in\Sigma} \vert f(a) \rangle \langle a \vert b \rangle
= \vert f(b)\rangle,
\]
for each $b\in\Sigma$, which is precisely what we require of the matrix $M$.

As we will discuss in greater detail later in a later lesson, $\langle a \vert
b \rangle$ may also be seen as an \emph{inner product} between the vectors
$\vert a\rangle$ and $\vert b\rangle$.
Inner products are critically important in quantum information, but we'll delay
a discussion of them until they are needed.

At this point the names ``bra'' and ``ket'' may be evident:
putting a ``bra'' $\langle a\vert$ together with a ``ket'' $\vert b\rangle$
yields a ``bracket'' $\langle a \vert b\rangle$.
This notation and terminology is due to Paul Dirac, and for this reason is
known as the \emph{Dirac notation}.

\subsubsection{Probabilistic operations and stochastic matrices}

In addition to deterministic operations, we have \emph{probabilistic
operations}.

For example, consider the following operation on a bit.
If the classical state of the bit is $0$, it is left alone; and if the
classical state of the bit is $1$, it is flipped, so that it becomes $0$ with
probability $1/2$ and $1$ with probability $1/2$.
This operation is represented by the matrix
\[
\begin{pmatrix}
  1 & \frac{1}{2}\\[1mm]
  0 & \frac{1}{2}
\end{pmatrix}.
\]
One can check that this matrix does the correct thing by multiplying the two
standard basis vectors by it.

For an arbitrary choice of a classical state set, we can describe the set of
all probabilistic operations in mathematical terms as those that are
represented by \emph{stochastic matrices}, which are matrices satisfying these
two properties:
\begin{enumerate}
\item
  All entries are nonnegative real numbers.
\item
  The entries in every column sum to $1$.
\end{enumerate}
Equivalently, stochastic matrices are matrices whose columns all form
probability vectors.

We can think about probabilistic operations at an intuitive level as ones where
randomness might somehow be used or introduced during the operation, just like
in the example above.
With respect to the stochastic matrix description of a probabilistic operation,
each column can be viewed as a vector representation of the probabilistic state
that is generated given the classical state input corresponding to that column.

We can also think about stochastic matrices as being exactly those matrices
that always map probability vectors to probability vectors.
That is, stochastic matrices always map probability vectors to probability
vectors, and any matrix that always maps probability vectors to probability
vectors must be a stochastic matrix.

Finally, a different way to think about probabilistic operations is that they
are random choices \emph{of} deterministic operations.
For instance, we can think about the operation in the example above as applying
either the identity function or the constant 0 function, each with probability
$1/2$.
This is consistent with the equation
\[
\begin{pmatrix}
  1 & \frac{1}{2}\\[1mm]
  0 & \frac{1}{2}
\end{pmatrix}
= \frac{1}{2}
\begin{pmatrix}
  1 & 0\\[1mm]
  0 & 1
\end{pmatrix}
+ \frac{1}{2}
\begin{pmatrix}
  1 & 1\\[1mm]
  0 & 0
\end{pmatrix}.
\]
Such an expression is always possible, for an arbitrary choice of a classical
state set and any stochastic matrix having rows and columns identified with it.

\subsubsection{Compositions of probabilistic operations}

Suppose that $\mathsf{X}$ is a system having classical state set $\Sigma$, and
$M_1,\ldots,M_n$ are stochastic matrices representing probabilistic operations
on the system $\mathsf{X}$.

If the first operation $M_1$ is applied to the probabilistic state represented
by a probability vector $u$, the resulting probabilistic state is represented
by the vector $M_1 u$.
If we then apply the second probabilistic operation $M_2$ to this new
probability vector, we obtain the probability vector
\[
M_2 (M_1 u) = (M_2 M_1) u.
\]
The equality follows from the fact that matrix multiplication, including
matrix-vector multiplication as a special case, is an associative operation.
Thus, the probabilistic operation obtained by composing the first and second
probabilistic operations, where we first apply $M_1$ and then apply $M_2$, is
represented by the matrix $M_2 M_1$, which is necessarily stochastic.

More generally, composing the probabilistic operations represented by the
matrices $M_1,\ldots,M_n$ in this order, meaning that $M_1$ is applied first,
$M_2$ is applied second, and so on, with $M_n$ applied last, is represented by
the matrix product
\[
M_n \,\cdots\, M_1.
\]
Note that the ordering is important here: although matrix multiplication is
associative, it is not a commutative operation. 
For example, if
\[
M_1 =
\begin{pmatrix}
  1 & 1\\[1mm]
  0 & 0
\end{pmatrix}
\quad\text{and}\quad
M_2 =
\begin{pmatrix}
  0 & 1\\[1mm]
  1 & 0
\end{pmatrix},
\]
then
\[
M_2 M_1 =
\begin{pmatrix}
  0 & 0 \\[1mm]
  1 & 1
\end{pmatrix}
\quad\text{and}\quad
M_1 M_2 =
\begin{pmatrix}
  1 & 1\\[1mm]
  0 & 0
\end{pmatrix}.
\]
That is, the order in which probabilistic operations are composed matters;
changing the order in which operations are applied in a composition can change
the resulting operation.

\section{Quantum information}

Now we're ready to move on to quantum information, where we make a different
choice for the type of vector that represents a state --- in this case a
\emph{quantum state} --- of the system being considered.
Like in the previous discussion of classical information, we'll be concerned
with systems having finite and nonempty sets of classical states, and we'll
make use of much of the same notation.

\subsection{Quantum state vectors}

A \emph{quantum state} of a system is represented by a column vector, similar
to a probabilistic state.
As before, the indices of the vector label the classical states of the system. 
Vectors representing quantum states are characterized by these two properties:
\begin{enumerate}
\item
  The entries of a quantum state vector are \emph{complex numbers}.
\item
  The sum of the \emph{absolute values squared} of the entries of a quantum
  state vector is $1$.
\end{enumerate}

Thus, in contrast to probabilistic states, vectors representing quantum states
need not have nonnegative real number entries, and it is the sum of the
absolute values squared of the entries (as opposed to the sum of the entries)
that must equal~$1$.
Simple as these changes are, they give rise to the differences between quantum
and classical information; any speedup from a quantum computer, or improvement
from a quantum communication protocol, is ultimately derived from these simple
mathematical changes.

The \emph{Euclidean norm} of a column vector
\[
v = \begin{pmatrix}
  \alpha_1\\
  \vdots\\
  \alpha_n
\end{pmatrix}
\]
is denoted and defined as follows:
\[
\| v \| = \sqrt{\sum_{k=1}^n |\alpha_k|^2}.
\]
The condition that the sum of the absolute values squared of a quantum state
vector equals $1$ is therefore equivalent to that vector having Euclidean norm
equal to $1$.
That is, quantum state vectors are \emph{unit vectors} with respect to the
Euclidean norm.

\subsubsection{Examples of qubit states}

The term \emph{qubit} refers to a quantum system whose classical state set
is $\{0,1\}$.
That is, a qubit is really just a bit --- but by using this name we explicitly
recognize that this bit can be in a quantum state.

These are examples of quantum states of a qubit:
\begin{gather}
  \begin{pmatrix}
    1\\[1mm]
    0
  \end{pmatrix}
  = \vert 0\rangle
  \quad\text{and}\quad
  \begin{pmatrix}
    0\\[1mm]
    1
  \end{pmatrix}
  = \vert 1\rangle,\nonumber\\[3mm]  
  \begin{pmatrix}
    \frac{1}{\sqrt{2}}\\[3mm]
    \frac{1}{\sqrt{2}}
  \end{pmatrix}
  = \frac{1}{\sqrt{2}}\,\vert 0\rangle + \frac{1}{\sqrt{2}}\,\vert 1\rangle,
  \label{eq:plus_state}\\[3mm]
  \begin{pmatrix}
    \frac{1+2i}{3}\\[1mm]
    -\frac{2}{3}
  \end{pmatrix}
  = \frac{1+2i}{3}\,\vert 0\rangle - \frac{2}{3}\,\vert 1\rangle.\nonumber
\end{gather}

The first two examples, $\vert 0\rangle$ and $\vert 1\rangle$, illustrate that
standard basis elements are valid quantum state vectors.
Their entries are complex numbers, where the imaginary part of these numbers
all happen to be $0$, and computing the sum of the absolute values squared of
the entries yields
\[
\vert 1\vert^2 + \vert 0\vert^2 = 1
\quad\text{and}\quad
\vert 0\vert^2 + \vert 1\vert^2 = 1,
\]
as required.
Similar to the classical setting, we associate the quantum state vectors $\vert
0\rangle$ and $\vert 1\rangle$ with a qubit being in the classical state $0$
and $1$, respectively.

For the other two examples, we again have complex number entries, and computing
the sum of the absolute value squared of the entries yields
\[
\biggl\vert\frac{1}{\sqrt{2}}\biggr\vert^2 +
\biggl\vert\frac{1}{\sqrt{2}}\biggr\vert^2 = \frac{1}{2} + \frac{1}{2} = 1
\]
and
\[
\biggl\vert \frac{1+2i}{3} \biggr\vert^2 +
\biggl\vert -\frac{2}{3} \biggr\vert^2 = \frac{5}{9} + \frac{4}{9} = 1.
\]

These are therefore valid quantum state vectors.
Note that they are linear combinations of the standard basis states $\vert 0
\rangle$ and $\vert 1 \rangle$, and for this reason we often say that they're
\emph{superpositions} of the states $0$ and $1$.
Within the context of quantum states, \emph{superposition} and \emph{linear
combination} are essentially synonymous.

The example \eqref{eq:plus_state} of a qubit state vector above is very
commonly encountered --- it is called the \emph{plus state} and is denoted as
follows:
\[
\vert {+} \rangle = \frac{1}{\sqrt{2}} \vert 0\rangle + \frac{1}{\sqrt{2}}
\vert 1\rangle.
\]
We also use the notation
\[
\vert {-} \rangle = \frac{1}{\sqrt{2}} \vert 0\rangle - \frac{1}{\sqrt{2}}
\vert 1\rangle
\]
to refer to a related quantum state vector where the second entry is negative
rather than positive, and we call this state the \emph{minus state}.

This sort of notation, where some symbol other than one referring to a
classical state appears inside of a ket, is common --- we can use whatever name
we wish inside of a ket to name a vector.
It is quite common to use the notation $\vert\psi\rangle$, or a different name
in place of $\psi$, to refer to an arbitrary vector that may not necessarily be
a standard basis vector.

Notice that, if we have a vector $\vert \psi \rangle$ whose indices correspond
to some classical state set $\Sigma$, and if $a\in\Sigma$ is an element of this
classical state set, then the matrix product $\langle a\vert \vert \psi\rangle$
is equal to the entry of the vector $\vert \psi \rangle$ whose index
corresponds to $a$.
As we did when $\vert \psi \rangle$ was a standard basis vector, we write
$\langle a \vert \psi \rangle$ rather than $\langle a\vert \vert \psi\rangle$
for the sake of readability.

For example, if $\Sigma = \{0,1\}$ and
\begin{equation}
\vert \psi \rangle =
\frac{1+2i}{3} \vert 0\rangle - \frac{2}{3} \vert 1\rangle
= \begin{pmatrix}
    \frac{1+2i}{3}\\[1mm]
    -\frac{2}{3}
  \end{pmatrix},
\label{eq:random-state}
\end{equation}
then
\[
\langle 0 \vert \psi \rangle = \frac{1+2i}{3}
\quad\text{and}\quad
\langle 1 \vert \psi \rangle = -\frac{2}{3}.
\]

In general, when using the Dirac notation for arbitrary vectors, the notation
$\langle \psi \vert$ refers to the row vector obtained by taking the
\emph{conjugate transpose} of the column vector $\vert\psi\rangle$, where the
vector is transposed from a column vector to a row vector and each entry is
replaced by its complex conjugate.
For example, if $\vert\psi\rangle$ is the vector defined in
\eqref{eq:random-state} then
\[
\langle\psi\vert
= \frac{1-2i}{3} \langle 0\vert - \frac{2}{3} \langle 1\vert
= \begin{pmatrix}
  \frac{1-2i}{3} &
  -\frac{2}{3}
\end{pmatrix}.
\]
The reason for taking the complex conjugate, in addition to the transpose, will
be made more clear later on when we discuss inner products.

\subsubsection{Quantum states of other systems}

We can consider quantum states of systems having arbitrary classical state sets.
For example, here is a quantum state vector for an electrical fan switch:
\[
\begin{pmatrix}
  \frac{1}{2}\\[1mm]
  0 \\[1mm]
  -\frac{i}{2}\\[1mm]
  \frac{1}{\sqrt{2}}
\end{pmatrix}
= \frac{1}{2} \vert\mathrm{high}\rangle
- \frac{i}{2} \vert\mathrm{low}\rangle
+ \frac{1}{\sqrt{2}} \vert\mathrm{off}\rangle.
\]
The assumption in place here is that the classical states are ordered as
\emph{high}, \emph{medium}, \emph{low}, \emph{off}.
There may be no particular reason why one would want to consider a quantum
state of an electrical fan switch, but it is possible in principle.

Here's another example, this time of a quantum decimal digit whose classical
states are $0, 1, \ldots, 9$:
\[
\frac{1}{\sqrt{385}}
\begin{pmatrix}
  1\\
  2\\
  3\\
  4\\
  5\\
  6\\
  7\\
  8\\
  9\\
  10
\end{pmatrix}
=
\frac{1}{\sqrt{385}}\sum_{k = 0}^9 (k+1) \vert k \rangle.
\]
This example illustrates the convenience of writing state vectors using the
Dirac notation.
For this particular example, the column vector representation is merely
cumbersome --- but if there were significantly more classical states it would
become unusable.
The Dirac notation, in contrast, supports precise descriptions of large and
complicated vectors in a compact form.

The Dirac notation also allows for the expression of vectors where different
aspects of the vectors are \emph{indeterminate}, meaning that they are unknown
or not yet established.
For example, for an arbitrary classical state set $\Sigma$, we can consider the
quantum state vector
\[
\frac{1}{\sqrt{|\Sigma|}} \sum_{a\in\Sigma} \vert a \rangle,
\]
where the notation $\vert\Sigma\vert$ refers to the number of elements in
$\Sigma$.
In words, this is a \emph{uniform superposition} over the classical states in
$\Sigma$.

We'll encounter much more complicated expressions of quantum state vectors in
later lessons, where the use of column vectors would be impractical or
impossible.
In fact, we'll mostly abandon the column vector representation of state
vectors, except for vectors having a small number of entries (often in the
context of examples), where it may be helpful to display and examine the
entries explicitly.

Here's one more reason why expressing state vectors using the Dirac notation is
convenient: it alleviates the need to explicitly specify an ordering of the
classical states (or, equivalently, the correspondence between classical states
and vector indices).

For example, a quantum state vector for a system having classical state set
$\{\clubsuit,\diamondsuit,\heartsuit,\spadesuit\}$, such as
\[
\frac{1}{2} \vert\clubsuit\rangle
+ \frac{i}{2} \vert\diamondsuit\rangle
- \frac{1}{2} \vert\heartsuit\rangle
- \frac{i}{2} \vert\spadesuit\rangle,
\]
is described by this expression without ambiguity, and there's really no need
to choose or specify an ordering of this classical state set to make sense of
the expression.
In this case, it's not difficult to specify an ordering of the standard card
suits --- for instance, we might choose to order them like this: $\clubsuit$,
$\diamondsuit$, $\heartsuit$, $\spadesuit$.
If we choose this particular ordering, the quantum state vector above would be
represented by the column vector
\[
\begin{pmatrix}
 \frac{1}{2}\\[1mm]
 \frac{i}{2}\\[1mm]
 -\frac{1}{2}\\[1mm]
 -\frac{i}{2}
\end{pmatrix}.
\]
In general, however, it is convenient to be able to simply ignore the issue of
how classical state sets are ordered.

\subsection{Measuring quantum states}

Next let us consider what happens when a quantum state is \emph{measured},
focusing on a simple type of measurement known as a
\emph{standard basis measurement}.
There are more general notions of measurement that we'll discuss later on.

Similar to the probabilistic setting, when a system in a quantum state is
measured, the hypothetical observer performing the measurement won't see a
quantum state vector, but rather will see some classical state.
In this sense, measurements act as an interface between quantum and classical
information, through which classical information is extracted from quantum
states.

The rule is simple: if a quantum state is measured, each classical state of the
system appears with probability equal to the \emph{absolute value squared} of
the entry in the quantum state vector corresponding to that classical state.
This is known as the \emph{Born rule} in quantum mechanics.
Notice that this rule is consistent with the requirement that the absolute
values squared of the entries in a quantum state vector sum to $1$, as it
implies that the probabilities of different classical state measurement
outcomes sum to $1$.

For example, measuring the plus state
\[
\vert {+} \rangle =
\frac{1}{\sqrt{2}} \vert 0 \rangle
+ \frac{1}{\sqrt{2}} \vert 1 \rangle
\]
results in the two possible outcomes, $0$ and $1$, with probabilities as
follows.
\begin{align*}
  \operatorname{Pr}(\text{outcome is 0})
  & = \bigl\vert \langle 0 \vert {+} \rangle \bigr\vert^2
  = \biggl\vert \frac{1}{\sqrt{2}} \biggr\vert^2
  = \frac{1}{2}\\[2mm]
  \operatorname{Pr}(\text{outcome is 1})
  & = \bigl\vert \langle 1 \vert {+} \rangle \bigr\vert^2
  = \biggl\vert \frac{1}{\sqrt{2}} \biggr\vert^2
  = \frac{1}{2}
\end{align*}

Interestingly, measuring the minus state
\[
\vert {-} \rangle =
\frac{1}{\sqrt{2}} \vert 0 \rangle
- \frac{1}{\sqrt{2}} \vert 1 \rangle
\]
results in exactly the same probabilities for the two outcomes.
\begin{align*}
  \operatorname{Pr}(\text{outcome is 0})
  & = \bigl\vert \langle 0 \vert {-} \rangle \bigr\vert^2
  = \biggl\vert \frac{1}{\sqrt{2}} \biggr\vert^2
  = \frac{1}{2}\\[2mm]
  \operatorname{Pr}(\text{outcome is 1})
  & = \bigl\vert \langle 1 \vert {-} \rangle \bigr\vert^2
  = \biggl\vert -\frac{1}{\sqrt{2}} \biggr\vert^2
  = \frac{1}{2}
\end{align*}
This suggests that, as far as standard basis measurements are concerned, the
plus and minus states are no different.
Why, then, would we care to make a distinction between them?
The answer is that these two states behave differently when operations are
performed on them, as we will discuss in the next subsection below.

Of course, measuring the quantum state $\vert 0\rangle$ results in the
classical state $0$ with certainty, and likewise measuring the quantum state
$\vert 1\rangle$ results in the classical state $1$ with certainty.
This is consistent with the identification of these quantum states with the
system \emph{being} in the corresponding classical state, as was suggested
previously.

As a final example, measuring the state
\[
\vert \psi \rangle = \frac{1+2i}{3} \vert 0\rangle - \frac{2}{3} \vert 1\rangle
\]
causes the two possible outcomes to appear with probabilities as follows:
\[
\operatorname{Pr}(\text{outcome is 0})
= \bigl\vert \langle 0 \vert \psi \rangle \bigr\vert^2
= \biggl\vert \frac{1+2i}{3} \biggr\vert^2
= \frac{5}{9},
\]
and
\[
\operatorname{Pr}(\text{outcome is 1})
= \bigl\vert \langle 1 \vert \psi \rangle \bigr\vert^2
= \biggl\vert -\frac{2}{3} \biggr\vert^2
= \frac{4}{9}.
\]

\subsection{Unitary operations}

Thus far, it may not be evident why quantum information is fundamentally
different from classical information.
That is, when a quantum state is measured, the probability to obtain each
classical state is given by the absolute value squared of the corresponding
vector entry --- so why not simply record these probabilities in a probability
vector?

The answer, at least in part, is that the set of allowable \emph{operations}
that can be performed on a quantum state is different than it is for classical
information.
Similar to the probabilistic setting, operations on quantum states are linear
mappings --- but rather than being represented by stochastic matrices, like in
the classical case, operations on quantum state vectors are represented by
\emph{unitary matrices}.

A square matrix $U$ having complex number entries is \emph{unitary} if it
satisfies the following two equations.
\begin{equation}
  \begin{aligned}
    U U^{\dagger} &= \mathbb{I} \\
    U^{\dagger} U &= \mathbb{I}
  \end{aligned}
  \label{eq:unitary-equations}
\end{equation}
Here, $\mathbb{I}$ is the identity matrix, and $U^{\dagger}$ is the
\emph{conjugate transpose} of $U$, meaning the matrix obtained by
transposing $U$ and taking the complex conjugate of each entry.
\[
U^{\dagger} = \overline{U^T}
\]
If either of the two equalities numbered \eqref{eq:unitary-equations} above is
true, then the other must also be true.
Both equalities are equivalent to $U^{\dagger}$ being the inverse of $U$:
\[
U^{-1} = U^{\dagger}.
\]
(Warning: if $M$ is not a square matrix, then it could be that $M^{\dagger} M =
\mathbb{I}$ while $M M^{\dagger} \neq \mathbb{I}$, for instance.
The equivalence of the two equalities \eqref{eq:unitary-equations} is only true
for square matrices.)

The condition that $U$ is unitary is equivalent to the condition that
multiplication by $U$ does not change the Euclidean norm of any vector.
That is, an $n\times n$ matrix $U$ is unitary if and only if
$\| U \vert \psi \rangle \| = \|\vert \psi \rangle \|$
for every $n$-dimensional column vector $\vert \psi \rangle$ with complex
number entries.
Thus, because the set of all quantum state vectors is the same as the set of
vectors having Euclidean norm equal to $1$, multiplying a unitary matrix to a
quantum state vector results in another quantum state vector.

Indeed, unitary matrices represent exactly the set of linear mappings that
always transform quantum state vectors to quantum state vectors.
Notice here a resemblance to the classical probabilistic case where operations
are associated with stochastic matrices, which are the ones that always
transform probability vectors into probability vectors.

\subsubsection{Examples of unitary operations on qubits}

The following list describes some commonly encountered unitary operations on
qubits.

\begin{description}[leftmargin=0mm]
\item[Pauli operations.]
  The four Pauli matrices are as follows:
  \[
  \mathbb{I} =
  \begin{pmatrix}
    1 & 0\\
    0 & 1
  \end{pmatrix},
  \quad
  \sigma_x =
  \begin{pmatrix}
    0 & 1\\
    1 & 0
  \end{pmatrix},
  \quad
  \sigma_y =
  \begin{pmatrix}
    0 & -i\\
    i & 0
  \end{pmatrix},
  \quad
  \sigma_z =
  \begin{pmatrix}
    1 & 0\\
    0 & -1
  \end{pmatrix}.
  \]
  A common alternative notation is $X = \sigma_x$, $Y = \sigma_y$, and $Z =
  \sigma_z$ (but be aware that the letters $X$, $Y$, and $Z$ are also commonly
  used for other purposes).
  The $X$ operation is also called a \emph{bit-flip} or a
  \emph{NOT operation} because it induces this action on bits:
  \[
  X \vert 0\rangle = \vert 1\rangle
  \quad \text{and} \quad
  X \vert 1\rangle = \vert 0\rangle.
  \]
  The $Z$ operation is also called a \emph{phase-flip}, and it has this action:
  \[
  Z \vert 0\rangle = \vert 0\rangle
  \quad \text{and} \quad
  Z \vert 1\rangle = - \vert 1\rangle.
  \]

\item[Hadamard operation.]
  The Hadamard operation is described by this matrix:
  \[
  H = 
.
  \]

\item[Phase operations.]
  A phase operation is one described by the matrix
  \[
  P_{\theta} =
  %
  \]
  for any choice of a real number $\theta$.
  The operations
  \[
  S = P_{\pi/2} =
  %
  \]
  are particularly important examples.
  Other examples include $\mathbb{I} = P_0$ and $Z = P_{\pi}$.
\end{description}

All of the matrices just defined are unitary, and therefore represent quantum
operations on a single qubit.
For example, here is a calculation that verifies that $H$ is unitary:
\begin{align*}
  %
.
\end{align*}
And here's the action of the Hadamard operation on a few commonly encountered
qubit state vectors.
\begingroup\allowdisplaybreaks
\begin{align*}
  H \vert 0 \rangle & =
  %
  = \vert 1 \rangle
\end{align*}
\endgroup
\noindent
More succinctly, we obtain these four equations.
\[
\begin{aligned}
  H \vert 0 \rangle = \vert {+} \rangle & \qquad H \vert {+} \rangle = \vert 0
  \rangle \\[1mm]
  H \vert 1 \rangle = \vert {-} \rangle & \qquad H \vert {-} \rangle = \vert 1
  \rangle
\end{aligned}
\]

It's worth pausing to consider the fact that
$H\vert {+} \rangle = \vert 0\rangle$ and
$H\vert {-} \rangle = \vert 1\rangle$, in light of the question
suggested in the previous section concerning the distinction between the states
$\vert {+} \rangle$ and $\vert {-} \rangle$.

Imagine a situation in which a qubit is prepared in one of the two quantum
states $\vert {+} \rangle$ and $\vert {-} \rangle$, but where it is not known
to us which one it is.
Measuring either state produces the same output distribution as the other, as
we already observed: $0$ and $1$ both appear with equal probability $1/2$,
which provides no information whatsoever about which of the two states was
prepared.

However, if we first apply a Hadamard operation and then measure, we obtain the
outcome $0$ with certainty if the original state was $\vert {+} \rangle$, and
we obtain the outcome $1$, again with certainty, if the original state was
$\vert {-} \rangle$.
The quantum states $\vert {+} \rangle$ and $\vert {-} \rangle$ can therefore be
discriminated \emph{perfectly}.
This reveals that sign changes, or more generally changes to the \emph{phases}
(which are also traditionally called \emph{arguments}) of the complex number
entries of a quantum state vector, can significantly change that state.

Here's another example, showing how a Hadamard operation acts on a state vector
that was mentioned previously.
\begin{multline*}
  \qquad
  H \biggl(\frac{1+2i}{3} \vert 0\rangle - \frac{2}{3} \vert 1\rangle\biggr)
  = \begin{pmatrix}
    \frac{1}{\sqrt{2}} & \frac{1}{\sqrt{2}} \\[2mm]
    \frac{1}{\sqrt{2}} & -\frac{1}{\sqrt{2}}
  \end{pmatrix}
  \begin{pmatrix}
    \frac{1+2i}{3}\\[2mm]
    -\frac{2}{3}
  \end{pmatrix}
  = \begin{pmatrix}
    \frac{-1+2i}{3\sqrt{2}}\\[3mm]
    \frac{3+2i}{3\sqrt{2}}
  \end{pmatrix}\\[2mm]
  = \frac{-1+2i}{3\sqrt{2}} | 0 \rangle
  + \frac{3+2i}{3\sqrt{2}} | 1 \rangle
  \qquad
\end{multline*}

Next, let's consider the action of a $T$ operation on a plus state.
\[
T \vert {+} \rangle
= T \biggl(\frac{1}{\sqrt{2}} \vert 0\rangle +
\frac{1}{\sqrt{2}} \vert 1\rangle\biggr)
= \frac{1}{\sqrt{2}} T\vert 0\rangle + \frac{1}{\sqrt{2}} T\vert 1\rangle
= \frac{1}{\sqrt{2}} \vert 0\rangle + \frac{1+i}{2} \vert 1\rangle
\]
Notice here that we did not bother to convert to the equivalent matrix/vector
forms, and instead used the linearity of matrix multiplication together with
the formulas
\[
T \vert 0\rangle = \vert 0\rangle
\quad\text{and}\quad
T \vert 1\rangle = \frac{1 + i}{\sqrt{2}} \vert 1\rangle.
\]
Along similar lines, we may compute the result of applying a Hadamard operation
to the quantum state vector just obtained.
\[
\begin{aligned}
H\, \biggl(\frac{1}{\sqrt{2}} \vert 0\rangle + \frac{1+i}{2} \vert
1\rangle\biggr)
& = \frac{1}{\sqrt{2}} H \vert 0\rangle + \frac{1+i}{2} H \vert 1\rangle\\[1mm]
& = \frac{1}{\sqrt{2}} \vert +\rangle + \frac{1+i}{2} \vert -\rangle \\[1mm]
& = \biggl(\frac{1}{2} \vert 0\rangle + \frac{1}{2} \vert 1\rangle\biggr)
+ \biggl(\frac{1+i}{2\sqrt{2}} \vert 0\rangle - \frac{1+i}{2\sqrt{2}} \vert
1\rangle\biggr)\\[1mm]
& = \biggl(\frac{1}{2} + \frac{1+i}{2\sqrt{2}}\biggr) \vert 0\rangle
+ \biggl(\frac{1}{2} - \frac{1+i}{2\sqrt{2}}\biggr) \vert 1\rangle
\end{aligned}
\]

The two approaches --- one where we explicitly convert to matrix
representations and the other where we use linearity and plug in the actions of
an operation on standard basis states --- are equivalent.
We can use whichever one is more convenient in the case at hand.

\subsubsection{Compositions of qubit unitary operations}

Compositions of unitary operations are represented by matrix multiplication,
just like we had in the probabilistic setting.

For example, suppose we first apply a Hadamard operation, followed by an $S$
operation, followed by another Hadamard operation.
The resulting operation, which we shall name $R$ for the sake of this example,
is as follows.
\[
R = H S H =
\begin{pmatrix}
  \frac{1}{\sqrt{2}} & \frac{1}{\sqrt{2}} \\[2mm]
  \frac{1}{\sqrt{2}} & -\frac{1}{\sqrt{2}}
\end{pmatrix}
\begin{pmatrix}
  1 & 0\\
  0 & i
\end{pmatrix}
\begin{pmatrix}
  \frac{1}{\sqrt{2}} & \frac{1}{\sqrt{2}} \\[2mm]
  \frac{1}{\sqrt{2}} & -\frac{1}{\sqrt{2}}
\end{pmatrix}
= \begin{pmatrix}
  \frac{1+i}{2} & \frac{1-i}{2} \\[2mm]
  \frac{1-i}{2} & \frac{1+i}{2}
\end{pmatrix}
\]
This unitary operation $R$ is an interesting example.
By applying this operation twice, which is equivalent to squaring its matrix
representation, we obtain a NOT operation:
\[
R^2 =
\begin{pmatrix}
  \frac{1+i}{2} & \frac{1-i}{2} \\[2mm]
  \frac{1-i}{2} & \frac{1+i}{2}
\end{pmatrix}^2
= \begin{pmatrix}
  0 & 1 \\[2mm]
  1 & 0
\end{pmatrix}.
\]
That is, $R$ is a \emph{square root of NOT} operation.
Such a behavior, where the same operation is applied twice to yield a NOT
operation, is not possible for a classical operation on a single bit.

\subsubsection{Unitary operations on larger systems}

In subsequent lessons, we will see many examples of unitary operations on
systems having more than two classical states.
An example of a unitary operation on a system having three classical states is
given by the following matrix.
\[
A =
\begin{pmatrix}
  {0} & {0} & {1} \\
  {1} & {0} & {0} \\
  {0} & {1} & {0}
\end{pmatrix}
\]
Assuming that the classical states of the system are $0$, $1$, and $2$, we can
describe this operation as addition modulo $3$.
\[
A \vert 0\rangle = \vert 1\rangle,
\quad
A \vert 1\rangle = \vert 2\rangle,
\quad\text{and}\quad
A \vert 2\rangle = \vert 0\rangle
\]

The matrix $A$ is an example of a \emph{permutation matrix}, which is a matrix
in which every row and column has exactly one $1$, with all other entries being
$0$.
Such matrices merely rearrange, or permute, the entries of the vectors they act
upon.
The identity matrix is perhaps the simplest example of a permutation matrix,
and another example is the NOT operation on a bit or qubit.
Every permutation matrix, in any positive integer dimension, is unitary.
These are the only examples of matrices that represent both classical and
quantum operations: a matrix is both stochastic and unitary if and only if it
is a permutation matrix.

Another example of a unitary matrix, this time being a $4\times 4$ matrix, is
this one:
\[
U =
\frac{1}{2}
\begin{pmatrix}
  1 & 1 & 1 & 1 \\[1mm]
  1 & i & -1 & -i \\[1mm]
  1 & -1 & 1 & -1 \\[1mm]
  1 & -i & -1 & i
\end{pmatrix}.
\]
This matrix describes an operation known as the
\emph{quantum Fourier transform}, specifically in the $4\times 4$ case. 
The quantum Fourier transform can be defined more generally, for any positive
integer dimension, and plays a key role in quantum algorithms.


\lesson{Multiple Systems}
\label{lesson:multiple-systems}

This lesson focuses on the basics of quantum information in the context of
\emph{multiple} systems.
This context arises both commonly and naturally in information processing,
classical and quantum;
information-carrying systems are typically constructed from collections of
smaller systems, such as bits or qubits.

A simple, yet critically important idea to keep in mind going into this lesson
is that we can always choose to view multiple systems \emph{together} as if
they form a single, compound system --- to which the discussion in the previous
lesson applies.
Indeed, this idea very directly leads to a description of how quantum states,
measurements, and operations work for multiple systems.

There is, however, more to understanding multiple quantum systems than simply
recognizing that they may be viewed collectively as single systems.
For instance, we may have multiple quantum systems that are collectively in a
particular quantum state, and then choose to measure some but not all of the
individual systems.
In general, this will affect the state of the systems that were not measured,
and it is important to understand exactly how when analyzing quantum algorithms
and protocols.
An understanding of the sorts of \emph{correlations} among multiple systems ---
and particularly a type of correlation known as \emph{entanglement} --- is also
important in quantum information and computation. 

\section{Classical information}

Like we did in the previous lesson, we'll begin this lesson with a discussion
of classical information.
Once again, the probabilistic and quantum descriptions are mathematically
similar, and recognizing how the mathematics works in the familiar setting of
classical information is helpful in understanding why quantum information is
described in the way that it is.

\subsection{Classical states via the Cartesian product}

We'll start at a very basic level, with classical states of multiple systems.
For simplicity, we'll begin by discussing just two systems, and then generalize
to more than two systems.

To be precise, let $\mathsf{X}$ be a system whose classical state set is
$\Sigma$, and let $\mathsf{Y}$ be a second system whose classical state set is
$\Gamma$.
Note that, because we have referred to these sets as
\emph{classical state sets}, our assumption is that $\Sigma$ and $\Gamma$ are
both finite and nonempty.
It could be that $\Sigma = \Gamma$, but this is not necessarily so --- and
regardless, it will be helpful to use different names to refer to these sets in
the interest of clarity.

Now imagine that the two systems, $\mathsf{X}$ and $\mathsf{Y}$, are placed
side-by-side, with $\mathsf{X}$ on the left and $\mathsf{Y}$ on the right.
If we so choose, we can view these two systems as if they form a single system,
which we can denote by $(\mathsf{X},\mathsf{Y})$ or $\mathsf{XY}$ depending on
our preference.
A natural question to ask about this compound system $(\mathsf{X},\mathsf{Y})$
is, ``What are its classical states?''

The answer is that the set of classical states of $(\mathsf{X},\mathsf{Y})$ is
the \emph{Cartesian product} of $\Sigma$ and $\Gamma$, which is the set defined
as
\[
\Sigma\times\Gamma
= \bigl\{(a,b)\,:\,a\in\Sigma\;\text{and}\;b\in\Gamma\bigr\}.
\]
In simple terms, the Cartesian product is precisely the mathematical notion
that captures the idea of viewing an element of one set and an element of a
second set together, as if they form a single element of a single set.

In the case at hand, to say that $(\mathsf{X},\mathsf{Y})$ is in the classical
state $(a,b)\in\Sigma\times\Gamma$ means that $\mathsf{X}$ is in the classical
state $a\in\Sigma$ and $\mathsf{Y}$ is in the classical state $b\in\Gamma$;
and if the classical state of $\mathsf{X}$ is $a\in\Sigma$ and the classical
state of $\mathsf{Y}$ is $b\in\Gamma$, then the classical state of the joint
system $(\mathsf{X},\mathsf{Y})$ is $(a,b)$.

For more than two systems, the situation generalizes in a natural way.
If we suppose that $\mathsf{X}_1,\ldots,\mathsf{X}_n$ are systems having
classical state sets $\Sigma_1,\ldots,\Sigma_n$, respectively, for any positive
integer $n$, the classical state set of the $n$-tuple
$(\mathsf{X}_1,\ldots,\mathsf{X}_n)$, viewed as a single joint system, is the
Cartesian product
\[
\Sigma_1\times\cdots\times\Sigma_n
= \bigl\{(a_1,\ldots,a_n)\,:\,
a_1\in\Sigma_1,\:\ldots,\:a_n\in\Sigma_n\bigr\}.
\]

Of course, we are free to use whatever names we wish for systems, and to order
them as we choose.
In particular, if we have $n$ systems like above, we could instead choose to
name them $\mathsf{X}_{0},\ldots,\mathsf{X}_{n-1}$ and arrange them from right
to left, so that the joint system becomes
$(\mathsf{X}_{n-1},\ldots,\mathsf{X}_0)$.
Following the same pattern for naming the associated classical states and
classical state sets, we might then refer to a classical state
\[
(a_{n-1},\ldots,a_0) \in \Sigma_{n-1}\times \cdots \times \Sigma_0
\]
of this compound system.

Indeed, this is the ordering convention used by Qiskit when naming multiple
qubits.
We'll come back to this convention and how it connects to quantum circuits in
the next lesson, but we'll start using it now to help to get used to it.

It is often convenient to write a classical state of the form
$(a_{n-1},\ldots,a_0)$ as a string $a_{n-1}\cdots a_0$ for the sake of brevity,
particularly in the very typical situation that the classical state sets
$\Sigma_0,\ldots,\Sigma_{n-1}$ are associated with sets of \emph{symbols} or
\emph{characters}.
In this context, the term \emph{alphabet} is commonly used to refer to sets of
symbols used to form strings, but the mathematical definition of an alphabet is
precisely the same as the definition of a classical state set: it is a finite
and nonempty set.

For example, suppose that $\mathsf{X}_0,\ldots,\mathsf{X}_9$ are bits, so that
the classical state sets of these systems are all the same.
\[
\Sigma_0 = \Sigma_1 = \cdots = \Sigma_9 = \{0,1\}
\]
There are then $2^{10} = 1024$ classical states of the joint system
$(\mathsf{X}_9,\ldots,\mathsf{X}_0)$, which are the elements of the set
\[
\Sigma_9\times\Sigma_8\times\cdots\times\Sigma_0 = \{0,1\}^{10}.
\]
Written as strings, these classical states look like this:
\[
\begin{array}{c}
  0000000000\\
  0000000001\\
  0000000010\\
  0000000011\\
  0000000100\\
  \vdots\\[1mm]
  1111111111
\end{array}
\]
For the classical state $0000000110$, for instance, we see that $\mathsf{X}_1$
and $\mathsf{X}_2$ are in the state~$1$, while all other systems are in the
state $0$.

\subsection{Probabilistic states}

Recall from the previous lesson that a \emph{probabilistic state} associates a
probability with each classical state of a system.
Thus, a probabilistic state of multiple systems --- viewed collectively as a
single system --- associates a probability with each element of the Cartesian
product of the classical state sets of the individual systems.

For example, suppose that $\mathsf{X}$ and $\mathsf{Y}$ are both bits, so that
their corresponding classical state sets are $\Sigma = \{0,1\}$ and
$\Gamma = \{0,1\}$, respectively. 
Here is a probabilistic state of the pair $(\mathsf{X},\mathsf{Y}):$
\[
\begin{aligned}
  \operatorname{Pr}\bigl( (\mathsf{X},\mathsf{Y}) = (0,0)\bigr)
  & = 1/2 \\[2mm]
  \operatorname{Pr}\bigl( (\mathsf{X},\mathsf{Y}) = (0,1)\bigr)
  & = 0\\[2mm]
  \operatorname{Pr}\bigl( (\mathsf{X},\mathsf{Y}) = (1,0)\bigr)
  & = 0\\[2mm]
  \operatorname{Pr}\bigl( (\mathsf{X},\mathsf{Y}) = (1,1)\bigr)
  & = 1/2
\end{aligned}
\]
This probabilistic state is one in which both $\mathsf{X}$ and $\mathsf{Y}$ are
random bits --- each is $0$ with probability $1/2$ and $1$ with probability
$1/2$ --- but the classical states of the two bits always agree.
This is an example of a \emph{correlation} between these systems.

\subsubsection{Ordering Cartesian product state sets}

Probabilistic states of systems can be represented by probability vectors, as
was discussed in the previous lesson.
In particular, the vector entries represent probabilities for the system to be
in the possible classical states of that system, and the understanding is that
a correspondence between the entries and the set of classical states has been
selected.

Choosing such a correspondence effectively means deciding on an ordering of the
classical states, which is often natural or determined by a standard
convention.
For example, the binary alphabet $\{0,1\}$ is naturally ordered with $0$ first
and $1$ second, so the first entry in a probability vector representing a
probabilistic state of a bit is the probability for it to be in the state $0$,
and the second entry is the probability for it to be in the state $1$.

None of this changes in the context of multiple systems, but there is a
decision to be made.
The classical state set of multiple systems together, viewed collectively as a
single system, is the Cartesian product of the classical state sets of the
individual systems --- so we must decide how the elements of Cartesian products
of classical state sets are to be ordered.

There is a simple convention that we follow for doing this, which is to start
with whatever orderings are already in place for the individual classical state
sets, and then to order the elements of the Cartesian product
\emph{alphabetically}.
Another way to say this is that the entries in each $n$-tuple (or,
equivalently, the symbols in each string) are treated as though they have
significance that \emph{decreases from left to right}.
For example, according to this convention, the Cartesian product
$\{1,2,3\}\times\{0,1\}$ is ordered like this:
\[
(1,0),\;
(1,1),\;
(2,0),\;
(2,1),\;
(3,0),\;
(3,1).
\]

When $n$-tuples are written as strings and ordered in this way, we observe
familiar patterns, such as $\{0,1\}\times\{0,1\}$ being ordered as $00, 01, 10,
11$, and the set $\{0,1\}^{10}$ being ordered as it was written earlier in the
lesson.
As another example, viewing the set $\{0, 1, \dots, 9\} \times \{0, 1, \dots,
9\}$ as a set of strings, we obtain the two-digit numbers $00$ through $99$,
ordered numerically.
This is obviously not a coincidence;
our decimal number system uses precisely this sort of alphabetical ordering,
where the word \emph{alphabetical} should be understood as having a broad
meaning that includes numerals in addition to letters.

Returning to the example of two bits from above, the probabilistic state
described previously is therefore represented by the following probability
vector, where the entries are labeled explicitly for the sake of clarity.
\begin{equation}
  \begin{pmatrix}
    \frac{1}{2}\\[1mm]
    0\\[1mm]
    0\\[1mm]
    \frac{1}{2}
  \end{pmatrix}
  \begin{array}{l}
    \leftarrow \text{probability of being in the state 00}\\[1mm]
    \leftarrow \text{probability of being in the state 01}\\[1mm]
    \leftarrow \text{probability of being in the state 10}\\[1mm]
    \leftarrow \text{probability of being in the state 11}
  \end{array}
  \label{eq:probability_vector}
\end{equation}

\subsubsection{Independence of two systems}

A special type of probabilistic state of two systems is one in which the
systems are \emph{independent}.
Intuitively speaking, two systems are independent if learning the classical
state of either system has no effect on the probabilities associated with the
other.
That is, learning what classical state one of the systems is in provides no
information at all about the classical state of the other.

To define this notion precisely, let us suppose once again that $\mathsf{X}$
and $\mathsf{Y}$ are systems having classical state sets $\Sigma$ and $\Gamma$,
respectively.
With respect to a given probabilistic state of these systems, they are said to
be \emph{independent} if it is the case that
\begin{equation}
  \operatorname{Pr}((\mathsf{X},\mathsf{Y}) = (a,b))
  = \operatorname{Pr}(\mathsf{X} = a) \operatorname{Pr}(\mathsf{Y} = b)
  \label{eq:independence_condition}
\end{equation}
for every choice of $a\in\Sigma$ and $b\in\Gamma$.

To express this condition in terms of probability vectors, assume that the
given probabilistic state of $(\mathsf{X},\mathsf{Y})$ is described by a
probability vector, written in the Dirac notation as
\[
\sum_{(a,b) \in \Sigma\times\Gamma} p_{ab} \vert a b\rangle.
\]
The condition \eqref{eq:independence_condition} for independence is then
equivalent to the existence of two probability vectors
\begin{equation}
  \vert \phi \rangle = \sum_{a\in\Sigma} q_a \vert a \rangle
  \quad\text{and}\quad
  \vert \psi \rangle = \sum_{b\in\Gamma} r_b \vert b \rangle,
  \label{eq:two_probability_vectors}
\end{equation}
representing the probabilities associated with the classical states of
$\mathsf{X}$ and $\mathsf{Y}$, respectively, such that
\begin{equation}
  p_{ab} = q_a r_b
  \label{eq:product_of_probabilities}
\end{equation}
for all $a\in\Sigma$ and $b\in\Gamma$.

For example, the probabilistic state of a pair of bits
$(\mathsf{X},\mathsf{Y})$ represented by the vector
\[
\frac{1}{6} \vert 00 \rangle
+ \frac{1}{12} \vert 01 \rangle
+ \frac{1}{2} \vert 10 \rangle
+ \frac{1}{4} \vert 11 \rangle
\]
is one in which $\mathsf{X}$ and $\mathsf{Y}$ are independent.
Specifically, the condition required for independence is true for the
probability vectors
\[
\vert \phi \rangle = \frac{1}{4} \vert 0 \rangle + \frac{3}{4} \vert 1 \rangle
\quad\text{and}\quad
\vert \psi \rangle = \frac{2}{3} \vert 0 \rangle + \frac{1}{3} \vert 1 \rangle.
\]
For instance, to make the probabilities for the $00$ state match, we need
$\frac{1}{6} = \frac{1}{4} \times \frac{2}{3}$, and indeed this is the
case. Other entries can be verified in a similar manner.

On the other hand, the probabilistic state \eqref{eq:probability_vector}, which
we may write as
\begin{equation}
  \frac{1}{2} \vert 00 \rangle + \frac{1}{2} \vert 11 \rangle,
  \label{eq:shared_random_bit}
\end{equation}
does not represent independence between the systems $\mathsf{X}$ and
$\mathsf{Y}$.
A simple way to argue this follows.

Suppose that there did exist probability vectors $\vert \phi\rangle$ and $\vert
\psi \rangle$, as in equation \eqref{eq:two_probability_vectors} above, for
which the condition \eqref{eq:product_of_probabilities} is satisfied for every
choice of $a$ and $b$.
It would then necessarily be that
\[
q_0 r_1 = \operatorname{Pr}\bigl((\mathsf{X},\mathsf{Y}) = (0,1)\bigr) = 0.
\]
This implies that either $q_0 = 0$ or $r_1 = 0$, because if both were nonzero,
the product $q_0 r_1$ would also be nonzero.
This leads to the conclusion that either $q_0 r_0 = 0$ (in case $q_0 = 0$) or
$q_1 r_1 = 0$ (in case $r_1 = 0$).
We see, however, that neither of those equalities can be true because we must
have $q_0 r_0 = 1/2$ and $q_1 r_1 = 1/2$.
Hence, there do not exist vectors $\vert\phi\rangle$ and $\vert\psi\rangle$
satisfying the property required for independence.

Having defined independence between two systems, we can now define what is
meant by \emph{correlation}: it is a \emph{lack of independence}.
For example, because the two bits in the probabilistic state represented by the
vector \eqref{eq:shared_random_bit} are not independent, they are, by
definition, correlated.

\subsubsection{Tensor products of vectors}

The condition of independence just described can be expressed succinctly
through the notion of a \emph{tensor product}.
Although tensor products are a very general notion, and can be defined quite
abstractly and applied to a variety of mathematical structures, we can adopt a
simple and concrete definition in the case at hand.

Given two vectors
\[
\vert \phi \rangle = \sum_{a\in\Sigma} \alpha_a \vert a \rangle
\quad\text{and}\quad
\vert \psi \rangle = \sum_{b\in\Gamma} \beta_b \vert b \rangle,
\]
the tensor product $\vert \phi \rangle \otimes \vert \psi \rangle$ is the
vector defined as
\[
\vert \phi \rangle \otimes \vert \psi \rangle
= \sum_{(a,b)\in\Sigma\times\Gamma} \alpha_a \beta_b \vert ab\rangle.
\]
The entries of this new vector correspond to the elements of the Cartesian
product $\Sigma\times\Gamma$, which are written as strings in the previous
equation.
Equivalently, the vector
$\vert \pi \rangle = \vert \phi \rangle \otimes \vert \psi \rangle$ is defined
by the equation
\[
\langle ab \vert \pi \rangle = \langle a \vert \phi \rangle \langle b \vert
\psi \rangle
\]
being true for every $a\in\Sigma$ and $b\in\Gamma$.

We can now recast the condition for independence:
for a joint system $(\mathsf{X}, \mathsf{Y})$ in a probabilistic state
represented by a probability vector $\vert \pi \rangle$, the systems
$\mathsf{X}$ and $\mathsf{Y}$ are independent if $\vert\pi\rangle$ is obtained
by taking a tensor product
\[
\vert \pi \rangle = \vert \phi \rangle \otimes \vert \psi \rangle
\]
of probability vectors $\vert \phi \rangle$ and $\vert \psi \rangle$ on each of
the subsystems $\mathsf{X}$ and $\mathsf{Y}$.
In this situation, $\vert \pi \rangle$ is said to be a \emph{product state} or
\emph{product vector}.

We often omit the symbol $\otimes$ when taking the tensor product of kets, such
as writing $\vert \phi \rangle \vert \psi \rangle$ rather than
$\vert \phi \rangle \otimes \vert \psi \rangle$. 
This convention captures the idea that the tensor product is, in this context,
the most natural or default way to take the product of two vectors.
Although it is less common, the notation $\vert \phi\otimes\psi\rangle$
is also sometimes used.

When we use the alphabetical convention for ordering elements of Cartesian
products, we obtain the following specification for the tensor product of two
column vectors.
\[
\begin{pmatrix}
  \alpha_1\\
  \vdots\\
  \alpha_m
\end{pmatrix}
\otimes
\begin{pmatrix}
  \beta_1\\
  \vdots\\
  \beta_k
\end{pmatrix}
= \begin{pmatrix}
  \alpha_1 \beta_1\\
  \vdots\\
  \alpha_1 \beta_k\\
  \alpha_2 \beta_1\\
  \vdots\\
  \alpha_2 \beta_k\\
  \vdots\\
  \alpha_m \beta_1\\
  \vdots\\
  \alpha_m \beta_k
\end{pmatrix}
\]

As an important aside, notice the following expression for tensor products of
standard basis vectors:
\[
\vert a \rangle \otimes \vert b \rangle = \vert ab \rangle.
\]
We could alternatively write $(a,b)$ as an ordered pair, rather than a string,
in which case we obtain
$\vert a \rangle \otimes \vert b \rangle = \vert (a,b) \rangle$.
It is, however, more common to omit the parentheses in this situation, instead
writing $\vert a \rangle \otimes \vert b \rangle = \vert a,b \rangle$.
This is typical in mathematics more generally; parentheses that don't add
clarity or remove ambiguity are often simply omitted.

The tensor product of two vectors has the important property that it is
\emph{bilinear}, which means that it is linear in each of the two arguments
separately, assuming that the other argument is fixed.
This property can be expressed through these equations:

\begin{enumerate}
\item
  Linearity in the first argument:
  \[
  \begin{aligned}
    \bigl(\vert\phi_1\rangle + \vert\phi_2\rangle\bigr)\otimes \vert\psi\rangle
    & = \vert\phi_1\rangle \otimes \vert\psi\rangle + \vert\phi_2\rangle
    \otimes \vert\psi\rangle \\[1mm]
    \bigl(\alpha \vert \phi \rangle\bigr) \otimes \vert \psi \rangle
    & = \alpha \bigl(\vert \phi \rangle \otimes \vert \psi \rangle \bigr)
  \end{aligned}
  \]

\item
  Linearity in the second argument:
  \[
  \begin{aligned}
    \vert \phi \rangle \otimes
    \bigl(\vert \psi_1 \rangle + \vert \psi_2 \rangle \bigr)
    & =
    \vert \phi \rangle \otimes \vert \psi_1 \rangle +
    \vert \phi \rangle \otimes \vert \psi_2 \rangle\\[1mm]
    \vert \phi \rangle \otimes
    \bigl(\alpha \vert \psi \rangle \bigr)
    & = \alpha \bigl(\vert\phi\rangle\otimes\vert\psi\rangle\bigr)
  \end{aligned}
  \]
\end{enumerate}

\noindent
Considering the second equation in each of these pairs of equations,
we see that scalars ``float freely'' within tensor products:
\[
\bigl(\alpha \vert \phi \rangle\bigr) \otimes \vert \psi \rangle
= \vert \phi \rangle \otimes \bigl(\alpha \vert \psi \rangle \bigr)
= \alpha \bigl(\vert \phi \rangle \otimes \vert \psi \rangle \bigr).
\]
There is therefore no ambiguity in simply writing
$\alpha\vert\phi\rangle\otimes\vert\psi\rangle$, or alternatively
$\alpha\vert\phi\rangle\vert\psi \rangle$ or
$\alpha\vert\phi\otimes\psi\rangle$, to refer to this vector.

\subsubsection{Independence and tensor products for three or more systems}

The notions of independence and tensor products generalize straightforwardly to
three or more systems.
If $\mathsf{X}_0,\ldots,\mathsf{X}_{n-1}$ are systems having classical state
sets $\Sigma_0,\ldots,\Sigma_{n-1}$, respectively, then a probabilistic state
of the combined system
\[
(\mathsf{X}_{n-1},\ldots,\mathsf{X}_0)
\]
is a \emph{product state} if the associated probability vector takes the form
\[
\vert \psi \rangle
= \vert \phi_{n-1} \rangle \otimes \cdots \otimes \vert \phi_0 \rangle
\]
for probability vectors $\vert \phi_0 \rangle,\ldots,\vert \phi_{n-1}\rangle$
describing probabilistic states of $\mathsf{X}_0,\ldots,\mathsf{X}_{n-1}$.
Here, the definition of the tensor product generalizes in a natural way: the
vector
\[
\vert \psi \rangle = \vert \phi_{n-1} \rangle \otimes \cdots \otimes \vert
\phi_0 \rangle
\]
is defined by the equation
\[
\langle a_{n-1} \cdots a_0 \vert \psi \rangle
= \langle a_{n-1} \vert \phi_{n-1} \rangle \cdots
\langle a_0 \vert \phi_0 \rangle
\]
being true for every $a_0\in\Sigma_0, \ldots a_{n-1}\in\Sigma_{n-1}$.

A different, but equivalent, way to define the tensor product of three or more
vectors is recursively in terms of tensor products of two vectors:
\[
\vert \phi_{n-1} \rangle \otimes \cdots \otimes \vert \phi_0 \rangle
= \vert \phi_{n-1} \rangle \otimes \bigl( \vert \phi_{n-2} \rangle
\otimes \cdots \otimes \vert \phi_0 \rangle \bigr).
\]

Similar to the tensor product of just two vectors, the tensor product of three
or more vectors is linear in each of the arguments individually, assuming that
all other arguments are fixed.
In this case it is said that the tensor product of three or more vectors is
\emph{multilinear}.

Like in the case of two systems, we could say that the systems
$\mathsf{X}_0,\ldots,\mathsf{X}_{n-1}$ are \emph{independent} when they are in
a product state, but the term \emph{mutually independent} is more precise.
There happen to be other notions of independence for three or more systems,
such as \emph{pairwise independence}, that are both interesting and important
--- but not in the context of this course.

Generalizing the observation earlier concerning tensor products of standard
basis vectors, for any positive integer $n$ and any classical states
$a_0,\ldots,a_{n-1}$, we have
\[
\vert a_{n-1} \rangle \otimes \cdots \otimes \vert a_0 \rangle
= \vert a_{n-1} \cdots a_0 \rangle.
\]

\subsection{Measurements of probabilistic states}

Now let us move on to measurements of probabilistic states of multiple systems.
By choosing to view multiple systems together as single systems, we immediately
obtain a specification of how measurements must work for multiple systems ---
provided that \emph{all} of the systems are measured.

For example, if the probabilistic state of two bits
$(\mathsf{X},\mathsf{Y})$ is described by the probability vector
\[
\frac{1}{2} \vert 00 \rangle + \frac{1}{2} \vert 11 \rangle,
\]
then the outcome $00$ --- meaning $0$ for the measurement of $\mathsf{X}$ and
$0$ for the measurement of $\mathsf{Y}$ --- is obtained with probability $1/2$
and the outcome $11$ is also obtained with probability $1/2$.
In each case we update the probability vector description of our knowledge
accordingly, so that the probabilistic state becomes $\vert 00\rangle$ or
$\vert 11\rangle$, respectively.

We could, however, choose to measure not \emph{every} system, but instead just
some of the systems.
This will result in a measurement outcome for each system that gets measured,
and will also (in general) affect our knowledge of the remaining systems that
we didn't measure.

To explain how this works, we'll focus on the case of two systems, one of which
is measured.
The more general situation --- in which some proper subset of three or more
systems is measured --- effectively reduces to the case of two systems when we
view the systems that are measured collectively as if they form one system and
the systems that are not measured as if they form a second system.

To be precise, let's suppose that $\mathsf{X}$ and $\mathsf{Y}$ are systems
whose classical state sets are $\Sigma$ and $\Gamma$, respectively, and that
the two systems together are in some probabilistic state.
We'll consider what happens when we measure just $\mathsf{X}$ and do nothing to
$\mathsf{Y}$.
The situation where just $\mathsf{Y}$ is measured and nothing happens to
$\mathsf{X}$ is handled symmetrically.

First, we know that the probability to observe a particular classical state
$a\in\Sigma$ when just $\mathsf{X}$ is measured must be consistent with the
probabilities we would obtain under the assumption that $\mathsf{Y}$ was also
measured.
That is, we must have
\[
\operatorname{Pr}(\mathsf{X} = a)
= \sum_{b\in\Gamma} \operatorname{Pr}\bigl( (\mathsf{X},\mathsf{Y})
= (a,b) \bigr).
\]
This is the formula for the so-called reduced (or marginal) probabilistic state
of $\mathsf{X}$ alone.

This formula makes perfect sense at an intuitive level, in the sense that
something very strange would have to happen for it to be wrong.
If it were wrong, that would mean that measuring $\mathsf{Y}$ could somehow
influence the probabilities associated with different outcomes of the
measurement of $\mathsf{X}$, irrespective of the actual outcome of the
measurement of $\mathsf{Y}$.
If $\mathsf{Y}$ happened to be in a distant location, such as somewhere in
another galaxy for instance, this would allow for faster-than-light signaling
--- which we reject based on our understanding of physics.

Another way to understand this comes from the interpretation of probability as
reflecting a degree of belief.
The mere fact that someone else might decide to look at $\mathsf{Y}$ cannot
change the classical state of $\mathsf{X}$, so without any information about
what they did or didn't see, one's beliefs about the state of $\mathsf{X}$
should not change as a result.

Now, given the assumption that only $\mathsf{X}$ is measured and $\mathsf{Y}$
is not, there may still exist uncertainty about the classical state of
$\mathsf{Y}$.
For this reason, rather than updating our description of the probabilistic
state of $(\mathsf{X},\mathsf{Y})$ to $\vert ab\rangle$ for some selection of
$a\in\Sigma$ and $b\in\Gamma$, we must update our description so that this
uncertainty about $\mathsf{Y}$ is properly reflected.

The following \emph{conditional probability} formula reflects this uncertainty.
\[
\operatorname{Pr}(\mathsf{Y} = b \,\vert\, \mathsf{X} = a)
= \frac{
  \operatorname{Pr}\bigl((\mathsf{X},\mathsf{Y}) = (a,b)\bigr)
}{
  \operatorname{Pr}(\mathsf{X} = a)
}
\]
Here, the expression
$\operatorname{Pr}(\mathsf{Y} = b \,\vert\, \mathsf{X} = a)$ denotes the
probability that $\mathsf{Y} = b$ \emph{conditioned} on (or \emph{given} that)
$\mathsf{X} = a$.
Technically speaking, this expression only makes sense if
$\operatorname{Pr}(\mathsf{X}=a)$ is nonzero, for if
$\operatorname{Pr}(\mathsf{X}=a) = 0$, then we're dividing by zero and we
obtain indeterminate form $\frac{0}{0}$.
This is not a problem, though, because if the probability associated with $a$
is zero, then we'll never obtain $a$ as an outcome of a measurement of
$\mathsf{X}$, so we don't need to be concerned with this possibility.

To express these formulas in terms of probability vectors, consider a
probability vector $\vert \psi \rangle$ describing a joint probabilistic state
of $(\mathsf{X},\mathsf{Y})$.
\[
\vert\psi\rangle = \sum_{(a,b)\in\Sigma\times\Gamma} p_{ab} \vert ab\rangle
\]
Measuring $\mathsf{X}$ alone yields each possible outcome $a\in\Sigma$ with
probability
\[
\operatorname{Pr}(\mathsf{X} = a) = \sum_{c\in\Gamma} p_{ac}.
\]
The vector representing the probabilistic state of $\mathsf{X}$ alone is
therefore given by
\[
\sum_{a\in\Sigma} \biggl(\sum_{c\in\Gamma} p_{ac}\biggr) \vert a\rangle.
\]
Having obtained a particular outcome $a\in\Sigma$ of the measurement of
$\mathsf{X}$, the probabilistic state of $\mathsf{Y}$ is updated according to
the formula for conditional probabilities, so that it is represented by this
probability vector:
\[
\vert \pi_a \rangle
= \frac{\sum_{b\in\Gamma}p_{ab}\vert b\rangle}{\sum_{c\in\Gamma} p_{ac}}.
\]
In the event that the measurement of $\mathsf{X}$ resulted in the classical
state $a$, we therefore update our description of the probabilistic state of
the joint system to $\vert a\rangle \otimes \vert\pi_a\rangle$.

One way to think about this definition of $\vert\pi_a\rangle$ is to see it as a
\emph{normalization} of the vector $\sum_{b\in\Gamma} p_{ab} \vert b\rangle$,
where we divide by the sum of the entries in this vector to obtain a
probability vector.
This normalization effectively accounts for a conditioning on the event that
the measurement of $\mathsf{X}$ has resulted in the outcome $a$.

For a specific example, suppose that classical state set of $\mathsf{X}$ is
$\Sigma = \{0,1\}$, the classical state set of $\mathsf{Y}$ is
$\Gamma = \{1,2,3\}$, and the probabilistic state of $(\mathsf{X},\mathsf{Y})$
is
\[
\vert \psi \rangle
= \frac{1}{2}  \vert 0,1 \rangle
+ \frac{1}{12} \vert 0,3 \rangle
+ \frac{1}{12} \vert 1,1 \rangle
+ \frac{1}{6}  \vert 1,2 \rangle
+ \frac{1}{6}  \vert 1,3 \rangle.
\]
Our goal will be to determine the probabilities of the two possible outcomes
($0$ and $1$), and to calculate what the resulting probabilistic state of
$\mathsf{Y}$ is for the two outcomes, assuming the system $\mathsf{X}$ is
measured.

Using the bilinearity of the tensor product, and specifically the fact that it
is linear in the \emph{second} argument, we may rewrite the vector $\vert \psi
\rangle$ as follows:
\[
\vert \psi \rangle
= \vert 0\rangle \otimes
\biggl( \frac{1}{2} \vert 1 \rangle + \frac{1}{12} \vert 3 \rangle\biggr)
+ \vert 1\rangle \otimes
\biggl( \frac{1}{12} \vert 1 \rangle + \frac{1}{6} \vert 2\rangle
+ \frac{1}{6} \vert 3 \rangle\biggr).
\]
In words, what we've done is to isolate the distinct standard basis vectors for
the first system (i.e., the one being measured), tensoring each with the linear
combination of standard basis vectors for the second system we get by picking
out the entries of the original vector that are consistent with the
corresponding classical state of the first system.
A moment's thought reveals that this is always possible, regardless of what
vector we started with.

Having expressed our probability vector in this way, the effects of measuring
the first system become easy to analyze.
The probabilities of the two outcomes can be obtained by summing the
probabilities in parentheses.
\[
\begin{aligned}
  \operatorname{Pr}(\mathsf{X} = 0)
  & = \frac{1}{2} + \frac{1}{12} = \frac{7}{12}\\[3mm]
  \operatorname{Pr}(\mathsf{X} = 1)
  & = \frac{1}{12} + \frac{1}{6} + \frac{1}{6} = \frac{5}{12}
\end{aligned}
\]
These probabilities sum to one, as expected --- but this is a useful check on
our calculations.

And now, the probabilistic state of $\mathsf{Y}$ conditioned on each possible
outcome can be inferred by normalizing the vectors in parentheses.
That is, we divide these vectors by the associated probabilities we just
calculated, so that they become probability vectors.
Thus, conditioned on $\mathsf{X}$ being $0$, the probabilistic state of
$\mathsf{Y}$ becomes
\[
\frac{\frac{1}{2} \vert 1 \rangle + \frac{1}{12} \vert 3 \rangle}{\frac{7}{12}}
= \frac{6}{7} \vert 1 \rangle + \frac{1}{7} \vert 3 \rangle,
\]
and conditioned on the measurement of $\mathsf{X}$ being $1$, the probabilistic
state of $\mathsf{Y}$ becomes
\[
\frac{\frac{1}{12} \vert 1 \rangle + \frac{1}{6} \vert 2\rangle
  + \frac{1}{6} \vert 3 \rangle}{\frac{5}{12}}
= \frac{1}{5} \vert 1 \rangle + \frac{2}{5} \vert 2 \rangle
+ \frac{2}{5} \vert 3 \rangle.
\]

\subsection{Operations on probabilistic states}

To conclude this discussion of classical information for multiple systems,
we'll consider \emph{operations} on multiple systems in probabilistic states.
Following the same idea as before, we can view multiple systems collectively as
single, compound systems, and then look to the previous lesson to see how this
works.

Returning to the typical set-up where we have two systems $\mathsf{X}$ and
$\mathsf{Y}$, let us consider classical operations on the compound system
$(\mathsf{X},\mathsf{Y})$.
Based on the previous lesson and the discussion above, we conclude that any
such operation is represented by a stochastic matrix whose rows and columns are
indexed by the Cartesian product $\Sigma\times\Gamma$.

For example, suppose that $\mathsf{X}$ and $\mathsf{Y}$ are bits, and consider
an operation with the following description.

\begin{callout}[leftrule=3pt]
  If $\mathsf{X} = 1$, then perform a NOT operation on $\mathsf{Y}$. \newline
  Otherwise do nothing.
\end{callout}

\noindent
This is a deterministic operation known as a \emph{controlled-NOT} operation,
where $\mathsf{X}$ is the \emph{control} bit that determines whether or not a
NOT operation should be applied to the \emph{target} bit $\mathsf{Y}$.
Here is the matrix representation of this operation:
\[
\begin{pmatrix}
  1 & 0 & 0 & 0\\
  0 & 1 & 0 & 0\\
  0 & 0 & 0 & 1\\
  0 & 0 & 1 & 0
\end{pmatrix}.
\]
Its action on standard basis states is as follows.
\[
\begin{aligned}
  \vert 00 \rangle & \mapsto \vert 00 \rangle\\
  \vert 01 \rangle & \mapsto \vert 01 \rangle\\
  \vert 10 \rangle & \mapsto \vert 11 \rangle\\
  \vert 11 \rangle & \mapsto \vert 10 \rangle
\end{aligned}
\]
If we were to exchange the roles of $\mathsf{X}$ and $\mathsf{Y}$, taking
$\mathsf{Y}$ to be the control bit and $\mathsf{X}$ to be the target bit, then
the matrix representation of the operation would become
\[
\begin{pmatrix}
  1 & 0 & 0 & 0\\
  0 & 0 & 0 & 1\\
  0 & 0 & 1 & 0\\
  0 & 1 & 0 & 0
\end{pmatrix}
\]
and its action on standard basis states would be as follows.
\[
\begin{aligned}
  \vert 00 \rangle & \mapsto \vert 00 \rangle\\
  \vert 01 \rangle & \mapsto \vert 11 \rangle\\
  \vert 10 \rangle & \mapsto \vert 10 \rangle\\
  \vert 11 \rangle & \mapsto \vert 01 \rangle
\end{aligned}
\]

Another example is the operation having this description:

\begin{callout}[leftrule=3pt]
  Perform one of the following two operations, each with probability $1/2:$
  \begin{enumerate}
  \item Set $\mathsf{Y}$ to be equal to $\mathsf{X}$.
  \item Set $\mathsf{X}$ to be equal to $\mathsf{Y}$.
  \end{enumerate}
\end{callout}

\noindent
The matrix representation of this operation is as follows:
\[
\begin{pmatrix}
  1 & \frac{1}{2} & \frac{1}{2} & 0\\[1mm]
  0 & 0 & 0 & 0\\[1mm]
  0 & 0 & 0 & 0\\[1mm]
  0 & \frac{1}{2} & \frac{1}{2} & 1
\end{pmatrix}
=
\frac{1}{2}
\begin{pmatrix}
  1 & 1 & 0 & 0\\
  0 & 0 & 0 & 0\\
  0 & 0 & 0 & 0\\
  0 & 0 & 1 & 1
\end{pmatrix}
+
\frac{1}{2}
\begin{pmatrix}
  1 & 0 & 1 & 0\\
  0 & 0 & 0 & 0\\
  0 & 0 & 0 & 0\\
  0 & 1 & 0 & 1
\end{pmatrix}.
\]
The action of this operation on standard basis vectors is as follows:
\[
\begin{aligned}
  \vert 00 \rangle & \mapsto \vert 00 \rangle\\[1mm]
  \vert 01 \rangle & \mapsto
  \frac{1}{2} \vert 00 \rangle + \frac{1}{2}\vert 11\rangle\\[3mm]
  \vert 10 \rangle & \mapsto \frac{1}{2} \vert 00 \rangle + \frac{1}{2}\vert
  11\rangle\\[2mm]
  \vert 11 \rangle & \mapsto \vert 11 \rangle
\end{aligned}
\]

In these examples, we are simply viewing two systems together as a single
system and proceeding as in the previous lesson.
The same thing can be done for any number of systems.

For example, imagine that we have three bits, and we increment the three bits
modulo $8$ --- meaning that we think about the three bits as encoding a number
between $0$ and $7$ using binary notation, add $1$, and then take the remainder
after dividing by $8$.
One way to express this operation is like this:
\[
\begin{aligned}
  & \vert 001 \rangle \langle 000 \vert
  + \vert 010 \rangle \langle 001 \vert
  + \vert 011 \rangle \langle 010 \vert
  + \vert 100 \rangle \langle 011 \vert\\[1mm]
  & \quad + \vert 101 \rangle \langle 100 \vert
  + \vert 110 \rangle \langle 101 \vert
  + \vert 111 \rangle \langle 110 \vert
  + \vert 000 \rangle \langle 111 \vert.
\end{aligned}
\]
Another way to express it is as
\[
\sum_{k = 0}^{7} \vert (k+1) \bmod 8 \rangle \langle k \vert,
\]
assuming we've agreed that numbers from $0$ to $7$ inside of kets refer to the
three-bit binary encodings of those numbers.
A third option is to express this operation as a matrix.
\[
\begin{pmatrix}
  0 & 0 & 0 & 0 & 0 & 0 & 0 & 1\\
  1 & 0 & 0 & 0 & 0 & 0 & 0 & 0\\
  0 & 1 & 0 & 0 & 0 & 0 & 0 & 0\\
  0 & 0 & 1 & 0 & 0 & 0 & 0 & 0\\
  0 & 0 & 0 & 1 & 0 & 0 & 0 & 0\\
  0 & 0 & 0 & 0 & 1 & 0 & 0 & 0\\
  0 & 0 & 0 & 0 & 0 & 1 & 0 & 0\\
  0 & 0 & 0 & 0 & 0 & 0 & 1 & 0
\end{pmatrix}
\]

\subsubsection{Independent operations}

Now suppose that we have multiple systems and we \emph{independently} perform
different operations on the systems separately.

For example, taking our usual set-up of two systems $\mathsf{X}$ and
$\mathsf{Y}$ having classical state sets $\Sigma$ and $\Gamma$, respectively,
let us suppose that we perform one operation on $\mathsf{X}$ and, completely
independently, another operation on $\mathsf{Y}$.
As we know from the previous lesson, these operations are represented by
stochastic matrices --- and to be precise, let us say that the operation on
$\mathsf{X}$ is represented by the matrix $M$ and the operation on $\mathsf{Y}$
is represented by the matrix $N$.
Thus, the rows and columns of $M$ have indices that are placed in
correspondence with the elements of $\Sigma$ and, likewise, the rows and
columns of $N$ correspond to the elements of $\Gamma$.

A natural question to ask is this: if we view $\mathsf{X}$ and $\mathsf{Y}$
together as a single, compound system $(\mathsf{X},\mathsf{Y})$, what is the
matrix that represents the combined action of the two operations on this
compound system?
To answer this question we must first introduce tensor products of matrices,
which are similar to tensor products of vectors and are defined analogously.

\subsubsection{Tensor products of matrices}

The tensor product $M\otimes N$ of the matrices
\[
M = \sum_{a,b\in\Sigma} \alpha_{ab} \vert a\rangle \langle b\vert
\]
and
\[
N = \sum_{c,d\in\Gamma} \beta_{cd} \vert c\rangle \langle d\vert
\]
is the matrix
\[
M \otimes N = \sum_{a,b\in\Sigma} \sum_{c,d\in\Gamma}
\alpha_{ab} \beta_{cd} \vert ac \rangle \langle bd \vert
\]

Equivalently, the tensor product of $M$ and $N$ is defined by the equation
\[
\langle ac \vert M \otimes N \vert bd\rangle
= \langle a \vert M \vert b\rangle \langle c \vert N \vert d\rangle
\]
being true for every selection of $a,b\in\Sigma$ and $c,d\in\Gamma$.

Another alternative, but equivalent, way to describe $M\otimes N$ is that it is
the unique matrix that satisfies the equation
\[
(M \otimes N)
\bigl( \vert \phi \rangle \otimes \vert \psi \rangle \bigr)
= \bigl(M \vert\phi\rangle\bigr) \otimes
\bigl(N \vert\psi\rangle\bigr)
\]
for every possible choice of vectors $\vert\phi\rangle$ and $\vert\psi\rangle$,
assuming that the indices of $\vert\phi\rangle$ correspond to the elements of
$\Sigma$ and the indices of $\vert\psi\rangle$ correspond to $\Gamma$.

Following the convention described previously for ordering the elements of
Cartesian products, we can also write the tensor product of two matrices
explicitly as follows.
\[
\begin{gathered}
  \begin{pmatrix}
    \alpha_{11} & \cdots & \alpha_{1m} \\
    \vdots & \ddots & \vdots \\
    \alpha_{m1} & \cdots & \alpha_{mm}
  \end{pmatrix}
  \otimes
  \begin{pmatrix}
    \beta_{11} & \cdots & \beta_{1k} \\
    \vdots & \ddots & \vdots\\
    \beta_{k1} & \cdots & \beta_{kk}
  \end{pmatrix}
  \hspace{6cm}\\[3mm]
  \hspace{1cm} =
  \begin{pmatrix}
    \alpha_{11}\beta_{11} & \cdots & \alpha_{11}\beta_{1k} & &
    \alpha_{1m}\beta_{11} & \cdots & \alpha_{1m}\beta_{1k} \\
    \vdots & \ddots & \vdots & \hspace{2mm}\cdots\hspace{2mm} & \vdots & \ddots
    & \vdots \\
    \alpha_{11}\beta_{k1} & \cdots & \alpha_{11}\beta_{kk} & &
    \alpha_{1m}\beta_{k1} & \cdots & \alpha_{1m}\beta_{kk} \\[2mm]
    & \vdots & & \ddots & & \vdots & \\[2mm]
    \alpha_{m1}\beta_{11} & \cdots & \alpha_{m1}\beta_{1k} & &
    \alpha_{mm}\beta_{11} & \cdots & \alpha_{mm}\beta_{1k} \\
    \vdots & \ddots & \vdots & \hspace{2mm}\cdots\hspace{2mm} & \vdots & \ddots
    & \vdots \\
    \alpha_{m1}\beta_{k1} & \cdots & \alpha_{m1}\beta_{kk} & &
    \alpha_{mm}\beta_{k1} & \cdots & \alpha_{mm}\beta_{kk}
  \end{pmatrix}
\end{gathered}
\]

Tensor products of three or more matrices are defined in an analogous way.
That is, if $M_0, \ldots, M_{n-1}$ are matrices whose indices correspond to
classical state sets $\Sigma_0,\ldots,\Sigma_{n-1}$, then the tensor product
$M_{n-1}\otimes\cdots\otimes M_0$ is defined by the condition that
\[
\langle a_{n-1}\cdots a_0 \vert M_{n-1}\otimes\cdots\otimes M_0 \vert
b_{n-1}\cdots b_0\rangle
= \langle a_{n-1} \vert M_{n-1} \vert b_{n-1} \rangle \cdots\langle a_0 \vert
M_0 \vert b_0 \rangle
\]
for every choice of classical states
$a_0,b_0\in\Sigma_0,\ldots,a_{n-1},b_{n-1}\in\Sigma_{n-1}$.
Alternatively, tensor products of three or more matrices can be defined
recursively, in terms of tensor products of two matrices, similar to what we
observed for vectors.

The tensor product of matrices is sometimes said to be \emph{multiplicative}
because the equation
\[
(M_{n-1}\otimes\cdots\otimes M_0)(N_{n-1}\otimes\cdots\otimes N_0)
= (M_{n-1} N_{n-1})\otimes\cdots\otimes (M_0 N_0)
\]
is always true, for any choice of matrices $M_0,\ldots,M_{n-1}$ and
$N_0\ldots,N_{n-1}$, provided that the products $M_0 N_0, \ldots, M_{n-1}
N_{n-1}$ make sense.

\subsubsection{Independent operations (continued)}

We can now answer the question asked previously:
if $M$ is a probabilistic operation on $\mathsf{X}$, $N$ is a probabilistic
operation on $\mathsf{Y}$, and the two operations are performed independently,
then the resulting operation on the compound system $(\mathsf{X},\mathsf{Y})$
is the tensor product $M\otimes N$.

So, for both probabilistic states and probabilistic operations, \emph{tensor
products represent independence}.
If we have two systems $\mathsf{X}$ and $\mathsf{Y}$ that are independently in
the probabilistic states $\vert\phi\rangle$ and $\vert\pi\rangle$, then the
compound system $(\mathsf{X},\mathsf{Y})$ is in the probabilistic state
$\vert\phi\rangle\otimes\vert\pi\rangle$;
and if we apply probabilistic operations $M$ and $N$ to the two systems
independently, then the resulting action on the compound system
$(\mathsf{X},\mathsf{Y})$ is described by the operation $M\otimes N$.

Let's take a look at an example, which recalls a probabilistic operation on a
single bit from the previous lesson:
if the classical state of the bit is $0$, it is left alone; and if the
classical state of the bit is $1$, it is flipped to 0 with probability $1/2$.
We observed that this operation is represented by the matrix
\[
\begin{pmatrix}
  1 & \frac{1}{2}\\[1mm]
  0 & \frac{1}{2}
\end{pmatrix}.
\]
If this operation is performed on a bit $\mathsf{X}$, and a NOT operation is
(independently) performed on a second bit $\mathsf{Y}$, then the joint
operation on the compound system $(\mathsf{X},\mathsf{Y})$ has the matrix
representation
\[
\begin{pmatrix}
  1 & \frac{1}{2}\\[1mm]
  0 & \frac{1}{2}
\end{pmatrix}
\otimes
\begin{pmatrix}
  0 & 1\\[1mm]
  1 & 0
\end{pmatrix}
= \begin{pmatrix}
  0 & 1 & 0 & \frac{1}{2} \\[1mm]
  1 & 0 & \frac{1}{2} & 0 \\[1mm]
  0 & 0 & 0 & \frac{1}{2} \\[1mm]
  0 & 0 & \frac{1}{2} & 0
\end{pmatrix}.
\]
By inspection, we see that this is a stochastic matrix.
This will always be the case: the tensor product of two or more stochastic
matrices is always stochastic.

A common situation that we encounter is one in which one operation is performed
on one system and \emph{nothing} is done to another.
In such a case, exactly the same prescription is followed, bearing in mind that
doing nothing is represented by the identity matrix.
For example, resetting the bit $\mathsf{X}$ to the $0$ state and doing nothing
to $\mathsf{Y}$ yields the probabilistic (and in fact deterministic) operation
on $(\mathsf{X},\mathsf{Y})$ represented by the matrix
\[
\begin{pmatrix}
  1 & 1\\[1mm]
  0 & 0
\end{pmatrix}
\otimes
\begin{pmatrix}
  1 & 0\\[1mm]
  0 & 1
\end{pmatrix}
= \begin{pmatrix}
  1 & 0 & 1 & 0 \\
  0 & 1 & 0 & 1 \\
  0 & 0 & 0 & 0 \\
  0 & 0 & 0 & 0
\end{pmatrix}.
\]

\section{Quantum information}

We're now prepared to move on to quantum information in the setting of multiple
systems.
Much like in the previous lesson on single systems, the mathematical
description of quantum information for multiple systems is quite similar to the
probabilistic case and makes use of similar concepts and techniques.

\subsection{Quantum states}

Multiple systems can be viewed collectively as single, compound systems.
We've already observed this in the probabilistic setting, and the quantum
setting is analogous.
Quantum states of multiple systems are therefore represented by column vectors
having complex number entries and Euclidean norm equal to $1,$ just like
quantum states of single systems.
In the multiple system case, the entries of these vectors are placed in
correspondence with the \emph{Cartesian product} of the classical state sets
associated with each of the individual systems, because that's the classical
state set of the compound system.

For instance, if $\mathsf{X}$ and $\mathsf{Y}$ are qubits, then the classical
state set of the pair of qubits $(\mathsf{X},\mathsf{Y}),$ viewed collectively
as a single system, is the Cartesian product $\{0,1\}\times\{0,1\}.$
By representing pairs of binary values as binary strings of length two, we
associate this Cartesian product set with the set $\{00,01,10,11\}.$
The following vectors are therefore all examples of quantum state vectors of
the pair $(\mathsf{X},\mathsf{Y})$:
\[
\frac{1}{\sqrt{2}} \vert 00 \rangle
- \frac{1}{\sqrt{6}} \vert 01\rangle
+ \frac{i}{\sqrt{6}} \vert 10\rangle
+ \frac{1}{\sqrt{6}} \vert 11\rangle, \quad
\frac{3}{5} \vert 00\rangle - \frac{4}{5} \vert 11\rangle,
\quad \text{and} \quad
\vert 01 \rangle.
\]

There are variations on how quantum state vectors of multiple systems are
expressed, and we can choose whichever variation suits our preferences.
Here are some examples for the first quantum state vector above.

\begin{enumerate}
\item
  We may use the fact that $\vert ab\rangle = \vert a\rangle \vert b\rangle$
  (for any classical states $a$ and $b$) to instead write
  \[
  \frac{1}{\sqrt{2}} \vert 0\rangle\vert 0 \rangle
  - \frac{1}{\sqrt{6}} \vert 0\rangle\vert 1\rangle
  + \frac{i}{\sqrt{6}} \vert 1\rangle\vert 0\rangle
  + \frac{1}{\sqrt{6}} \vert 1\rangle\vert 1\rangle.
  \]

\item
  We may choose to write the tensor product symbol explicitly like this:
  \[
  \frac{1}{\sqrt{2}} \vert 0\rangle\otimes\vert 0 \rangle
  - \frac{1}{\sqrt{6}} \vert 0\rangle\otimes\vert 1\rangle
  + \frac{i}{\sqrt{6}} \vert 1\rangle\otimes\vert 0\rangle
  + \frac{1}{\sqrt{6}} \vert 1\rangle\otimes\vert 1\rangle.
  \]

\item
  We may subscript the kets to indicate how they correspond to the systems
  being considered, like this:
  \[
  \frac{1}{\sqrt{2}} \vert 0\rangle_{\mathsf{X}}\vert 0 \rangle_{\mathsf{Y}}
  - \frac{1}{\sqrt{6}} \vert 0\rangle_{\mathsf{X}}\vert 1\rangle_{\mathsf{Y}}
  + \frac{i}{\sqrt{6}} \vert 1\rangle_{\mathsf{X}}\vert 0\rangle_{\mathsf{Y}}
  + \frac{1}{\sqrt{6}} \vert 1\rangle_{\mathsf{X}}\vert 1\rangle_{\mathsf{Y}}.
  \]
\end{enumerate}

\noindent
Of course, we may also write quantum state vectors explicitly as column
vectors:
\[
\begin{pmatrix}
  \frac{1}{\sqrt{2}}\\[2mm]
  - \frac{1}{\sqrt{6}}\\[2mm]
  \frac{i}{\sqrt{6}}\\[2mm]
  \frac{1}{\sqrt{6}}
\end{pmatrix}.
\]
Depending upon the context in which it appears, one of these variations may be
preferred --- but they are all equivalent in the sense that they describe the
same vector.

\subsubsection{Tensor products of quantum state vectors}

Similar to what we have for probability vectors, tensor products of quantum
state vectors are also quantum state vectors --- and again they represent
\emph{independence} among systems.

In greater detail, and beginning with the case of two systems, suppose that
$\vert \phi \rangle$ is a quantum state vector of a system $\mathsf{X}$ and
$\vert \psi \rangle$ is a quantum state vector of a system~$\mathsf{Y}.$
The tensor product $\vert \phi \rangle \otimes \vert \psi \rangle,$ which may
alternatively be written as
$\vert \phi \rangle \vert \psi \rangle$ or as
$\vert \phi \otimes \psi \rangle,$ is then a quantum state vector of the joint
system $(\mathsf{X},\mathsf{Y}).$ 
Again we refer to a state of this form as a being a \emph{product state}.

Intuitively speaking, when a pair of systems $(\mathsf{X},\mathsf{Y})$ is in a
product state $\vert \phi \rangle \otimes \vert \psi \rangle,$ we may interpret
this as meaning that $\mathsf{X}$ is in the quantum state $\vert \phi \rangle,$
$\mathsf{Y}$ is in the quantum state $\vert \psi \rangle,$ and the states of
the two systems have nothing to do with one another.

The fact that the tensor product vector $\vert \phi \rangle \otimes \vert \psi
\rangle$ is indeed a quantum state vector is consistent with the Euclidean norm
being \emph{multiplicative} with respect to tensor products:
\[
\begin{aligned}
  \bigl\| \vert \phi \rangle \otimes \vert \psi \rangle \bigr\|
  & = \sqrt{
    \sum_{(a,b)\in\Sigma\times\Gamma}
    \bigl\vert\langle ab \vert \phi\otimes\psi \rangle \bigr\vert^2
  }\\[1mm]
  & = \sqrt{
    \sum_{a\in\Sigma} \sum_{b\in\Gamma}
    \bigl\vert\langle a \vert \phi \rangle
    \langle b \vert \psi \rangle \bigr\vert^2
  }\\[1mm]
  & = \sqrt{
    \biggl(\sum_{a\in\Sigma}
    \bigl\vert \langle a \vert \phi \rangle \bigr\vert^2
    \biggr)
    \biggl(\sum_{b\in\Gamma}
    \bigl\vert \langle b \vert \psi \rangle \bigr\vert^2
    \biggr)
  }\\[1mm]
  & = \bigl\|
  \vert \phi \rangle \bigr\| \bigl\| \vert \psi \rangle
  \bigr\|.
\end{aligned}
\]
Because $\vert \phi \rangle$ and $\vert \psi \rangle$ are quantum state
vectors, we have $\|\vert \phi \rangle\| = 1$ and $\|\vert \psi \rangle\| = 1,$
and therefore $\|\vert \phi \rangle \otimes \vert \psi \rangle\| = 1,$ so
$\vert \phi \rangle \otimes \vert \psi \rangle$ is also a quantum state vector.

This generalizes to more than two systems.
If $\vert \psi_0 \rangle,\ldots,\vert \psi_{n-1} \rangle$ are quantum state
vectors of systems $\mathsf{X}_0,\ldots,\mathsf{X}_{n-1},$ then $\vert
\psi_{n-1} \rangle\otimes\cdots\otimes \vert \psi_0 \rangle$ is a quantum state
vector representing a \emph{product state} of the joint system
$(\mathsf{X}_{n-1},\ldots,\mathsf{X}_0).$
Again, we know that this is a quantum state vector because
\[
\bigl\|
\vert \psi_{n-1} \rangle\otimes\cdots\otimes \vert \psi_0 \rangle
\bigr\|
= \bigl\|\vert \psi_{n-1} \rangle\bigl\| \cdots
\bigl\|\vert \psi_0 \rangle \bigr\| = 1^n = 1.
\]

\subsubsection{Entangled states}

Not all quantum state vectors of multiple systems are product states.
For example, the quantum state vector
\begin{equation}
  \frac{1}{\sqrt{2}} \vert 00\rangle + \frac{1}{\sqrt{2}} \vert 11\rangle
  \label{eq:entangled-bits}
\end{equation}
of two qubits is not a product state.
To reason this, we may follow exactly the same argument that we used in the
previous section for a probabilistic state.
That is, if \eqref{eq:entangled-bits} were a product state, there would exist
quantum state vectors $\vert\phi\rangle$ and $\vert\psi\rangle$ for which
\[
\vert\phi\rangle\otimes\vert\psi\rangle
= \frac{1}{\sqrt{2}} \vert 00\rangle
+ \frac{1}{\sqrt{2}} \vert 11\rangle.
\]
But then it would necessarily be the case that
\[
\langle 0 \vert \phi\rangle
\langle 1 \vert \psi\rangle
= \langle 01 \vert \phi\otimes\psi\rangle
= 0
\]
implying that $\langle 0 \vert \phi\rangle = 0$ or
$\langle 1 \vert \psi\rangle = 0$ (or both).
That contradicts the fact that
\[
\langle 0 \vert \phi\rangle \langle 0 \vert \psi\rangle
= \langle 00 \vert \phi\otimes\psi\rangle
= \frac{1}{\sqrt{2}}
\]
and
\[
\langle 1 \vert \phi\rangle \langle 1 \vert \psi\rangle
= \langle 11 \vert \phi\otimes\psi\rangle
= \frac{1}{\sqrt{2}}
\]
are both nonzero.
Thus, the quantum state vector \eqref{eq:entangled-bits} represents a
\emph{correlation} between two systems, and specifically we say that the
systems are \emph{entangled}.

Notice that the specific value $1/\sqrt{2}$ is not important to this argument
--- all that is important is that this value is nonzero.
Thus, for instance, the quantum state
\[
\frac{3}{5} \vert 00\rangle + \frac{4}{5} \vert 11\rangle
\]
is also not a product state, by the same argument.

Entanglement is a quintessential feature of quantum information that will be
discussed in greater detail in a later lesson.
Entanglement can be complicated, particularly for the sorts of noisy quantum
states that can be described by density matrices, which are discussed later
in the course in Lesson~\ref{lesson:density-matrices}
\emph{(Density Matrices)}.
For quantum state vectors, however, entanglement is equivalent to correlation:
any quantum state vector that is not a product state represents an entangled
state.

\pagebreak

In contrast, the quantum state vector
\[
\frac{1}{2} \vert 00\rangle
+ \frac{i}{2} \vert 01\rangle
- \frac{1}{2} \vert 10\rangle
- \frac{i}{2} \vert 11\rangle
\]
is an example of a product state.
\begin{multline*}
  \qquad\qquad
\frac{1}{2} \vert 00\rangle
+ \frac{i}{2} \vert 01\rangle
- \frac{1}{2} \vert 10\rangle
- \frac{i}{2} \vert 11\rangle\\[2mm]
=
\biggl(
\frac{1}{\sqrt{2}}\vert 0\rangle - \frac{1}{\sqrt{2}}\vert 1\rangle
\biggr)
\otimes
\biggl(
\frac{1}{\sqrt{2}}\vert 0\rangle + \frac{i}{\sqrt{2}}\vert 1\rangle
\biggr)
\qquad\qquad
\end{multline*}
Hence, this state is not entangled.

\subsubsection{Bell states}

We'll now take a look as some important examples of multiple-qubit quantum
states, beginning with the \emph{Bell states}.
These are the following four two-qubit states.
\[
\begin{aligned}
  \vert \phi^+ \rangle & = \frac{1}{\sqrt{2}} \vert 00 \rangle +
  \frac{1}{\sqrt{2}} \vert 11 \rangle \\[2mm]
  \vert \phi^- \rangle & = \frac{1}{\sqrt{2}} \vert 00 \rangle -
  \frac{1}{\sqrt{2}} \vert 11 \rangle \\[2mm]
  \vert \psi^+ \rangle & = \frac{1}{\sqrt{2}} \vert 01 \rangle +
  \frac{1}{\sqrt{2}} \vert 10 \rangle \\[2mm]
  \vert \psi^- \rangle & = \frac{1}{\sqrt{2}} \vert 01 \rangle -
  \frac{1}{\sqrt{2}} \vert 10 \rangle
\end{aligned}
\]

The Bell states are so-named in honor of John Bell.
Notice that the same argument that establishes that $\vert\phi^+\rangle$ is not
a product state reveals that none of the other Bell states are product states
either: all four of the Bell states represent entanglement between two qubits.

The collection of all four Bell states
\[
\bigl\{\vert \phi^+ \rangle, \vert \phi^- \rangle, \vert \psi^+ \rangle, \vert
\psi^- \rangle\bigr\}
\]
is known as the \emph{Bell basis}.
True to its name, this is a basis; any quantum state vector of two qubits, or
indeed any complex vector at all having entries corresponding to the four
classical states of two bits, can be expressed as a linear combination of the
four Bell states.
For example,
\[
\vert 0 0 \rangle
= \frac{1}{\sqrt{2}} \vert \phi^+\rangle
+ \frac{1}{\sqrt{2}} \vert \phi^-\rangle.
\]

\subsubsection{GHZ and W states}

Next we will consider two interesting examples of states of three qubits.
The first example is the \emph{GHZ state} (so named in honor of Daniel
Greenberger, Michael Horne, and Anton Zeilinger, who first studied some of its
properties):
\[
\frac{1}{\sqrt{2}} \vert 000\rangle +
\frac{1}{\sqrt{2}} \vert 111\rangle.
\]
The second example is the so-called W state:
\[
\frac{1}{\sqrt{3}} \vert 001\rangle +
\frac{1}{\sqrt{3}} \vert 010\rangle +
\frac{1}{\sqrt{3}} \vert 100\rangle.
\]
Neither of these states is a product state, meaning that they cannot be written
as a tensor product of three qubit quantum state vectors.
We'll examine both of these states later when we discuss partial measurements
of quantum states of multiple systems.

\subsubsection{Additional examples}

The examples of quantum states of multiple systems we've seen so far are states
of two or three qubits, but we can also consider quantum states of multiple
systems having different classical state sets.

For example, here's a quantum state of three systems, $\mathsf{X},$
$\mathsf{Y},$ and $\mathsf{Z},$ where the classical state set of $\mathsf{X}$
is the binary alphabet (so $\mathsf{X}$ is a qubit) and the classical state set
of $\mathsf{Y}$ and $\mathsf{Z}$ is
$\{\clubsuit,\diamondsuit,\heartsuit,\spadesuit\}:$
\[
\frac{1}{2} \vert 0 \rangle \vert \heartsuit\rangle
\vert \heartsuit \rangle
+ \frac{1}{2} \vert 1 \rangle \vert \spadesuit\rangle
\vert \heartsuit \rangle
- \frac{1}{\sqrt{2}} \vert 0 \rangle \vert \heartsuit\rangle
\vert \diamondsuit \rangle.
\]

And here's an example of a quantum state of three systems, $\mathsf{X},$
$\mathsf{Y},$ and $\mathsf{Z},$ that all share the same classical state set
$\{0,1,2\}:$
\[
\frac{
  \vert 012 \rangle
  - \vert 021 \rangle
  + \vert 120 \rangle
  - \vert 102 \rangle
  + \vert 201 \rangle
  - \vert 210 \rangle
}{\sqrt{6}}.
\]
Systems having the classical state set $\{0,1,2\}$ are often called
\emph{trits} or (assuming that they can be in a quantum state) \emph{qutrits}.
The term \emph{qudit} refers to a system having classical state set
$\{0,\ldots,d-1\}$ for an arbitrary choice of $d.$

\subsection{Measurements of quantum states}

Standard basis measurements of quantum states of single systems were discussed
in the previous lesson: if a system having classical state set $\Sigma$ is in a
quantum state represented by the vector $\vert \psi \rangle,$ and that system
is measured (with respect to a standard basis measurement), then each classical
state $a\in\Sigma$ appears with probability $\vert \langle a \vert \psi
\rangle\vert^2.$
This tells us what happens when we have a quantum state of multiple systems and
choose to measure the entire compound system, which is equivalent to measuring
\emph{all} of the systems.

To state this precisely, let us suppose that
$\mathsf{X}_0,\ldots,\mathsf{X}_{n-1}$ are systems having classical state sets
$\Sigma_0,\ldots,\Sigma_{n-1},$ respectively.
We may then view $(\mathsf{X}_{n-1},\ldots,\mathsf{X}_0)$ collectively as a
single system whose classical state set is the Cartesian product
$\Sigma_{n-1}\times\cdots\times\Sigma_0.$
If a quantum state of this system is represented by the quantum state vector
$\vert\psi\rangle,$ and all of the systems are measured, then each possible
outcome $(a_{n-1},\ldots,a_0)\in\Sigma_{n-1}\times\cdots\times\Sigma_0$ appears
with probability $\vert\langle a_{n-1}\cdots a_0\vert \psi\rangle\vert^2.$

For example, if systems $\mathsf{X}$ and $\mathsf{Y}$ are jointly in the
quantum state
\[
\frac{3}{5} \vert 0\rangle \vert \heartsuit \rangle
- \frac{4i}{5} \vert 1\rangle \vert \spadesuit \rangle,
\]
then measuring both systems with standard basis measurements yields the outcome
$(0,\heartsuit)$ with probability $9/25$ and the outcome $(1,\spadesuit)$ with
probability $16/25.$

\subsubsection{Partial measurements}

Now let us consider the situation in which we have multiple systems in some
quantum state, and we measure a proper subset of the systems.
As before, we will begin with two systems $\mathsf{X}$ and $\mathsf{Y}$ having
classical state sets $\Sigma$ and $\Gamma,$ respectively.

In general, a quantum state vector of $(\mathsf{X},\mathsf{Y})$ takes the form
\[
\vert \psi \rangle
= \sum_{(a,b)\in\Sigma\times\Gamma} \alpha_{ab} \vert ab\rangle,
\]
where $\{\alpha_{ab} : (a,b)\in\Sigma\times\Gamma\}$ is a collection of complex
numbers satisfying
\[
\sum_{(a,b)\in\Sigma\times\Gamma} \vert \alpha_{ab} \vert^2 = 1,
\]
which is equivalent to $\vert \psi \rangle$ being a unit vector.

We already know, from the discussion above, that if both $\mathsf{X}$ and
$\mathsf{Y}$ are measured, then each possible outcome
$(a,b)\in\Sigma\times\Gamma$ appears with probability
\[
\bigl\vert \langle ab \vert \psi \rangle \bigr\vert^2 = \vert\alpha_{ab}\vert^2.
\]
If we suppose instead that just the first system $\mathsf{X}$ is measured, the
probability for each outcome $a\in\Sigma$ to appear must therefore be equal to
\[
\sum_{b\in\Gamma} \bigl\vert \langle ab \vert \psi \rangle \bigr\vert^{2} =
\sum_{b\in\Gamma} \vert\alpha_{ab}\vert^2.
\]
This is consistent with what we already saw in the probabilistic setting, as
well as our current understanding of physics:
the probability for each outcome to appear when $\mathsf{X}$ is measured can't
possibly depend on whether or not $\mathsf{Y}$ was also measured, as that would
allow for faster-than-light communication.

Having obtained a particular outcome $a\in\Sigma$ of a standard basis
measurement of $\mathsf{X},$ we naturally expect that the quantum state of
$\mathsf{X}$ changes so that it is equal to $\vert a\rangle,$ just like we had
for single systems.
But what happens to the quantum state of $\mathsf{Y}$?

To answer this question, we can first express the vector $\vert\psi\rangle$ as
\[
\vert\psi\rangle
= \sum_{a\in\Sigma}
\vert a \rangle
\otimes \vert \phi_a \rangle,
\]
where
\[
\vert \phi_a \rangle = \sum_{b\in\Gamma} \alpha_{ab} \vert b\rangle
\]
for each $a\in\Sigma.$
Here we're following the same methodology as in the probabilistic case, of
isolating the standard basis states of the system being measured.
The probability for the standard basis measurement of $\mathsf{X}$ to give each
outcome $a$ is then as follows
\[
\sum_{b\in\Gamma} \vert\alpha_{ab}\vert^2 = \bigl\| \vert \phi_a \rangle
\bigr\|^2
\]
And, as a result of the standard basis measurement of $\mathsf{X}$ giving the
outcome $a,$ the quantum state of the pair $(\mathsf{X},\mathsf{Y})$ together
becomes
\[
\vert a \rangle \otimes \frac{\vert \phi_a \rangle}{\|\vert \phi_a \rangle\|}.
\]
That is, the state ``collapses'' like in the single-system case, but only as far
as is required for the state to be consistent with the measurement of
$\mathsf{X}$ having produced the outcome $a.$

Informally speaking, $\vert a \rangle \otimes \vert \phi_a\rangle$ represents
the component of $\vert \psi\rangle$ that is consistent with the a measurement
of $\mathsf{X}$ resulting in the outcome $a.$
We then \emph{normalize} this vector --- by dividing it by its Euclidean norm,
which is equal to $\|\vert\phi_a\rangle\|$ --- to obtain a valid quantum state
vector having Euclidean norm equal to $1.$
This normalization step is analogous to what we did in the probabilistic
setting when we divided vectors by the sum of their entries to obtain a
probability vector.

As an example, consider the state of two qubits $(\mathsf{X},\mathsf{Y})$ from
the beginning of the section:
\[
\vert \psi \rangle
= \frac{1}{\sqrt{2}} \vert 00 \rangle
- \frac{1}{\sqrt{6}} \vert 01 \rangle
+ \frac{i}{\sqrt{6}} \vert 10 \rangle
+ \frac{1}{\sqrt{6}} \vert 11 \rangle.
\]
To understand what happens when the first system $\mathsf{X}$ is measured, we
begin by writing
\[
\vert \psi \rangle
= \vert 0 \rangle \otimes \biggl(
\frac{1}{\sqrt{2}}  \vert 0 \rangle
- \frac{1}{\sqrt{6}} \vert 1 \rangle \biggr)
+ \vert 1 \rangle \otimes \biggl(
\frac{i}{\sqrt{6}} \vert 0 \rangle
+ \frac{1}{\sqrt{6}} \vert 1 \rangle \biggr).
\]
We now see, based on the description above, that the probability for the
measurement to result in the outcome $0$ is
\[
\biggl\|\frac{1}{\sqrt{2}}  \vert 0 \rangle
-\frac{1}{\sqrt{6}} \vert 1 \rangle\biggr\|^2
= \frac{1}{2} + \frac{1}{6}
= \frac{2}{3},
\]
in which case the state of $(\mathsf{X},\mathsf{Y})$ becomes
\[
\vert 0\rangle \otimes
\frac{\frac{1}{\sqrt{2}} \vert 0 \rangle
  -\frac{1}{\sqrt{6}} \vert 1 \rangle}{\sqrt{\frac{2}{3}}}
= \vert 0\rangle \otimes
\Biggl( \frac{\sqrt{3}}{2} \vert 0 \rangle - \frac{1}{2} \vert 1\rangle\Biggr);
\]
and the probability for the measurement to result in the outcome $1$ is
\[
\biggl\|\frac{i}{\sqrt{6}}  \vert 0 \rangle
+ \frac{1}{\sqrt{6}} \vert 1 \rangle\biggr\|^2
= \frac{1}{6} + \frac{1}{6}
= \frac{1}{3},
\]
in which case the state of $(\mathsf{X},\mathsf{Y})$ becomes
\[
\vert 1\rangle \otimes
\frac{\frac{i}{\sqrt{6}} \vert 0 \rangle
  +\frac{1}{\sqrt{6}} \vert 1 \rangle}{\sqrt{\frac{1}{3}}}
= \vert 1\rangle \otimes
\Biggl( \frac{i}{\sqrt{2}} \vert 0 \rangle
+\frac{1}{\sqrt{2}} \vert 1\rangle\Biggr).
\]

The same technique, used in a symmetric way, describes what happens if the
second system $\mathsf{Y}$ is measured rather than the first.
This time we rewrite the vector $\vert \psi \rangle$ as
\[
\vert \psi \rangle
= \biggl(
\frac{1}{\sqrt{2}} \vert 0 \rangle
+ \frac{i}{\sqrt{6}} \vert 1 \rangle
\biggr) \otimes \vert 0\rangle
+ \biggl(
-\frac{1}{\sqrt{6}} \vert 0 \rangle
+\frac{1}{\sqrt{6}} \vert 1\rangle
\biggr) \otimes \vert 1\rangle.
\]
The probability that the measurement of $\mathsf{Y}$ gives the outcome $0$ is
\[
\biggl\| \frac{1}{\sqrt{2}} \vert 0 \rangle
+ \frac{i}{\sqrt{6}} \vert 1 \rangle \biggr\|^2
= \frac{1}{2} + \frac{1}{6} = \frac{2}{3},
\]
in which case the state of $(\mathsf{X},\mathsf{Y})$ becomes
\[
\frac{\frac{1}{\sqrt{2}} \vert 0 \rangle
  + \frac{i}{\sqrt{6}} \vert 1 \rangle}{\sqrt{\frac{2}{3}}} \otimes \vert 0
\rangle = \biggl(\frac{\sqrt{3}}{2} \vert 0 \rangle + \frac{i}{2} \vert 1
\rangle\biggr) \otimes\vert 0 \rangle;
\]
and the probability that the measurement outcome is $1$ is
\[
\biggl\|
-\frac{1}{\sqrt{6}} \vert 0 \rangle
+\frac{1}{\sqrt{6}} \vert 1\rangle
\biggr\|^2
= \frac{1}{6} + \frac{1}{6} = \frac{1}{3},
\]
in which case the state of $(\mathsf{X},\mathsf{Y})$ becomes
\[
\frac{
  -\frac{1}{\sqrt{6}} \vert 0 \rangle
  +\frac{1}{\sqrt{6}} \vert 1\rangle }{\frac{1}{\sqrt{3}}}
\otimes \vert 1\rangle
= \biggl(-\frac{1}{\sqrt{2}} \vert 0\rangle
+ \frac{1}{\sqrt{2}} \vert 1\rangle\biggr) \otimes \vert 1\rangle.
\]

\subsubsection{Remark on reduced quantum states}

The previous example shows a limitation of the description of quantum
information we've been using, which we'll later refer to as the
\emph{simplified formulation} when contrasting it with one based on density
matrices later on starting in Lesson~\ref{lesson:density-matrices}
\emph{(Density Matrices)}.
The limitation is that the simplified formulation doesn't offer us a way to
describe the reduced (or marginal) quantum state of just one of two systems (or
of a proper subset of any number of systems) like in the probabilistic case.

Specifically, for a probabilistic state of two systems
$(\mathsf{X},\mathsf{Y})$ described by a probability vector
\[
\sum_{(a,b)\in\Sigma\times\Gamma} p_{ab} \vert ab\rangle,
\]
we can write the \emph{reduced} or \emph{marginal} probabilistic state of
$\mathsf{X}$ alone as
\[
\sum_{a\in\Sigma} \biggl( \sum_{b\in\Gamma} p_{ab}\biggr) \vert a\rangle =
\sum_{(a,b)\in\Sigma\times\Gamma} p_{ab} \vert a\rangle.
\]
For quantum state vectors, there isn't an analogous way to do this.
In particular, for a quantum state vector
\[
\vert \psi \rangle = \sum_{(a,b)\in\Sigma\times\Gamma} \alpha_{ab} \vert
ab\rangle,
\]
the vector
\[
\sum_{(a,b)\in\Sigma\times\Gamma} \alpha_{ab} \vert a\rangle
\]
is not a quantum state vector in general, and does not properly represent the
concept of a reduced or marginal state.

Density matrices do, in fact, provide us with a meaningful way to define
reduced quantum states in an analogous way to the probabilistic setting.

\subsubsection{Partial measurements for three or more systems}

Partial measurements for three or more systems, where some proper subset of the
systems are measured, can be reduced to the case of two systems by dividing the
systems into two collections, those that are measured and those that are not.

Here is a specific example that illustrates how this can be done.
It demonstrates specifically how subscripting kets by the names of the systems
they represent can be useful --- in this case because it gives us a simple way
to describe permutations of the systems.

For this example, consider a quantum state of a 5-tuple of systems
$(\mathsf{X}_4,\ldots,\mathsf{X}_0),$ where all five of these systems share the
same classical state set $\{\clubsuit,\diamondsuit,\heartsuit,\spadesuit\}:$
\[
\begin{gathered}
  \sqrt{\frac{1}{7}}
  \vert\heartsuit\rangle \vert\clubsuit\rangle \vert\diamondsuit\rangle
  \vert\spadesuit\rangle \vert\spadesuit\rangle
  + \sqrt{\frac{2}{7}}
  \vert\diamondsuit\rangle \vert\clubsuit\rangle \vert\diamondsuit\rangle
  \vert\spadesuit\rangle \vert\clubsuit\rangle
  + \sqrt{\frac{1}{7}}
  \vert\spadesuit\rangle \vert\spadesuit\rangle \vert\clubsuit\rangle
  \vert\diamondsuit\rangle \vert\clubsuit\rangle \\
  -i \sqrt{\frac{2}{7}}
  \vert\heartsuit\rangle \vert\clubsuit\rangle \vert\diamondsuit\rangle
  \vert\heartsuit\rangle \vert\heartsuit\rangle
  - \sqrt{\frac{1}{7}}
  \vert\spadesuit\rangle \vert\heartsuit\rangle \vert\clubsuit\rangle
  \vert\spadesuit\rangle \vert\clubsuit\rangle.
\end{gathered}
\]
We'll examine the situation in which the first and third systems are measured,
and the remaining systems are left alone.

Conceptually speaking, there's no fundamental difference between this situation
and one in which one of two systems is measured.
Unfortunately, because the measured systems are interspersed with the
unmeasured systems, we face a hurdle in writing down the expressions needed to
perform these calculations.

One way to proceed, as suggested above, is to subscript the kets to indicate
which systems they refer to.
This gives us a way to keep track of the systems as we permute the ordering of
the kets, which makes the mathematics simpler.

First, the quantum state vector above can alternatively be written as
\[
\begin{gathered}
  \sqrt{\frac{1}{7}}
  \vert\heartsuit\rangle_4 \vert\clubsuit\rangle_3 \vert\diamondsuit\rangle_2
  \vert\spadesuit\rangle_1 \vert\spadesuit\rangle_0
  + \sqrt{\frac{2}{7}}
  \vert\diamondsuit\rangle_4 \vert\clubsuit\rangle_3 \vert\diamondsuit\rangle_2
  \vert\spadesuit\rangle_1 \vert\clubsuit\rangle_0\\
  + \sqrt{\frac{1}{7}}
  \vert\spadesuit\rangle_4 \vert\spadesuit\rangle_3 \vert\clubsuit\rangle_2
  \vert\diamondsuit\rangle_1 \vert\clubsuit\rangle_0
  -i \sqrt{\frac{2}{7}}
  \vert\heartsuit\rangle_4 \vert\clubsuit\rangle_3 \vert\diamondsuit\rangle_2
  \vert\heartsuit\rangle_1 \vert\heartsuit\rangle_0\\
  - \sqrt{\frac{1}{7}}
  \vert\spadesuit\rangle_4 \vert\heartsuit\rangle_3 \vert\clubsuit\rangle_2
  \vert\spadesuit\rangle_1 \vert\clubsuit\rangle_0.
\end{gathered}
\]
Nothing has changed, except that each ket now has a subscript indicating which
system it corresponds to.
Here we've used the subscripts $0,\ldots,4,$ but the names of the systems
themselves could also be used (in a situation where we have system names such
as $\mathsf{X},$ $\mathsf{Y},$ and $\mathsf{Z},$ for instance).

We can now re-order the kets and collect terms as follows:
\[
\begin{aligned}
  &
  \sqrt{\frac{1}{7}}
  \vert\heartsuit\rangle_4 \vert\diamondsuit\rangle_2 \vert\clubsuit\rangle_3
  \vert\spadesuit\rangle_1 \vert\spadesuit\rangle_0
  +
  \sqrt{\frac{2}{7}}
  \vert\diamondsuit\rangle_4 \vert\diamondsuit\rangle_2 \vert\clubsuit\rangle_3
  \vert\spadesuit\rangle_1 \vert\clubsuit\rangle_0\\
  & \quad +
  \sqrt{\frac{1}{7}}
  \vert\spadesuit\rangle_4 \vert\clubsuit\rangle_2 \vert\spadesuit\rangle_3
  \vert\diamondsuit\rangle_1 \vert\clubsuit\rangle_0
  -i
  \sqrt{\frac{2}{7}}
  \vert\heartsuit\rangle_4 \vert\diamondsuit\rangle_2 \vert\clubsuit\rangle_3
  \vert\heartsuit\rangle_1 \vert\heartsuit\rangle_0\\
  & \quad -\sqrt{\frac{1}{7}}
  \vert\spadesuit\rangle_4 \vert\clubsuit\rangle_2 \vert\heartsuit\rangle_3
  \vert\spadesuit\rangle_1 \vert\clubsuit\rangle_0\\[2mm]
  & \hspace{1.5cm} = \vert\heartsuit\rangle_4 \vert\diamondsuit\rangle_2
  \biggl(
  \sqrt{\frac{1}{7}} \vert\clubsuit\rangle_3 \vert\spadesuit\rangle_1
  \vert\spadesuit\rangle_0
  -i \sqrt{\frac{2}{7}} \vert\clubsuit\rangle_3 \vert\heartsuit\rangle_1
  \vert\heartsuit\rangle_0
  \biggr)\\
  & \hspace{1.5cm} \quad
  + \vert\diamondsuit\rangle_4 \vert\diamondsuit\rangle_2
  \biggl(
  \sqrt{\frac{2}{7}} \vert\clubsuit\rangle_3 \vert\spadesuit\rangle_1
  \vert\clubsuit\rangle_0
  \biggr)\\
  & \hspace{1.5cm} \quad + \vert\spadesuit\rangle_4 \vert\clubsuit\rangle_2
  \biggl(
  \sqrt{\frac{1}{7}} \vert\spadesuit\rangle_3 \vert\diamondsuit\rangle_1
  \vert\clubsuit\rangle_0
  - \sqrt{\frac{1}{7}} \vert\heartsuit\rangle_3 \vert\spadesuit\rangle_1
  \vert\clubsuit\rangle_0\biggr).
\end{aligned}
\]
The tensor products are still implicit, even when parentheses are used, as in
this example.

To be clear about permuting the kets, tensor products are not commutative: if
$\vert \phi\rangle$ and $\vert \pi \rangle$ are vectors, then, in general,
$\vert \phi\rangle\otimes\vert \pi \rangle$ is different from
$\vert \pi\rangle\otimes\vert \phi \rangle,$ and likewise for tensor products
of three or more vectors.
For instance,
$\vert\heartsuit\rangle \vert\clubsuit\rangle \vert\diamondsuit\rangle
\vert\spadesuit\rangle \vert\spadesuit\rangle$
is a different vector than
$\vert\heartsuit\rangle \vert\diamondsuit\rangle \vert\clubsuit\rangle
\vert\spadesuit\rangle \vert\spadesuit\rangle.$
Re-ordering the kets as we have just done should not be interpreted as
suggesting otherwise.

Rather, for the sake of performing calculations, we're simply making a decision
that it's more convenient to collect the systems together as
$(\mathsf{X}_4,\mathsf{X}_2,\mathsf{X}_3,\mathsf{X}_1,\mathsf{X}_0)$ rather
than $(\mathsf{X}_4,\mathsf{X}_3,\mathsf{X}_2,\mathsf{X}_1,\mathsf{X}_0).$
The subscripts on the kets serve to keep this all straight, and we're free to
revert back to the original ordering later if we wish to do that.

We now see that, if the systems $\mathsf{X}_4$ and $\mathsf{X}_2$ are measured,
the (nonzero) probabilities of the different outcomes are as follow:
\begin{itemize}
\item
  The measurement outcome $(\heartsuit,\diamondsuit)$ occurs with probability
  \[
  \biggl\|
  \sqrt{\frac{1}{7}} \vert\clubsuit\rangle_3 \vert\spadesuit\rangle_1
  \vert\spadesuit\rangle_0
  -i \sqrt{\frac{2}{7}} \vert\clubsuit\rangle_3 \vert\heartsuit\rangle_1
  \vert\heartsuit\rangle_0
  \biggr\|^2 = \frac{1}{7} + \frac{2}{7} = \frac{3}{7}
  \]

\item
  The measurement outcome $(\diamondsuit,\diamondsuit)$ occurs with probability
  \[
  \biggl\|
  \sqrt{\frac{2}{7}} \vert\clubsuit\rangle_3 \vert\spadesuit\rangle_1
  \vert\clubsuit\rangle_0 \biggr\|^2 = \frac{2}{7}
  \]

\item
  The measurement outcome $(\spadesuit,\clubsuit)$ occurs with probability
  \[
  \biggl\|
  \sqrt{\frac{1}{7}} \vert\spadesuit\rangle_3 \vert\diamondsuit\rangle_1
  \vert\clubsuit\rangle_0
  - \sqrt{\frac{1}{7}} \vert\heartsuit\rangle_3 \vert\spadesuit\rangle_1
  \vert\clubsuit\rangle_0
  \biggr\|^2 = \frac{1}{7} + \frac{1}{7} = \frac{2}{7}.
  \]

\end{itemize}

\noindent
If the measurement outcome is $(\heartsuit,\diamondsuit),$ for instance, the
resulting state of our five systems becomes
\[
\begin{aligned}
  & \vert \heartsuit\rangle_4 \vert \diamondsuit \rangle_2
  \otimes
  \frac{
    \sqrt{\frac{1}{7}}
    \vert\clubsuit\rangle_3 \vert\spadesuit\rangle_1 \vert\spadesuit\rangle_0
    - i
    \sqrt{\frac{2}{7}}
    \vert\clubsuit\rangle_3 \vert\heartsuit\rangle_1 \vert\heartsuit\rangle_0
  }{\sqrt{\frac{3}{7}}}\\
  & \qquad
  =
  \sqrt{\frac{1}{3}}
  \vert \heartsuit\rangle_4 \vert\clubsuit\rangle_3 \vert \diamondsuit
  \rangle_2\vert\spadesuit\rangle_1 \vert\spadesuit\rangle_0
  -i
  \sqrt{\frac{2}{3}}
  \vert \heartsuit\rangle_4 \vert\clubsuit\rangle_3 \vert \diamondsuit
  \rangle_2\vert\heartsuit\rangle_1 \vert\heartsuit\rangle_0.
\end{aligned}
\]
Here, for the final answer, we've reverted back to our original ordering of the
systems, just to illustrate that we can do this.
For the other possible measurement outcomes, the state can be determined in a
similar way.

Finally, here are two examples promised earlier, beginning with the GHZ state
\[
\frac{1}{\sqrt{2}} \vert 000\rangle + \frac{1}{\sqrt{2}} \vert 111\rangle.
\]
If just the first system is measured, we obtain the outcome $0$ with
probability $1/2,$ in which case the state of the three qubits becomes $\vert
000\rangle;$ and we also obtain the outcome $1$ with probability $1/2,$ in
which case the state of the three qubits becomes $\vert 111\rangle.$

For a W state, on the other hand, assuming again that just the first system is
measured, we begin by writing this state like this:
\[
\begin{aligned}
  &
  \frac{1}{\sqrt{3}} \vert 001\rangle +
  \frac{1}{\sqrt{3}} \vert 010\rangle +
  \frac{1}{\sqrt{3}} \vert 100\rangle \\
  & \qquad
  = \vert 0 \rangle \biggl(
  \frac{1}{\sqrt{3}} \vert 01\rangle +
\frac{1}{\sqrt{3}} \vert 10\rangle\biggr)
+ \vert 1 \rangle \biggl(\frac{1}{\sqrt{3}}\vert 00\rangle\biggr).
\end{aligned}
\]
The probability that a measurement of the first qubit results in the outcome 0
is therefore equal to
\[
\biggl\|
\frac{1}{\sqrt{3}} \vert 01\rangle +
\frac{1}{\sqrt{3}} \vert 10\rangle
\biggr\|^2 = \frac{2}{3},
\]
and conditioned upon the measurement producing this outcome, the quantum state
of the three qubits becomes
\[
\vert 0\rangle\otimes
\frac{
  \frac{1}{\sqrt{3}} \vert 01\rangle +
  \frac{1}{\sqrt{3}} \vert 10\rangle
}{
  \sqrt{\frac{2}{3}}
}
= \vert 0\rangle \biggl(\frac{1}{\sqrt{2}} \vert 01\rangle
+ \frac{1}{\sqrt{2}} \vert 10\rangle \biggr)
= \vert 0\rangle\vert \psi^+\rangle.
\]
The probability that the measurement outcome is 1 is $1/3,$ in which case the
state of the three qubits becomes $\vert 100\rangle.$

The W state is symmetric, in the sense that it does not change if we permute
the qubits.
We therefore obtain a similar description for measuring the second or third
qubit rather than the first.

\subsection{Unitary operations}

In principle, any unitary matrix whose rows and columns correspond to the
classical states of a system represents a valid quantum operation on that
system.
This, of course, remains true for compound systems, whose classical state sets
happen to be Cartesian products of the classical state sets of the individual
systems.

Focusing in on two systems, if $\mathsf{X}$ is a system having classical state
set $\Sigma,$ and $\mathsf{Y}$ is a system having classical state set $\Gamma,$
then the classical state set of the joint system $(\mathsf{X},\mathsf{Y})$ is
$\Sigma\times\Gamma.$ Therefore, quantum operations on this joint system are
represented by unitary matrices whose rows and columns are placed in
correspondence with the set $\Sigma\times\Gamma.$
The ordering of the rows and columns of these matrices is the same as the
ordering used for quantum state vectors of the system
$(\mathsf{X},\mathsf{Y}).$

For example, let us suppose that $\Sigma = \{1,2,3\}$ and $\Gamma = \{0,1\},$
and recall that the standard convention for ordering the elements of the
Cartesian product $\{1,2,3\}\times\{0,1\}$ is this:
\[
(1,0),\;(1,1),\;(2,0),\;(2,1),\;(3,0),\; (3,1).
\]

Here's an example of a unitary matrix representing an operation on
$(\mathsf{X},\mathsf{Y}):$
\[
U =
\begin{pmatrix}
  \frac{1}{2} & \frac{1}{2} & \frac{1}{2} & 0 & 0 & \frac{1}{2} \\[2mm]
  \frac{1}{2} & \frac{i}{2} & -\frac{1}{2} & 0 & 0 & -\frac{i}{2} \\[2mm]
  \frac{1}{2} & -\frac{1}{2} & \frac{1}{2} & 0 & 0 & -\frac{1}{2} \\[2mm]
  0 & 0 & 0 & \frac{1}{\sqrt{2}} & \frac{1}{\sqrt{2}} & 0\\[2mm]
  \frac{1}{2} & -\frac{i}{2} & -\frac{1}{2} & 0 & 0 & \frac{i}{2} \\[2mm]
  0 & 0 & 0 &  -\frac{1}{\sqrt{2}} & \frac{1}{\sqrt{2}} & 0
\end{pmatrix}.
\]
This unitary matrix isn't special, it's just an example.
To check that $U$ is unitary, it suffices to compute and check that
$U^{\dagger} U = \mathbb{I},$ for instance.
Alternatively, we can check that the rows (or the columns) are orthonormal,
which is made simpler in this case given the particular form of the matrix $U.$

The action of $U$ on the standard basis vector $\vert 1, 1 \rangle,$ for
instance, is
\[
U \vert 1, 1\rangle =
\frac{1}{2} \vert 1, 0 \rangle
+ \frac{i}{2} \vert 1, 1 \rangle
- \frac{1}{2} \vert 2, 0 \rangle
- \frac{i}{2} \vert 3, 0\rangle,
\]
which we can see by examining the second column of $U,$ considering our
ordering of the set $\{1,2,3\}\times\{0,1\}.$

As with any matrix, it is possible to express $U$ using Dirac notation, which
would require 20 terms for the 20 nonzero entries of $U.$
If we did write down all of these terms, however, rather than writing a
$6\times 6$ matrix, it would be messy and the patterns that are evident from
the matrix expression would not likely be as clear.
Simply put, Dirac notation is not always the best choice.

Unitary operations on three or more systems work in a similar way, with the
unitary matrices having rows and columns corresponding to the Cartesian product
of the classical state sets of the systems.
We've already seen one example in this lesson: the three-qubit operation
\[
\sum_{k = 0}^{7} \vert (k+1) \bmod 8 \rangle \langle k \vert,
\]
where numbers in bras and kets mean their $3$-bit binary encodings.
In addition to being a deterministic operation, this is also a unitary
operation.
Operations that are both deterministic and unitary
are sometimes called \emph{reversible} operations, and are represented by
permutation matrices, which we encountered in the previous lesson.
The conjugate transpose of this particular matrix can be written like this:
\[
\sum_{k = 0}^{7} \vert k \rangle \langle (k+1) \bmod 8 \vert
= \sum_{k = 0}^{7} \vert (k-1) \bmod 8 \rangle \langle k \vert.
\]
This represents the \emph{reverse}, or in mathematical terms the
\emph{inverse}, of the original operation --- which is what we expect from the
conjugate transpose of a unitary matrix.
We'll see other examples of unitary operations on multiple systems as the
lesson continues.

\subsubsection{Independent unitary operations}

When unitary operations are performed independently on a collection of
individual systems, the combined action of these independent operations is
described by the tensor product of the unitary matrices that represent them.
That is, if $\mathsf{X}_{0},\ldots,\mathsf{X}_{n-1}$ are quantum systems,
$U_0,\ldots, U_{n-1}$ are unitary matrices representing operations on these
systems, and the operations are performed independently on the systems, the
combined action on $(\mathsf{X}_{n-1},\ldots,\mathsf{X}_0)$ is represented by
the matrix $U_{n-1}\otimes\cdots\otimes U_0.$
Once again, we find that the probabilistic and quantum settings are analogous
in this regard.

One would naturally expect, from reading the previous paragraph, that the
tensor product of any collection of unitary matrices is unitary.
Indeed this is true, and we can verify it as follows.

Notice first that the conjugate transpose operation satisfies
\[
(M_{n-1} \otimes \cdots \otimes M_0)^{\dagger} = M_{n-1}^{\dagger} \otimes
\cdots \otimes M_0^{\dagger}
\]
for any chosen matrices $M_0,\ldots,M_{n-1}.$
This can be checked by going back to the definition of the tensor product and
of the conjugate transpose, and checking that each entry of the two sides of
the equation are in agreement.
This means that
\[
(U_{n-1} \otimes \cdots \otimes U_0)^{\dagger} (U_{n-1}\otimes\cdots\otimes U_0)
= (U_{n-1}^{\dagger} \otimes \cdots \otimes U_0^{\dagger})
(U_{n-1}\otimes\cdots\otimes U_0).
\]

Next, because the tensor product of matrices is multiplicative, we find that
\begin{multline*}
  \qquad\qquad
  (U_{n-1}^{\dagger} \otimes \cdots \otimes U_0^{\dagger})
  (U_{n-1}\otimes\cdots\otimes U_0)\\[1mm]
  = (U_{n-1}^{\dagger} U_{n-1}) \otimes \cdots \otimes (U_0^{\dagger} U_0)
  = \mathbb{I}_{n-1} \otimes \cdots \otimes \mathbb{I}_0.
  \qquad\qquad
\end{multline*}
Here we have written $\mathbb{I}_0,\ldots,\mathbb{I}_{n-1}$ to refer to the
matrices representing the identity operation on the systems
$\mathsf{X}_0,\ldots,\mathsf{X}_{n-1},$ which is to say that these are identity
matrices whose sizes agree with the number of classical states of
$\mathsf{X}_0,\ldots,\mathsf{X}_{n-1}.$

Finally, the tensor product $\mathbb{I}_{n-1} \otimes \cdots \otimes
\mathbb{I}_0$ is equal to the identity matrix for which we have a number of
rows and columns that agrees with the product of the number of rows and columns
of the matrices $\mathbb{I}_{n-1},\ldots,\mathbb{I}_0.$
This larger identity matrix represents the identity operation on the joint
system $(\mathsf{X}_{n-1},\ldots,\mathsf{X}_0).$

In summary, we have the following sequence of equalities.
\[
\begin{aligned}
  & (U_{n-1} \otimes \cdots \otimes U_0)^{\dagger} (U_{n-1}\otimes\cdots\otimes
  U_0) \\
  & \quad = (U_{n-1}^{\dagger} \otimes \cdots \otimes U_0^{\dagger})
  (U_{n-1}\otimes\cdots\otimes U_0) \\
  & \quad = (U_{n-1}^{\dagger} U_{n-1}) \otimes \cdots \otimes (U_0^{\dagger}
  U_0)\\
  & \quad = \mathbb{I}_{n-1} \otimes \cdots \otimes \mathbb{I}_0 =
  \mathbb{I}
\end{aligned}
\]
We therefore conclude that $U_{n-1} \otimes \cdots \otimes U_0$ is unitary.

An important situation that often arises is one in which a unitary operation is
applied to just one system --- or a proper subset of systems --- within a
larger joint system.
For instance, suppose that $\mathsf{X}$ and $\mathsf{Y}$ are systems that we
can view together as forming a single, compound system
$(\mathsf{X},\mathsf{Y}),$ and we perform an operation just on the system
$\mathsf{X}.$
To be precise, let us suppose that $U$ is a unitary matrix representing an
operation on $\mathsf{X},$ so that its rows and columns have been placed in
correspondence with the classical states of $\mathsf{X}.$

To say that we perform the operation represented by $U$ just on the system
$\mathsf{X}$ implies that we do nothing to $\mathsf{Y},$ meaning that we
independently perform $U$ on $\mathsf{X}$ and the \emph{identity operation} on
$\mathsf{Y}.$
That is, ``doing nothing'' to $\mathsf{Y}$ is equivalent to performing the
identity operation on $\mathsf{Y},$ which is represented by the identity matrix
$\mathbb{I}_\mathsf{Y}.$
(Here, by the way, the subscript $\mathsf{Y}$ tells us that
$\mathbb{I}_\mathsf{Y}$ refers to the identity matrix having a number of rows
and columns in agreement with the classical state set of $\mathsf{Y}.$)
The operation on $(\mathsf{X},\mathsf{Y})$ that is obtained when we perform $U$
on $\mathsf{X}$ and do nothing to $\mathsf{Y}$ is therefore represented by the
unitary matrix $U \otimes \mathbb{I}_{\mathsf{Y}}$.

For example, if $\mathsf{X}$ and $\mathsf{Y}$ are qubits, performing a Hadamard
operation on $\mathsf{X}$ and doing nothing to $\mathsf{Y}$ is equivalent to
performing the operation
\[
H \otimes \mathbb{I}_{\mathsf{Y}} =
\begin{pmatrix}
  \frac{1}{\sqrt{2}} & \frac{1}{\sqrt{2}}\\[2mm]
  \frac{1}{\sqrt{2}} & -\frac{1}{\sqrt{2}}
\end{pmatrix}
\otimes
\begin{pmatrix}
  1 & 0\\
  0 & 1
\end{pmatrix}
= \begin{pmatrix}
  \frac{1}{\sqrt{2}} & 0 & \frac{1}{\sqrt{2}} & 0\\[2mm]
  0 & \frac{1}{\sqrt{2}} & 0 & \frac{1}{\sqrt{2}}\\[2mm]
  \frac{1}{\sqrt{2}} & 0 & -\frac{1}{\sqrt{2}} & 0\\[2mm]
  0 & \frac{1}{\sqrt{2}} & 0 & -\frac{1}{\sqrt{2}}
\end{pmatrix}
\]
on the joint system $(\mathsf{X},\mathsf{Y}).$

Along similar lines, if an operation represented by a unitary matrix $U$ is
applied to $\mathsf{Y}$ and nothing is done to $\mathsf{X},$ the resulting
operation on $(\mathsf{X},\mathsf{Y})$ is represented by the unitary matrix
$\mathbb{I}_{\mathsf{X}} \otimes U$.
For example, if we again consider the situation in which both $\mathsf{X}$ and
$\mathsf{Y}$ are qubits and $U$ is a Hadamard operation, the resulting
operation on $(\mathsf{X},\mathsf{Y})$ is represented by the matrix
\[
\begin{pmatrix}
  1 & 0\\
  0 & 1
\end{pmatrix}
\otimes
\begin{pmatrix}
  \frac{1}{\sqrt{2}} & \frac{1}{\sqrt{2}}\\[2mm]
  \frac{1}{\sqrt{2}} & -\frac{1}{\sqrt{2}}
\end{pmatrix}
= \begin{pmatrix}
  \frac{1}{\sqrt{2}} & \frac{1}{\sqrt{2}} & 0 & 0\\[2mm]
  \frac{1}{\sqrt{2}} & -\frac{1}{\sqrt{2}} & 0 & 0\\[2mm]
  0 & 0 & \frac{1}{\sqrt{2}} & \frac{1}{\sqrt{2}}\\[2mm]
  0 & 0 & \frac{1}{\sqrt{2}} & -\frac{1}{\sqrt{2}}
\end{pmatrix}.
\]

Not every unitary operation on a collection of systems can be written as a
tensor product of unitary operations like this, just as not every quantum state
vector of these systems is a product state.
For example, neither the swap operation nor the controlled-NOT operation on two
qubits, which are described next, can be expressed as a tensor product of
unitary operations.

\subsubsection{The swap operation}

To conclude the lesson, let's take a look at two classes of examples of unitary
operations on multiple systems, beginning with the \emph{swap operation}.

Suppose that $\mathsf{X}$ and $\mathsf{Y}$ are systems that share the same
classical state set $\Sigma.$
The \emph{swap} operation on the pair $(\mathsf{X},\mathsf{Y})$ is the
operation that exchanges the contents of the two systems, but otherwise leaves
the systems alone --- so that $\mathsf{X}$ remains on the left and $\mathsf{Y}$
remains on the right.
We'll denote this operation as $\operatorname{SWAP},$ and it operates like this
for every choice of classical states $a,b\in\Sigma:$
\[
\operatorname{SWAP} \vert a \rangle \vert b \rangle = \vert b \rangle \vert a
\rangle.
\]
One way to write the matrix associated with this operation using the Dirac
notation is as follows: 
\[
\mathrm{SWAP} = \sum_{c,d\in\Sigma} \vert c \rangle \langle d \vert \otimes
\vert d \rangle \langle c \vert.
\]
It may not be immediately clear that this matrix represents
$\operatorname{SWAP},$ but we can check it satisfies the condition
$\operatorname{SWAP} \vert a \rangle \vert b \rangle = \vert b \rangle \vert a
\rangle$ for every choice of classical states $a,b\in\Sigma.$
As a simple example, when $\mathsf{X}$ and $\mathsf{Y}$ are qubits, we find that
\[
\operatorname{SWAP} =
\begin{pmatrix}
  1 & 0 & 0 & 0\\
  0 & 0 & 1 & 0\\
  0 & 1 & 0 & 0\\
  0 & 0 & 0 & 1
\end{pmatrix}.
\]

\subsubsection{Controlled-unitary operations}

Now let us suppose that $\mathsf{Q}$ is a qubit and $\mathsf{R}$ is an
arbitrary system, having whatever classical state set we wish.
For every unitary operation $U$ acting on the system $\mathsf{R},$ a
\emph{controlled}-$U$ operation is a unitary operation on the pair
$(\mathsf{Q},\mathsf{R})$ defined as follows.
\[
\vert 0\rangle \langle 0\vert \otimes \mathbb{I}_{\mathsf{R}} + \vert 1\rangle
\langle 1\vert \otimes U
\]

For example, if $\mathsf{R}$ is also a qubit, and we consider the Pauli $X$
operation on $\mathrm{R},$ then a controlled-$X$ operation is given by
\[
\vert 0\rangle \langle 0\vert \otimes \mathbb{I}_{\mathsf{R}} + \vert 1\rangle
\langle 1\vert \otimes X =
\begin{pmatrix}
  1 & 0 & 0 & 0\\
  0 & 1 & 0 & 0\\
  0 & 0 & 0 & 1\\
  0 & 0 & 1 & 0
\end{pmatrix}.
\]
We already encountered this operation in the context of classical information
and probabilistic operations earlier in the lesson.
Replacing the Pauli $X$ operation on $\mathsf{R}$ with a $Z$ operation gives
this operation:
\[
\vert 0\rangle \langle 0\vert \otimes \mathbb{I}_{\mathsf{R}} + \vert 1\rangle
\langle 1\vert \otimes Z =
\begin{pmatrix}
  1 & 0 & 0 & 0\\
  0 & 1 & 0 & 0\\
  0 & 0 & 1 & 0\\
  0 & 0 & 0 & -1
\end{pmatrix}.
\]

If instead we take $\mathsf{R}$ to be two qubits, and we take $U$ to be the
\emph{swap operation} between these two qubits, we obtain this operation:
\[
\operatorname{CSWAP} =
\begin{pmatrix}
  1 & 0 & 0 & 0 & 0 & 0 & 0 & 0 \\
  0 & 1 & 0 & 0 & 0 & 0 & 0 & 0 \\
  0 & 0 & 1 & 0 & 0 & 0 & 0 & 0 \\
  0 & 0 & 0 & 1 & 0 & 0 & 0 & 0 \\
  0 & 0 & 0 & 0 & 1 & 0 & 0 & 0 \\
  0 & 0 & 0 & 0 & 0 & 0 & 1 & 0 \\
  0 & 0 & 0 & 0 & 0 & 1 & 0 & 0 \\
  0 & 0 & 0 & 0 & 0 & 0 & 0 & 1
\end{pmatrix}.
\]
This operation is also known as a \emph{Fredkin operation}, or more commonly, a
\emph{Fredkin gate}.
Its action on standard basis states can be described as follows:
\[
\begin{aligned}
  \operatorname{CSWAP} \vert 0 b c \rangle
  & = \vert 0 b c \rangle \\[1mm]
  \operatorname{CSWAP} \vert 1 b c \rangle
  & = \vert 1 c b \rangle
\end{aligned}
\]

Finally, a \emph{controlled-controlled-NOT operation} is called a \emph{Toffoli
operation} or \emph{Toffoli gate}.
Its matrix representation looks like this:
\[
\begin{pmatrix}
  1 & 0 & 0 & 0 & 0 & 0 & 0 & 0\\
  0 & 1 & 0 & 0 & 0 & 0 & 0 & 0\\
  0 & 0 & 1 & 0 & 0 & 0 & 0 & 0\\
  0 & 0 & 0 & 1 & 0 & 0 & 0 & 0\\
  0 & 0 & 0 & 0 & 1 & 0 & 0 & 0\\
  0 & 0 & 0 & 0 & 0 & 1 & 0 & 0\\
  0 & 0 & 0 & 0 & 0 & 0 & 0 & 1\\
  0 & 0 & 0 & 0 & 0 & 0 & 1 & 0
\end{pmatrix}.
\]
We may alternatively express it using the Dirac notation as follows:
\[
\bigl(
\vert 00 \rangle \langle 00 \vert
+ \vert 01 \rangle \langle 01 \vert
+ \vert 10 \rangle \langle 10 \vert \bigr) \otimes \mathbb{I}
+ \vert 11 \rangle \langle 11 \vert \otimes X.
\]


\lesson{Quantum Circuits}
\label{lesson:quantum-circuits}

This lesson introduces the \emph{quantum circuit} model of computation, which
provides a standard way to describe quantum computations.

The lesson also introduces a few important mathematical concepts, including
\emph{inner products} between vectors, the notions of \emph{orthogonality} and
\emph{orthonormality}, and \emph{projections} and \emph{projective
measurements}, which generalize standard basis measurements.
Through these concepts, we'll derive fundamental limitations on quantum
information, including the \emph{no-cloning theorem} and the impossibility to
perfectly discriminate non-orthogonal quantum states.

\section{Circuits}

In computer science, \emph{circuits} are models of computation in which
information is carried by wires through a network of \emph{gates}, which
represent operations on the information carried by the wires.
\emph{Quantum circuits} are a specific model of computation based on this more
general concept.

Although the word ``circuit'' often refers to a circular path, circular paths
aren't actually allowed in the circuit models of computation that are most
commonly studied.
That is to say, we usually consider \emph{acyclic circuits} when we're thinking
about circuits as computational models.
Quantum circuits follow this pattern; a quantum circuit represents a finite
sequence of operations that cannot contain feedback loops.

\subsection{Boolean circuits}

\begin{figure}[!ht]
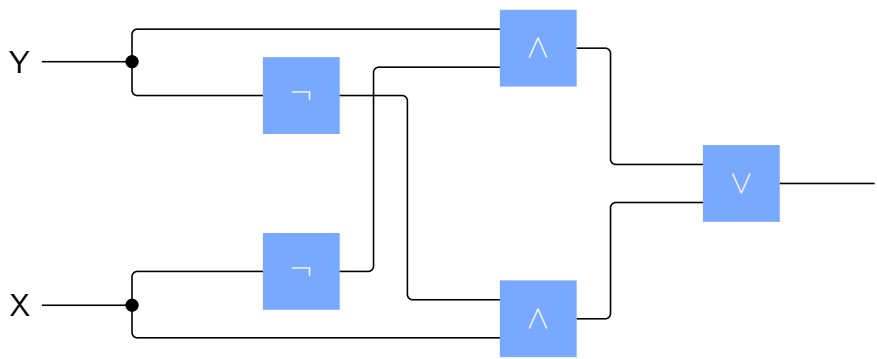

  \begin{center}

  \end{center}
  \caption{A Boolean circuit for computing the exclusive-OR of two bits.}
  \label{fig:Boolean-circuit-XOR}
\end{figure}

Figure~\ref{fig:Boolean-circuit-XOR} shows an example of a (classical) Boolean
circuit, where the wires carry binary values and the gates represent Boolean
logic operations.
The flow of information along the wires goes from left to right: the wires on
the left-hand side of the figure labeled $\mathsf{X}$ and $\mathsf{Y}$ are
input bits, which can be set to whatever binary values we choose, and the wire
on the right-hand side is the output.
The intermediate wires take values determined by the gates, which are evaluated
from left to right.

The gates are AND gates (labeled $\wedge$), OR gates (labeled $\vee$), and NOT
gates (labeled $\neg$).
The functions computed by these gates will likely be familiar to many readers,
but here they are represented by tables of values:
\[
\begin{array}{c}
  \begin{array}{c|c}
    a & \neg a\\
    \hline
    0 & 1\\
    1 & 0\\
  \end{array}\\
  \\
  \\
\end{array}
\qquad\quad
\begin{array}{c|c}
  ab & a \wedge b\\
  \hline
  00 & 0\\
  01 & 0\\
  10 & 0\\
  11 & 1
\end{array}
\qquad\quad
\begin{array}{c|c}
  ab & a \vee b\\
  \hline
  00 & 0\\
  01 & 1\\
  10 & 1\\
  11 & 1
\end{array}
\]

The two small, solid circles on the wires just to the right of the names
$\mathsf{X}$ and $\mathsf{Y}$ represent \emph{fan-out} operations, which simply
create a copy of whatever value is carried on the wire on which they appear,
allowing this value to be input into multiple gates.
Fan-out operations are not always considered to be gates in the classical
setting; sometimes they're treated as if they're ``free'' in some sense.
When Boolean circuits are converted into equivalent quantum circuits, however,
we do need to classify fan-out operations explicitly as gates to handle and
account for them correctly.

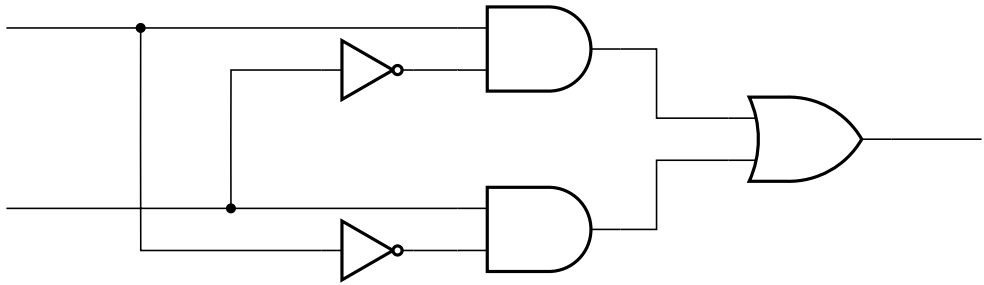
\begin{figure}[!ht]
  \begin{center}
    \begin{circuitikz}[scale=1.2,line width=0.6pt]
      \ctikzset{logic ports=ieee}
      
      \draw[color=black]
      (0,-1) node[and port] (lower_and) {}
      (0,1) node[and port] (upper_and) {}
      (3,0) node[or port] (or) {}
      (upper_and.in 1) -| ++(-5,0) node[anchor=east] (X) {}
      (lower_and.in 1) -| ++(-5,0) node[anchor=east] (Y) {}
      (upper_and.in 2) ++(-1,0) node[scale=0.7,not port] (upper_not) {}
      (lower_and.in 2) ++(-1,0) node[scale=0.7,not port] (lower_not) {}
      (upper_not.out) -| (upper_and.in 2)
      (lower_not.out) -| (lower_and.in 2)
      (upper_and.out) -| ++(0.4,0) |- (or.in 1)
      (lower_and.out) -| ++(0.4,0) |- (or.in 2)
      (upper_not.in) to ++(-1,0) coordinate (lower_fanout)
      to[short, -*] (lower_fanout |- Y)
      (lower_not.in) to ++(-2,0) coordinate (upper_fanout)
      to[short, -*] (upper_fanout |- X)
      (or.out) -| ++(1,0) node[anchor=west] (out) {};
      
    \end{circuitikz}
  \end{center}
  \caption{The same Boolean circuit as in Figure~\ref{fig:Boolean-circuit-XOR}
    expressed using standardized symbols in electrical engineering.}
  \label{fig:Boolean-circuit-IEEE}
\end{figure}

The same circuit is illustrated in Figure~\ref{fig:Boolean-circuit-IEEE}
using a style more common in electrical engineering, which uses conventional
symbols for the AND, OR, and NOT gates.
We won't use this style or these particular gate symbols further, but we will
use different symbols to represent gates in quantum circuits, which we'll
explain as we encounter them.

The particular circuit in this example computes the \emph{exclusive-OR} (or XOR
for short), which is denoted by the symbol $\oplus$:
\[
\begin{array}{c|c}
  ab & a \oplus b\\
  \hline
  00 & 0\\
  01 & 1\\
  10 & 1\\
  11 & 0
\end{array}
\]

\begin{figure}[!ht]
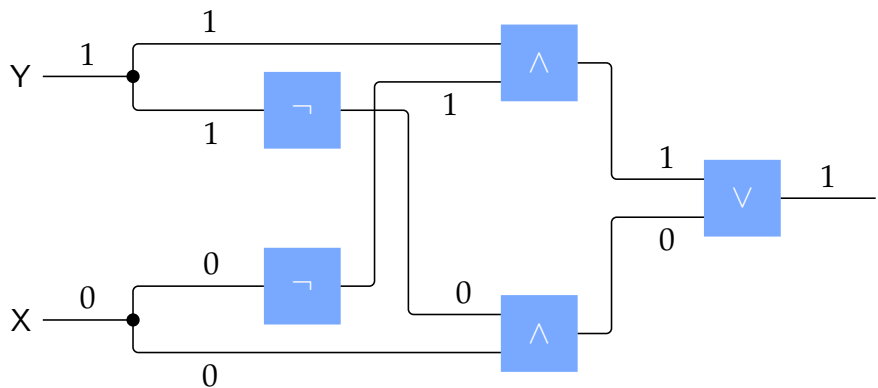

\begin{center}
  %
\end{center}
\caption{The same Boolean circuit as in Figure~\ref{fig:Boolean-circuit-XOR}
  evaluated on the inputs $\mathsf{X}=0$ and $\mathsf{Y}=1$.}
\label{fig:Boolean-circuit-XOR-evaluated}
\end{figure}

Figure~\ref{fig:Boolean-circuit-XOR-evaluated} illustrates the evaluation
of our circuit on just one choice for the inputs: $\mathsf{X}=0$
and $\mathsf{Y}=1.$
Each wire is labeled by value it carries so you can follow the operations.
The output value is $1$ in this case, which is the correct value for the XOR:
$0 \oplus 1 = 1.$
The other three possible input settings can be checked in a similar way.

\subsection{Other types of circuits}

As was suggested above, the notion of a circuit in computer science is very
general.
For example, circuits whose wires carry values other than $0$ and $1$ are
sometimes analyzed, as are gates representing different choices of operations.

\begin{figure}[!ht]
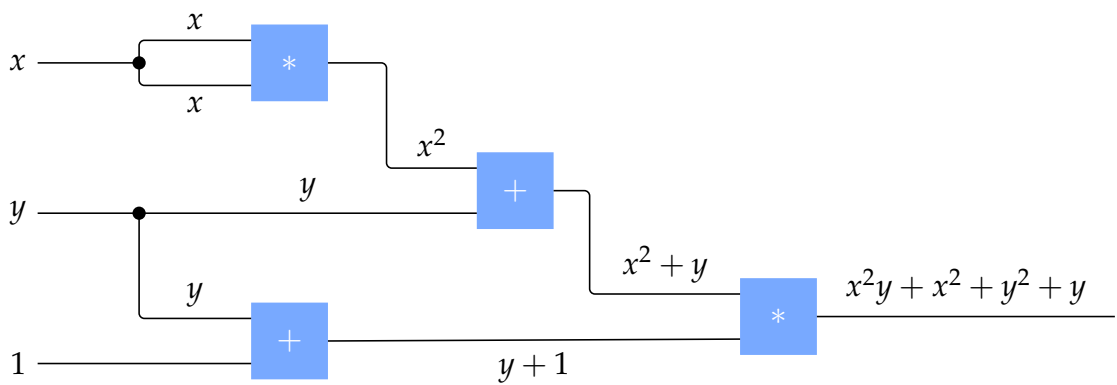

  \begin{center}

    
  \end{center}
  \caption{An arithmetic circuit.}
  \label{fig:arithmetic_circuit}
\end{figure}

In \emph{arithmetic circuits}, for instance, the wires may carry integer values
while the gates represent arithmetic operations, such as addition and
multiplication.
Figure~\ref{fig:arithmetic_circuit} depicts an arithmetic circuit that takes
two variable input values ($x$~and~$y$) as well as a third input set to the
value $1.$
The values carried by the wires, as functions of the values $x$ and $y,$ are
shown in the figure.

We can also consider circuits that incorporate randomness, such as ones where
gates represent probabilistic operations.

\subsection{Quantum circuits}

In the quantum circuit model, wires represent qubits and gates represent
operations on these qubits.
We'll focus for now on operations we've encountered so far, namely
\emph{unitary operations} and \emph{standard basis measurements}.
As we learn about other sorts of quantum operations and measurements, we can
enhance our model accordingly.

\begin{figure}[!ht]
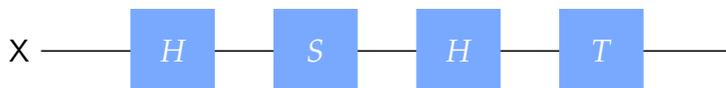

  \begin{center}

  \end{center}
  \caption{A simple quantum circuit on one qubit.}
  \label{fig:one-qubit-circuit}
\end{figure}

A simple example of a quantum circuit is shown in
Figure~\ref{fig:one-qubit-circuit}.
In this circuit, we have a single qubit named $\mathsf{X},$ which is
represented by the horizontal line, and a sequence of gates representing
unitary operations on this qubit.
Just like in the examples above, the flow of information goes from left to
right --- so the first operation performed is a Hadamard operation, the second
is an $S$ operation, the third is another Hadamard operation, and the final
operation is a $T$ operation.
Applying the entire circuit therefore applies the composition of these
operations, $T H S H,$ to the qubit $\mathsf{X}.$

\begin{figure}[!ht]
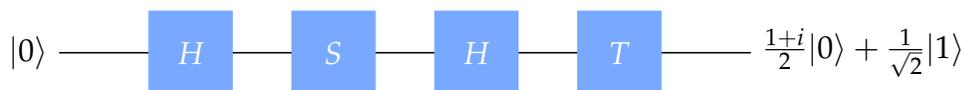

  \begin{center}
    %
  \end{center}
  \caption{The circuit from Figure~\ref{fig:one-qubit-circuit} evaluated
    on the input $\vert 0\rangle$.}
  \label{fig:one-qubit-circuit-evaluated}
\end{figure}

Sometimes we may wish to explicitly indicate the input or output states of
circuits.
For example, if we apply the operation $T H S H$ to the state $\vert 0\rangle,$
we obtain the state
$\frac{1+i}{2}\vert 0\rangle + \frac{1}{\sqrt{2}} \vert 1 \rangle.$
This can be indicated as is shown in
Figure~\ref{fig:one-qubit-circuit-evaluated}.
Quantum circuits often start out with all qubits initialized to $\vert
0\rangle,$ as we have in this case, but there are also situations where the
input qubits are initially set to different states.

\begin{figure}[!ht]
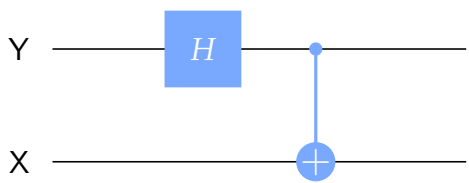

  \begin{center}
    \scalebox{1}{%
      %
    }
  \end{center}  
  \caption{A simple quantum circuit on two qubits.}
  \label{fig:ebit-circuit}
\end{figure}

Another example of a quantum circuit, this time with two qubits, is shown in
Figure~\ref{fig:ebit-circuit}.
As always, the gate labeled $H$ refers to a Hadamard operation, while the
second gate is a \emph{controlled-NOT} operation: the solid circle represents
the \emph{control qubit} and the circle resembling the symbol $\oplus$ denotes
the \emph{target qubit.}

Before examining this circuit in greater detail and explaining what it does, it
is imperative that we first clarify how qubits are ordered in quantum circuits.
This connects with the convention that Qiskit uses for naming and ordering
systems that was mentioned briefly in the previous lesson.

\begin{callout}[title={Qiskit's qubit ordering convention for circuits}]
  In Qiskit, the \emph{topmost} qubit in a circuit diagram has index $0$ and
  corresponds to the \emph{rightmost} position in a tuple of qubits (or in a
  string, Cartesian product, or tensor product corresponding to this tuple),
  the \emph{second-from-top} qubit has index~$1$ and corresponds to the
  position \emph{second-from-right} in a tuple, and so on. The
  \emph{bottommost} qubit, which has the highest index, therefore corresponds
  to the \emph{leftmost} position in a tuple.\vspace{2mm}

  In particular, Qiskit's default
  names for the qubits in an $n$-qubit circuit are represented by the $n$-tuple
  $(\mathsf{q_{n-1}},\ldots,\mathsf{q_{0}}),$ with $\mathsf{q_{0}}$ being the
  qubit on the top and $\mathsf{q_{n-1}}$ on the bottom in quantum circuit
  diagrams.
\end{callout}

\noindent
Please be aware that this is a reversal of a more common convention for
ordering qubits in circuits, and is a frequent source of confusion.

Although we sometimes deviate from the specific default names
$\mathsf{q_{0}},\ldots,\mathsf{q_{n-1}}$ used for qubits by Qiskit, we will
always follow the ordering convention described above when interpreting circuit
diagrams throughout this course.
Thus, our interpretation of the circuit above is that it describes an operation
on a pair of qubits $(\mathsf{X},\mathsf{Y}).$
If the input to the circuit is a quantum state $\vert\psi\rangle \otimes
\vert\phi\rangle,$ for instance, then this means that the lower qubit
$\mathsf{X}$ starts in the state $\vert\psi\rangle$ and the upper qubit
$\mathsf{Y}$ starts in the state $\vert\phi\rangle.$

\pagebreak

Now, to understand what the circuit in Figure~\ref{fig:ebit-circuit} does, we
can go from left to right through its operations.

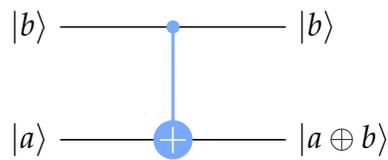
\begin{figure}[t]
  \begin{center}
    \begin{tikzpicture}[
        scale=1,
        line width=0.6pt,
        gate/.style={%
          inner sep = 0,
          fill = CircuitBlue,
          draw = CircuitBlue,
          text = white,
          minimum size = 10mm},
        control/.style={%
          circle,
          fill=CircuitBlue,
          minimum size = 5pt,
          inner sep=0mm},
        not/.style={%
          circle,
          fill = CircuitBlue,
          draw = CircuitBlue,
          text = white,
          inner sep=0,
          minimum size=5mm,
          label = {center:\textcolor{white}{\large $+$}}
        }
      ]
      
      \node[anchor = east] (In1) at (-1.5,1.5) {%
        \makebox(0,0)[r]{$\vert b\rangle$}};
      
      \node[anchor = east] (In2) at (-1.5,0) {%
        \makebox(0,0)[r]{$\vert a\rangle$}};
      
      \node[anchor = west] (Out1) at (1.5,1.5) {%
        \makebox(0,0)[l]{$\vert b\rangle$}};
      
      \node[anchor = west] (Out2) at (1.5,0) {%
        \makebox(0,0)[l]{$\vert a\oplus b\rangle$}};
      
      \draw (In1) -- (Out1);
      \draw (In2) -- (Out2);
      
      \node[control] (Control1) at (0,1.5) {};
      
      \node[not] (Not) at (0,0) {};
      
      \draw[very thick,draw=CircuitBlue] (Control1.center) --
      ([yshift=-1pt]Not.north);
      
    \end{tikzpicture}
    
  \end{center}
  \caption{The action of a controlled-NOT gate on standard basis states.}
  \label{fig:CNOT}
\end{figure}

\begin{enumerate}
\item
  The first operation is a Hadamard operation on $\mathsf{Y}$.
  When applying a gate to a single qubit like this, nothing happens to the
  other qubits (which is just one other qubit in this case). Nothing happening
  is equivalent to the identity operation being performed.
  The effect of the Hadamard operation on the two qubits together is
  therefore represented by this matrix:
  \[
  \mathbb{I}\otimes H
  = \begin{pmatrix}
    \frac{1}{\sqrt{2}} & \frac{1}{\sqrt{2}} & 0 & 0\\[2mm]
    \frac{1}{\sqrt{2}} & -\frac{1}{\sqrt{2}} & 0 & 0\\[2mm]
    0 & 0 & \frac{1}{\sqrt{2}} & \frac{1}{\sqrt{2}}\\[2mm]
    0 & 0 & \frac{1}{\sqrt{2}} & -\frac{1}{\sqrt{2}}
  \end{pmatrix}.
  \]
  Note that the identity matrix is on the left of the tensor product and $H$ is
  on the right, which is consistent with Qiskit's ordering convention.

\item
  The second operation is the controlled-NOT operation, where $\mathsf{Y}$ is
  the control and $\mathsf{X}$ is the target.
  The controlled-NOT gate's action on standard basis states is
  illustrated in Figure~\ref{fig:CNOT}.
  
  Given that we order the qubits as $(\mathsf{X}, \mathsf{Y}),$ with
  $\mathsf{X}$ being on the bottom and $\mathsf{Y}$ being on the top of our
  circuit, the matrix representation of the controlled-NOT gate is this:
  \[
  \begin{pmatrix}
    1 & 0 & 0 & 0\\
    0 & 0 & 0 & 1\\
    0 & 0 & 1 & 0\\
    0 & 1 & 0 & 0
  \end{pmatrix}.
  \]
\end{enumerate}

The unitary operation implemented by the entire circuit, which we'll give the
name $U,$ is the composition of the operations:
\[
U = \begin{pmatrix}
  1 & 0 & 0 & 0\\
  0 & 0 & 0 & 1\\
  0 & 0 & 1 & 0\\
  0 & 1 & 0 & 0
\end{pmatrix}
\begin{pmatrix}
  \frac{1}{\sqrt{2}} & \frac{1}{\sqrt{2}} & 0 & 0\\[2mm]
  \frac{1}{\sqrt{2}} & -\frac{1}{\sqrt{2}} & 0 & 0\\[2mm]
  0 & 0 & \frac{1}{\sqrt{2}} & \frac{1}{\sqrt{2}}\\[2mm]
  0 & 0 & \frac{1}{\sqrt{2}} & -\frac{1}{\sqrt{2}}
\end{pmatrix}
= \begin{pmatrix}
  \frac{1}{\sqrt{2}} & \frac{1}{\sqrt{2}} & 0 & 0\\[2mm]
  0 & 0 & \frac{1}{\sqrt{2}} & -\frac{1}{\sqrt{2}}\\[2mm]
  0 & 0 & \frac{1}{\sqrt{2}} & \frac{1}{\sqrt{2}}\\[2mm]
  \frac{1}{\sqrt{2}} & -\frac{1}{\sqrt{2}} & 0 & 0
\end{pmatrix}.
\]
In particular, recalling our notation for the Bell states,
\[
\begin{aligned}
  \vert \phi^+ \rangle & = \frac{1}{\sqrt{2}} \vert 0 0 \rangle
  + \frac{1}{\sqrt{2}} \vert 1 1 \rangle \\[2mm]
  \vert \phi^- \rangle & = \frac{1}{\sqrt{2}} \vert 0 0 \rangle
  - \frac{1}{\sqrt{2}} \vert 1 1 \rangle \\[2mm]
  \vert \psi^+ \rangle & = \frac{1}{\sqrt{2}} \vert 0 1 \rangle
  + \frac{1}{\sqrt{2}} \vert 1 0 \rangle \\[2mm]
  \vert \psi^- \rangle & = \frac{1}{\sqrt{2}} \vert 0 1 \rangle
  - \frac{1}{\sqrt{2}} \vert 1 0 \rangle,
\end{aligned}
\]
we find that
\[
\begin{aligned}
  U \vert 00\rangle & = \vert \phi^+\rangle\\
  U \vert 01\rangle & = \vert \phi^-\rangle\\
  U \vert 10\rangle & = \vert \psi^+\rangle\\
  U \vert 11\rangle & = -\vert \psi^-\rangle.
\end{aligned}
\]

This circuit therefore gives us a way to create the state $\vert\phi^+\rangle$
if we run it on two qubits initialized to $\vert 00\rangle.$
More generally, it provides us with a way to convert the standard basis to the
Bell basis.

(Note that, while it is not important for this example, the $-1$ phase factor
on the last state, $-\vert \psi^{-} \rangle,$ could be eliminated if we wanted
by making a small addition to the circuit.
For instance, we could add a controlled-$Z$ gate at the beginning, which is
similar to a controlled-NOT gate except that a $Z$ operation is applied to the
target qubit rather than a NOT operation when the control is set to $1.$
Alternatively, we could add a swap gate at the end. Either choice eliminates
the minus sign without affecting the circuit's action on the other three
standard basis states.)

\begin{figure}[t]
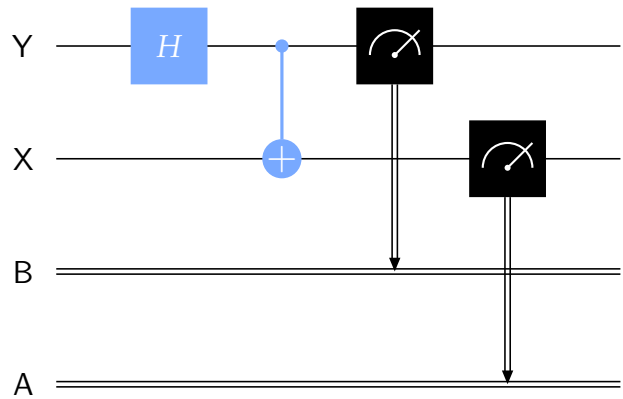

  \begin{center}

  \end{center}
  
  \caption{A quantum circuit including measurements and classical bit wires.}
  \label{fig:ebit-circuit-measure}
\end{figure}

In general, quantum circuits can contain any number of qubit wires. We may also
include \emph{classical bit} wires, which are indicated by double lines, like
in the example in Figure~\ref{fig:ebit-circuit-measure}.
\begin{figure}[t]
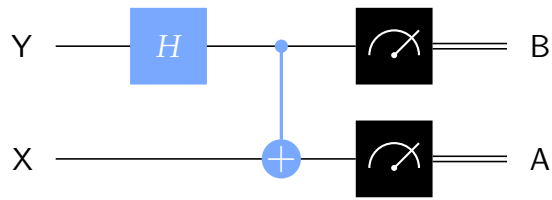

  \vspace{3mm}
  \begin{center}
    %
  \end{center}
  \caption{A compact representation of the circuit in
    Figure~\ref{fig:ebit-circuit-measure}.}
  \label{fig:ebit-circuit-measured-compact}
\end{figure}%
Here we have a Hadamard gate and a controlled-NOT gate on two qubits,
$\mathsf{X}$ and $\mathsf{Y},$ just like in the previous example.
We also have two \emph{classical} bits, $\mathsf{A}$ and $\mathsf{B},$ as well
as two measurement gates.
The measurement gates represent standard basis measurements:
the qubits are changed into their post-measurement states, while the
measurement outcomes are \emph{overwritten} onto the classical bits to which
the arrows point.

It's often convenient to depict a measurement as a gate that takes a qubit as
input and outputs a classical bit (as opposed to outputting the qubit in its
post-measurement state and writing the result to a separate classical bit).
This means the measured qubit has been discarded and can safely be ignored
thereafter, its state having changed into $\vert 0\rangle$ or $\vert 1\rangle$
depending upon the measurement outcome.
For example, the circuit diagram in
Figure~\ref{fig:ebit-circuit-measured-compact} represents the same process as
the one in Figure~\ref{fig:ebit-circuit-measure}, but where we disregard
$\mathsf{X}$ and $\mathsf{Y}$ after measuring them.

\pagebreak

As the course continues, we'll see more examples of quantum circuits, which are
usually more complicated than the simple examples above.
Here are some examples of symbols used to denote gates that commonly appear in
circuit diagrams.

\begin{itemize}
\item
  Single-qubit gates are generally shown as squares with a letter indicating
  which operation it is, like in Figure~\ref{fig:single-qubit-gates}.
  \begin{figure}[t]
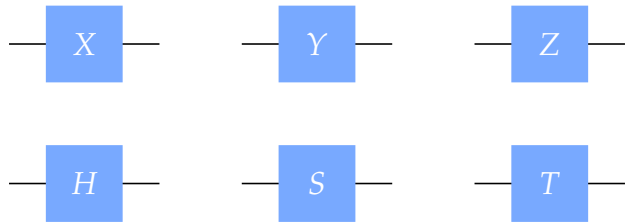

    \begin{center}

    \end{center}
    \caption{Six common single-qubit gates.}
    \label{fig:single-qubit-gates}
  \end{figure}%
  NOT gates (or $X$ gates) are also sometimes denoted by a circle
  around a plus sign, as shown in Figure~\ref{fig:not-gate}.

  \begin{figure}[t]
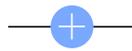

    \begin{center}
      %
    \end{center}

  \caption{An alternative symbol for a NOT (or $X$) gate.}
  \label{fig:not-gate}
\end{figure}%

\item
  Swap gates are denoted as is shown in Figure~\ref{fig:swap-gate}.

  \begin{figure}[t]
    \begin{center}
        %
    \end{center}
  \caption{A swap gate.}
  \label{fig:swap-gate}
\end{figure}

\item
  Controlled-gates, meaning gates that describe controlled-unitary operations,
  are denoted by a filled-in circle (indicating the control) connected by a
  vertical line to whatever operation is being controlled. For instance,
  controlled-NOT gates, controlled-controlled-NOT (or Toffoli) gates, and
  controlled-swap (or Fredkin) gates are denoted as illustrated in
  Figure~\ref{fig:controlled-gates}.

  \begin{figure}[t]
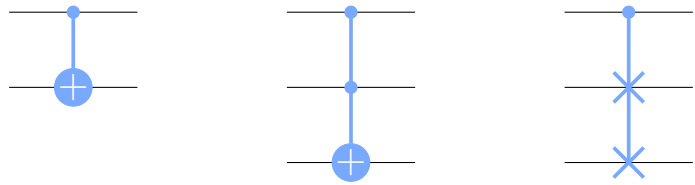

   \begin{center}
     %
   \end{center}
  \caption{A controlled-NOT gate, a controlled-controlled NOT (or Toffoli)
    gate, and a controlled-SWAP (or Fredkin) gate.}
  \label{fig:controlled-gates}
\end{figure}

\item
  Arbitrary unitary operations on multiple qubits may be viewed as gates. They
  are depicted by rectangles labeled by the name of the unitary operation.
  Figure~\ref{fig:uncontrolled-and-controlled-unitary} depicts an (unspecified)
  unitary operation $U$ as a gate, along with a controlled version of this
  gate.

  \begin{figure}[!ht]
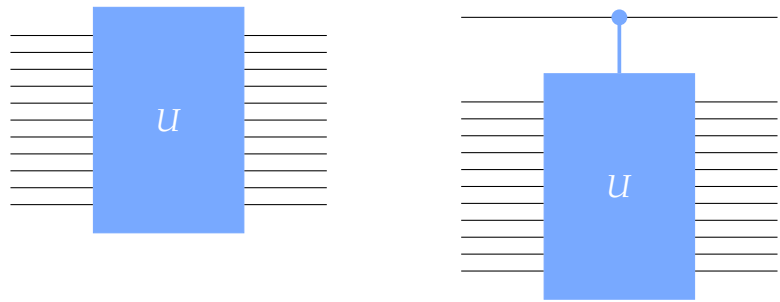

    \begin{center}
      \vspace{1mm}
      
      %
    \end{center}
  \caption{A unitary operation $U$ as a quantum gate along with a controlled
    version of this gate.}
  \label{fig:uncontrolled-and-controlled-unitary}
\end{figure}

\end{itemize}
  
\section{Inner products and projections}

To better prepare ourselves to explore the capabilities and limitations of
quantum circuits, we'll now introduce some additional mathematical concepts ---
namely the \emph{inner product} between vectors (and its connection to the
Euclidean norm), the notions of \emph{orthogonality} and \emph{orthonormality}
for sets of vectors, and \emph{projection} matrices, which will allow us to
introduce generalizations of standard basis measurements.

\subsection{Inner products}

Recall from Lesson~\ref{lesson:single-systems} \emph{(Single Systems)} that
when we use the Dirac notation to refer to an arbitrary column vector as a ket,
such as
\[
\vert \psi \rangle =
\begin{pmatrix}
  \alpha_1\\
  \alpha_2\\
  \vdots\\
  \alpha_n
\end{pmatrix},
\]
the corresponding bra vector is the \emph{conjugate transpose} of this vector:
\begin{equation}
  \langle \psi \vert = \bigl(\vert \psi \rangle \bigr)^{\dagger}
  =
  \begin{pmatrix}
    \overline{\alpha_1} & \overline{\alpha_2} & \cdots & \overline{\alpha_n}
  \end{pmatrix}.
  \label{eq:vector-dagger}
\end{equation}
Alternatively, if we have some classical state set $\Sigma$ in mind, and we
express a column vector as a ket such as
\[
\vert \psi \rangle = \sum_{a\in\Sigma} \alpha_a \vert a \rangle,
\]
then the corresponding row (or bra) vector is the conjugate transpose
\begin{equation}
  \langle \psi \vert = \sum_{a\in\Sigma} \overline{\alpha_a} \langle a \vert.
  \label{eq:vector-dagger-dirac}
\end{equation}

We also have that the product of a bra vector and a ket vector, viewed as
matrices either having a single row or a single column, results in a scalar.
Specifically, if we have two column vectors
\[
\vert \psi \rangle =
\begin{pmatrix}
  \alpha_1\\
  \alpha_2\\
  \vdots\\
  \alpha_n
\end{pmatrix}
\quad\text{and}\quad
\vert \phi \rangle =
\begin{pmatrix}
  \beta_1\\
  \beta_2\\
  \vdots\\
  \beta_n
\end{pmatrix},
\]
so that the row vector $\langle \psi \vert$ is as in equation
\eqref{eq:vector-dagger}, then
\[
\langle \psi \vert \phi \rangle = \langle \psi \vert \vert \phi \rangle
= \begin{pmatrix}
  \overline{\alpha_1} & \overline{\alpha_2} & \cdots & \overline{\alpha_n}
\end{pmatrix}
\begin{pmatrix}
  \beta_1\\
  \beta_2\\
  \vdots\\
  \beta_n
\end{pmatrix}
= \overline{\alpha_1} \beta_1 + \cdots + \overline{\alpha_n}\beta_n.
\]
Alternatively, if we have two column vectors that we have written as
\[
\vert \psi \rangle = \sum_{a\in\Sigma} \alpha_a \vert a \rangle
\quad\text{and}\quad
\vert \phi \rangle = \sum_{b\in\Sigma} \beta_b \vert b \rangle,
\]
so that $\langle \psi \vert$ is the row vector \eqref{eq:vector-dagger-dirac},
we find that
\[
\langle \psi \vert \phi \rangle = \langle \psi \vert \vert \phi \rangle
= \Biggl(\sum_{a\in\Sigma} \overline{\alpha_a} \langle a \vert\Biggr)
\Biggl(\sum_{b\in\Sigma} \beta_b \vert b\rangle\Biggr)
= \sum_{a\in\Sigma}\sum_{b\in\Sigma} \overline{\alpha_a} \beta_b \langle a
\vert b \rangle
= \sum_{a\in\Sigma} \overline{\alpha_a} \beta_a,
\]
where the last equality follows from the observation that $\langle a \vert a
\rangle = 1$ and $\langle a \vert b \rangle = 0$ for classical states $a$ and
$b$ satisfying $a\neq b.$

The value $\langle \psi \vert \phi \rangle$ is called the \emph{inner product}
between the vectors $\vert \psi\rangle$ and $\vert \phi \rangle.$
Inner products are critically important in quantum information and computation;
we would not get far in understanding quantum information at a mathematical
level without them.
Some basic facts about inner products of vectors follow.

\begin{trivlist}
\item
  \textbf{Relationship to the Euclidean norm.}
  The inner product of any vector
  \[
  \vert \psi \rangle = \sum_{a\in\Sigma} \alpha_a \vert a \rangle
  \]
  with itself is
  \[
  \langle \psi \vert \psi \rangle
  = \sum_{a\in\Sigma} \overline{\alpha_a} \alpha_a
  = \sum_{a\in\Sigma} \vert\alpha_a\vert^2
  = \bigl\| \vert \psi \rangle \bigr\|^2.
  \]
  Thus, the Euclidean norm of a vector may alternatively be expressed as
  \[
  \bigl\| \vert \psi \rangle \bigr\| = \sqrt{ \langle \psi \vert \psi \rangle }.
  \]

  Notice that the Euclidean norm of a vector must always be a nonnegative real
  number.
  Moreover, the only way the Euclidean norm of a vector can be equal to zero
  is if every one of the entries is equal to zero, which is to say that the
  vector is the zero vector.

  We can summarize these observations like this: for every vector
  $\vert \psi \rangle$ we have 
  \[
  \langle \psi \vert \psi \rangle \geq 0,
  \]
  with $\langle \psi \vert \psi \rangle = 0$ if and only if
  $\vert \psi \rangle = 0.$
  This property of the inner product is sometimes referred to as
  \emph{positive definiteness}.

\item
  \textbf{Conjugate symmetry.}
  For any two vectors
  \[
  \vert \psi \rangle = \sum_{a\in\Sigma} \alpha_a \vert a \rangle
  \quad\text{and}\quad
  \vert \phi \rangle = \sum_{b\in\Sigma} \beta_b \vert b \rangle,
  \]
  we have
  \[
  \langle \psi \vert \phi \rangle = \sum_{a\in\Sigma} \overline{\alpha_a}
  \beta_a
  \quad\text{and}\quad
  \langle \phi \vert \psi \rangle = \sum_{a\in\Sigma} \overline{\beta_a}
  \alpha_a,
  \]
  and therefore
  \[
  \overline{\langle \psi \vert \phi \rangle} = \langle \phi \vert \psi \rangle.
  \]

\item
  \textbf{Linearity in the second argument
    (and conjugate linearity in the first).}
  Let us suppose that $\vert \psi \rangle,$ $\vert \phi_1 \rangle,$ and $\vert
  \phi_2 \rangle$ are vectors and $\alpha_1$ and $\alpha_2$ are complex
  numbers.
  If we define a new vector
  \[
  \vert \phi\rangle = \alpha_1 \vert \phi_1\rangle + \alpha_2 \vert
  \phi_2\rangle,
  \]
  then
  \[
  \langle \psi \vert \phi \rangle
  = \langle \psi \vert \bigl( \alpha_1\vert \phi_1 \rangle
  + \alpha_2\vert \phi_2 \rangle\bigr)
  = \alpha_1 \langle \psi \vert \phi_1 \rangle
  + \alpha_2 \langle \psi \vert \phi_2 \rangle.
  \]
  That is to say, the inner product is \emph{linear} in the second argument.
  This can be verified either through the formulas above or simply by noting
  that matrix multiplication is linear in each argument (and specifically in
  the second argument).
  
  Combining this fact with conjugate symmetry reveals that the inner product is
  \emph{conjugate linear} in the first argument.
  That is, if $\vert \psi_1 \rangle,$ $\vert \psi_2 \rangle,$ and $\vert \phi
  \rangle$ are vectors and $\alpha_1$ and $\alpha_2$ are complex numbers, and
  we define
  \[
  \vert \psi \rangle = \alpha_1 \vert \psi_1\rangle
  + \alpha_2 \vert \psi_2 \rangle,
  \]
  then
  \[
  \langle \psi \vert \phi \rangle
  =
  \bigl( \overline{\alpha_1} \langle \psi_1 \vert
  + \overline{\alpha_2} \langle \psi_2 \vert \bigr) \vert\phi\rangle
  = \overline{\alpha_1} \langle \psi_1 \vert \phi \rangle
  + \overline{\alpha_2} \langle \psi_2 \vert \phi \rangle.
  \]

\item
  \textbf{The Cauchy--Schwarz inequality.}
  For every choice of vectors $\vert \phi \rangle$ and $\vert \psi \rangle$
  having the same number of entries, we have
  \[
  \bigl\vert \langle \psi \vert \phi \rangle\bigr| \leq \bigl\| \vert\psi
  \rangle \bigr\| \bigl\| \vert \phi \rangle \bigr\|.
  \]
  This is an incredibly handy inequality that gets used quite extensively in
  quantum information (and in many other topics of study).

\end{trivlist}

\subsection{Orthogonal and orthonormal sets}

Two vectors $\vert \phi \rangle$ and $\vert \psi \rangle$ are said to be
\emph{orthogonal} if their inner product is zero:
\[
\langle \psi \vert \phi \rangle = 0.
\]
Geometrically, we can think about orthogonal vectors as vectors at right angles
to each other.

A set of vectors $\{ \vert \psi_1\rangle,\ldots,\vert\psi_m\rangle\}$ is called
an \emph{orthogonal set} if every vector in the set is orthogonal to every
other vector in the set.
That is, this set is orthogonal if
\[
\langle \psi_j \vert \psi_k\rangle = 0
\]
for all choices of $j,k\in\{1,\ldots,m\}$ for which $j\neq k.$

A set of vectors $\{ \vert \psi_1\rangle,\ldots,\vert\psi_m\rangle\}$ is called
an \emph{orthonormal} set if it is an orthogonal set and, in addition, every
vector in the set is a unit vector.
Alternatively, this set is an orthonormal set if we have
\begin{equation}
  \langle \psi_j \vert \psi_k\rangle =
  \begin{cases}
    1 & j = k\\[1mm]
    0 & j\neq k
  \end{cases}
  \label{eq:orthonormal}
\end{equation}
for all choices of $j,k\in\{1,\ldots,m\}.$

Finally, a set $\{ \vert \psi_1\rangle,\ldots,\vert\psi_m\rangle\}$ is an
\emph{orthonormal basis} if, in addition to being an orthonormal set, it forms
a basis.
This is equivalent to $\{ \vert \psi_1\rangle,\ldots,\vert\psi_m\rangle\}$
being an orthonormal set and $m$ being equal to the dimension of the space from
which $\vert \psi_1\rangle,\ldots,\vert\psi_m\rangle$ are drawn.

For example, for any classical state set $\Sigma,$ the set of all standard
basis vectors $\big\{ \vert a \rangle \,:\, a\in\Sigma\bigr\}$ is an
orthonormal basis.
The set $\{\vert+\rangle,\vert-\rangle\}$ is an orthonormal basis for the
$2$-dimensional space corresponding to a single qubit, and the Bell basis
\[
\bigl\{\vert\phi^+\rangle, \vert\phi^-\rangle, \vert\psi^+\rangle,
\vert\psi^-\rangle\bigr\}
\]
is an orthonormal basis for the $4$-dimensional space
corresponding to two qubits.

\subsubsection{Extending orthonormal sets to orthonormal bases}

Suppose that $\vert\psi_1\rangle,\ldots,\vert\psi_m\rangle$ are vectors that
live in an $n$-dimensional space, and assume moreover that
$\{\vert\psi_1\rangle,\ldots,\vert\psi_m\rangle\}$ is an orthonormal set.
Orthonormal sets are always linearly independent sets, so these vectors
necessarily span a subspace of dimension $m.$
From this we conclude that $m\leq n$ because the dimension of the subspace
spanned by these vectors cannot be larger than the dimension of the entire
space from which they're drawn.

If it is the case that $m<n,$ then it is always possible to choose an
additional $n-m$ vectors $\vert \psi_{m+1}\rangle,\ldots,\vert\psi_n\rangle$ so
that $\{\vert\psi_1\rangle,\ldots,\vert\psi_n\rangle\}$ forms an orthonormal
basis.
A procedure known as the \emph{Gram--Schmidt orthogonalization process} can be
used to construct these vectors.

\subsubsection{Orthonormal sets and unitary matrices}

Orthonormal bases are closely connected with unitary matrices.
One way to express this connection is to say that the following three
statements are logically equivalent (meaning that they are all true or all
false) for any choice of a square matrix $U$.

\begin{enumerate}
\item
  The matrix $U$ is unitary (i.e., $U^{\dagger} U = \mathbb{I} = U
  U^{\dagger}$).
\item
  The rows of $U$ form an orthonormal basis.
\item
  The columns of $U$ form an orthonormal basis.
\end{enumerate}

This equivalence is actually pretty straightforward when we think about how
matrix multiplication and the conjugate transpose work.
Suppose, for instance, that we have a $3\times 3$ matrix like this:
\[
U = 

\]
The conjugate transpose of $U$ looks like this:
\[
U^{\dagger} = %
\]
Multiplying the two matrices, with the conjugate transpose on the left-hand
side, gives us this matrix:\vspace{1mm}
\[
\begin{aligned}
  &%
$}
\end{aligned}\vspace{2mm}
\]
If we form three vectors from the columns of $U,$
\[
\vert \psi_1\rangle = %
,
\]
then we can alternatively express the product above as follows:
\[
U^{\dagger} U =
%
\]
Referring to the equation \eqref{eq:orthonormal}, we see that this matrix is
equal to the identity matrix if and only if the set
$\{\vert\psi_1\rangle,\vert\psi_2\rangle,\vert\psi_3\rangle\}$
is orthonormal.
This argument generalizes to unitary matrices of any size.

The fact that the rows of a square matrix form an orthonormal basis if and only
if the matrix is unitary follows from the fact that a matrix is unitary if and
only if its transpose is unitary.

Given the equivalence described above, together with the fact that every
orthonormal set can be extended to form an orthonormal basis, we conclude the
following useful fact:
Given any orthonormal set of vectors
$\{\vert\psi_1\rangle,\ldots,\vert\psi_m\rangle\}$ drawn from an
$n$-dimensional space, there exists a unitary matrix $U$ whose first $m$
columns are the vectors $\vert\psi_1\rangle,\ldots,\vert\psi_m\rangle.$
Pictorially, we can always find a unitary matrix having this form:
\[
U =
\left(
\begin{array}{ccccccc}
  \rule{0.4pt}{16pt} & \rule{0.4pt}{16pt} & & \rule{0.4pt}{16pt} &
  \rule{0.4pt}{16pt} & & \rule{0.4pt}{16pt}\\
  \vert\psi_1\rangle & \vert\psi_2\rangle & \cdots & \vert\psi_m\rangle &
  \vert\psi_{m+1}\rangle &
  \cdots & \vert\psi_n\rangle\\[2mm]
  \rule{0.4pt}{16pt} & \rule{0.4pt}{16pt} & & \rule{0.4pt}{16pt} &
  \rule{0.4pt}{16pt} & & \rule{0.4pt}{16pt}
\end{array}
\right).
\]
The last $n-m$ columns ban be filled in with any choice of vectors
$\vert\psi_{m+1}\rangle,\ldots,\vert\psi_n\rangle$ that make
$\{\vert\psi_1\rangle,\ldots,\vert\psi_n\rangle\}$ an orthonormal basis.

\subsection{Projections and projective measurements}

\subsubsection{Projection matrices}

A square matrix $\Pi$ is called a \emph{projection} if it satisfies two
properties:
\begin{enumerate}
\item
  $\Pi = \Pi^{\dagger}.$
\item
  $\Pi^2 = \Pi.$
\end{enumerate}
Matrices that satisfy the first condition --- that they are equal to their own
conjugate transpose --- are called \emph{Hermitian matrices}, and matrices that
satisfy the second condition --- that squaring them leaves them unchanged ---
are called \emph{idempotent} matrices.

As a word of caution, the word \emph{projection} is sometimes used to refer to
any matrix that satisfies just the second condition but not necessarily the
first, and when this is done the term \emph{orthogonal projection} is typically
used to refer to matrices satisfying both properties.
In the context of quantum information and computation, however, the terms
\emph{projection} and \emph{projection matrix} more typically refer to matrices
satisfying both conditions.
\pagebreak

An example of a projection is the matrix
\begin{equation}
  \Pi = \vert \psi \rangle \langle \psi \vert
  \label{eq:pure-state}
\end{equation}
for any unit vector $\vert \psi\rangle.$
We can see that this matrix is Hermitian as follows:
\[
\Pi^{\dagger} = \bigl( \vert \psi \rangle \langle \psi \vert \bigr)^{\dagger}
= \bigl( \langle \psi \vert \bigr)^{\dagger}\bigl( \vert \psi \rangle
\bigr)^{\dagger}
= \vert \psi \rangle \langle \psi \vert = \Pi.
\]
Here, to obtain the second equality, we have used the formula
\[
(A B)^{\dagger} = B^{\dagger} A^{\dagger},
\]
which is always true, for any two matrices $A$ and $B$ for which the product
$AB$ makes sense.

To see that the matrix $\Pi$ in \eqref{eq:pure-state} is idempotent, we can use
the assumption that $\vert\psi\rangle$ is a unit vector, so that it satisfies
$\langle \psi \vert \psi\rangle = 1.$
Thus, we have
\[
\Pi^2
= \bigl( \vert\psi\rangle\langle \psi\vert \bigr)^2
= \vert\psi\rangle\langle \psi\vert\psi\rangle\langle\psi\vert
= \vert\psi\rangle\langle\psi\vert = \Pi.
\]

More generally, if $\{\vert \psi_1\rangle,\ldots,\vert \psi_m\rangle\}$ is any
orthonormal set of vectors, then the matrix
\begin{equation}
  \Pi = \sum_{k = 1}^m \vert \psi_k\rangle \langle \psi_k \vert
  \label{eq:projection}
\end{equation}
is a projection.
Specifically, we have
\[
\Pi^{\dagger}
= \biggl(\sum_{k = 1}^m \vert \psi_k\rangle \langle \psi_k
\vert\biggr)^{\dagger}
= \sum_{k = 1}^m \bigl(\vert\psi_k\rangle\langle\psi_k\vert\bigr)^{\dagger}
= \sum_{k = 1}^m \vert \psi_k\rangle \langle \psi_k \vert
= \Pi,
\]
and
\begin{multline*}
  \quad\qquad  \Pi^2
  = \biggl( \sum_{j = 1}^m \vert \psi_j\rangle \langle \psi_j
  \vert\Bigr)\Bigl(\sum_{k = 1}^m \vert \psi_k\rangle \langle \psi_k
  \vert\biggr)\\ 
  = \sum_{j = 1}^m\sum_{k = 1}^m \vert \psi_j\rangle \langle \psi_j \vert
  \psi_k\rangle \langle \psi_k \vert 
  = \sum_{k = 1}^m \vert \psi_k\rangle \langle \psi_k \vert
  = \Pi,\qquad\quad
\end{multline*}
where the orthonormality of
$\{\vert \psi_1\rangle,\ldots,\vert\psi_m\rangle\}$ implies the second-to-last
equality.

In fact, this exhausts all of the possibilities: \emph{every} projection $\Pi$
can be written in the form \eqref{eq:projection} for some choice of an
orthonormal set $\{\vert \psi_1\rangle,\ldots,\vert \psi_m\rangle\}.$
(Technically speaking, the zero matrix $\Pi=0,$ which is a projection, is a
special case.
To fit it into the general form \eqref{eq:projection} we must allow the
possibility that the sum is empty, resulting in the zero matrix.)

\subsubsection{Projective measurements}

The notion of a measurement of a quantum system is more general than just
standard basis measurements.
\emph{Projective measurements} are measurements that are described by a
collection of projections whose sum is equal to the identity matrix.
In symbols, a collection $\{\Pi_0,\ldots,\Pi_{m-1}\}$ of projection matrices
describes a projective measurement if
\[
\Pi_0 + \cdots + \Pi_{m-1} = \mathbb{I}.
\]
When such a measurement is performed on a system $\mathsf{X}$ while it is in
some state $\vert\psi\rangle,$ two things happen:
\begin{enumerate}
\item
  For each $k\in\{0,\ldots,m-1\},$ the outcome of the measurement is $k$ with
  probability equal to
  \[
  \operatorname{Pr}\bigl(\text{outcome is $k$}\bigr) = \bigl\| \Pi_k \vert \psi
  \rangle \bigr\|^2.
  \]

\item
  For whichever outcome $k$ the measurement produces, the state of $\mathsf{X}$
  becomes
  \[
  \frac{\Pi_k \vert\psi\rangle}{\bigl\|\Pi_k \vert\psi\rangle\bigr\|}.
  \]
\end{enumerate}

We can also choose outcomes other than $\{0,\ldots,m-1\}$ for projective
measurements if we wish.
More generally, for any finite and nonempty set $\Sigma,$ if we have a
collection of projection matrices $\{\Pi_a:a\in\Sigma\}$ that satisfies the
condition
\[
\sum_{a\in\Sigma} \Pi_a = \mathbb{I},
\]
then this collection describes a projective measurement whose possible outcomes
coincide with the set $\Sigma,$ where the rules are the same as before:
\begin{enumerate}
\item
  For each $a\in\Sigma,$ the outcome of the measurement is $a$ with probability
  equal to
  \[
  \operatorname{Pr}\bigl(\text{outcome is $a$}\bigr) = \bigl\| \Pi_a \vert \psi
  \rangle \bigr\|^2.
  \]

\item
  For whichever outcome $a$ the measurement produces, the state of $\mathsf{X}$
  becomes
  \[
  \frac{\Pi_a \vert\psi\rangle}{\bigl\|\Pi_a \vert\psi\rangle\bigr\|}.
  \]
\end{enumerate}

For example, standard basis measurements are equivalent to projective
measurements for which $\Sigma$ is the set of classical states of whatever
system $\mathsf{X}$ we're talking about and our set of projection matrices is
$\{\vert a\rangle\langle a\vert:a\in\Sigma\}.$

Another example of a projective measurement, this time on two qubits
$(\mathsf{X},\mathsf{Y}),$ is given by the set $\{\Pi_0,\Pi_1\},$ where
\[
\Pi_0 = \vert \phi^+\rangle\langle \phi^+ \vert + \vert \phi^-\rangle\langle
\phi^- \vert + \vert \psi^+\rangle\langle \psi^+ \vert
\quad\text{and}\quad
\Pi_1 = \vert\psi^-\rangle\langle\psi^-\vert.
\]

If we have multiple systems that are jointly in some quantum state and a
projective measurement is performed on just one of the systems, the action is
similar to what we had for standard basis measurements --- and in fact we can
now describe this action in much simpler terms than we could before.

To be precise, let us suppose that we have two systems
$(\mathsf{X},\mathsf{Y})$ in a quantum state $\vert\psi\rangle,$ and a
projective measurement described by a collection $\{\Pi_a : a\in\Sigma\}$ is
performed on the system $\mathsf{X},$ while nothing is done to $\mathsf{Y}.$
Doing this is then equivalent to performing the projective measurement
described by the collection
\[
\bigl\{ \Pi_a \otimes \mathbb{I} \,:\, a\in\Sigma\bigr\}
\]
on the joint system $(\mathsf{X},\mathsf{Y}).$
Each measurement outcome $a$ results with probability
\[
\bigl\| (\Pi_a \otimes \mathbb{I})\vert \psi\rangle \bigr\|^2,
\]
and conditioned on the result $a$ appearing, the state of the joint system
$(\mathsf{X},\mathsf{Y})$ becomes
\[
\frac{(\Pi_a \otimes \mathbb{I})\vert \psi\rangle}{\bigl\| (\Pi_a \otimes
  \mathbb{I})\vert \psi\rangle \bigr\|}.
\]

\subsubsection{Implementing projective measurements}

Arbitrary projective measurements can be implemented using unitary operations,
standard basis measurements, and an extra workspace system, as will now be
explained.

Let us suppose that $\mathsf{X}$ is a system and $\{\Pi_0,\ldots,\Pi_{m-1}\}$
is a projective measurement on $\mathsf{X}.$ We can easily generalize this
discussion to projective measurements having different sets of outcomes, but in
the interest of convenience and simplicity we'll assume the set of possible
outcomes for our measurement is $\{0,\ldots,m-1\}.$

Let us note explicitly that $m$ is not necessarily equal to the number of
classical states of $\mathsf{X}$ --- we'll let $n$ be the number of classical
states of $\mathsf{X},$ which means that each matrix $\Pi_k$ is an $n\times n$
projection matrix.

Because we assume that $\{\Pi_0\ldots,\Pi_{m-1}\}$ represents a projective
measurement, it is necessarily the case that
\[
\sum_{k = 0}^{m-1} \Pi_k = \mathbb{I}.
\]
Our goal is to perform a process that has the same effect as performing this
projective measurement on $\mathsf{X},$ but to do this using only unitary
operations and standard basis measurements.

We will make use of an extra workspace system $\mathsf{Y}$ to do this, and
specifically we'll take the classical state set of $\mathsf{Y}$ to be
$\{0,\ldots,m-1\},$ which is the same as the set of outcomes of the projective
measurement.
The idea is that we will perform a standard basis measurement on $\mathsf{Y},$
and interpret the outcome of this measurement as being equivalent to the
outcome of the projective measurement on $\mathsf{X}.$
We'll need to assume that $\mathsf{Y}$ is initialized to some fixed state,
which we'll choose to be $\vert 0\rangle.$
(Any other choice of fixed quantum state vector could be made to work, but
choosing $\vert 0\rangle$ makes the explanation to follow much simpler.)

Of course, in order for a standard basis measurement of $\mathsf{Y}$ to tell us
anything about $\mathsf{X},$ we will need to allow $\mathsf{X}$ and
$\mathsf{Y}$ to interact somehow before measuring $\mathsf{Y},$ by performing a
unitary operation on the system $(\mathsf{Y},\mathsf{X}).$
First consider this matrix:
\[
M = \sum_{k = 0}^{m-1} \vert k \rangle \langle 0 \vert \otimes \Pi_k.
\]
Expressed explicitly as a so-called \emph{block matrix}, which is a matrix of
matrices that we interpret as a single, larger matrix, $M$ looks like this:
\[
M =
\begin{pmatrix}
  \Pi_0 & 0 & \cdots & 0\\[1mm]
  \Pi_1 & 0 & \cdots & 0\\[1mm]
  \vdots & \vdots & \ddots & \vdots\\[1mm]
  \Pi_{m-1} & 0 & \cdots & 0
\end{pmatrix}.
\]
Here, each $0$ represents an $n\times n$ matrix filled entirely with zeros, so
that the entire matrix $M$ is an $nm\times nm$ matrix.

Now, $M$ is certainly not a unitary matrix (unless $m=1,$ in which case $\Pi_0
= \mathbb{I},$ giving $M = \mathbb{I}$ in this trivial case) because unitary
matrices cannot have any columns (or rows) that are entirely $0;$ unitary
matrices have columns that form orthonormal bases, and the all-zero vector is
not a unit vector.

However, it is the case that the first $n$ columns of $M$ are orthonormal, and
we get this from the assumption that $\{\Pi_0,\ldots,\Pi_{m-1}\}$ is a
measurement.
To verify this claim, notice that for each $j\in\{0,\ldots,n-1\},$ the vector
formed by column number $j$ of $M$ is as follows:
\[
\vert \psi_j\rangle = M \vert 0, j\rangle = \sum_{k = 0}^{m-1} \vert k \rangle
\otimes \Pi_k \vert j\rangle.
\]
Note that here we're numbering the columns starting from column $0.$ Taking the
inner product of column $i$ with column $j$ when $i,j\in\{0,\ldots,n-1\}$ gives
\begin{multline*}
  \langle \psi_i \vert \psi_j \rangle
  =
  \biggl(\sum_{k = 0}^{m-1} \vert k \rangle \otimes \Pi_k \vert
  i\rangle\biggr)^{\dagger}
  \biggl(\sum_{l = 0}^{m-1} \vert l \rangle \otimes \Pi_l \vert j\rangle\biggr)
  =
  \sum_{k = 0}^{m-1} \sum_{l = 0}^{m-1}
  \langle k \vert l \rangle \langle i \vert \Pi_k \Pi_l \vert j\rangle\\
  =
  \sum_{k = 0}^{m-1}
  \langle i \vert \Pi_k \Pi_k \vert j\rangle
  =
  \sum_{k = 0}^{m-1}
  \langle i \vert \Pi_k \vert j\rangle
  = \langle i \vert \mathbb{I} \vert j \rangle
  = \begin{cases}
    1 & i = j\\
    0 & i\neq j,
  \end{cases}
\end{multline*}
which is what we needed to show.

Thus, because the first $n$ columns of the matrix $M$ are orthonormal, we can
replace all of the remaining zero entries by some different choice of complex
number entries so that the entire matrix is unitary.
\[
U = \begin{pmatrix}
  \Pi_0 & \fbox{?} & \cdots & \fbox{?}\\[1mm]
  \Pi_1 & \fbox{?} & \cdots & \fbox{?}\\[1mm]
  \vdots & \vdots & \ddots & \vdots\\[1mm]
  \Pi_{m-1} & \fbox{?} & \cdots & \fbox{?}
\end{pmatrix}
\]
If we're given the matrices $\Pi_0,\ldots,\Pi_{m-1},$ we can compute suitable
matrices to fill in for the blocks marked $\fbox{?}$ --- using
the Gram--Schmidt process --- but it does not matter specifically what these
matrices are for the sake of this discussion.

Finally we can describe the measurement process: we first perform $U$ on the
joint system $(\mathsf{Y},\mathsf{X})$ and then measure $\mathsf{Y}$ with
respect to a standard basis measurement.
For an arbitrary state $\vert \phi \rangle$ of $\mathsf{X},$ we obtain the
state
\[
U \bigl( \vert 0\rangle \vert \phi\rangle\bigr)
= M \bigl( \vert 0\rangle \vert \phi\rangle\bigr)
= \sum_{k = 0}^{m-1} \vert k\rangle \otimes \Pi_k \vert\phi\rangle,
\]
where the first equality follows from the fact that $U$ and $M$ agree on their
first $n$ columns.
When we perform a projective measurement on $\mathsf{Y},$ we obtain each
outcome $k$ with probability
\[
\bigl\| \Pi_k \vert \phi\rangle \bigr\|^2,
\]
in which case the state of $(\mathsf{Y},\mathsf{X})$ becomes
\[
\vert k\rangle \otimes \frac{\Pi_k \vert \phi\rangle}{\bigl\| \Pi_k \vert
  \phi\rangle \bigr\|}.
\]
Thus, $\mathsf{Y}$ stores a copy of the measurement outcome and $\mathsf{X}$
changes precisely as it would had the projective measurement described by
$\{\Pi_0,\ldots,\Pi_{m-1}\}$ been performed directly on $\mathsf{X}.$

\section{Limitations on quantum information}

Despite sharing a common underlying mathematical structure, quantum and
classical information have key differences.
As a result, there are many examples of tasks that quantum information allows
but classical information does not.

Before exploring some of these examples, however, we'll take note of some
important limitations on quantum information.
Understanding things quantum information can't do helps us identify the things
it can do.

\subsection{Irrelevance of global phases}

The first limitation we'll cover --- which is really more of a slight
degeneracy in the way that quantum states are represented by quantum state
vectors, as opposed to an actual limitation --- concerns the notion of a
\emph{global phase}.

What we mean by a global phase is this.
Let $\vert \psi \rangle$ and $\vert \phi \rangle$ be unit vectors representing
quantum states of some system, and suppose that there exists a complex number
$\alpha$ on the unit circle, meaning that $\vert \alpha \vert = 1,$ or
alternatively $\alpha = e^{i\theta}$ for some real number $\theta,$ such that
\[
\vert \phi \rangle = \alpha \vert \psi \rangle.
\]
The vectors $\vert \psi \rangle$ and $\vert \phi \rangle$ are then said to
\emph{differ by a global phase}.
We also sometimes refer to $\alpha$ as a \emph{global phase}, although this is
context-dependent;
any number on the unit circle can be thought of as a global phase when
multiplied to a unit vector.

Consider what happens when a system is in one of the two quantum states
$\vert\psi\rangle$ and $\vert\phi\rangle,$ and the system undergoes a standard
basis measurement.
In the first case, in which the system is in the state $\vert\psi\rangle,$ the
probability of measuring any given classical state $a$ is
\[
\bigl\vert \langle a \vert \psi \rangle \bigr\vert^2.
\]
In the second case, in which the system is in the state $\vert\phi\rangle,$ the
probability of measuring any classical state $a$ is
\[
\bigl\vert \langle a \vert \phi \rangle \bigr\vert^2
= \bigl\vert \alpha \langle a \vert \psi \rangle \bigr\vert^2
= \vert \alpha \vert^2 \bigl\vert \langle a \vert \psi \rangle \bigr\vert^2
= \bigl\vert \langle a \vert \psi \rangle \bigr\vert^2,
\]
because $\vert\alpha\vert = 1.$
That is, the probability of an outcome appearing is the same for both states.

Now consider what happens when we apply an arbitrary unitary operation $U$ to
both states.
In the first case, in which the initial state is $\vert \psi \rangle,$ the
state becomes
\[
U \vert \psi \rangle,
\]
and in the second case, in which the initial state is $\vert \phi\rangle,$ it
becomes
\[
U \vert \phi \rangle = \alpha U \vert \psi \rangle.
\]
That is, the two resulting states still differ by the same global phase
$\alpha.$

Consequently, two quantum states $\vert\psi\rangle$ and $\vert\phi\rangle$ that
differ by a global phase are completely indistinguishable;
no matter what operation, or sequence of operations, we apply to the two
states, they will always differ by a global phase, and performing a standard
basis measurement will produce outcomes with precisely the same probabilities
as the other.
For this reason, two quantum state vectors that differ by a global phase are
considered to be equivalent, and are effectively viewed as being the same
state.

For example, the quantum states
\[
\vert - \rangle = \frac{1}{\sqrt{2}} \vert 0 \rangle - \frac{1}{\sqrt{2}} \vert
1 \rangle
\quad\text{and}\quad
-\vert - \rangle = -\frac{1}{\sqrt{2}} \vert 0 \rangle + \frac{1}{\sqrt{2}}
\vert 1 \rangle
\]
differ by a global phase (which is $-1$ in this example), and are therefore
considered to be the same state.

On the other hand, the quantum states
\[
\vert + \rangle = \frac{1}{\sqrt{2}} \vert 0 \rangle + \frac{1}{\sqrt{2}} \vert
1 \rangle
\quad\text{and}\quad
\vert - \rangle = \frac{1}{\sqrt{2}} \vert 0 \rangle - \frac{1}{\sqrt{2}} \vert
1 \rangle
\]
do not differ by a global phase.
Although the only difference between the two states is that a plus sign turns
into a minus sign, this is not a \emph{global} phase difference, it is a
\emph{relative} phase difference because it does not affect every vector entry,
but only a proper subset of the entries.
This is consistent with what we have already observed previously, which is that
the states $\vert{+} \rangle$ and $\vert{-}\rangle$ can be discriminated
perfectly.
In particular, performing a Hadamard operation and then measuring yields
outcome probabilities as follows:
\[
\begin{aligned}
  \bigl\vert \langle 0 \vert H \vert {+} \rangle \bigr\vert^2 = 1 & \hspace{1cm}
  \bigl\vert \langle 0 \vert H \vert {-} \rangle \bigr\vert^2 = 0 \\[1mm]
  \bigl\vert \langle 1 \vert H \vert {+} \rangle \bigr\vert^2 = 0 & \hspace{1cm}
  \bigl\vert \langle 1 \vert H \vert {-} \rangle \bigr\vert^2 = 1.
\end{aligned}
\]

\subsection{No-cloning theorem}

The \emph{no-cloning theorem} shows that it is impossible to create a perfect
copy of an unknown quantum state.

\begin{callout}[title={No-cloning theorem}]
  Let $\Sigma$ be a classical state set having at least two elements, and let
  $\mathsf{X}$ and $\mathsf{Y}$ be systems sharing the same classical state set
  $\Sigma.$ There does not exist a quantum state $\vert \phi\rangle$ of
  $\mathsf{Y}$ and a unitary operation $U$ on the pair
  $(\mathsf{X},\mathsf{Y})$ such that
  \[
  U \bigl( \vert \psi \rangle \otimes \vert\phi\rangle\bigr) = \vert \psi
  \rangle \otimes \vert\psi\rangle
  \]
  for every state $\vert \psi \rangle$ of $\mathsf{X}.$
\end{callout}

That is, there is no way to initialize the system $\mathsf{Y}$ (to any state
$\vert\phi\rangle$ whatsoever) and perform a unitary operation $U$ on the joint
system $(\mathsf{X},\mathsf{Y})$ so that the effect is for the state
$\vert\psi\rangle$ of $\mathsf{X}$ to be \emph{cloned} --- resulting in
$(\mathsf{X},\mathsf{Y})$ being in the state
$\vert \psi \rangle \otimes \vert\psi\rangle.$

The proof of this theorem is actually quite simple: it boils down to the
observation that the mapping
\[
\vert\psi\rangle \otimes \vert \phi\rangle\mapsto\vert\psi\rangle \otimes \vert
\psi\rangle
\]
is not linear in $\vert\psi\rangle.$

In detail, because $\Sigma$ has at least two elements, we may choose
$a,b\in\Sigma$ with $a\neq b.$
If there did exist a quantum state $\vert \phi\rangle$ of $\mathsf{Y}$ and a
unitary operation $U$ on the pair
$(\mathsf{X},\mathsf{Y})$ for which $ U \bigl( \vert \psi \rangle \otimes
\vert\phi\rangle\bigr) = \vert \psi \rangle \otimes \vert\psi\rangle $ for
every quantum state $\vert\psi\rangle$ of $\mathsf{X},$ then it would be the
case that
\[
U \bigl( \vert a \rangle \otimes \vert\phi\rangle\bigr)
= \vert a \rangle \otimes \vert a\rangle
\quad\text{and}\quad
U \bigl( \vert b \rangle \otimes \vert\phi\rangle\bigr)
= \vert b \rangle \otimes \vert b\rangle.
\]
By linearity, meaning specifically the linearity of the tensor product in the
first argument and the linearity of matrix-vector multiplication in the second
(vector) argument, we must therefore have
\[
U \biggl(\biggl( \frac{1}{\sqrt{2}}\vert a \rangle + \frac{1}{\sqrt{2}} \vert
b\rangle \biggr) \otimes \vert\phi\rangle\biggr)
= \frac{1}{\sqrt{2}} \vert a \rangle \otimes \vert a\rangle
+ \frac{1}{\sqrt{2}} \vert b \rangle \otimes \vert b\rangle.
\]
However, the requirement that
$U \bigl( \vert \psi \rangle \otimes \vert\phi\rangle\bigr) = \vert \psi
\rangle \otimes \vert\psi\rangle$
for every quantum state $\vert\psi\rangle$ demands that
\[
\begin{aligned}
  & U \biggl(\biggl( \frac{1}{\sqrt{2}}\vert a \rangle + \frac{1}{\sqrt{2}}
  \vert b\rangle \biggr)
  \otimes \vert\phi\rangle\biggr)\\
  & \qquad = \biggl(\frac{1}{\sqrt{2}} \vert a \rangle + \frac{1}{\sqrt{2}}
  \vert b \rangle\biggr)
  \otimes \biggl(\frac{1}{\sqrt{2}} \vert a \rangle + \frac{1}{\sqrt{2}} \vert
  b \rangle\biggr)\\
  & \qquad = \frac{1}{2} \vert a \rangle \otimes \vert a\rangle
  + \frac{1}{2} \vert a \rangle \otimes \vert b\rangle
  + \frac{1}{2} \vert b \rangle \otimes \vert a\rangle
  + \frac{1}{2} \vert b \rangle \otimes \vert b\rangle\\
  & \qquad \neq \frac{1}{\sqrt{2}} \vert a \rangle \otimes \vert a\rangle
  + \frac{1}{\sqrt{2}} \vert b \rangle \otimes \vert b\rangle.
\end{aligned}
\]
Therefore there cannot exist a state $\vert \phi\rangle$ and a unitary
operation $U$ for which
\[
U \bigl( \vert \psi \rangle \otimes \vert\phi\rangle\bigr)
= \vert \psi \rangle \otimes \vert\psi\rangle
\]
for every quantum state vector $\vert \psi\rangle.$

A few remarks concerning the no-cloning theorem are in order.
The first one is that the statement of the no-cloning theorem above is
absolute, in the sense that it states that \emph{perfect} cloning is impossible
--- but it does not say anything about possibly cloning with limited accuracy,
where we might succeed in producing an approximate clone (with respect to some
way of measuring how similar two different quantum states might be).
There are, in fact, statements of the no-cloning theorem that place limitations
on approximate cloning, as well as methods to achieve approximate cloning with
limited accuracy.

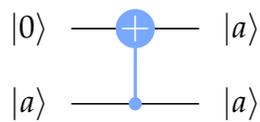
\begin{figure}[!ht]
  \begin{center}
    \begin{tikzpicture}[
        scale=1,
        line width=0.6pt,          
        gate/.style={%
          inner sep = 0,
          fill = CircuitBlue,
          draw = CircuitBlue,
          text = white,
          minimum size = 10mm},
        control/.style={%
          circle,
          fill=CircuitBlue,
          minimum size = 5pt,
          inner sep=0mm},
        not/.style={%
          circle,
          fill = CircuitBlue,
          draw = CircuitBlue,
          text = white,
          inner sep=0,
          minimum size=5mm,
          label = {center:\textcolor{white}{\large $+$}}
        }
      ]
      
      \node (In1) at (-1,1) {};
      \node (In2) at (-1,0) {};
      \node (Out1) at (1,1) {};
      \node (Out2) at (1,0) {};
      
      \draw (In1) -- (Out1);
      \draw (In2) -- (Out2);
      
      \node[control] (Control) at (0,0) {};
      \node[not] (Not) at (0,1) {};
      \draw[very thick,draw=CircuitBlue] (Control.center) --
      ([yshift=0.1pt]Not.south);
      
      \node[anchor=east] at (In1) {$\vert 0 \rangle$};
      \node[anchor=east] at (In2) {$\vert a \rangle$};
      \node[anchor=west] at (Out1) {$\vert a \rangle$};
      \node[anchor=west] at (Out2) {$\vert a \rangle$};
      
    \end{tikzpicture}
  \end{center}
  \caption{A quantum circuit for copying a standard basis state.}
  \label{fig:cNOT-copy}
\end{figure}

The second remark is that the no-cloning theorem is a statement about the
impossibility of cloning an \emph{arbitrary} state $\vert\psi\rangle.$
In contrast, we can easily create a clone of any standard basis state, for
instance.
For example, we can clone a qubit standard basis state using a controlled-NOT
operation as shown in Figure~\ref{fig:cNOT-copy}.
While there is no difficulty in creating a clone of a standard basis state,
this does not contradict the no-cloning theorem.
This approach of using a controlled-NOT gate would not succeed in creating a
clone of the state $\vert + \rangle,$ for instance.

One final remark about the no-cloning theorem is that it really isn't unique to
quantum information --- it's also impossible to clone an arbitrary
probabilistic state using a classical (deterministic or probabilistic) process.
Imagine someone hands you a system in some probabilistic state, but you're not
sure what that probabilistic state is.
For example, maybe they randomly generated a number between $1$ and $10,$ but
they didn't tell you how they generated that number.
There's certainly no physical process through which you can obtain two
\emph{independent} copies of that same probabilistic state: all you have in
your hands is a number between $1$ and $10,$ and there just isn't enough
information present for you to somehow reconstruct the probabilities for all of
the other outcomes to appear.

Mathematically speaking, a version of the no-cloning theorem for probabilistic
states can be proved in exactly the same way as the regular no-cloning theorem
(for quantum states).
That is, cloning an arbitrary probabilistic state is a non-linear process, so
it cannot possibly be represented by a stochastic matrix.

\subsection{Non-orthogonal states cannot be perfectly discriminated}

For the final limitation to be covered in this lesson, we'll show that if we
have two quantum states $\vert\psi\rangle$ and $\vert\phi\rangle$ that are not
orthogonal, which means that $\langle \phi\vert\psi\rangle \neq 0,$ then it's
impossible to discriminate them (or, in other words, to tell them apart)
perfectly.
In fact, we'll show something logically equivalent: if we do have a way to
discriminate two states perfectly, without any error, then they must be
orthogonal.

\begin{figure}[!ht]
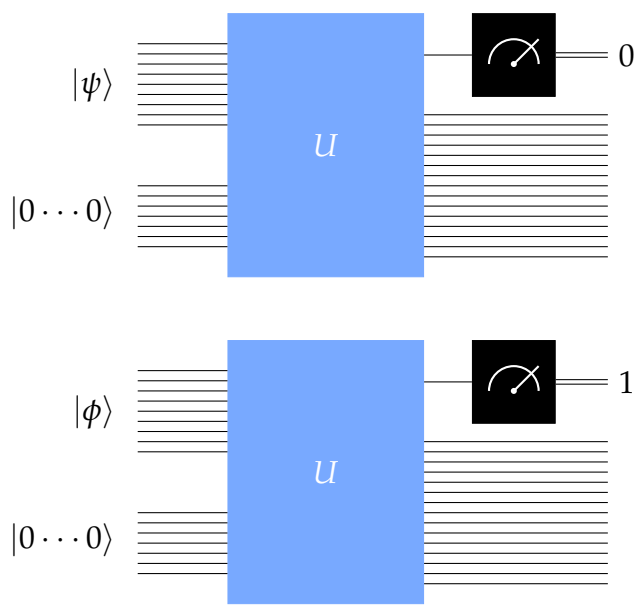

  \begin{center}
    \vspace{1mm}
%
  \end{center}

  \caption{A quantum circuit $U$ perfectly discriminates the states
    $\vert\psi\rangle$ and $\vert\phi\rangle$.}
  \label{fig:discriminate}
\end{figure}

We'll restrict our attention to quantum circuits that consist of any number of
unitary gates, followed by a single standard basis measurement of the top
qubit.
What we require of a quantum circuit, to say that it perfectly discriminates
the states $\vert\psi\rangle$ and $\vert\phi\rangle,$ is that the measurement
always yields the value $0$ for one of the two states and always yields $1$ for
the other state.
To be precise, we shall assume that we have a quantum circuit that operates as
Figure~\ref{fig:discriminate} suggests.

The box labeled $U$ denotes the unitary operation representing the combined
action of all of the unitary gates in our circuit, but not including the final
measurement.
There is no loss of generality in assuming that the measurement outputs $0$ for
$\vert\psi\rangle$ and $1$ for $\vert\phi\rangle;$ the analysis would not
differ fundamentally if these output values were reversed.

Notice that, in addition to the qubits that initially store either
$\vert\psi\rangle$ or $\vert\phi\rangle,$ the circuit is free to make use of
any number of additional \emph{workspace} qubits.
These qubits are initially each set to the $\vert 0\rangle$ state --- so their
combined state is denoted $\vert 0\cdots 0\rangle$ in the figures --- and these
qubits can be used by the circuit in any way that might be beneficial.
It is very common to make use of workspace qubits in quantum circuits like
this.

Now, consider what happens when we run our circuit on the state
$\vert\psi\rangle$ (along with the initialized workspace qubits).
The resulting state, immediately prior to the measurement being performed, can
be written as
\[
U \bigl(  \vert 0\cdots 0 \rangle \vert \psi \rangle\bigr)
= \vert \gamma_0\rangle\vert 0 \rangle + \vert \gamma_1 \rangle\vert 1 \rangle
\]
for two vectors $\vert \gamma_0\rangle$ and $\vert \gamma_1\rangle$ that
correspond to all of the qubits except the top qubit.
In general, for such a state the probabilities that a measurement of the top
qubit yields the outcomes $0$ and $1$ are as follows:
\[
\operatorname{Pr}(\text{outcome is $0$}) = \bigl\| \vert\gamma_0\rangle
\bigr\|^2
\qquad\text{and}\qquad
\operatorname{Pr}(\text{outcome is $1$}) = \bigl\| \vert\gamma_1\rangle
\bigr\|^2.
\]
Because our circuit always outputs $0$ for the state $\vert\psi\rangle,$ it
must be that $\vert\gamma_1\rangle = 0,$ and so
\[
U \bigl( \vert 0\cdots 0\rangle\vert \psi \rangle  \bigr)
= \vert\gamma_0\rangle\vert 0 \rangle.
\]
Multiplying both sides of this equation by $U^{\dagger}$ yields this equation:
\begin{equation}
  \vert 0\cdots 0\rangle\vert \psi \rangle
  = U^{\dagger} \bigl( \vert \gamma_0\rangle\vert 0 \rangle \bigr).
  \label{eq:distinguish-equation-1}
\end{equation}

Reasoning similarly for $\vert\phi\rangle$ in place of $\vert\psi\rangle,$ we
conclude that
\[
U \bigl( \vert 0\cdots 0\rangle\vert \phi \rangle  \bigr)
=  \vert \delta_1\rangle\vert 1 \rangle
\]
for some vector $\vert\delta_1\rangle,$ and therefore
\begin{equation}
  \vert 0\cdots 0\rangle\vert \phi \rangle
  = U^{\dagger} \bigl(  \vert \delta_1\rangle\vert 1 \rangle\bigr).
  \label{eq:distinguish-equation-2}
\end{equation}

Now let us take the inner product of the vectors represented by the equations
\eqref{eq:distinguish-equation-1} and \eqref{eq:distinguish-equation-2},
starting with the representations on the right-hand side of each equation.
We have
\[
\bigl(U^{\dagger} \bigl( \vert \gamma_0\rangle\vert 0 \rangle
\bigr)\bigr)^{\dagger}
= \bigl( \langle\gamma_0\vert\langle 0\vert \bigr)U,
\]
so the inner product of the vector \eqref{eq:distinguish-equation-1}
with the vector \eqref{eq:distinguish-equation-2} is
\[
\bigl( \langle\gamma_0\vert\langle 0\vert \bigr)U U^{\dagger} \bigl(  \vert
\delta_1\rangle\vert 1 \rangle\bigr)
= \bigl( \langle\gamma_0\vert\langle 0\vert \bigr) \bigl(  \vert
\delta_1\rangle\vert 1 \rangle\bigr)
=  \langle \gamma_0 \vert \delta_1\rangle \langle 0 \vert 1 \rangle = 0.
\]
Here we have used the fact that $U U^{\dagger} = \mathbb{I},$ as well as the
fact that the inner product of tensor products is the product of the inner
products:
\[
\langle u \otimes v \vert w \otimes x\rangle = \langle u \vert w\rangle \langle
v \vert x\rangle
\]
for any choices of these vectors (assuming $\vert u\rangle$ and $\vert
w\rangle$ have the same number of entries
and $\vert v\rangle$ and $\vert x\rangle$ have the same number of entries, so
that it makes sense to form the inner products $\langle u\vert w\rangle$ and
$\langle v\vert x \rangle$).
Notice that the value of the inner product $\langle \gamma_0 \vert
\delta_1\rangle$ is irrelevant because it is multiplied by $\langle 0 \vert 1
\rangle = 0.$

Finally, taking the inner product of the vectors on the left-hand sides of the
equations \eqref{eq:distinguish-equation-1} and
\eqref{eq:distinguish-equation-2} must result in the same zero value that we've
already calculated, so
\[
0 = \bigl(  \vert 0\cdots 0\rangle \vert \psi\rangle\bigr)^{\dagger}
\bigl(\vert 0\cdots 0\rangle\vert \phi\rangle\bigr)
=  \langle 0\cdots 0 \vert 0\cdots 0 \rangle \langle \psi \vert \phi \rangle =
\langle \psi \vert \phi \rangle.
\]
We have therefore concluded what we wanted, which is that $\vert \psi\rangle$
and $\vert\phi\rangle$ are orthogonal:
$\langle \psi \vert \phi \rangle = 0.$

It is possible, by the way, to perfectly discriminate any two states that are
orthogonal, which is the converse to the statement we just proved.
Suppose that the two states to be discriminated are $\vert \phi\rangle$ and
$\vert \psi\rangle,$ where $\langle \phi\vert\psi\rangle = 0.$
We can then perfectly discriminate these states by performing the projective
measurement described by these matrices, for instance:
\[
\bigl\{
\vert\phi\rangle\langle\phi\vert,\,\mathbb{I} - \vert\phi\rangle\langle\phi\vert
\bigr\}.
\]
For the state $\vert\phi\rangle,$ the first outcome is always obtained:
\[
\begin{aligned}
  & \bigl\| \vert\phi\rangle\langle\phi\vert \vert\phi\rangle \bigr\|^2 =
  \bigl\| \vert\phi\rangle\langle\phi\vert\phi\rangle \bigr\|^2 =
  \bigl\| \vert\phi\rangle \bigr\|^2 = 1,\\[1mm]
  & \bigl\| (\mathbb{I} - \vert\phi\rangle\langle\phi\vert) \vert\phi\rangle
  \bigr\|^2 =
  \bigl\| \vert\phi\rangle - \vert\phi\rangle\langle\phi\vert\phi\rangle
  \bigr\|^2 =
  \bigl\| \vert\phi\rangle - \vert\phi\rangle \bigr\|^2 = 0.
\end{aligned}
\]
And, for the state $\vert\psi\rangle,$ the second outcome is always obtained:
\[
\begin{aligned}
  & \bigl\| \vert\phi\rangle\langle\phi\vert \vert\psi\rangle \bigr\|^2 =
  \bigl\| \vert\phi\rangle\langle\phi\vert\psi\rangle \bigr\|^2 =
  \bigl\| 0 \bigr\|^2 = 0,\\[1mm]
  & \bigl\| (\mathbb{I} - \vert\phi\rangle\langle\phi\vert) \vert\psi\rangle
  \bigr\|^2 =
  \bigl\| \vert\psi\rangle - \vert\phi\rangle\langle\phi\vert\psi\rangle
  \bigr\|^2 =
  \bigl\| \vert\psi\rangle \bigr\|^2 = 1.
\end{aligned}
\]
More generally, any orthogonal collection of quantum state vectors can be
discriminated perfectly.


\lesson{Entanglement in Action}
\label{lesson:entanglement-in-action}

In this lesson we'll take a look at three fundamentally important examples.
The first two are the \emph{quantum teleportation} and \emph{superdense coding}
protocols, which are principally concerned with the transmission of information
from a sender to a receiver.
The third example is an abstract game, called the \emph{CHSH game}, which
illustrates a phenomenon in quantum information that is sometimes referred to
as \emph{nonlocality}.
(The CHSH game is not always described as a game.
It is often described instead as an experiment --- specifically, it is an
example of a \emph{Bell test} --- and is referred to as the \emph{CHSH
inequality}.)

Quantum teleportation, superdense coding, and the CHSH game are not merely
examples meant to illustrate how quantum information works, although they do
serve well in this regard.
Rather, they are stones in the foundation of quantum information.
Entanglement plays a key role in all three examples, so this lesson provides
the first opportunity in this course to see entanglement in action, and to
begin to explore what it is that makes entanglement such an interesting and
important concept.

Before proceeding to the examples themselves, a few preliminary comments that
connect to all three examples are in order.

\subsection{Alice and Bob}

\emph{Alice} and \emph{Bob} are names traditionally given to hypothetical
entities or agents in systems, protocols, games, and other interactions that
involve the exchange of information.
While these are human names, it should be understood that they represent
abstractions and not necessarily actual human beings --- so Alice and Bob might
be expected to perform complex computations, for instance.

These names were first used in this way in the 1970s in the context of
cryptography, but the convention has become common more broadly since then.
The idea is simply that these are common names (at least in some parts of the
world) that start with the letters A and B.
It is also quite convenient to refer to Alice with the pronoun \emph{her} and
Bob with the pronoun \emph{him} for the sake of brevity.

By default, we imagine that Alice and Bob are in different locations.
They may have different goals and behaviors depending on the context in which
they arise.
For example, in the context of \emph{communication}, meaning the transmission
of information, we might decide to use the name Alice to refer to the sender
and Bob to refer to the receiver of whatever information is transmitted.
In general, it may be that Alice and Bob cooperate, which is typical of a wide
range of settings --- but in other settings they may be in competition, or they
may have different goals that may or may not be consistent or harmonious.
These things must be made clear in the situation at hand.

We can also introduce additional characters, such as \emph{Charlie} and
\emph{Diane}, as needed.
Other names that represent different personas, such as \emph{Eve} for an
eavesdropper or \emph{Mallory} for someone behaving maliciously, are also
sometimes used.

\subsection{Entanglement as a resource}

Recall this example of an entangled quantum state of two qubits:
\begin{equation}
  \vert \phi^+ \rangle =
  \frac{1}{\sqrt{2}} \vert 00\rangle + \frac{1}{\sqrt{2}} \vert 11\rangle.
  \label{eq:e-bit}
\end{equation}
It is one of the four Bell states, and is often viewed as the archetypal
example of an entangled quantum state.

We also previously encountered this example of a probabilistic state of two
bits:
\begin{equation}
  \frac{1}{2} \vert 00 \rangle + \frac{1}{2} \vert 11 \rangle.
  \label{eq:sr-bit}
\end{equation}
It is, in some sense, analogous to the entangled quantum state
\eqref{eq:e-bit}.
It represents a probabilistic state in which two bits are correlated, but it is
not entangled.
Entanglement is a uniquely quantum phenomenon, essentially by definition:
in simplified terms, entanglement refers to \emph{non-classical} quantum
correlations.

Unfortunately, defining entanglement as non-classical quantum correlation is
somewhat unsatisfying at an intuitive level, because it's a definition of what
entanglement is in terms of what it is not.
This may be why it's actually rather challenging to explain precisely what
entanglement is, and what makes it special, in intuitive terms.

Typical explanations of entanglement often fail to distinguish the two states
\eqref{eq:e-bit} and \eqref{eq:sr-bit} in a meaningful way.
For example, it is sometimes said that if one of two entangled qubits is
measured, then the state of the other qubit is somehow instantaneously
affected; or that the state of the two qubits together cannot be described
separately; or that the two qubits somehow maintain a memory of each other.
These statements are not false, but why are they not also true for the
(unentangled) probabilistic state \eqref{eq:sr-bit} above?
The two bits represented by this state are intimately connected: each one has a
perfect memory of the other in a literal sense.
But the state is nevertheless not entangled.

One way to explain what makes entanglement special, and what makes the quantum
state \eqref{eq:e-bit} different from the probabilistic state
\eqref{eq:sr-bit}, is to explain what can be done with entanglement, or what we
can see happening because of entanglement, that goes beyond the decisions we
make about how to represent our knowledge of states using vectors.
All three of the examples to be discussed in this lesson have this nature, in
that they illustrate things that can be done with the state \eqref{eq:e-bit}
that cannot be done with \emph{any} classically correlated state, including the
state \eqref{eq:sr-bit}.

Indeed, it is typical in the study of quantum information and computation that
entanglement is viewed as a resource through which different tasks can be
accomplished.
When this is done, the state \eqref{eq:e-bit} is viewed as representing one
\emph{unit} of entanglement, which we refer to as an \emph{e-bit}. 
The ``e'' stands for ``entangled'' or ``entanglement.''
While it is true that the state \eqref{eq:e-bit} is a state of two qubits, the
quantity of entanglement that it represents is one e-bit.

Incidentally, we can also view the probabilistic state \eqref{eq:sr-bit} as a
resource, which is one bit of \emph{shared randomness}.
It can be very useful in cryptography, for instance, to share a random bit with
somebody (presuming that nobody else knows what the bit is), so that it can be
used as a private key, or part of a private key, for the sake of encryption.
But in this lesson the focus is on entanglement and a few things we can do with
it.

As a point of clarification regarding terminology, when we say that Alice and
Bob \emph{share an e-bit}, what we mean is that Alice has a qubit named
$\mathsf{A}$, Bob has a qubit named $\mathsf{B}$, and together the pair
$(\mathsf{A},\mathsf{B})$ is in the quantum state \eqref{eq:e-bit}.
Different names could, of course, be chosen for the qubits, but throughout this
lesson we will stick with these names in the interest of clarity.

\section{Quantum teleportation}

Quantum teleportation, or just teleportation for short, is a protocol where a
sender (Alice) transmits a qubit to a receiver (Bob) by making use of a shared
entangled quantum state (one e-bit, to be specific) along with two bits of
classical communication.
The name \emph{teleportation} is meant to be suggestive of the concept in
science fiction where matter is transported from one location to another by a
futuristic process, but it must be understood that matter is not teleported in
quantum teleportation --- what is actually teleported is quantum information.

The set-up for teleportation is as follows.
We assume that Alice and Bob share an e-bit: Alice holds a qubit $\mathsf{A}$,
Bob holds a qubit $\mathsf{B}$, and together the pair $(\mathsf{A},\mathsf{B})$
is in the state $\vert\phi^+\rangle.$
It could be, for instance, that Alice and Bob were in the same location in the
past, they prepared the qubits $\mathsf{A}$ and $\mathsf{B}$ in the state
$\vert \phi^+ \rangle$, and then each went their own way with their qubit in
hand.
Or, it could be that a different process, such as one involving a third party
or a complex distributed process, was used to establish this shared e-bit.
These details are not part of the teleportation protocol itself.

Alice then comes into possession of a third qubit $\mathsf{Q}$ that she wishes
to transmit to Bob.
The state of the qubit $\mathsf{Q}$ is considered to be \emph{unknown} to Alice
and Bob, and no assumptions are made about it.
For example, the qubit $\mathsf{Q}$ might be entangled with one or more other
systems that neither Alice nor Bob can access.
To say that Alice wishes to transmit the qubit $\mathsf{Q}$ to Bob means that
Alice would like Bob to be holding a qubit that is in the same state that
$\mathsf{Q}$ was in at the start of the protocol, having whatever correlations
that $\mathsf{Q}$ had with other systems, as if Alice had physically handed
$\mathsf{Q}$ to Bob.

We could imagine that Alice physically sends the qubit $\mathsf{Q}$ to Bob, and
if it reaches Bob without being altered or disturbed in transit, then Alice and
Bob's task will be accomplished.
In the context of teleportation, however, it is our assumption that this is not
feasible; Alice cannot send qubits directly to Bob.
She may, however, send classical information to Bob.

These are reasonable assumptions in a variety of settings.
For example, if Alice doesn't know Bob's exact location, or the distance
between them is large, physically sending a qubit using the technology of
today, or the foreseeable future, would be challenging to say the least.
However, as we know from everyday experiences, classical information
transmission under these circumstances is quite straightforward.

At this point, one might ask whether it is possible for Alice and Bob to
accomplish their task without even needing to make use of a shared e-bit.
In other words, is there any way to transmit a qubit using classical
communication alone?

The answer is no, it is not possible to transmit quantum information using
classical communication alone.
This is not too difficult to prove mathematically using basic quantum
information theory, but we can alternatively rule out the possibility of
transmitting qubits using classical communication alone by thinking about the
no-cloning theorem.

Imagine that there was a way to send quantum information using classical
communication alone.
Classical information can easily be copied and broadcast, which means that any
classical transmission from Alice to Bob might also be received by a second
receiver (Charlie, let us say).
But if Charlie receives the same classical communication that Bob received,
then would he not also be able to obtain a copy of the qubit $\mathsf{Q}?$
This would suggest that $\mathsf{Q}$ was cloned, which we already know is
impossible by the no-cloning theorem, and so we conclude that there is no way
to send quantum information using classical communication alone.

When the assumption that Alice and Bob share an e-bit is in place, however, it
is possible for Alice and Bob to accomplish their task.
This is precisely what the quantum teleportation protocol does.

\subsection{Protocol}

\begin{figure}[!ht]
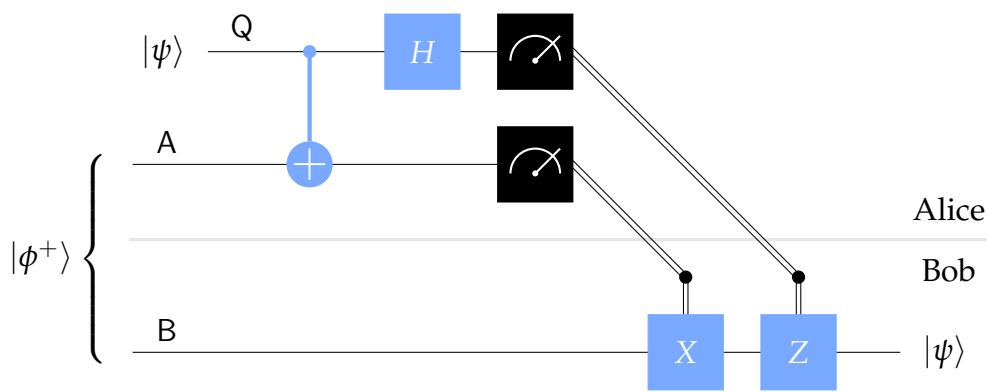

\begin{center}

  
\end{center}
\caption{The quantum teleportation protocol expressed as a quantum circuit.}
\label{fig:teleportation}
\end{figure}

Figure~\ref{fig:teleportation} describes the teleportation protocol as a
quantum circuit.
The diagram is slightly stylized in that it depicts the separation between
Alice and Bob, with two diagonal wires representing classical bits that are
sent from Alice to Bob, but otherwise it is an ordinary quantum circuit
diagram.
The qubit names are shown above the wires rather than to the left so that the
initial states can be shown as well (which we will commonly do when it is
convenient).
It should also be noted that the $X$ and $Z$ gates have \emph{classical}
controls, which simply means that the gates are not applied or applied
depending on whether these classical control bits are $0$ or $1$, respectively.

In words, the teleportation protocol is as follows:
\begin{enumerate}
\item
  Alice performs a controlled-NOT operation on the pair
  $(\mathsf{A},\mathsf{Q})$, with $\mathsf{Q}$ the control and
  $\mathsf{A}$ the target, and then performs a Hadamard operation on
  $\mathsf{Q}.$

\item
  Alice then measures both $\mathsf{A}$ and $\mathsf{Q}$, with respect to a
  standard basis measurement in both cases, and transmits the classical
  outcomes to Bob. Let us refer to the outcome of the measurement of
  $\mathsf{A}$ as $a$ and the outcome of the measurement of $\mathsf{Q}$ as
  $b.$

\item
  Bob receives $a$ and $b$ from Alice, and depending on the values of these
  bits he performs these operations:
  \begin{itemize}
  \item
    If $a = 1$, then Bob performs a bit-flip (or $X$ gate) on his qubit
    $\mathsf{B}.$
  \item
    If $b = 1$, then Bob performs a phase-flip (or $Z$ gate) on his qubit
    $\mathsf{B}.$
  \end{itemize}
  \noindent
  That is, conditioned on $ab$ being $00$, $01$, $10$, or $11$, Bob performs
  one of the operations $\mathbb{I}$, $Z$, $X$, or $ZX$ on the qubit
  $\mathsf{B}.$
\end{enumerate}

This is the complete description of the teleportation protocol.
The analysis that appears below reveals that when it is run, the qubit
$\mathsf{B}$ will be in whatever state $\mathsf{Q}$ was in prior to the
protocol being executed, including whatever correlations it had with any other
systems --- which is to say that the protocol has effectively implemented a
perfect qubit communication channel, where the state of $\mathsf{Q}$ has been
``teleported'' into $\mathsf{B}.$

Before proceeding to the analysis, notice that this protocol does not succeed
in cloning the state of $\mathsf{Q}$, which we already know is impossible by
the no-cloning theorem.
Rather, when the protocol is finished, the state of the qubit $\mathsf{Q}$ will
have changed from its original value to $\vert b\rangle$ as a result of the
measurement performed on it.
Also notice that the e-bit has effectively been ``burned'' in the process: the
state of $\mathsf{A}$ has changed to $\vert a\rangle$ and is no longer
entangled with $\mathsf{B}$ (or any other system).
This is the cost of teleportation.

\subsection{Analysis}

To analyze the teleportation protocol, we'll examine the behavior of the
circuit described above, one step at a time, beginning with the situation in
which $\mathsf{Q}$ is initially in the state $\alpha\vert 0\rangle + \beta\vert
1\rangle.$
This is not the most general situation, as it does not capture the possibility
that $\mathsf{Q}$ is entangled with other systems, but starting with this
simpler case will add clarity to the analysis.
The more general case is addressed below, following the analysis of the simpler
case.

\begin{figure}[t]
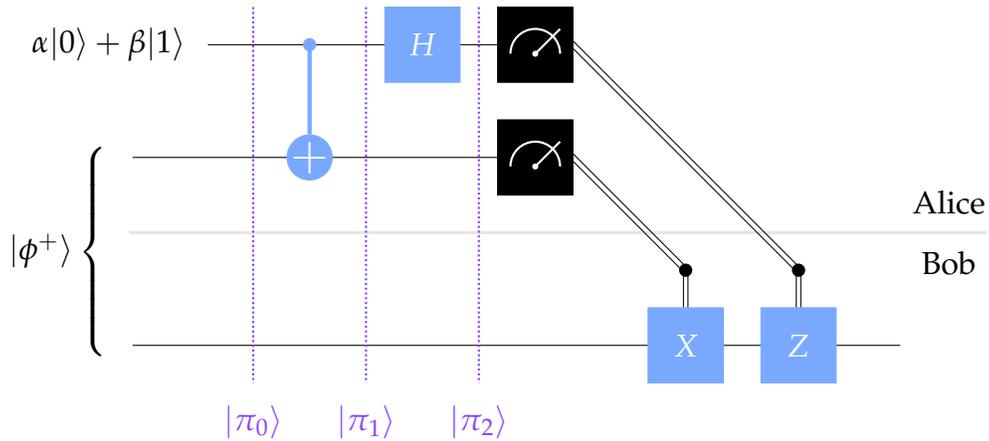

\begin{center}

  
\end{center}
\caption{Three states $\ket{\pi_0}$, $\ket{\pi_1}$, and $\ket{\pi_2}$ relevant
  to the analysis of the teleportation protocol.}
\label{fig:teleportation-time-steps}
\end{figure}

Consider the states of the qubits $(\mathsf{B},\mathsf{A},\mathsf{Q})$ at the
times suggested by Figure~\ref{fig:teleportation-time-steps}.
Under the assumption that the qubit $\mathsf{Q}$ begins the protocol in the
state $\alpha\vert 0\rangle + \beta\vert 1\rangle$, the state of the three
qubits $(\mathsf{B},\mathsf{A},\mathsf{Q})$ together at the start of the
protocol is therefore
\[
\vert \pi_0 \rangle
= \vert \phi^+\rangle \otimes
\bigl(\alpha\vert 0\rangle + \beta\vert 1\rangle \bigr)
= \frac{\alpha \vert 000 \rangle + \alpha \vert 110\rangle + \beta \vert
  001\rangle + \beta \vert 111\rangle}{\sqrt{2}}.
\]
The first gate that is performed is the controlled-NOT gate, which transforms
the state $\vert\pi_0\rangle$ into
\[
\vert \pi_1 \rangle  = \frac{\alpha \vert 000 \rangle + \alpha \vert 110\rangle
  + \beta \vert 011\rangle + \beta \vert 101\rangle}{\sqrt{2}}.
\]
Then the Hadamard gate is applied, which transforms the state
$\vert\pi_1\rangle$ into
\[
\begin{aligned}
\vert\pi_2\rangle
& =
\frac{\alpha \vert 00\rangle \vert + \rangle + \alpha \vert 11\rangle\vert
  +\rangle + \beta \vert 01\rangle\vert -\rangle + \beta \vert 10\rangle\vert
  -\rangle}{\sqrt{2}}\\[2mm]
& = \frac{\alpha \vert 000 \rangle
  + \alpha \vert 001 \rangle
  + \alpha \vert 110 \rangle
  + \alpha \vert 111 \rangle
  + \beta \vert 010 \rangle
  - \beta \vert 011 \rangle
  + \beta \vert 100 \rangle
  - \beta \vert 101 \rangle}{2}.
\end{aligned}
\]
Using the multilinearity of the tensor product, we may alternatively write this
state as follows.
\begin{align*}
  \vert\pi_2\rangle = \quad
  & \frac{1}{2} \bigl(\alpha\vert 0 \rangle + \beta \vert 1\rangle \bigr)\vert
  00\rangle \\[1mm]
  + & \frac{1}{2} \bigl(\alpha\vert 0 \rangle - \beta \vert 1\rangle
  \bigr)\vert 01\rangle \\[1mm]
  + & \frac{1}{2} \bigl(\alpha\vert 1 \rangle + \beta \vert 0\rangle \bigr)\vert
  10\rangle \\[1mm]
  + & \frac{1}{2} \bigl(\alpha\vert 1 \rangle - \beta \vert 0\rangle
  \bigr)\vert 11\rangle
\end{align*}

At first glance, it might look like something magical has happened, because the
leftmost qubit $\mathsf{B}$ now seems to depend on the numbers $\alpha$ and
$\beta$, even though there has not yet been any communication from Alice to
Bob.
This is an illusion.
Scalars float freely through tensor products, so $\alpha$ and $\beta$ are
neither more nor less associated with the leftmost qubit than they are with the
other qubits, and all we have done is to use algebra to express the state in a
way that facilitates an analysis of the measurements.

Now let us consider the four possible outcomes of Alice's standard basis
measurements, together with the actions that Bob performs as a result.

\subsubsection{Possible outcomes}

\begin{itemize}
\item
  The outcome of Alice's measurement is $ab = 00$ with probability
  \[
  \Biggl\| \frac{1}{2}\bigl(\alpha \vert 0\rangle + \beta\vert 1\rangle\bigr)
  \Biggr\|^2
  = \frac{\vert\alpha\vert^2 + \vert\beta\vert^2}{4} = \frac{1}{4},
  \]
  in which case the state of $(\mathsf{B},\mathsf{A},\mathsf{Q})$ becomes
  \[
  \bigl( \alpha \vert 0 \rangle + \beta \vert 1 \rangle \bigr) \vert 00
  \rangle.
  \]
  Bob does nothing in this case, and so this is the final state of these
  three qubits.

\item
  The outcome of Alice's measurement is $ab = 01$ with probability
  \[
  \Biggl\| \frac{1}{2}\bigl(\alpha \vert 0\rangle - \beta\vert 1\rangle\bigr)
  \Biggr\|^2
  = \frac{\vert\alpha\vert^2 + \vert{-\beta}\vert^2}{4} = \frac{1}{4},
  \]
  in which case the state of $(\mathsf{B},\mathsf{A},\mathsf{Q})$ becomes
  \[
  \bigl( \alpha \vert 0 \rangle - \beta \vert 1 \rangle \bigr) \vert 01
  \rangle.
  \]
  In this case Bob applies a $Z$ gate to $\mathsf{B}$, leaving
  $(\mathsf{B},\mathsf{A},\mathsf{Q})$ in the state
  \[
  \bigl( \alpha \vert 0 \rangle + \beta \vert 1 \rangle \bigr) \vert 01
  \rangle.
  \]

\item
  The outcome of Alice's measurement is $ab = 10$ with probability
  \[
  \Biggl\| \frac{1}{2}\bigl(\alpha \vert 1\rangle + \beta\vert 0\rangle\bigr)
  \Biggr\|^2
  = \frac{\vert\alpha\vert^2 + \vert\beta\vert^2}{4} = \frac{1}{4},
  \]
  in which case the state of $(\mathsf{B},\mathsf{A},\mathsf{Q})$ becomes
  \[
  \bigl( \alpha \vert 1 \rangle + \beta \vert 0 \rangle \bigr) \vert 10
  \rangle.
  \]
  In this case, Bob applies an $X$ gate to the qubit $\mathsf{B}$, leaving
  $(\mathsf{B},\mathsf{A},\mathsf{Q})$ in the state
  \[
  \bigl( \alpha \vert 0 \rangle + \beta \vert 1 \rangle \bigr) \vert 10
  \rangle.
  \]

\item
  The outcome of Alice's measurement is $ab = 11$ with probability
  \[
  \Biggl\| \frac{1}{2}\bigl(\alpha \vert 1\rangle - \beta\vert 0\rangle\bigr)
  \Biggr\|^2
  = \frac{\vert\alpha\vert^2 + \vert{-\beta}\vert^2}{4} = \frac{1}{4},
  \]
  in which case the state of $(\mathsf{B},\mathsf{A},\mathsf{Q})$ becomes
  \[
  \bigl( \alpha \vert 1 \rangle - \beta \vert 0 \rangle \bigr) \vert 11
  \rangle.
  \]
  In this case, Bob performs the operation $ZX$ on the qubit $\mathsf{B}$,
  leaving $(\mathsf{B},\mathsf{A},\mathsf{Q})$ in the state
  \[
  \bigl( \alpha \vert 0 \rangle + \beta \vert 1 \rangle \bigr) \vert 11
  \rangle.
  \]
\end{itemize}

We now see, in all four cases, that Bob's qubit $\mathsf{B}$ is left in the
state $\alpha\vert 0\rangle + \beta\vert 1\rangle$ at the end of the protocol,
which is the initial state of the qubit $\mathsf{Q}.$
This is what we wanted to show: the teleportation protocol has worked
correctly.

We also see that the qubits $\mathsf{A}$ and $\mathsf{Q}$ are left in one of
the four states $\vert 00\rangle$, $\vert 01\rangle$, $\vert 10\rangle$, or
$\vert 11\rangle$, each with probability $1/4$, depending upon the measurement
outcomes that Alice obtained.
Thus, as was already suggested above, at the end of the protocol Alice no
longer has the state $\alpha \vert 0\rangle + \beta \vert 1\rangle$, which is
consistent with the no-cloning theorem.

Notice that Alice's measurements yield absolutely no information about the
state $\alpha \vert 0\rangle + \beta \vert 1\rangle.$
That is, the probability for each of the four possible measurement outcomes is
$1/4$, irrespective of $\alpha$ and $\beta.$
This is also essential for teleportation to work correctly.
Extracting information from an unknown quantum state necessarily disturbs it in
general, but here Bob obtains the state without it being disturbed.

\subsubsection{General case}

Now let's consider the more general situation in which the qubit $\mathsf{Q}$
is initially entangled with another system, which we'll name $\mathsf{R}.$
A similar analysis to the one above reveals that the teleportation protocol
functions correctly in this more general case:
at the end of the protocol, the qubit $\mathsf{B}$ held by Bob is entangled
with $\mathsf{R}$ in the same way that $\mathsf{Q}$ was at the start of the
protocol, as if Alice had simply handed $\mathsf{Q}$ to Bob.

To prove this, let us suppose that the state of the pair
$(\mathsf{Q},\mathsf{R})$ is initially given by a quantum state vector of the
form
\[
\alpha \vert 0 \rangle_{\mathsf{Q}} \vert \gamma_0\rangle_{\mathsf{R}}
+ \beta \vert 1 \rangle_{\mathsf{Q}} \vert \gamma_1\rangle_{\mathsf{R}},
\]
where $\vert\gamma_0\rangle$ and $\vert\gamma_1\rangle$ are quantum state
vectors for the system $\mathsf{R}$ and $\alpha$ and $\beta$ are complex
numbers satisfying $\vert \alpha \vert^2 + \vert\beta\vert^2 = 1.$
Any quantum state vector of the pair $(\mathsf{Q},\mathsf{R})$ can be expressed
in this way.

\begin{figure}[!ht]
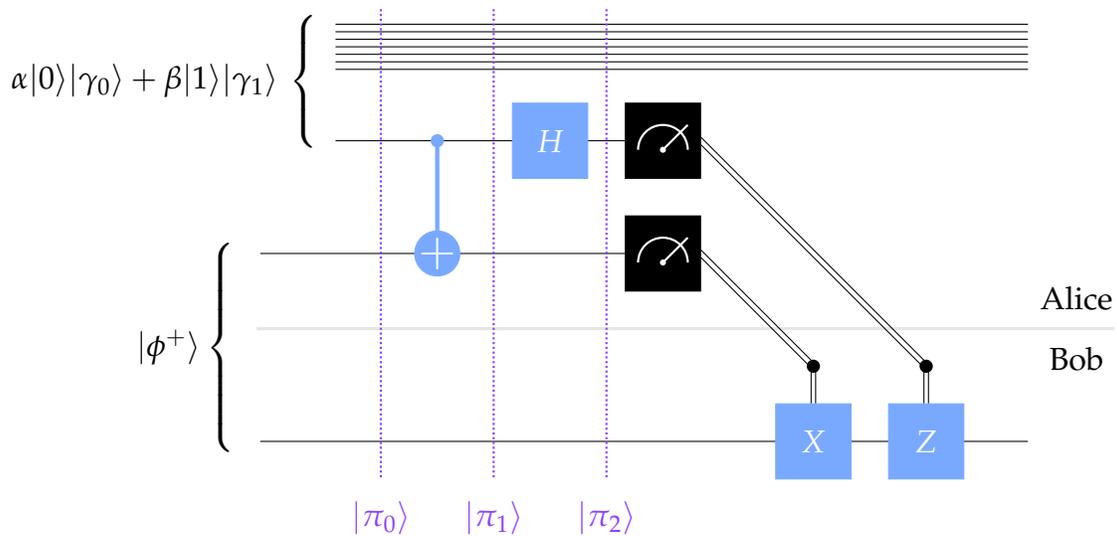

\begin{center}

  
\end{center}
  \caption{The three states $\ket{\pi_0}$, $\ket{\pi_1}$, and $\ket{\pi_2}$
    in the general case where there may be an additional system.}
  \label{fig:teleportation-with-entanglement}
\end{figure}

Figure~\ref{fig:teleportation-with-entanglement} depicts the same circuit as
before, with the addition of the system $\mathsf{R}$ (represented by a
collection of qubits on the top of the diagram that nothing happens to).

To analyze what happens when the teleportation protocol is run in this
situation, it is helpful to permute the systems, along the same lines that was
described in the previous lesson.
Specifically, we'll consider the state of the systems in the order
$(\mathsf{B},\mathsf{R},\mathsf{A},\mathsf{Q})$ rather than
$(\mathsf{B},\mathsf{A},\mathsf{Q},\mathsf{R}).$
The names of the various systems are included as subscripts in the expressions
that follow for clarity.

At the start of the protocol, the state of these systems is as follows:
\[
\scalebox{0.95}{$\displaystyle
  \begin{aligned}
    \vert \pi_0\rangle
    & = \vert \phi^+\rangle_{\mathsf{BA}} \otimes \bigl(
    \alpha \vert 0\rangle_{\mathsf{Q}} \vert\gamma_0\rangle_{\mathsf{R}}
    + \beta \vert
    1\rangle_{\mathsf{Q}}\vert\gamma_1\rangle_{\mathsf{R}}\bigr)\\[1mm]
    & =
    \frac{
      \alpha \vert 0\rangle_{\mathsf{B}} \vert \gamma_0 \rangle_{\mathsf{R}}
      \vert 00 \rangle_{\mathsf{AQ}}
      + \alpha \vert 1\rangle_{\mathsf{B}} \vert \gamma_0 \rangle_{\mathsf{R}}
      \vert 10 \rangle_{\mathsf{AQ}}
      + \beta \vert 0\rangle_{\mathsf{B}} \vert \gamma_1 \rangle_{\mathsf{R}}
      \vert 01 \rangle_{\mathsf{AQ}}
      + \beta \vert 1\rangle_{\mathsf{B}} \vert \gamma_1 \rangle_{\mathsf{R}}
      \vert 11 \rangle_{\mathsf{AQ}}}{\sqrt{2}}.
  \end{aligned}$}
\]
First the controlled-NOT gate is applied, which transforms this state to
\[
\scalebox{0.95}{$\displaystyle
\vert\pi_1\rangle =
\frac{
  \alpha \vert 0\rangle_{\mathsf{B}} \vert\gamma_0 \rangle_{\mathsf{R}} \vert
  00\rangle_{\mathsf{AQ}}
  + \alpha \vert 1\rangle_{\mathsf{B}} \vert\gamma_0 \rangle_{\mathsf{R}} \vert
  10\rangle_{\mathsf{AQ}}
  + \beta \vert 0\rangle_{\mathsf{B}} \vert\gamma_1 \rangle_{\mathsf{R}} \vert
  11\rangle_{\mathsf{AQ}}
  + \beta \vert 1\rangle_{\mathsf{B}} \vert\gamma_1 \rangle_{\mathsf{R}} \vert
  01\rangle_{\mathsf{AQ}}}{\sqrt{2}}.$}
\]
Then the Hadamard gate is applied.
After expanding and simplifying the resulting state, along similar lines to the
analysis of the simpler case above, we obtain this expression of the resulting
state:
\[
\begin{aligned}
\vert \pi_2 \rangle = \quad
  & \frac{1}{2} \bigl(
   \alpha \vert 0\rangle_{\mathsf{B}} \vert\gamma_0\rangle_{\mathsf{R}}
  + \beta \vert 1\rangle_{\mathsf{B}} \vert\gamma_1\rangle_{\mathsf{R}}
  \bigr) \vert 00\rangle_{\mathsf{AQ}}\\[2mm]
  + & \frac{1}{2} \bigl(
   \alpha \vert 0\rangle_{\mathsf{B}} \vert\gamma_0\rangle_{\mathsf{R}}
  - \beta \vert 1\rangle_{\mathsf{B}} \vert\gamma_1\rangle_{\mathsf{R}}
  \bigr) \vert 01\rangle_{\mathsf{AQ}}\\[2mm]
  + & \frac{1}{2} \bigl(
   \alpha \vert 1\rangle_{\mathsf{B}} \vert\gamma_0\rangle_{\mathsf{R}}
  + \beta \vert 0\rangle_{\mathsf{B}} \vert\gamma_1\rangle_{\mathsf{R}}
  \bigr) \vert 10\rangle_{\mathsf{AQ}}\\[2mm]
  + & \frac{1}{2} \bigl(
   \alpha \vert 1\rangle_{\mathsf{B}} \vert\gamma_0\rangle_{\mathsf{R}}
  - \beta \vert 0\rangle_{\mathsf{B}} \vert\gamma_1\rangle_{\mathsf{R}}
  \bigr) \vert 11\rangle_{\mathsf{AQ}}.
\end{aligned}
\]

Proceeding exactly as before, where we consider the four different possible
outcomes of Alice's measurements along with the corresponding actions performed
by Bob, we find that at the end of the protocol, the state of
$(\mathsf{B},\mathsf{R})$ is always
\[
\alpha \vert 0 \rangle \vert \gamma_0\rangle + \beta \vert 1 \rangle \vert
\gamma_1\rangle.
\]
Informally speaking, the analysis does not change in a significant way as
compared with the simpler case above;
$\vert\gamma_0\rangle$ and $\vert\gamma_1\rangle$ essentially just ``come along
for the ride.''
So, teleportation succeeds in creating a perfect quantum communication channel,
effectively transmitting the contents of the qubit $\mathsf{Q}$ into
$\mathsf{B}$ and preserving all correlations with other systems.

This is actually not surprising at all, given the analysis of the simpler case
above.
As that analysis revealed, we have a physical process that acts like the
identity operation on a qubit in an arbitrary quantum state, and there's only
one way that can happen: the operation implemented by the protocol must
\emph{be} the identity operation.
That is, once we know that teleportation works correctly for a single qubit in
isolation, we can conclude that the protocol effectively implements a perfect,
noiseless quantum channel, and so it must work correctly even if the input
qubit is entangled with another system.

\subsection{Further discussion}

Here are a few brief, concluding remarks on teleportation, beginning with the
clarification that teleportation is not an \emph{application} of quantum
information, it's a \emph{protocol} for performing quantum communication.
It is therefore useful only insofar as quantum communication is useful.

Indeed, it is reasonable to speculate that teleportation could one day become a
standard way to communicate quantum information, perhaps through a process
known as \emph{entanglement distillation}.
This is a process that converts a larger number of noisy (or imperfect) e-bits
into a smaller number of high quality e-bits, that could then be used for
noiseless or near-noiseless teleportation.
The idea is that the process of entanglement distillation is not as delicate as
direct quantum communication.
We could accept losses, for instance, and if the process doesn't work out, we
can just try again.
In contrast, the actual qubits we hope to communicate might be much more
precious.

Finally, it should be understood that the idea behind teleportation and the way
that it works is quite fundamental in quantum information and computation.
It really is a cornerstone of quantum information theory, and variations of it
arise.
For example, quantum gates can be implemented through a closely related process
known as \emph{quantum gate teleportation}, which uses teleportation to apply
\emph{operations} to qubits rather than communicating them.

\section{Superdense coding}

Superdense coding is a protocol that, in some sense, achieves a complementary
aim to teleportation.
Rather than allowing for the transmission of one qubit using two classical bits
of communication (at the cost of one e-bit of entanglement), it allows for the
transmission of two classical bits using one qubit of quantum communication
(again, at the cost of one e-bit of entanglement).

In greater detail, we have a sender (Alice) and a receiver (Bob) that share one
e-bit of entanglement.
According to the conventions in place for the lesson, this means that Alice
holds a qubit $\mathsf{A}$, Bob holds a qubit $\mathsf{B}$, and together the
pair $(\mathsf{A},\mathsf{B})$ is in the state $\vert\phi^+\rangle.$
Alice wishes to transmit two classical bits to Bob, which we'll denote by $c$
and $d$, and she will accomplish this by sending him one qubit.

It is reasonable to view this feat as being less interesting than the one that
teleportation accomplishes.
Sending qubits is likely to be so much more difficult than sending classical
bits for the foreseeable future that trading one qubit of quantum communication
for two bits of classical communication, at the cost of an e-bit no less,
hardly seems worth it.
However, this does not imply that superdense coding is not interesting, for it
most certainly is.

Fitting the theme of the lesson, one reason why superdense coding is
interesting is that it demonstrates a concrete and (in the context of
information theory) rather striking use of entanglement.
A famous theorem in quantum information theory, known as \emph{Holevo's
theorem}, implies that without the use of a shared entangled state, it is
impossible to communicate more than one bit of classical information by sending
a single qubit.
(Holevo's theorem is more general than this.
Its precise statement is technical and requires explanation, but this is one
consequence of it.)
So, through superdense coding, shared entanglement effectively allows for the
\emph{doubling} of the classical information-carrying capacity of sending
qubits.

\subsection{Protocol}

\begin{figure}[!ht]
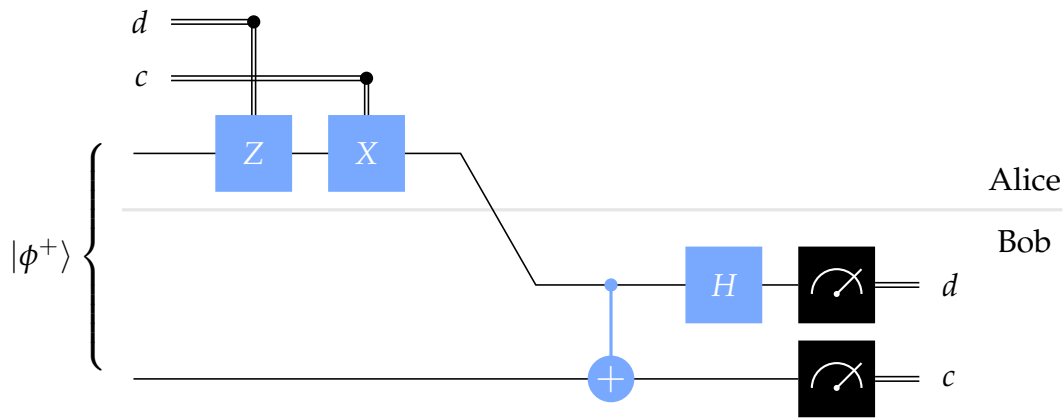

\begin{center}

      
\end{center}
\caption{The superdense coding protocol described as a quantum circuit.}
\label{fig:superdense-coding}
\end{figure}

The superdense coding protocol is described as a quantum circuit in
Figure~\ref{fig:superdense-coding}.
In words, here is what Alice does:
\begin{enumerate}
\item
  If $d=1$, Alice performs a $Z$ gate on her qubit $\mathsf{A}$ (and if $d=0$
  she does not).
\item
  If $c=1$, Alice performs an $X$ gate on her qubit $\mathsf{A}$ (and if $c=0$
  she does not).
\end{enumerate}
Alice then sends her qubit $\mathsf{A}$ to Bob.

What Bob does when he receives the qubit $\mathsf{A}$ is to first perform a
controlled-NOT gate, with $\mathsf{A}$ being the control and $\mathsf{B}$ being
the target, and then he applies a Hadamard gate to $\mathsf{A}.$
He then measures $\mathsf{B}$ to obtain $c$ and $\mathsf{A}$ to obtain $d$,
with standard basis measurements in both cases.

\subsection{Analysis}

The idea behind this protocol is pretty simple:
Alice effectively chooses which Bell state she would like to be sharing with
Bob, she sends Bob her qubit, and Bob measures to determine which Bell state
Alice chose.

That is, they initially share $\vert\phi^+\rangle$, and depending upon the bits
$c$ and $d$, Alice either leaves this state alone or shifts it to one of the
other Bell states by applying $\mathbb{I}$, $X$, $Z$, or $XZ$ to her qubit
$\mathsf{A}.$
\[
\begin{aligned}
  (\mathbb{I} \otimes \mathbb{I}) \vert \phi^+ \rangle & = \vert \phi^+\rangle
  \\
  (\mathbb{I} \otimes Z) \vert \phi^+ \rangle & = \vert \phi^-\rangle \\
  (\mathbb{I} \otimes X) \vert \phi^+ \rangle & = \vert \psi^+\rangle \\
  (\mathbb{I} \otimes XZ) \vert \phi^+ \rangle & = \vert \psi^-\rangle
\end{aligned}
\]
Bob's actions have the following effects on the four Bell states.
\[
\begin{aligned}
  \vert \phi^+\rangle & \mapsto \vert 00\rangle\\
  \vert \phi^-\rangle & \mapsto \vert 01\rangle\\
  \vert \psi^+\rangle & \mapsto \vert 10\rangle\\
  \vert \psi^-\rangle & \mapsto -\vert 11\rangle\\
\end{aligned}
\]
This can be checked directly, by computing the results of Bob's operations on
these states one at a time.

\pagebreak

So, when Bob performs his measurements, he is able to determine which Bell
state Alice chose.
To verify that the protocol works correctly is a matter of checking each case:
\begin{itemize}
\item
  If $cd = 00$, then the state of $(\mathsf{B},\mathsf{A})$ when Bob receives
  $\mathsf{A}$ is $\vert \phi^+\rangle.$ He transforms this state into $\vert
  00\rangle$ and obtains $cd = 00.$

\item
  If $cd = 01$, then the state of $(\mathsf{B},\mathsf{A})$ when Bob receives
  $\mathsf{A}$ is $\vert \phi^-\rangle.$ He transforms this state into $\vert
  01\rangle$ and obtains $cd = 01.$

\item
  If $cd = 10$, then the state of $(\mathsf{B},\mathsf{A})$ when Bob receives
  $\mathsf{A}$ is $\vert \psi^+\rangle.$ He transforms this state into $\vert
  10\rangle$ and obtains $cd = 10.$

\item
  If $cd = 11$, then the state of $(\mathsf{B},\mathsf{A})$ when Bob receives
  $\mathsf{A}$ is $\vert \psi^-\rangle.$ He transforms this state into $-\vert
  11\rangle$ and obtains $cd = 11.$ (The negative-one phase factor has no
  effect here.)
\end{itemize}

\section{The CHSH game}

The last example to be discussed in this lesson is not a protocol, but a
\emph{game} known as the \emph{CHSH game}.
When we speak of a game in this context, we're not talking about something
that's meant to be played for fun or sport, but rather a mathematical
abstraction in the sense of \emph{game theory}.
Mathematical abstractions of games are studied in economics and computer
science, for instance, and they are both fascinating and useful.

The letters CHSH refer to the authors --- John Clauser, Michael Horne, Abner
Shimony, and Richard Holt --- of a 1969 paper where the example was first
described.
They did not describe the example as a game, but rather as an experiment.
Its description as a game, however, is both natural and intuitive.

The CHSH game falls within a class of games known as \emph{nonlocal games}.
Nonlocal games are incredibly interesting and have deep connections to physics,
computer science, and mathematics --- holding mysteries that still remain
unsolved.
We'll begin the section by explaining what nonlocal games are, and then we'll
focus in on the CHSH game and what makes it interesting.

\subsection{Nonlocal games}

A nonlocal game is a \emph{cooperative game} where two players, Alice and Bob,
work together to achieve a particular outcome.
The game is run by a \emph{referee}, who behaves according to strict guidelines
that are known to Alice and Bob.

Alice and Bob can prepare for the game however they choose, but once the game
starts they are \emph{forbidden from communicating}.
We might imagine the game taking place in a secure facility of some sort --- as
if the referee is playing the role of a detective and Alice and Bob are
suspects being interrogated in different rooms.
But another way to think about the set-up is that Alice and Bob are separated
by a vast distance, and communication is prohibited because the speed of light
doesn't allow for it within the running time of the game.
That is to say, if Alice tries to send a message to Bob, the game will be over
by the time he receives it, and vice versa.

The way a nonlocal game works is that the referee first asks each of Alice and
Bob a question.
We'll use the letter $x$ to refer to Alice's question and $y$ to refer to Bob's
question.
Here we're thinking of $x$ and $y$ as being classical states, and in the CHSH
game $x$ and $y$ are bits.
The referee uses \emph{randomness} to select these questions.
To be precise, there is some probability $p(x,y)$ associated with each possible
pair $(x,y)$ of questions, and the referee has vowed to choose the questions
randomly, at the time of the game, in this way.
Everyone, including Alice and Bob, knows these probabilities --- but nobody
knows specifically which pair $(x,y)$ will be chosen until the game begins.

\begin{figure}[!ht]
\begin{center}
  \begin{tikzpicture}[
      >=latex,
      scale=2,
      line width = 0.6pt,
      gate/.style={%
        inner sep = 0,
        fill = CircuitBlue,
        draw = CircuitBlue,
        text = white
      },
      blackgate/.style={%
        inner sep = 0,
        fill = black,
        draw = black,
        text = white
      }
      ]
    
    \node[gate, minimum width = 20mm, minimum height = 15mm]
    (Alice) at (-1.375,1) {Alice};
    
    \node[gate, minimum width = 20mm, minimum height = 15mm]
    (Bob) at (1.375,1) {Bob};
    
    \node[gate, minimum width = 20mm, minimum height = 15mm]
    (Referee) at (0,-0.65) {Referee};
    
    \node[blackgate, minimum width = 4mm, minimum height = 30mm]
    (wall) at (0,1) {};
    
    \node[anchor = south] at (0,1.9) {%
      \begin{tabular}{c}
        No communication\\
        between Alice and Bob
      \end{tabular}
    };
    
    \draw[->,rounded corners = 2pt] ([yshift=-2mm]Referee.west) -|
    ([xshift=-2mm]Alice.south)
    node [pos = 0.75, left] {$x$};
    
    \draw[->,rounded corners = 2pt] ([yshift=-2mm]Referee.east) -|
    ([xshift=2mm]Bob.south)
    node [pos = 0.725, right] {$y$};
    
    \draw[->,rounded corners = 2pt]
    ([xshift=2mm]Alice.south) |-
    ([yshift=2mm]Referee.west)             
    node [pos = 0.25, right] {$a$};
    
    \draw[->,rounded corners = 2pt]
    ([xshift=-2mm]Bob.south) |-
    ([yshift=2mm]Referee.east)             
    node [pos = 0.25, left] {$b$};
    
  \end{tikzpicture}
  
\end{center}
\caption{The interactions between the Referee and Alice and Bob in a nonlocal
  game.}
\label{fig:nonlocal-game}
\end{figure}
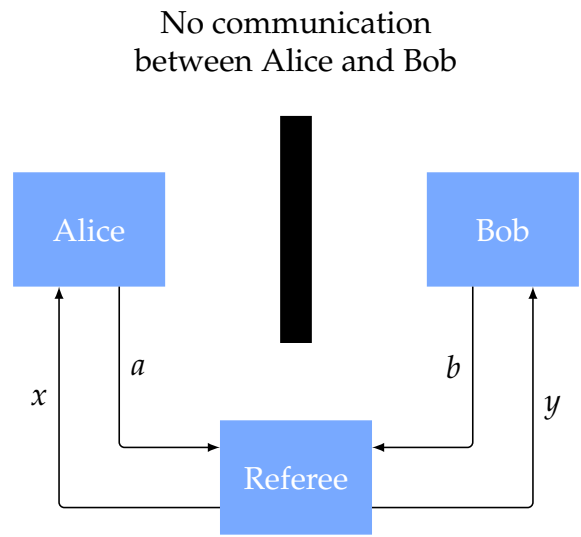

After Alice and Bob receive their questions, they must then provide answers:
Alice's answer is $a$ and Bob's answer is $b.$
Again, these are classical states in general, and bits in the CHSH game.
Upon receiving these answers, the referee makes a decision: Alice and Bob
either \emph{win} or \emph{lose} depending on whether or not the pair of
answers $(a,b)$ is deemed correct for the pair of questions $(x,y)$ according
to some fixed set of rules.
Different rules mean different games, and the rules for the CHSH game
specifically are described in the section following this one.
As was already suggested, the rules are known to everyone.
Figure~\ref{fig:nonlocal-game} provides a graphic representation of the
interactions just described.

It is the uncertainty about which questions will be asked, and specifically the
fact that each player doesn't know the other player's question, that makes
nonlocal games challenging for Alice and Bob --- just like colluding suspects
in different rooms trying to keep their story straight.

A precise description of the referee defines an instance of a nonlocal game.
This includes a specification of the probabilities $p(x,y)$ for each question
pair along with the rules that determine whether each pair of answers $(a,b)$
wins or loses for each possible question pair $(x,y).$

We'll take a look at the CHSH game momentarily, but before that let us briefly
acknowledge that it's also interesting to consider other nonlocal games.
It's extremely interesting, in fact, and there are some nonlocal games for
which it's currently not known how well Alice and Bob can play using
entanglement.
The set-up is simple, but there's complexity at work --- and for some games it
can be impossibly difficult to compute best or near-best strategies for Alice
and Bob.
This is the mind-blowing nature of the non-local games model.

\subsection{CHSH game description}

Here is the precise description of the CHSH game, where (as above) $x$ is
Alice's question, $y$ is Bob's question, $a$ is Alice's answer, and $b$ is
Bob's answer:
\begin{itemize}
\item
  The questions and answers are all bits: $x,y,a,b\in\{0,1\}.$

\item
  The referee chooses the questions $(x,y)$ \emph{uniformly at random}:
  each of the four possibilities, $(0,0)$, $(0,1)$, $(1,0)$, and $(1,1)$,
  is selected with probability~$1/4.$

\item
  The answers $(a,b)$ \emph{win} for the questions $(x,y)$ if $a\oplus b =
  x\wedge y$ and \emph{lose} otherwise.
  The following table expresses this rule by listing the winning and losing
  conditions on the answers $(a,b)$ for each pair of questions $(x,y).$
  \[
  \begin{array}{c@{\hspace{15pt}}c@{\hspace{15pt}}c}
    (x,y) & \text{win} & \text{lose} \\[1mm]\hline
    \rule{0mm}{4mm}(0,0) & a = b & a \neq b \\[1mm]
    (0,1) & a = b & a \neq b \\[1mm]
    (1,0) & a = b & a \neq b \\[1mm]
    (1,1) & a \neq b & a = b
  \end{array}
  \]
\end{itemize}

\subsection{Limitations of classical strategies}

Now let's consider strategies for Alice and Bob in the CHSH game, beginning
with \emph{classical} strategies.

\subsubsection{Deterministic strategies}

We'll start with \emph{deterministic} strategies, where Alice's answer $a$ is a
function of the question $x$ that she receives, and likewise Bob's answer $b$
is a function of the question $y$ he receives.
So, for instance, we may write $a(0)$ to represent Alice's answer when her
question is $0$, and $a(1)$ to represent Alice's answer when her question
is~$1.$

No deterministic strategy can possibly win the CHSH game every time.
One way to reason this is simply to go one-by-one through all of the possible
deterministic strategies and check that every one of them loses for at least
one of the four possible question pairs.
Alice and Bob can each choose from four possible functions from one bit to one
bit --- which we encountered back in the first lesson of the course --- so
there are $16$ different deterministic strategies in total to check.

We can also reason this analytically.
If Alice and Bob's strategy wins when $(x,y) = (0,0)$, then it must be that
$a(0) = b(0);$ if their strategy wins when $(x,y) = (0,1)$, then $a(0) = b(1);$
and similarly, if the strategy wins for $(x,y)=(1,0)$ then $a(1) = b(0).$
So, if their strategy wins for all three possibilities, then
\[
b(1) = a(0) = b(0) = a(1).
\]
This implies that the strategy loses in the final case $(x,y) = (1,1)$, for
here winning requires that $a(1) \neq b(1).$
Thus, there can be no deterministic strategy that wins every time.

On the other hand, it is easy to find deterministic strategies that win in
three of the four cases, such as $a(0)=a(1)=b(0)=b(1)=0.$
From this we conclude that the maximum probability for Alice and Bob to win
using a deterministic strategy is~$3/4.$

\subsubsection{Probabilistic strategies}

As we just concluded, Alice and Bob cannot do better than winning the CHSH game
75\% of the time using a deterministic strategy. 
But what about a probabilistic strategy?
Could it help Alice and Bob to use randomness --- including the possibility of
\emph{shared randomness}, where their random choices are correlated?

It turns out that probabilistic strategies don't help at all to increase the
probability that Alice and Bob win.
This is because every probabilistic strategy can alternatively be viewed as a
random selection of a deterministic strategy, just like probabilistic
operations can be viewed as random selections of deterministic operations.
The average is never larger than the maximum, and so it follows that
probabilistic strategies don't offer any advantage in terms of their overall
winning probability.

Thus, winning with probability $3/4$ is the best that Alice and Bob can do
using any classical strategy, whether deterministic or probabilistic.

\subsection{CHSH game strategy}

\begin{figure}[!ht]
  \begin{center}
    \begin{tikzpicture}[
        >=latex,
        scale=2,
        line width = 0.6pt,
        gate/.style={%
          inner sep = 0,
          fill = CircuitBlue,
          draw = CircuitBlue,
          text = white
        },
        blackgate/.style={%
          inner sep = 0,
          fill = black,
          draw = black,
          text = white
        }
      ]
      
      \node[gate, minimum width = 20mm, minimum height = 15mm]
      (Alice) at (-1.375,1) {Alice};
      
      \node[gate, minimum width = 20mm, minimum height = 15mm]
      (Bob) at (1.375,1) {Bob};
      
      \node[gate, minimum width = 20mm, minimum height = 15mm]
      (Referee) at (0,-0.65) {Referee};
      
      \node[blackgate, minimum width = 4mm, minimum height = 30mm]
      (wall) at (0,1) {};
      
      \draw[->,rounded corners = 2pt] ([yshift=-2mm]Referee.west) -|
      ([xshift=-2mm]Alice.south)
      node [pos = 0.75, left] {$x$};
      
      \draw[->,rounded corners = 2pt] ([yshift=-2mm]Referee.east) -|
      ([xshift=2mm]Bob.south)
      node [pos = 0.725, right] {$y$};
      
      \draw[->,rounded corners = 2pt]
      ([xshift=2mm]Alice.south) |-
      ([yshift=2mm]Referee.west)             
      node [pos = 0.25, right] {$a$};
      
      \draw[->,rounded corners = 2pt]
      ([xshift=-2mm]Bob.south) |-
      ([yshift=2mm]Referee.east)             
      node [pos = 0.25, left] {$b$};

      \node[text=Highlight, minimum width = 4mm]
      (psi) at (0,2.25) {$\ket{\psi}$};

      \draw[color=Highlight, ->, rounded corners = 2pt]
      (psi.west) -| (Alice.north);
      
      \draw[color=Highlight, ->, rounded corners = 2pt]
      (psi.east) -| (Bob.north);
      
    \end{tikzpicture}
  \end{center}
  \caption{A quantum strategy in which Alice and Bob make use of a shared
    entangled state $\ket{\psi}$.}
  \label{fig:nonlocal-game-entanglement}
\end{figure}
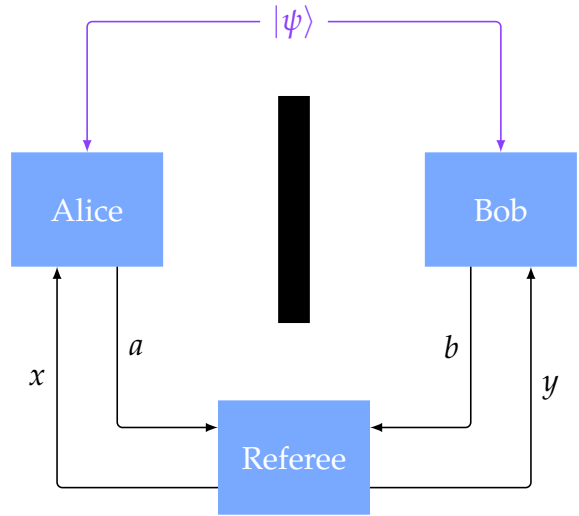

A natural question to ask at this point is whether Alice and Bob can do any
better using a \emph{quantum} strategy.
In particular, if they share an entangled quantum state as
Figure~\ref{fig:nonlocal-game-entanglement} suggests, which they could have
prepared prior to playing the game, can they increase their winning
probability?

The answer is yes, and this is the main point of the example and why it's so
interesting.
So let's see exactly how Alice and Bob can do better in this game using
entanglement.

\subsubsection{Required vectors and matrices}

The first thing we need to do is to define a qubit state vector $\vert
\psi_{\theta}\rangle$, for each real number $\theta$ (which we'll think of as
an angle measured in radians) as follows.
\[
\vert\psi_{\theta}\rangle = \cos(\theta)\vert 0\rangle + \sin(\theta) \vert
1\rangle
\]
Here are some simple examples.
\[
\vert\psi_{0}\rangle = \vert 0\rangle \qquad
\vert\psi_{\pi/2}\rangle = \vert 1\rangle \qquad
\vert\psi_{\pi/4}\rangle = \vert + \rangle \qquad
\vert\psi_{-\pi/4}\rangle = \vert - \rangle
\]
We also have the following examples, which arise in the analysis below.
\[
\begin{aligned}
  \vert\psi_{-\pi/8}\rangle & = \frac{\sqrt{2 + \sqrt{2}}}{2}\vert 0\rangle
  -\frac{\sqrt{2 - \sqrt{2}}}{2}\vert 1\rangle \\[1mm]
  \vert\psi_{\pi/8}\rangle & = \frac{\sqrt{2 + \sqrt{2}}}{2}\vert 0\rangle +
  \frac{\sqrt{2 - \sqrt{2}}}{2}\vert 1\rangle \\[1mm]
  \vert\psi_{3\pi/8}\rangle & = \frac{\sqrt{2 - \sqrt{2}}}{2}\vert 0\rangle +
  \frac{\sqrt{2 + \sqrt{2}}}{2}\vert 1\rangle \\[1mm]
  \vert\psi_{5\pi/8}\rangle & = -\frac{\sqrt{2 - \sqrt{2}}}{2}\vert 0\rangle +
  \frac{\sqrt{2 + \sqrt{2}}}{2}\vert 1\rangle
\end{aligned}
\]

Looking at the general form, we see that the inner product between any two of
these vectors has this formula:
\begin{equation}
\langle \psi_{\alpha} \vert \psi_{\beta} \rangle
= \cos(\alpha)\cos(\beta) + \sin(\alpha)\sin(\beta)
= \cos(\alpha-\beta).
\label{eq:angle-inner-product-1}
\end{equation}
In detail, there are only real number entries in these vectors, so there are no
complex conjugates to worry about:
the inner product is the product of the cosines plus the product of the sines.
Using one of the \emph{angle addition formulas} from trigonometry leads to the
simplification above.
This formula reveals the geometric interpretation of the inner product between
real unit vectors as the cosine of the angle between them.

If we compute the inner product of the \emph{tensor product} of any two of
these vectors with the $\vert \phi^+\rangle$ state, we obtain a similar
expression, except that it has a $\sqrt{2}$ in the denominator:
\begin{equation}
  \langle \psi_{\alpha} \otimes \psi_{\beta} \vert \phi^+ \rangle
  = \frac{\cos(\alpha)\cos(\beta) + \sin(\alpha)\sin(\beta)}{\sqrt{2}}
  = \frac{\cos(\alpha-\beta)}{\sqrt{2}}.
  \label{eq:angle-inner-product-2}
\end{equation}
Our interest in this particular inner product will become clear shortly, but
for now we're simply observing this as a formula.

Next, define a unitary matrix $U_{\theta}$ for each angle $\theta$ as follows.
\[
U_{\theta} = \vert 0 \rangle \langle \psi_{\theta} \vert + \vert
1\rangle\langle \psi_{\theta+\pi/2} \vert
\]
Intuitively speaking, this matrix transforms $\vert\psi_{\theta}\rangle$ into
$\vert 0\rangle$ and $\vert \psi_{\theta + \pi/2}\rangle$ into $\vert
1\rangle.$
To check that this is a unitary matrix, a key observation is that the vectors
$\vert\psi_{\theta}\rangle$ and $\vert\psi_{\theta + \pi/2}\rangle$ are
orthogonal for every angle $\theta$:
\[
\langle \psi_{\theta} \vert \psi_{\theta + \pi/2} \rangle = \cos(\pi/2) = 0.
\]
Thus, we find that
\[
\begin{aligned}
  U_{\theta} U_{\theta}^{\dagger}
  & = \bigl(\vert 0 \rangle \langle \psi_{\theta} \vert + \vert 1\rangle\langle
  \psi_{\theta+\pi/2} \vert\bigr)
  \bigl(\vert \psi_{\theta} \rangle \langle 0 \vert + \vert
  \psi_{\theta+\pi/2}\rangle\langle 1 \vert\bigr) \\[1mm]
  & =
  \vert 0 \rangle \langle \psi_{\theta} \vert \psi_{\theta} \rangle \langle 0
  \vert
  + \vert 0 \rangle \langle \psi_{\theta} \vert \psi_{\theta+\pi/2} \rangle
  \langle 1 \vert\\
  & \qquad
  + \vert 1 \rangle \langle \psi_{\theta+\pi/2} \vert \psi_{\theta} \rangle
  \langle 0 \vert
  + \vert 1 \rangle \langle \psi_{\theta+\pi/2} \vert \psi_{\theta+\pi/2}
  \rangle \langle 1 \vert \\[1mm]
  & =
  \vert 0 \rangle \langle 0 \vert + \vert 1 \rangle \langle 1 \vert\\[1mm]
  & = \mathbb{I}.
\end{aligned}
\]
We may alternatively write this matrix explicitly as
\[
U_{\theta}
= \begin{pmatrix}
\cos(\theta) & \sin(\theta)\\[1mm]
\cos(\theta+ \pi/2) & \sin(\theta + \pi/2)
\end{pmatrix}
= \begin{pmatrix}
\cos(\theta) & \sin(\theta)\\[1mm]
-\sin(\theta) & \cos(\theta)
\end{pmatrix}.
\]

This is an example of a \emph{rotation matrix}, and specifically it rotates
two-dimensional vectors with real number entries by an angle of $-\theta$ about
the origin.
If we follow a standard convention for naming and parameterizing rotations of
various forms, we have $U_{\theta} = R_y(-2\theta)$ where
\[
R_y(\theta) =
\begin{pmatrix}
  \cos(\theta/2) & -\sin(\theta/2)\\[1mm]
  \sin(\theta/2) & \cos(\theta/2)
\end{pmatrix}.
\]

\subsubsection{Strategy description}

Now we can describe the quantum strategy.
\begin{trivlist}
\item
  \textbf{Set-up:} Alice and Bob start the game sharing an e-bit: Alice holds a
  qubit $\mathsf{A}$, Bob holds a qubit $\mathsf{B}$, and together the two
  qubits $(\mathsf{X},\mathsf{Y})$ are in the $\vert\phi^+\rangle$ state.

\item
  \textbf{Alice's actions:}
  \begin{itemize}
  \item
    If Alice receives the question $x=0$, she applies $U_{0}$ to her qubit
    $\mathsf{A}.$
  \item
    If Alice receives the question  $x=1$, she applies $U_{\pi/4}$ to her qubit
    $\mathsf{A}.$
  \end{itemize}
  \noindent
  The operation Alice performs on $\mathsf{A}$ may alternatively be described
  like this:
  \[
  \begin{cases}
    U_0 & \text{if $x = 0$}\\
    U_{\pi/4} & \text{if $x = 1$}.
  \end{cases}
  \]
  After Alice applies this operation, she measures $\mathsf{A}$ with a standard
  basis measurement and sets her answer $a$ to be the measurement outcome.

\item
  \textbf{Bob's actions:}
  \begin{itemize}
  \item
    If Bob receives the question $y=0$, he applies $U_{\pi/8}$ to his qubit
    $\mathsf{B}.$
  \item
    If Bob receives the question $y=1$, he applies $U_{-\pi/8}$ to his qubit
    $\mathsf{B}.$
  \end{itemize}

  \noindent
  Like we did for Alice, we can express Bob's operation on $\mathsf{B}$ like
  this:
  \[
  \begin{cases}
    U_{\pi/8} & \text{if $y = 0$}\\
    U_{-\pi/8} & \text{if $y = 1$}.
  \end{cases}
  \]
  After Bob applies this operation, he measures $\mathsf{B}$ with a standard
  basis measurement and sets his answer $b$ to be the measurement outcome.
\end{trivlist}

\begin{figure}[!ht]
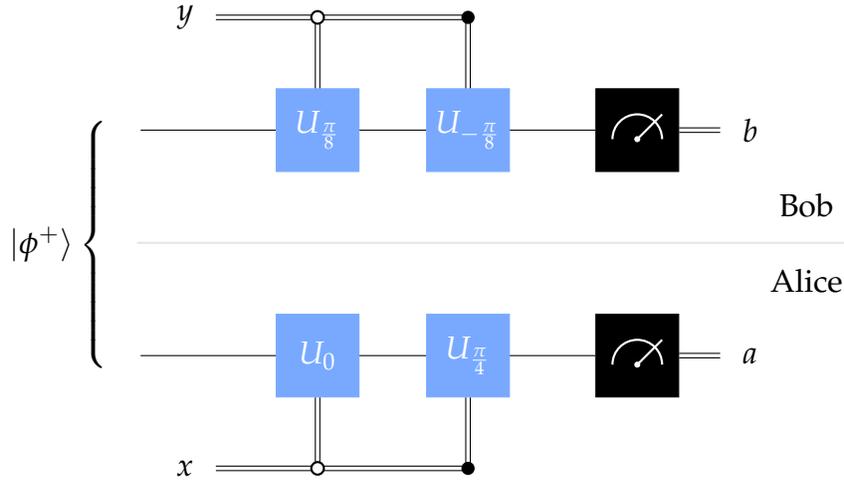

  \begin{center}

  \end{center}
  \caption{A quantum circuit description of Alice and Bob's strategy.}
  \label{fig:CHSH}
\end{figure}

Figure~\ref{fig:CHSH} describes this strategy as a quantum circuit diagram.
In this diagram we see two ordinary controlled gates, one for $U_{-\pi/8}$ on
the top and one for $U_{\pi/4}$ on the bottom.
We also have two gates that look like controlled gates, one for $U_{\pi/8}$ on
the top and one for $U_{0}$ on the bottom, except that the circle representing
the control is not filled in.
This denotes a different type of controlled gate where the gate is performed if
the control is set to $0$ (rather than $1$ like an ordinary controlled gate).
So, effectively, Bob performs $U_{\pi/8}$ on his qubit if $y=0$ and
$U_{-\pi/8}$ if $y=1;$
and Alice performs $U_0$ on her qubit if $x=0$ and $U_{\pi/4}$ if $x=1$, which
is consistent with the description of the protocol in words above.

It remains to figure out how well this strategy for Alice and Bob works.
We'll do this by going through the four possible question pairs individually.

\subsubsection{Case-by-case analysis}

\begin{description}[leftmargin=0mm]
\item[Case 1:  $(x,y) = (0,0).$]
  In this case Alice performs $U_{0}$ on her qubit and Bob performs $U_{\pi/8}$
  on his, so the state of the two qubits $(\mathsf{A},\mathsf{B})$ after they
  perform their operations is
  \[
  \begin{aligned}
    \bigl(U_0 \otimes U_{\pi/8}\bigr) \vert \phi^+\rangle
    & =
    \vert 00 \rangle \langle \psi_0 \otimes \psi_{\pi/8}\vert \phi^+\rangle
    + \vert 01 \rangle \langle \psi_0 \otimes\psi_{5\pi/8}\vert \phi^+\rangle \\
    & \qquad + \vert 10 \rangle \langle \psi_{\pi/2} \otimes \psi_{\pi/8}\vert
    \phi^+\rangle
   + \vert 11 \rangle \langle \psi_{\pi/2} \otimes \psi_{5\pi/8}\vert
   \phi^+\rangle\\[2mm]
   & = \frac{
     \cos\bigl(-\frac{\pi}{8}\bigr) \vert 00\rangle
     + \cos\bigl(-\frac{5\pi}{8}\bigr) \vert 01\rangle
     + \cos\bigl(\frac{3\pi}{8}\bigr) \vert 10\rangle
     + \cos\bigl(-\frac{\pi}{8}\bigr) \vert 11\rangle}{\sqrt{2}}.
  \end{aligned}
  \]
  The probabilities for the four possible answer pairs $(a,b)$ are therefore as
  follows.
  \[
  \begin{aligned}
    \operatorname{Pr}\bigl((a,b)=(0,0)\bigr) & =
    \frac{1}{2}\cos^2\Bigl(-\frac{\pi}{8}\Bigr) = \frac{2+\sqrt{2}}{8} \\[2mm]
   \operatorname{Pr}\bigl((a,b)=(0,1)\bigr) & =
   \frac{1}{2}\cos^2\Bigl(-\frac{5\pi}{8}\Bigr) = \frac{2-\sqrt{2}}{8}\\[2mm]
   \operatorname{Pr}\bigl((a,b)=(1,0)\bigr) & =
   \frac{1}{2}\cos^2\Bigl(\frac{3\pi}{8}\Bigr) = \frac{2-\sqrt{2}}{8}\\[2mm]
   \operatorname{Pr}\bigl((a,b)=(1,1)\bigr) & =
   \frac{1}{2}\cos^2\Bigl(-\frac{\pi}{8}\Bigr) = \frac{2+\sqrt{2}}{8}
  \end{aligned}
  \]
  We can then obtain the probabilities that $a=b$ and $a\neq b$ by summing.
  \[
  \operatorname{Pr}(a = b) = \frac{2 + \sqrt{2}}{4} \qquad
    \operatorname{Pr}(a \neq b) = \frac{2 - \sqrt{2}}{4}
  \]
  For the question pair $(0,0)$, Alice and Bob win if $a=b$, and therefore they
  win in this case with probability
  \[
  \frac{2 + \sqrt{2}}{4}.
  \]
  
\item[Case 2:  $(x,y) = (0,1).$] 
  In this case Alice performs $U_{0}$ on her qubit and Bob performs
  $U_{-\pi/8}$ on his, so the state of the two qubits $(\mathsf{A},\mathsf{B})$
  after they perform their operations is
  \[
  \begin{aligned}
    \bigl(U_0 \otimes U_{-\pi/8}\bigr) \vert \phi^+\rangle
    & =
    \vert 00 \rangle \langle \psi_0 \otimes \psi_{-\pi/8}\vert \phi^+\rangle
    + \vert 01 \rangle \langle \psi_0 \otimes\psi_{3\pi/8}\vert \phi^+\rangle
    \\
    & \qquad + \vert 10 \rangle \langle \psi_{\pi/2} \otimes \psi_{-\pi/8}\vert
    \phi^+\rangle
    + \vert 11 \rangle \langle \psi_{\pi/2} \otimes \psi_{3\pi/8}\vert
    \phi^+\rangle\\[2mm]
    & = \frac{
      \cos\bigl(\frac{\pi}{8}\bigr) \vert 00\rangle
      + \cos\bigl(-\frac{3\pi}{8}\bigr) \vert 01\rangle
      + \cos\bigl(\frac{5\pi}{8}\bigr) \vert 10\rangle
      + \cos\bigl(\frac{\pi}{8}\bigr) \vert 11\rangle}{\sqrt{2}}.
  \end{aligned}
  \]
  The probabilities for the four possible answer pairs $(a,b)$ are therefore as
  follows.
  \[
  \begin{aligned}
    \operatorname{Pr}\bigl((a,b)=(0,0)\bigr) & =
    \frac{1}{2}\cos^2\Bigl(\frac{\pi}{8}\Bigr) = \frac{2+\sqrt{2}}{8} \\[2mm]
    \operatorname{Pr}\bigl((a,b)=(0,1)\bigr) & =
    \frac{1}{2}\cos^2\Bigl(-\frac{3\pi}{8}\Bigr) = \frac{2-\sqrt{2}}{8}\\[2mm]
    \operatorname{Pr}\bigl((a,b)=(1,0)\bigr) & =
    \frac{1}{2}\cos^2\Bigl(\frac{5\pi}{8}\Bigr) = \frac{2-\sqrt{2}}{8}\\[2mm]
    \operatorname{Pr}\bigl((a,b)=(1,1)\bigr) & =
    \frac{1}{2}\cos^2\Bigl(\frac{\pi}{8}\Bigr) = \frac{2+\sqrt{2}}{8}
  \end{aligned}
  \]
  Again, we can obtain the probabilities that $a=b$ and $a\neq b$ by summing.
  \[
  \operatorname{Pr}(a = b) = \frac{2 + \sqrt{2}}{4} \qquad
  \operatorname{Pr}(a \neq b) = \frac{2 - \sqrt{2}}{4}
  \]
  For the question pair $(0,1)$, Alice and Bob win if $a=b$, and therefore they
  win in this case with probability
  \[
  \frac{2 + \sqrt{2}}{4}.
  \]

\item[Case 3:  $(x,y) = (1,0).$]
  In this case Alice performs $U_{\pi/4}$ on her qubit and Bob performs
  $U_{\pi/8}$ on his, so the state of the two qubits $(\mathsf{A},\mathsf{B})$
  after they perform their operations is
  \[
  \begin{aligned}
    \bigl(U_{\pi/4} \otimes U_{\pi/8}\bigr) \vert \phi^+\rangle
    & =
    \vert 00 \rangle \langle \psi_{\pi/4} \otimes \psi_{\pi/8}\vert
    \phi^+\rangle
    + \vert 01 \rangle \langle \psi_{\pi/4} \otimes\psi_{5\pi/8}\vert
    \phi^+\rangle \\ 
    & \qquad + \vert 10 \rangle \langle \psi_{3\pi/4} \otimes \psi_{\pi/8}\vert
    \phi^+\rangle
    + \vert 11 \rangle \langle \psi_{3\pi/4} \otimes \psi_{5\pi/8}\vert
    \phi^+\rangle\\[2mm]
    & = \frac{
      \cos\bigl(\frac{\pi}{8}\bigr) \vert 00\rangle
      + \cos\bigl(-\frac{3\pi}{8}\bigr) \vert 01\rangle
      + \cos\bigl(\frac{5\pi}{8}\bigr) \vert 10\rangle
      + \cos\bigl(\frac{\pi}{8}\bigr) \vert 11\rangle}{\sqrt{2}}.
  \end{aligned}
  \]
  The probabilities for the four possible answer pairs $(a,b)$ are therefore as
  follows.
  \[
  \begin{aligned}
    \operatorname{Pr}\bigl((a,b)=(0,0)\bigr) & =
    \frac{1}{2}\cos^2\Bigl(\frac{\pi}{8}\Bigr) = \frac{2+\sqrt{2}}{8} \\[2mm]
    \operatorname{Pr}\bigl((a,b)=(0,1)\bigr) & =
    \frac{1}{2}\cos^2\Bigl(-\frac{3\pi}{8}\Bigr) = \frac{2-\sqrt{2}}{8}\\[2mm]
    \operatorname{Pr}\bigl((a,b)=(1,0)\bigr) & =
    \frac{1}{2}\cos^2\Bigl(\frac{5\pi}{8}\Bigr) = \frac{2-\sqrt{2}}{8}\\[2mm]
    \operatorname{Pr}\bigl((a,b)=(1,1)\bigr) & =
    \frac{1}{2}\cos^2\Bigl(\frac{\pi}{8}\Bigr) = \frac{2+\sqrt{2}}{8}
  \end{aligned}
  \]
  We find, once again, that probabilities that $a=b$ and $a\neq b$ are as
  follows.
  \[
  \operatorname{Pr}(a = b) = \frac{2 + \sqrt{2}}{4}\qquad
  \operatorname{Pr}(a \neq b) = \frac{2 - \sqrt{2}}{4}
  \]
  For the question pair $(1,0)$, Alice and Bob win if $a=b$, so they win in
  this case with probability
  \[
  \frac{2 + \sqrt{2}}{4}.
  \]

\item[Case 4:  $(x,y) = (1,1).$]  
  The last case is a little bit different, as we might expect because the
  winning condition is different in this case. When $x$ and $y$ are both $1$,
  Alice and Bob win when $a$ and $b$ are \emph{different}. In this case Alice
  performs $U_{\pi/4}$ on her qubit and Bob performs $U_{-\pi/8}$ on his, so
  the state of the two qubits $(\mathsf{A},\mathsf{B})$ after they perform
  their operations is
  \[
  \begin{aligned}
    \bigl(U_{\pi/4} \otimes U_{-\pi/8}\bigr) \vert
    \phi^+\rangle\hspace{-0.6pt}
    & =
    \vert 00 \rangle \langle \psi_{\pi/4} \otimes \psi_{-\pi/8}\vert
    \phi^+\rangle
    + \vert 01 \rangle \langle \psi_{\pi/4} \otimes\psi_{3\pi/8}\vert
    \phi^+\rangle \\
    & \qquad + \vert 10 \rangle \langle \psi_{3\pi/4} \otimes
    \psi_{-\pi/8}\vert \phi^+\rangle
    + \vert 11 \rangle \langle \psi_{3\pi/4} \otimes \psi_{3\pi/8}\vert
    \phi^+\rangle\\[2mm]
    & = \frac{
      \cos\bigl(\frac{3\pi}{8}\bigr) \vert 00\rangle
      + \cos\bigl(-\frac{\pi}{8}\bigr) \vert 01\rangle
      + \cos\bigl(\frac{7\pi}{8}\bigr) \vert 10\rangle
      + \cos\bigl(\frac{3\pi}{8}\bigr) \vert 11\rangle}{\sqrt{2}}.
  \end{aligned}
  \]
  The probabilities for the four possible answer pairs $(a,b)$ are therefore as
  follows.
  \[
  \begin{aligned}
    \operatorname{Pr}\bigl((a,b)=(0,0)\bigr) & =
    \frac{1}{2}\cos^2\Bigl(\frac{3\pi}{8}\Bigr) = \frac{2-\sqrt{2}}{8} \\[2mm]
    \operatorname{Pr}\bigl((a,b)=(0,1)\bigr) & =
    \frac{1}{2}\cos^2\Bigl(-\frac{\pi}{8}\Bigr) = \frac{2+\sqrt{2}}{8}\\[2mm]
    \operatorname{Pr}\bigl((a,b)=(1,0)\bigr) & =
    \frac{1}{2}\cos^2\Bigl(\frac{7\pi}{8}\Bigr) = \frac{2+\sqrt{2}}{8}\\[2mm]
    \operatorname{Pr}\bigl((a,b)=(1,1)\bigr) & =
    \frac{1}{2}\cos^2\Bigl(\frac{3\pi}{8}\Bigr) = \frac{2-\sqrt{2}}{8}
  \end{aligned}
  \]
  The probabilities have effectively swapped places from in the three other
  cases.
  We obtain the probabilities that $a=b$ and $a\neq b$ by summing.
  \[
  \operatorname{Pr}(a = b) = \frac{2 - \sqrt{2}}{4}\qquad
  \operatorname{Pr}(a \neq b) = \frac{2 + \sqrt{2}}{4}
  \]
  For the question pair $(1,1)$, Alice and Bob win if $a\neq b$, and therefore
  they win in this case with probability
  \[
  \frac{2 + \sqrt{2}}{4}.
  \]
\end{description}

They win in every case with the same probability:
\[
\frac{2 + \sqrt{2}}{4} \approx 0.85.
\]
This is therefore the probability that they win overall.
That's significantly better than any classical strategy can do for this game;
classical strategies have winning probability bounded by $3/4.$ And that makes
this a very interesting example.

This happens to be the \emph{optimal} winning probability for quantum
strategies.
That is, we can't do any better than this, no matter what entangled state or
measurements we choose.
This fact is known as \emph{Tsirelson's inequality}, named for Boris Tsirelson
who first proved it --- and who first described the CHSH experiment as a game.

\subsubsection{Geometric picture}

It is possible to think about the strategy described above geometrically, which
may be helpful for understanding the relationships among the various angles
chosen for Alice and Bob's operations.

What Alice effectively does is to choose an angle $\alpha$, depending on her
question~$x$, and then to apply $U_{\alpha}$ to her qubit and measure.
Similarly, Bob chooses an angle~$\beta$, depending on $y$, and then he applies
$U_{\beta}$ to his qubit and measures.
We've chosen~$\alpha$ and~$\beta$ like so.
\[
\begin{aligned}
  \alpha & = \begin{cases}
    0 & x=0\\
    \pi/4 & x=1
  \end{cases}\\[2mm]
  \beta & = \begin{cases}
    \pi/8 & y = 0\\
    -\pi/8 & y = 1
  \end{cases}
\end{aligned}
\]
\pagebreak

\begin{figure}[!ht]
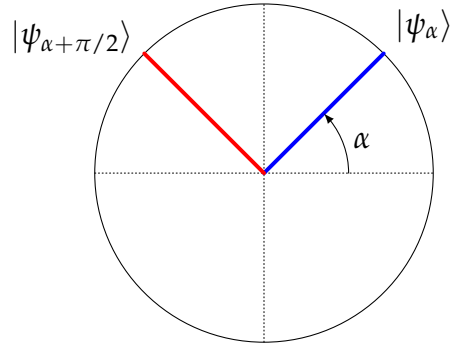

  \begin{center}

  \end{center}
  \caption{Alice's basis is determined by the angle $\alpha$.}
  \label{fig:alpha-basis}
\end{figure}

\begin{figure}[!ht]
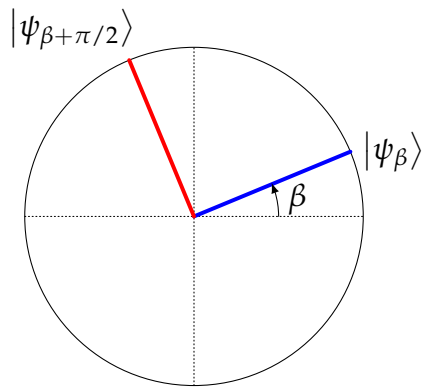

  \begin{center}
    %
    
  \end{center}
  \caption{Bob's basis is determined by the angle $\beta$.}
  \label{fig:beta-basis}
\end{figure}

For the moment, though, let's take $\alpha$ and $\beta$ to be arbitrary.
By choosing $\alpha$, Alice effectively defines an orthonormal basis of vectors
as is shown in Figure~\ref{fig:alpha-basis}.
Bob does likewise, except that his angle is $\beta$, as illustrated in
Figure~\ref{fig:beta-basis}.
The colors of the vectors correspond to Alice and Bob's answers: blue for $0$
and red for $1.$

Now, if we combine together \eqref{eq:angle-inner-product-1}
and \eqref{eq:angle-inner-product-2} we get the formula
\[
\langle \psi_{\alpha} \otimes\psi_{\beta} \vert \phi^+ \rangle
= \frac{1}{\sqrt{2}} \langle \psi_{\alpha} \vert \psi_{\beta} \rangle,
\]
which works for all real numbers $\alpha$ and $\beta.$

Following the same sort of analysis that we went through above, but with
$\alpha$ and $\beta$ being variables, we find this:
\[
\begin{aligned}
  \bigl(U_{\alpha} \otimes U_{\beta}\bigr) \vert \phi^+\rangle
  \hspace{-24mm}\\[1mm]
  & =
  \vert 00 \rangle \langle \psi_{\alpha} \otimes \psi_{\beta}\vert \phi^+\rangle
  + \vert 01 \rangle \langle \psi_{\alpha} \otimes\psi_{\beta + \pi/2}\vert
  \phi^+\rangle \\
  & \qquad + \vert 10 \rangle \langle \psi_{\alpha+\pi/2} \otimes
  \psi_{\beta}\vert \phi^+\rangle
  + \vert 11 \rangle \langle \psi_{\alpha+\pi/2} \otimes
  \psi_{\beta+\pi/2}\vert \phi^+\rangle\\[2mm]
  & = \frac{
    \langle \psi_\alpha \vert \psi_\beta \rangle \vert 00\rangle
    + \langle \psi_\alpha \vert \psi_{\beta+\pi/2} \rangle \vert 01\rangle
    + \langle \psi_{\alpha+\pi/2} \vert \psi_\beta \rangle \vert 10\rangle
    + \langle \psi_{\alpha+\pi/2} \vert \psi_{\beta+\pi/2} \rangle \vert
    11\rangle
  }{\sqrt{2}}.
\end{aligned}
\]
We conclude these two formulas:
\[
\begin{aligned}
  \operatorname{Pr}(a = b)  & = \frac{1}{2} \vert \langle \psi_\alpha \vert
  \psi_\beta \rangle \vert^2
  + \frac{1}{2} \vert \langle \psi_{\alpha+\pi/2} \vert \psi_{\beta+\pi/2}
  \rangle \vert^2
  = \cos^2(\alpha - \beta)\\[2mm]
  \operatorname{Pr}(a \neq b) & = \frac{1}{2} \vert \langle \psi_\alpha \vert
  \psi_{\beta+\pi/2} \rangle \vert^2
  + \frac{1}{2} \vert \langle \psi_{\alpha+\pi/2} \vert \psi_\beta \rangle
  \vert^2
  = \sin^2(\alpha - \beta).
\end{aligned}
\]

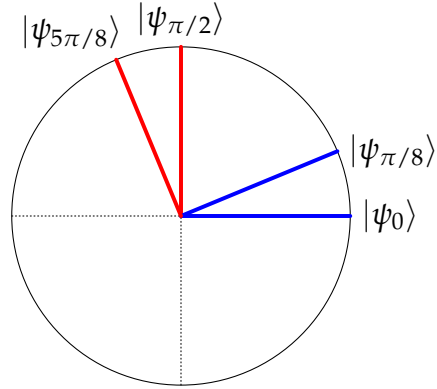
\begin{figure}[t]
  \begin{center}
    \begin{tikzpicture}[scale=2.25,cap=round,>=latex]
      
      \coordinate (a00) at (1,0);
      \coordinate (a01) at (0,1);
      \coordinate (a10) at (0.7071, 0.7071);
      \coordinate (a11) at (-0.7071, 0.7071);
      
      \coordinate (b00) at (0.9239,0.3827);
      \coordinate (b01) at (-0.3827,0.9239);
      \coordinate (b10) at (0.9239,-0.3827);
      \coordinate (b11) at (0.3827,0.9239);
      
      \draw (0,0) circle(1);
     
      \node at (2,0) {};
      \node at (-2,0) {};
      
      \draw[densely dotted] (-1,0) -- (1,0) {};
      \draw[densely dotted] (0,-1) -- (0,1) {};

      \node[anchor = west] at (a00) {$\ket*{\psi_0}$};
      \node[anchor = south] at (a01) {%
        $\ket*{\psi_{\pi/2}}$
      };
      
      \draw[line width = 1.5pt, draw = blue] (0,0) -- (a00);
      \draw[line width = 1.5pt, draw = red] (0,0) -- (a01);
      
      \node[anchor = west] at (b00) {
        $\ket*{\psi_{\pi/8}}$};
      \node[anchor = south east] at ([xshift=1mm]b01.center) {
        $\ket*{\psi_{5\pi/8}}$};
      \draw[line width = 1.5pt, draw = blue] (0,0) -- (b00);
      \draw[line width = 1.5pt, draw = red] (0,0) -- (b01);
      
    \end{tikzpicture}
  \end{center}
  \caption{Alice and Bob's bases when $x=0$ and $y=0$.}
  \label{fig:strategy00}
\end{figure}

These equations can be connected to the figures above by imagining that we
superimpose the bases chosen by Alice and Bob.
In particular, when $(x,y) = (0,0)$, Alice and Bob choose $\alpha = 0$ and
$\beta = \pi/8$, resulting in the bases shown in Figure~\ref{fig:strategy00}.
The angle between the red vectors is $\pi/8$, which is the same as the angle
between the two blue vectors.
The probability that Alice and Bob's outcomes agree is the cosine-squared of
this angle,
\[
\cos^2\Bigl(\frac{\pi}{8}\Bigr) = \frac{2 + \sqrt{2}}{4},
\]
while the probability they disagree is the sine-squared of this angle,
\[
\sin^2\Bigl(\frac{\pi}{8}\Bigr) = \frac{2 - \sqrt{2}}{4}.
\]

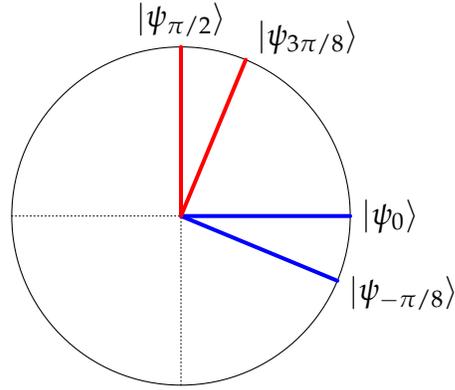
\begin{figure}[t]
  \begin{center}
    \begin{tikzpicture}[scale=2.25,cap=round,>=latex]
      
      \coordinate (a00) at (1,0);
      \coordinate (a01) at (0,1);
      \coordinate (a10) at (0.7071, 0.7071);
      \coordinate (a11) at (-0.7071, 0.7071);
      
      \coordinate (b00) at (0.9239,0.3827);
      \coordinate (b01) at (-0.3827,0.9239);
      \coordinate (b10) at (0.9239,-0.3827);
      \coordinate (b11) at (0.3827,0.9239);
      
      \draw (0,0) circle(1);
           
      \node at (2,0) {};
      \node at (-2,0) {};

      \draw[densely dotted] (-1,0) -- (1,0) {};
      \draw[densely dotted] (0,-1) -- (0,1) {};

      \node[anchor = west] at (a00) {$\ket*{\psi_0}$};
      \node[anchor = south] at (a01) {%
        $\ket*{\psi_{\pi/2}}$
      };
      
      \draw[line width = 1.5pt, draw = blue] (0,0) -- (a00);
      \draw[line width = 1.5pt, draw = red] (0,0) -- (a01);

      \node[anchor = 170] at (b10) {$\ket*{\psi_{-\pi/8}}$};
      \node[anchor = 200] at (b11.center) {%
        $\ket*{\psi_{3\pi/8}}$};
      \draw[line width = 1.5pt, draw = blue] (0,0) -- (b10);
      \draw[line width = 1.5pt, draw = red] (0,0) -- (b11);
    \end{tikzpicture}
  \end{center}
  \caption{Alice and Bob's bases when $x=0$ and $y=1$.}
  \label{fig:strategy01}
\end{figure}

When $(x,y) = (0,1)$, Alice and Bob choose $\alpha = 0$ and $\beta = -\pi/8$,
resulting in the bases shown in Figure~\ref{fig:strategy01}.
The angle between the red vectors is again $\pi/8$, as is the angle between the
blue vectors.
The probability that Alice and Bob's outcomes agree is again the cosine-squared
of this angle,
\[
\cos^2\Bigl(\frac{\pi}{8}\Bigr) = \frac{2 + \sqrt{2}}{4},
\]
while the probability they disagree is the sine-squared of this angle,
\[
\sin^2\Bigl(\frac{\pi}{8}\Bigr) = \frac{2 - \sqrt{2}}{4}.
\]

\begin{figure}[t]
  \begin{center}
    \begin{tikzpicture}[scale=2.25,cap=round,>=latex]
      
      \coordinate (a00) at (1,0);
      \coordinate (a01) at (0,1);
      \coordinate (a10) at (0.7071, 0.7071);
      \coordinate (a11) at (-0.7071, 0.7071);
      
      \coordinate (b00) at (0.9239,0.3827);
      \coordinate (b01) at (-0.3827,0.9239);
      \coordinate (b10) at (0.9239,-0.3827);
      \coordinate (b11) at (0.3827,0.9239);
      
      \draw (0,0) circle(1);
           
      \node at (2,0) {};
      \node at (-2,0) {};

      \draw[densely dotted] (-1,0) -- (1,0) {};
      \draw[densely dotted] (0,-1) -- (0,1) {};
      
      \node[anchor = south west] at (a10) {%
        $\ket*{\psi_{\pi/4}}$};
      \node[anchor = 350] at (a11.center) {%
        $\ket*{\psi_{3\pi/4}}$};
      \draw[line width = 1.5pt, draw = blue] (0,0) -- (a10);
      \draw[line width = 1.5pt, draw = red] (0,0) -- (a11);
      
      \node[anchor = west] at (b00) {%
        $\ket*{\psi_{\pi/8}}$};
      \node[anchor = 320] at (b01.center) {
        $\ket*{\psi_{5\pi/8}}$};
      \draw[line width = 1.5pt, draw = blue] (0,0) -- (b00);
      \draw[line width = 1.5pt, draw = red] (0,0) -- (b01);
      
    \end{tikzpicture}
  \end{center}
  \caption{Alice and Bob's bases when $x=1$ and $y=0$.}
  \label{fig:strategy10}
\end{figure}
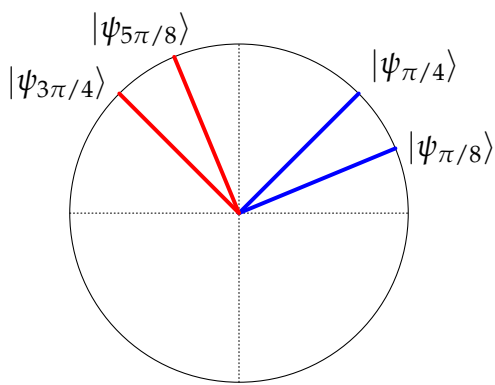

When $(x,y) = (1,0)$, Alice and Bob choose $\alpha = \pi/4$ and $\beta =
\pi/8$, resulting in the bases shown in Figure~\ref{fig:strategy10}.
The bases have changed but the angles haven't --- once again the angle between
vectors of the same color is $\pi/8.$
The probability that Alice and Bob's outcomes agree is
\[
\cos^2\Bigl(\frac{\pi}{8}\Bigr) = \frac{2 + \sqrt{2}}{4},
\]
and the probability they disagree is
\[
\sin^2\Bigl(\frac{\pi}{8}\Bigr) = \frac{2 - \sqrt{2}}{4}.
\]

\begin{figure}[t]
  \begin{center}

    \begin{tikzpicture}[scale=2.25,cap=round,>=latex]
      
      \coordinate (a00) at (1,0);
      \coordinate (a01) at (0,1);
      \coordinate (a10) at (0.7071, 0.7071);
      \coordinate (a11) at (-0.7071, 0.7071);
      
      \coordinate (b00) at (0.9239,0.3827);
      \coordinate (b01) at (-0.3827,0.9239);
      \coordinate (b10) at (0.9239,-0.3827);
      \coordinate (b11) at (0.3827,0.9239);
      
      \draw (0,0) circle(1);
      
      \draw[densely dotted] (-1,0) -- (1,0) {};
      \draw[densely dotted] (0,-1) -- (0,1) {};

      \node[anchor = 190] at (a10) {%
        $\ket*{\psi_{\pi/4}}$};
      \node[anchor = 340] at (a11.center) {%
        $\ket*{\psi_{3\pi/4}}$};
      \draw[line width = 1.5pt, draw = blue] (0,0) -- (a10);
      \draw[line width = 1.5pt, draw = red] (0,0) -- (a11);
      
      \node[anchor = 170] at (b10) {%
        $\ket*{\psi_{-\pi/8}}$};
      \node[anchor = 220] at (b11.center) {%
        $\ket*{\psi_{3\pi/8}}$};
      \draw[line width = 1.5pt, draw = blue] (0,0) -- (b10);
      \draw[line width = 1.5pt, draw = red] (0,0) -- (b11);
      
    \end{tikzpicture}
    
  \end{center}
  \caption{Alice and Bob's bases when $x=1$ and $y=1$.}
  \label{fig:strategy11}
\end{figure}
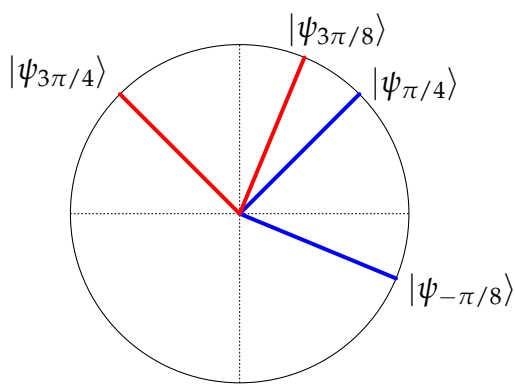

When $(x,y) = (1,1)$, Alice and Bob choose $\alpha = \pi/4$ and $\beta =
-\pi/8.$ This results in the bases shown in Figure~\ref{fig:strategy11},
which reveals that something different has happened.
By the way the angles were chosen, this time the angle between vectors having
the same color is $3\pi/8$ rather than $\pi/8.$
The probability that Alice and Bob's outcomes agree is still the cosine-squared
of this angle, but this time the value is
\[
\cos^2\Bigl(\frac{3\pi}{8}\Bigr) = \frac{2 - \sqrt{2}}{4}.
\]
The probability the outcomes disagree is the sine-squared of this angle, which
in this case is this:
\[
\sin^2\Bigl(\frac{3\pi}{8}\Bigr) = \frac{2 + \sqrt{2}}{4}.
\]

\subsection{Remarks}

The basic idea of an experiment like the CHSH game, where entanglement leads to
statistical results that are inconsistent with purely classical reasoning, is
due to John Bell, the namesake of the Bell states.
For this reason, people often refer to experiments of this sort as \emph{Bell
tests}.
Sometimes people also refer to \emph{Bell's theorem}, which can be formulated
in different ways --- but the essence of it is that quantum mechanics is not
compatible with so-called \emph{local hidden variable theories}.
The CHSH game is a particularly clean and simple example of a Bell test, and
can be viewed as a proof, or demonstration, of Bell's theorem.

The CHSH game offers a way to experimentally test the theory of quantum
information.
Experiments can be performed that implement the CHSH game, and test the sorts
of strategies based on entanglement described above.
This provides us with a high degree of confidence that entanglement is real ---
and unlike the sometimes vague or poetic ways that we come up with to explain
entanglement, the CHSH game gives us a concrete and testable way to
\emph{observe} entanglement.
The 2022 Nobel Prize in Physics acknowledges the importance of this line of
work: the prize was awarded to Alain Aspect, John Clauser (the C in CHSH) and
Anton Zeilinger for observing entanglement through Bell tests on entangled
photons.

\stopcontents[part]


\unit[Fundamentals of Quantum Algorithms]{Fundamentals of\\[-1mm]
  Quantum Algorithms}
\label{unit:fundamentals-of-quantum-algorithms}

This unit explores computational advantages of quantum information, including
what we can do with quantum computers and their advantages over classical
computers.
It begins with quantum query algorithms, which offer simple proof-of-concept
demonstrations for quantum algorithms, and then moves on to quantum algorithms
for problems including integer factorization and unstructured search.

\begin{trivlist}
  \setlength{\parindent}{0mm}
  \setlength{\parskip}{2mm}
  \setlength{\itemsep}{1mm}
\item
  \textbf{Lesson 5: Quantum Query Algorithms}

  This lesson is on the quantum query model of computation. It describes a
  progression of quantum algorithms that offer advantages over classical
  algorithms within this model, including Deutsch's algorithm, the
  Deutsch--Jozsa algorithm, and Simon's algorithm.

  Lesson video URL: \url{https://youtu.be/2wticzHE1vs}

\item
  \textbf{Lesson 6: Quantum Algorithmic Foundations}

  This lesson discusses a notion of computational cost for both classical and
  quantum computations, and describes a method through which classical
  computations can be performed by quantum circuits at roughly the same
  cost. This opens up many interesting possibilities for quantum algorithms by
  allowing them to use classical computations as subroutines.

  Lesson video URL: \url{https://youtu.be/2wxxvwRGANQ}
  
\item
  \textbf{Lesson 7: Phase Estimation and Factoring}

  This lesson discusses the phase estimation problem and a quantum algorithm to
  solve it. By applying this algorithm to a number-theoretic problem known as
  the order finding problem, we obtain Shor's algorithm, which is an efficient
  quantum algorithm for the integer factorization problem.

  Lesson video URL: \url{https://youtu.be/4nT0BTUxhJY}
  
\item
  \textbf{Lesson 8: Grover's Algorithm}

  This lesson is about Grover's algorithm, which is a quantum algorithm for
  so-called unstructured search problems that offers a quadratic improvement
  over classical algorithms --- meaning that it requires a number
  of operations on the order of the square-root of the number required to solve
  unstructured search classically.

  Lesson video URL: \url{https://youtu.be/hnpjC8WQVrQ}

\end{trivlist}


\lesson{Quantum Query Algorithms}
\label{lesson:quantum-query-algorithms}

In this first lesson of the unit, we'll formulate a simple algorithmic
framework --- known as the \emph{query model} --- and explore the advantages
that quantum computers offer within this framework.

The query model of computation is like a Petri dish for quantum algorithmic
ideas.
It's rigid and unnatural in the sense that it doesn't accurately represent the
sorts of computational problems we generally care about in practice, but it has
nevertheless proved to be incredibly useful as a tool for developing quantum
algorithmic techniques.
This includes the ones that power the most well-known quantum algorithms, such
as Shor's algorithm for integer factorization.
The query model also happens to be a very useful framework for
\emph{explaining} quantum algorithmic techniques.

After introducing the query model itself, we'll discuss the very first quantum
algorithm that was discovered, which is \emph{Deutsch's algorithm,} along with
an extension of Deutsch's algorithm known as the \emph{Deutsch--Jozsa
algorithm}.
These algorithms demonstrate quantifiable advantages of quantum over classical
computers within the context of the query model.
We'll then discuss a quantum algorithm known as \emph{Simon's algorithm,} which
offers a more robust and satisfying advantage of quantum over classical
computations, for reasons that will be explained when we get to it.

\section{The query model of computation}

\begin{figure}[!ht]
  \begin{center}
    \begin{tikzpicture}[
        >=latex,
        line width=0.6pt,
        scale=1.5,
        box/.style={%
          inner sep = 0,
          fill = CircuitBlue,
          draw = CircuitBlue,
          text = white,
          minimum width = 32mm,
          minimum height = 20mm}
      ]
      
      \node (in) at (-2.5,0) {input};
      \node[box] (computation) at (0,0) {computation};
      \node (out) at (2.5,0) {output};

      \draw[->] (in.east) -- (computation.west);
      \draw[->] (computation.east) -- (out.west);

    \end{tikzpicture}    
  \end{center}
  \caption{A simple abstraction of a standard model of computation.}
  \label{fig:standard-computation}
\end{figure}
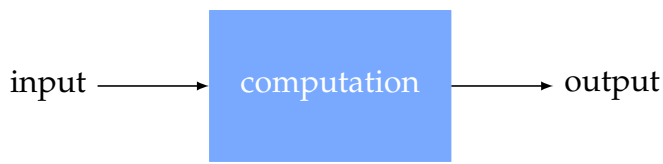

When we model computations in mathematical terms, we typically have in mind the
sort of process represented in Figure~\ref{fig:standard-computation}, where
information is provided as input, a computation takes place, and output is
produced.

While it is true that the computers we use today continuously receive input and
produce output, essentially interacting with both us and with other computers
in a way not reflected by the figure, the intention is not to represent the
ongoing operation of computers.
Rather, it is to create a simple abstraction of computation, focusing on
isolated computational tasks.
For example, the input might encode a number, a vector, a matrix, a graph, a
description of a molecule, or something more complicated, while the output
encodes a solution to the computational task we have in mind.

The key point is that the input is provided to the computation, usually in the
form of a binary string, with no part of it being hidden.

\subsection{Description of the model}

\begin{figure}[!ht]
  \begin{center}
    \begin{tikzpicture}[
        >=latex,
        line width=0.6pt,
        scale=1.5,
        box/.style={%
          inner sep = 0,
          fill = CircuitBlue,
          draw = CircuitBlue,
          text = white,
          minimum width = 32mm,
          minimum height = 20mm},
        smallbox/.style={%
          inner sep = 0,
          fill = CircuitBlue,
          draw = CircuitBlue,
          text = white,
          minimum width = 28mm,
          minimum height = 16mm}
      ]
      
      \node[smallbox] (in) at (0,2) {input};
      \node[box] (computation) at (0,0) {computation};
      \node (out) at (2.5,0) {output};

      \foreach \x in {-8,-4,6} {
        \draw[->] ([xshift=\x mm]computation.north)
        -- ([xshift=\x mm]in.south);
      }

      \foreach \x in {-6,-2,8} {
        \draw[->] ([xshift=\x mm]in.south)
        -- ([xshift=\x mm]computation.north);
      }
      
      \draw[->] (computation.east) -- (out.west);

      \node at (0.2,1.05) {$\cdots$};
      \node at (-1.5,1.05) {queries};

    \end{tikzpicture}
  \end{center}
  \caption{An abstraction of the query model of computation}
  \label{fig:query-computation}
\end{figure}
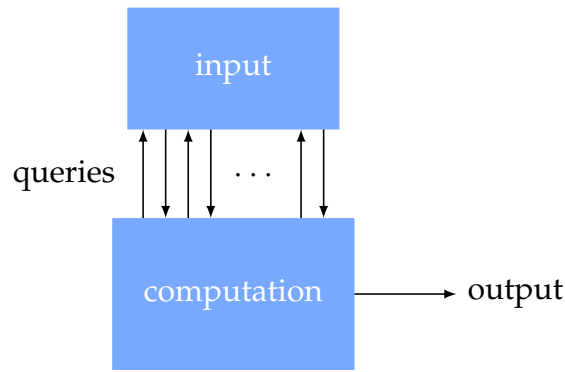

In the \emph{query model} of computation, the entire input is not provided to
the computation like in a more standard model suggested above.
Rather, the input is made available in the form of a \emph{function}, which the
computation accesses by making \emph{queries}, as is depicted in
Figure~\ref{fig:query-computation}.
Alternatively, we may view computations in the query model as having
random access to bits (or segments of bits) of the input.

We often refer to the input as being provided by an \emph{oracle} or
\emph{black box} in the context of the query model.
Both terms suggest that a complete description of the input is hidden from the
computation, with the only way to access it being to ask questions.
It is as if we're consulting the Oracle at Delphi about the input: she won't
tell us everything she knows, she only answers specific questions.
The term \emph{black box} makes sense especially when we think about the input
as being represented by a function;
we cannot look inside the function and understand how it works, we can only
evaluate it on arguments we select.

We're going to be working exclusively with binary strings in this lesson, as
opposed to strings containing different symbols, so let's write
$\Sigma = \{0,1\}$ hereafter to refer to the binary alphabet for convenience.
We'll be thinking about different computational problems, with some simple
examples described shortly, but for all of them the input will be represented
by a function taking the form
\[
f:\Sigma^n \rightarrow \Sigma^m
\]
for two positive integers $n$ and $m.$
Naturally, we could choose a different name in place of $f,$ but we'll stick
with $f$ throughout the lesson.

To say that a computation makes a \emph{query} means that some string
$x \in \Sigma^n$ is selected, and then the string $f(x)\in\Sigma^m$ is made
available to the computation by the oracle.
The precise way that this works for quantum algorithms will be discussed
shortly --- we need to make sure that this is possible to do with a unitary
quantum operation allowing queries to be made in superposition --- but for now
we can think about it intuitively at a high level.

Finally, the way that we'll measure efficiency of query algorithms is simple:
we'll count the \emph{number of queries} they require.
This is related to the time required to perform a computation, but it's not
exactly the same because we're ignoring the time for operations other than the
queries, and we're also treating the queries as if they each have unit cost.
We can take the operations besides the queries into account if we wish (and
this is sometimes done), but restricting our attention just to the number of
queries helps to keep things simple.

\pagebreak

\subsection{Examples of query problems}

Here are a few simple examples of query problems.
\begin{description}[leftmargin=0mm]

\item[Or:]
  The input function takes the form $f:\Sigma^n \rightarrow\Sigma$
  (so $m=1$ for this problem). The task is to output $1$ if there
  exists a string $x\in\Sigma^n$ for which $f(x) = 1,$ and to output $0$ if
  there is no such string. If we think about the function $f$ as representing a
  sequence of $2^n$ bits to which we have random access, the problem is to
  compute the OR of these bits.

\item[Parity:]
  The input function again takes the form $f:\Sigma^n \rightarrow \Sigma.$ The
  task is to determine whether the number of strings $x\in\Sigma^n$ for which
  $f(x) = 1$ is \emph{even} or \emph{odd}.
  To be precise, the required output is $0$ if the set
  $\{x\in\Sigma^n : f(x) = 1\}$ has an even number of elements and $1$ if it
  has an odd number of elements. If we think about the function $f$ as
  representing a sequence of $2^n$ bits to which we have random access, the
  problem is to compute the parity (or exclusive-OR) of these bits.

\item[Minimum:]
  The input function takes the form $f:\Sigma^n \rightarrow \Sigma^m$ for any
  choices of positive integers $n$ and $m.$ The required output is the string
  $y \in \{f(x) : x\in\Sigma^n\}$ that comes first in the lexicographic (i.e.,
  dictionary) ordering of $\Sigma^m.$
  If we think about the function $f$ as representing a sequence of $2^n$
  integers encoded as strings of length $m$ in binary notation to which we have
  random access, the problem is to compute the minimum of these integers.
\end{description}

We also sometimes consider query problems where we have a \emph{promise} on the
input.
What this means is that we're given some sort of guarantee on the input, and
we're not responsible for what happens when this guarantee is not met.
Another way to describe this type of problem is to say that some input
functions (the ones for which the promise is not satisfied) are considered as
``don't care'' inputs.
No requirements at all are placed on algorithms when they're given
``don't care'' inputs. 
Here's one example of a problem with a promise:
\begin{description}[leftmargin=0mm]
\item[Unique search.]
  The input function takes the form $f:\Sigma^n \rightarrow \Sigma,$ and we are
  promised that there is exactly one string $z\in\Sigma^n$ for which
  $f(z) = 1,$ with $f(x) = 0$ for all strings $x\neq z.$ The task is to find
  this unique string $z.$
\end{description}

All four of the examples just described are natural, in the sense that they're
easy to describe and we can imagine a variety of situations or contexts in
which they might arise.
In contrast, some query problems aren't ``natural'' like this at all.
In fact, in the study of the query model, we sometimes come up with very
complicated and highly contrived problems where it's difficult to imagine that
anyone would ever actually want to solve them in practice.
This doesn't mean that the problems aren't interesting, though!
Things that might seem contrived or unnatural at first can provide unexpected
clues or inspire new ideas.
Shor's quantum algorithm for factoring, which was inspired by Simon's
algorithm, is a great example.
It's also an important part of the study of the query model to look for
extremes, which can shed light on both the potential advantages and the
limitations of quantum computing.

\subsection{Query gates}

When we're describing computations with circuits, queries are made by special
gates called \emph{query gates}.

\begin{figure}[!ht]
  \begin{center}
    \begin{tikzpicture}[
        baseline=(current bounding box.north),
        scale=1.5,
        line width=0.6pt,
        control/.style={%
          circle,
          fill=CircuitBlue,
          minimum size = 5pt,
          inner sep=0mm},
        gate/.style={%
          inner sep = 0,
          fill = CircuitBlue,
          draw = CircuitBlue,
          text = white,
          minimum size = 6mm}
      ]
      
      \node[gate, minimum height=32mm, minimum width=20mm]
      (U) at (0,0) {$f$};
      
      \node (In) at (-1.5,0) {};
      \node (Out) at (1.5,0) {};
      
      \foreach \y in {-7.5,-6,...,7.5} {
        \draw ([yshift=\y mm]In.east) -- ([yshift=\y mm]U.west) {};
      }
      
      \foreach \y in {-6,-4.5,...,6} {
        \draw ([yshift=\y mm]Out.west) -- ([yshift=\y mm]U.east) {};
      }
      
      \node[anchor=east] at (In) {$x\; \left\{ \rule{0mm}{14mm} \right.$};
      
      \node[anchor=west] at (Out) {$\left.\rule{0mm}{12mm}\right\}\;f(x)$};
      
    \end{tikzpicture}
  \end{center}
  \caption{A classical query gate.}
  \label{fig:classical-query-gate}
\end{figure}
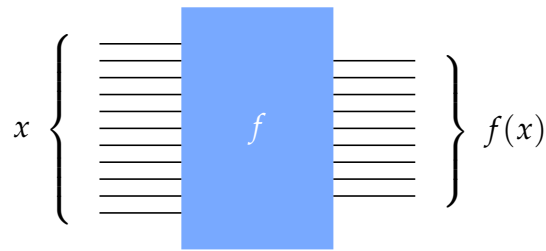

The simplest way to define query gates for classical Boolean circuits is to
simply allow them to compute the input function $f$ directly, as
Figure~\ref{fig:classical-query-gate} suggests.

When a Boolean circuit is created for a query problem, the input function $f$
is accessed through these gates, and the number of queries that the circuit
makes is simply the number of query gates that appear in the circuit.
The input wires of the Boolean circuit itself are initialized to fixed values,
which should be considered as part of the algorithm (as opposed to being inputs
to the problem).

\begin{figure}[!ht]
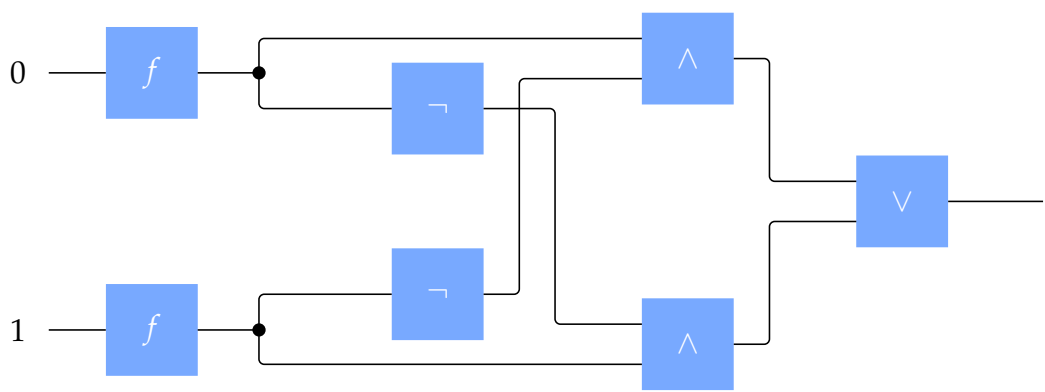

  \begin{center}

  \end{center}
  \caption{A Boolean circuit that solves the parity problem for a function
    $f:\Sigma\rightarrow\Sigma$.}
  \label{fig:classical-parity}
\end{figure}

For example, Figure~\ref{fig:classical-parity} describes a Boolean circuit with
classical query gates that solves the parity problem described above for a
function of the form $f:\Sigma\rightarrow\Sigma$.
This algorithm makes two queries because there are two query gates.
The way it works is that the function $f$ is queried on the two possible
inputs, $0$ and $1,$ and the results are plugged into a Boolean circuit that
computes the XOR.
This particular circuit appeared as an example of a Boolean circuit in
Lesson~\ref{lesson:quantum-circuits}
\emph{(Quantum Circuits)}.

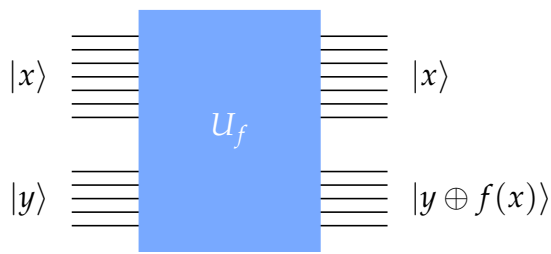
\begin{figure}[!ht]
  \begin{center}
    \begin{tikzpicture}[
        scale=1.8,
        line width=0.6pt,        
        control/.style={%
          circle,
          fill=CircuitBlue,
          minimum size = 5pt,
          inner sep=0mm},
        gate/.style={%
          inner sep = 0,
          fill = CircuitBlue,
          draw = CircuitBlue,
          text = white,
          minimum size = 6mm}
      ]

      \node (in_top) at (-1.25,0.4) {};
      \node (in_bottom) at (-1.25,-0.5) {};
      
      \node (out_top) at (1.25,0.4) {};
      \node (out_bottom) at (1.25,-0.5) {};

      \node[anchor=east] at (in_top) {$\ket{x}$};
      \node[anchor=east] at (in_bottom) {$\ket{y}$};
      
      \node[anchor=west] at (out_top) {$\ket{x}$};        
      \node[anchor=west] at (out_bottom) {$\ket{y \oplus f(x)}$};

      \foreach \y in {-3,-2,...,3} {
        \draw ([yshift=\y mm]in_top.east) --
        ([yshift=\y mm]out_top.west) {};
      }

      \foreach \y in {-2,-1,...,2} {
        \draw ([yshift=\y mm]in_bottom.east) --
        ([yshift=\y mm]out_bottom.west) {};
      }
      
      \node[gate, minimum height=32mm, minimum width=24mm]
      (U) at (0,0) {$U_f$};

    \end{tikzpicture}
  \end{center}
  \caption{The action of a unitary query gate $U_f$ on standard basis inputs.}
  \label{fig:unitary-query-gate}
\end{figure}

For quantum circuits, this definition of query gates doesn't work, because
these gates will be non-unitary for some choices of the function $f.$
So, what we do instead is to define \emph{unitary query gates} that operate
on standard basis states as shown in Figure~\ref{fig:unitary-query-gate}.

Here, our assumption is that $x\in\Sigma^n$ and $y\in\Sigma^m$ are arbitrary
strings.
The notation $y\oplus f(x)$ refers to the \emph{bitwise exclusive-OR} of two
strings, which have length~$m$ in this case.
For example, $001 \oplus 101 = 100.$

Intuitively speaking, what the gate $U_f$ does (for any chosen function $f$) is
to echo the top input string $x$ and XOR the function value $f(x)$ onto the
bottom input string $y,$ which is a unitary operation for every choice for the
function $f.$
In fact, it's a deterministic operation, and it is its own inverse.
This implies that, as a matrix, $U_f$ is always a permutation matrix,
meaning a matrix with a single $1$ in each row and each column, with all other
entries being $0.$
Applying a permutation matrix to a vector simply shuffles the entries of the
vector (hence the term \emph{permutation matrix}), and therefore does not
change that vector's Euclidean norm --- revealing that permutation matrices are
always unitary.

It should be highlighted that, when we analyze query algorithms by simply
counting the number of queries that a query algorithm makes, we're completely
ignoring the difficulty of physically constructing the query gates --- for both
the classical and quantum versions just described.
Intuitively speaking, the construction of the query gates is part of the
preparation of the input, not part of finding a solution.

That might seem unreasonable, but we must keep in mind that we're not trying to
describe practical computing or fully account for the resources required.
Rather, we're defining a theoretical model that helps to shed light on the
potential advantages of quantum computing.
We'll have more to say about this point in the lesson following this one when
we turn our attention to a more standard model of computation where inputs are
given explicitly to circuits as binary strings.

\section{Deutsch's algorithm}

Deutsch's algorithm solves the parity problem for the special case that
$n = 1.$ 
In the context of quantum computing this problem is sometimes referred to as
\emph{Deutsch's problem}, and we'll follow that nomenclature in this lesson.

To be precise, the input is represented by a function
$f:\Sigma \rightarrow \Sigma$ from one bit to one bit. 
There are four such functions, which we encountered earlier in the course:
\[
\begin{array}{c|c}
  a & f_1(a)\\
  \hline
  0 & 0\\
  1 & 0
\end{array}
\qquad
\begin{array}{c|c}
  a & f_2(a)\\
  \hline
  0 & 0\\
  1 & 1
\end{array}
\qquad
\begin{array}{c|c}
  a & f_3(a)\\
  \hline
  0 & 1\\
  1 & 0
\end{array}
\qquad
\begin{array}{c|c}
  a & f_4(a)\\
  \hline
  0 & 1\\
  1 & 1
\end{array}
\]
The first and last of these functions are \emph{constant} and the middle two
are \emph{balanced}, meaning that the two possible output values for the
function occur the same number of times as we range over the inputs.
Deutsch's problem is to determine which of these two categories the input
function belongs to: constant or balanced.

\begin{callout}[title={Deutsch's problem}]
  \begin{problem}
    \Input{A function $f:\{0,1\}\rightarrow\{0,1\}$.}
    \Output{$0$ if $f$ is constant, $1$ if $f$ is balanced.}
  \end{problem}
\end{callout}

If we view the input function $f$ in Deutsch's problem as representing random
access to a string, we're thinking about a two-bit string: $f(0)f(1).$
\[
\begin{array}{cc}
  \text{function} & \text{string}\\
  \hline
  f_1 & 00 \\
  f_2 & 01 \\
  f_3 & 10 \\
  f_4 & 11
\end{array}
\]
When viewed in this way, Deutsch's problem is to compute the parity (or,
equivalently, the exclusive-OR) of the two bits.

Every classical query algorithm that correctly solves this problem must query
both bits: $f(0)$ and $f(1).$
If we learn that $f(1) = 1,$ for instance, the answer could still be $0$ or
$1,$ depending on whether $f(0) = 1$ or $f(0) = 0,$ respectively.
Every other case is similar; knowing just one of two bits doesn't provide any
information at all about their parity.
So, the Boolean circuit described in the previous section is the best we can do
in terms of the number of queries required to solve this problem.

\subsection{Quantum circuit description}

Deutsch's algorithm solves Deutsch's problem using a single query, therefore
providing a quantifiable advantage of quantum over classical computations.
This may be a modest advantage --- one query as opposed to two --- but we have
to start somewhere.
Scientific advances sometimes have seemingly humble origins.
Figure~\ref{fig:Deutsch-circuit} describes Deutsch's algorithm
as a quantum circuit.

\begin{figure}[b]
  \vspace{3mm}
  \begin{center}

  \end{center}
  \caption{Deutsch's algorithm.}
  \label{fig:Deutsch-circuit}
\end{figure}

\subsection{Analysis}

To analyze Deutsch's algorithm, we will trace through the action of the circuit
above and identify the states of the qubits at the times suggested by
Figure~\ref{fig:Deutsch-circuit-states}.

\begin{figure}[!ht]
  \begin{center}
    %
  \end{center}
  \caption{Three states $\ket{\pi_1}$, $\ket{\pi_2}$, and $\ket{\pi_3}$
    considered in the analysis of Deutsch's algorithm.}
  \label{fig:Deutsch-circuit-states}
\end{figure}

The initial state is $\vert 1\rangle \vert 0 \rangle,$ and the two Hadamard
operations on the left-hand side of the circuit transform this state to
\[
\vert \pi_1 \rangle = \vert - \rangle \vert + \rangle
= \frac{1}{2} \bigl( \vert 0\rangle - \vert 1\rangle \bigr) \vert 0\rangle
+ \frac{1}{2} \bigl( \vert 0\rangle - \vert 1\rangle \bigr) \vert 1\rangle.
\]
(As always, we're following Qiskit's qubit ordering convention, which puts the
top qubit to the right and the bottom qubit to the left.)

Next, the $U_f$ gate is performed.
According to the definition of the $U_f$ gate, the value of the function $f$
for the classical state of the top/rightmost qubit is XORed onto the
bottom/leftmost qubit, which transforms $\vert \pi_1\rangle$ into the state
\[
\vert \pi_2 \rangle
= \frac{1}{2} \bigl( \vert 0 \oplus f(0) \rangle - \vert 1 \oplus f(0) \rangle
\bigr) \vert 0 \rangle
+ \frac{1}{2} \bigl( \vert 0 \oplus f(1) \rangle - \vert 1 \oplus f(1) \rangle
\bigr) \vert 1 \rangle.
\]
We can simplify this expression by observing that the formula
\[
\vert 0 \oplus a\rangle - \vert 1 \oplus a\rangle = (-1)^a \bigl( \vert
0\rangle - \vert 1\rangle \bigr)
\]
works for both possible values $a\in\Sigma.$
More explicitly, the two cases are as follows.
\[
\begin{aligned}
  \vert 0 \oplus 0\rangle - \vert 1 \oplus 0\rangle
  & = \vert 0 \rangle - \vert 1 \rangle
  = (-1)^0 \bigl( \vert 0\rangle - \vert 1\rangle \bigr)\\
  \vert 0 \oplus 1\rangle - \vert 1 \oplus 1\rangle & = \vert 1 \rangle - \vert
  0\rangle
  = (-1)^1 \bigl( \vert 0\rangle - \vert 1\rangle \bigr)
\end{aligned}
\]
Thus, we can alternatively express $\vert\pi_2\rangle$ like this:
\[
\begin{aligned}
  \vert\pi_2\rangle
  & = \frac{1}{2} (-1)^{f(0)} \bigl( \vert 0 \rangle - \vert 1 \rangle \bigr)
  \vert 0 \rangle
  + \frac{1}{2} (-1)^{f(1)} \bigl( \vert 0 \rangle - \vert 1 \rangle \bigr)
  \vert 1 \rangle \\
  & = \vert - \rangle \biggl( \frac{(-1)^{f(0)} \vert 0\rangle + (-1)^{f(1)}
    \vert 1\rangle}{\sqrt{2}}\biggr).
\end{aligned}
\]

Something interesting just happened!
Although the action of the $U_f$ gate on standard basis states leaves the
top/rightmost qubit alone and XORs the function value onto the bottom/leftmost
qubit, here we see that the state of the top/rightmost qubit has changed (in
general) while the state of the bottom/leftmost qubit remains the same ---
specifically being in the $\vert - \rangle$ state before and after the $U_f$
gate is performed.
This phenomenon is known as the \emph{phase kickback}, and we will have more to
say about it shortly.

With one final simplification, which is to pull the factor of $(-1)^{f(0)}$
outside of the sum, we obtain this expression of the state $\vert\pi_2\rangle$:
\[
\begin{aligned}
  \vert\pi_2\rangle
  & = (-1)^{f(0)} \vert - \rangle
  \biggl( \frac{\vert 0\rangle + (-1)^{f(0) \oplus f(1)} \vert
    1\rangle}{\sqrt{2}}\biggr) \\
  & = \begin{cases}
    (-1)^{f(0)} \vert - \rangle \vert + \rangle & \text{if $f(0) \oplus f(1) =
      0$}\\[1mm]
    (-1)^{f(0)} \vert - \rangle \vert - \rangle & \text{if $f(0) \oplus f(1) =
      1$}.
  \end{cases}
\end{aligned}
\]
Notice that in this expression, we have $f(0) \oplus f(1)$ in the exponent of
$-1$ as opposed to $f(1) - f(0),$ which is what we might expect from a purely
algebraic viewpoint, but we obtain the same result either way.
This is because the value $(-1)^k$ for any integer $k$ depends only on whether
$k$ is even or odd.

Applying the final Hadamard gate to the top qubit leaves us with the state
\[
\vert \pi_3 \rangle =
\begin{cases}
  (-1)^{f(0)} \vert - \rangle \vert 0 \rangle & \text{if $f(0) \oplus f(1) =
    0$}\\[1mm]
  (-1)^{f(0)} \vert - \rangle \vert 1 \rangle & \text{if $f(0) \oplus f(1) =
    1$},
\end{cases}
\]
which leads to the correct outcome with probability $1$ when the right/topmost
qubit is measured.

\subsection{Further remarks on the phase kickback}

Before moving on, let's look at the analysis above from a slightly different
angle that may shed some light on the phase kickback phenomenon.

First, notice that the following formula works for all choices of bits
$b,c\in\Sigma.$
\[
\vert b \oplus c\rangle = X^c \vert b \rangle
\]
This can be verified by checking it for the two possible values $c = 0$ and $c
= 1$:
\[
\begin{aligned}
  \vert b \oplus 0 \rangle & = \vert b\rangle = \mathbb{I} \vert b \rangle =
  X^0 \vert b \rangle\\
  \vert b \oplus 1 \rangle & = \vert \neg b\rangle = X \vert b \rangle = X^1
  \vert b \rangle.
\end{aligned}
\]

Using this formula, we see that
\[
U_f \bigl(\vert b\rangle \vert a \rangle\bigr)
= \vert b \oplus f(a) \rangle \vert a \rangle
= \bigl(X^{f(a)}\vert b \rangle\bigr) \vert a \rangle
\]
for every choice of bits $a,b\in\Sigma.$
Because this formula is true for $b=0$ and $b=1,$ we see by linearity that
\[
U_f \bigl( \vert \psi \rangle \vert a \rangle \bigr) = \bigl(X^{f(a)}\vert \psi
\rangle\bigr) \vert a \rangle
\]
for all qubit state vectors $\vert \psi\rangle,$ and therefore
\[
U_f \bigl( \vert - \rangle \vert a \rangle \bigr) = \bigl(X^{f(a)} \vert -
\rangle \bigr) \vert a \rangle
= (-1)^{f(a)} \vert - \rangle \vert a \rangle.
\]

The key that makes this work is that
\[
X\vert - \rangle = - \vert - \rangle.
\]
In mathematical terms, the vector $\vert - \rangle$ is an \emph{eigenvector} of
the matrix $X$ having \emph{eigenvalue} $-1.$
We'll discuss eigenvectors and eigenvalues in greater detail in
Lesson~\ref{lesson:phase-estimation-and-factoring}
\emph{(Phase Estimation and Factoring)}, where the phase kickback phenomenon is
generalized to other unitary operations.

Keeping in mind that scalars float freely through tensor products, we find an
alternative way of reasoning how the operation $U_f$ transforms $\vert
\pi_1\rangle$ into $\vert \pi_2\rangle$ in the analysis above:
\begin{align*}
  \vert \pi_2 \rangle
  & = U_f \bigl( \vert - \rangle \vert + \rangle \bigr)\\
  & = \frac{1}{\sqrt{2}} U_f \bigl(\vert - \rangle \vert 0\rangle \bigr)
  + \frac{1}{\sqrt{2}} U_f \bigl(\vert - \rangle \vert 1\rangle \bigr)\\
  & = \vert - \rangle \biggl( \frac{(-1)^{f(0)} \vert 0\rangle + (-1)^{f(1)}
    \vert 1\rangle}{\sqrt{2}}\biggr).
\end{align*}

\section{The Deutsch--Jozsa algorithm}

Deutsch's algorithm outperforms all classical algorithms for a query problem,
but the advantage is quite modest: one query versus two.
The Deutsch--Jozsa algorithm extends this advantage --- and, in fact, it can be
used to solve a couple of different query problems.

\begin{figure}[!ht]
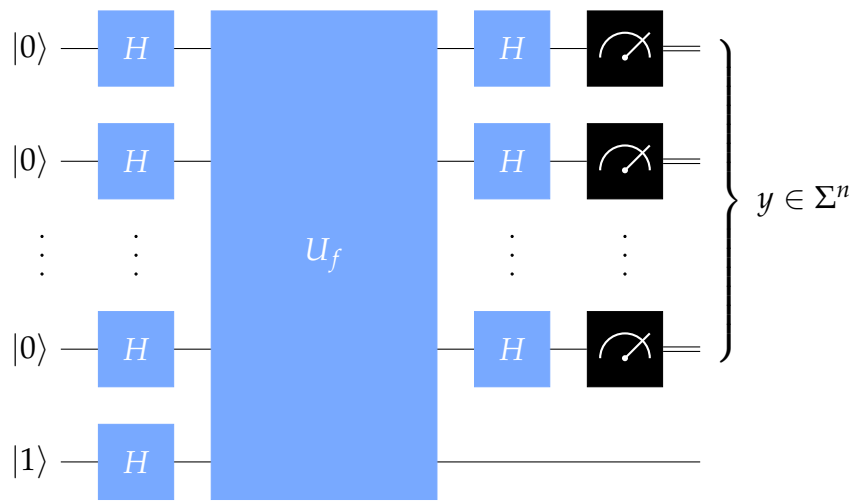

  \begin{center}
    \vspace{3mm}


  \end{center}
  \caption{The Deutsch--Jozsa algorithm as a quantum circuit.}
  \label{fig:Deutsch-Jozsa}
\end{figure}

A quantum circuit description of the Deutsch--Jozsa algorithm appears in
Figure~\ref{fig:Deutsch-Jozsa}.
An additional classical post-processing step, not shown in the figure, may also
be required depending on the specific problem being solved.
Of course, we haven't actually discussed what problems this algorithm solves;
this is done in the two sections that follow.

\subsection{The Deutsch--Jozsa problem}

We'll begin with the query problem the Deutsch--Jozsa algorithm was originally
intended to solve, which is known as the \emph{Deutsch--Jozsa problem}.

The input function for this problem takes the form $f:\Sigma^n \rightarrow
\Sigma$ for an arbitrary positive integer $n.$
Like Deutsch's problem, the task is to output $0$ if $f$ is constant and $1$ if
$f$ is balanced, which again means that the number of input strings on which
the function takes the value $0$ is equal to the number of input strings on
which the function takes the value $1$.

Notice that, when $n$ is larger than $1,$ there are functions of the form
$f:\Sigma^n \rightarrow \Sigma$ that are neither constant nor balanced.
For example, the function $f:\Sigma^2\rightarrow\Sigma$ defined as
\[
\begin{aligned}
  f(00) & = 0 \\
  f(01) & = 0 \\
  f(10) & = 0 \\
  f(11) & = 1
\end{aligned}
\]
falls into neither of these two categories.
For the Deutsch--Jozsa problem, we simply don't worry about functions like this
--- they're considered to be ``don't care'' inputs.
That is, for this problem we have a \emph{promise} that $f$ is either constant
or balanced.

\begin{callout}[title={The Deutsch--Jozsa problem}]
  \begin{problem}
    \Input{A function $f:\{0,1\}^n\rightarrow\{0,1\}$.}
    \Promise{$f$ is either constant or balanced.}
    \Output{$0$ if $f$ is constant, $1$ if $f$ is balanced.}
  \end{problem}
\end{callout}

The Deutsch--Jozsa algorithm, with its single query, solves this problem in the
following sense:
if every one of the $n$ measurement outcomes is $0,$ then the function $f$ is
constant;
and otherwise, if at least one of the measurement outcomes is $1,$ then the
function $f$ is balanced.
Another way to say this is that the circuit described above is followed by a
classical post-processing step in which the OR of the measurement outcomes is
computed to produce the output.

\subsubsection{Algorithm analysis}

To analyze the performance of the Deutsch--Jozsa algorithm for the
Deutsch--Jozsa problem, it's helpful to begin by thinking about the action of a
single layer of Hadamard gates.
A Hadamard operation can be expressed as a matrix in the usual way,
\[
H = \begin{pmatrix}
  \frac{1}{\sqrt{2}} & \frac{1}{\sqrt{2}} \\[2mm]
  \frac{1}{\sqrt{2}} & -\frac{1}{\sqrt{2}}
\end{pmatrix},
\]
but we can also express this operation in terms of its action on standard basis
states:
\[
\begin{aligned}
  H \vert 0\rangle & = \frac{1}{\sqrt{2}} \vert 0 \rangle + \frac{1}{\sqrt{2}}
  \vert 1 \rangle\\[3mm]
  H \vert 1\rangle & = \frac{1}{\sqrt{2}} \vert 0 \rangle - \frac{1}{\sqrt{2}}
  \vert 1 \rangle.
\end{aligned}
\]
These two equations can be combined into a single formula,
\[
H \vert a \rangle = \frac{1}{\sqrt{2}} \vert 0 \rangle + \frac{1}{\sqrt{2}}
(-1)^a \vert 1 \rangle
= \frac{1}{\sqrt{2}} \sum_{b\in\{0,1\}} (-1)^{ab} \vert b\rangle,
\]
which is true for both choices of $a\in\Sigma.$

Now suppose that instead of just a single qubit we have $n$ qubits, and a
Hadamard operation is performed on each.
The combined operation on the $n$ qubits is described by the tensor product
$H\otimes \cdots \otimes H$ ($n$ times), which we write as $H^{\otimes n}$ for
conciseness and clarity.
Using the formula from above, followed by expanding and then simplifying, we
can express the action of this combined operation on the standard basis states
of $n$ qubits like this:
\[
\begin{aligned}
  & H^{\otimes n} \vert x_{n-1} \cdots x_1 x_0 \rangle \\
  & \qquad = \bigl(H \vert x_{n-1} \rangle \bigr) \otimes \cdots \otimes
  \bigl(H \vert x_{0} \rangle \bigr) \\
  & \qquad = \Biggl( \frac{1}{\sqrt{2}} \sum_{y_{n-1}\in\Sigma} (-1)^{x_{n-1}
    y_{n-1}} \vert y_{n-1} \rangle \Biggr)
  \otimes \cdots \otimes
  \Biggl( \frac{1}{\sqrt{2}} \sum_{y_{0}\in\Sigma} (-1)^{x_{0} y_{0}} \vert
  y_{0} \rangle \Biggr) \\
  & \qquad = \frac{1}{\sqrt{2^n}} \sum_{y_{n-1}\cdots y_0 \in \Sigma^n}
  (-1)^{x_{n-1}y_{n-1} + \cdots + x_0 y_0} \vert y_{n-1} \cdots y_0 \rangle.
\end{aligned}
\]
Here, by the way, we're writing binary strings of length $n$ as $x_{n-1}\cdots
x_0$ and $y_{n-1}\cdots y_0,$ following Qiskit's indexing convention.
This formula provides us with a useful tool for analyzing the quantum circuit
above.

After the first layer of Hadamard gates is performed, the state of the $n+1$
qubits (including the leftmost/bottom qubit, which is treated separately from
the rest) is
\[
\bigl( H \vert 1 \rangle \bigr) \bigl( H^{\otimes n} \vert 0 \cdots 0 \rangle
\bigr)
= \vert - \rangle \otimes \frac{1}{\sqrt{2^n}} \sum_{x_{n-1}\cdots x_0 \in
  \Sigma^n} \vert x_{n-1} \cdots x_0 \rangle.
\]
When the $U_f$ operation is performed, this state is transformed into
\[
\vert - \rangle \otimes \frac{1}{\sqrt{2^n}}
\sum_{x_{n-1}\cdots x_0 \in \Sigma^n} (-1)^{f(x_{n-1}\cdots x_0)} \vert x_{n-1}
\cdots x_0 \rangle
\]
through exactly the same phase kickback phenomenon that we saw in the analysis
of Deutsch's algorithm.
Then the second layer of Hadamard gates is performed, which (by the formula
above) transforms this state into
\[
\vert - \rangle \otimes \frac{1}{2^n}
\sum_{x_{n-1}\cdots x_0 \in \Sigma^n}
\sum_{y_{n-1}\cdots y_0 \in \Sigma^n}
(-1)^{f(x_{n-1}\cdots x_0) + x_{n-1}y_{n-1} + \cdots + x_0 y_0}
\vert y_{n-1} \cdots y_0 \rangle.
\]

This expression looks somewhat complicated, and not too much can be concluded
about the probabilities to obtain different measurement outcomes without
knowing more about the function $f.$
Fortunately, all we need to know is the probability that every one of the
measurement outcomes is $0$ --- because that's the probability that the
algorithm determines that $f$ is constant.
This probability has a simple formula.
\[
\Biggl\vert
\frac{1}{2^n}
\sum_{x_{n-1}\cdots x_0 \in \Sigma^n}
(-1)^{f(x_{n-1}\cdots x_0)}
\Biggr\vert^2
= \begin{cases}
  1 & \text{if $f$ is constant}\\[1mm]
  0 & \text{if $f$ is balanced}
\end{cases}
\]

In greater detail, if $f$ is constant, then either $f(x_{n-1}\cdots x_0) = 0$
for every string $x_{n-1}\cdots x_0,$ in which case the value of the sum is
$2^n,$ or $f(x_{n-1}\cdots x_0) = 1$ for every string $x_{n-1}\cdots x_0,$
in which case the value of the sum is $-2^n.$
Dividing by $2^n$ and taking the square of the absolute value yields $1.$
If, on the other hand, $f$ is balanced, then $f$ takes the value $0$ on half of
the strings $x_{n-1}\cdots x_0$ and the value $1$ on the other half, so the
$+1$ terms and $-1$ terms in the sum cancel and we're left with the value $0.$

We conclude that the algorithm operates correctly provided that the promise is
fulfilled.

\subsubsection{Classical difficulty}

The Deutsch--Jozsa algorithm works every time, always giving us the correct
answer when the promise is met, and requires a single query.
How does this compare with classical query algorithms for the Deutsch--Jozsa
problem?

First, any \emph{deterministic} classical algorithm that correctly solves the
Deutsch--Jozsa problem must make exponentially many queries:
$2^{n-1} + 1$ queries are required in the worst case.
The reasoning is that, if a deterministic algorithm queries $f$ on $2^{n-1}$ or
fewer different strings, and obtains the same function value every time, then
both answers are still possible.
The function might be constant, or it might be balanced but through bad luck
the queries all happen to return the same function value.

The second possibility might seem unlikely --- but for deterministic algorithms
there's no randomness or uncertainty, so they will fail systematically on
certain functions.
We therefore have a significant advantage of quantum over classical algorithms
in this regard.

There is a catch, however, which is that \emph{probabilistic} classical
algorithms can solve the Deutsch--Jozsa problem with very high probability using
just a few queries.
In particular, if we simply choose a few different strings of length $n$
randomly, and query $f$ on those strings, it's unlikely that we'll get the same
function value for all of them when $f$ is balanced.

To be specific, if we choose $k$ input strings $x^1,\ldots,x^k \in \Sigma^n$
uniformly at random, evaluate $f(x^1),\ldots,f(x^k),$ and answer $0$ if the
function values are all the same, and~$1$ if not, then we'll always be correct
when $f$ is constant, and wrong in the case that $f$ is balanced with
probability just $2^{-k + 1}.$
If we take $k = 11,$ for instance, this algorithm will answer correctly with
probability greater than $99.9$

For this reason, we do still have a rather modest advantage of quantum over
classical algorithms --- but it is nevertheless a quantifiable advantage
representing an improvement over Deutsch's algorithm.

\subsection{The Bernstein--Vazirani problem}

Next, we'll discuss a problem known as the \emph{Bernstein--Vazirani problem}.
It's also called the \emph{Fourier sampling problem}, although there are more
general formulations of this problem that also go by that name.

First, let's introduce some notation.
For any two binary strings $x = x_{n-1} \cdots x_0$ and $y = y_{n-1}\cdots y_0$
of length $n,$ we define
\[
x \cdot y = x_{n-1} y_{n-1} \oplus \cdots \oplus x_0 y_0.
\]
We'll refer to this operation as the \emph{binary dot product}.
An alternative way to define it is like so.
\[
x \cdot y =
\begin{cases}
1 & x_{{n-1}} y_{n-1} + \cdots + x_0 y_0 \text{ is odd}\\[0.5mm]
0 & x_{{n-1}} y_{n-1} + \cdots + x_0 y_0 \text{ is even}
\end{cases}
\]

Notice that this is a symmetric operation, meaning that the result doesn't
change if we swap $x$ and $y,$ so we're free to do that whenever it's
convenient.
Sometimes it's useful to think about the binary dot product $x \cdot y$ as
being the parity of the bits of $x$ in positions where the string $y$ has a
$1,$ or equivalently, the parity of the bits of $y$ in positions where the
string $x$ has a $1.$

With this notation in hand we can now define the Bernstein--Vazirani problem.

\begin{callout}[title={Bernstein--Vazirani problem}]
  \begin{problem}
    \Input{A function $f:\{0,1\}^n\rightarrow\{0,1\}$.}
    \Promise{There exists a binary string $s = s_{n-1} \cdots s_0$ for which
      $f(x) = s\cdot x$ for all $x\in\Sigma^n$.}
    \Output{The string $s$.}
  \end{problem}
\end{callout}

\noindent
We don't actually need a new quantum algorithm for this problem; the
Deutsch--Jozsa algorithm solves it.
In the interest of clarity, let's refer to the quantum circuit from above,
which doesn't include the classical post-processing step of computing the OR,
as the \emph{Deutsch--Jozsa circuit}.

\subsubsection{Algorithm analysis}

To analyze how the Deutsch--Jozsa circuit works for a function satisfying the
promise for the Bernstein--Vazirani problem, we'll begin with a quick
observation.
Using the binary dot product, we can alternatively describe the action of $n$
Hadamard gates on the standard basis states of $n$ qubits as follows.
\[
H^{\otimes n} \vert x \rangle = \frac{1}{\sqrt{2^n}} \sum_{y\in\Sigma^n}
(-1)^{x\cdot y} \vert y\rangle
\]
Similar to what we saw when analyzing Deutsch's algorithm, this is because the
value $(-1)^k$ for any integer $k$ depends only on whether $k$ is even or odd.

Turning to the Deutsch--Jozsa circuit, after the first layer of Hadamard gates
is performed, the state of the $n+1$ qubits is
\[
\vert {-} \rangle \otimes \frac{1}{\sqrt{2^n}} \sum_{x \in \Sigma^n} \vert x
\rangle.
\]
The query gate is then performed, which (through the phase kickback phenomenon)
transforms the state into
\[
\vert - \rangle \otimes \frac{1}{\sqrt{2^n}} \sum_{x \in \Sigma^n} (-1)^{f(x)}
\vert x \rangle.
\]
Using our formula for the action of a layer of Hadamard gates, we see that the
second layer of Hadamard gates then transforms this state into
\[
\vert {-} \rangle \otimes \frac{1}{2^n}
\sum_{x \in \Sigma^n} \sum_{y \in \Sigma^n} (-1)^{f(x) + x \cdot y} \vert y
\rangle.
\]

Now we can make some simplifications, in the exponent of $-1$ inside the sum.
We're promised that $f(x) = s\cdot x$ for some string $s = s_{n-1} \cdots s_0,$
so we can express the state as
\[
\vert - \rangle \otimes \frac{1}{2^n}
\sum_{x \in \Sigma^n} \sum_{y \in \Sigma^n} (-1)^{s\cdot x + x \cdot y} \vert y
\rangle.
\]
Because $s\cdot x$ and $x\cdot y$ are binary values, we can replace the
addition with the exclusive-OR --- again because the only thing that matters
for an integer in the exponent of $-1$ is whether it is even or odd.
Making use of the symmetry of the binary dot product, we obtain this expression
for the state:
\[
\vert - \rangle \otimes \frac{1}{2^n}
\sum_{x \in \Sigma^n} \sum_{y \in \Sigma^n} (-1)^{(s\cdot x) \oplus (y \cdot
  x)} \vert y \rangle.
\]
Parentheses have been added for clarity, though they aren't really necessary
because it's conventional to treat the binary dot product as having higher
precedence than the exclusive-OR.

At this point we will make use of the following formula.
\[
(s\cdot x) \oplus (y \cdot x) = (s \oplus y) \cdot x
\]
We can obtain the formula through a similar formula for bits,
\[
(a c) \oplus (b c) = (a \oplus b) c,
\]
together with an expansion of the binary dot product and bitwise exclusive-OR.
\[
\begin{aligned}
  (s\cdot x) \oplus (y \cdot x)
  & = (s_{n-1} x_{n-1}) \oplus \cdots \oplus (s_{0} x_{0}) \oplus
  (y_{n-1} x_{n-1})  \oplus \cdots \oplus (y_{0} x_{0}) \\
  & = (s_{n-1} \oplus y_{n-1}) x_{n-1}  \oplus \cdots \oplus (s_{0} \oplus
  y_{0}) x_{0} \\
  & = (s \oplus y) \cdot x
\end{aligned}
\]
This allows us to express the state of the circuit immediately prior to the
measurements like this:
\[
\vert {-} \rangle \otimes \frac{1}{2^n}
\sum_{x \in \Sigma^n} \sum_{y \in \Sigma^n} (-1)^{(s\oplus y)\cdot x} \vert y
\rangle.
\]

The final step is to make use of yet another formula, which works for every
binary string $z = z_{n-1}\cdots z_0.$
\[
\frac{1}{2^n}
\sum_{x \in \Sigma^n} (-1)^{z \cdot x}
= \begin{cases}
  1 & \text{if $z = 0^n$}\\
  0 & \text{if $z\neq 0^n$}
\end{cases}
\]
Here we're using a simple notation for strings that we'll use throughout the
remainder of course: $0^n$ is the all-zero string of length $n.$

A simple way to argue that this formula works is to consider the two cases
separately.
If $z = 0^n,$ then $z\cdot x = 0$ for every string $x\in\Sigma^n,$ so the value
of each term in the sum is $1,$ and we obtain $1$ by summing and dividing by
$2^n.$
On the other hand, if any one of the bits of $z$ is equal to $1,$ then the
binary dot product $z\cdot x$ is equal to $0$ for exactly half of the possible
choices for $x\in\Sigma^n$ and $1$ for the other half --- because the value of
the binary dot product $z\cdot x$ flips (from $0$ to $1$ or from $1$ to $0$) if
we flip any bit of $x$ in a position where $z$ has a $1.$

If we now apply this formula to simplify the state of the circuit prior to the
measurements, we obtain
\[
\vert - \rangle \otimes \frac{1}{2^n}
\sum_{x \in \Sigma^n} \sum_{y \in \Sigma^n} (-1)^{(s\oplus y)\cdot x} \vert y
\rangle
= \vert - \rangle \otimes \vert s \rangle,
\]
owing to the fact that $s\oplus y = 0^n$ if and only if $y = s.$
Thus, the measurements reveal precisely the string $s$ we're looking for.

\subsubsection{Classical difficulty}

While the Deutsch--Jozsa circuit solves the Bernstein--Vazirani problem with a
single query, any classical query algorithm must make at least $n$ queries to
solve this problem.
This can be reasoned through a so-called \emph{information theoretic} argument,
which is very simple in this case:
each classical query reveals a single bit of information about the solution,
and there are $n$ bits of information that need to be uncovered, so at least
$n$ queries are needed.

It is, in fact, possible to solve the Bernstein--Vazirani problem classically by
querying the function on each of the $n$ strings having a single $1,$ in each
possible position, and $0$ for all other bits, which reveals the bits of $s$
one at a time.
So, the advantage of quantum over classical algorithms for this problem is $1$
query versus $n$ queries.

\subsubsection{Remark on nomenclature}

In the context of the Bernstein--Vazirani problem, it is common that the
Deutsch--Jozsa algorithm is referred to as the ``Bernstein--Vazirani
algorithm.''
This is slightly misleading, because the algorithm \emph{is} the Deutsch--Jozsa
algorithm, as Bernstein and Vazirani were very clear about in their work.

What Bernstein and Vazirani did after showing that the Deutsch--Jozsa algorithm
solves the Bernstein--Vazirani problem (as it is stated above) was to define a
much more complicated problem, known as the \emph{recursive Fourier sampling
problem}.
This is a highly contrived problem where solutions to different instances of
the problem effectively unlock new levels of the problem arranged in a
tree-like structure.
The Bernstein--Vazirani problem is essentially just the base case of this more
complicated problem.

The recursive Fourier sampling problem was the first known example of a query
problem where quantum algorithms have a so-called \emph{super-polynomial}
advantage over probabilistic algorithms, thereby surpassing the advantage of
quantum over classical offered by the Deutsch--Jozsa algorithm.
Intuitively speaking, the recursive version of the problem amplifies the $1$
versus $n$ advantage of quantum algorithms to something much larger.
The most challenging aspect of the mathematical analysis establishing this
advantage is showing that classical query algorithms can't solve the problem
without making lots of queries.
This is quite typical; for many problems it can be very difficult to rule out
creative classical approaches that solve them efficiently.

Simon's problem, and the algorithm for it described in the next section, does
provide a much simpler example of a super-polynomial (and, in fact,
exponential) advantage of quantum over classical algorithms, and for this
reason the recursive Fourier sampling problem is less often discussed.
It is, nevertheless, an interesting computational problem in its own right.

\section{Simon's algorithm}

Simon's algorithm is a quantum query algorithm for a problem known as
\emph{Simon's problem}.
This is a promise problem with a flavor similar to the Deutsch--Jozsa and
Bernstein--Vazirani problems, but the specifics are different.

Simon's algorithm is significant because it provides an \emph{exponential}
advantage of quantum over classical (including probabilistic) algorithms, and
the technique it uses inspired Peter Shor's discovery of an efficient quantum
algorithm for integer factorization.

\subsection{Simon's problem}

The input function for Simon's problem takes the form
\[
f:\Sigma^n \rightarrow \Sigma^m
\]
for positive integers $n$ and $m.$
We could restrict our attention to the case $m = n$ in the interest of
simplicity, but there's little to be gained in making this assumption ---
Simon's algorithm and its analysis are basically the same either way.

\begin{callout}[title={Simon's problem}]
  \begin{problem}
    \Input{A function $f:\Sigma^n \rightarrow \Sigma^m$.}
    \Promise{There exists a string $s\in\Sigma^n$ such that
      \[
      [f(x) = f(y)]
      \Leftrightarrow [(x = y) \vee (x \oplus s = y)]
      \]
      for all $x,y\in\Sigma^n$.
    }
    \Output{The string $s$.}
  \end{problem}
\end{callout}

We'll unpack the promise to better understand what it says momentarily, but
first let's be clear that it requires that $f$ has a very special structure ---
so most functions won't satisfy this promise.
It's also fitting to acknowledge that this problem isn't intended to have
practical importance.
Rather, it's a somewhat artificial problem tailor-made to be easy for quantum
computers and hard for classical computers.

There are two main cases: the first case is that $s$ is the all-zero string
$0^n,$ and the second case is that $s$ is not the all-zero string.
\begin{description}[leftmargin=0mm]
\item[Case 1: $s=0^n.$]
  If $s$ is the all-zero string, then we can simplify the if and only if
  statement in the promise so that it reads
  $[f(x) = f(y)] \Leftrightarrow [x = y].$
  This is equivalent to $f$ being a one-to-one function.
\item[Case 2: $s\neq 0^n.$]
  If $s$ is not the all-zero string, then the promise being satisfied for this
  string implies that $f$ is \emph{two-to-one}, meaning that for every possible
  output string of $f,$ there are exactly two input strings that cause $f$ to
  output that string. Moreover, these two input strings must take the form $w$
  and $w \oplus s$ for some string $w.$
\end{description}

It's important to recognize that there can only be one string $s$ that works if
the promise is met, so there's always a unique correct answer for functions
that satisfy the promise.

Here's an example of a function taking the form $f:\Sigma^3 \rightarrow
\Sigma^5$ that satisfies the promise for the string $s = 011.$
\[
\begin{aligned}
  f(000) & = 10011 \\
  f(001) & = 00101 \\
  f(010) & = 00101 \\
  f(011) & = 10011 \\
  f(100) & = 11010 \\
  f(101) & = 00001 \\
  f(110) & = 00001 \\
  f(111) & = 11010
\end{aligned}
\]
There are $8$ different input strings and $4$ different output strings, each of
which occurs twice --- so this is a two-to-one function.
Moreover, for any two different input strings that produce the same output
string, we see that the bitwise XOR of these two input strings is equal to
$011,$ which is equivalent to saying that either one of them equals the other
XORed with $s.$

Notice that the only thing that matters about the actual output strings is
whether they're the same or different for different choices of input strings.
For instance, in the example above, there are four strings $(10011,$ $00101,$
$00001,$ and $11010)$ that appear as outputs of $f.$ We could replace these
four strings with different strings, so long as they're all distinct, and the
correct solution $s = 011$ would not change.

\subsection{Algorithm description}

\begin{figure}[!ht]
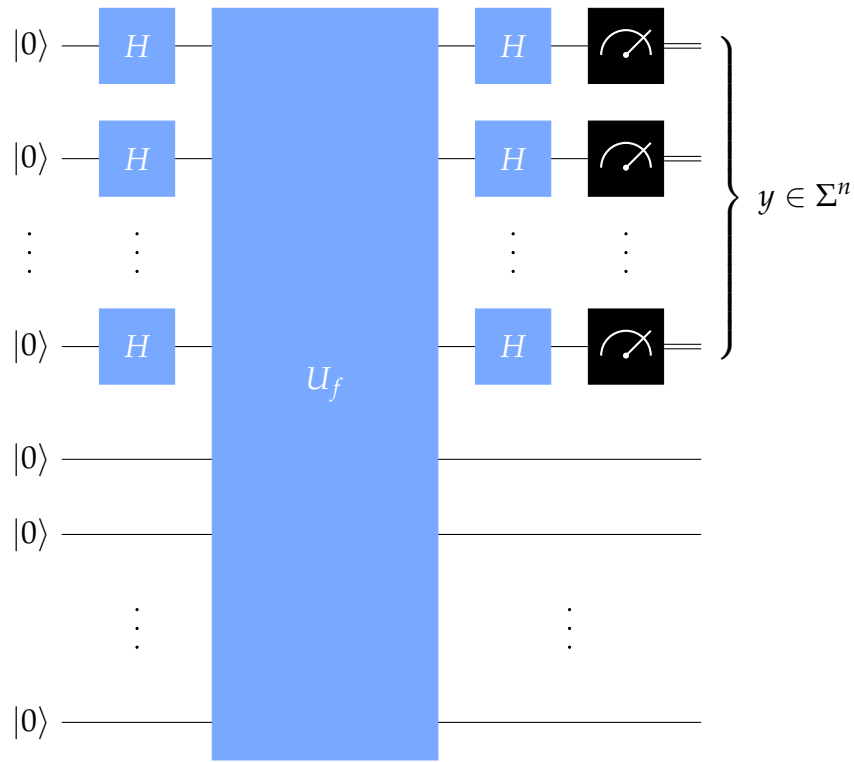

  \begin{center}

    
  \end{center}
  \caption{The quantum circuit portion of Simon's algorithm.}
  \label{fig:Simon}
\end{figure}

Figure~\ref{fig:Simon} describes the quantum circuit portion of Simon's
algorithm.
To be clear, there are $n$ qubits on the top that are acted upon by Hadamard
gates and $m$ qubits on the bottom that go directly into the query gate.
It looks very similar to the algorithms we've already discussed in the lesson,
but this time there's no phase kickback; the bottom $m$ qubits all go into the
query gate in the state $\vert 0\rangle.$

To solve Simon's problem using this circuit requires several independent runs
of it followed by a classical post-processing step, which will be described
later after the behavior of the circuit is analyzed.

\subsection{Analysis}

The analysis of Simon's algorithm begins along similar lines to the
Deutsch--Jozsa algorithm.
After the first layer of Hadamard gates is performed on the top $n$ qubits, the
state becomes
\[
\frac{1}{\sqrt{2^n}} \sum_{x\in\Sigma^n} \vert 0^m \rangle \vert x\rangle.
\]
When the $U_f$ is performed, the output of the function $f$ is XORed onto the
all-zero state of the bottom $m$ qubits, so the state becomes
\[
\frac{1}{\sqrt{2^n}} \sum_{x\in\Sigma^n} \vert f(x) \rangle \vert x\rangle.
\]
When the second layer of Hadamard gates is performed, we obtain the following
state by using the same formula for the action of a layer of Hadamard gates as
before.
\[
\frac{1}{2^n} \sum_{x\in\Sigma^n} \sum_{y\in\Sigma^n} (-1)^{x\cdot y} \vert
f(x) \rangle \vert y\rangle
\]

At this point, the analysis diverges from the ones for the previous algorithms
in this lesson.
We're interested in the probability for the measurements to result in each
possible string $y\in\Sigma^n.$
Through the rules for analyzing measurements described in
Lesson~\ref{lesson:multiple-systems} \emph{(Multiple Systems)}, we find
that the probability $p(y)$ to obtain the string $y$ is equal to
\[
p(y) = \left\|\frac{1}{2^n} \sum_{x\in\Sigma^n} (-1)^{x\cdot y} \vert f(x)
\rangle \right\|^2.
\]

To get a better handle on these probabilities, we'll need just a bit more
notation and terminology.
First, the \emph{range} of the function $f$ is the set containing all of its
output strings.
\[
\operatorname{range}(f) = \{ f(x) : x\in \Sigma^n \}
\]
Second, for each string $z\in\operatorname{range}(f),$ we can express the set
of all input strings that cause the function to evaluate to this output string
$z$ as $f^{-1}(\{z\}).$
\[
f^{-1}(\{z\}) = \{ x\in\Sigma^n : f(x) = z \}
\]
The set $f^{-1}(\{z\})$ is known as the \emph{preimage} of $\{z\}$ under $f.$
We can define the preimage under $f$ of any set in place of $\{z\}$ in an
analogous way --- it's the set of all elements that $f$ maps to that set.
(This notation should not be confused with the \emph{inverse} of the function
$f,$ which may not exist.
The fact that the argument on the left-hand side is the set $\{z\}$ rather than
the element $z$ is the clue that allows us to avoid this confusion.)

Using this notation, we can split up the sum in our expression for the
probabilities above to obtain
\[
p(y) =
\left\|
\frac{1}{2^n}
\sum_{z\in\operatorname{range}(f)}
\Biggl(\sum_{x\in f^{-1}(\{z\})}  (-1)^{x\cdot y}\Biggr)
\vert z \rangle
\right\|^2.
\]
Every string $x\in\Sigma^n$ is represented exactly once by the two summations
--- we're basically just putting these strings into separate buckets depending
on which output string $z = f(x)$ they produce when we evaluate the function
$f,$ and then summing separately over all the buckets.

We can now evaluate the Euclidean norm squared to obtain
\[
p(y) = \frac{1}{2^{2n}}
\sum_{z\in\operatorname{range}(f)}
\left\vert \sum_{x\in f^{-1}(\{z\})}  (-1)^{x\cdot y} \right\vert^2.
\]

To simplify these probabilities further, let's take a look at the value
\begin{equation}
  \left\vert \sum_{x\in f^{-1}(\{z\})}  (-1)^{x\cdot y} \right\vert^2
  \label{eq:Simon-absolute-sum}
\end{equation}
for an arbitrary selection of $z\in\operatorname{range}(f).$
If it happens to be the case that $s = 0^n,$ then $f$ is a one-to-one function
and there's always just a single element $x\in f^{-1}(\{z\}),$ for every
$z\in\operatorname{range}(f).$
The value of the expression \eqref{eq:Simon-absolute-sum} is $1$ in this case.

If, on the other hand, $s\neq 0^n,$ then there are exactly two strings in the
set $f^{-1}(\{z\}).$
To be precise, if we choose $w\in f^{-1}(\{z\})$ to be any one of these two
strings, then the other string must be $w \oplus s$ by the promise in Simon's
problem.
Using this observation we can simplify \eqref{eq:Simon-absolute-sum} as
follows.
\[
\begin{aligned}
  \left\vert \sum_{x\in f^{-1}(\{z\})}  (-1)^{x\cdot y} \right\vert^2
  & = \Bigl\vert (-1)^{w\cdot y} + (-1)^{(w\oplus s)\cdot y} \Bigr\vert^2 \\
  & = \Bigl\vert (-1)^{w\cdot y} \Bigl(1 + (-1)^{s\cdot y}\Bigr) \Bigr\vert^2 \\
  & = \Bigl\vert 1 + (-1)^{y\cdot s} \Bigr\vert^2 \\
  & = \begin{cases}
    4 & y \cdot s = 0\\[1mm]
    0 & y \cdot s = 1
  \end{cases}
\end{aligned}
\]
So, it turns out that the value \eqref{eq:Simon-absolute-sum} is independent of
the specific choice of $z\in\operatorname{range}(f)$ in both cases.

We can now finish off the analysis by looking at the same two cases as before
separately.
\begin{description}[leftmargin=0mm]
\item[Case 1: $s = 0^n.$]
  In this case the function $f$ is one-to-one, so there are $2^n$ strings
  $z\in\operatorname{range}(f),$ and we obtain
  \[
  p(y) = \frac{1}{2^{2n}} \cdot 2^n = \frac{1}{2^n}.
  \]
  In words, the measurements result in a string $y\in\Sigma^n$ chosen uniformly
  at random.

\item[Case 2: $s \neq 0^n.$]
  In this case $f$ is two-to-one, so there are $2^{n-1}$ elements in
  $\operatorname{range}(f).$
  Using the formula from above we conclude that the probability to measure each
  $y\in\Sigma^n$ is
  \[
  p(y)
  = \frac{1}{2^{2n}} \sum_{z\in\operatorname{range}(f)}
  \Biggl\vert \sum_{x\in f^{-1}(\{z\})} (-1)^{x\cdot y} \Biggr\vert^2
  =
  \begin{cases}
    \frac{1}{2^{n-1}} & y \cdot s = 0\\[1mm]
    0 & y \cdot s = 1.
  \end{cases}
  \]
  In words, we obtain a string chosen uniformly at random from the set
  \[
  \{y\in\Sigma^n : y \cdot s = 0\},
  \]
  which contains $2^{n-1}$ strings.
  This is because, when $s\neq 0^n$, exactly half of the binary strings of
  length $n$ have binary dot product $1$ with $s$ and the other have binary dot
  product $0$ with $s,$ as we already observed in the analysis of the
  Deutsch--Jozsa algorithm for the Bernstein--Vazirani problem.
\end{description}

\subsubsection{Classical post-processing}

We now know what the probabilities are for the possible measurement outcomes
when we run the quantum circuit for Simon's algorithm.
Is this enough information to determine $s$?

The answer is yes, provided that we're willing to repeat the process several
times and accept that it could fail with some probability, which we can make
very small by running the circuit enough times.
The essential idea is that each execution of the circuit provides us with
statistical evidence concerning $s,$ and we can use that evidence to find $s$
with very high probability if we run the circuit sufficiently many times.

Let's suppose that we run the circuit independently $k$ times, for
$k = n + 10.$ 
There's nothing special about this particular number of iterations --- we could
take $k$ to be larger (or smaller) depending on the probability of failure
we're willing to tolerate, as we will see.
Choosing $k = n + 10$ will ensure that we have greater than a $99.9$\% chance
of recovering $s.$

By running the circuit $k$ times, we obtain strings
$y^1,...,y^{k} \in \Sigma^n.$ 
To be clear, the superscripts here are part of the names of these strings, not
exponents or indexes to their bits, so we have
\[
\begin{aligned}
  y^1 & = y^1_{n-1} \cdots y^1_{0}\\[1mm]
  y^2 & = y^2_{n-1} \cdots y^2_{0}\\[1mm]
  & \;\; \vdots\\[1mm]
  y^{k} & = y^{k}_{n-1} \cdots y^{k}_{0}
\end{aligned}
\]
We then form a matrix $M$ having $k$ rows and $n$ columns by taking the bits of
these strings as binary-valued entries.
\[
M = \begin{pmatrix}
  y^1_{n-1} & \cdots & y^1_{0}\\[1mm]
  y^2_{n-1} & \cdots & y^2_{0}\\[1mm]
  \vdots & \ddots & \vdots \\[1mm]
  y^{k}_{n-1} & \cdots & y^{k}_{0}
\end{pmatrix}
\]

Now, we don't know what $s$ is at this point --- our goal is to find this
string.
But imagine for a moment that we do know the string $s,$ and we form a column
vector $v$ from the bits of the string $s = s_{n-1} \cdots s_0$ as follows.
\[
v = \begin{pmatrix}
  s_{n-1}\\
  \vdots\\
  s_0
\end{pmatrix}
\]
If we perform the matrix-vector multiplication $M v$ modulo $2$ --- meaning
that we perform the multiplication as usual and then take the remainder of the
entries of the result after dividing by $2$ --- we obtain the all-zero vector.
\[
M v = \begin{pmatrix}
  y^1 \cdot s\\
  y^2 \cdot s\\
  \vdots\\[1mm]
  y^{k} \cdot s
\end{pmatrix}
= \begin{pmatrix}
  0\\
  0\\
  \vdots\\[1mm]
  0
\end{pmatrix}
\]
That is, treated as a column vector $v$ as just described, the string $s$ will
always be an element of the \emph{null space} of the matrix $M,$ provided that
we do the arithmetic modulo $2.$
This is true in both the case that $s = 0^n$ and $s\neq 0^n.$
To be more precise, the all-zero vector is always in the null space of $M,$ and
it's joined by the vector whose entries are the bits of $s$ in case
$s\neq 0^n.$

The question remaining is whether there will be any other vectors in the null
space of $M$ besides the ones corresponding to $0^n$ and $s.$
The answer is that this becomes increasingly unlikely as $k$ increases --- and
if we choose $k = n + 10,$ the null space of $M$ will contain no other vectors
in addition to those corresponding to $0^n$ and $s$ with greater than a
$99.9$\% chance.
More generally, if we replace $k = n + 10$ with $k = n + r$ for an arbitrary
choice of a positive integer $r,$ the probability that the vectors
corresponding to $0^n$ and $s$ are alone in the null space of $M$ is at least
$1 - 2^{-r}.$

Using linear algebra, it is possible to efficiently calculate a description of
the null space of $M$ modulo $2.$
Specifically, it can be done using \emph{Gaussian elimination}, which works the
same way when arithmetic is done modulo $2$ as it does with real or complex
numbers.
So long as the vectors corresponding to $0^n$ and $s$ are alone in the null
space of $M,$ which happens with high probability, we can deduce $s$ from the
results of this computation.

\subsubsection{Classical difficulty}

How many queries does a \emph{classical} query algorithm need to solve Simon's
problem?
The answer is: a lot, in general.

There are different precise statements that can be made about the classical
difficulty of this problem, and here's just one of them.
If we have any probabilistic query algorithm, and that algorithm makes fewer
than $2^{n/2 - 1} - 1$ queries, which is a number of queries that's
\emph{exponential} in $n,$ then that algorithm will fail to solve Simon's
problem with probability at least $1/2.$

Sometimes, proving impossibility results like this can be very challenging, but
this one isn't too difficult to prove through an elementary probabilistic
analysis.
Here, however, we'll only briefly examine the basic intuition behind it.

We're trying to find the hidden string $s,$ but so long as we don't query the
function on two strings having the same output value, we'll get very limited
information about $s.$
Intuitively speaking, all we'll learn is that the hidden string $s$ is
\emph{not} the exclusive-OR of any two distinct strings we've queried.
And if we query fewer than $2^{n/2 - 1} - 1$ strings, then there will still be
a lot of choices for $s$ that we haven't ruled out because there aren't enough
pairs of strings to cover all the possibilities.
This isn't a formal proof, it's just the basic idea.

So, in summary, Simon's algorithm provides us with a striking advantage of
quantum over classical algorithms within the query model.
In particular, Simon's algorithm solves Simon's problem with a number of
queries that's \emph{linear} in the number of input bits $n$ of our function,
whereas any classical algorithm, even if it's probabilistic, needs to make a
number of queries that's \emph{exponential} in $n$ in order to solve Simon's
problem with a reasonable probability of success.


\lesson{Quantum Algorithmic Foundations}
\label{lesson:quantum-algorithmic-foundations}

Quantum algorithms offer provable advantages over classical algorithms in the
query model of computation.
But what about a more standard model of computation, where problem inputs are
given explicitly rather than in the form of an oracle or black box?
This turns out to be a much more difficult question to answer, and to address
it we must first establish a solid foundation on which to base our
investigation.
This is the primary purpose of this lesson.

We'll begin by discussing \emph{computational cost}, for both classical and
quantum computations, and how it can be measured.
This is a general notion that can be applied to a wide range of computational
problems --- but to keep things simple we'll mainly examine it through the lens
of \emph{computational number theory}, which addresses computational tasks that
are likely to be familiar to most readers, including basic arithmetic,
computing greatest common divisors, and integer factorization.
While computational number theory is a narrow application domain, these
problems serve well to illustrate the basic issues (and they also happen to be
highly relevant to the lesson following this one).

Our focus is on \emph{algorithms}, as opposed to the ever-improving hardware on
which they're run.
Correspondingly, we'll be more concerned with how the cost of running an
algorithm scales as the specific problem instances it's run on grow in size,
rather than how many seconds, minutes, or hours some particular computation
requires.
We focus on this aspect of computational cost in recognition of the fact that
algorithms have fundamental importance, and will naturally be deployed against
larger and larger problem instances using faster and more reliable hardware as
technology develops.

Finally, we'll turn to a critically important task, which is running
\emph{classical} computations on quantum computers.
The reason this task is important is not because we hope to replace classical
computers with quantum computers --- which seems extremely unlikely to happen
any time soon, if ever --- but rather because it opens up many interesting
possibilities for quantum algorithms.
Specifically, classical computations running on quantum computers become
available as \emph{subroutines,} effectively leveraging decades of research and
development on classical algorithms in pursuit of quantum computational
advantages.

\section{Two examples: factoring and GCDs}

The classical computers that exist today are incredibly fast, and their speed
seems to be ever increasing.
For this reason, some might be inclined to believe that computers are so fast
that no computational problem is beyond their reach.

This belief is false.
Some computational problems are so inherently complex that, although there
exist algorithms to solve them, no computer on the planet Earth today is fast
enough to run these algorithms to completion on even moderately sized inputs
within the lifetime of a human --- or even within the lifetime of the Earth
itself.

To explain further, let's introduce the \emph{integer factorization} problem.

\begin{callout}[title={Integer factorization}]
  \begin{problem}
    \Input{An integer $N\geq 2$.}
    \Output{The prime factorization of $N$.}
  \end{problem}
\end{callout}

\noindent
By the \emph{prime factorization} of $N$ we mean a list of the prime factors of
$N$ and the powers to which they must be raised to obtain $N$ by
multiplication.
For example, the prime factors of $12$ are $2$ and $3,$ and to obtain $12$ we
must take the product of $2$ to the power $2$ and $3$ to the power $1.$
\[
12 = 2^2 \cdot 3
\]
Up to the ordering of the prime factors, there is only one prime factorization
for each positive integer $N\geq 2,$ which is a fact known as the
\emph{fundamental theorem of arithmetic}.

A few simple code demonstrations in Python will be helpful for further
explaining integer factorization and other concepts that relate to this
discussion. The following imports are needed for these demonstrations.
\vspace{2mm}

\begin{code}
  \begin{lstlisting}
import math
from sympy.ntheory import factorint\end{lstlisting}
\end{code}
\vspace{2mm}

The \texttt{factorint} function from the \texttt{SymPy} symbolic mathematics
package for Python solves the integer factorization problem for whatever input
$N$ we choose.
For example, we can obtain the prime factorization for $12$, which naturally
agrees with the factorization above.
\vspace{2mm}

\begin{code}
  \begin{lstlisting}
N = 12
print(factorint(N))\end{lstlisting}
\end{code}
\vspace{1mm}

\begin{code}
  \begin{lstlisting}
{2: 2, 3: 1}\end{lstlisting}
\end{code}
\vspace{2mm}

Factoring small numbers like $12$ is easy, but when the number $N$ to be
factored gets larger, the problem becomes more difficult.
For example, running \texttt{factorint} on a significantly larger number causes
a short but noticeable delay on a typical personal computer.
\vspace{2mm}

\begin{code}
  \begin{lstlisting}
N = 3402823669209384634633740743176823109843098343
print(factorint(N))\end{lstlisting}
\end{code}
\vspace{1mm}

\begin{code}
  \begin{lstlisting}
{3: 2, 74519450661011221: 1, 5073729280707932631243580787: 1}\end{lstlisting}
\end{code}
\vspace{2mm}

For even larger values of $N,$ things become impossibly difficult, at least as far as we know.
For example, the \emph{RSA Factoring Challenge}, which was run by RSA
Laboratories from 1991 to 2007, offered a cash prize of \$100,000 to factor the
following number, which has 309 decimal digits (or 1024 bits when written in
binary).
The prize for this number was never collected and its prime factors remain
unknown.
\pagebreak

\begin{code}
  \vspace*{5mm}
  
  \ttfamily\small
  RSA1024 = \seqsplit{135066410865995223349603216278805969938881475605667027524485143851526510604859533833940287150571909441798207282164471551373680419703964191743046496589274256239341020864383202110372958725762358509643110564073501508187510676594629205563685529475213500852879416377328533906109750544334999811150056977236890927563}

  \vspace*{5mm}
\end{code}
\vspace{2mm}

\noindent
Don't bother running \texttt{factorint} on RSA1024, it would not finish within
our lifetimes.

The fastest known algorithm for factoring large integers is known as the
\emph{number field sieve}.
As an example of this algorithm's use, the RSA challenge number RSA250, which
has 250 decimal digits (or 829 bits when written in binary), was factored using
the number field sieve in 2020.
The computation required thousands of CPU core-years, distributed across tens
of thousands of machines around the world.
Here we can appreciate this effort by checking the solution.
\vspace{2mm}

\begin{code}
  \vspace*{5mm}
  
  \ttfamily\small
  RSA250 = \seqsplit{2140324650240744961264423072839333563008614715144755017797754920881418023447140136643345519095804679610992851872470914587687396261921557363047454770520805119056493106687691590019759405693457452230589325976697471681738069364894699871578494975937497937}

  \vspace{1mm}

  p = \seqsplit{64135289477071580278790190170577389084825014742943447208116859632024532344630238623598752668347708737661925585694639798853367}

  \vspace{1mm}

  q = \seqsplit{33372027594978156556226010605355114227940760344767554666784520987023841729210037080257448673296881877565718986258036932062711}
  
  \vspace{1mm}

  print(RSA250 == p * q)

  \vspace*{5mm}  
\end{code}

\vspace{1mm}

\begin{code}
  \begin{lstlisting}
True\end{lstlisting}
\end{code}

\vspace{2mm}

The security of the RSA public-key cryptosystem is based on the computational
difficulty of integer factoring, in the sense that an efficient algorithm for
integer factoring would break it.

Next let's consider a related but very different problem, which is computing
the greatest common divisor (or GCD) of two integers.

\begin{callout}[title={Greatest common divisor (GCD)}]
  \begin{problem}
    \Input{Nonnegative integers $N$ and $M,$ at least one of which is
      positive.}
    \Output{The greatest common divisor of $N$ and $M$.}
  \end{problem}
\end{callout}

\noindent
The greatest common divisor of two numbers is the largest integer that evenly
divides both of them.

This problem is easy to solve with a computer --- it has roughly the same
computational cost as multiplying the two input numbers together.
The \texttt{gcd} function from the Python \texttt{math} module computes the
greatest common divisor of numbers that are considerably larger than RSA1024 in
the blink of an eye. (In fact, RSA1024 is the GCD of the two numbers in this
example.) \vspace{2mm}

\begin{code}
  \vspace*{5mm}
  
  \ttfamily\small
  N = \seqsplit{4636759690183918349682239573236686632636353319755818421393667064929987310592347460711767784882455889983961546491666129915628431549982893638464243493812487979530329460863532041588297885958272943021122033997933550246447236884738870576045537199814804920281890355275625050796526864093092006894744790739778376848205654332434378295899591539239698896074}

  \vspace{1mm}
  
  M = \seqsplit{5056714874804877864225164843977749374751021379176083540426461360945653967249306494545888621353613218518084414930846655066495767441010526886803458300440345782982127522212209489410315422285463057656809702949608368597012967321172325810519806487247195259818074918082416290513738155834341957254558278151385588990304622183174568167973121179585331770773}

  \vspace{1mm}

  print(math.gcd(N, M))

  \vspace*{5mm}
\end{code}

\vspace{1mm}

\begin{code}
  \vspace*{5mm}

  \ttfamily\small
  \seqsplit{135066410865995223349603216278805969938881475605667027524485143851526510604859533833940287150571909441798207282164471551373680419703964191743046496589274256239341020864383202110372958725762358509643110564073501508187510676594629205563685529475213500852879416377328533906109750544334999811150056977236890927563}

  \vspace*{5mm}
\end{code}

\noindent
This is possible because we have very efficient algorithms for computing GCDs,
the most well-known of which is \emph{Euclid's algorithm}, discovered over
2,000 years ago.

Could there be a fast algorithm for integer factorization that we just haven't
discovered yet, allowing large numbers like RSA1024 to be factored in the blink
of an eye?
The answer is yes.
Although we might expect that an efficient algorithm for factoring as simple
and elegant as Euclid's algorithm for computing GCDs would have been discovered
by now, there is nothing that rules out the existence of a very fast classical
algorithm for integer factorization, beyond the fact that we've failed to find
one thus far.
One could be discovered tomorrow --- but don't hold your breath.
Generations of mathematicians and computer scientists have searched, and
factoring numbers like RSA1024 remains beyond our reach.

\section{Measuring computational cost}

Next, we'll discuss a mathematical framework through which computational cost
can be measured, narrowly focused on the needs of this course.
The \emph{analysis of algorithms} and \emph{computational complexity} are
entire subjects onto themselves, and have much more to say about these notions.

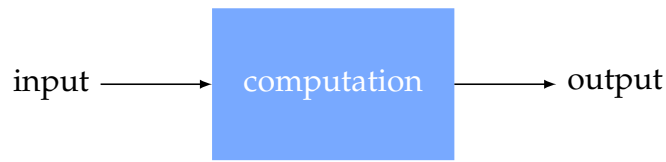
\begin{figure}[!ht]
  \begin{center}
    \begin{tikzpicture}[
        >=latex,
        line width=0.6pt,
        scale=1.5,
        box/.style={%
          inner sep = 0,
          fill = CircuitBlue,
          draw = CircuitBlue,
          text = white,
          minimum width = 32mm,
          minimum height = 20mm}
      ]
      
      \node (in) at (-2.5,0) {input};
      \node[box] (computation) at (0,0) {computation};
      \node (out) at (2.5,0) {output};

      \draw[->] (in.east) -- (computation.west);
      \draw[->] (computation.east) -- (out.west);

    \end{tikzpicture}    
  \end{center}
  \caption{A simple abstraction of a standard model of computation.}
  \label{fig:standard-computation-2}
\end{figure}

As a starting point, consider Figure~\ref{fig:standard-computation-2}, which
also appeared in the previous lesson, which represents a very high level
abstraction of a computation.
The computation itself could be modeled or described in a variety of ways, such
as by a computer program written in Python, a Turing machine, a Boolean
circuit, or a quantum circuit.
Our focus will be on circuits (both Boolean and quantum).

\subsection{Encodings and input length}

Let's begin with the input and output of a computational problem, which we'll
assume are \emph{binary strings}.
Different symbols could be used, but we'll keep things simple for the purposes
of this discussion by restricting our attention to binary string inputs and
outputs.
Through binary strings, we can \emph{encode} a variety of interesting objects
that the problems we're interested in solving might concern, such as numbers,
vectors, matrices, and graphs, as well as lists of these and other objects.

For example, to encode nonnegative integers, we can use \emph{binary notation}.
The following table lists the binary encoding of the first nine nonnegative
integers, along with the \emph{length} (meaning the total number of bits) of
each encoding.
\begin{center}
\begin{tabular}{lll}
  Number & Binary encoding & Length \\\hline
  0 & 0 & 1 \\
  1 & 1 & 1 \\
  2 & 10 & 2 \\
  3 & 11 & 2 \\
  4 & 100 & 3 \\
  5 & 101 & 3 \\
  6 & 110 & 3 \\
  7 & 111 & 3 \\
  8 & 1000 & 4 
\end{tabular}
\end{center}
We can easily extend this encoding to handle both positive and negative
integers by appending a \emph{sign bit} to the representations if we choose.
Sometimes it's also convenient to allow binary representations of nonnegative
integers to have leading zeros, which don't change the value being encoded but
can allow representations to fill up a string or word of a fixed size.

Using binary notation to represent nonnegative integers is both common and
efficient, but if we wanted to we could choose a different way to represent
nonnegative integers using binary strings, such as the ones suggested in the
following table.
The specifics of these alternatives are not important to this discussion ---
the point is only to clarify that we do have choices for the encodings we use.
\begin{center}
\begin{tabular}{lll}
 Number & Unary encoding & Lexicographic encoding \\\hline
 0 & $\varepsilon$ & $\varepsilon$ \\
 1 & 0 & 0 \\
 2 & 00 & 1 \\
 3 & 000 & 00 \\
 4 & 0000 & 01 \\
 5 & 00000 & 10 \\
 6 & 000000 & 11 \\
 7 & 0000000 & 000 \\
 8 & 00000000 & 001 
\end{tabular}
\end{center}

\noindent
(In this table, the symbol $\varepsilon$ represents the \emph{empty string},
which has no symbols in it and length equal to zero. Naturally, to avoid an
obvious source of confusion, we use a special symbol such as $\varepsilon$ to
represent the empty string rather than literally writing nothing.)

Other types of inputs, such as vectors and matrices, or more complicated
objects like descriptions of molecules, can also be encoded as binary strings.
Just like we have for nonnegative integers, a variety of different encoding
schemes can be selected or invented.
For whatever scheme we come up with to encode inputs to a given problem, we
interpret the \emph{length} of an input string as representing the size of the
problem instance being solved.

For example, the number of bits required to express a nonnegative integer $N$
in binary notation, which is sometimes denoted $\operatorname{lg}(N),$ is given
by the following formula.
\[
\operatorname{lg}(N) =
\begin{cases}
  1 & N = 0\\
  1 + \lfloor \log_2 (N) \rfloor & N \geq 1
\end{cases}
\]
Assuming that we use binary notation to encode the input to the integer
factoring problem, the \emph{input length} for the number $N$ is therefore
$\operatorname{lg}(N).$
Note, in particular, that the length (or size) of the input $N$ is not $N$
itself; when $N$ is large we don't need nearly this many bits to express $N$ in
binary notation.

From a strictly formal viewpoint, whenever we consider a computational problem
or task, it should be understood that a specific scheme has been selected for
encoding whatever objects are given as input or produced as output.
This allows computations that solve interesting problems to be viewed
abstractly as transformations from binary string inputs to binary string
outputs.

The details of how objects are encoded as binary strings must necessarily be
important to these computations at some level.
Usually, though, we don't worry all that much about these details when we're
analyzing computational cost, so that we can avoid getting into details of
secondary importance.
The basic reasoning is that we expect the computational cost of converting back
and forth between ``reasonable'' encoding schemes to be insignificant compared
with the cost of solving the actual problem.
In those situations in which this is not the case, the details can (and should) be clarified.

For example, a very simple computation converts between the binary
representation of a nonnegative integer and its lexicographic encoding (which
we have not explained in detail, but it can be inferred from the table
above). For this reason, the computational cost of integer factoring wouldn't
differ significantly if we decided to switch from using one of these encodings
to the other for the input $N.$
On the other hand, encoding nonnegative integers in unary notation incurs an
exponential blow-up in the total number of symbols required, and we would not
consider it to be a ``reasonable'' encoding scheme for this reason.

\subsection{Elementary operations}

Now let's consider the computation itself, which is represented by the blue
rectangle in Figure~\ref{fig:standard-computation-2}.
The way that we'll measure computational cost is to count the number of
\emph{elementary operations} that each computation requires.
Intuitively speaking, an elementary operation is one involving a small, fixed
number of bits or qubits, that can be performed quickly and easily --- such as
computing the AND of two bits.
In contrast, running the \texttt{factorint} function is not reasonably viewed as
being an elementary operation.

Formally speaking, there are different choices for what constitutes an
elementary operation depending on the computational model being used.
Our focus will be on circuit models, and specifically quantum and Boolean
circuits.

\subsubsection{Universal gate sets}

For circuit-based models of computation, it's typical that each \emph{gate} is
viewed as an elementary operation.
This leads to the question of precisely which gates we permit in our circuits.
Focusing for the moment on quantum circuits, we've seen several gates thus far
in this course, including $X,$ $Y,$ $Z,$ $H,$ $S,$ and $T$ gates, \emph{swap}
gates, controlled versions of gates (including \emph{controlled-NOT},
\emph{Toffoli}, and \emph{Fredkin} gates), as well as gates that represent
standard basis measurements.
In the context of the CHSH game we also saw a few additional \emph{rotation}
gates.

We also discussed \emph{query gates} in the context of the query model, and we
also saw that any unitary operation $U,$ acting on any number of qubits, can be
viewed as being a gate if we so choose --- but we'll disregard both of these
options for the sake of this discussion.
We won't be working in the query model (although the implementation of query
gates using elementary operations is discussed later in the lesson), and
viewing arbitrary unitary gates, potentially acting on millions of qubits, as
being elementary operations does not lead to meaningful or realistic notions of
computational cost.

Sticking with quantum gates that operate on small numbers of qubits, one
approach to deciding which ones are to be considered elementary is to tease out
a precise criterion --- but this is not the standard approach or the one we'll
take.
Rather, we simply make a choice.

Here's one standard choice, which we shall adopt as the \emph{default} gate set
for quantum circuits:
\begin{itemize}
\item
  Single-qubit unitary gates from this list: $X,$ $Y,$ $Z,$ $H,$ $S,$
  $S^{\dagger},$ $T,$ and $T^{\dagger}$.
\item
  Controlled-NOT gates.
\item
  Single-qubit standard basis measurements.
\end{itemize}

A common alternative is to view Toffoli, Hadamard, and $S$ gates as being
elementary, in addition to standard basis measurements.
Sometimes all single-qubit gates are viewed as being elementary, though this
does lead to an unrealistically powerful model when the accuracy with which
gates are performed is not properly taken into account.

The unitary gates in our default collection form what's called a
\emph{universal} gate set.
This means that we can approximate any unitary operation on any number of
qubits to any degree of accuracy we wish, using circuits composed of these
gates alone.
To be clear, the definition of universality places no requirements on the cost
of such approximations, meaning the number of gates from our set that we need.
Indeed, a fairly simple argument based on the mathematical notion of measure
reveals that most unitary operations must have extremely high cost.
Proving the universality of quantum gate sets is not a simple matter and won't
be covered in this course.

For Boolean circuits, we'll take AND, OR, NOT, and FANOUT gates to be the ones
representing elementary operations.
We don't actually need both AND gates and OR gates --- we can use
\emph{De~Morgan's laws} to convert from either one to the other by placing NOT
gates on all three input/output wires --- but nevertheless it is both typical
and convenient to allow both AND and OR gates.
AND, OR, NOT, and FANOUT gates form a universal set for deterministic
computations, meaning that any function from any fixed number of input bits to
any fixed number of output bits can be implemented with these gates.

\subsubsection{The principle of deferred measurement}

\begin{figure}[b]
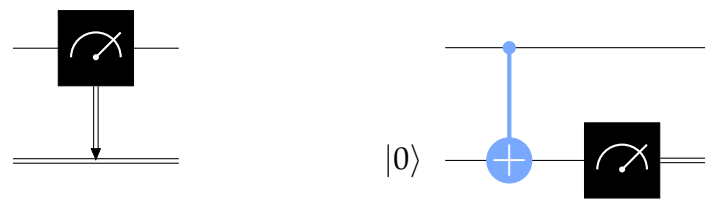

  \begin{center}


  \end{center}
  \caption{The standard basis measurement on the left can be deferred
    through the introduction of a workspace qubit and a controlled-NOT gate, as
    shown on the right.}
  \label{fig:deferred-measurement}
\end{figure}

Standard basis measurement gates can appear within quantum circuits, but
sometimes it's convenient to delay them until the end.
This allows us to view quantum computations as consisting of a unitary part
(representing the computation itself), followed by a simple read-out phase
where qubits are measured and the results are output.
This can always be done, provided that we're willing to add an additional qubit
for each standard basis measurement.
Figure~\ref{fig:deferred-measurement} illustrates how this can be done.

Specifically, the classical bit in the circuit on the left is replaced by a
qubit on the right (initialized to the $\vert 0\rangle$ state), and the
standard basis measurement is replaced by a controlled-NOT gate, followed by a
standard basis measurement on the bottom qubit.
The point is that the standard basis measurement in the right-hand circuit can
be pushed all the way to the end of the circuit.
If the classical bit in the circuit on the left is later used as a control bit,
we can use the bottom qubit in the circuit on the right as a control instead,
and the overall effect will be the same.
(We are assuming that the classical bit in the circuit on the left doesn't get
overwritten after the measurement takes place by another measurement --- but if
it did we could always just use a new classical bit rather than overwriting one
that was used for a previous measurement.)

\subsection{Circuit size and depth}

\subsubsection{Circuit size}

The total number of gates in a circuit is referred to as that circuit's
\emph{size}.
Thus, presuming that the gates in our circuits represent elementary operations,
a circuit's size represents the number of elementary operations it requires ---
or, in other words, its \emph{computational cost}.
We write $\operatorname{size}(C)$ to refer to the size of a given circuit $C.$

\begin{figure}[!ht]
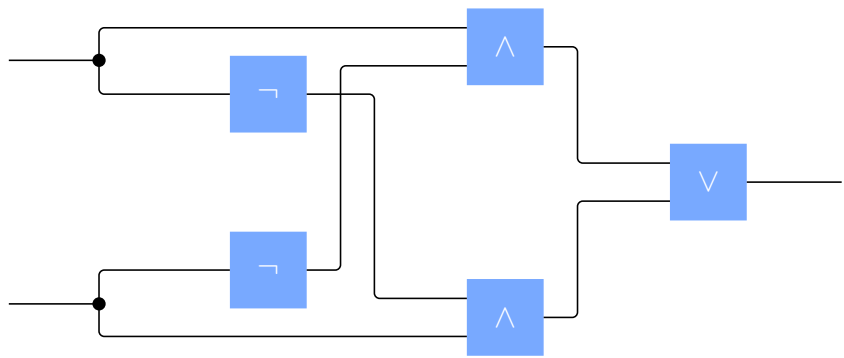

  \begin{center}

  \end{center}
  \caption{A Boolean circuit for computing the exclusive-OR of two bits.}
  \label{fig:Boolean-circuit-XOR-2}
\end{figure}

For example, consider the Boolean circuit for computing the XOR of
two bits shown in Figure~\ref{fig:Boolean-circuit-XOR-2}, which we've now
encountered a few times.
The size of this circuit is 7 because there are 7 gates in total.
(Fanout operations are not always counted as being gates, but for the purposes
of this lesson we will count them as being gates.)

\subsubsection{Time, cost, and circuit depth}

\emph{Time} is a critically important resource, or a limiting constraint, for
computations.
The examples above, such as the task of factoring RSA1024, reinforce this
viewpoint.
The \texttt{factorint} function doesn't fail to factor RSA1024 per se, it's
just that we don't have enough time to let it finish.

The notion of computational cost, as the number of elementary operations
required to perform a computation, is intended to be an abstract proxy for the
time required to implement a computation.
Each elementary operation requires a certain amount of time to perform, and the
more of them a computation needs, the longer it's going to take, at least in
general.
In the interest of simplicity, we'll continue to make this association between
computational cost and the time required to run algorithms.

But notice that the size of a circuit doesn't necessarily correspond directly
to how long it takes to run.
In our Boolean circuit for computing the XOR of two bits, for instance, the two
FANOUT gates could be performed simultaneously, as could the two NOT gates, as
well as the two AND gates.
A different way to measure the efficiency of circuits, which takes this
possibility of \emph{parallelization} into account, is by their \emph{depth}.
This is the minimum number of \emph{layers} of gates needed within the circuit,
where the gates within each layer operate on different wires.
Equivalently, the depth of a circuit is the maximum number of gates encountered
on any path from an input wire to an output wire.
For the circuit above, for instance, the depth is 4.

Circuit depth is one way to formalize the running time of parallel
computations.
It's an advanced topic, and there exist very sophisticated circuit
constructions known to minimize the depth required for certain computations.
There are also some fascinating unanswered questions concerning circuit depth.
For example, much remains unknown about the circuit depth required to compute
GCDs.

We won't have too much more to say about circuit depth in this course, aside
from including a few interesting facts concerning circuit depth as we go along,
but it should be clearly acknowledged that parallelization is a potential
source of computational advantages.

\subsubsection{Assigning costs to different gates}

One final note concerning circuit size and computational cost is that it is
possible to assign different costs to gates, rather than viewing every gate as
contributing equally to the total cost.

For example, as was already mentioned, FANOUT gates are often viewed as being
free for Boolean circuits --- which is to say that we could choose that FANOUT
gates have zero cost.
As another example, when we're working in the query model and we count the
number of queries that a circuit makes to an input function (in the form of a
black box), we're effectively assigning unit cost to query gates and zero cost
to other gates, such as Hadamard gates.
A final example is that we sometimes assign different costs to gates depending
on how difficult they are to implement, which could vary depending upon the
hardware being considered.

While all of these options are sensible in different contexts, for this lesson
we'll keep things simple and stick with circuit size as a representation of
computational cost.

\subsection{Cost as a function of input length}

We're primarily interested in how computational cost scales as inputs become
larger and larger.
This leads us to represent the costs of algorithms as \emph{functions} of the
input length.

\subsubsection{Families of circuits}

Inputs to a given computational problem can vary in length, potentially
becoming arbitrarily large.
Every circuit, on the other hand, has a fixed number of gates and wires.
For this reason, when we think about algorithms in terms of circuits, we
generally need infinitely large \emph{families} of circuits to represent
algorithms.
By a family of circuits, we mean a sequence of circuits that grow in size,
allowing larger and larger inputs to be accommodated.

For example, imagine that we have a classical algorithm for integer
factorization, such as the one used by \texttt{factorint}.
Like all classical algorithms, this algorithm can be implemented using Boolean
circuits --- but to do it we'll need a separate circuit for each possible input
length.
If we looked at the resulting circuits for different input lengths, we would
see that their sizes naturally grow as the input length grows --- reflecting
the fact that factoring 4-bit integers is much easier and requires far fewer
elementary operations than factoring 1024-bit integers, for instance.

This leads us to represent the computational cost of an algorithm by a
function~$t,$ defined so that $t(n)$ is the number of gates in the circuit that
implements the algorithm for $n$ bit inputs.
In more formal terms, an algorithm in the Boolean circuit model is described by
a sequence of circuits
\[
\{C_1, C_2, C_3,\ldots\},
\]
where $C_n$ solves whatever
problem we're talking about for $n$-bit inputs (or, more generally, for inputs
whose size is parameterized in some way by $n$), and the function $t$
representing the cost of this algorithm is defined as
\[
t(n) = \operatorname{size}(C_n).
\]

For quantum circuits the situation is similar, where larger and larger circuits
are needed to accommodate longer and longer input strings.

\subsubsection{Example: integer addition}

To explain further, let's take a moment to consider the problem of integer
addition, which is much simpler than integer factoring or even computing GCDs.

\begin{callout}[title={Integer addition}]
  \begin{problem}
    \Input{Integers $N$ and $M$.}
    \Output{$N+M$.}
  \end{problem}
\end{callout}

\noindent
How might we design Boolean circuits for solving this problem?

To keep things simple, let's restrict our attention to the case where $N$ and
$M$ are both nonnegative integers represented by $n$-bit strings using binary
notation.
We'll allow for any number of leading zeros in these encodings, so that
\[
0\leq N,M\leq 2^n - 1.
\]
The output will be an $(n+1)$-bit binary string representing the sum, which is
the maximum number of bits we need to express the result.

We begin with an algorithm --- the \emph{standard} algorithm for addition of
binary representations --- which is the base $2$ analogue to the way addition
is taught in elementary/primary schools around the world.
This algorithm can be implemented with Boolean circuits as follows.

Starting from the least significant bits, we can compute their XOR to determine
the least significant bit for the sum.
Then we compute the carry bit, which is the AND of the two least significant
bits of $N$ and $M.$
Sometimes these two operations together are known as a \emph{half adder}.

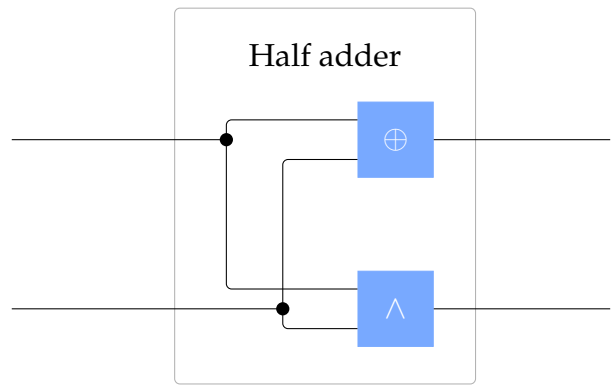
\begin{figure}[!ht]
  \begin{center}

    \begin{tikzpicture}[scale=1.5,
        gate/.style={%
          inner sep = 0,
          fill = CircuitBlue,
          draw = CircuitBlue,
          text = white,
          minimum size = 10mm},
        control/.style={%
          circle,
          fill,
          minimum size = 5pt,
          inner sep=0mm}        
      ]
      
      \node (In0) at (-3.5,0.75) {};
      \node (In1) at (-3.5,-0.75) {};
      
      \node (Out0) at (2,0.75) {};
      \node (Out1) at (2,-0.75) {};

      \node[
        draw = black!30,
        rounded corners=2pt,
        minimum width = 40mm,
        minimum height = 50mm
      ] at (-0.625,0.25) {}; 
      
      \node[control] (Fanout0) at (-1.5,0.75) {};
      
      \node[control] (Fanout1) at (-1,-0.75) {};
      
      \node [gate] (XOR) at (0,0.75) {$\oplus$};
      \node [gate] (AND) at (0,-0.75) {$\wedge$};
      
      \draw (In0.east) -- (Fanout0.center)
      node [midway, above] {};
      
      \draw (In1.east) -- (Fanout1.center)
      node [midway, above] {};
      
      \draw[rounded corners = 2pt] (Fanout0.center)
      |- ([yshift=5]AND.west) node [pos = 0.605, above] {};
      
      \draw[rounded corners = 2pt] (Fanout0.center)
      |- ([yshift=5]XOR.west) node [pos = 0.605, above] {};
      
      \draw[rounded corners = 2pt] (Fanout1.center)
      |- ([yshift=-5]AND.west) node [pos = 0.605, above] {};
      
      \draw[rounded corners = 2pt] (Fanout1.center)
      |- ([yshift=-5]XOR.west) node [pos = 0.605, above] {};
      
      \draw (XOR.east) -- (Out0.west);
      \draw (AND.east) -- (Out1.west);
      
      \node at (-0.625,1.5) {Half adder};
      
    \end{tikzpicture}
    
  \end{center}
  \caption{A Boolean circuit implementing a half adder using two FANOUT gates,
    an XOR gate, and an AND gate.}
  \label{fig:half-adder}
\end{figure}

Using the XOR circuit we've now seen a few times together with an AND gate and
two FANOUT gates, we can build a half adder with 10 gates.
If for some reason we changed our minds and decided to include XOR gates in our
set of elementary operations, we would need 1 AND gate, 1 XOR gate, and 2
FANOUT gates to build a half adder.

\begin{figure}[!ht]
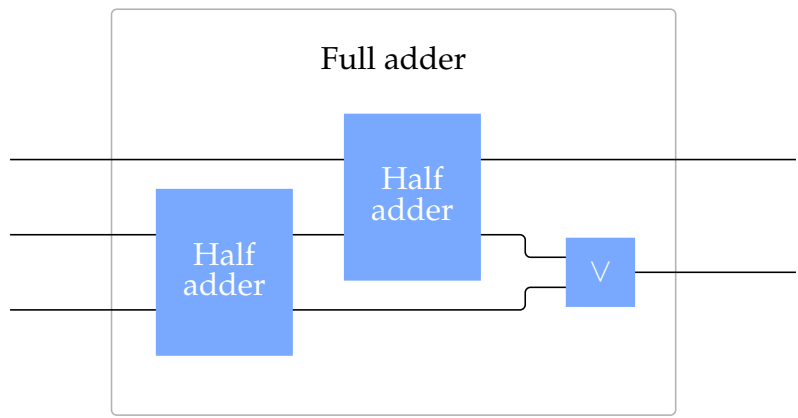

  \begin{center}

    
  \end{center}
  \caption{A full adder constructed from two half adders and an OR gate.}
  \label{fig:full-adder}
\end{figure}

Moving on to the more significant bits, we can use a similar procedure, but
this time including the carry bit from each previous position into our
calculation.
By cascading two half adders and taking the OR of the carry bits they produce,
we can create what's known as a \emph{full adder}.
Figure~\ref{fig:full-adder} illustrates this construction.
This requires 21 gates in total: 2 AND gates, 2 XOR gates (each requiring 7
gates to implement), one OR gate, and 4 FANOUT gates.

\begin{figure}[!ht]
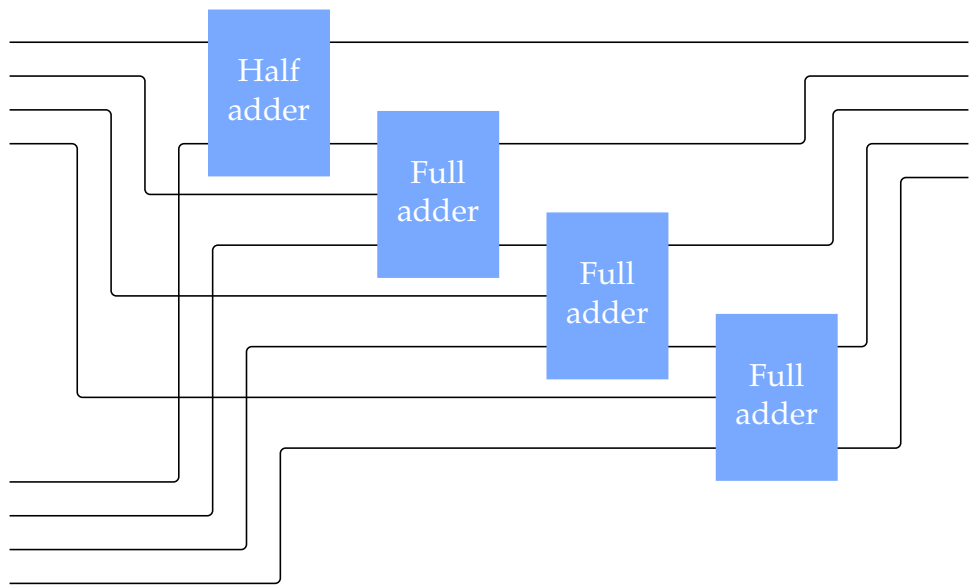

  \begin{center}
    %
  \end{center}
  \caption{Cascading a half adder and three full adders creates a Boolean
    circuit for adding two $4$-bit integers.}
  \label{fig:addition-circuit}
\end{figure}

Finally, by cascading a half adder along with however many full adders as
needed, we obtain a Boolean circuit for nonnegative integer addition.
For example, Figure~\ref{fig:addition-circuit} illustrates how this is done
when computing the sum of two 4-bit integers.

In general, this requires
\[
21 (n-1) + 10 = 21 n - 11
\]
gates.
Had we decided to include XOR gates in our set of elementary operations, we
would need $2n-1$ AND gates, $2n-1$ XOR gates, $n-1$ OR gates, and $4n-2$
FANOUT gates, for a total of $9n-5$ gates.
If in addition we decide not to count FANOUT gates, it's $5n-3$ gates.

\subsubsection{Asymptotic notation}

On the one hand, it's good to know precisely how many gates are needed to
perform various computations, like in the example of integer addition above.
These details are important for actually building the circuits.

On the other hand, if we perform analyses at this level of detail for all the
computations we're interested in, including ones for tasks that are much more
complicated than addition, we'll very quickly be buried in details.
To keep things manageable, and to intentionally suppress details of secondary
importance, we typically use \emph{Big-O} notation when analyzing algorithms.
Through this notation we can make useful statements about the rate at which
functions grow.

Formally speaking, if we have two functions $g(n)$ and $h(n),$ we write that
$g(n) = O(h(n))$ if there exists a positive real number $c > 0$ and a positive
integer $n_0$ such that
\[
g(n) \leq c\cdot h(n)
\]
for all $n \geq n_0.$
Typically $h(n)$ is chosen to be as simple an expression as possible, so that
the notation can be used to reveal the limiting behavior of a function in
simple terms.
For example, $17 n^3 - 257 n^2 + 65537 = O(n^3).$

This notation can be extended to functions having multiple arguments in a
fairly straightforward way.
For instance, if we have two functions $g(n,m)$ and $h(n,m)$ defined on
positive integers $n$ and $m,$ we write that $g(n,m) = O(h(n,m))$ if there
exists a positive real number $c > 0$ and a positive integer $k_0$ such that
\[
g(n,m) \leq c\cdot h(n,m)
\]
whenever $n+m \geq k_0.$

Connecting this notation to the example of nonnegative integer addition, we
conclude that there exists a family of Boolean circuits $\{C_1, C_2,\ldots,\},$
where $C_n$ adds two $n$-bit nonnegative integers together, such that
$\operatorname{size}(C_n) = O(n).$
This reveals the most essential feature of how the cost of addition scales with
the input size: it scales \emph{linearly}.

Notice also that it doesn't depend on the specific detail of whether we
consider XOR gates to have unit cost or cost $7.$
In general, using Big-O notation allows us to make statements about
computational costs that aren't sensitive to such low-level details.

\subsubsection{More examples}

Here are a few more examples of problems from computational number theory,
beginning with \emph{multiplication}.

\begin{callout}[title={Integer multiplication}]
  \begin{problem}
    \Input{Integers $N$ and $M$.}
    \Output{$NM$.}
  \end{problem}
\end{callout}

\noindent
Creating Boolean circuits for this problem is more difficult than creating
circuits for addition --- but by thinking about the
\emph{standard multiplication algorithm}, we can come up with circuits having
size $O(n^2)$ for this problem (assuming $N$ and $M$ are both represented by
$n$-bit binary representations).
More generally, if $N$ has $n$ bits and $M$ has $m$ bits, there are Boolean
circuits of size $O(nm)$ for multiplying $N$ and~$M.$

There are, in fact, other ways to multiply that scale better.
For instance, the Sch\"{o}nhage--Strassen multiplication algorithm can be used
to create Boolean circuits for multiplying two $n$-bit integers at cost
$O(n \operatorname{lg}(n) \operatorname{lg}(\operatorname{lg}(n))).$
The intricacy of this method causes a lot of overhead, however, making it only
practical for numbers having tens of thousands of bits or more.

Another basic problem is \emph{division}, which we interpret to mean computing
both the quotient and remainder given an integer divisor and dividend.

\begin{callout}[title={Integer division}]
  \begin{problem}
    \Input{Integers $N$ and $M\neq 0$.}
    \Output{Integers $q$ and $r$ satisfying $0\leq r < |M|$
      and $N = q M + r$.}
  \end{problem}
\end{callout}

\noindent
The cost of integer division is similar to multiplication: if $N$ has $n$ bits
and $M$ has $m$ bits, there are Boolean circuits of size $O(nm)$ for solving
this problem.
And like multiplication, asymptotically superior methods are known.

We can now compare known algorithms for computing GCDs with those for addition
and multiplication.
Euclid's algorithm for computing the GCD of an $n$-bit number $N$ and an
$m$-bit number $M$ requires Boolean circuits of size $O(nm),$ similar to the
standard algorithms for multiplication and division.
Also similar to multiplication and division, there are asymptotically faster
GCD algorithms --- including ones requiring $O(n(\operatorname{lg}(n))^2
\operatorname{lg}(\operatorname{lg}(n)))$ elementary operations to compute the
GCD of two $n$-bit numbers.

A somewhat more expensive computation that arises in number theory is
\emph{modular exponentiation}.

\begin{callout}[title={Integer modular exponentiation}]
  \begin{problem}
    \Input{Integers $N,$ $K,$ and $M$ with $K\geq 0$ and $M\geq 1$.}
    \Output{$N^K \hspace{1mm} (\text{mod }M)$}
  \end{problem}
\end{callout}

\noindent
By $N^K\hspace{1mm} (\text{mod }M)$ we mean the remainder when $N^K$ is divided
by $M,$ meaning the unique integer $r$ satisfying $0\leq r < M$ and
$N^K = q M + r$ for some integer $q.$

If $N$ has $n$ bits, $M$ has $m$ bits, and $K$ has $k$ bits, this problem can
be solved by Boolean circuits having size $O(k m^2 + nm).$
This is not at all obvious.
The solution is not to first compute $N^K$ and then take the remainder, which
would necessitate using exponentially many bits just to store the number $N^K.$
Rather, we can use the \emph{power algorithm} (known alternatively as the
\emph{binary method} and \emph{repeated squaring}), which makes use of the
binary representation of $K$ to perform the entire computation modulo $M.$
Assuming $N,$ $M,$ and $K$ are all $n$-bit numbers, we obtain an $O(n^3)$
algorithm --- or a \emph{cubic} time algorithm.
And once again, there are known algorithms that are more complicated but
asymptotically faster.

\subsubsection{Cost of integer factorization}

In contrast to the algorithms just discussed, known algorithms for integer
factorization are much more expensive --- as we might expect from the
discussion earlier in the lesson.

One simple approach to factoring is \emph{trial division,} where an algorithm
searches through the list $2,\ldots,\sqrt{N}$ to find a prime factor of an
input number $N.$
This requires $O(2^{n/2})$ iterations in the worst case when $N$ is an $n$-bit
number.
Each iteration requires a trial division, which means $O(n^2)$ elementary
operations for each iteration (using a standard algorithm for integer
division).
We end up with circuits of size $O(n^2 2^{n/2}),$ which is \emph{exponential}
in the input size $n.$

There are algorithms for integer factorization having better scaling.
The number field sieve mentioned earlier, for instance, which is an algorithm
that makes use of randomness, is generally believed (but not rigorously proven)
to require
\[
2^{O(n^{1/3} (\operatorname{lg}(n))^{2/3})}
\]
elementary operations to factor $n$-bit integers with high probability.
While it is quite significant that $n$ is raised to the power $1/3$ in the
exponent of this expression, the fact it appears in the exponent is still a
problem that causes poor scaling --- and explains in part why RSA1024 remains
outside of its domain of applicability.

\subsubsection{Polynomial versus exponential cost}

Classical algorithms for integer addition, multiplication, division, and
computing greatest common divisors allow us to solve these problems in the
blink of an eye for inputs with thousands of bits.
Addition has \emph{linear} cost while the other three problems have
\emph{quadratic} cost (or \emph{subquadratic} cost using asymptotically fast
algorithms).
Modular exponentiation is more expensive but can still be done pretty
efficiently, with \emph{cubic} cost (or sub-cubic cost using asymptotically
fast algorithms).

These are all examples of algorithms having \emph{polynomial} cost, meaning
that they have cost $O(n^c)$ for some choice of a fixed constant $c > 0.$
As a rough, first-order approximation, algorithms having polynomial cost are
abstractly viewed as representing \emph{efficient} algorithms.

In contrast, known classical algorithms for integer factoring have
\emph{exponential} cost.
Sometimes the cost of the number field sieve is described as
\emph{sub-exponential} because $n$ is raised to the power $1/3$ in the
exponent, but in complexity theory it is more typical to reserve this term for
algorithms whose cost is
\[
O\bigl(2^{n^{\varepsilon}}\bigr)
\]
for \emph{every} $\varepsilon > 0.$
The so-called \emph{NP-complete} problems are a class of problems not known to
(and widely conjectured not to) have polynomial-cost algorithms.
A circuit-based formulation of the \emph{exponential-time hypothesis} posits
something even stronger, which is that no NP-complete problem can have a
sub-exponential cost algorithm.

The association of polynomial-cost algorithms with efficient algorithms must be
understood as being a loose abstraction.
Of course, if an algorithm's cost scales as $n^{1000}$ or $n^{1000000}$ for
inputs of size $n,$ then it's a stretch to describe that algorithm as being
efficient.
However, even an algorithm having cost that scales as $n^{1000000}$ must be
doing something clever to avoid having \emph{exponential} cost, which is
generally what we expect of algorithms based in some way on ``brute force'' or
``exhaustive search.''
Even the sophisticated refinements of the number field sieve, for instance,
fail to avoid this exponential scaling in cost.
Polynomial-cost algorithms, on the other hand, manage to take advantage of the
problem structure in some way that avoids an exponential scaling.

In practice, the identification of a polynomial-cost algorithm for a problem is
just a first step toward actual efficiency.
Through algorithmic refinements, polynomial-cost algorithms with large
exponents can sometimes be improved dramatically, lowering the cost to a more
``reasonable'' polynomial scaling.
Sometimes things become easier when they're known to be possible --- so the
identification of a polynomial-cost algorithm for a problem can also have the
effect of inspiring new, even more efficient algorithms.

As we consider advantages of quantum computing over classical computing, our
eyes are generally turned first toward \emph{exponential} advantages, or at
least \emph{super-polynomial} advantages --- ideally finding polynomial-cost
quantum algorithms for problems not known to have polynomial-cost classical
algorithms.
Theoretical advantages on this order have the greatest chances to lead to
actual practical advantages --- but identifying such advantages is an extremely
difficult challenge.
Only a few examples are currently known, but the search continues.

Polynomial (but not super-polynomial) advantages in computational cost of
quantum over classical are also interesting and should not be dismissed --- but
given the current gap between quantum and classical computing technology, they
do seem rather less compelling at the present time.
One day, though, they could become significant.
\emph{Grover's algorithm,} for instance, which is covered in the last lesson of
this unit, offers a \emph{quadratic} advantage of quantum over classical for
so-called \emph{unstructured searching,} and has a potential for broad
applications.

\subsubsection{A hidden cost of circuit computation}

There is one final issue that's worth mentioning, although we will not concern
ourselves with it further in this course.
There's a ``hidden'' computational cost when we're working with circuits, and
it concerns the specifications of the circuits themselves.
As inputs get longer and longer, larger and larger circuits are required ---
but we need to get our hands on the descriptions of these circuits somehow if
we're going to implement them.

For all of the examples we've discussed, or will discuss in subsequent lessons,
there's an underlying algorithm from which the circuits are derived.
Usually the circuits in a family follow some basic pattern that's easy to
extrapolate to larger and larger inputs, such as cascading full adders to
create Boolean circuits for addition or performing layers of Hadamard gates and
other gates in some simple-to-describe pattern.

But what happens if there are prohibitive computational costs associated with
the patterns in the circuits themselves?
For instance, the description of each member $C_n$ in a circuit family could,
in principle, be determined by some extremely difficult to compute function of
$n.$

The answer is that this is indeed a problem --- and so we must place additional
restrictions on families of circuits beyond having polynomial cost in order for
them to truly represent efficient algorithms.
The property of \emph{uniformity} for circuits does this by stipulating that,
in various precise formulations, it must be computationally easy to obtain the
description of each circuit in a family.
All of the circuit families we'll discuss do have this property --- but this is
nevertheless an important issue to be aware of in general when studying circuit
models of computation from a formal viewpoint.

\section{Classical computations on quantum computers}

We'll now turn our attention to implementing classical algorithms on quantum
computers.
We'll see that any computation that can be performed with a classical Boolean
circuit can also be performed by a quantum circuit with a similar asymptotic
computational cost.
Moreover, this can be done in a ``clean'' manner to be described shortly, which
is an important requirement for using these computations as subroutines inside
of larger quantum computations.

\subsection{Simulating Boolean circuits with quantum circuits}

Boolean circuits are composed of AND, OR, NOT, and FANOUT gates.
To simulate Boolean circuits with quantum circuits, we'll begin by showing how
each of these four gates can be simulated by quantum gates.
Once that's done, converting a given Boolean circuit to a quantum circuit is a
simple matter of simulating one gate at a time.
We'll only need NOT gates, controlled-NOT gates, and Toffoli gates to do this,
which are all deterministic operations in addition to being unitary.

\subsubsection{Toffoli gates}

\begin{figure}[!ht]
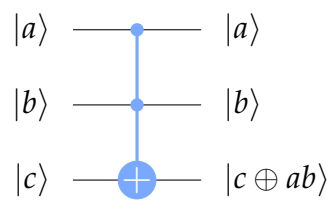

  \begin{center}

    
  \end{center}
  \caption{The action of a Toffoli gate on a standard basis state.}
  \label{fig:Toffoli-gate}
\end{figure}

Toffoli gates can alternatively be described as controlled-controlled-NOT
gates, whose action on standard basis states is as shown in
Figure~\ref{fig:Toffoli-gate}.

Bearing in mind that we're using Qiskit's ordering convention, where the qubits
are ordered in increasing significance from top to bottom, the matrix
representation of this gate is as follows.
\[
%
\]

Another way to think about Toffoli gates is that they're essentially query
gates for the AND function, in the sense that they follow the pattern we saw in
the previous lesson for unitary query gate implementations of arbitrary
functions having binary string inputs and outputs.

\begin{figure}[!ht]
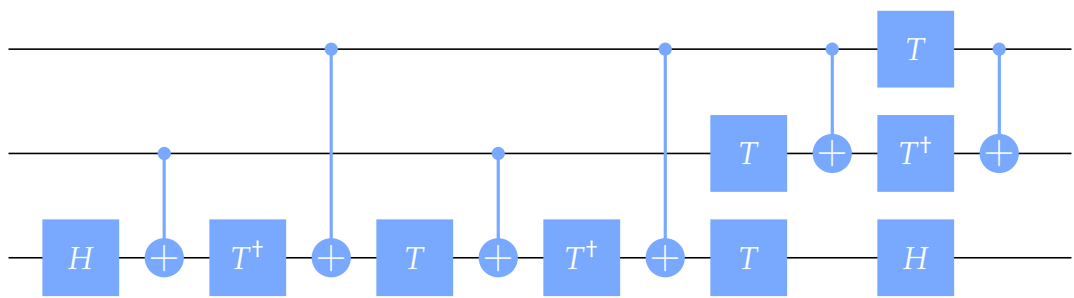

  \begin{center}
    %
    
  \end{center}
  \caption{A quantum circuit implementation of a Toffoli gate.}
  \label{fig:build-Toffoli}
\end{figure}

Toffoli gates are not included in the default gate set discussed earlier in the
lesson, but it is possible to construct a Toffoli gate from $H,$ $T,$
$T^{\dagger},$ and CNOT gates as shown in Figure~\ref{fig:build-Toffoli}.

\subsubsection{Simulating Boolean gates with Toffoli, controlled-NOT, and NOT
  gates}

\begin{figure}[!ht]
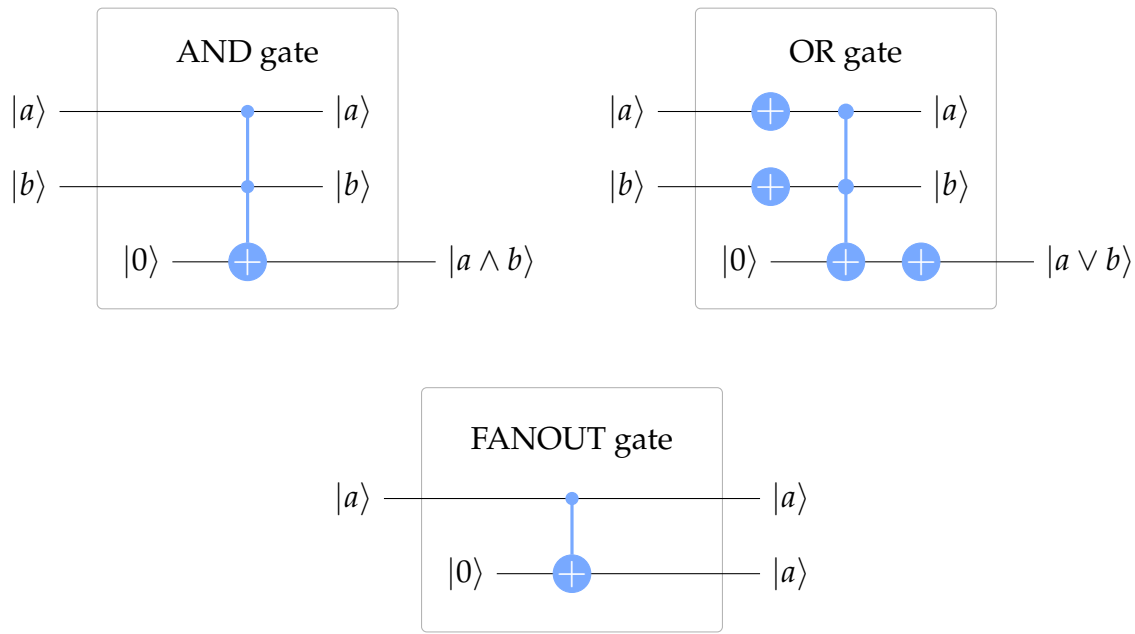

  \begin{center}
    %
      
  \end{center}
  \caption{Implementations of AND, OR, and FANOUT gates using Toffoli and NOT
    gates along with an initialized workspace qubit.}
  \label{fig:AND-OR-FANOUT-with-Toffoli}
\end{figure}

A single Toffoli gate, used in conjunction with a few NOT gates, can implement
an AND and OR gate, and FANOUT gates can easily be implemented using
controlled-NOT gates, as Figure~\ref{fig:AND-OR-FANOUT-with-Toffoli}
illustrates.

In all three cases, the qubits that the AND, OR, and FANOUT gates act upon come
in from the left as inputs, and we also require one \emph{workspace} qubit
initialized to the zero state for each one.
These workspace qubits appear inside of the boxes representing the gate
implementations to suggest that they're new, and therefore part of the cost of
these implementations.

For the AND and OR gates we also have two qubits left over, in addition to the
output qubit.
For example, inside the box in the diagram representing the simulation of an
AND gate, the top two qubits are left in the states $\vert a\rangle$ and $\vert
b\rangle.$
These qubits are illustrated as remaining inside of the boxes because they're
no longer needed and are not part of the output.
They can be ignored for now, though we will turn our attention back to them
shortly.

The remaining Boolean gate, the NOT gate, is included in our default set of
quantum gates, so we don't require a simulation for this one.

\subsubsection{Gate by gate simulation of Boolean circuits}

Now suppose that we have an ordinary Boolean circuit named $C,$ composed of
AND, OR, NOT, and FANOUT gates, and having $n$ input bits and $m$ of output
bits.
Let $t = \operatorname{size}(C)$ be the number of gates in $C,$ and let's give
the name $f$ to the function that $C$ computes, which takes the form
\[
f:\Sigma^n\rightarrow\Sigma^m
\]
for $\Sigma = \{0,1\}.$

\begin{figure}[!ht]
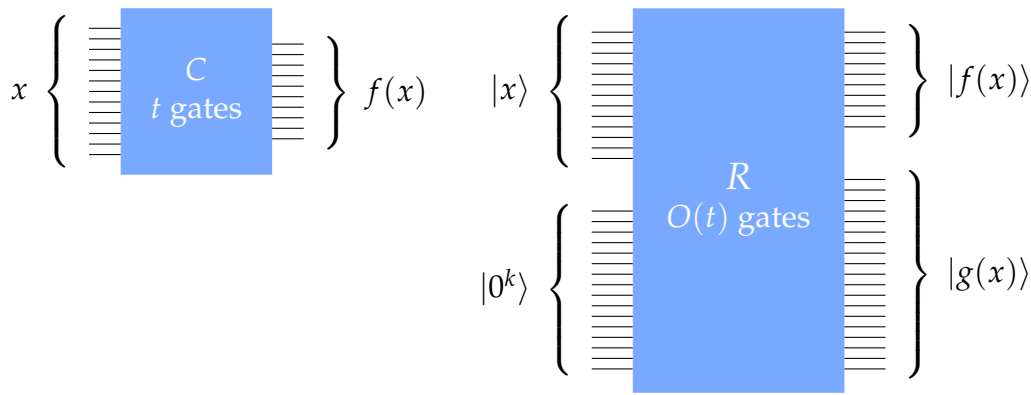

  \begin{center}
    \begin{minipage}[t][5cm]{6.125cm}
      \begin{center}

      \end{center}
    \end{minipage}
  \end{center}
  \caption{For a given Boolean circuit $C$, a circuit $R$ is obtained by
    replacing each AND, OR, and FANOUT gate with its Toffoli gate simulation.
    The action of $R$ on standard basis states is as shown.}
  \label{fig:reversible-circuit-simulation}
\end{figure}

Now consider what happens when we go one at a time through the AND, OR, and
FANOUT gates of $C,$ replacing each one by the corresponding simulation
described above, including the addition of the required workspace qubits.
Let's name the resulting circuit $R,$ and let's order the qubits of $R$ so that
the $n$ input bits of $C$ correspond to the top $n$ qubits of $R$ and the
workspace qubits are on the bottom.
Figure~\ref{fig:reversible-circuit-simulation} depicts the actions of the
circuits $C$ and $R$ side-by-side.

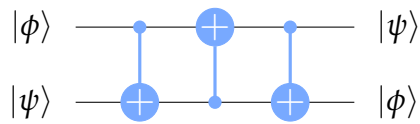
\begin{figure}[!ht]
  \begin{center}
    \begin{tikzpicture}[
        baseline=(current bounding box.north),
        scale=1,
        control/.style={%
          circle,
          fill=CircuitBlue,
          minimum size = 5pt,
          inner sep=0mm},
        not/.style={%
          circle,
          fill = CircuitBlue,
          draw = CircuitBlue,
          text = white,
          minimum size = 5mm,
          inner sep=0mm,
          label = {center:\textcolor{white}{\large $+$}}
        }
      ]
      
      \node (In1) at (-2,1) {};
      \node (In2) at (-2,0) {};
      
      \node (Out1) at (2,1) {};
      \node (Out2) at (2,0) {};
      
      \draw (In1) -- (Out1);
      \draw (In2) -- (Out2);
      
      \node[control] (Control1) at (-1,1) {};
      \node[control] (Control2) at (0,0) {};
      \node[control] (Control3) at (1,1) {};
      
      \node[not] (Not1) at (-1,0) {};
      \node[not] (Not2) at (0,1) {};
      \node[not] (Not3) at (1,0) {};
      
      \draw[very thick,draw=CircuitBlue]
      (Control1.center) -- ([yshift=-0.1pt]Not1.north);
      
      \draw[very thick,draw=CircuitBlue]
      (Control2.center) -- ([yshift=0.1pt]Not2.south);
      
      \draw[very thick,draw=CircuitBlue]
      (Control3.center) -- ([yshift=-0.1pt]Not3.north);

      \node[anchor = east] at (In1) {$\ket{\phi}$};
      \node[anchor = east] at (In2) {$\ket{\psi}$};
      
      \node[anchor = west] at (Out1) {$\ket{\psi}$};
      \node[anchor = west] at (Out2) {$\ket{\phi}$};
      
    \end{tikzpicture}
  \end{center}
  \caption{An implementation of a SWAP gate using three CNOT gates.}
  \label{fig:swap}
\end{figure}

Here, $k$ is the number of workspace qubits required --- one for each AND, OR,
and FANOUT gate of $C$ --- and $g$ is a function of the form
$g:\Sigma^n \rightarrow \Sigma^{n+k-m}$ that describes the states of the
leftover qubits created by the gate simulations after $R$ is run.
In the figure, the qubits corresponding to the output $f(x)$ are on the top and
the remaining, leftover qubits storing $g(x)$ are on the bottom.
We can force this to happen if we wish by rearranging the qubits using SWAP
gates, which can be implemented with three controlled-NOT gates as shown in
Figure~\ref{fig:swap}.
As we'll see in the next section, it's not really essential to rearrange the
output qubits like this, but it's easy enough to do it if we choose.

The function $g$ that describes the classical states of the leftover qubits is
determined by the circuit $C,$ but we actually don't need to worry all that
much about it;
we don't care specifically what state these qubits are in when the computation
finishes.
The letter $g$ comes after $f,$ so it's a reasonable name for this function on
that account, but there's a better reason to pick the name $g$ --- it's short
for \emph{garbage.}

\subsection{Cleaning up the garbage}

If our only interest is in evaluating the function $f$ computed by a given
Boolean circuit $C$ with a quantum circuit, we don't need to proceed any
further than the gate-by-gate simulation just described.
This means that, in addition to the output of the function, we'll have a bunch
of garbage left over.

However, this is not good enough if we want to perform classical computations
as subroutines within larger quantum computations, because those garbage qubits
will cause problems.
The phenomenon of \emph{interference} is critically important to quantum
algorithms, and garbage qubits can ruin the interference patterns needed to
make quantum algorithms work.

Fortunately, it's not difficult to clean up the garbage, so to speak.
The key is to use the fact that because $R$ is a quantum circuit, we can run it
in reverse, by simply replacing each gate with its inverse and applying them in
the reverse order, thereby obtaining a quantum circuit for the operation
$R^{\dagger}.$
Toffoli gates, CNOT gates, and NOT gates are actually their own inverses, so
running $R$ in reverse is really just a matter of applying the gates in the
reverse order --- but more generally any quantum circuit can be reversed as
just described.

\begin{figure}[p]
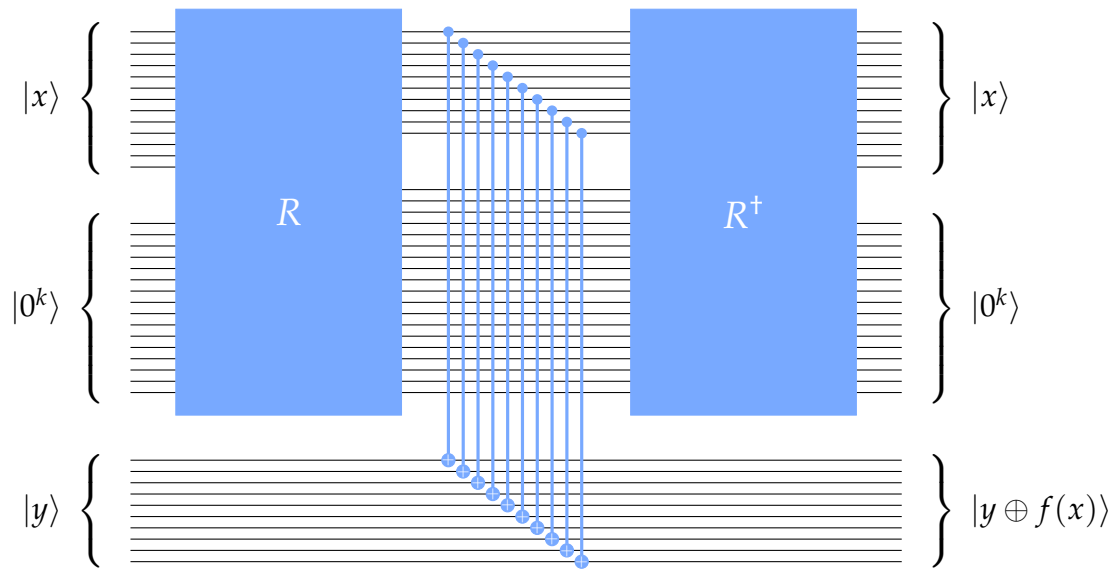

  \begin{center}

  \end{center}
  \caption{A garbage-free implementation of the original Boolean circuit $C$ is
    obtained by applying $R$, XORing the classical output to a new system, and
    then running $R$ in reverse.}
  \label{fig:garbage-free-computation}
\end{figure}

\begin{figure}[p]
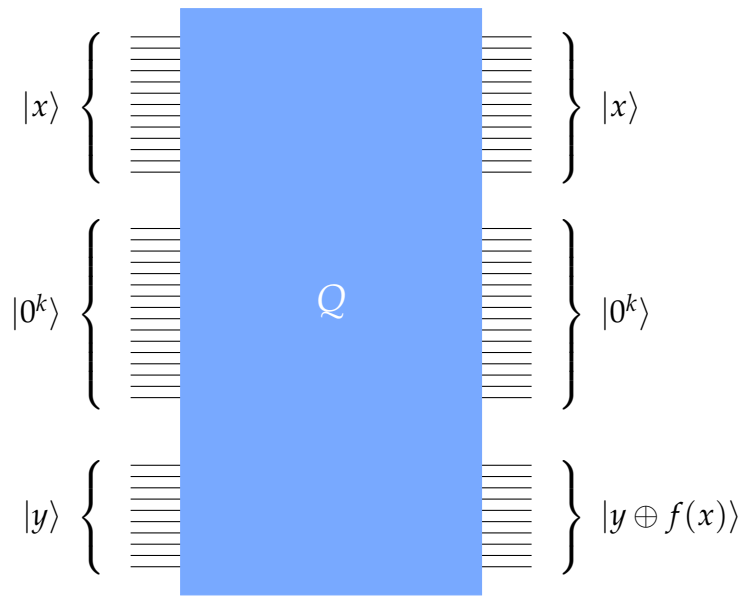

  \begin{center}
    %

  \end{center}
  \caption{The circuit in Figure~\ref{fig:garbage-free-computation}
    depicted as a single operation $Q$.}
  \label{fig:simulation-as-query-gate}
\end{figure}

Specifically, what we can do is to add $m$ more qubits (recalling that the
function $f$ has $m$ output bits), use CNOT gates to copy the output of $R$
onto these qubits, and reverse $R$ to clean up the garbage.
Figure~\ref{fig:garbage-free-computation} illustrates the resulting circuit and
describes its action on standard basis states.
Figure~\ref{fig:simulation-as-query-gate} depicts the result as a single
quantum circuit $Q.$
Given that $C$ has $t$ gates, the circuit $Q$ has $O(t)$ gates.

If we disregard the $k$ additional workspace qubits, what we have is a circuit
$Q$ that functions exactly like a query gate for the function $f.$
If we simply want to compute the function $f$ on some string $x,$ we can set $y
= 0^m$ and the resulting value $f(x)$ will appear on the bottom $m$ qubits ---
or we can feed in a different state to the bottom $m$ qubits if we wish
(perhaps to make use of a phase kickback, like in Deutsch's or the
Deutsch--Jozsa algorithm).

This means that for any query algorithm, if we have a Boolean circuit that
computes the input function, we can replace each query gate with a circuit
implementation of it, and the query algorithm will function correctly.

Note that the workspace qubits are needed to make this process work, but they
are returned to their initial states once the combined circuit is executed.
This allows these qubits to be used again as workspace qubits for other
purposes.
There are also known strategies to reduce the number of workspace qubits
required (which come at a cost of making the circuits larger), but we won't
discuss those strategies here.

\subsubsection{Implementing invertible functions}

The construction just described allows us to simulate any Boolean circuit with
a quantum circuit in a garbage-free manner.
If $C$ is a Boolean circuit implementing a function $f:\Sigma^n \rightarrow
\Sigma^m,$ then we obtain a quantum circuit $Q$ that operates as follows on
standard basis states.
\[
Q \bigl( \vert y \rangle \vert 0^k \rangle \vert x\rangle\bigr)
= \vert y \oplus f(x) \rangle \vert 0^k \rangle \vert x\rangle
\]
The number $k$ indicates how many workspace qubits are required in total.

It is possible to take this methodology one step further when the function $f$
itself is invertible.
To be precise, suppose that the function $f$ takes the form $f:\Sigma^n
\rightarrow \Sigma^n,$ and also suppose that there exists a function $f^{-1}$
such that $f^{-1}(f(x)) = x$ for every $x\in\Sigma^n$ (which is necessarily
unique when it exists).
This means that the operation that transforms $\vert x \rangle$ into $\vert
f(x) \rangle$ for every $x\in\Sigma^n$ is unitary, so we might hope to build a
quantum circuit that implements the unitary operation defined by
\[
U \vert x \rangle = \vert f(x) \rangle
\]
for every $x\in\Sigma^n.$

To be clear, the fact that this is a unitary operation relies on $f$ being
invertible --- it's not unitary when $f$ isn't invertible.
Disregarding the workspace qubits, $U$ is different from the operation that the
circuit $Q$ implements because we're not keeping a copy of the input around and
XORing it to an arbitrary string, we're \emph{replacing} $x$ by $f(x).$

The question is: when $f$ is invertible, can we do this?

\begin{figure}[!ht]
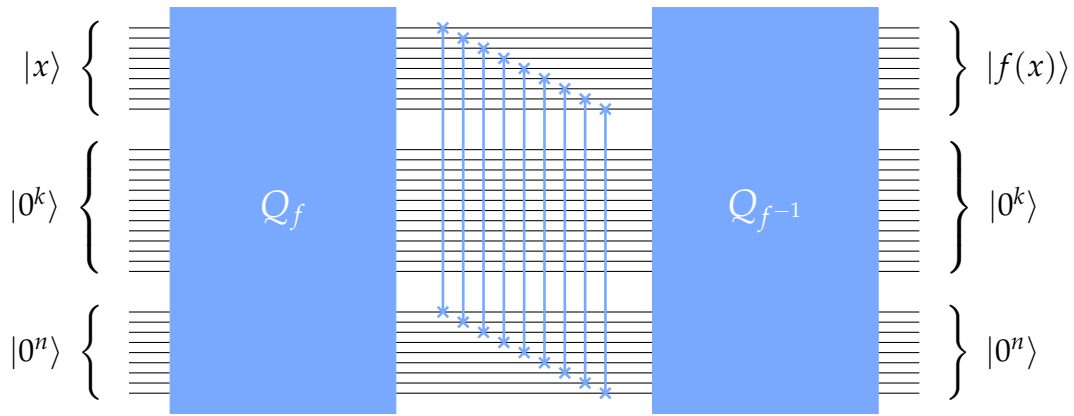

  \begin{center}

  \end{center}
  \caption{A unitary implementation of an invertible function $f$
    using garbage-free implementations of $f$ and $f^{-1}$ along with swap
    gates.
  }
  \label{fig:fully-reversible-simulation}
\end{figure}

The answer is yes, provided that we're allowed to use workspace qubits and, in
addition to having a Boolean circuit that computes $f,$ we also have one that
computes $f^{-1}.$
So, this isn't a shortcut for computationally inverting functions when we don't
already know how to do that!
Figure~\ref{fig:fully-reversible-simulation} illustrates how it can be done by
composing two quantum circuits, $Q_f$ and $Q_{f^{-1}},$ which are obtained
individually for the functions $f$ and $f^{-1}$ through the method described
above, along with $n$ swap gates, taking $k$ to be the maximum of the numbers
of workspace qubits required by $Q_f$ and $Q_{f^{-1}}.$

\lesson{Phase Estimation and Factoring}
\label{lesson:phase-estimation-and-factoring}

In this lesson, we'll discuss the phase estimation problem and how to solve it
with a quantum computer.
We'll then use this solution to obtain \emph{Shor's algorithm} --- an efficient
quantum algorithm for the integer factorization problem.
Along the way, we'll encounter the quantum Fourier transform, and we'll see how
it can be implemented efficiently by a quantum circuit.

\section{The phase estimation problem}

This section of the lesson explains the \emph{phase estimation problem}.
We'll begin with a short discussion of the \emph{spectral theorem} from linear
algebra, and then move on to a statement of the phase estimation problem
itself.

\subsection{Spectral theorem}

The \emph{spectral theorem} is an important fact from linear algebra that
states that matrices of a certain type, called \emph{normal matrices,} can be
expressed in a simple and useful way.
We'll only need this theorem for unitary matrices in this lesson, but later
in the course we'll apply it to Hermitian matrices as well.

\subsubsection{Normal matrices}

A square matrix $M$ with complex number entries is said to be a \emph{normal}
matrix if it commutes with its conjugate transpose:
$M M^{\dagger} = M^{\dagger} M.$
Every unitary matrix $U$ is normal because
\[
U U^{\dagger} = \mathbb{I} = U^{\dagger} U.
\]
Hermitian matrices, which are matrices that equal their own conjugate
transpose, are another important class of normal matrices.
If $M$ is a Hermitian matrix, then
\[
M M^{\dagger} = M^2 = M^{\dagger} M,
\]
so $M$ is normal.

Not every square matrix is normal.
For instance, this matrix isn't normal:
\[
\begin{pmatrix}
0 & 1\\
0 & 0
\end{pmatrix}
\]
(This is a simple but great example of a matrix that's often very helpful to
consider.)
It isn't normal because
\[
\begin{pmatrix}
  0 & 1\\
  0 & 0
\end{pmatrix}
\begin{pmatrix}
  0 & 1\\
  0 & 0
\end{pmatrix}^{\dagger}
=
\begin{pmatrix}
  0 & 1\\
  0 & 0
\end{pmatrix}
\begin{pmatrix}
  0 & 0\\
  1 & 0
\end{pmatrix}
=
\begin{pmatrix}
  1 & 0\\
  0 & 0
\end{pmatrix}
\]
while
\[
\begin{pmatrix}
  0 & 1\\
  0 & 0
\end{pmatrix}^{\dagger}
\begin{pmatrix}
  0 & 1\\
  0 & 0
\end{pmatrix}
=
\begin{pmatrix}
  0 & 0\\
  1 & 0
\end{pmatrix}
\begin{pmatrix}
  0 & 1\\
  0 & 0
\end{pmatrix}
=
\begin{pmatrix}
  0 & 0\\
  0 & 1
\end{pmatrix}.
\]

\subsubsection{Theorem statement}

Now here's a statement of the spectral theorem.

\begin{callout}[title = {Spectral theorem}]
  Let $M$ be a normal $N\times N$ complex matrix.
  There exists an orthonormal basis of $N$-dimensional complex vectors
  $\bigl\{\vert\psi_0\rangle,\ldots,\vert\psi_{N-1}\rangle \bigr\}$ along with
  complex numbers $\lambda_0,\ldots,\lambda_{N-1}$ such that
  \[
  M = \lambda_0 \vert \psi_0\rangle\langle \psi_0\vert + \cdots +
  \lambda_{N-1} \vert \psi_{N-1}\rangle\langle \psi_{N-1}\vert.
  \]
\end{callout}

\noindent
The expression of a matrix in the form
\begin{equation}
  M = \sum_{k = 0}^{N-1} \lambda_k \vert \psi_k\rangle\langle \psi_k\vert
  \label{eq:spectral-decomposition}
\end{equation}
is commonly called a \emph{spectral decomposition}.
Notice that if $M$ is a normal matrix expressed in the form
\eqref{eq:spectral-decomposition}, then the equation
\[
M \vert \psi_j \rangle = \lambda_j \vert \psi_j \rangle
\]
must be true for every $j = 0,\ldots,N-1.$
This is a consequence of the orthonormality of the set
$\bigl\{\vert\psi_0\rangle,\ldots,\vert\psi_{N-1}\rangle\bigr\}$.
\[
M \vert \psi_j \rangle
= \left(\sum_{k = 0}^{N-1} \lambda_k \vert \psi_k\rangle\langle
\psi_k\vert\right)\vert \psi_j\rangle
= \sum_{k = 0}^{N-1} \lambda_k \vert \psi_k\rangle\langle \psi_k\vert\psi_j
\rangle
= \lambda_j \vert\psi_j \rangle
\]
That is, each number $\lambda_j$ is an \emph{eigenvalue} of $M$ and
$\vert\psi_j\rangle$ is an \emph{eigenvector} corresponding to that eigenvalue.

\begin{description}
\item[Example 1.]
  Let
  \[
  \mathbb{I} = \begin{pmatrix}1 & 0\\0 & 1\end{pmatrix},
  \]
  which is normal.
  The theorem implies that $\mathbb{I}$ can be written in the form
  \eqref{eq:spectral-decomposition} for some choice of $\lambda_0,$
  $\lambda_1,$ $\vert\psi_0\rangle,$ and $\vert\psi_1\rangle.$
  There are multiple choices that work, including
  \[
  \lambda_0 = 1, \hspace{5pt}
  \lambda_1 = 1, \hspace{5pt}
  \vert\psi_0\rangle = \vert 0\rangle, \hspace{5pt}
  \vert\psi_1\rangle = \vert 1\rangle.
  \]

  Notice that the theorem does not say that the complex numbers
  $\lambda_0,\ldots,\lambda_{N-1}$ are distinct --- we can have the same
  complex number repeated, which is necessary for this example.
  These choices work because
  \[
  \mathbb{I} = \vert 0\rangle\langle 0\vert + \vert 1\rangle\langle 1\vert.
  \]
  Indeed, we could choose $\{\vert\psi_0\rangle,\vert\psi_1\rangle\}$ to be
  \emph{any} orthonormal basis and the equation will be true.
  For instance,
  \[
  \mathbb{I} = \vert +\rangle\langle +\vert + \vert -\rangle\langle -\vert.
  \]

\item[Example 2.]
  Consider a Hadamard operation.
  \[
  H = \frac{1}{\sqrt{2}}
  \begin{pmatrix}
    1 & 1\\[1mm]
    1 & -1
  \end{pmatrix}
  \]
  This is a unitary matrix, so it is normal.
  The spectral theorem implies that $H$ can be written in the form
  \eqref{eq:spectral-decomposition}, and in particular we have
  \[
  H =
  \vert\psi_{\pi/8}\rangle \langle \psi_{\pi/8}\vert
  - \vert\psi_{5\pi/8}\rangle \langle \psi_{5\pi/8}\vert
  \]
  where
  \[
  \vert\psi_{\theta}\rangle
  = \cos(\theta)\vert 0\rangle + \sin(\theta) \vert 1\rangle.
  \]

  More explicitly,
  \[
  \begin{aligned}
    \vert\psi_{\pi/8}\rangle & = \frac{\sqrt{2 + \sqrt{2}}}{2}\vert 0\rangle
    + \frac{\sqrt{2 - \sqrt{2}}}{2}\vert 1\rangle, \\[3mm]
    \vert\psi_{5\pi/8}\rangle & = -\frac{\sqrt{2 - \sqrt{2}}}{2}\vert 0\rangle
    + \frac{\sqrt{2 + \sqrt{2}}}{2}\vert 1\rangle.
  \end{aligned}
  \]
  We can check that this decomposition is correct by performing the required
  calculations:
  \[
  \vert\psi_{\pi/8}\rangle \langle \psi_{\pi/8}\vert
  - \vert\psi_{5\pi/8}\rangle \langle \psi_{5\pi/8}\vert
  = \begin{pmatrix}
    \frac{2 + \sqrt{2}}{4} & \frac{\sqrt{2}}{4}\\[2mm]
    \frac{\sqrt{2}}{4} & \frac{2 - \sqrt{2}}{4}
  \end{pmatrix}
  -
  \begin{pmatrix}
    \frac{2 - \sqrt{2}}{4} & -\frac{\sqrt{2}}{4}\\[2mm]
    -\frac{\sqrt{2}}{4} & \frac{2 + \sqrt{2}}{4}
  \end{pmatrix}
  = H.
  \]
\end{description}

As the first example above reveals, there can be some freedom in how
eigenvectors are selected.
There is, however, no freedom at all in how the eigenvalues are chosen, except
for their ordering:
the same $N$ complex numbers $\lambda_0,\ldots,\lambda_{N-1},$ which can
include repetitions of the same complex number, will always occur in the
equation \eqref{eq:spectral-decomposition} for a given choice of a matrix $M.$

Now let's focus in on unitary matrices.
Suppose $U$ is unitary and we have a complex number $\lambda$ and a nonzero
vector $\vert\psi\rangle$ that satisfy the equation
\begin{equation}
U\vert\psi\rangle = \lambda\vert\psi\rangle.
\label{eq:eigen}
\end{equation}
That is, $\lambda$ is an eigenvalue of $U$ and $\vert\psi\rangle$ is an
eigenvector corresponding to this eigenvalue.

Unitary matrices preserve Euclidean norm, and so we conclude the following
from \eqref{eq:eigen}.
\[
\bigl\| \vert\psi\rangle \bigr\|
= \bigl\| U \vert\psi\rangle \bigr\|
= \bigl\| \lambda \vert\psi\rangle \bigr\|
= \vert \lambda \vert \bigl\| \vert\psi\rangle \bigr\|
\]
The condition that $\vert\psi\rangle$ is nonzero implies that $\bigl\|
\vert\psi\rangle \bigr\|\neq 0,$ so we can cancel it from both sides to obtain
\[
\vert \lambda \vert = 1.
\]
This reveals that eigenvalues of unitary matrices must always have absolute
value equal to one, so they lie on the \emph{unit circle.}
\[
\mathbb{T} = \{\alpha\in\mathbb{C} : \vert\alpha\vert = 1\}
\]
(The symbol $\mathbb{T}$ is a common name for the complex unit circle. The name
$S^1$ is also common.)

\subsection{Phase estimation problem statement}

In the \emph{phase estimation problem}, we're given a quantum state $\vert
\psi\rangle$ of $n$ qubits, along with a unitary quantum circuit that acts on
$n$ qubits.
We're \emph{promised} that $\vert \psi\rangle$ is an eigenvector of the unitary
matrix $U$ that describes the action of the circuit, and our goal is to compute
or approximate the eigenvalue $\lambda$ to which $\vert \psi\rangle$
corresponds.
More precisely, because $\lambda$ lies on the complex unit circle, we can write
\[
\lambda = e^{2\pi i \theta}
\]
for a unique real number $\theta$ satisfying $0\leq\theta<1.$
The goal of the problem is to compute or approximate this real number $\theta.$

\begin{callout}[title={Phase estimation problem}]
  \begin{problem}
    \Input{A unitary quantum circuit for an $n$-qubit operation $U$ along with
      an $n$-qubit quantum state $\vert\psi\rangle$.}
    \Promise{$\vert\psi\rangle$ is an eigenvector of $U$.}
    \Output{An approximation to the number $\theta\in[0,1)$ satisfying
        \[
        U\vert\psi\rangle = e^{2\pi i \theta}\vert\psi\rangle.
        \]}
  \end{problem}
\end{callout}

\noindent
Here are a few remarks about this problem statement:
\begin{enumerate}
\item
  The phase estimation problem is different from other problems we've seen so
  far in the course in that the input includes a quantum state.
  Typically we focus on problems having classical inputs and outputs, but
  nothing prevents us from considering quantum state inputs like this.
  In terms of its practical relevance, the phase estimation problem is
  typically encountered as a \emph{subproblem} inside of a larger computation,
  like we'll see in the context of integer factorization later in the lesson.

\item
  The statement of the phase estimation problem above isn't specific about what
  constitutes an approximation of $\theta,$ but we can formulate more precise
  problem statements depending on our needs and interests.
  In the context of integer factorization, we'll demand a very precise
  approximation to $\theta,$ but in other cases we might be satisfied with a
  very rough approximation.
  We'll discuss shortly how the precision we require affects the computational
  cost of a solution.

\item
  Notice that as we go from $\theta = 0$ toward $\theta = 1$ in the phase
  estimation problem, we're going all the way around the unit circle, starting
  from $e^{2\pi i \cdot 0} = 1$ and moving counter-clockwise toward
  $e^{2\pi i \cdot 1} = 1.$
  That is, when we reach $\theta = 1$ we're back where we started at
  $\theta = 0.$
  So, as we consider the accuracy of approximations, choices of $\theta$ near
  $1$ should be considered as being near $0.$
  For example, an approximation $\theta = 0.999$ should be considered as being
  within $1/1000$ of $\theta = 0.$
\end{enumerate}

\section{Phase estimation procedure}

Next, we'll discuss the \emph{phase estimation procedure}, which is a quantum
algorithm for solving the phase estimation problem.

We'll begin with a low-precision warm-up, which explains some of the basic
intuition behind the method.
We'll then talk about the \emph{quantum Fourier transform}, which is an
important quantum operation used in the phase estimation procedure, as well as
its quantum circuit implementation.
Once we have the quantum Fourier transform in hand, we'll describe the
phase estimation procedure in full generality and analyze its performance.

\subsection{Warm-up: approximating phases with low precision}

We'll begin with a couple of simple versions of the phase estimation procedure
that provide low-precision solutions to the phase estimation problem.
This is helpful for explaining the intuition behind the general procedure that
we'll see a bit later in the lesson.

\subsubsection{Using the phase kickback}

A simple approach to the phase estimation problem, which allows us to learn
something about the value $\theta$ we seek, is based on the \emph{phase
kickback} phenomenon.
As we'll see, this is essentially a single control-qubit version of the general
phase estimation procedure to be discussed later in the lesson.

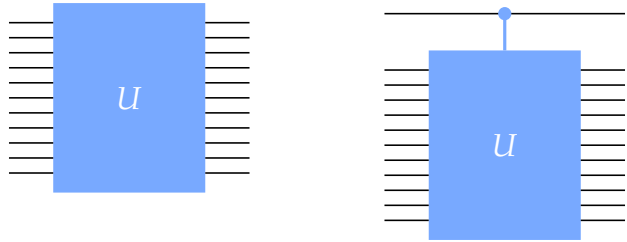
\begin{figure}[!ht]
  \begin{center}
    \begin{tikzpicture}[
        baseline=(current bounding box.north),
        scale=1,
        line width = 0.6pt,
        control/.style={%
          circle,
          fill=CircuitBlue,
          minimum size = 5pt,
          inner sep=0mm},
        gate/.style={%
          inner sep = 0,
          fill = CircuitBlue,
          draw = CircuitBlue,
          text = white,
          minimum size = 10mm}
      ]
      
      \node[gate, minimum height=25mm, minimum width=20mm] (U) at (0,0) {$U$};
      
      \node (In) at (-1.75,0) {};
      \node (Out) at (1.75,0) {};
      
      \foreach \y in {-10,-8,...,10} {
        \draw ([yshift=\y mm]In.east) -- ([yshift=\y mm]U.west) {};
        \draw ([yshift=\y mm]Out.west) -- ([yshift=\y mm]U.east) {};
      }
      
    \end{tikzpicture}
    \hspace{10mm}
    \begin{tikzpicture}[
        baseline=(current bounding box.north),
        scale=1,
        line width = 0.6pt,
        control/.style={%
          circle,
          fill=CircuitBlue,
          minimum size = 5pt,
          inner sep=0mm},
        gate/.style={%
          inner sep = 0,
          fill = CircuitBlue,
          draw = CircuitBlue,
          text = white,
          minimum size = 10mm}
      ]
          
      \node (ControlIn) at (-1.75,1.75) {};
      \node (ControlOut) at (1.75,1.75) {};
      
      \draw (ControlIn.east) -- (ControlOut.west);
      
      \node[gate, minimum height=25mm, minimum width=20mm] (U) at (0,0) {$U$};
          
      \node (In) at (-1.75,0) {};
      \node (Out) at (1.75,0) {};
      
      \foreach \y in {-10,-8,...,10} {
        \draw ([yshift=\y mm]In.east) -- 
        ([yshift=\y mm]U.west) {};
        \draw ([yshift=\y mm]Out.west) -- 
        ([yshift=\y mm]U.east) {};
      }
      
      \node[control] (Control) at (0,1.75) {};
      
      \draw[very thick,draw=CircuitBlue] (Control.center) -- (U.north);
          
    \end{tikzpicture}
  \end{center}
  \caption{A unitary operation $U$ (viewed as a quantum gate) on the left
    and a controlled-$U$ operation on the right.}
  \label{fig:uncontrolled-and-controlled-unitary-2}
\end{figure}

As part of the input to the phase estimation problem, we have a unitary quantum
circuit for the operation $U.$
We can use the description of this circuit to create a circuit for a
\emph{controlled}-$U$ operation, which can be depicted as
Figure~\ref{fig:uncontrolled-and-controlled-unitary-2} suggests.

We can create a quantum circuit for a controlled-$U$ operation by first adding
a control qubit to the circuit for $U,$ and then replacing every gate in the
circuit for $U$ with a controlled version of that gate --- so our one new
control qubit effectively controls every single gate in the circuit for $U.$
This requires that we have a controlled version of every gate in our circuit,
but we can always build circuits for these controlled operations in case
they're not included in our gate set.

\begin{figure}[!ht]
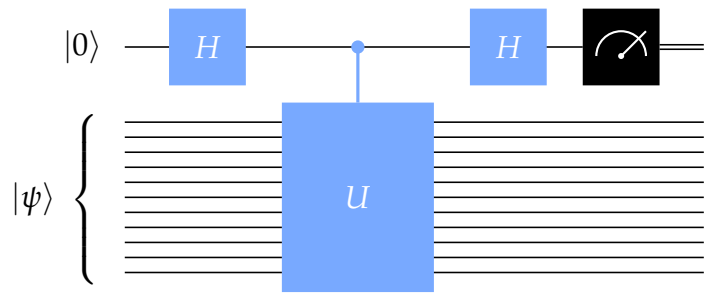

  \begin{center}

   
  \end{center}
  \caption{A phase estimation circuit with a single control qubit.}
  \label{fig:estimate-phase-with-kickback}
\end{figure}

\begin{figure}[!ht]
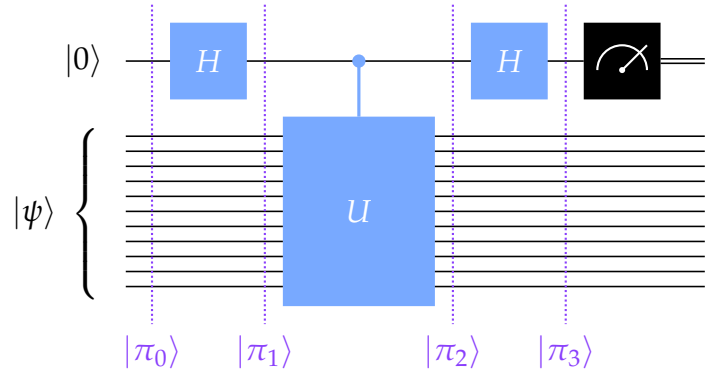

  \begin{center}
    %

  \end{center}
  \caption{The states $\ket{\pi_0},\ldots,\ket{\pi_3}$ considered in the
    analysis of the single control qubit phase estimation procedure.}
  \label{fig:estimate-phase-with-kickback-analysis}
\end{figure}

Now consider the circuit in Figure~\ref{fig:estimate-phase-with-kickback},
where the input state $\vert\psi\rangle$ of all of the qubits except the top
one is the quantum state eigenvector of $U.$
The measurement outcome probabilities for this circuit depend on the eigenvalue
of $U$ corresponding to the eigenvector $\vert\psi\rangle.$
Let's analyze the circuit in detail to determine exactly how by considering the
states indicated in Figure~\ref{fig:estimate-phase-with-kickback-analysis}.

The initial state of the circuit is
\[
\vert\pi_0\rangle = \vert\psi\rangle \vert 0\rangle
\]
and the first Hadamard gate transforms this state to
\[
\vert\pi_1\rangle = \vert\psi\rangle \vert +\rangle
= \frac{1}{\sqrt{2}} \vert\psi\rangle \vert 0\rangle + \frac{1}{\sqrt{2}}
\vert\psi\rangle \vert 1\rangle.
\]

Next, the controlled-$U$ operation is performed, which results in the state
\[
\vert\pi_2\rangle
= \frac{1}{\sqrt{2}} \vert\psi\rangle \vert 0\rangle + \frac{1}{\sqrt{2}}
\bigl(U \vert\psi\rangle\bigr) \vert 1\rangle.
\]
Using the assumption that $\vert\psi\rangle$ is an eigenvector of $U$ having
eigenvalue $\lambda = e^{2\pi i\theta},$ we can alternatively express this
state as follows.
\[
\vert\pi_2\rangle
= \frac{1}{\sqrt{2}} \vert\psi\rangle \vert 0\rangle
+ \frac{e^{2\pi i \theta}}{\sqrt{2}} \vert\psi\rangle \vert 1\rangle
= \vert\psi\rangle \otimes \left( \frac{1}{\sqrt{2}} \vert 0\rangle
+ \frac{e^{2\pi i \theta}}{\sqrt{2}} \vert 1\rangle\right)
\]
Here we observe the phase kickback phenomenon.
It is slightly different this time than it was for Deutsch's algorithm and the
Deutsch--Jozsa algorithm because we're not working with a query gate --- but the
idea is similar.

Finally, the second Hadamard gate is performed. After just a bit of
simplification, we obtain this expression for this state.
\[
\vert\pi_3\rangle
= \vert\psi\rangle \otimes \left( \frac{1+ e^{2\pi i \theta}}{2} \vert 0\rangle
+ \frac{1 - e^{2\pi i \theta}}{2} \vert 1\rangle\right)
\]
The measurement therefore yields the outcomes $0$ and $1$ with these
probabilities:
\[
\begin{aligned}
  p_0 &= \left\vert \frac{1+ e^{2\pi i \theta}}{2} \right\vert^2 =
  \cos^2(\pi\theta)\\[1mm]
  p_1 &= \left\vert \frac{1- e^{2\pi i \theta}}{2} \right\vert^2 =
  \sin^2(\pi\theta).
\end{aligned}
\]

\begin{figure}[!ht]
  \begin{center}
    \begin{tikzpicture}
      \begin{axis}[
          samples=\samples,
          domain = 0:1,
          y = 5cm,
          x = 8cm,
          width=8cm,
          xtick = {0.25, 0.5, 0.75, 1},
          axis lines=center,
          axis line style={-},
          x label style={%
            at={(axis description cs:.5,-0.15)},anchor=north},
          y label style={%
            at={(axis description cs:-0.15,.5)},rotate=90,
            anchor=south},
          xlabel={$\theta$},
          ylabel={probability},
          clip=false,
          legend style={at={(1.125,0.5)},anchor=west}
        ]
        
        \addplot[thick, DataColor0] {cos(deg(pi*x))*cos(deg(pi*x))};
        \addplot[thick, DataColor1] {sin(deg(pi*x))*sin(deg(pi*x))};
        
        \legend{$0$,$1$}
        
      \end{axis}
      
    \end{tikzpicture}
  \end{center}
  \caption{Output probabilities for phase estimation with a single control
    qubit.}
  \label{fig:kickback-probabilities}
\end{figure}
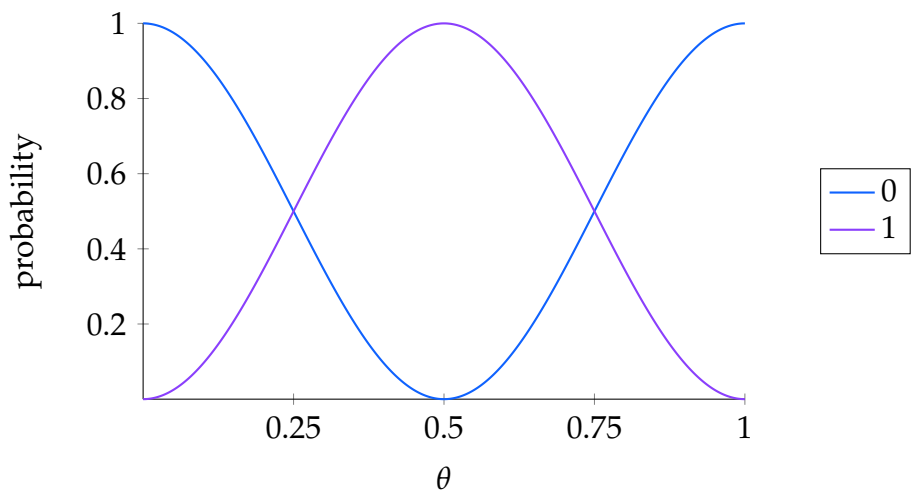

Figure~\ref{fig:kickback-probabilities} shows a plot of the probabilities for
the two possible outcomes, $0$ and $1,$ as functions of $\theta.$
Naturally, the two probabilities always sum to $1.$
Notice that when $\theta = 0,$ the measurement outcome is always $0,$ and when
$\theta = 1/2,$ the measurement outcome is always $1.$
So, although the measurement result doesn't reveal exactly what $\theta$ is, it
does provide us with some information about it --- and if we were promised that
either $\theta = 0$ or $\theta = 1/2,$ we could learn from the circuit which
one is correct without error.

Intuitively speaking, we can think of the circuit's measurement outcome as
being a guess for $\theta$ to ``one bit of accuracy.''
In other words, if we were to write $\theta$ in binary notation and round it
off to one bit, we'd have a number like this:
\[
0.a = \begin{cases}
  0 & a = 0\\
  \frac{1}{2} & a = 1.
\end{cases}
\]

The measurement outcome can be viewed as a guess for the bit $a.$
When $\theta$ is neither $0$ nor $1/2,$ there's a nonzero probability that the
guess will be wrong --- but the probability of making an error becomes smaller
and smaller as we get closer to $0$ or $1/2.$

It's natural to ask what role the two Hadamard gates play in this procedure:
\begin{itemize}
\item
  The first Hadamard gate sets the control qubit to a uniform superposition of
  $\vert 0\rangle$ and $\vert 1\rangle,$ so that when the phase kickback
  occurs, it happens for the $\vert 1\rangle$ state and not the $\vert
  0\rangle$ state, creating a \emph{relative} phase difference that affects the
  measurement outcomes.
  If we didn't do this and the phase kickback produced a \emph{global} phase,
  it would have no effect on the probabilities of obtaining different
  measurement outcomes.

\item
  The second Hadamard gate allows us to learn something about the number
  $\theta$ through the phenomenon of \emph{interference}.
  Prior to the second Hadamard gate, the state of the top qubit is
  \[
  \frac{1}{\sqrt{2}} \vert 0\rangle + \frac{e^{2\pi i \theta}}{\sqrt{2}} \vert
  1\rangle,
  \]
  and if we were to measure this state, we would obtain $0$ and $1$ each with
  probability $1/2,$ telling us nothing about $\theta.$
  By performing the second Hadamard gate, however, we cause the number $\theta$
  to affect the output probabilities.
\end{itemize}

\subsubsection{Doubling the phase}

The circuit above uses the phase kickback phenomenon to approximate $\theta$ to
a single bit of accuracy.
One bit of accuracy may be all we need in some situations --- but for factoring
we're going to need a lot more accuracy than that.
A natural question is, how can we learn more about $\theta?$

\begin{figure}[!ht]
  \begin{center}
    \begin{tikzpicture}[
        baseline=(current bounding box.north),
        scale=1,
        line width = 0.6pt,
        control/.style={%
          circle,
          fill=CircuitBlue,
          minimum size = 5pt,
          inner sep=0mm},
        gate/.style={%
          inner sep = 0,
          fill = CircuitBlue,
          draw = CircuitBlue,
          text = white,
          minimum size = 10mm},
        blackgate/.style={%
          inner sep = 0,
          fill = black,
          draw = black,
          text = white,
          minimum size = 10mm}
      ]
      
      \node (In) at (-3.25,0) {};
      \node (Out) at (7.25,0) {};
      
      \node (ControlIn) at (-3.25,2) {};
      \node (ControlOut) at (7.25,2) {};
      
      \node[blackgate] (M) at (6,2) {};
      \readout{M}
      
      \draw (ControlIn.east) -- (M.west);
      
      \draw ([yshift=0.3mm]M.east) -- ([yshift=0.3mm]ControlOut.west);
      \draw ([yshift=-0.3mm]M.east) -- ([yshift=-0.3mm]ControlOut.west);
      
      \node[gate] (H1) at (-2,2) {$H$};
      \node[gate] (H2) at (4.5,2) {$H$};
      
      \node[anchor=east] at (In.center) {$\ket{\psi} \;
        \left\{\rule{0mm}{13mm}\right.$};
      \node[anchor=east] at (ControlIn) {$\ket{0}$};

      \foreach \y in {-10,-8,...,10} {
        \draw ([yshift=\y mm]In.east) -- ([yshift=\y mm]Out.west) {};
      }
      
      \node[gate, minimum height=25mm, minimum width=20mm]
      (U1) at (0,0) {$U$};

      \node[gate, minimum height=25mm, minimum width=20mm]
      (U2) at (2.5,0) {$U$};
      
      \node[control] (Control1) at (0,2) {};
      \node[control] (Control2) at (2.5,2) {};
      
      \draw[very thick,draw=CircuitBlue] (Control1.center) -- (U1.north);
      \draw[very thick,draw=CircuitBlue] (Control2.center) -- (U2.north);
      
    \end{tikzpicture}
  \end{center}
  \caption{A modified version of the circuit in
    Figure~\ref{fig:estimate-phase-with-kickback} with two controlled-$U$ gates
    in place of one.}
  \label{fig:double-phase-kickback}
\end{figure}
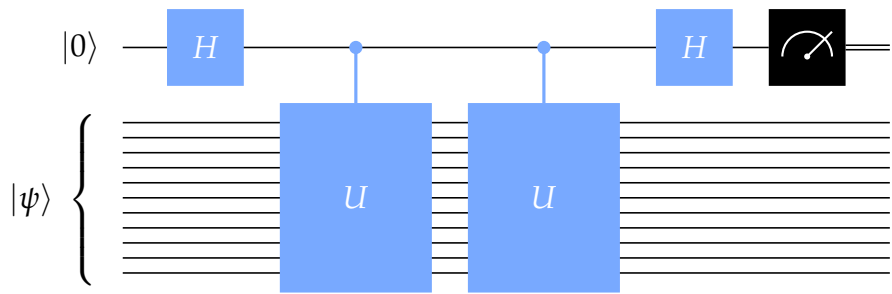

One very simple thing we can do is to replace the controlled-$U$ operation in
our circuit with \emph{two copies} of this operation, like in
Figure~\ref{fig:double-phase-kickback}.
Two copies of a controlled-$U$ operation is equivalent to a controlled-$U^2$
operation.
If $\vert\psi\rangle$ is an eigenvector of $U$ having eigenvalue
$\lambda = e^{2\pi i \theta},$ then this state is also an eigenvector of
$U^2,$ this time having eigenvalue $\lambda^2 = e^{2\pi i (2\theta)}.$

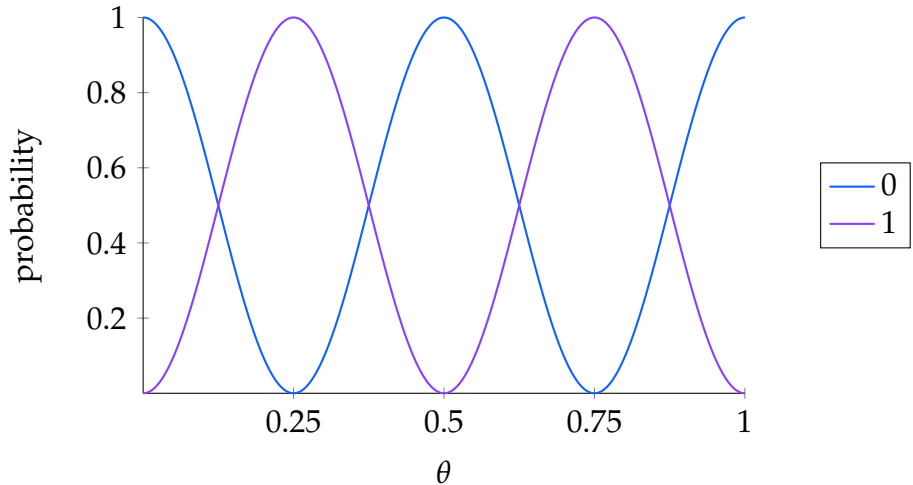
\begin{figure}[!ht]
  \begin{center}
    \begin{tikzpicture}
      \begin{axis}[
          samples=\samples,
          domain = 0:1,
          y = 5cm,
          x = 8cm,
          width=8cm,
          xtick = {0.25, 0.5, 0.75, 1},
          axis lines=center,
          axis line style={-},
          x label style={%
            at={(axis description cs:.5,-0.15)},anchor=north},
          y label style={%
            at={(axis description cs:-0.15,.5)},rotate=90,
            anchor=south},
          xlabel={$\theta$},
          ylabel={probability},
          clip=false,
          legend style={at={(1.125,0.5)},anchor=west}
        ]
        
        \addplot[thick, DataColor0] {cos(2*deg(pi*x))*cos(2*deg(pi*x))};
        \addplot[thick, DataColor1] {sin(2*deg(pi*x))*sin(2*deg(pi*x))};
        
        \legend{$0$,$1$}
        
      \end{axis}
      
    \end{tikzpicture}

  \end{center}
  \caption{Output probabilities for phase estimation with a single control
    qubit and two controlled-unitary gates.}
  \label{fig:double-kickback-probabilities}
\end{figure}

So, if we run this version of the circuit, we're effectively performing the
same computation as before, except that the number $\theta$ is replaced by
$2\theta.$
Figure~\ref{fig:double-kickback-probabilities} shows a plot illustrating the
output probabilities as $\theta$ ranges from $0$ to $1.$

Doing this can indeed provide us with some additional information about
$\theta.$
If the binary representation of $\theta$ is
\[
\theta = 0.a_1 a_2 a_3\cdots
\]
then doubling $\theta$ effectively shifts the binary point one position to the
right.
\[
2\theta = a_1. a_2 a_3\cdots
\]
And because we're equating $\theta = 1$ with $\theta = 0$ as we move around the
unit circle, we see that the bit $a_1$ has no influence on our probabilities,
and we're effectively obtaining a guess for the \emph{second} bit after the
binary point if we round $\theta$ to two bits.
For instance, if we knew in advance that $\theta$ was either $0$ or $1/4,$ then
we could fully trust the measurement outcome to tell us which.

It's not immediately clear, though, how this estimation should be reconciled
with what we learned from the original (non-doubled) phase kickback circuit to
give us the most accurate information possible about $\theta.$
So let's take a step back and consider how to proceed.

\subsubsection{Two-qubit phase estimation}

\begin{figure}[!ht]
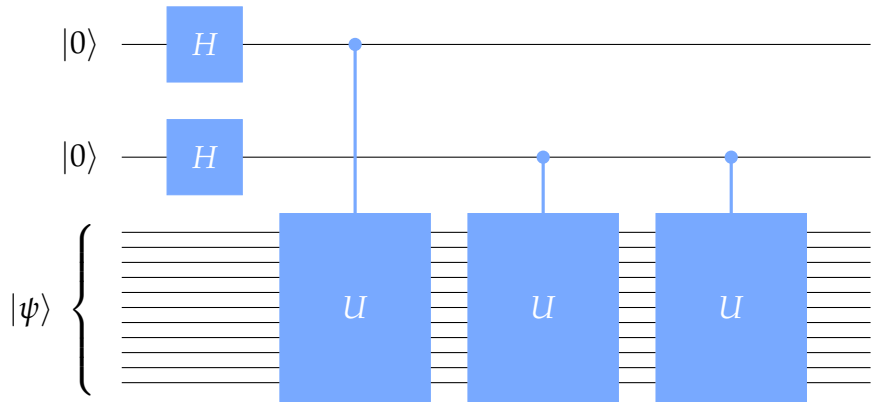

  \begin{center}

  \end{center}
  \caption{The initial portion of a quantum circuit for phase estimation with
    two control qubits.}
  \label{fig:two-bit-phase-estimation-initial}
\end{figure}

Rather than considering the two options described above separately, let's
combine them into a single circuit, like in
Figure~\ref{fig:two-bit-phase-estimation-initial}.
The Hadamard gates after the controlled operations have been removed and there
are no measurements here yet.
We'll add more to the circuit as we consider our options for learning as much
as we can about~$\theta.$

If we run this circuit when $\vert\psi\rangle$ is an eigenvector of $U,$ the
state of the bottom qubits will remain $\vert\psi\rangle$ throughout the entire
circuit, and phases will be kicked into the state of the top two qubits.
Let's analyze the circuit carefully by considering the states indicated in
Figure~\ref{fig:two-bit-phase-estimation-states}.

\begin{figure}[!ht]
  \begin{center}


  \end{center}
  \caption{The states $\ket{\pi_0},\ldots,\ket{\pi_3}$ considered in the
    analysis of two-qubit phase estimation.}
  \label{fig:two-bit-phase-estimation-states}
\end{figure}

We can write the state $\vert\pi_1\rangle$ like this:
\[
\vert\pi_1\rangle = \vert \psi\rangle \otimes \frac{1}{2} \sum_{a_0 = 0}^1
\sum_{a_1 = 0}^1 \vert a_1 a_0 \rangle.
\]

When the first controlled-$U$ operation is performed, the eigenvalue
$\lambda = e^{2\pi i\theta}$ gets kicked into the phase when $a_0$ (the top
qubit) is equal to $1,$ but not when it's $0.$
So, we can express the resulting state like this:
\[
\vert\pi_2\rangle
= \vert\psi\rangle \otimes \frac{1}{2} \sum_{a_0=0}^1 \sum_{a_1=0}^1
e^{2 \pi i a_0 \theta} \vert a_1 a_0 \rangle. 
\]

The second and third controlled-$U$ gates do something similar, except for
$a_1$ rather than $a_0,$ and with $\theta$ replaced by $2\theta.$
We can express the resulting state like this:
\[
\vert\pi_3\rangle
= \vert\psi\rangle\otimes\frac{1}{2}\sum_{a_0 = 0}^1 \sum_{a_1 = 0}^1
e^{2\pi i (2 a_1 + a_0)\theta} \vert a_1 a_0 \rangle.
\]
If we think about the binary string $a_1 a_0$ as representing an integer $x \in
\{0,1,2,3\}$ in binary notation, which is $x = 2 a_1 + a_0,$ we can
alternatively express this state as follows.
\[
\vert\pi_3\rangle = \vert \psi\rangle \otimes \frac{1}{2} \sum_{x = 0}^3
e^{2\pi i x \theta} \vert x \rangle
\]
Our goal is to extract as much information about $\theta$ as we can from this
state.

At this point we'll consider a special case, where we're promised that $\theta
= \frac{y}{4}$ for some integer $y\in\{0,1,2,3\}.$
In other words, we have $\theta\in \{0, 1/4, 1/2, 3/4\},$ so we can express
this number exactly using binary notation with two bits, as .$00,$ .$01,$
.$10,$ or .$11.$
In general, $\theta$ might not be one of these four values, but thinking about
this special case will help us to figure out how to most effectively extract
information about $\theta$ in general.

First we'll define a two-qubit state vector for each possible value $y \in \{0,
1, 2, 3\}.$
\[
\vert \phi_y\rangle = \frac{1}{2} \sum_{x = 0}^3 e^{2\pi i x (\frac{y}{4})}
\vert x \rangle
= \frac{1}{2} \sum_{x = 0}^3 e^{2\pi i \frac{x y}{4}} \vert x \rangle
\]
After simplifying the exponentials, we can write these vectors as follows.
\[
\begin{aligned}
  \vert\phi_0\rangle & = \frac{1}{2} \vert 0 \rangle + \frac{1}{2} \vert 1
  \rangle + \frac{1}{2} \vert 2 \rangle + \frac{1}{2} \vert 3 \rangle \\[3mm]
  \vert\phi_1\rangle & = \frac{1}{2} \vert 0 \rangle + \frac{i}{2} \vert 1
  \rangle - \frac{1}{2} \vert 2 \rangle - \frac{i}{2} \vert 3 \rangle \\[3mm]
  \vert\phi_2\rangle & = \frac{1}{2} \vert 0 \rangle - \frac{1}{2} \vert 1
  \rangle + \frac{1}{2} \vert 2 \rangle - \frac{1}{2} \vert 3 \rangle \\[3mm]
  \vert\phi_3\rangle & = \frac{1}{2} \vert 0 \rangle - \frac{i}{2} \vert 1
  \rangle - \frac{1}{2} \vert 2 \rangle + \frac{i}{2} \vert 3 \rangle
\end{aligned}
\]

These vectors are orthogonal: if we choose any pair of them and compute their
inner product, we get $0.$
Each one is also a unit vector, so $\{\vert\phi_0\rangle, \vert\phi_1\rangle,
\vert\phi_2\rangle, \vert\phi_3\rangle\}$ is an orthonormal basis.
We therefore know right away that there is a measurement that can discriminate
them perfectly --- meaning that, if we're given one of them but we don't know
which, then we can figure out which one it is without error.

To perform such a discrimination with a quantum circuit, we can first define a
unitary operation $V$ that transforms standard basis states into the four
states listed above.
\[
\begin{aligned}
  V \vert 00 \rangle & = \vert\phi_0\rangle \\
  V \vert 01 \rangle & = \vert\phi_1\rangle \\
  V \vert 10 \rangle & = \vert\phi_2\rangle \\
  V \vert 11 \rangle & = \vert\phi_3\rangle
\end{aligned}
\]
To write down $V$ as a $4\times 4$ matrix, it's just a matter of taking the
columns of $V$ to be the states $\vert\phi_0\rangle,\ldots,\vert\phi_3\rangle.$
\[
V =
\frac{1}{2}

\]

This is a special matrix, and it's likely that some readers will have
encountered it before: it's the matrix associated with the $4$-dimensional
\emph{discrete Fourier transform}.
In light of this fact, let us call it by the name $\mathrm{QFT}_4$ rather than
$V.$
The name $\mathrm{QFT}$ is short for \emph{quantum Fourier transform} --- which
is essentially just the discrete Fourier transform, viewed as a unitary
operation.
We'll discuss the quantum Fourier transform in greater detail and generality
shortly.
\[
\mathrm{QFT}_4 =
\frac{1}{2}
%
\]

\begin{figure}[!ht]
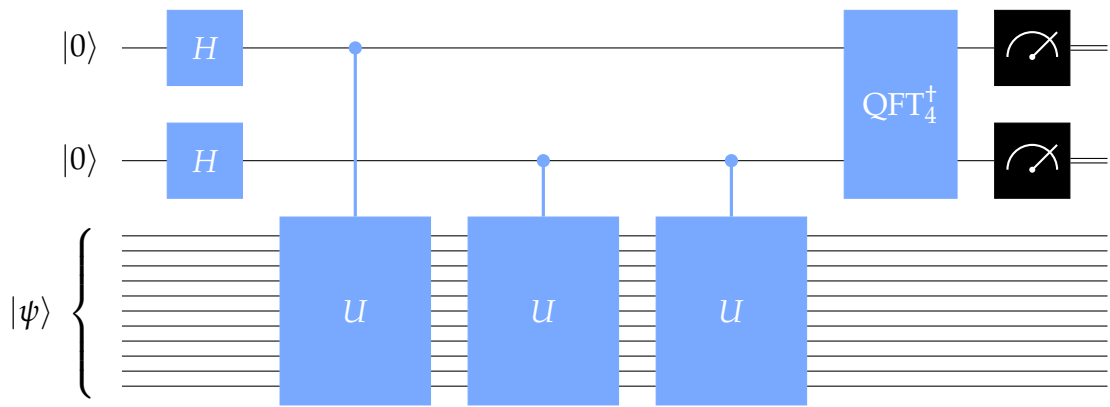

  \begin{center}
    %
    

  \end{center}
  \caption{The complete quantum circuit for phase estimation with two control
    qubits.}
  \label{fig:two-bit-phase-estimation}
\end{figure}

We can perform the inverse of this operation to go the other way, to transform
the states $\vert\phi_0\rangle,\ldots,\vert\phi_3\rangle$ into the standard
basis states $\vert 0\rangle,\ldots,\vert 3\rangle.$
If we do this, then we can measure to learn which value $y\in\{0,1,2,3\}$
describes $\theta$ as $\theta = y/4.$
Figure~\ref{fig:two-bit-phase-estimation} depicts a quantum circuit that does
this.

To summarize, if we run this circuit when $\theta = y/4$ for $y\in\{0,1,2,3\},$
the state immediately before the measurements take place will be $\vert
\psi\rangle \vert y\rangle$ (for $y$ encoded as a two-bit binary string), so
the measurements will reveal the value $y$ without error.

\begin{figure}[!ht]
  \begin{center}
    \begin{tikzpicture}
      \begin{axis}[
          samples=\samples,
          domain = 0:1,
          y = 5cm,
          x = 8cm,
          width=8cm,
          xtick = {0.25, 0.5, 0.75, 1},
          axis lines=center,
          axis line style={-},
          x label style={%
            at={(axis description cs:.5,-0.15)},anchor=north},
          y label style={%
            at={(axis description cs:-0.15,.5)},rotate=90,
            anchor=south},
          xlabel={$\theta$},
          ylabel={probability},
          clip=false,
          legend style={at={(1.125,0.5)},anchor=west}
        ]

        \addplot[thick, DataColor0] {%
          ( 4
          + 6*cos(deg(2*pi*x))
          + 4*cos(deg(4*pi*x))
          + 2*cos(deg(6*pi*x))
          )/16};

        \addplot[thick, DataColor1] {%
          ( 4
          + 6*cos(deg(2*pi*(x-0.25)))
          + 4*cos(deg(4*pi*(x-0.25)))
          + 2*cos(deg(6*pi*(x-0.25)))
          )/16};

        \addplot[thick, DataColor2] {%
          ( 4
          + 6*cos(deg(2*pi*(x+0.5)))
          + 4*cos(deg(4*pi*(x+0.5)))
          + 2*cos(deg(6*pi*(x+0.5)))
          )/16};
        
        \addplot[thick, DataColor3] {%
          ( 4
          + 6*cos(deg(2*pi*(x+0.25)))
          + 4*cos(deg(4*pi*(x+0.25)))
          + 2*cos(deg(6*pi*(x+0.25)))
          )/16};
        
        \legend{$0$, $1$, $2$, $3$}
        
      \end{axis}

    \end{tikzpicture}
  \end{center}
  \caption{Output probabilities for phase estimation with two control qubits.}
  \label{fig:two-bit-probabilities}
\end{figure}
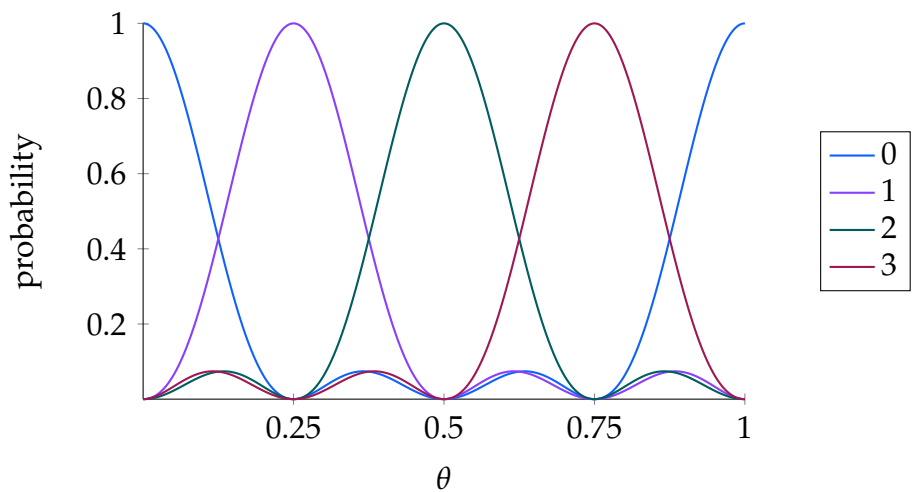

This circuit is motivated by the special case that $\theta \in
\{0,1/4,1/2,3/4\}$ --- but we can run it for any choice of $U$ and $\vert
\psi\rangle,$ and hence any value of $\theta,$ that we wish.
Figure~\ref{fig:two-bit-probabilities} shows a plot of the output probabilities
the circuit produces for arbitrary choices of~$\theta.$

This is a clear improvement over the single-qubit variant described earlier in
the lesson.
It's not perfect --- it can give us the wrong answer --- but the answer is
heavily skewed toward values of $y$ for which $y/4$ is close to $\theta.$
In particular, the most likely outcome always corresponds to the closest value
of $y/4$ to $\theta$ (equating $\theta = 0$ and $\theta = 1$ as before), and
from the plot it looks like this closest value for $x$ always appears with
probability just above $40\%.$
When $\theta$ is exactly halfway between two such values, like $\theta = 0.375$
for instance, the two equally close values of $y$ are equally likely.

\subsubsection{Preparing to generalize to many qubits}

Given the improvement we've just obtained by using two control qubits rather
than one, in conjunction with the inverse of the $4$-dimensional quantum
Fourier transform, it's natural to consider generalizing it further --- by
adding more control qubits.
When we do this, we obtain the general \emph{phase estimation procedure}.
We'll see how this works shortly, but in order to describe it precisely we're
going to need to discuss the quantum Fourier transform in greater generality,
to see how it's defined for other dimensions and to see how we can implement it
(or its inverse) with a quantum circuit.

\subsection{Quantum Fourier transform}

The quantum Fourier transform is a unitary operation that can be defined for
any positive integer dimension $N.$
In this section, we'll see how this operation is defined and how it can be
implemented with a quantum circuit on $m$ qubits with cost $O(m^2)$ when
$N = 2^m.$

The matrices that describe the quantum Fourier transform are derived from an
analogous operation on $N$-dimensional vectors known as the
\emph{discrete Fourier transform.}
This operation can be thought about in different ways.
For instance, we can think about the discrete Fourier transform in purely
abstract, mathematical terms as a linear mapping.
Or we can think about it in computational terms, where we're given an
$N$-dimensional vector of complex numbers (using binary notation to encode the
real and imaginary parts of the entries, let us suppose) and the goal is to
calculate the $N$-dimensional vector obtained by applying the discrete Fourier
transform.
Our focus is on third way, which is viewing this transformation as a unitary
operation that can be performed on a quantum system.

There's an efficient algorithm for computing the discrete Fourier transform on
a given input vector known as the \emph{fast Fourier transform.}
It has applications in signal processing and many other areas, and is
considered by many to be one of the most important algorithms ever discovered.
As it turns out, the implementation of the quantum Fourier transform when $N$
is a power of 2 that we'll study is based on precisely the same underlying
structure that makes the fast Fourier transform possible.

\subsubsection{Definition of the quantum Fourier transform}

To define the quantum Fourier transform, we'll first define a complex number
$\omega_N,$ for each positive integer $N,$ like this:
\[
\omega_N = e^{\frac{2\pi i}{N}} = \cos\left(\frac{2\pi}{N}\right) + i
\sin\left(\frac{2\pi}{N}\right).
\]
This is the number on the complex unit circle we obtain if we start at $1$ and
move counter-clockwise by an angle of $2\pi/N$ radians, or a fraction of $1/N$
of the circumference of the circle. Here are a few examples.
\begingroup\allowdisplaybreaks
\begin{gather*}
  \omega_1 = 1\\[1mm]
  \omega_2 = -1\\[1mm]
  \omega_3 = -\frac{1}{2} + \frac{\sqrt{3}}{2} i\\[2mm]
  \omega_4 = i\\[1mm]
  \omega_8 = \frac{1+i}{\sqrt{2}}\\[3mm]
  \omega_{16} = \frac{\sqrt{2 + \sqrt{2}}}{2} + \frac{\sqrt{2 - \sqrt{2}}}{2}
  i\\[2mm]
  \omega_{100} \approx 0.998 + 0.063 i
\end{gather*}
\endgroup

Now we can define the $N$-dimensional quantum Fourier transform, which is
described by an $N\times N$ matrix whose rows and columns are associated with
the standard basis states $\vert 0\rangle,\ldots,\vert N-1\rangle.$
We're only going to need this operation for when $N = 2^m$ is a power of $2$
for phase estimation, but the operation can be defined for any positive integer
$N.$
\[
\mathrm{QFT}_N = \frac{1}{\sqrt{N}} \sum_{x = 0}^{N-1} \sum_{y = 0}^{N-1}
\omega_N^{xy} \vert x \rangle\langle y\vert
\]

As was already stated, this is the matrix associated with the $N$-dimensional
\emph{discrete Fourier transform}.
Often the leading factor of $1/\sqrt{N}$ is not included in the definition of
this matrix, but we need to include it to obtain a unitary matrix.

Here's the quantum Fourier transform, written as a matrix, for some small
values of $N.$
\begingroup\allowdisplaybreaks
\begin{gather*}
  \mathrm{QFT}_1 = \begin{pmatrix} 1 \end{pmatrix}\\[4mm]
  \mathrm{QFT}_2 =
  \frac{1}{\sqrt{2}} \begin{pmatrix} 1 & 1\\[1mm] 1 & -1 \end{pmatrix}\\[4mm]
  \mathrm{QFT}_3 =
  \frac{1}{\sqrt{3}}
  \begin{pmatrix}
    1 & 1 & 1\\[2mm]
    1 & \frac{-1 + i\sqrt{3}}{2} & \frac{-1 - i\sqrt{3}}{2}\\[2mm]
    1 & \frac{-1 - i\sqrt{3}}{2} & \frac{-1 + i\sqrt{3}}{2}
  \end{pmatrix}\\[4mm]
  \mathrm{QFT}_4 =
  \frac{1}{2}
  \begin{pmatrix}
    1 & 1 & 1 & 1\\[1mm]
    1 & i & -1 & -i\\[1mm]
    1 & -1 & 1 & -1\\[1mm]
    1 & -i & -1 & i
  \end{pmatrix}\\[4mm]
  \mathrm{QFT}_8 =
  \frac{1}{2\sqrt{2}}
  \begin{pmatrix}
    1 & 1 & 1 & 1 & 1 & 1 & 1 & 1\\[2mm]
    1 & \frac{1+i}{\sqrt{2}} & i & \frac{-1+i}{\sqrt{2}} & -1 &
    \frac{-1-i}{\sqrt{2}} & -i & \frac{1-i}{\sqrt{2}}\\[2mm]
    1 & i & -1 & -i & 1 & i & -1 & -i\\[2mm]
    1 & \frac{-1+i}{\sqrt{2}} & -i & \frac{1+i}{\sqrt{2}} & -1 &
    \frac{1-i}{\sqrt{2}} & i & \frac{-1-i}{\sqrt{2}}\\[2mm]
    1 & -1 & 1 & -1 & 1 & -1 & 1 & -1\\[2mm]
    1 & \frac{-1-i}{\sqrt{2}} & i & \frac{1-i}{\sqrt{2}} & -1 &
    \frac{1+i}{\sqrt{2}} & -i & \frac{-1+i}{\sqrt{2}}\\[2mm]
    1 & -i & -1 & i & 1 & -i & -1 & i\\[2mm]
    1 & \frac{1-i}{\sqrt{2}} & -i & \frac{-1-i}{\sqrt{2}} & -1 &
    \frac{-1+i}{\sqrt{2}} & i & \frac{1+i}{\sqrt{2}}
  \end{pmatrix}
\end{gather*}
\endgroup
Notice, in particular, that $\mathrm{QFT}_2$ is another name for a Hadamard
operation.

\subsubsection{Unitarity}

Let's check that $\mathrm{QFT}_N$ is unitary, for any selection of $N.$
One way to do this is to show that its columns form an orthonormal basis.
We can define a vector corresponding to column number $y,$ starting from $y =
0$ and going up to $y = N-1,$ like this:
\[
\vert\phi_y\rangle = \frac{1}{\sqrt{N}} \sum_{x = 0}^{N-1} \omega_N^{xy} \vert
x \rangle.
\]
Taking the inner product between any two of these vectors gives us this
expression:
\[
\langle \phi_z \vert \phi_y \rangle = \frac{1}{N} \sum_{x = 0}^{N-1}
\omega_N^{x (y - z)}
\]

We can evaluate sums like this using the following formula for the sum of the
first $N$ terms of a geometric series.
\[
1 + \alpha + \alpha^2 + \cdots + \alpha^{N-1} =
\begin{cases}
  \frac{\alpha^N - 1}{\alpha - 1} & \text{if } \alpha\neq 1\\[2mm]
  N & \text{if } \alpha=1
\end{cases}
\]
Specifically, we can use this formula when $\alpha = \omega_N^{y-z}.$
When $y = z,$ we have $\alpha = 1,$ so using the formula and dividing by $N$
gives
\[
\langle \phi_y \vert \phi_y \rangle = 1.
\]
When $y\neq z,$ we have $\alpha \neq 1,$ so the formula reveals this:
\[
\langle \phi_z \vert \phi_y \rangle
= \frac{1}{N} \frac{\omega_N^{N(y-z)} - 1}{\omega_N^{y-z} - 1}
= \frac{1}{N} \frac{1 - 1}{\omega_N^{y-z} - 1} = 0.
\]
This happens because $\omega_N^N = e^{2\pi i} = 1,$ so $\omega_N^{N(y-z)} =
1^{y-z} = 1,$ making numerator zero, while the denominator is nonzero because
$\omega_N^{y-z} \neq 1.$
Intuitively speaking, what we're doing is summing a bunch of points that are
distributed around the unit circle, and they cancel out and leave $0$ when
summed.

We have therefore established that
$\{\vert\phi_0\rangle,\ldots,\vert\phi_{N-1}\rangle\}$ is an orthonormal set,
\[
\langle \phi_z \vert \phi_y \rangle =
\begin{cases}
  1 & y=z\\[1mm]
  0 & y\neq z,
\end{cases}
\]
which reveals that $\mathrm{QFT}_N$ is unitary.

\subsubsection{Controlled-phase gates}

\begin{figure}[!ht]
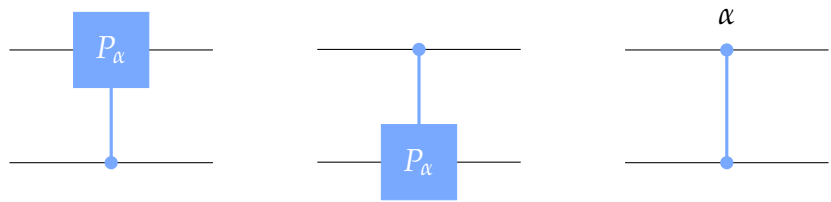

  \begin{center}
    %
    
  \end{center}
  \caption{Three equivalent ways to denote controlled-phase gates.}
  \label{fig:controlled-phase-gates}
\end{figure}

To implement the quantum Fourier transform with a quantum circuit, we'll need
to make use of \emph{controlled-phase} gates.
Recall that a \emph{phase operation} is a single-qubit unitary operation of the
form
\[
P_{\alpha} =

\]
for any real number $\alpha.$
A controlled version of this gate has the following matrix.
\[
%
\]
For this controlled gate, it doesn't actually matter which qubit is the control
and which is the target because the two possibilities are equivalent.
We can use any of the symbols shown in Figure~\ref{fig:controlled-phase-gates}
to represent this gate in quantum circuit diagrams.
For the third form, the number $\alpha$ is also sometimes placed on the side of
the control line or under the lower control when that's convenient.

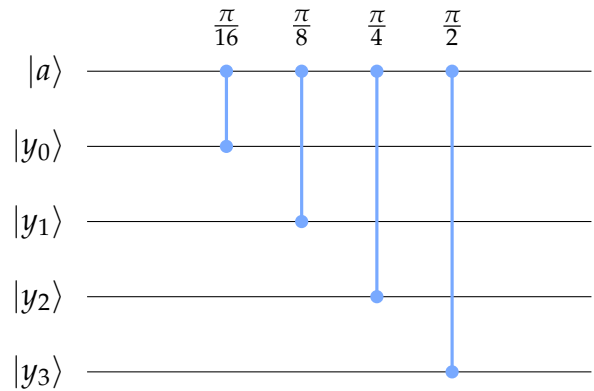
\begin{figure}[!ht]
  \begin{center}
    \begin{tikzpicture}[
        baseline=(current bounding box.north),
        scale=1,
        control/.style={%
          circle,
          fill=CircuitBlue,
          minimum size = 5pt,
          inner sep=0mm}
      ]
      
      \node (y3_in) at (-4,-2) {};
      \node (y2_in) at (-4,-1) {};
      \node (y1_in) at (-4,-0) {};
      \node (y0_in) at (-4,1) {};
      \node (a_in) at (-4,2) {};
      
      \node (y3_out) at (3,-2) {};
      \node (y2_out) at (3,-1) {};
      \node (y1_out) at (3,-0) {};
      \node (y0_out) at (3,1) {};
      \node (a_out) at (3,2) {};

      \draw (y3_in) -- (y3_out);
      \draw (y2_in) -- (y2_out);
      \draw (y1_in) -- (y1_out);
      \draw (y0_in) -- (y0_out);
      \draw (a_in) -- (a_out);

      \node[control] (c32) at (-2,2) {};
      \node[control] (c16) at (-1,2) {};
      \node[control] (c8) at (0,2) {};
      \node[control] (c4) at (1,2) {};

      \node[control] (d32) at (-2,1) {};
      \node[control] (d16) at (-1,0) {};
      \node[control] (d8) at (0,-1) {};
      \node[control] (d4) at (1,-2) {};
      
      \draw[very thick,draw=CircuitBlue] (c32.center) -- (d32.center);
      \draw[very thick,draw=CircuitBlue] (c16.center) -- (d16.center);
      \draw[very thick,draw=CircuitBlue] (c8.center) -- (d8.center);
      \draw[very thick,draw=CircuitBlue] (c4.center) -- (d4.center);

      \node[anchor=south] at ([yshift=2mm]c32.center) {$\frac{\pi}{16}$};
      \node[anchor=south] at ([yshift=2mm]c16.center) {$\frac{\pi}{8}$};
      \node[anchor=south] at ([yshift=2mm]c8.center) {$\frac{\pi}{4}$};
      \node[anchor=south] at ([yshift=2mm]c4.center) {$\frac{\pi}{2}$};

      \node[anchor = east] at (a_in) {$\ket{a}$};
      \node[anchor = east] at (y0_in) {$\ket{y_0}$};
      \node[anchor = east] at (y1_in) {$\ket{y_1}$};
      \node[anchor = east] at (y2_in) {$\ket{y_2}$};
      \node[anchor = east] at (y3_in) {$\ket{y_3}$};

    \end{tikzpicture}
    
  \end{center}
  \caption{A quantum circuit for performing the operation
    \eqref{eq:phase-injection} when $m=5$.}
  \label{fig:phase-injection}
\end{figure}

To perform the quantum Fourier transform when $N = 2^m$ and $m\geq 2,$ we're
going to need to perform an operation on $m$ qubits whose action on standard
basis states can be described as
\begin{equation}
\vert y \rangle \vert a \rangle \mapsto \omega_{2^m}^{ay} \vert y \rangle \vert
a \rangle,
\label{eq:phase-injection}
\end{equation}
where $a$ is a bit and $y \in \{0,\ldots,2^{m-1} - 1\}$ is a number encoded in
binary notation as a string of $m-1$ bits.
This can be done using controlled-phase gates by generalizing the example in
Figure~\ref{fig:phase-injection}, for which $m=5.$

In general, for an arbitrary choice of $m\geq 2,$ the top qubit corresponding
to the bit $a$ can be viewed as the control, with the phase gates $P_{\alpha}$
ranging from $\alpha = \pi/2^{m-1}$ on the qubit corresponding to the least
significant bit of $y$ to $\alpha = \frac{\pi}{2}$ on the qubit corresponding
to the most significant bit of $y.$
These controlled-phase gates all commute with one another and could be
performed in any order.

\subsubsection{Circuit implementation of the QFT}

Now we'll see how we can implement the quantum Fourier transform with a circuit
when the dimension $N = 2^m$ is a power of $2.$
There are, in fact, multiple ways to implement the quantum Fourier transform,
but this is arguably the simplest method known.
Once we know how to implement the quantum Fourier transform with a quantum
circuit, it's straightforward to implement its inverse: we can replace each
gate with its inverse (or, equivalently, conjugate transpose) and apply the
gates in the reverse order.
Every quantum circuit composed of unitary gates alone can be inverted in this
way.

The implementation is recursive in nature, so that's how it's most naturally
described.
The base case is $m=1,$ in which case the quantum Fourier transform is a
Hadamard operation.

To perform the quantum Fourier transform on $m$ qubits when $m \geq 2,$ we can
perform the following steps, whose actions we'll describe for standard basis
states of the form $\vert x \rangle \vert a\rangle,$ where
$x\in\{0,\ldots,2^{m-1} - 1\}$ is an integer encoded as $m-1$ bits using binary
notation and $a$ is a single bit.

\begin{enumerate}
\item
  First apply the $2^{m-1}$-dimensional quantum Fourier transform to the
  bottom/leftmost $m-1$ qubits to obtain this state:
  \[
  \Bigl(\mathrm{QFT}_{2^{m-1}} \vert x \rangle\Bigr) \vert a\rangle
  = \frac{1}{\sqrt{2^{m-1}}} \sum_{y = 0}^{2^{m-1} - 1} \omega_{2^{m-1}}^{xy}
  \vert y \rangle \vert a \rangle.
  \]
  This is done by recursively applying the method being described for one fewer
  qubit, using the Hadamard operation on a single qubit as the base case.

\item
  Use the top/rightmost qubit as a control to inject the phase $\omega_{2^m}^y$
  for each standard basis state $\vert y\rangle$ of the remaining $m-1$ qubits
  (as is described above) to obtain this state:
  \[
  \frac{1}{\sqrt{2^{m-1}}} \sum_{y = 0}^{2^{m-1} - 1} \omega_{2^{m-1}}^{xy}
  \omega_{2^m}^{ay} \vert y \rangle \vert a \rangle.
  \]

\item
  Perform a Hadamard gate on the top/rightmost qubit to obtain this state:
  \[
  \frac{1}{\sqrt{2^{m}}} \sum_{y = 0}^{2^{m-1} - 1} \sum_{b=0}^1
  (-1)^{ab} \omega_{2^{m-1}}^{xy} \omega_{2^m}^{ay}
  \vert y \rangle \vert b \rangle.
  \]

\item
  Permute the order of the qubits so that the least significant bit becomes the
  most significant bit, with all others shifted up/right:
  \[
  \frac{1}{\sqrt{2^{m}}} \sum_{y = 0}^{2^{m-1} - 1} \sum_{b=0}^1
  (-1)^{ab} \omega_{2^{m-1}}^{xy} \omega_{2^m}^{ay}
  \vert b \rangle \vert y \rangle.
  \]
\end{enumerate}

For example, Figure~\ref{fig:QFT32} shows the circuit we obtain for
$N = 32 = 2^5.$ 
In this diagram, the qubits are given names that correspond to the standard
basis vectors $\vert x\rangle \vert a\rangle$ (for the input) and $\vert
b\rangle \vert y\rangle$ (for the output) for clarity.

\begin{figure}[!ht]
  \begin{center}



  \end{center}
  \caption{A quantum circuit for $\mathrm{QFT}_{32}$ using an operation for
    $\mathrm{QFT}_{16}$.}
  \label{fig:QFT32}
\end{figure}

\subsubsection{Analysis}

The key formula we need to verify that the circuit just described implements
the $2^m$-dimensional quantum Fourier transform is this one:
\[
(-1)^{ab}
\omega_{2^{m-1}}^{xy}
\omega_{2^m}^{ay}
=
\omega_{2^m}^{(2x+ a)(2^{m-1}b + y)}.
\]
This formula works for any choice of integers $a,$ $b,$ $x,$ and $y,$ but we'll
only need it for $a,b\in\{0,1\}$ and $x,y\in\{0,\ldots,2^{m-1}-1\}.$
It can be checked by expanding the product in the exponent on the right-hand
side,
\[
\omega_{2^m}^{(2x+ a)(2^{m-1}b + y)}
= \omega_{2^m}^{2^m xb} \omega_{2^m}^{2xy} \omega_{2^m}^{2^{m-1}ab}
\omega_{2^m}^{ay}
= (-1)^{ab} \omega_{2^{m-1}}^{xy} \omega_{2^m}^{ay},
\]
where the second equality makes use of the observation that
\[
\omega_{2^m}^{2^m xb} = \bigl(\omega_{2^m}^{2^m}\bigr)^{xb} = 1^{xb} = 1.
\]

The $2^m$-dimensional quantum Fourier transform is defined as follows for every
$u\in\{0,\ldots,2^m - 1\}.$
\[
\mathrm{QFT}_{2^m} \vert u\rangle = \frac{1}{\sqrt{2^m}}
\sum_{v = 0}^{2^m - 1} \omega_{2^m}^{uv} \vert v\rangle
\]
If we write $u$ and $v$ as
\[
\begin{aligned}
u & = 2x + a\\
v & = 2^{m-1}b + y
\end{aligned}
\]
for $a,b\in\{0,1\}$ and $x,y\in\{0,\ldots,2^{m-1} - 1\},$ we obtain
\[
\begin{aligned}
\mathrm{QFT}_{2^m} \vert 2x + a\rangle
& =
\frac{1}{\sqrt{2^m}}
\sum_{y = 0}^{2^{m-1} - 1}
\sum_{b=0}^1
\omega_{2^m}^{(2x+ a)(2^{m-1}b + y)} \vert b 2^{m-1} + y\rangle\\[2mm]
& =
\frac{1}{\sqrt{2^m}}
\sum_{y = 0}^{2^{m-1} - 1}
\sum_{b=0}^1
(-1)^{ab}
\omega_{2^{m-1}}^{xy}
\omega_{2^m}^{ay}
\vert b 2^{m-1} + y\rangle.
\end{aligned}
\]
Finally, by thinking about the standard basis states
$\vert x \rangle \vert a\rangle$ and $\vert b \rangle \vert y \rangle$ as
binary encodings of integers in the range $\{0,\ldots,2^m-1\},$
\[
\begin{aligned}
\vert x \rangle \vert a\rangle & = \vert 2x + a \rangle\\
\vert b \rangle \vert y \rangle & = \vert 2^{m-1}b + y\rangle,
\end{aligned}
\]
we see that the circuit above implements the required operation.

If this method for performing the quantum Fourier transform seems remarkable,
it's because it is: it's essentially the fast Fourier transform in the form of
a quantum circuit.

Finally, let's count how many gates are used in the circuit just described.
The controlled-phase gates aren't in the standard gate set that we discussed in
the previous lesson, but to begin we'll ignore this and count each of them as a
single gate.
Let's let $s_m$ denote the number of gates we need for each possible choice of
$m.$
If $m=1,$ the quantum Fourier transform is just a Hadamard operation, so
\[
s_1 = 1.
\]
If $m\geq 2,$ then in the circuit above we need $s_{m-1}$ gates for the quantum
Fourier transform on $m-1$ qubits, plus $m-1$ controlled-phase gates, plus a
Hadamard gate, plus $m-1$ swap gates, so
\[
s_m = s_{m-1} + (2m - 1).
\]
We can obtain a closed-form expression by summing:
\[
s_m = \sum_{k = 1}^m (2k - 1) = m^2.
\]

We don't actually need as many swap gates as the method describes.
If we rearrange the gates just a bit, we can push all of the swap gates out to
the right and reduce the number of swap gates required to $\lfloor m/2\rfloor.$
Asymptotically speaking this isn't a major improvement: we still obtain
circuits with size $O(m^2)$ for performing $\mathrm{QFT}_{2^m}.$

If we wish to implement the quantum Fourier transform using only gates from our
standard gate set, we need to either build or approximate each of the
controlled-phase gates with gates from our set.
The number required depends on how much accuracy we require, but as a function
of $m$ the total cost remains quadratic.

It is, in fact, possible to approximate the quantum Fourier transform quite
closely with a sub-quadratic number of gates by using the fact that
$P_{\alpha}$ is very close to the identity operation when $\alpha$ is very
small --- which means that we can simply leave out most of the controlled-phase
gates without suffering too much of a loss in terms of accuracy.

\subsection{General procedure and analysis}

\begin{figure}[!ht]
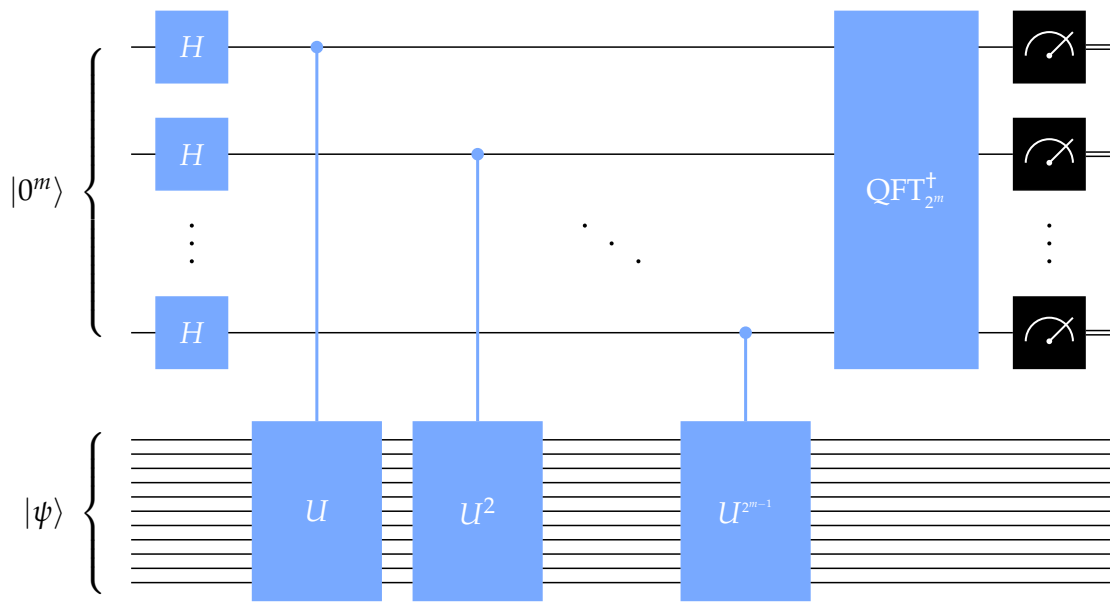

  \begin{center}
    \scalebox{0.95}{
%
    }
    
  \end{center}
  \caption{A quantum circuit for the general phase estimation procedure.}
  \label{fig:phase-estimation-procedure}
\end{figure}

Now we'll examine the phase estimation procedure in general.
The idea is to extend the two-qubit version of phase estimation that we
considered above in the natural way suggested by
Figure~\ref{fig:phase-estimation-procedure}.

Notice that, for each new control qubit added on the top, we \emph{double} the
number of times the unitary operation $U$ is performed.
This is indicated in the diagram by the powers on $U$ for each of the
controlled-unitary operations.

The most straightforward way to implement a controlled-$U^k$ operation for some
choice of $k$ is simply to repeat a controlled-$U$ operation $k$ times.
If this is indeed the methodology that is used, it must be recognized that the
addition of control qubits contributes significantly to the size of the
circuit: if we have $m$ control qubits, like the diagram depicts, a total of
$2^m - 1$ copies of the controlled-$U$ operation are required.
This means that a significant computational cost is incurred as $m$ is
increased --- but as we will see, it also leads to a significantly more
accurate approximation of $\theta.$

It is important to note, however, that for \emph{some} choices of $U$ it may be
possible to create a circuit that implements the operation $U^k$ for large
values of $k$ in a more efficient way than simply repeating $k$ times the
circuit for $U.$
We'll see a specific example of this in the context of integer factorization
later in the lesson, where the efficient algorithm for \emph{modular
exponentiation} discussed in the previous lesson comes to the rescue.

Now let us analyze the circuit just described.
The state immediately prior to the inverse quantum Fourier transform looks like
this:
\[
\frac{1}{\sqrt{2^m}} \sum_{x = 0}^{2^m - 1} \bigl( U^x \vert\psi\rangle \bigr)
\vert x\rangle
= \vert\psi\rangle \otimes \frac{1}{\sqrt{2^m}} \sum_{x = 0}^{2^m - 1} e^{2\pi i x\theta}  \vert x\rangle.
\]

\subsubsection{A special case}

Along similar lines to what we did in the $m=2$ case, we'll first consider the
special case that $\theta = y/2^m$ for $y\in\{0,\ldots,2^m-1\}.$
In this case the state prior to the inverse quantum Fourier transform can
alternatively be written like this:
\[
\vert\psi\rangle \otimes \frac{1}{\sqrt{2^m}} \sum_{x = 0}^{2^m - 1} e^{2\pi i
  \frac{xy}{2^m}}  \vert x\rangle
= \vert\psi\rangle \otimes \frac{1}{\sqrt{2^m}} \sum_{x = 0}^{2^m - 1}
\omega_{2^m}^{xy}  \vert x\rangle
= \vert\psi\rangle \otimes \mathrm{QFT}_{2^m} \vert y\rangle.
\]
So, when the inverse quantum Fourier transform is applied, the state becomes
\[
\vert\psi\rangle \vert y\rangle
\]
and the measurements reveal $y$ (encoded in binary).

\subsubsection{Bounding the probabilities}

For other values of $\theta,$ meaning ones that don't take the form $y/2^m$ for
an integer~$y,$ the measurement outcomes won't be certain, but we can prove
bounds on the probabilities for different outcomes.
Going forward, let's consider an arbitrary choice of $\theta$ satisfying
$0\leq \theta < 1.$

After the inverse quantum Fourier transform is performed, the state of the
circuit is this:
\[
\vert \psi \rangle \otimes
\frac{1}{2^m} \sum_{y=0}^{2^m - 1} \sum_{x=0}^{2^m-1}
e^{2\pi i x (\theta - y/2^m)} \vert y\rangle.
\]
So, when the measurements on the top $m$ qubits are performed, we see each
outcome $y$ with probability
\[
p_y =
\left\vert \frac{1}{2^m} \sum_{x=0}^{2^m - 1} e^{2\pi i x (\theta - y/2^m)}
\right\vert^2.
\]

To get a better handle on these probabilities, we'll make use of the same
formula that we saw before, for the sum of the initial portion of a geometric
series.
\[
1 + \alpha + \alpha^2 + \cdots + \alpha^{N-1} =
\begin{cases}
\frac{\alpha^N - 1}{\alpha - 1} & \text{if } \alpha\neq 1\\[2mm]
N & \text{if } \alpha=1
\end{cases}
\]
We can simplify the sum appearing in the formula for $p_y$ by taking
$\alpha = e^{2\pi i (\theta - y/2^m)}.$
Here's what we obtain.
\[
\sum_{x=0}^{2^m - 1} e^{2\pi i x (\theta - y/2^m)}
=
\begin{cases}
2^m & \theta = y/2^m\\[2mm]
\frac{e^{2\pi (2^m \theta - y)} - 1}{e^{2\pi (\theta - y/2^m)} - 1}
& \theta\neq y/2^m
\end{cases}
\]

So, in the case that $\theta = y/2^m,$ we find that $p_y = 1$ (as we already
knew from considering this special case), and in the case that
$\theta \neq y/2^m,$ we find that
\[
p_y = \frac{1}{2^{2m}} \left\vert \frac{e^{2\pi i (2^m \theta - y)} -
  1}{e^{2\pi i (\theta - y/2^m)} - 1}\right\vert^2.
\]

\begin{figure}[!ht]
  \begin{center}
    \begin{tikzpicture}          
      
      \newcommand\CircleRadius{3cm}

      \draw[Gray] (-3cm,0cm) -- (3cm,0cm);
      \draw[Gray] (0cm,-3cm) -- (0cm,3cm);
      
      \draw (0,0) circle (\CircleRadius);

      \node (P) at (125:\CircleRadius) {};
      \node (1) at (0:\CircleRadius) {};

      \node[anchor = south east] at (P) {$e^{2\pi i \delta}$};
      \node[anchor = west] at (1) {$1$};
      
      \draw[very thick, DataColor0]
      (1.center) -- node[above,midway,rotate=-27.5, text=black] {
        $\bigl\vert e^{2\pi i\delta} - 1\bigr\vert$
      } (P.center);

      \draw[very thick, DataColor1] (1.center) arc (0:125:\CircleRadius)
      node[midway,above,rotate=-27.5, text=black] {$2\pi\vert\delta\vert$};
      
      \filldraw (1) circle (0.75pt);
      \filldraw (P) circle (0.75pt);

    \end{tikzpicture}
  \end{center}
  \caption{Arc and chord lengths on the complex unit circle.}
  \label{fig:arc-and-chord}
\end{figure}
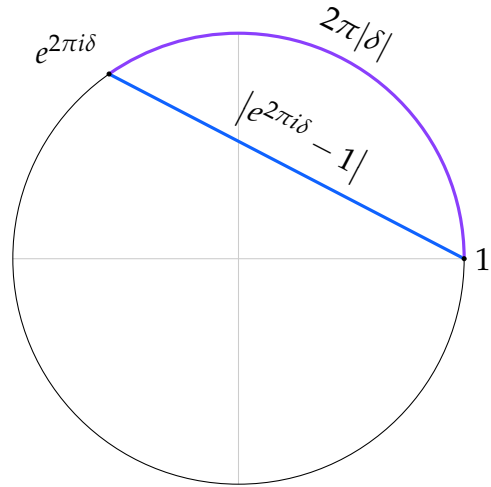

We can learn more about these probabilities by thinking about how arc lengths
and chord lengths on the unit circle are related.
Figure~\ref{fig:arc-and-chord} illustrates the relationships we need for any
real number $\delta\in \bigl[ -\frac{1}{2},\frac{1}{2}\bigr].$

First, the chord length (drawn in blue) can't possibly be larger than the arc
length (drawn in purple):
\[
\bigl\vert e^{2\pi i \delta} - 1\bigr\vert \leq 2\pi\vert\delta\vert.
\]
Relating these lengths in the other direction, we see that the ratio of the arc
length to the chord length is greatest when $\delta = \pm 1/2,$ and in this
case the ratio is half the circumference of the circle divided by the diameter,
which is $\pi/2.$
Thus, we have
\[
\frac{2\pi\vert\delta\vert}{\bigl\vert e^{2\pi i \delta} - 1\bigr\vert} \leq
\frac{\pi}{2},
\]
and so
\[
\bigl\vert e^{2\pi i \delta} - 1\bigr\vert \geq 4\vert\delta\vert.
\]
An analysis based on these relations reveals the following two facts.

\begin{enumerate}
  \setlength\parskip{5pt}
\item
  Suppose that $\theta$ is a real number and $y\in \{0,\ldots,2^m-1\}$
  satisfies
  \[
  \Bigl\vert \theta - \frac{y}{2^m}\Bigr\vert \leq 2^{-(m+1)}.
  \]
  This means that $y/2^m$ is either the best $m$-bit approximation to $\theta,$
  or it's exactly halfway between $y/2^m$ and either $(y-1)/2^m$ or
  $(y+1)/2^m,$ so it's one of the two best approximations to $\theta.$ 

   We'll prove that $p_y$ has to be pretty large in this case.
   By the assumption we're considering, it follows that
   $\vert 2^m \theta - y \vert \leq 1/2,$ so we can use the second observation
   above relating arc and chord lengths to conclude that
   \[
   \left\vert e^{2\pi i (2^m \theta - y)} - 1\right\vert
   \geq 4 \vert 2^m \theta - y \vert = 4 \cdot 2^m \cdot
   \Bigl\vert \theta - \frac{y}{2^m}\Bigr\vert.
   \]
   We can also use the first observation about arc and chord lengths to
   conclude
   \[
   \left\vert e^{2\pi i (\theta - y/2^m)} - 1\right\vert \leq 2\pi \Bigl\vert
   \theta - \frac{y}{2^m}\Bigr\vert.
   \]
   Putting these two inequalities to use on $p_y$ reveals
   \[
   p_y \geq \frac{1}{2^{2m}} \frac{16 \cdot 2^{2m}}{4 \pi^2}
   = \frac{4}{\pi^2} \approx 0.405.
   \]

   This explains our observation that the best outcome occurs with probability
   greater than $40\%$ in the $m=2$ version of phase estimation discussed
   earlier.
   It's not really 40\%, it's $4/\pi^2,$ and this bound holds for every
   choice of~$m.$

 \item
   Now suppose that $y\in \{0,\ldots,2^m-1\}$ satisfies
   \[
   2^{-m} \leq \Bigl\vert \theta - \frac{y}{2^m}\Bigr\vert \leq \frac{1}{2}.
   \]
   This means that there's a better approximation $z/2^m$ to $\theta$
   between $\theta$ and~$y/2^m.$
   This time we'll prove that $p_y$ can't be too big.
   
   We can start with the simple observation that
   \[
   \left\vert e^{2\pi i (2^m \theta - y)} - 1\right\vert \leq 2,
   \]
   which follows from the fact that any two points on the unit circle can
   differ in absolute value by at most $2.$

   We can also use the second observation about arc and chord lengths from
   above, this time working with the denominator of $p_y$ rather than the
   numerator, to conclude
   \[
   \left\vert e^{2\pi i (\theta - y/2^m)} - 1\right\vert
   \geq  4\Bigl\vert \theta - \frac{y}{2^m}\Bigr\vert
   \geq 4 \cdot 2^{-m}.
   \]
   Putting the two inequalities together reveals
   \[
   p_y \leq \frac{1}{2^{2m}} \frac{4}{16 \cdot 2^{-2m}} = \frac{1}{4}.
   \]

   Note that, while this bound is good enough for our purposes, it is fairly
   crude --- the probability is usually much lower than $1/4.$

\end{enumerate}

The important take-away from this analysis is that very close approximations to
$\theta$ are likely to occur --- we'll get a best $m$-bit approximation with
probability greater than $40\%$ --- whereas approximations off by more than
$2^{-m}$ are less likely to occur, with probability upper bounded by $25\%.$

Given these guarantees, it is possible to boost our confidence by repeating the
phase estimation procedure several times, to gather statistical evidence about
$\theta.$
It is important to note that the state $\vert\psi\rangle$ of the bottom
collection of qubits is unchanged by the phase estimation procedure, so it can
be used to run the procedure as many times as we like.
In particular, each time we run the circuit, we get a best $m$-bit
approximation to $\theta$ with probability greater than $40\%,$ while the
probability of being off by more than $2^{-m}$ is bounded by $25\%.$
If we run the circuit several times and take the most commonly appearing
outcome of the runs, it's therefore exceedingly likely that the outcome that
appears most commonly will not be one that occurs at most $25\%$ of the time.
As a result, we'll be very likely to obtain an approximation $y/2^m$ that's
within $1/2^m$ of the value $\theta.$
Indeed, the unlikely chance that we're off by more than $1/2^m$ decreases
exponentially in the number of times the procedure is run.

\begin{figure}[p]
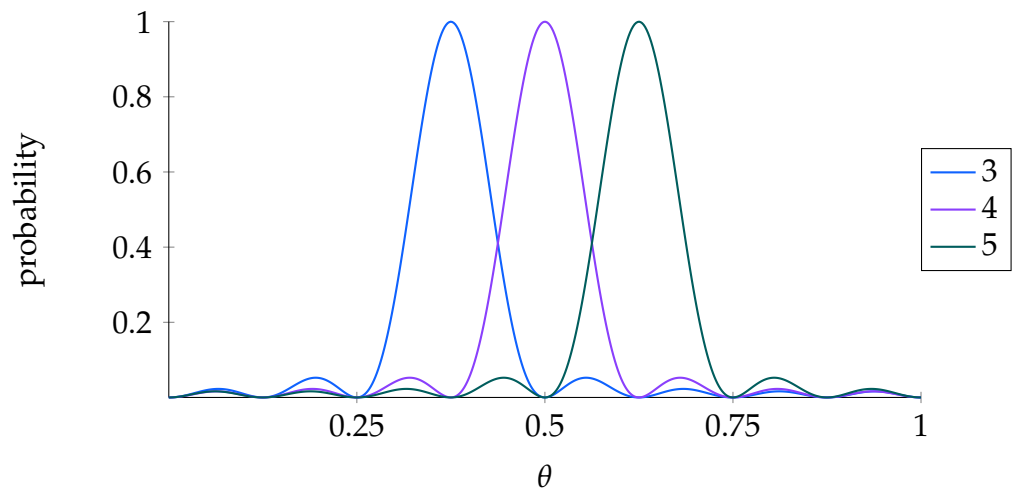

  \begin{center}


  \end{center}
  \caption{Output probabilities for the outcomes 3, 4, and 5 in the
    phase estimation procedure using $m=3$ control qubits.}
  \label{fig:three-qubit-probabilities}
\end{figure}

\begin{figure}[p]
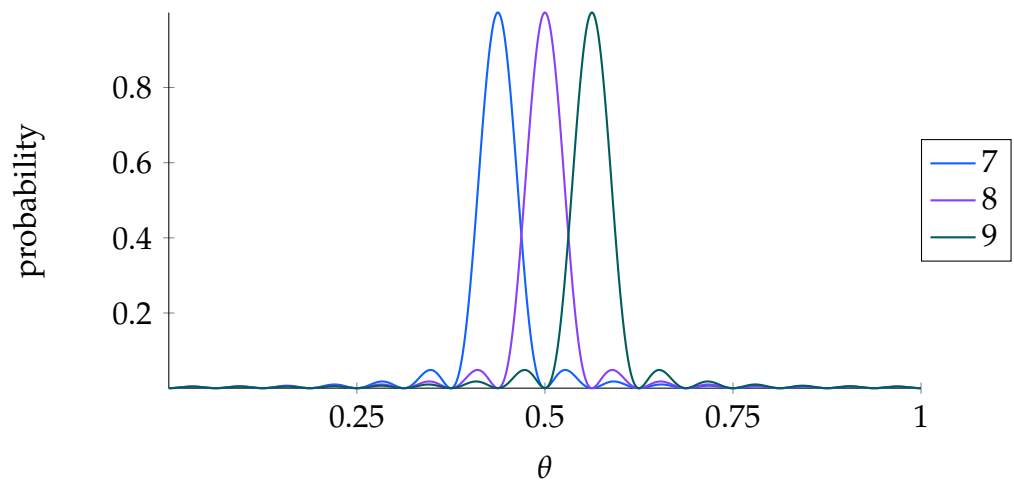

  \vspace*{6mm}
  \begin{center}
    %
  \end{center}
  \caption{Output probabilities for the outcomes 7, 8, and 9 in the
    phase estimation procedure using $m=4$ control qubits.}
  \label{fig:four-qubit-probabilities}
\end{figure}

Figures~\ref{fig:three-qubit-probabilities} and
\ref{fig:four-qubit-probabilities} show plots of the probabilities for three
consecutive values for $y$ when $m = 3$ and $m=4$ as functions of $\theta.$
(Only three outcomes are shown for clarity.
Probabilities for other outcomes are obtained by cyclically shifting the same
underlying function.)

\section{Shor's algorithm}

Now we'll turn our attention to the integer factorization problem, and see how
it can be solved efficiently on a quantum computer using phase estimation.
The algorithm we'll obtain is
\emph{Shor's algorithm} for integer factorization.
Shor didn't describe his algorithm specifically in terms of phase estimation,
but it is a natural and intuitive way to explain how it works.

We'll begin by discussing an intermediate problem known as the
\emph{order finding problem} and see how phase estimation provides a solution
to this problem.
We'll then see how an efficient solution to the order finding problem gives us
an efficient solution to the integer factorization problem.
(When a solution to one problem provides a solution to another problem like
this, we say that the second problem \emph{reduces} to the first --- so in this
case we're reducing integer factorization to order finding.)
This second part of Shor's algorithm doesn't make use of quantum computing at
all; it's completely classical.
Quantum computing is only needed to solve order finding.

\subsection{The order finding problem}

\subsubsection{Some basic number theory}

To explain the order finding problem and how it can be solved using phase
estimation, it will be helpful to begin with a couple of basic number theory
concepts, and to introduce some handy notation along the way.

To begin, for any given positive integer $N,$ define the set $\mathbb{Z}_N$
like this.
\[
\mathbb{Z}_N = \{0,1,\ldots,N-1\}
\]
For instance,
$\mathbb{Z}_1 = \{0\},\;$
$\mathbb{Z}_2 = \{0,1\},\;$
$\mathbb{Z}_3 = \{0,1,2\},\;$
and so on.

These are sets of numbers, but we can think of them as more than sets.
In particular, we can think about \emph{arithmetic operations} on
$\mathbb{Z}_N$ such as addition and multiplication --- and if we agree to
always take our answers modulo $N$
(i.e., divide by $N$ and take the remainder as the result), we'll always stay
within this set when we perform these operations.
The two specific operations of addition and multiplication, both taken modulo
$N,$ turn $\mathbb{Z}_N$ into a \emph{ring}, which is a fundamentally important
type of object in algebra.

For example, $3$ and $5$ are elements of $\mathbb{Z}_7,$ and if we multiply
them together we get $3\cdot 5 = 15,$ which leaves a remainder of $1$ when
divided by $7.$
Sometimes we express this as follows.
\[
3 \cdot 5 \equiv 1 \; (\textrm{mod } 7)
\]
But we can also simply write $3 \cdot 5 = 1,$ provided that it's been made
clear that we're working in $\mathbb{Z}_7,$ just to keep our notation as simple
as possible.

As an example, here are the addition and multiplication tables for
$\mathbb{Z}_6.$
\[
\begin{array}{c|cccccc}
    + & 0 & 1 & 2 & 3 & 4 & 5 \\\hline
    0 & 0 & 1 & 2 & 3 & 4 & 5 \\
    1 & 1 & 2 & 3 & 4 & 5 & 0 \\
    2 & 2 & 3 & 4 & 5 & 0 & 1 \\
    3 & 3 & 4 & 5 & 0 & 1 & 2 \\
    4 & 4 & 5 & 0 & 1 & 2 & 3 \\
    5 & 5 & 0 & 1 & 2 & 3 & 4 \\
\end{array}
\qquad
\begin{array}{c|cccccc}
\cdot & 0 & 1 & 2 & 3 & 4 & 5 \\\hline
    0 & 0 & 0 & 0 & 0 & 0 & 0 \\
    1 & 0 & 1 & 2 & 3 & 4 & 5 \\
    2 & 0 & 2 & 4 & 0 & 2 & 4 \\
    3 & 0 & 3 & 0 & 3 & 0 & 3 \\
    4 & 0 & 4 & 2 & 0 & 4 & 2 \\
    5 & 0 & 5 & 4 & 3 & 2 & 1 \\
\end{array}
\]
Among the $N$ elements of $\mathbb{Z}_N,$ the elements $a\in\mathbb{Z}_N$ that
satisfy $\gcd(a,N) = 1$ are special.
Frequently the set containing these elements is denoted with a star like so.
\[
\mathbb{Z}_N^{\ast} = \{a\in \mathbb{Z}_N : \gcd(a,N) = 1\}
\]

If we focus our attention on the operation of multiplication, the set
$\mathbb{Z}_N^{\ast}$ forms a \emph{group} --- specifically an
\emph{abelian group} --- which is another important type of object in algebra.
It's a basic fact about these sets (and finite groups in general), that if we
pick any element $a\in\mathbb{Z}_N^{\ast}$ and repeatedly multiply $a$ to
itself, we'll always eventually get the number $1.$

For a first example, let's take $N=6.$
We have that $5\in\mathbb{Z}_6^{\ast}$ because $\gcd(5,6) = 1,$ and if we
multiply $5$ to itself we get $1,$ as the table above confirms.
\[
5^2 = 1 \quad \text{(working within $\mathbb{Z}_6$)}
\]

As a second example, let's take $N = 21.$
If we go through the numbers from $0$ to $20,$ the ones having GCD equal to $1$
with $21$ are as follows.
\[
\mathbb{Z}_{21}^{\ast} = \{1,2,4,5,8,10,11,13,16,17,19,20\}
\]
For each of these elements, it is possible to raise that number to a positive
integer power to get $1.$
Here are the smallest powers for which this works:
\[
\begin{array}{c@{\hspace{8mm}}c@{\hspace{8mm}}c}
  1^{1} = 1  &
  8^{2} = 1  &
  16^{3} = 1 \\[1mm]
  2^{6} = 1  &
  10^{6} = 1  &
  17^{6} = 1 \\[1mm]
  4^{3} = 1  &
  11^{6} = 1  &
  19^{6} = 1 \\[1mm]
  5^{6} = 1  &
  13^{2} = 1  &
  20^{2} = 1
\end{array}
\]

Naturally we're working within $\mathbb{Z}_{21}$ for all of these equations,
which we haven't bothered to write --- we take it to be implicit to avoid
cluttering things up.
We'll continue to do that throughout the rest of the lesson.

\subsubsection{Problem statement and connection to phase estimation}

Now we can state the order finding problem.

\begin{callout}[title={Order finding}]
  \begin{problem}
    \Input{Positive integers $N$ and $a$ satisfying $\gcd(N,a) = 1$.}
    \Output{The smallest positive integer $r$ such that
      $a^r \equiv 1$ $(\textrm{mod } N)$.}
  \end{problem}
\end{callout}

Alternatively, in terms of the notation we just introduced above, we're given
$a \in \mathbb{Z}_N^{\ast},$ and we're looking for the smallest positive
integer $r$ such that $a^r = 1.$
This number $r$ is called the \emph{order} of $a$ modulo $N.$

To connect the order finding problem to phase estimation, let's think about the
operation defined on a system whose classical states correspond to
$\mathbb{Z}_N,$ where we multiply by a fixed element $a\in\mathbb{Z}_N^{\ast}.$
\[
M_a \vert x\rangle = \vert ax \rangle \qquad
\text{(for each $x\in\mathbb{Z}_N$)}
\]
To be clear, we're doing the multiplication in $\mathbb{Z}_N,$ so it's implicit
that we're taking the product modulo $N$ inside of the ket on the right-hand
side of the equation.

For example, if we take $N = 15$ and $a=2,$ then the action of $M_2$ on the
standard basis $\{\vert 0\rangle,\ldots,\vert 14\rangle\}$ is as follows.
\[
\begin{array}{ccc}
  M_{2} \vert 0 \rangle = \vert 0\rangle \quad &
  M_{2} \vert 5 \rangle = \vert 10\rangle \quad &
  M_{2} \vert 10 \rangle = \vert 5\rangle \\[1mm]
  M_{2} \vert 1 \rangle = \vert 2\rangle \quad &
  M_{2} \vert 6 \rangle = \vert 12\rangle \quad &
  M_{2} \vert 11 \rangle = \vert 7\rangle \\[1mm]
  M_{2} \vert 2 \rangle = \vert 4\rangle \quad &
  M_{2} \vert 7 \rangle = \vert 14\rangle \quad &
  M_{2} \vert 12 \rangle = \vert 9\rangle \\[1mm]
  M_{2} \vert 3 \rangle = \vert 6\rangle \quad &
  M_{2} \vert 8 \rangle = \vert 1\rangle \quad &
  M_{2} \vert 13 \rangle = \vert 11\rangle \\[1mm]
  M_{2} \vert 4 \rangle = \vert 8\rangle \quad &
  M_{2} \vert 9 \rangle = \vert 3\rangle \quad &
  M_{2} \vert 14 \rangle = \vert 13\rangle
\end{array}
\]

This is a unitary operation provided that $\gcd(a,N)=1;$ it shuffles the
elements of the standard basis $\{\vert 0\rangle,\ldots,\vert N-1\rangle\},$ so
as a matrix it's a permutation matrix.
It's evident from its definition that this operation is deterministic, and a
simple way to see that it's invertible is to think about the order $r$ of $a$
modulo $N,$ and to recognize that the inverse of $M_a$ is $M_a^{r-1}.$
\[
M_a^{r-1} M_a = M_a^r = M_{a^r} = M_1 = \mathbb{I}
\]

There's another way to think about the inverse that doesn't require any
knowledge of $r$ (which, after all, is what we're trying to compute).
For every element $a\in\mathbb{Z}_N^{\ast}$ there's always a unique element
$b\in\mathbb{Z}_N^{\ast}$ that satisfies $ab=1.$
We denote this element $b$ by $a^{-1},$ and it can be computed efficiently;
an extension of Euclid's GCD algorithm does it with cost quadratic in
$\operatorname{lg}(N).$
And thus
\[
M_{a^{-1}} M_a = M_{a^{-1}a} = M_1 = \mathbb{I}.
\]

So, the operation $M_a$ is both deterministic and invertible.
That implies that it's described by a permutation matrix, and is therefore
unitary.

Now let's think about the eigenvectors and eigenvalues of the operation $M_a,$
assuming that $a\in\mathbb{Z}_N^{\ast}.$
As was just argued, this assumption tells us that $M_a$ is unitary.

There are $N$ eigenvalues of $M_a,$ possibly including the same eigenvalue
repeated multiple times, and in general there's some freedom in selecting
corresponding eigenvectors --- but we won't need to worry about all of the
possibilities.
Let's start simply and identify just one eigenvector of $M_a.$
\[
\vert \psi_0 \rangle
= \frac{\vert 1 \rangle + \vert a \rangle + \cdots
  + \vert a^{r-1} \rangle}{\sqrt{r}}
\]
The number $r$ is the order of $a$ modulo $N,$ here and throughout the
remainder of the lesson.
The eigenvalue associated with this eigenvector is $1$ because it isn't changed
when we multiply by $a.$
\[
M_a \vert \psi_0 \rangle
= \frac{\vert a \rangle + \cdots + \vert a^{r-1} \rangle
  + \vert a^r \rangle}{\sqrt{r}}
= \frac{\vert a \rangle + \cdots + \vert a^{r-1} \rangle
  + \vert 1 \rangle}{\sqrt{r}}
= \vert \psi_0 \rangle
\]
This happens because $a^r = 1,$ so each standard basis state $\vert a^k
\rangle$ gets shifted to $\vert a^{k+1} \rangle$ for $k\leq r-1,$ and $\vert
a^{r-1} \rangle$ gets shifted back to $\vert 1\rangle.$
Informally speaking, it's like we're slowly stirring $\vert \psi_0 \rangle,$
but it's already completely stirred so nothing changes.

Here's another example of an eigenvector of $M_a.$
This one happens to be more interesting in the context of order finding and
phase estimation.
\[
\vert \psi_1 \rangle = \frac{\vert 1 \rangle + \omega_r^{-1} \vert a \rangle +
  \cdots + \omega_r^{-(r-1)}\vert a^{r-1} \rangle}{\sqrt{r}}
\]
Alternatively, we can write this vector using a summation as follows.
\[
\vert \psi_1 \rangle = \frac{1}{\sqrt{r}}
\sum_{k = 0}^{r-1} \omega_r^{-k} \vert a^k \rangle
\]

Here we're seeing the complex number $\omega_r = e^{2\pi i/r}$ showing up
naturally, due to the way that multiplication by $a$ works modulo $N.$
This time the corresponding eigenvalue is $\omega_r.$
To see this, we can first compute as follows.
\[
M_a \vert \psi_1 \rangle
= \sum_{k = 0}^{r-1} \omega_r^{-k} M_a\vert a^k \rangle
= \sum_{k = 0}^{r-1} \omega_r^{-k} \vert a^{k+1} \rangle
= \sum_{k = 1}^{r} \omega_r^{-(k - 1)} \vert a^{k} \rangle
= \omega_r \sum_{k = 1}^{r} \omega_r^{-k} \vert a^{k} \rangle
\]
Then, because $\omega_r^{-r} = 1 = \omega_r^0$ and
$\vert a^r \rangle = \vert 1\rangle = \vert a^0\rangle,$ we see that
\[
\sum_{k = 1}^{r} \omega_r^{-k} \vert a^{k} \rangle
= \sum_{k = 0}^{r-1} \omega_r^{-k} \vert a^k \rangle
= \vert\psi_1\rangle,
\]
so $M_a \vert\psi_1\rangle = \omega_r \vert\psi_1\rangle.$

Using the same reasoning, we can identify additional eigenvector/eigenvalue
pairs for $M_a.$
For any choice of $j\in\{0,\ldots,r-1\}$ we have that
\[
\vert \psi_j \rangle = \frac{1}{\sqrt{r}}
\sum_{k = 0}^{r-1} \omega_r^{-jk} \vert a^k \rangle
\]
is an eigenvector of $M_a$ whose corresponding eigenvalue is $\omega_r^j.$
\[
M_a \vert \psi_j \rangle = \omega_r^j \vert \psi_j \rangle
\]

There are other eigenvectors of $M_a,$ but we don't need to concern ourselves
with them --- we'll focus solely on the eigenvectors
$\vert\psi_0\rangle,\ldots,\vert\psi_{r-1}\rangle$ that we've just identified.

\subsection{Order finding through phase estimation}

To solve the order finding problem for a given choice of
$a\in\mathbb{Z}_N^{\ast},$ we can apply the phase estimation procedure to the
operation $M_a.$

To do this, we need to implement not only $M_a$ efficiently with a quantum
circuit, but also $M_a^2,$ $M_a^4,$ $M_a^8,$ and so on, going as far as needed
to obtain a precise enough estimate from the phase estimation procedure.
Here we'll explain how this can be done, and we'll figure out exactly how much
precision is needed later.

Let's start with the operation $M_a$ by itself.
Naturally, because we're working with the quantum circuit model, we'll use
binary notation to encode the numbers between $0$ and $N-1.$
The largest number we need to encode is $N-1,$ so the number of bits we need is
\[
n = \operatorname{lg}(N-1) = \lfloor \log(N-1) \rfloor + 1.
\]

For example, if $N = 21$ we have $n = \operatorname{lg}(N-1) = 5.$
Here's what the encoding of elements of $\mathbb{Z}_{21}$ as binary strings of
length $5$ looks like.
\[
\begin{gathered}
  0  \mapsto 00000\\[1mm]
  1  \mapsto 00001\\[1mm]
  \vdots\\[1mm]
  20 \mapsto 10100
\end{gathered}
\]

And now, here's a precise definition of how $M_a$ is defined as an $n$-qubit
operation.
\[
M_a \vert x\rangle =
\begin{cases}
  \vert ax \; (\textrm{mod}\;N)\rangle & 0\leq x < N\\[1mm]
  \vert x\rangle & N\leq x < 2^n
\end{cases}
\]
The point is that although we only care about how $M_a$ works for
$\vert 0\rangle,\ldots,\vert N-1\rangle,$ we do have to specify how it works
for the remaining $2^n - N$ standard basis states --- and we need to do this in
a way that still gives us a unitary operation.
Defining $M_a$ so that it does nothing to the remaining standard basis states
accomplishes this.

Using the algorithms for integer multiplication and division discussed in the
previous lesson, together with the methodology for reversible, garbage-free
implementations of them, we can build a quantum circuit that performs $M_a,$
for any choice of $a\in\mathbb{Z}_N^{\ast},$ at cost $O(n^2).$
Here is one way this can be done, which mirrors the method described at the end
of Lesson~\ref{lesson:quantum-algorithmic-foundations}
\emph{(Quantum Algorithmic Foundations)} for implementing reversible functions
with quantum circuits.
\begin{enumerate}
\item
  Build a circuit for performing the operation
  \[
  \vert x \rangle \vert y \rangle \mapsto \vert x \rangle \vert y
  \oplus f_a(x)\rangle
  \]
  where
  \[
  f_a(x) =
  \begin{cases}
    ax \; (\textrm{mod}\;N) & 0\leq x < N\\[1mm]
    x & N\leq x < 2^n
  \end{cases}
  \]
  using the method described in the previous lesson.
  This gives us a circuit of size $O(n^2).$

\item
  Swap the two $n$-qubit systems qubit-by-qubit using $n$ swap gates.

\item
  Along similar lines to the first step, build a circuit for the operation
  \[
  \vert x \rangle \vert y \rangle \mapsto \vert x \rangle \bigl\vert y \oplus
  f_{a^{-1}}(x)\bigr\rangle
  \]
  where $a^{-1}$ is the inverse of $a$ in $\mathbb{Z}_N^{\ast}.$
\end{enumerate}

\noindent
By initializing the bottom $n$ qubits and composing the three steps, we obtain
this transformation:
\[
\vert x \rangle \vert 0^n \rangle
\stackrel{\text{step 1}}{\mapsto}
\vert x \rangle \vert f_a(x)\rangle
\stackrel{\text{step 2}}{\mapsto}
\vert f_a(x)\rangle \vert x \rangle
\stackrel{\text{step 3}}{\mapsto}
\vert f_a(x)\rangle \bigl\vert x \oplus f_{a^{-1}}(f_a(x)) \bigr\rangle
= \vert f_a(x)\rangle\vert 0^n \rangle
\]
The method requires workspace qubits, but they're returned to their initialized
state at the end, which allows us to use these circuits for phase estimation.
The total cost of the circuit we obtain is $O(n^2).$

To perform $M_a^2,$ $M_a^4,$ $M_a^8,$ and so on, we can use exactly the same
method, except that we replace $a$ with $a^2,$ $a^4,$ $a^8,$ and so on, as
elements of $\mathbb{Z}_N^{\ast}.$
That is, for any power $k$ we choose, we can create a circuit for $M_a^k$ not
by iterating $k$ times the circuit for $M_a,$ but instead by computing $b = a^k
\in \mathbb{Z}_N^{\ast}$ and then using the circuit for $M_b.$

The computation of powers $a^k \in \mathbb{Z}_N$ is the
\emph{modular exponentiation} problem mentioned in the previous lesson. 
This computation can be done \emph{classically}, using the
\emph{power algorithm} for modular exponentiation mentioned in the previous
lesson.
In fact, we only require \emph{power-of-2} powers of $a,$ in particular
$a^2, a^4, \ldots a^{2^{m-1}} \in \mathbb{Z}_N^{\ast},$ and we can obtain these
powers by iteratively squaring $m-1$ times.
Each squaring can be performed by a Boolean circuit of size $O(n^2).$

In essence, what we're effectively doing here is offloading the problem of
iterating $M_a$ as many as $2^{m-1}$ times to an efficient classical
computation.
And it's good fortune that this is possible!
For an arbitrary choice of a quantum circuit in the phase estimation problem,
this is not likely to be possible --- and in that case the resulting cost for
phase estimation grows \emph{exponentially} in the number of control qubits
$m.$

\subsubsection{Solution given a convenient eigenvector}

To understand how we can solve the order finding problem using phase
estimation, let's start by supposing that we run the phase estimation procedure
on the operation $M_a$ using the eigenvector $\vert\psi_1\rangle.$
Getting our hands on this eigenvector isn't easy, as it turns out, so this
won't be the end of the story --- but it's helpful to start here.

The eigenvalue of $M_a$ corresponding to the eigenvector $\vert \psi_1\rangle$
is
\[
\omega_r = e^{2\pi i \frac{1}{r}}.
\]
That is, $\omega_r = e^{2\pi i \theta}$ for $\theta = 1/r.$
So, if we run the phase estimation procedure on $M_a$ using the eigenvector
$\vert\psi_1\rangle,$ we'll get an approximation to $1/r.$
By computing the reciprocal we'll be able to learn $r$ --- provided that our
approximation is good enough.

In more detail, when we run the phase estimation procedure using $m$ control
qubits, what we obtain is a number $y\in\{0,\ldots,2^m-1\}.$
We then take $y/2^m$ as a guess for $\theta,$ which is $1/r$ in the case at
hand.
To figure out what $r$ is from this approximation, the natural thing to do is
to compute the reciprocal of our approximation and round to the nearest
integer.
\[
\left\lfloor \frac{2^m}{y} + \frac{1}{2} \right\rfloor
\]

For example, let's suppose $r = 6$ and we perform phase estimation on $M_a$
with the eigenvector $\vert\psi_1\rangle$ using $m = 5$ control bits.
The best $5$-bit approximation to $1/r = 1/6$ is $5/32,$ and we have a pretty
good chance (about $68\%$ in this case) to obtain the outcome $y=5$ from phase
estimation.
We have
\[
\frac{2^m}{y} = \frac{32}{5} = 6.4,
\]
and rounding to the nearest integer gives $6,$ which is the correct answer.

On the other hand, if we don't use enough precision, we might not get the right
answer.
For instance, if we take $m = 4$ control qubits in phase estimation, we might
obtain the best $4$-bit approximation to $1/r = 1/6,$ which is $3/16.$
Taking the reciprocal yields
\[
\frac{2^m}{y} = \frac{16}{3} = 5.333 \cdots
\]
and rounding to the nearest integer gives an incorrect answer of $5.$

How much precision do we need to get the right answer?
We know that the order $r$ is an integer, and intuitively speaking what we need
is enough precision to distinguish $1/r$ from nearby possibilities, including
$1/(r+1)$ and $1/(r-1).$
The closest number to $1/r$ that we need to be concerned with is $1/(r+1),$ and
the distance between these two numbers is
\[
\frac{1}{r} - \frac{1}{r+1} = \frac{1}{r(r+1)}.
\]
So, if we want to make sure that we don't mistake $1/r$ for $1/(r+1),$ it
suffices to use enough precision to guarantee that a best approximation $y/2^m$
to $1/r$ is closer to $1/r$ than it is to $1/(r+1).$
If we use enough precision so that
\[
\left\vert
\frac{y}{2^m} - \frac{1}{r}
\right\vert
< \frac{1}{2 r (r+1)},
\]
so that the error is less than half of the distance between $1/r$ and
$1/(r+1),$ then $y/2^m$ will be closer to $1/r$ than to any other possibility,
including $1/(r+1)$ and $1/(r-1).$

We can double-check this as follows.
Suppose that
\[
\frac{y}{2^m} = \frac{1}{r} + \varepsilon
\]
for $\varepsilon$ satisfying
\[
\vert\varepsilon\vert < \frac{1}{2 r (r+1)}.
\]
When we take the reciprocal we obtain
\[
\frac{2^m}{y} = \frac{1}{\frac{1}{r} + \varepsilon}
= \frac{r}{1+\varepsilon r} = r - \frac{\varepsilon r^2}{1+\varepsilon r}.
\]
By maximizing in the numerator and minimizing in the denominator, we can bound
how far away we are from $r$ as follows.
\[
\left\vert
\frac{\varepsilon r^2}{1+\varepsilon r}
\right\vert
\leq \frac{ \frac{r^2}{2 r(r+1)}}{1 - \frac{r}{2r(r+1)}}
= \frac{r}{2 r + 1}
< \frac{1}{2}
\]
We're less than $1/2$ away from $r,$ so as expected we'll get $r$ when we
round.

Unfortunately, because we don't yet know what $r$ is, we can't use it to tell
us how much accuracy we need.
What we can do instead is to use the fact that $r$ must be smaller than $N$ to
ensure that we use enough precision.
In particular, if we use enough accuracy to guarantee that the best
approximation $y/2^m$ to $1/r$ satisfies
\[
\left\vert \frac{y}{2^m} - \frac{1}{r} \right\vert \leq \frac{1}{2N^2},
\]
then we'll have enough precision to correctly determine $r$ when we take the
reciprocal.
Taking $m = 2\operatorname{lg}(N)+1$ ensures that we have a high chance to
obtain an estimation with this precision using the method described previously.
(Taking $m = 2\operatorname{lg}(N)$ is good enough if we're comfortable with a
lower bound of 40\% on the probability of success.)

\subsubsection{General solution}

As we just saw, if we have the eigenvector $\vert \psi_1 \rangle$ of $M_a,$ we
can learn $r$ through phase estimation, so long as we use enough control qubits
to do this with sufficient precision.
Unfortunately, it's not easy to get our hands on the eigenvector
$\vert\psi_1\rangle,$ so we need to figure out how to proceed.

Let's suppose momentarily that we proceed just like above, except with the
eigenvector $\vert\psi_k\rangle$ in place of $\vert\psi_1\rangle,$ for any
choice of $k\in\{0,\ldots,r-1\}$ that we choose to think about.
The result we get from the phase estimation procedure will be an approximation
\[
\frac{y}{2^m} \approx \frac{k}{r}.
\]

Working under the assumption that we don't know either $k$ or $r,$ this might
or might not allow us to identify $r.$
For example, if $k = 0$ we'll get an approximation $y/2^m$ to $0,$ which
unfortunately tells us nothing.
This, however, is an unusual case; for other values of $k,$ we'll at least be
able to learn something about $r.$

We can use an algorithm known as the \emph{continued fraction algorithm} to
turn our approximation $y/2^m$ into nearby fractions --- including $k/r$ if the
approximation is good enough.
We won't explain the continued fraction algorithm here.
Instead, here's a statement of a known fact about this algorithm.

\begin{callout}[title = {Finding fractions with the continued fraction
      algorithm}]
  Given an integer $N\geq 2$ and a real number $\alpha\in(0,1),$ there is at
  most one choice of integers $u,v\in\{0,\ldots,N-1\}$ with $v\neq 0$ and
  $\gcd(u,v)=1$ satisfying
  \[
  \vert \alpha - u/v\vert < \frac{1}{2N^2}.
  \]
  Given $\alpha$ and $N,$ the \emph{continued fraction algorithm} finds $u$ and
  $v,$ or reports that they don't exist.
  This algorithm can be implemented as a Boolean circuit having size
  $O((\operatorname{lg}(N))^3).$
\end{callout}

If we have a very close approximation $y/2^m$ to $k/r,$ and we run the
continued fraction algorithm for $N$ and $\alpha = y/2^m,$ we'll get $u$ and
$v,$ as they're described in the fact.
An analysis of the fact allows us to conclude that
\[
\frac{u}{v} = \frac{k}{r}.
\]
Notice in particular that we don't necessarily learn $k$ and $r,$ we only learn
$k/r$ in lowest terms.

For example, and as we've already noticed, we're not going to learn anything
from $k=0.$
But that's the only value of $k$ where that happens.
When $k$ is nonzero, it might have common factors with $r,$ but the number $v$
we obtain from the continued fraction algorithm must at least divide $r.$

It's far from obvious, but it is true that if we have the ability to learn $u$
and $v$ for $u/v = k/r$ for $k\in\{0,\ldots,r-1\}$ chosen
\emph{uniformly at random}, then we're very likely to be able to recover $r$
after just a few samples.
In particular, if our guess for $r$ is the \emph{least common multiple} of all
the values for the denominator $v$ that we observe, we'll be right with high
probability.
Intuitively speaking, some values of $k$ aren't good because they share common
factors with $r,$ and those common factors are hidden from us when we learn $u$
and $v.$
But \emph{random} choices of $k$ aren't likely to hide factors of $r$ for long,
and the probability that we don't guess $r$ correctly by taking the least
common multiple of the denominators we observe drops exponentially in the
number of samples.

It remains to address the issue of how we get our hands on an eigenvector
$\vert\psi_k\rangle$ of $M_a$ on which to run the phase estimation procedure.
As it turns out, we don't actually need to create them!

What we will do instead is to run the phase estimation procedure on the state
$\vert 1\rangle,$ by which we mean the $n$-bit binary encoding of the number
$1,$ in place of an eigenvector $\vert\psi\rangle$ of $M_a.$
So far, we've only talked about running the phase estimation procedure on a
particular eigenvector, but nothing prevents us from running the procedure on
an input state that isn't an eigenvector of $M_a,$ and that's what we're doing
here with the state $\vert 1\rangle.$
(This isn't an eigenvector of $M_a$ unless $a=1,$ which isn't a choice we'll be
interested in.)

The rationale for choosing the state $\vert 1\rangle$ in place of an
eigenvector of $M_a$ is that the following equation is true.
\[
\vert 1\rangle = \frac{1}{\sqrt{r}} \sum_{k = 0}^{r-1} \vert \psi_k\rangle
\]
One way to verify this equation is to compare the inner products of the two
sides with each standard basis state, using formulas mentioned previously in
the lesson to help to evaluate the results for the right-hand side.
As a consequence, we will obtain precisely the same measurement results as if
we had chosen $k\in\{0,\ldots,r-1\}$ uniformly at random and used
$\vert\psi_k\rangle$ as an eigenvector.

In greater detail, let's imagine that we run the phase estimation procedure
with the state $\vert 1\rangle$ in place of one of the eigenvectors
$\vert\psi_k\rangle.$
After the inverse quantum Fourier transform is performed, this leaves us with
the state
\[
\frac{1}{\sqrt{r}} \sum_{k = 0}^{r-1} \vert \psi_k\rangle \vert \gamma_k\rangle,
\]
where
\[
\vert\gamma_k\rangle =
\frac{1}{2^m} \sum_{y=0}^{2^m - 1}
\sum_{x=0}^{2^m-1} e^{2\pi i x (k/r - y/2^m)} \vert y\rangle.
\]
The vector $\vert\gamma_k\rangle$ represents the state of the top $m$ qubits
after the inverse of the quantum Fourier transform has been performed on them.

So, by virtue of the fact that
$\{\vert\psi_0\rangle,\ldots,\vert\psi_{r-1}\rangle\}$ is an orthonormal set,
we find that a measurement of the top $m$ qubits
yields an approximation $y/2^m$ to the value $k/r$ where $k\in\{0,\ldots,r-1\}$
is chosen uniformly at random.
As we've already discussed, this allows us to learn $r$ with a high degree of
confidence after several independent runs, which was our goal.

\subsubsection{Total cost}

The cost to implement each $M_a^k$, and hence each controlled version of
these unitary operations, is $O(n^2).$
There are $m$ controlled-unitary operations, and we have $m = O(n),$ so the
total cost for the controlled-unitary operations is $O(n^3).$
In addition, we have $m$ Hadamard gates (which contribute $O(n)$ to the cost),
and the inverse quantum Fourier transform contributes $O(n^2)$ to the cost.
Thus, the cost of the controlled-unitary operations dominates the cost of the
entire procedure --- which is therefore $O(n^3).$

In addition to the quantum circuit itself, there are a few classical
computations that need to be performed along the way.
This includes computing the powers $a^k$ in $\mathbb{Z}_N$ for
$k = 2, 4, 8, \ldots, 2^{m-1},$ which are needed to create the
controlled-unitary gates, as well as the continued fraction algorithm that
converts approximations of $\theta$ into fractions.
These computations can be performed by Boolean circuits with a total cost of
$O(n^3).$

As is typical, all of these bounds can be improved using asymptotically fast
algorithms; these bounds assume we're using standard algorithms for basic
arithmetic operations.

\subsection{Factoring by order finding}

The very last thing we need to discuss is how solving the order finding problem
helps us to factor.
This part is completely classical --- it has nothing specifically to do with
quantum computing.

Here's the basic idea.
We want to factorize the number $N,$ and we can do this \emph{recursively}.
Specifically, we can focus on the task of \emph{splitting} $N,$ which means
finding any two integers $b,c\geq 2$ for which $N = bc.$
This isn't possible if $N$ is a prime number, but we can efficiently test to
see if $N$ is prime using a primality testing algorithm first, and if $N$ isn't
prime we'll try to split it.
Once we split $N,$ we can simply recurse on $b$ and $c$ until all of our
factors are prime and we obtain the prime factorization of $N.$

Splitting even integers is easy: we just output $2$ and $N/2.$

It's also easy to split perfect powers, meaning numbers of the form $N = s^j$
for integers $s,j\geq 2,$ just by approximating the roots
$N^{1/2},$ $N^{1/3},$ $N^{1/4},$ etc., and checking nearby integers as suspects
for $s.$
We don't need to go further than $\log(N)$ steps into this sequence, because at
that point the root drops below $2$ and won't reveal additional candidates.

It's good that we can do both of these things because order finding won't help
us to factor even numbers or \emph{prime} powers, where the number $s$
happens to be prime.
If $N$ is odd and not a prime power, however, order finding allows us to
split~$N$ through the following algorithm.

\begin{callout}[title={Probabilistic algorithm to split an odd, composite,
      non-prime-power integer}]
  \begin{enumerate}
  \item
    Randomly choose $a\in\{2,\ldots,N-1\}.$
  \item
    Compute $d=\gcd(a,N).$
  \item
    If $d > 1$ then output $b = d$ and $c = N/d$ and stop. Otherwise continue
    to the next step knowing that $a\in\mathbb{Z}_N^{\ast}.$
  \item
    Let $r$ be the order of $a$ modulo $N.$ (Here's where we need order
    finding.)
  \item
    If $r$ is even:
    \begin{enumerate}
    \item[5.1] Compute $x = a^{r/2} - 1$ modulo $N$.
    \item[5.2] Compute $d = \gcd(x,N).$
    \item[5.3] If $d>1$ then output $b=d$ and $c = N/d$ and stop.
    \end{enumerate}
  \item
    If this point is reached, the algorithm has failed to find a factor
    of $N.$
  \end{enumerate}
\end{callout}

A run of this algorithm may fail to find a factor of $N.$
Specifically, this happens in two situations:
\begin{itemize}
  \item The order of $a$ modulo $N$ is odd.
  \item The order of $a$ modulo $N$ is even and
    $\gcd\bigl(a^{r/2} - 1, N\bigr) = 1.$
\end{itemize}
Using basic number theory it can be proved that, for a random choice of $a,$
with probability at least $1/2$ neither of these events happen.
In fact, the probability that either event happens is at most $2^{-(m-1)}$ for
$m$ being the number of distinct prime factors of $N,$ which is why the
assumption that $N$ is not a prime power is needed.
(The assumption that $N$ is odd is also required for this fact to be true.)

This means that each run has at least a 50\% chance to split $N.$
Therefore, if we run the algorithm $t$ times, randomly choosing $a$ each time,
we'll succeed in splitting $N$ with probability at least $1 - 2^{-t}.$

The basic idea behind the algorithm is as follows.
If we have a choice of $a$ for which the order $r$ of $a$ modulo $N$ is even,
then $r/2$ is an integer and we can consider the numbers
\[
a^{r/2} - 1\; (\textrm{mod}\; N) \quad \text{and} \quad a^{r/2} + 1\;
(\textrm{mod}\; N).
\]
Using the formula $Z^2 - 1 = (Z+1)(Z-1),$ we conclude that
\[
\bigl(a^{r/2} - 1\bigr) \bigl(a^{r/2} + 1\bigr) = a^r - 1.
\]
Now, we know that $a^r \; (\textrm{mod}\; N) = 1$ by the definition of the
order --- which is another way of saying that $N$ evenly divides $a^r - 1.$
That means that $N$ evenly divides the product
\[
\bigl(a^{r/2} - 1\bigr) \bigl(a^{r/2} + 1\bigr).
\]
For this to be true, all of the prime factors of $N$ must also be prime factors
of $a^{r/2} - 1$ or $a^{r/2} + 1$ (or both) --- and for a random selection of
$a$ it turns out to be unlikely that all of the prime factors of $N$ will
divide one of the terms and none will divide the other.
Otherwise, so long as some of the prime factors of $N$ divide the first term
and some divide the second term, we'll be able to find a non-trivial factor of
$N$ by computing the GCD with the first term.


\lesson{Grover's Algorithm}
\label{lesson:grover-algorithm}

Grover's algorithm is a quantum algorithm for so-called \emph{unstructured
search} problems that offers a \emph{quadratic} improvement over classical
algorithms.
What this means is that Grover's algorithm requires a number of operations on
the order of the \emph{square-root} of the number of operations required to
solve unstructured search classically --- which is equivalent to saying that
classical algorithms for unstructured search must have a cost at least on the
order of the \emph{square} of the cost of Grover's algorithm.

Grover's algorithm, together with its extensions and underlying methodology,
turn out to be broadly applicable, leading to a quadratic advantage for many
interesting computational tasks that may not initially look like unstructured
search problems on the surface.

While the broad applicability of Grover's searching technique is compelling, it
should be acknowledged here at the start of the lesson that the quadratic
advantage it offers seems unlikely to lead to a practical advantage of
quantum over classical computing any time soon.
Classical computing hardware is much more advanced than quantum computing
hardware --- and the quadratic quantum-over-classical advantage offered by
Grover's algorithm is sure to be washed away by the staggering clock speeds of
modern classical computers for any unstructured search problem that could
feasibly be run any time soon.

As quantum computing technology advances, however, Grover's algorithm could
have potential.
Indeed, some of the most important and impactful classical algorithms ever
discovered, including the fast Fourier transform and fast sorting (e.g.,
quicksort and merge sort), offer slightly less than a quadratic advantage over
naive approaches to the problems they solve.
The key difference here, of course, is that an entirely new technology (meaning
quantum computing) is required to run Grover's algorithm.
While this technology is still very much in its infancy in comparison to
classical computing, we should not be so quick to underestimate the potential
of technological advances that could allow a quadratic advantage of quantum
computing to one day offer tangible practical benefits.

\section{Unstructured search}

We'll begin with a description of the problem that Grover's algorithm solves.
As usual, we'll let $\Sigma = \{0,1\}$ denote the binary alphabet throughout
this discussion.

Suppose that
\[
f:\Sigma^n \rightarrow \Sigma
\]
is a function from binary strings of length $n$ to bits.
We'll assume that we can compute this function efficiently, but otherwise it's
arbitrary and we can't rely on it having a special structure or specific
implementation that suits our needs.

What Grover's algorithm does is to search for a string $x\in\Sigma^n$ for which
$f(x) = 1.$
We'll refer to strings like this as \emph{solutions} to the searching problem.
If there are multiple solutions, then any one of them is considered to be a
correct output, and if there are no solutions, then a correct answer requires
that we report that there are no solutions.

This task is described as an \emph{unstructured} search problem because we
can't rely on $f$ having any particular structure to make it easy.
We're not searching an ordered list, or within some data structure specifically
designed to facilitate searching, we're essentially looking for a needle in a
haystack.

Intuitively speaking, we might imagine that we have an extremely complicated
Boolean circuit that computes $f,$ and we can easily run this circuit on a
selected input string if we choose.
But because the circuit is so complicated, we have no hope of making sense of
the circuit by examining it (beyond having the ability to evaluate it on
selected input strings).

One way to perform this searching task classically is to simply iterate through
all of the strings $x\in\Sigma^n,$ evaluating $f$ on each one to check whether
or not it is a solution.
Hereafter, let's write
\[
N = 2^n
\]
for the sake of convenience.
There are $N$ strings in $\Sigma^n,$ so iterating through all of them requires
$N$ evaluations of $f.$
Operating under the assumption that we're limited to evaluating $f$ on chosen
inputs, this is the best we can do with a deterministic algorithm if we want to
guarantee success.

With a probabilistic algorithm, we might hope to save time by randomly choosing
input strings to $f,$ but we'll still require $O(N)$ evaluations of $f$ if we
want this method to succeed with high probability.

Grover's algorithm solves this search problem with high probability with just
$O(\sqrt{N})$ evaluations of $f.$
To be clear, these function evaluations must happen \emph{in superposition},
similar to the query algorithms discussed in
Lesson~\ref{lesson:quantum-query-algorithms}
\emph{(Quantum Query Algorithms)}, including Deutsch's algorithm, the
Deutsch--Jozsa algorithm, and Simon's algorithm.
Unlike those algorithms, Grover's algorithm takes an iterative approach: it
evaluates $f$ on superpositions of input strings and intersperses these
evaluations with other operations that have the effect of creating interference
patterns, leading to a solution with high probability (if one exists) after
$O(\sqrt{N})$ iterations.

\subsection{Formal problem statement}

We'll formalize the problem that Grover's algorithm solves using the query
model of computation.
That is, we'll assume that we have access to the function
$f:\Sigma^n\rightarrow\Sigma$ through a query gate defined in the usual way:
\[
U_f \bigl( \vert a\rangle \vert x\rangle \bigr) = \vert a \oplus f(x) \rangle
\vert x \rangle
\]
for every $x\in\Sigma^n$ and $a\in\Sigma.$
This is the action of $U_f$ on standard basis states, and its action in general
is determined by linearity.

As was discussed in
Lesson~\ref{lesson:quantum-algorithmic-foundations}
\emph{(Quantum Algorithmic Foundations)}, if we
have a Boolean circuit for computing $f,$ we can transform that Boolean circuit
description into a quantum circuit implementing $U_f$ (using some number of
workspace qubits that start and end the computation in the $\vert 0\rangle$
state).
So, although we're using the query model to formalize the problem that Grover's
algorithm solves, it is not limited to this model; we can run Grover's
algorithm on any function $f$ for which we have a Boolean circuit.

Here's a precise statement of the problem, which is named \emph{search} because
we're searching for a solution, meaning a string $x$ that causes $f$ to
evaluate to $1.$

\begin{callout}[title={Search}]
  \begin{problem}
    \Input{A function $f:\Sigma^n\rightarrow\Sigma$.}
    \Output{A string $x\in\Sigma^n$ satisfying $f(x) = 1,$ or
      ``no solution'' if no such string $x$ exists.}
  \end{problem}
\end{callout}

\noindent
Notice that this is \emph{not} a promise problem --- the function $f$ is
arbitrary.
It will, however, be helpful to consider the following promise variant of the
problem, where we're guaranteed that there's exactly one solution.
This problem appeared as an example of a promise problem in
Lesson~\ref{lesson:quantum-query-algorithms}
\emph{(Quantum Query Algorithms)}.

\begin{callout}[title={Unique search}]
  \begin{problem}
    \Input{A function of the form $f:\Sigma^n \rightarrow \Sigma$.}
    \Promise{There is exactly one string $z\in\Sigma^n$ for which
      $f(z) = 1,$ with\linebreak
      $f(x) = 0$ for all strings $x\neq z$.}
    \Output{The string $z$.}
  \end{problem}
\end{callout}

Also notice that the \emph{or} problem mentioned in the same lesson is closely
related to search.
For that problem, the goal is simply to determine whether or not a solution
exists, as opposed to actually finding a solution.

\section{Description of Grover's algorithm}

In this section, we'll describe Grover's algorithm.
We'll begin by discussing \emph{phase query gates} and how to build them,
followed by the description of Grover's algorithm itself.
Finally, we'll briefly discuss how this algorithm is naturally applied to
searching.

\subsection{Phase query gates}

Grover's algorithm makes use of operations known as \emph{phase query gates}.
In contrast to an ordinary query gate $U_f,$ defined for a given function $f$
in the usual way described previously, a phase query gate for the function $f$
is defined as
\[
Z_f \vert x\rangle = (-1)^{f(x)} \vert x\rangle
\]
for every string $x\in\Sigma^n.$

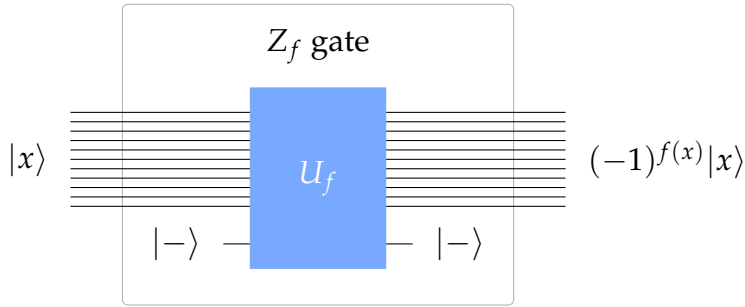
\begin{figure}[!ht]
  \begin{center}
    \begin{tikzpicture}[
        scale=1.25,
        gate/.style={%
          inner sep = 0,
          fill = CircuitBlue,
          draw = CircuitBlue,
          text = white,
          minimum size = 10mm}
      ]
      
      \node (in_top) at (-2.75,0.4) {};
      \node (in_bottom) at (-1.125,-0.5) {};
      
      \node (out_top) at (2.75,0.4) {};
      \node (out_bottom) at (1.125,-0.5) {};

      \node[
        draw = black!30,
        rounded corners=2pt,
        minimum width = 52mm,
        minimum height = 40mm
      ] at (0,0.45) {}; 

      \node[anchor=east] at (in_top) {$\ket{x}$};
      \node[anchor=east] at (in_bottom) {$\ket{-}$};
      
      \node[anchor=west] at (out_top) {$(-1)^{f(x)}\ket{x}$};
      \node[anchor=west] at (out_bottom) {$\ket{-}$};
      
      \foreach \y in {-5,-4,...,5} {
        \draw ([yshift=\y mm]in_top.east) --
        ([yshift=\y mm]out_top.west) {};
      }
      
      \draw (in_bottom.east) -- (out_bottom.west) {};
      
      \node[gate, minimum height=24mm, minimum width=18mm]
      (U) at (0,0.2) {$U_f$};
      
      \node at (0,1.6) {$Z_f$ gate};

    \end{tikzpicture}
  \end{center}
  \caption{An implementation of a phase query gate $Z_f$ using a standard
    query gate $U_f$.}
  \label{fig:Zf}
\end{figure}

The operation $Z_f$ can be implemented using one query gate $U_f$ as
Figure~\ref{fig:Zf} suggests.
The implementation makes use of the phase kickback phenomenon, and requires
that one workspace qubit, initialized to a $\vert -\rangle$ state, is made
available.
This qubit remains in the $\vert - \rangle$ state after the implementation has
completed, and can be reused (to implement subsequent $Z_f$ gates, for
instance) or simply discarded.

In addition to the operation $Z_f,$ we will also make use of a phase query gate
for the $n$-bit OR function, which is defined as follows for each string
$x\in\Sigma^n.$
\[
\mathrm{OR}(x) =
\begin{cases}
  0 & x = 0^n\\[0.5mm]
  1 & x \neq 0^n
\end{cases}
\]
Explicitly, the phase query gate for the $n$-bit OR function operates like
this:
\[
Z_{\mathrm{OR}} \vert x\rangle
= \begin{cases}
  \vert x\rangle & x = 0^n \\[0.5mm]
- \vert x\rangle & x \neq 0^n.
\end{cases}
\]
To be clear, this is how $Z_{\mathrm{OR}}$ operates on standard basis states;
its behavior on arbitrary states is determined from this expression by
linearity.

The operation $Z_{\mathrm{OR}}$ can be implemented as a quantum circuit by
beginning with a Boolean circuit for the OR function, then constructing a
$U_{\mathrm{OR}}$ operation (i.e., a standard query gate for the $n$-bit OR
function) using the procedure described in
Lesson~\ref{lesson:quantum-algorithmic-foundations}
\emph{(Quantum Algorithmic Foundations)}, and finally a $Z_{\mathrm{OR}}$
operation using the phase kickback phenomenon as described above.
Notice that the operation $Z_{\mathrm{OR}}$ has no dependence on the function
$f$ and can therefore be implemented by a quantum circuit having no query
gates.

\subsection{Description of the algorithm}

Now that we have the two operations $Z_f$ and $Z_{\mathrm{OR}},$ we can
describe Grover's algorithm.

The algorithm refers to a number $t,$ which is the number of \emph{iterations}
it performs (as well as the number of \emph{queries} to the function $f$ it
requires).
This number $t$ isn't specified by Grover's algorithm as we're describing it,
and we'll discuss later in the lesson how it can be chosen.

\begin{callout}[title={Grover's algorithm}]
  \begin{enumerate}
  \item
    Initialize an $n$ qubit register $\mathsf{Q}$ to the all-zero state
    $\vert 0^n \rangle$ and then apply a Hadamard operation to each qubit of
    $\mathsf{Q}.$
  \item
    Apply $t$ times the unitary operation
    $G = H^{\otimes n} Z_{\mathrm{OR}} H^{\otimes n} Z_f$ to the
    register~$\mathsf{Q}$.
  \item
    Measure the qubits of $\mathsf{Q}$ with respect to standard basis
    measurements and output the resulting string.
  \end{enumerate}
\end{callout}

\begin{figure}[b]
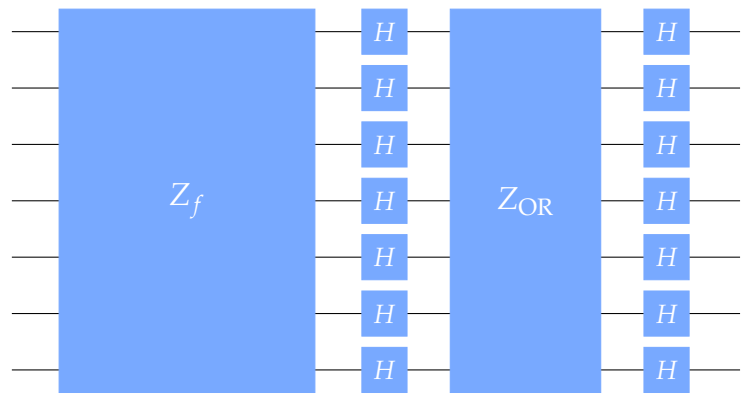

  \begin{center}

  \end{center}
  \caption{A quantum circuit implementation of the Grover operation
    on $7$ qubits.}
  \label{fig:Grover-operation}
\end{figure}

\noindent
The operation $G = H^{\otimes n} Z_{\mathrm{OR}} H^{\otimes n} Z_f$ iterated in
step 2 will be called the \emph{Grover operation} throughout the remainder of
this lesson.
Figure~\ref{fig:Grover-operation} shows a quantum circuit representation of the
Grover operation when $n=7$.

In this figure, the $Z_f$ operation is depicted as being larger than
$Z_{\mathrm{OR}}$ as an informal visual clue to suggest that it is likely to be
the more costly operation.
In particular, when we're working within the query model, $Z_f$ requires one
query while $Z_{\mathrm{OR}}$ requires no queries.
If instead we have a Boolean circuit for the function $f,$ and then convert it
to a quantum circuit for $Z_f,$ we can reasonably expect that the resulting
quantum circuit will be larger and more complicated than one for
$Z_{\mathrm{OR}}.$

\begin{figure}[t]
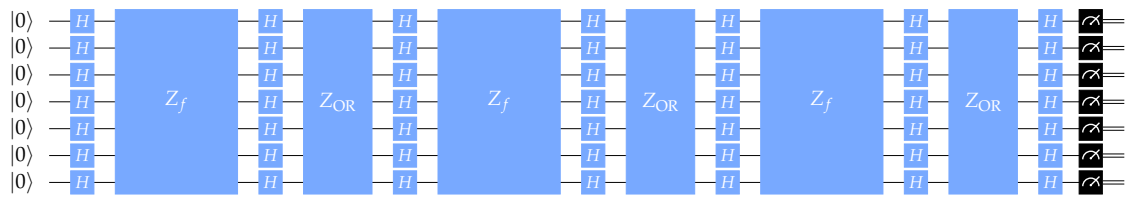

  \begin{center}
   \scalebox{0.65}{%

   }
  \end{center}
  \caption{A quantum circuit running Grover's algorithm for $3$ iterations on
    $7$ qubits.}
  \label{fig:Grover-circuit}
\end{figure}

Figure~\ref{fig:Grover-circuit} shows a quantum circuit for the entire
algorithm when $n=7$ and $t=3.$
For larger values of $t$ we can simply insert additional instances of the
Grover operation immediately before the measurements.

\subsection{Application to search}

Grover's algorithm can be applied to the search problem as follows:

\begin{itemize}
\item
  Choose the number $t$ in step 2. (This is discussed later in the lesson.)
\item
  Run Grover's algorithm on the function $f,$ using whatever choice we made for
  $t,$ to obtain a string $x\in\Sigma^n.$
\item
  Query the function $f$ on the string $x$ to see if it's a valid solution:
  \begin{itemize}
  \item
    If $f(x) = 1,$ then we have found a solution, so we can stop and
    output $x.$
  \item
    Otherwise, if $f(x) = 0,$ then we can either run the procedure again,
    possibly with a different choice for $t,$ or we can decide to give up and
    output ``no solution.''
  \end{itemize}
\end{itemize}

Once we've analyzed how Grover's algorithm works, we'll see that by taking
$t = O(\sqrt{N}),$ we obtain a solution to our search problem (if one exists)
with high probability.

\pagebreak
\section{Analysis}

Now we'll analyze Grover's algorithm to understand how it works.
We'll start with what could be described as a \emph{symbolic} analysis, where
we calculate how the Grover operation $G$ acts on certain states, and then
we'll tie this symbolic analysis to a \emph{geometric} picture that's helpful
for visualizing how the algorithm works.

\subsection{Solutions and non-solutions}

Let's start by defining two sets of strings.
\[
\begin{aligned}
  A_0 &= \bigl\{ x\in\Sigma^n : f(x) = 0\bigr\} \\
  A_1 &= \bigl\{ x\in\Sigma^n : f(x) = 1\bigr\}
\end{aligned}
\]
The set $A_1$ contains all of the solutions to our search problem while $A_0$
contains the strings that aren't solutions (which we can refer to as
\emph{non-solutions} when it's convenient).
These two sets satisfy $A_0 \cap A_1 = \varnothing$ and
$A_0 \cup A_1 = \Sigma^n,$ which is to say that this is a \emph{bipartition} of
$\Sigma^n.$

Next we'll define two unit vectors representing uniform superpositions over the
sets of solutions and non-solutions.
\[
\begin{aligned}
  \vert A_0\rangle &=
  \frac{1}{\sqrt{\vert A_0\vert}} \sum_{x\in A_0} \vert x\rangle \\
  \vert A_1\rangle &=
  \frac{1}{\sqrt{\vert A_1\vert}} \sum_{x\in A_1} \vert x\rangle
\end{aligned}
\]
Formally speaking, each of these vectors is only defined when its corresponding
set is nonempty, but hereafter we're going to focus on the case that neither
$A_0$ nor $A_1$ is empty.
The cases that $A_0 = \varnothing$ and $A_1 = \varnothing$ are easily handled
separately, and we'll do that later.

As an aside, the notation being used here is common: any time we have a finite
and nonempty set $S,$ we can write $\vert S\rangle$ to denote the quantum state
vector that's uniform over the elements of $S.$

Let's also define $\vert u \rangle$ to be a \emph{uniform} quantum state over
all $n$-bit strings:
\[
\vert u\rangle = \frac{1}{\sqrt{N}} \sum_{x\in\Sigma^n} \vert x\rangle.
\]
Notice that
\[
\vert u\rangle
= \sqrt{\frac{\vert A_0 \vert}{N}} \vert A_0\rangle
+ \sqrt{\frac{\vert A_1 \vert}{N}} \vert A_1\rangle.
\]
We also have that $\vert u\rangle = H^{\otimes n} \vert 0^n \rangle,$ so $\vert
u\rangle$ represents the state of the register $\mathsf{Q}$ after the
initialization in step 1 of Grover's algorithm.

This implies that just before the iterations of $G$ happen in step 2, the state
of $\mathsf{Q}$ is contained in the two-dimensional vector space spanned by
$\vert A_0\rangle$ and $\vert A_1\rangle,$ and moreover the coefficients of
these vectors are real numbers.
As we will see, the state of $\mathsf{Q}$ will always have these properties ---
meaning that the state is a real linear combination of $\vert A_0\rangle$ and
$\vert A_1\rangle$ --- after any number of iterations of the operation $G$ in
step 2.

\subsection{An observation about the Grover operation}

We'll now turn our attention to the Grover operation
\[
G = H^{\otimes n} Z_{\mathrm{OR}} H^{\otimes n} Z_f,
\]
beginning with an interesting observation about it.

Imagine for a moment that we replaced the function $f$ by the composition of
$f$ with the NOT function --- or, in other words, the function we get by
flipping the output bit of $f.$
We'll call this new function $g,$ and we can express it using symbols in a few
alternative ways.
\[
g(x) = \neg f(x) = 1 \oplus f(x) = 1 - f(x) =
\begin{cases}
1 & f(x) = 0\\[1mm]
0 & f(x) = 1
\end{cases}
\]

Notice that
\[
(-1)^{g(x)} = (-1)^{1 \oplus f(x)} = - (-1)^{f(x)}
\]
for every string $x\in\Sigma^n,$ and therefore
\[
Z_g = - Z_f.
\]
This means that if we were to substitute the function $f$ with the function
$g,$ Grover's algorithm wouldn't function any differently --- because the
states we obtain from the algorithm in the two cases are necessarily equivalent
up to a global phase.

This isn't a problem!
Intuitively speaking, the algorithm doesn't care which strings are solutions
and which are non-solutions --- it only needs to be able to \emph{distinguish}
solutions and non-solutions to operate correctly.

\subsection{Action of the Grover operation}

Now let's consider the action of $G$ on the quantum state vectors
$\vert A_0\rangle$ and $\vert A_1\rangle.$
First, let's observe that the operation $Z_f$ has a simple action on
$\vert A_0\rangle$ and $\vert A_1\rangle.$
\[
\begin{aligned}
  Z_f \vert A_0\rangle & = \vert A_0\rangle \\[1mm]
  Z_f \vert A_1\rangle & = -\vert A_1\rangle
\end{aligned}
\]

Second, we have the operation $H^{\otimes n} Z_{\mathrm{OR}} H^{\otimes n}.$
The operation $Z_{\mathrm{OR}}$ is defined as
\[
Z_{\mathrm{OR}} \vert x\rangle
= \begin{cases}
  \vert x\rangle & x = 0^n \\[1mm]
  -\vert x\rangle & x \neq 0^n,
\end{cases}
\]
again for every string $x\in\Sigma^n,$ and a convenient alternative way to
express this operation is like this:
\[
Z_{\mathrm{OR}} = 2 \vert 0^n \rangle \langle 0^n \vert - \mathbb{I}.
\]
A simple way to verify that this expression agrees with the definition of
$Z_{\mathrm{OR}}$ is to evaluate its action on standard basis states.
The operation $H^{\otimes n} Z_{\mathrm{OR}} H^{\otimes n}$ can therefore be
written like this:
\[
H^{\otimes n} Z_{\mathrm{OR}} H^{\otimes n} = 2 H^{\otimes n} \vert 0^n \rangle
\langle 0^n \vert H^{\otimes n} - \mathbb{I} = 2 \vert u \rangle \langle u
\vert - \mathbb{I},
\]
using the same notation $\vert u \rangle$ that we used above for the uniform
superposition over all $n$-bit strings.

And now we have what we need to compute the action of $G$ on
$\vert A_0\rangle$ and $\vert A_1\rangle.$
First let's compute the action of $G$ on $\vert A_0\rangle.$
\[
\begin{aligned}
  G \vert A_0 \rangle
  & = \bigl( 2 \vert u\rangle \langle u \vert - \mathbb{I}\bigr)
  Z_f \vert A_0\rangle \\
  & = \bigl( 2 \vert u\rangle \langle u \vert - \mathbb{I}\bigr)
  \vert A_0\rangle \\
  & = 2 \sqrt{\frac{\vert A_0\vert}{N}} \vert u\rangle -\vert A_0 \rangle\\
  & = 2 \sqrt{\frac{\vert A_0\vert}{N}} \biggl(
  \sqrt{\frac{\vert A_0\vert}{N}} \vert A_0\rangle
  + \sqrt{\frac{\vert A_1\vert}{N}} \vert A_1\rangle\biggr)
  -\vert A_0 \rangle \\
  & = \biggl( \frac{2\vert A_0\vert}{N} - 1\biggr) \vert A_0 \rangle
  + \frac{2 \sqrt{\vert A_0\vert \cdot \vert A_1\vert}}{N} \vert A_1\rangle\\
  & = \frac{\vert A_0\vert - \vert A_1\vert}{N} \vert A_0 \rangle
  + \frac{2 \sqrt{\vert A_0\vert \cdot \vert A_1\vert}}{N} \vert A_1\rangle
\end{aligned}
\]
And second, let's compute the action of $G$ on $\vert A_1\rangle.$
\[
\begin{aligned}
  G \vert A_1 \rangle
  & = \bigl( 2 \vert u\rangle \langle u \vert - \mathbb{I} \bigr) Z_f \vert
  A_1\rangle \\
  & = - \bigl( 2 \vert u\rangle \langle u \vert - \mathbb{I} \bigr) \vert
  A_1\rangle \\
  & = - 2 \sqrt{\frac{\vert A_1\vert}{N}} \vert u\rangle + \vert A_1 \rangle\\
  & = - 2 \sqrt{\frac{\vert A_1\vert}{N}}
  \biggl(\sqrt{\frac{\vert A_0\vert}{N}} \vert A_0\rangle
  + \sqrt{\frac{\vert A_1\vert}{N}} \vert A_1\rangle\biggr) + \vert A_1
  \rangle \\
  & = - \frac{2 \sqrt{\vert A_1\vert \cdot \vert A_0\vert}}{N} \vert A_0
  \rangle + \biggl( 1 - \frac{2\vert A_1\vert}{N} \biggr) \vert A_1 \rangle \\
  & = - \frac{2 \sqrt{\vert A_1\vert \cdot \vert A_0\vert}}{N} \vert A_0
  \rangle + \frac{\vert A_0\vert - \vert A_1\vert}{N} \vert A_1 \rangle
\end{aligned}
\]
In both cases we're using the equation
\[
\vert u\rangle
= \sqrt{\frac{\vert A_0 \vert}{N}} \vert A_0\rangle
+ \sqrt{\frac{\vert A_1 \vert}{N}} \vert A_1\rangle
\]
along with the expressions
\[
\langle u \vert A_0\rangle = \sqrt{\frac{\vert A_0 \vert}{N}}
\qquad\text{and}\qquad
\langle u \vert A_1\rangle = \sqrt{\frac{\vert A_1 \vert}{N}}
\]
that follow.
In summary, we have
\[
\begin{aligned}
  G \vert A_0 \rangle
  & = \frac{\vert A_0\vert - \vert A_1\vert}{N} \vert A_0 \rangle
  + \frac{2 \sqrt{\vert A_0\vert \cdot \vert A_1\vert}}{N} \vert A_1
  \rangle\\[2mm]
  G \vert A_1 \rangle
  & = - \frac{2 \sqrt{\vert A_1\vert \cdot \vert A_0\vert}}{N} \vert A_0
  \rangle + \frac{\vert A_0\vert - \vert A_1\vert}{N} \vert A_1 \rangle.
\end{aligned}
\]

As we already noted, the state of $\mathsf{Q}$ just prior to step 2 is
contained in the two-dimensional space spanned by $\vert A_0\rangle$ and $\vert
A_1\rangle,$ and we have just established that $G$ maps any vector in this
space to another vector in the same space.
This means that, for the sake of the analysis, we can focus our attention
exclusively on this subspace.

To better understand what's happening within this two-dimensional space, let's
express the action of $G$ on this space as a matrix,
\[
M = 
,
\]
whose first and second rows/columns correspond to $\vert A_0\rangle$ and
$\vert A_1\rangle,$ respectively. 
So far in this course, we've always connected the rows and columns of matrices
with the classical states of a system, but matrices can also be used to
describe the actions of linear mappings on different bases like we have here.

While it isn't at all obvious at first glance, the matrix $M$ is what we obtain
by \emph{squaring} a simpler-looking matrix.
\[
%
\]
is a \emph{rotation matrix}, which we can alternatively express as
\[
%
\]
for
\[
\theta = \sin^{-1}\biggl(\sqrt{\frac{\vert A_1\vert}{N}}\biggr).
\]
This angle $\theta$ is going to play a very important role in the analysis that
follows, so it's worth stressing its importance here as we see it for the first
time.

In light of this expression of this matrix, we observe that
\[
M = %
.
\]
This is because rotating by the angle $\theta$ two times is equivalent to
rotating by the angle $2\theta.$
Another way to see this is to make use of the alternative expression
\[
\theta = \cos^{-1}\biggl(\sqrt{\frac{\vert A_0\vert}{N}}\biggr),
\]
together with the \emph{double angle} formulas from trigonometry:
\[
\begin{aligned}
  \cos(2\theta) & = \cos^2(\theta) - \sin^2(\theta)\\[1mm]
  \sin(2\theta) & = 2 \sin(\theta)\cos(\theta).
\end{aligned}
\]

In summary, the state of the register $\mathsf{Q}$ at the start of step 2 is
\[
\vert u\rangle
= \sqrt{\frac{\vert A_0\vert}{N}} \vert A_0\rangle
+ \sqrt{\frac{\vert A_1\vert}{N}} \vert A_1\rangle
= \cos(\theta) \vert A_0\rangle + \sin(\theta) \vert A_1\rangle,
\]
and the effect of applying $G$ to this state is to rotate it by an angle
$2\theta$ within the space spanned by $\vert A_0\rangle$ and $\vert
A_1\rangle.$
So, for example, we have
\[
\begin{aligned}
  G \vert u \rangle &= \cos(3\theta) \vert A_0\rangle
  + \sin(3\theta) \vert A_1\rangle\\[1mm]
  G^2 \vert u \rangle &= \cos(5\theta) \vert A_0\rangle
  + \sin(5\theta) \vert A_1\rangle\\[1mm]
  G^3 \vert u \rangle &= \cos(7\theta) \vert A_0\rangle
  + \sin(7\theta) \vert A_1\rangle
\end{aligned}
\]
and in general
\[
G^t \vert u \rangle
= \cos\bigl((2t + 1)\theta\bigr) \vert A_0\rangle
+ \sin\bigl((2t + 1)\theta\bigr) \vert A_1\rangle.
\]

\subsection{Geometric picture}

Now let's connect the analysis we just went through to a geometric picture.
The idea is that the operation $G$ is the product of two \emph{reflections},
$Z_f$ and $H^{\otimes n} Z_{\mathrm{OR}} H^{\otimes n}.$
And the net effect of performing two reflections is to perform a
\emph{rotation}.

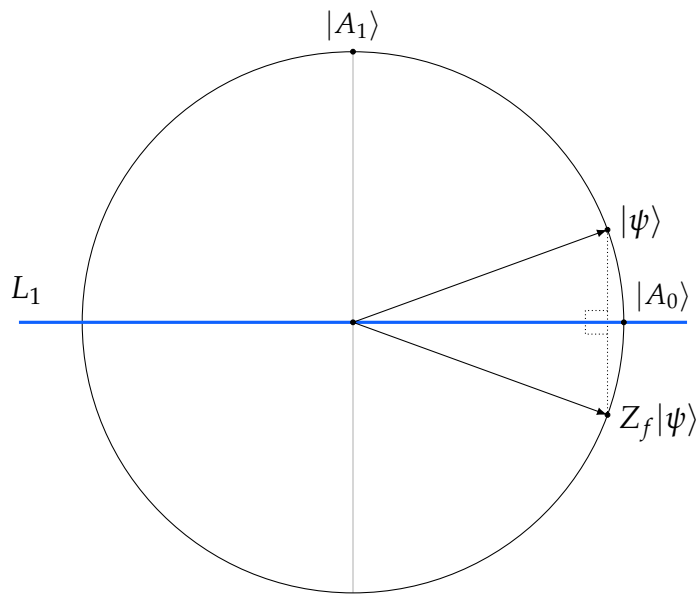
\begin{figure}[p]
  \begin{center}
   \begin{tikzpicture}[>=latex, scale=1.2]
          
     \newcommand\CircleRadius{3cm}

     \draw[black!30] (-3cm,0cm) -- (3cm,0cm);
     \draw[black!30] (0cm,-3cm) -- (0cm,3cm);                    
     
     \node (Origin) at (0,0) {};
     \draw (Origin) circle (\CircleRadius);
     
     \node (L0) at (0:3.7) {};
     \node (L1) at (180:3.7) {};
     
     \node (P) at (20:3cm) {};
     \node (Q) at (-20:3cm) {};
     
     \node (A0) at (0:3) {};
     \node (A1) at (90:3) {};
     
     \node[anchor = south, yshift = 1mm, xshift = 1mm] at (L1) {$L_1$};
          
     \node (angle1) at (0:2.82cm) {};
          
     \node[
       anchor = east,
       minimum width=.15cm,
       minimum height=.3cm,
       rotate=0
     ] (S1) at (angle1) {};
     
     \draw[densely dotted] (P.center) -- (Q.center);
     \draw[densely dotted] (S1.north east) -- (S1.north west)
     -- (S1.south west) -- (S1.south east);
     
     \draw[very thick, DataColor0] (L0.center) -- (L1.center);

     \filldraw (Origin) circle (0.75pt);
     \node[anchor = west, yshift=1mm] at (P) {$\ket{\psi}$};
     \filldraw (P) circle (0.75pt);
     \draw[->] (Origin.center) -- (P.center);
    
     \node[anchor = west, yshift = -1mm] at (Q) {$Z_f\ket{\psi}$};
     \filldraw (Q) circle (0.75pt);
     \draw[->] (Origin.center) -- (Q.center);

     \node[anchor = south west] at (A0) {\small $\ket{A_0}$};
     \filldraw (A0) circle (0.75pt);
     
     \node[anchor = south] at (A1) {\small $\ket{A_1}$};
     \filldraw (A1) circle (0.75pt);

     \node at (5,0) {};
     \node at (-5,0) {};
     
   \end{tikzpicture}
  \end{center}
  \caption{The action of $Z_f$, which reflects about the line $L_1$,
    on a vector~$\ket{\psi}$ that is a real linear combination of $\ket{A_0}$
    and $\ket{A_1}$.}
  \label{fig:reflection1}
\end{figure}

Let's start with $Z_f.$
As we already observed previously, we have
\[
\begin{aligned}
  Z_f \vert A_0\rangle & = \vert A_0\rangle \\[1mm]
  Z_f \vert A_1\rangle & = -\vert A_1\rangle.
\end{aligned}
\]
Within the two-dimensional vector space spanned by $\vert A_0\rangle$ and
$\vert A_1\rangle,$ this is a \emph{reflection} about the line parallel to
$\vert A_0\rangle,$ which we'll call $L_1.$
Figure~\ref{fig:reflection1} illustrates the action of this reflection on a
hypothetical unit vector $\vert\psi\rangle,$ which we're assuming is a real
linear combination of $\vert A_0\rangle$ and $\vert A_1\rangle.$

\begin{figure}[p]
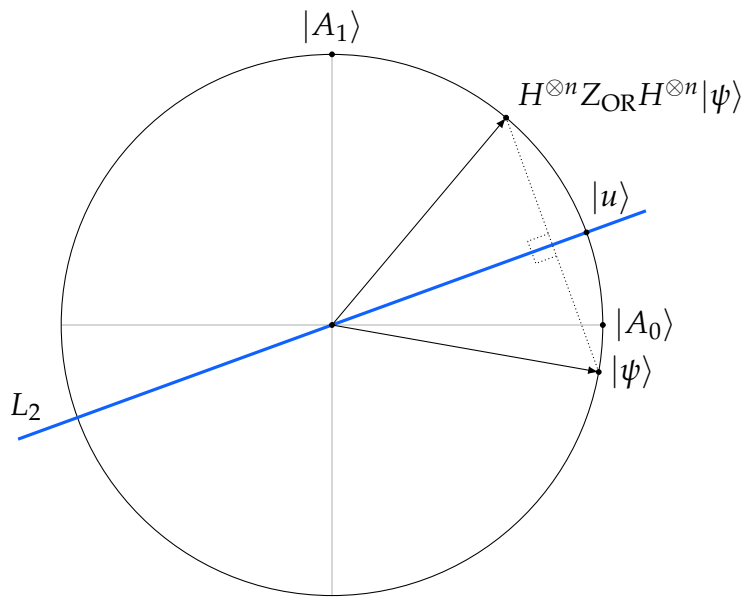

  \begin{center}

  \end{center}
  \caption{The action of $H^{\otimes n} Z_{\mathrm{OR}} H^{\otimes n}$, which
    reflects about the line $L_2$, on a vector~$\ket{\psi}$ that is a real
    linear combination of $\ket{A_0}$ and $\ket{A_1}$.}
  \label{fig:reflection2}
\end{figure}

Second we have the operation $H^{\otimes n} Z_{\mathrm{OR}} H^{\otimes n},$
which we've already seen can be written as
\[
H^{\otimes n} Z_{\mathrm{OR}} H^{\otimes n} = 2 \vert u \rangle \langle u \vert
- \mathbb{I}.
\]
This is also a reflection, this time about the line $L_2$ parallel to the
vector $\vert u\rangle.$
Figure~\ref{fig:reflection2} depicts the action of this reflection on a unit
vector $\vert\psi\rangle.$

\begin{figure}[!ht]
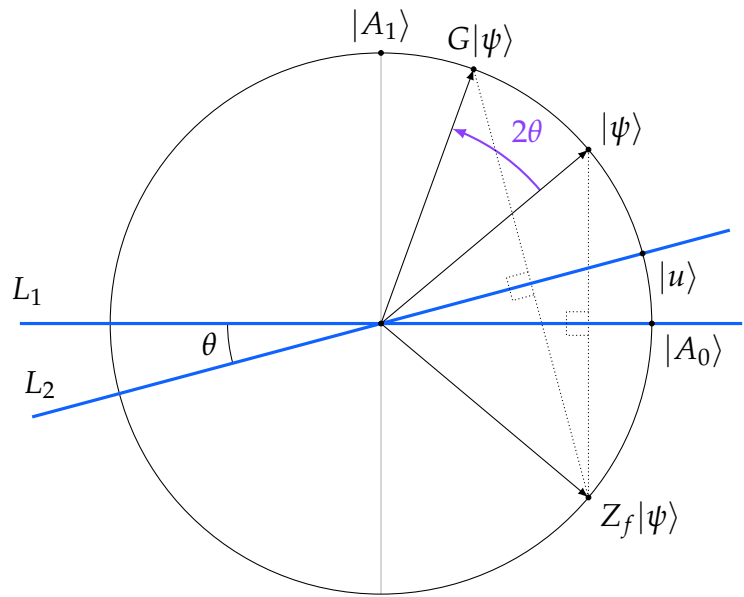

  \begin{center}
    %
  \end{center}
  \caption{The Grover operation $G$ is a composition of the reflections about
    the lines $L_1$ and $L_2$. Its action on real linear combinations of
    $\ket{A_0}$ and $\ket{A_1}$ is to rotate by twice the angle between $L_1$
    and $L_2$.}
  \label{fig:Grover-rotation}
\end{figure}

When we compose these two reflections, we obtain a rotation --- by twice the
angle between the lines of reflection --- as Figure~\ref{fig:Grover-rotation}
illustrates.
This explains, in geometric terms, why the effect of the Grover operation is to
rotate linear combinations of $\vert A_0\rangle$ and $\vert A_1\rangle$ by an
angle of $2\theta.$

\section{Choosing the number of iterations}

We have established that the state vector of the register $\mathsf{Q}$ in
Grover's algorithm remains in the two-dimensional subspace spanned by $\vert
A_0\rangle$ and $\vert A_1\rangle$ once the initialization step has been
performed.

The goal is to find an element $x\in A_1,$ and this goal will be accomplished
if we can obtain the state $\vert A_1\rangle$ --- for if we measure this state,
we're guaranteed to get a measurement outcome $x\in A_1.$
Given that the state of $\mathsf{Q}$ after $t$ iterations in step 2 is
\[
G^t \vert u \rangle
= \cos\bigl((2t + 1)\theta\bigr) \vert A_0\rangle
+ \sin\bigl((2t + 1)\theta\bigr) \vert A_1\rangle,
\]
we should choose $t$ so that
\[
\langle A_1 \vert G^t \vert u \rangle =
\sin((2t + 1)\theta)
\]
is as close to $1$ as possible in absolute value, to maximize the probability
to obtain $x\in A_1$ from the measurement.
For any angle $\theta \in (0,2\pi),$ the value $\sin((2t + 1)\theta)$
\emph{oscillates} as $t$ increases, though it is not necessarily periodic ---
there's no guarantee that we'll ever get the same value twice.

Naturally, in addition to making the probability of obtaining an element
$x\in A_1$ from the measurement large, we would also like to choose $t$ to be
as small as possible, because $t$ applications of the operation $G$ requires
$t$ queries to the function~$f.$
Because we're aiming to make $\sin( (2t + 1) \theta)$ close to $1$ in absolute
value, a natural way to do this is to choose $t$ so that
\[
(2t + 1) \theta \approx \frac{\pi}{2}.
\]
Solving for $t$ yields
\[
t \approx \frac{\pi}{4\theta} - \frac{1}{2}.
\]
Of course, $t$ must be an integer, so we won't necessarily be able to hit this
value exactly --- but what we can do is to take the closest integer to this
value, which is
\[
t = \Bigl\lfloor \frac{\pi}{4\theta} \Bigr\rfloor.
\]
This is the recommended number of iterations for Grover's algorithm.
As we proceed with the analysis, we'll see that the closeness of this integer
to the target value naturally affects the performance of the algorithm.

(As an aside, if the target value $\pi/(4\theta) - 1/2$ happens to be exactly
half-way between two integers, this expression of $t$ is what we get by
rounding up. We could alternatively round down, which makes sense to do because
it means one fewer query --- but this is secondary and unimportant for the sake
of the lesson.)

Recalling that the value of the angle $\theta$ is given by the formula
\[
\theta = \sin^{-1}\biggl(\sqrt{\frac{\vert A_1\vert}{N}}\biggr),
\]
we see that the recommended number of iterations $t$ depends on the number of
strings in $A_1.$
This presents a challenge if we don't know how many solutions we have, as we'll
discuss later.

\subsection{Unique search}

First, let's focus on the situation in which there's a single string $x$ such
that $f(x)=1.$
Another way to say this is that we're considering an instance of the
unique search problem.
In this case we have
\[
\theta = \sin^{-1}\biggl( \sqrt{\frac{1}{N}} \biggr),
\]
which can be conveniently approximated as
\[
\theta = \sin^{-1}\biggl( \sqrt{\frac{1}{N}} \biggr) \approx \sqrt{\frac{1}{N}}
\]
when $N$ is large.
If we substitute $\theta = 1/\sqrt{N}$ into the expression
\[
t = \Bigl\lfloor \frac{\pi}{4\theta} \Bigr\rfloor
\]
we obtain
\[
t = \Bigl\lfloor \frac{\pi}{4}\sqrt{N} \Bigr\rfloor.
\]
Recalling that $t$ is not only the number of times the operation $G$ is
performed, but also the number of queries to the function $f$ required by the
algorithm, we see that we're on track to obtaining an algorithm that requires
$O(\sqrt{N})$ queries.

Now we'll investigate how well the recommended choice of $t$ works.
The probability that the final measurement results in the unique solution can
be expressed explicitly as
\[
p(N,1) = \sin^2 \bigl( (2t + 1) \theta \bigr).
\]
The first argument, $N,$ refers to the number of items we're searching over,
and the second argument, which is $1$ in this case, refers to the number of
solutions.
A bit later we'll use the same notation more generally, where there are
multiple solutions.

Here's a table of the probabilities of success for increasing values of
$N = 2^n.$
\[
\begin{array}{ll}
N & p(N,1)\\ \hline
2	& 0.5000000000\\
4	& 1.0000000000\\
8	& 0.9453125000\\
16	& 0.9613189697\\
32	& 0.9991823155\\
64	& 0.9965856808\\
128	& 0.9956198657\\
256	& 0.9999470421
\end{array}\qquad\qquad
\begin{array}{ll}
N & p(N,1)\\ \hline
512	& 0.9994480262\\
1024 &	 0.9994612447\\
2048 &	 0.9999968478\\
4096 &	 0.9999453461\\
8192 &	 0.9999157752\\
16384 &	 0.9999997811\\
32768 &	 0.9999868295\\
65536 &	 0.9999882596
\end{array}
\]
Notice that these probabilities are not strictly increasing.
In particular, we have an interesting anomaly when $N=4,$ where we get a
solution with certainty.
It can, however, be proved in general that
\[
p(N,1) \geq 1 - \frac{1}{N}
\]
for all $N,$ so the probability of success goes to $1$ in the limit as $N$
becomes large, as the values above seem to suggest.
This is good!

But notice, however, that even a weak bound such as $p(N,1) \geq 1/2$
establishes the utility of Grover's algorithm.
For whatever measurement outcome $x$ we obtain from running the procedure, we
can always check to see if $f(x) = 1$ using a single query to $f.$
And if we fail to obtain the unique string $x$ for which $f(x) = 1$ with
probability at most $1/2$ by running the procedure once, then after $m$
independent runs of the procedure we will have failed to obtain this unique
string $x$ with probability at most $2^{-m}.$
That is, using $O(m \sqrt{N})$ queries to $f,$ we'll obtain the unique solution
$x$ with probability at least $1 - 2^{-m}.$
Using the better bound $p(N,1) \geq 1 - 1/N$ reveals that the probability to
find $x\in A_1$ using this method is actually at least $1 - N^{-m}.$

\subsection{Multiple solutions}

As the number of elements in $A_1$ varies, so too does the angle $\theta,$
which can have a significant effect on the algorithm's probability of success.
For the sake of brevity, let's write $s = \vert A_1 \vert$ to denote the number
of solutions, and as before we'll assume that $s\geq 1.$

As a motivating example, let's imagine that we have $s = 4$ solutions rather
than a single solution, as we considered above.
This means that
\[
\theta = \sin^{-1}\biggl( \sqrt{\frac{4}{N}} \biggr),
\]
which is approximately double the angle we had in the $\vert A_1 \vert = 1$
case when $N$ is large.
Suppose that we didn't know any better, and selected the same value of $t$ as
in the unique solution setting:
\[
t = \Biggl\lfloor \frac{\pi}{4\sin^{-1}\bigl(1/\sqrt{N}\bigr)}\Biggr\rfloor.
\]
The effect will be catastrophic as the following table of probabilities reveals.
\[
\begin{array}{ll}
N & \text{Success probability}\\ \hline
4	& 1.0000000000\\
8	& 0.5000000000\\
16	& 0.2500000000\\
32	& 0.0122070313\\
64	& 0.0203807689\\
128	& 0.0144530758\\
256	& 0.0000705058\\
512	& 0.0019310741
\end{array}\qquad\qquad
\begin{array}{ll}
N & \text{Success probability}\\ \hline
1024	& 0.0023009083\\
2048	& 0.0000077506\\
4096	& 0.0002301502\\
8192	& 0.0003439882\\
16384	& 0.0000007053\\
32768	& 0.0000533810\\
65536	& 0.0000472907\\
\phantom{0}
\end{array}
\]
This time the probability of success goes to $0$ as $N$ goes to infinity.
This happens because we're effectively rotating twice as fast as we did when
there was a unique solution, so we end up zooming past the target $\vert
A_1\rangle$ and landing near $-\vert A_0\rangle.$

However, if instead we use the recommended choice of $t,$ which is
\[
t = \Bigl\lfloor \frac{\pi}{4\theta}\Bigr\rfloor
\]
for
\[
\theta = \sin^{-1}\biggl( \sqrt{\frac{s}{N}} \biggr),
\]
then the performance will be better.
To be more precise, using this choice of $t$ leads to success with high
probability.
\[
\begin{array}{ll}
N & p(N,4)\\ \hline
4	 & 1.0000000000\\
8	 & 0.5000000000\\
16	 & 1.0000000000\\
32	 & 0.9453125000\\
64	 & 0.9613189697\\
128	 & 0.9991823155\\
256	 & 0.9965856808\\
512	 & 0.9956198657
\end{array}\qquad\qquad
\begin{array}{ll}
N & p(N,4)\\ \hline
1024	& 0.9999470421\\
2048	& 0.9994480262\\
4096	& 0.9994612447\\
8192	& 0.9999968478\\
16384	& 0.9999453461\\
32768	& 0.9999157752\\
65536	& 0.9999997811\\
\phantom{0}
\end{array}
\]

Generalizing what was claimed earlier, it can be proved that
\[
p(N,s) \geq 1 - \frac{s}{N},
\]
where we're using the notation suggested earlier: $p(N,s)$ denotes the
probability that Grover's algorithm run for $t$ iterations reveals a solution
when there are $s$ solutions in total out of $N$ possibilities.

This lower bound of $1 - s/N$ on the probability of success is slightly
peculiar in that more solutions implies a worse lower bound --- but under the
assumption that $s$ is significantly smaller than $N,$ we nevertheless conclude
that the probability of success is reasonably high.
As before, the mere fact that $p(N,s)$ is reasonably large implies the
algorithm's usefulness.

It also happens to be the case that
\[
p(N,s) \geq \frac{s}{N}.
\]
This lower bound describes the probability that a string $x\in\Sigma^n$
selected uniformly at random is a solution --- so Grover's algorithm always
does at least as well as random guessing.
(In fact, when $t=0,$ Grover's algorithm \emph{is} random guessing.)

Now let's take a look at the number of iterations (and hence the number of
queries)
\[
t = \Bigl\lfloor \frac{\pi}{4\theta}\Bigr\rfloor,
\]
for
\[
\theta = \sin^{-1}\biggl(\sqrt{\frac{s}{N}}\biggr).
\]
For every $\alpha \in [0,1],$ it is the case that
$\sin^{-1}(\alpha)\geq\alpha,$ and so
\[
\theta = \sin^{-1}\left(\sqrt{\frac{s}{N}}\right) \geq \sqrt{\frac{s}{N}}.
\]
This implies that
\[
t \leq \frac{\pi}{4\theta} \leq \frac{\pi}{4}\sqrt{\frac{N}{s}},
\]
which translates to a savings in the number of queries as $s$ grows.
In particular, the number of queries required is
\[
O\biggl(\sqrt{\frac{N}{s}}\biggr).
\]

\subsection{Unknown number of solutions}

If the number of solutions $s = \vert A_1 \vert$ is \emph{unknown}, then a
different approach is required, for in this situation we have no knowledge of
$s$ to inform our choice of $t.$
There are, in fact, multiple approaches.

One simple approach is to choose
\[
t \in \Bigl\{ 1,\ldots,\Bigl\lfloor\pi\sqrt{N}/4\Bigr\rfloor \Bigr\}
\]
\emph{uniformly at random}.
Selecting $t$ in this way always finds a solution (assuming one exists) with
probability greater than 40\%, though this is not obvious and requires an
analysis that will not be included here.
It does makes sense, however, particularly when we think about the geometric
picture:
rotating the state of $\mathsf{Q}$ a random number of times like this is not
unlike choosing a random unit vector in the space spanned by $\vert A_0\rangle$
and $\vert A_1\rangle,$ for which it is likely that the coefficient of $\vert
A_1\rangle$ is reasonably large.
By repeating this procedure and checking the outcome in the same way as
described before, the probability to find a solution can be made very close to
$1.$

There is a refined method that finds a solution when one exists using
$O(\sqrt{N/s})$ queries, even when the number of solutions $s$ is not known,
and requires $O(\sqrt{N})$ queries to determine that there are no solutions
when $s=0.$

The basic idea is to choose $t$ uniformly at random from the set
$\{1,\ldots,T\}$ iteratively, for increasing values of $T.$
In particular, we can start with $T = 1$ and increase it exponentially, always
terminating the process as soon as a solution is found and capping $T$ so as
not to waste queries when there isn't a solution.
The process takes advantage of the fact that fewer queries are required when
more solutions exist.
Some care is required, however, to balance the rate of growth of $T$ with the
probability of success for each iteration.
(Taking $T \leftarrow \lceil \frac{5}{4}T\rceil$ works, for instance, as an
analysis reveals.
Doubling $T,$ however, does not --- this turns out to be too fast of an
increase.)

\subsection{The trivial cases}

Throughout the analysis we've just gone through, we've assumed that the number
of solutions is nonzero.
Indeed, by simply referring to the vectors $\ket{A_0}$ and $\ket{A_1}$ we have
implicitly assumed that $A_0$ and $A_1$ are both nonempty.
Here we will briefly consider what happens when one of these sets is empty.

Before we bother with an analysis, let's observe the obvious:
if every string $x\in\Sigma^n$ is a solution, then we'll see a solution when we
measure; and when there aren't any solutions, we won't see one.
In some sense there's no need to go deeper than this.

We can, however, quickly verify the mathematics for these trivial cases.
The situation where one of $A_0$ and $A_1$ is empty happens when $f$ is
constant;
$A_1$ is empty when $f(x) = 0$ for every $x\in\Sigma^n,$ and $A_0$ is empty
when $f(x) = 1$ for every $x\in\Sigma^n.$
This means that
\[
Z_f \vert u\rangle = \pm \vert u\rangle,
\]
and therefore
\[
\begin{aligned}
  G \vert u \rangle
  & = \bigl( 2 \vert u\rangle \langle u \vert - \mathbb{I}\bigr)
  Z_f\vert u\rangle \\
  & = \pm \bigl( 2 \vert u\rangle \langle u \vert - \mathbb{I}\bigr)
  \vert u\rangle \\
  & = \pm \vert u\rangle.
\end{aligned}
\]
So, irrespective of the number of iterations $t$ we perform in these cases, the
measurements will always reveal a uniform random string $x\in\Sigma^n.$

\section{Concluding remarks}

Within the query model, Grover's algorithm is \emph{asymptotically optimal}.
What this means is that it's not possible to come up with a query algorithm for
solving the search problem, or even the unique search problem specifically,
that uses asymptotically less than $O(\sqrt{N})$ queries in the worst case.
This is something that has been proved rigorously in multiple ways.
Interestingly, this was known even before Grover's algorithm was discovered ---
Grover's algorithm matched an already-known lower bound.

Grover's algorithm is also broadly applicable, in the sense that the
square-root speed-up that it offers can be obtained in a variety of different
settings.
For example, sometimes it's possible to use Grover's algorithm in conjunction
with another algorithm to get an improvement.
Grover's algorithm is also quite commonly used as a subroutine inside of other
quantum algorithms to obtain speed-ups.

Finally, the technique used in Grover's algorithm, where two reflections are
composed and iterated to rotate a quantum state vector, can be generalized.
An example is a technique known as \emph{amplitude amplification}, where a
process similar to Grover's algorithm can be applied to another quantum
algorithm to boost its success probability quadratically faster than what is
possible classically.
Amplitude amplification has broad applications in quantum algorithms.

So, although Grover's algorithm may not lead to a practical quantum advantage
for searching any time soon, it is a fundamentally important quantum algorithm,
and it is representative of a more general technique that finds many
applications in quantum algorithms.

\stopcontents[part]


\unit[General Formulation of Quantum Information]{General Formulation of\\[-1mm]
  Quantum Information}
\label{unit:general-formulation-of-quantum-information}

This unit describes the general formulation of quantum information, where
quantum states are represented by density matrices, changes in states
are described by channels, and a more general class of measurements can be
considered than those discussed previously.
The unit also discusses mathematical ways of formalizing the distance or
similarity between quantum states, and how they relate to channels and
measurements in operational ways.

\begin{trivlist}
  \setlength{\parindent}{0mm}
  \setlength{\parskip}{2mm}
  \setlength{\itemsep}{1mm}
\item
  \textbf{Lesson 9: Density Matrices}

  This lesson describes the basics of how density matrices work and explains
  how they relate to quantum state vectors.
  It also introduces the Bloch sphere, which provides a useful geometric
  representation of qubit states.
  
  Lesson video URL: \url{https://youtu.be/CeK9ry8G8HQ}

\item
  \textbf{Lesson 10: Quantum Channels}

  This lesson begins with a discussion of basic aspects of channels along with
  some examples. It then moves on to different ways that channels can be
  described in mathematical terms --- including the so-called Stinespring,
  Kraus, and Choi representations of channels --- and explains why these
  different representations offer equivalent characterizations of channels.
  
  Lesson video URL: \url{https://youtu.be/cMl-xIDSmXI}

\item
  \textbf{Lesson 11: General Measurements}

  This lesson explains quantum measurements in full generality, including
  different ways that general measurements can be described in mathematical
  terms. It also describes quantum state discrimination and quantum state
  tomography, which are important notions connected with measurements.
  
  Lesson video URL: \url{https://youtu.be/Xi9YTYzQErY}

\item
  \textbf{Lesson 12: Purifications and Fidelity}

  This lesson explores the incredibly useful concept of a purification in
  quantum information, where an arbitrary quantum state is represented by a
  pure state of a larger system that leaves the original state when the rest
  of the larger system is discarded.
  It also introduces fidelity, a measure of similarity between two quantum
  states, which plays a key role in quantum computing.
  
  Lesson video URL: \url{https://youtu.be/jemWEdnJTnI}

\end{trivlist}


\lesson{Density Matrices}
\label{lesson:density-matrices}

In Unit~\ref{unit:basics-of-quantum-information}
\emph{(Basics of Quantum Information)}, we discussed a framework for quantum
information in which quantum states are represented by quantum state vectors,
operations are represented by unitary matrices, and so on.
We then used this framework in
Unit~\ref{unit:fundamentals-of-quantum-algorithms} \emph{(Fundamentals of
Quantum Algorithms)} to describe and analyze quantum algorithms.

There are actually two common mathematical descriptions of quantum information,
with the one introduced in Unit~\ref{unit:basics-of-quantum-information} being
the simpler of the two.
For this reason we'll refer to it as the \emph{simplified formulation of
quantum information.}

In this lesson, we'll begin our exploration of the second description, which is
the \emph{general formulation of quantum information.}
It is, naturally, consistent with the simplified formulation, but offers
noteworthy advantages.
For instance, it can be used to describe uncertainty in quantum states and
model the effects of noise on quantum computations.
It provides the foundation for quantum information theory, quantum
cryptography, and other topics connected with quantum information, and also
happens to be quite beautiful from a mathematical perspective.

In the general formulation of quantum information, quantum states are not
represented by vectors like in the simplified formulation, but instead are
represented by a special class of matrices called \emph{density matrices}. Here
are a few key points that motivate their use.

\begin{itemize}
\item
  Density matrices can represent a broader class of quantum states than quantum
  state vectors. This includes states that arise in practical settings, such as
  states of quantum systems that have been subjected to noise, as well as
  random choices of quantum states.

\item
  Density matrices allow us to describe states of isolated parts of systems,
  such as the state of one system that happens to be entangled with another
  system that we wish to ignore. This isn't easily done in the simplified
  formulation of quantum information.

\item
  Classical (probabilistic) states can also be represented by density matrices,
  specifically ones that are diagonal.
  This is important because it allows quantum and classical information to be
  described together within a single mathematical framework, with classical
  information essentially being a special case of quantum information.
\end{itemize}

At first glance, it may seem peculiar that quantum states are represented by
matrices, which more typically represent actions or operations, as opposed to
states. For example, unitary matrices describe quantum operations in the
simplified formulation of quantum information and stochastic matrices describe
probabilistic operations in the context of classical information.
In contrast, although density matrices are indeed matrices, they represent
states --- not actions or operations.

Despite this, the fact that density matrices can (like all matrices) be
associated with linear mappings is a critically important aspect of them.
For example, the \emph{eigenvalues} of density matrices describe the randomness
or uncertainty inherent to the states they represent.

\section{Density matrix basics}

We'll begin by describing what density matrices are in mathematical terms, and
then we'll take a look at some examples.
After that, we'll discuss a few basic aspects of how density matrices work and
how they relate to quantum state vectors in the simplified formulation of
quantum information.

\subsection{Definition}

Suppose that we have a quantum system named $\mathsf{X},$ and let $\Sigma$ be
the (finite and nonempty) classical state set of this system.
Here we're mirroring the naming conventions used in
Unit~\ref{unit:basics-of-quantum-information}, which we'll continue to do when
the opportunity arises.

In the general formulation of quantum information, a quantum state of the
system $\mathsf{X}$ is described by a \emph{density matrix} $\rho$ whose
entries are complex numbers and whose indices (for both its rows and columns)
have been placed in correspondence with the classical state set $\Sigma.$
The lowercase Greek letter $\rho$ is a conventional first choice for the name
of a density matrix, although $\sigma$ and $\xi$ are also common choices.
Here are a few examples of density matrices that describe states of qubits:
\[
\begin{pmatrix}
  1 & 0\\
  0 & 0
\end{pmatrix},
\quad
\begin{pmatrix}
  \frac{1}{2} & \frac{1}{2}\\[2mm]
  \frac{1}{2} & \frac{1}{2}
\end{pmatrix},
\quad
\begin{pmatrix}
  \frac{3}{4} & \frac{i}{8}\\[2mm]
  -\frac{i}{8} & \frac{1}{4}
\end{pmatrix},
\quad\text{and}\quad
\begin{pmatrix}
  \frac{1}{2} & 0\\[2mm]
  0 & \frac{1}{2}
\end{pmatrix}.
\]

To say that $\rho$ is a density matrix means that these two conditions, which
will be explained momentarily, are both satisfied:
\begin{enumerate}
\item
  Unit trace: $\operatorname{Tr}(\rho) = 1.$
\item
  Positive semidefiniteness: $\rho \geq 0.$
\end{enumerate}

\subsubsection{The trace of a matrix}

The first condition on density matrices refers to the \emph{trace} of a matrix.
This is a function that is defined, for all square matrices, as the sum of the
diagonal entries:
\[
\operatorname{Tr}
\begin{pmatrix}
\alpha_{0,0} & \alpha_{0,1} & \cdots & \alpha_{0,n-1}\\[1.5mm]
\alpha_{1,0} & \alpha_{1,1} & \cdots & \alpha_{1,n-1}\\[1.5mm]
\vdots & \vdots & \ddots & \vdots\\[1.5mm]
\alpha_{n-1,0} & \alpha_{n-1,1} & \cdots & \alpha_{n-1,n-1}
\end{pmatrix}
= \alpha_{0,0} + \alpha_{1,1} + \cdots + \alpha_{n-1,n-1}.
\]

The trace is a \emph{linear} function: for any two square matrices $A$ and $B$
of the same size, and any two complex numbers $\alpha$ and $\beta,$ the
following equation is always true.
\[
\operatorname{Tr}(\alpha A + \beta B) = \alpha \operatorname{Tr}(A) +
\beta\operatorname{Tr}(B)
\]

The trace is an extremely important function and there's a lot more that can be
said about it, but we'll wait until the need arises to say more.

\subsubsection{Positive semidefinite matrices}

The second condition refers to the property of a matrix being
\emph{positive semidefinite}, which is a fundamental concept in quantum
information theory and in many other subjects.
A matrix $P$ is \emph{positive semidefinite} if there exists a matrix $M$ such
that
\[
P = M^{\dagger} M.
\]
Here we can either demand that $M$ is a square matrix of the same size as $P$
or allow it to be non-square --- we obtain the same class of matrices either
way.

There are several alternative (but equivalent) ways to define this condition,
including these:
\begin{itemize}
\item
  A matrix $P$ is positive semidefinite if and only if $P$ is Hermitian (i.e.,
  equal to its own conjugate transpose) and all of its eigenvalues are
  nonnegative real numbers. Checking that a matrix is Hermitian and all of its
  eigenvalues are nonnegative is a simple computational way to verify that it's
  positive semidefinite.

\item
  A matrix $P$ is positive semidefinite if and only if
  $\langle \psi \vert P \vert \psi \rangle \geq 0$ for every complex vector
  $\vert\psi\rangle$ having the same indices as the rows and columns of $P.$
\end{itemize}

An intuitive way to think about positive semidefinite matrices is that they're
like matrix analogues of nonnegative real numbers.
That is, positive semidefinite matrices are to complex square matrices as
nonnegative real numbers are to complex numbers.
For example, a complex number $\alpha$ is a nonnegative real number if and only
if
\[
\alpha = \overline{\beta} \beta
\]
for some complex number $\beta,$ which matches the definition of positive
semidefiniteness when we replace matrices with scalars.
While matrices are more complicated objects than scalars in general, this is
nevertheless a helpful way to think about positive semidefinite matrices.

This also explains the common notation $P\geq 0,$ which indicates that $P$ is
positive semidefinite.
Notice in particular that $P\geq 0$ does \emph{not} mean that each entry of $P$
is nonnegative in this context;
there are positive semidefinite matrices having negative entries, as well as
matrices whose entries are all positive that are not positive semidefinite.

\subsubsection{Interpretation of density matrices}

At this point, the definition of density matrices may seem rather arbitrary and
abstract, as we have not yet associated any meaning with these matrices or
their entries.
The way density matrices work and can be interpreted will be clarified as the
lesson continues, but for now it may be helpful to think about the entries of
density matrices in the following (somewhat informal) way.
\begin{itemize}
\item
  The \emph{diagonal} entries of a density matrix give us the probabilities for
  each classical state to appear if we perform a standard basis measurement ---
  so we can think about these entries as describing the ``weight'' or
  ``likelihood'' associated with each classical state. 
\item
  The \emph{off-diagonal} entries of a density matrix describe the degree to
  which the two classical states corresponding to that entry (meaning the one
  corresponding to the row and the one corresponding to the column) are in
  quantum superposition, as well as the relative phase between them.
\end{itemize}

It is certainly not obvious \emph{a priori} that quantum states should be
represented by density matrices.
Indeed, there is a sense in which the choice to represent quantum states by
density matrices leads naturally to the entire mathematical description of
quantum information.
Everything else about quantum information actually follows pretty logically
from this one choice!

\subsection{Connection to quantum state vectors}

Recall that a quantum state vector $\vert\psi\rangle$ describing a quantum
state of $\mathsf{X}$ is a column vector having Euclidean norm equal to $1$
whose entries have been placed in correspondence with the classical state set
$\Sigma.$
The density matrix representation $\rho$ of the same state is defined as
follows.
\[
\rho = \vert\psi\rangle\langle\psi\vert
\]
To be clear, we're multiplying a column vector to a row vector, so the result
is a square matrix whose rows and columns correspond to $\Sigma.$
Matrices of this form, in addition to being density matrices, are always
projections and have rank equal to $1.$

For example, let's define two qubit state vectors.
\[
\begin{aligned}
  \vert {+i} \rangle & = \frac{1}{\sqrt{2}} \vert 0 \rangle +
  \frac{i}{\sqrt{2}} \vert 1 \rangle
  = 

\end{aligned}
\]
The density matrices corresponding to these two vectors are as follows.
\[
\begin{aligned}
\vert {+i} \rangle\langle{+i}\vert
& = %
\end{aligned}
\]
Here's a table listing these states along with a few other basic examples:
$\vert 0\rangle,$ $\vert 1\rangle,$ $\vert {+}\rangle,$ and
$\vert {-}\rangle.$
We'll see these six states again later in the lesson.

\begin{center}
  \begin{tabular}{l@{\hspace{20pt}}l}
    State vector & Density matrix \\\hline
    \rule{0pt}{30pt}
    $\vert 0\rangle = \begin{pmatrix} 1 \\[1mm] 0 \end{pmatrix}$
    &
    $\vert 0\rangle\langle 0\vert =
    \begin{pmatrix} 1 & 0\\[1mm] 0 & 0 \end{pmatrix}$
    \\[5mm]
    \rule{0pt}{30pt}
    $\vert 1\rangle = \begin{pmatrix} 0 \\[1mm] 1 \end{pmatrix}$
    &
    $\vert 1\rangle\langle 1\vert = \begin{pmatrix} 0 & 0\\[1mm] 0 &
      1 \end{pmatrix}$
    \\[5mm]
    \rule{0pt}{30pt}
    $\vert {+}\rangle = \begin{pmatrix} \frac{1}{\sqrt{2}} \\[2mm]
      \frac{1}{\sqrt{2}} \end{pmatrix}$
    & $\vert {+}\rangle\langle {+}\vert
    = \begin{pmatrix} \frac{1}{2} & \frac{1}{2}\\[2mm] \frac{1}{2} &
      \frac{1}{2} \end{pmatrix}$
    \\[5mm]
    \rule{0pt}{30pt}
    $\vert {-} \rangle = \begin{pmatrix} \frac{1}{\sqrt{2}} \\[2mm]
      -\frac{1}{\sqrt{2}}\end{pmatrix}$ &
    $\vert {-}\rangle\langle {-}\vert
    = \begin{pmatrix} \frac{1}{2} & -\frac{1}{2}\\[2mm] -\frac{1}{2} &
      \frac{1}{2} \end{pmatrix}$
    \\[5mm]
    \rule{0pt}{30pt}
    $\vert {+i} \rangle = \begin{pmatrix} \frac{1}{\sqrt{2}} \\[2mm]
      \frac{i}{\sqrt{2}} \end{pmatrix}$
    & $\vert {+i} \rangle\langle {+i} \vert
    = \begin{pmatrix} \frac{1}{2} & -\frac{i}{2}\\[2mm] \frac{1}{2} &
      \frac{1}{2} \end{pmatrix}$
    \\[5mm]
    \rule{0pt}{30pt}
    $\vert {-i} \rangle = \begin{pmatrix} \frac{1}{\sqrt{2}} \\[2mm]
      -\frac{i}{\sqrt{2}}\end{pmatrix}$ &
    $\vert {-i} \rangle\langle {-i} \vert
    = \begin{pmatrix} \frac{1}{2} & \frac{i}{2}\\[2mm] -\frac{i}{2} &
      \frac{1}{2} \end{pmatrix}$
  \end{tabular}
\end{center}

For one more example, here's a state from Lesson~\ref{lesson:single-systems}
(\emph{Single Systems}), including both its state vector and density matrix
representations.
\[
\vert v\rangle =
\frac{1 + 2 i}{3}\,\vert 0\rangle - \frac{2}{3}\,\vert 1\rangle
\qquad
\vert v\rangle\langle v\vert =
%
\]

Density matrices that take the form
$\rho = \vert \psi \rangle \langle \psi \vert$ for a quantum state vector
$\vert \psi \rangle$ are known as \emph{pure states}. 
Not every density matrix can be written in this form; some states are not pure.

As density matrices, pure states always have one eigenvalue equal to $1$ and
all other eigenvalues equal to $0.$
This is consistent with the interpretation that the eigenvalues of a density
matrix describe the randomness or uncertainty inherent to that state.
In essence, there's no uncertainty for a pure state
$\rho = \vert \psi \rangle \langle \psi \vert$ --- the state is definitely
$\vert \psi \rangle.$

In general, for a quantum state vector
\[
\vert\psi\rangle =
\begin{pmatrix}
  \alpha_0\\
  \alpha_1\\
  \vdots\\
  \alpha_{n-1}
\end{pmatrix}
\]
for a system with $n$ classical states, the density matrix representation of
the same state is as follows.
\[
\begin{aligned}
  \vert\psi\rangle\langle\psi\vert
  & =
  \begin{pmatrix}
    \alpha_0 \overline{\alpha_0} & \alpha_0 \overline{\alpha_1} & \cdots &
    \alpha_0 \overline{\alpha_{n-1}}\\[1mm]
    \alpha_1 \overline{\alpha_0} & \alpha_1 \overline{\alpha_1} & \cdots &
    \alpha_1 \overline{\alpha_{n-1}}\\[1mm]
    \vdots & \vdots & \ddots & \vdots\\[1mm]
    \alpha_{n-1} \overline{\alpha_0} & \alpha_{n-1} \overline{\alpha_1} &
    \cdots & \alpha_{n-1} \overline{\alpha_{n-1}}
  \end{pmatrix}\\[3mm]
  & = \begin{pmatrix}
    \vert\alpha_0\vert^2 & \alpha_0 \overline{\alpha_1} & \cdots & \alpha_0
    \overline{\alpha_{n-1}}\\[1mm]
    \alpha_1 \overline{\alpha_0} & \vert\alpha_1\vert^2 & \cdots & \alpha_1
    \overline{\alpha_{n-1}}\\[1mm]
    \vdots & \vdots & \ddots & \vdots\\[1mm]
    \alpha_{n-1} \overline{\alpha_0} & \alpha_{n-1} \overline{\alpha_1} &
    \cdots & \vert\alpha_{n-1}\vert^2
  \end{pmatrix}
\end{aligned}
\]
So, for the special case of pure states, we can verify that the diagonal
entries of a density matrix describe the probabilities that a standard basis
measurement would output each possible classical state.

A final remark about pure states is that density matrices eliminate the
degeneracy concerning global phases found for quantum state vectors.
Suppose we have two quantum state vectors that differ by a global phase:
$\vert \psi \rangle$ and
$\vert \phi \rangle = e^{i \theta} \vert \psi \rangle,$ for some real number
$\theta.$
Because they differ by a global phase, these vectors represent exactly the same
quantum state, despite the fact that the vectors may be different.
The density matrices that we obtain from these two state vectors, on the other
hand, are identical.
\[
\vert \phi \rangle \langle \phi \vert =
\bigl( e^{i\theta} \vert \psi \rangle \bigr) \bigl( e^{i\theta} \vert \psi
\rangle \bigr)^{\dagger}
= e^{i(\theta - \theta)} \vert \psi \rangle \langle \psi \vert
= \vert \psi \rangle \langle \psi \vert
\]

In general, density matrices provide a unique representation of quantum states:
two quantum states are identical, generating exactly the same outcome
statistics for every possible measurement that can be performed on them, if and
only if their density matrix representations are equal.
Using mathematical parlance, we can express this by saying that density
matrices offer a \emph{faithful} representation of quantum states.

\section{Convex combinations of density matrices}

\subsection{Probabilistic selections of density matrices}

A key aspect of density matrices is that \emph{probabilistic selections} of
quantum states are represented by \emph{convex combinations} of their
associated density matrices.

For example, if we have two density matrices, $\rho$ and $\sigma,$ representing
quantum states of a system $\mathsf{X},$ and we prepare the system in the state
$\rho$ with probability $p$ and $\sigma$ with probability $1 - p,$ then the
resulting quantum state is represented by the density matrix
\[
p \rho + (1 - p) \sigma.
\]

More generally, if we have $m$ quantum states represented by density matrices
$\rho_0,\ldots,\rho_{m-1},$ and a system is prepared in the state $\rho_k$ with
probability $p_k$ for some probability vector $(p_0,\ldots,p_{m-1}),$ the
resulting state is represented by the density matrix
\[
\sum_{k = 0}^{m-1} p_k \rho_k.
\]
This is a \emph{convex combination} of the density matrices
$\rho_0,\ldots,\rho_{m-1}.$

It follows that if we have $m$ quantum state vectors
$\vert\psi_0\rangle,\ldots,\vert\psi_{m-1}\rangle,$ and we prepare a system in
the state $\vert\psi_k\rangle$ with probability $p_k$ for each
$k\in\{0,\ldots,m-1\},$ the state we obtain is represented by the density
matrix
\[
\sum_{k = 0}^{m-1} p_k \vert\psi_k\rangle\langle\psi_k\vert.
\]

For example, if a qubit is prepared in the state $\vert 0\rangle$ with
probability $1/2$ and in the state $\vert + \rangle$ with probability $1/2,$
the density matrix representation of the state we obtain is given by
\[
\frac{1}{2} \vert 0\rangle\langle 0 \vert + \frac{1}{2} \vert +\rangle\langle +
\vert
= \frac{1}{2}
\begin{pmatrix}
  1 & 0\\[1mm]
  0 & 0
\end{pmatrix} +
\frac{1}{2}
\begin{pmatrix}
  \frac{1}{2} & \frac{1}{2}\\[2mm]
  \frac{1}{2} & \frac{1}{2}
\end{pmatrix}
= \begin{pmatrix}
  \frac{3}{4} & \frac{1}{4}\\[2mm]
  \frac{1}{4} & \frac{1}{4}
\end{pmatrix}.
\]
In the simplified formulation of quantum information, averaging quantum state
vectors like this doesn't work.
For instance, the vector
\[
\frac{1}{2} \vert 0\rangle + \frac{1}{2} \vert + \rangle
= \frac{1}{2}
\begin{pmatrix}
  1\\[1mm] 0
\end{pmatrix} +
\frac{1}{2}
\begin{pmatrix}
  \frac{1}{\sqrt{2}}\\[2mm]
  \frac{1}{\sqrt{2}}
\end{pmatrix}
= \begin{pmatrix}
  \frac{2 + \sqrt{2}}{4}\\[2mm]
  \frac{\sqrt{2}}{4}
\end{pmatrix}
\]
is not a valid quantum state vector because its Euclidean norm is not equal to
$1.$
A more extreme example that shows that this doesn't work for quantum state
vectors is that we fix any quantum state vector $\vert\psi\rangle$ that we
wish, and then we take our state to be $\vert\psi\rangle$ with probability
$1/2$ and $-\vert\psi\rangle$ with probability $1/2.$
These states differ by a global phase, so they're actually the same state ---
but averaging gives us the zero vector, which is not a valid quantum state
vector.

\subsection{The completely mixed state}

Suppose we set the state of a qubit to be $\vert 0\rangle$ or $\vert 1\rangle$
randomly, each with probability $1/2.$
The density matrix representing the resulting state is as follows.
\[
\frac{1}{2} \vert 0\rangle\langle 0\vert
+ \frac{1}{2} \vert 1\rangle\langle 1\vert
= \frac{1}{2}
\begin{pmatrix}
  1 & 0\\[1mm]
  0 & 0
\end{pmatrix} + \frac{1}{2}
\begin{pmatrix}
  0 & 0\\[1mm]
  0 & 1
\end{pmatrix}
= \begin{pmatrix}
  \frac{1}{2} & 0\\[1mm]
  0 & \frac{1}{2}
\end{pmatrix}
= \frac{1}{2} \mathbb{I}
\]
(In this equation the symbol $\mathbb{I}$ denotes the $2\times 2$ identity
matrix.)
This is a special state known as the \emph{completely mixed state}.
It represents complete uncertainty about the state of a qubit, similar to a
uniform random bit in the probabilistic setting.

Now suppose that we change the procedure: in place of the states $\vert
0\rangle$ and $\vert 1\rangle$ we'll use the states $\vert + \rangle$ and
$\vert - \rangle.$
We can compute the density matrix that describes the resulting state in a
similar way.
\[
\frac{1}{2} \vert +\rangle\langle {+} \vert
+ \frac{1}{2} \vert -\rangle\langle {-} \vert
= \frac{1}{2}
\begin{pmatrix}
  \frac{1}{2} & \frac{1}{2}\\[2mm]
  \frac{1}{2} & \frac{1}{2}
\end{pmatrix}
+ \frac{1}{2}
\begin{pmatrix}
  \frac{1}{2} & -\frac{1}{2}\\[2mm]
  -\frac{1}{2} & \frac{1}{2}
\end{pmatrix}
=
\begin{pmatrix}
  \frac{1}{2} & 0\\[2mm]
  0 & \frac{1}{2}
\end{pmatrix}
= \frac{1}{2} \mathbb{I}
\]
It's the same density matrix as before, even though we changed the states.
In fact, we would again obtain the same result --- the completely mixed state
--- by substituting \emph{any} two orthogonal qubit state vectors for $\vert
0\rangle$ and $\vert 1\rangle.$

This is a feature, not a bug!
We do in fact obtain exactly the same state either way.
That is, there's no way to distinguish the two procedures by measuring the
qubit they produce, even in a statistical sense.
Our two different procedures are simply different ways to prepare this state.

We can verify that this makes sense by thinking about what we could hope to
learn given a random selection of a state from one of the two possible state
sets $\{\vert 0\rangle,\vert 1\rangle\}$ and
$\{\vert {+}\rangle,\vert{-}\rangle\}.$
To keep things simple, let's suppose that we perform a unitary operation $U$ on
our qubit and then measure in the standard basis.

In the first scenario, the state of the qubit is chosen uniformly from the set
$\{\vert 0\rangle,\vert 1\rangle\}.$
If the state is $\vert 0\rangle,$ we obtain the outcomes $0$ and $1$ with
probabilities
\[
\vert \langle 0 \vert U \vert 0 \rangle \vert^2
\quad\text{and}\quad
\vert \langle 1 \vert U \vert 0 \rangle \vert^2
\]
respectively.
If the state is $\vert 1\rangle,$ we obtain the outcomes $0$ and $1$ with
probabilities
\[
\vert \langle 0 \vert U \vert 1 \rangle \vert^2
\quad\text{and}\quad
\vert \langle 1 \vert U \vert 1 \rangle \vert^2.
\]
Because the two possibilities each happen with probability $1/2,$ we obtain the
outcome $0$ with probability
\[
\frac{1}{2}\vert \langle 0 \vert U \vert 0 \rangle \vert^2
+ \frac{1}{2}\vert \langle 0 \vert U \vert 1 \rangle \vert^2
\]
and the outcome $1$ with probability
\[
\frac{1}{2}\vert \langle 1 \vert U \vert 0 \rangle \vert^2
+ \frac{1}{2}\vert \langle 1 \vert U \vert 1 \rangle \vert^2.
\]
Both of these expressions are equal to $1/2.$
One way to argue this is to use a fact from linear algebra that can be seen as
a generalization of the Pythagorean theorem.

\begin{callout}[title = {Parseval's identity}]
  Suppose $\{\vert\psi_0\rangle,\ldots,\vert\psi_{N-1}\rangle\}$ is an
  orthonormal basis of a (real or complex) vector space $\mathcal{V}.$ For
  every vector $\vert \phi\rangle \in \mathcal{V}$ we have
  \[
  \vert \langle \psi_0\vert\phi\rangle\vert^2 + \cdots + \vert \langle
  \psi_{N-1} \vert \phi \rangle\vert^2 = \| \vert\phi\rangle \|^2.
  \]
\end{callout}

We can apply this theorem to determine the probabilities as follows.
The probability to get $0$ is
\[
\begin{aligned}
\frac{1}{2}\vert \langle 0 \vert U \vert 0 \rangle \vert^2
+ \frac{1}{2}\vert \langle 0 \vert U \vert 1 \rangle \vert^2
& = \frac{1}{2} \Bigl(
\vert \langle 0 \vert U \vert 0 \rangle \vert^2
+ \vert \langle 0 \vert U \vert 1 \rangle \vert^2
\Bigr) \\[2mm]
& = \frac{1}{2} \Bigl(
\vert \langle 0 \vert U^{\dagger} \vert 0 \rangle \vert^2
+ \vert \langle 1 \vert U^{\dagger} \vert 0 \rangle \vert^2
\Bigr)\\[2mm]
& = \frac{1}{2} \bigl\| U^{\dagger} \vert 0 \rangle \bigr\|^2
\end{aligned}
\]
and the probability to get $1$ is
\[
\begin{aligned}
\frac{1}{2}\vert \langle 1 \vert U \vert 0 \rangle \vert^2
+ \frac{1}{2}\vert \langle 1 \vert U \vert 1 \rangle \vert^2
& = \frac{1}{2} \Bigl(
\vert \langle 1 \vert U \vert 0 \rangle \vert^2
+ \vert \langle 1 \vert U \vert 1 \rangle \vert^2
\Bigr) \\[2mm]
& = \frac{1}{2} \Bigl(
\vert \langle 0 \vert U^{\dagger} \vert 1 \rangle \vert^2
+ \vert \langle 1 \vert U^{\dagger} \vert 1 \rangle \vert^2
\Bigr)\\[2mm]
& = \frac{1}{2} \bigl\| U^{\dagger} \vert 1 \rangle \bigr\|^2.
\end{aligned}
\]
Because $U$ is unitary, we know that $U^{\dagger}$ is unitary as well, implying
that both $U^{\dagger} \vert 0 \rangle$ and $U^{\dagger} \vert 1 \rangle$ are
unit vectors.
Both probabilities are therefore equal to $1/2.$
This means that no matter how we choose $U,$ we're just going to get a uniform
random bit from the measurement.

We can perform a similar verification for any other pair of orthonormal states
in place of $\vert 0\rangle$ and $\vert 1\rangle.$
For example, because $\{\vert + \rangle, \vert - \rangle\}$ is an orthonormal
basis, the probability to obtain the measurement outcome $0$ in the second
procedure is
\[
\frac{1}{2}\vert \langle 0 \vert U \vert + \rangle \vert^2
+ \frac{1}{2}\vert \langle 0 \vert U \vert - \rangle \vert^2
= \frac{1}{2} \bigl\| U^{\dagger} \vert 0 \rangle \bigr\|^2 = \frac{1}{2}
\]
and the probability to get $1$ is
\[
\frac{1}{2}\vert \langle 1 \vert U \vert + \rangle \vert^2
+ \frac{1}{2}\vert \langle 1 \vert U \vert - \rangle \vert^2
= \frac{1}{2} \bigl\| U^{\dagger} \vert 1 \rangle \bigr\|^2 = \frac{1}{2}.
\]
In particular, we obtain exactly the same output statistics as we did for the
states $\vert 0\rangle$ and $\vert 1\rangle.$

\subsection{Probabilistic states}

Classical states can be represented by density matrices.
In particular, for each classical state $a$ of a system $\mathsf{X},$ the
density matrix
\[
\rho = \vert a\rangle \langle a \vert
\]
represents $\mathsf{X}$ being definitively in the classical state $a.$
For qubits we have
\[
\vert 0\rangle \langle 0 \vert = \begin{pmatrix}1 & 0 \\ 0 & 0\end{pmatrix}
\quad\text{and}\quad
\vert 1\rangle \langle 1 \vert = \begin{pmatrix}0 & 0 \\ 0 & 1\end{pmatrix},
\]
and in general we have a single $1$ on the diagonal in the position
corresponding to the classical state we have in mind, with all other entries
zero.

We can take convex combinations of these density matrices to represent
probabilistic states.
Supposing for simplicity that our classical state set is $\{0,\ldots,n-1\},$ if
$\mathsf{X}$ is in the state $a$ with probability $p_a$ for each
$a\in\{0,\ldots,n-1\},$ then the density matrix we obtain is
\[
\rho = \sum_{a = 0}^{n-1} p_a \vert a\rangle \langle a \vert
= \begin{pmatrix}
p_0 & 0 & \cdots & 0\\
0 & p_1 & \ddots & \vdots\\
\vdots & \ddots & \ddots & 0\\
0 & \cdots & 0 & p_{n-1}
\end{pmatrix}.
\]

Going in the other direction, any diagonal density matrix can naturally be
identified with the probabilistic state we obtain by simply reading the
probability vector off from the diagonal.

To be clear, when a density matrix is diagonal, it's not necessarily the case
that we're talking about a classical system, or that the system must have been
prepared through the random selection of a classical state, but rather that the
state \emph{could} have been obtained through the random selection of a
classical state.

The fact that probabilistic states are represented by diagonal density matrices
is consistent with the intuition suggested at the start of the lesson that
off-diagonal entries describe the degree to which the two classical states
corresponding to the row and column of that entry are in quantum superposition.
Here, all of the off-diagonal entries are zero, so we just have classical
randomness and nothing is in quantum superposition.

\subsection{Density matrices and the spectral theorem}

We've seen that if we take a convex combination of pure states,
\[
\rho = \sum_{k = 0}^{m-1} p_k \vert \psi_k\rangle \langle \psi_k \vert,
\]
we obtain a density matrix.
Every density matrix $\rho,$ in fact, can be expressed as a convex combination
of pure states like this.
That is, there will always exist a collection of unit vectors
$\{\vert\psi_0\rangle,\ldots,\vert\psi_{m-1}\rangle\}$ and a probability vector
$(p_0,\ldots,p_{m-1})$ for which the equation above is true.

We can, moreover, always choose the number $m$ so that it agrees with the
number of classical states of the system being considered, and we can select
the quantum state vectors to be orthogonal.
The spectral theorem, which we encountered in
Lesson~\ref{lesson:phase-estimation-and-factoring}
\emph{(Phase Estimation and Factoring)}, allows us to conclude this.
Here's a restatement of the spectral theorem for convenience.

\begin{callout}[title = {Spectral theorem}]
  Let $M$ be a normal $N\times N$ complex matrix.
  There exists an orthonormal basis of $N$-dimensional complex vectors
  $\bigl\{\vert\psi_0\rangle,\ldots,\vert\psi_{N-1}\rangle \bigr\}$ along with
  complex numbers $\lambda_0,\ldots,\lambda_{N-1}$ such that
  \[
  M = \lambda_0 \vert \psi_0\rangle\langle \psi_0\vert + \cdots +
  \lambda_{N-1} \vert \psi_{N-1}\rangle\langle \psi_{N-1}\vert.
  \]
\end{callout}

\noindent
(Recall that a matrix $M$ is \emph{normal} if it satisfies $M^{\dagger} M = M
M^{\dagger}.$ In words, normal matrices are matrices that commute with their
own conjugate transpose.)

We can apply the spectral theorem to any given density matrix $\rho$ because
density matrices are always Hermitian and therefore normal.
This allows us to write
\[
\rho = \lambda_0 \vert \psi_0\rangle\langle \psi_0\vert + \cdots +
\lambda_{n-1} \vert \psi_{n-1}\rangle\langle \psi_{n-1}\vert
\]
for an orthonormal basis
$\{\vert\psi_0\rangle,\ldots,\vert\psi_{n-1}\rangle\}.$
It remains to verify that $(\lambda_0,\ldots,\lambda_{n-1})$ is a probability
vector, which we can then rename to $(p_0,\ldots,p_{n-1})$ if we wish.

The numbers $\lambda_0,\ldots,\lambda_{n-1}$ are the eigenvalues of $\rho,$ and
because $\rho$ is positive semidefinite, these numbers must therefore be
nonnegative real numbers.
We can conclude that $\lambda_0 + \cdots + \lambda_{n-1} = 1$ from the fact
that $\rho$ has trace equal to $1.$
Going through the details will give us an opportunity to point out the
following important and very useful property of the trace.

\begin{callout}[title = {Cyclic property of the trace}]
  For any two matrices $A$ and $B$ that give us a square matrix $AB$ by
  multiplying, the equality $\operatorname{Tr}(AB) = \operatorname{Tr}(BA)$ is
  true.
\end{callout}

\noindent
Note that this works even if $A$ and $B$ are not themselves square matrices.
That is, we may have that $A$ is $n\times m$ and $B$ is $m\times n,$ for some
choice of positive integers $n$ and $m,$ so that $AB$ is an $n\times n$ square
matrix and $BA$ is $m\times m.$

In particular, if we let $A$ be a column vector $\vert\phi\rangle$ and let $B$
be the row vector $\langle \phi\vert,$ then we see that
\[
\operatorname{Tr}\bigl(\vert\phi\rangle\langle\phi\vert\bigr)
= \operatorname{Tr}\bigl(\langle\phi\vert\phi\rangle\bigr)
= \langle\phi\vert\phi\rangle.
\]
The second equality follows from the fact that $\langle\phi\vert\phi\rangle$ is
a scalar, which we can also think of as a $1\times 1$ matrix whose trace is its
single entry.
Using this fact, we can conclude that $\lambda_0 + \cdots + \lambda_{n-1} = 1$
by the linearity of the trace function.
\[
\begin{gathered}
  1 = \operatorname{Tr}(\rho) =
  \operatorname{Tr}\bigl(\lambda_0 \vert \psi_0\rangle\langle \psi_0\vert +
  \cdots + \lambda_{n-1} \vert \psi_{n-1}\rangle\langle
  \psi_{n-1}\vert\bigr)\\[2mm]
  = \lambda_0 \operatorname{Tr}\bigl(\vert \psi_0\rangle\langle
  \psi_0\vert\bigr) + \cdots + \lambda_{n-1} \operatorname{Tr}\bigl(\vert
  \psi_{n-1}\rangle\langle \psi_{n-1}\vert\bigr)
  = \lambda_0 + \cdots + \lambda_{n-1}
\end{gathered}
\]
Alternatively, we can reach the same conclusion by using the fact that the
trace of a square matrix (even one that isn't normal) is equal to the sum of
its eigenvalues.

We have therefore concluded that any given density matrix $\rho$ can be
expressed as a convex combination of pure states.
We also see that we can, moreover, take the pure states to be
\emph{orthogonal}.
This means, in particular, that we never need the number $n$ to be larger than
the size of the classical state set of $\mathsf{X}.$

In general, it must be understood that there will be different ways to write a
density matrix as a convex combination of pure states, not just the ways that
the spectral theorem provides.
A previous example illustrates this.
\[
\frac{1}{2} \vert 0\rangle\langle 0 \vert + \frac{1}{2} \vert +\rangle\langle +
\vert
= \begin{pmatrix}
\frac{3}{4} & \frac{1}{4}\\[2mm]
\frac{1}{4} & \frac{1}{4}
\end{pmatrix}
\]
This is not a spectral decomposition of this matrix because $\vert 0\rangle$
and $\vert + \rangle$ are not orthogonal.
Here's a spectral decomposition:
\[
\begin{pmatrix}
  \frac{3}{4} & \frac{1}{4}\\[2mm]
  \frac{1}{4} & \frac{1}{4}
\end{pmatrix}
= \cos^2(\pi/8) \vert \psi_{\pi/8} \rangle \langle \psi_{\pi/8}\vert +
\sin^2(\pi/8) \vert \psi_{5\pi/8} \rangle \langle \psi_{5\pi/8}\vert,
\]
where
$\vert \psi_{\theta} \rangle = \cos(\theta)\vert 0\rangle
+ \sin(\theta)\vert 1\rangle.$
The eigenvalues are numbers that will likely look familiar:
\[
\cos^2(\pi/8)
= \frac{2+\sqrt{2}}{4}
\approx 0.85 \quad\text{and}\quad \sin^2(\pi/8)
= \frac{2-\sqrt{2}}{4}
\approx 0.15.
\]
The eigenvectors can be written explicitly like this.
\[
\begin{aligned}
  \vert\psi_{\pi/8}\rangle & =
  \frac{\sqrt{2 + \sqrt{2}}}{2}\vert 0\rangle + \frac{\sqrt{2 -
      \sqrt{2}}}{2}\vert 1\rangle \\[3mm]
  \vert\psi_{5\pi/8}\rangle & =
  -\frac{\sqrt{2 - \sqrt{2}}}{2}\vert 0\rangle + \frac{\sqrt{2 +
      \sqrt{2}}}{2}\vert 1\rangle
\end{aligned}
\]

As another, more general example, suppose $\vert \phi_0\rangle,\ldots,\vert
\phi_{99} \rangle$ are quantum state vectors representing states of a single
qubit, chosen arbitrarily --- so we're not assuming any particular
relationships among these vectors.
We could then consider the state we obtain by choosing one of these $100$
states uniformly at random:
\[
\rho = \frac{1}{100} \sum_{k = 0}^{99} \vert \phi_k\rangle\langle \phi_k \vert.
\]
Because we're talking about a qubit, the density matrix $\rho$ is $2\times 2,$
so by the spectral theorem we could alternatively write
\[
\rho = p \vert\psi_0\rangle\langle\psi_0\vert + (1 - p)
\vert\psi_1\rangle\langle\psi_1\vert
\]
for some real number $p\in[0,1]$ and an orthonormal basis
$\{\vert\psi_0\rangle,\vert\psi_1\rangle\}$ --- but naturally the existence of
this expression doesn't prohibit us from writing $\rho$ as an average of 100
pure states if we choose to do that.

\section{Bloch sphere}

There's a useful geometric way to represent qubit states known as the
\emph{Bloch sphere}.
It's very convenient, but unfortunately it only works for qubits --- the
analogous representation no longer corresponds to a spherical object once we
have three or more classical states of our system.

\subsection{Qubit states as points on a sphere}

Let's start by thinking about a quantum state vector of a qubit: $\alpha \vert
0\rangle + \beta \vert 1\rangle.$
We can restrict our attention to vectors for which $\alpha$ is a nonnegative
real number because every qubit state vector is equivalent up to a global phase
to one for which $\alpha \geq 0.$
This allows us to write
\[
\vert\psi\rangle = \cos\bigl(\theta/2\bigr) \vert 0\rangle + e^{i\phi}
\sin\bigl(\theta/2\bigr) \vert 1\rangle
\]
for two real numbers $\theta \in [0,\pi]$ and $\phi\in[0,2\pi).$
Here, we're allowing $\theta$ to range from $0$ to $\pi$ and dividing by $2$ in
the argument of sine and cosine because this is a conventional way to
parameterize vectors of this sort, and it will make things simpler a bit later
on.

Now, it isn't quite the case that the numbers $\theta$ and $\phi$ are uniquely
determined by a given quantum state vector $\alpha \vert 0\rangle + \beta \vert
1\rangle,$ but it is nearly so.
In particular, if $\beta = 0,$ then $\theta = 0$ and it doesn't make any
difference what value $\phi$ takes, so it can be chosen arbitrarily.
Similarly, if $\alpha = 0,$ then $\theta = \pi,$ and once again $\phi$ is
irrelevant (as our state is equivalent to $e^{i\phi}\vert 1\rangle$ for any
$\phi$ up to a global phase).
If, however, neither $\alpha$ nor $\beta$ is zero, then there's a unique choice
for the pair $(\theta,\phi)$ for which $\vert\psi\rangle$ is equivalent to
$\alpha\vert 0\rangle + \beta\vert 1\rangle$ up to a global phase.

Next, let's consider the density matrix representation of this state.
\[
\vert\psi\rangle\langle\psi\vert
= \begin{pmatrix}
\cos^2(\theta/2) & e^{-i\phi}\cos(\theta/2)\sin(\theta/2)\\[2mm]
e^{i\phi}\cos(\theta/2)\sin(\theta/2) & \sin^2(\theta/2)
\end{pmatrix}
\]
We can use some trigonometric identities,
\[
\begin{gathered}
\cos^2(\theta/2) = \frac{1 + \cos(\theta)}{2},\\[2mm]
\sin^2(\theta/2) = \frac{1 - \cos(\theta)}{2},\\[2mm]
\cos(\theta/2) \sin(\theta/2) = \frac{\sin(\theta)}{2},
\end{gathered}
\]
as well as the formula $e^{i\phi} = \cos(\phi) + i\sin(\phi),$ to simplify the
density matrix as follows.
\[
\vert\psi\rangle\langle\psi\vert
= \frac{1}{2}
\begin{pmatrix}
  1 + \cos(\theta) & (\cos(\phi) - i \sin(\phi)) \sin(\theta)\\[1mm]
  (\cos(\phi) + i \sin(\phi)) \sin(\theta) & 1 - \cos(\theta)
\end{pmatrix}
\]
This makes it easy to express this density matrix as a linear combination of
the Pauli matrices:
\[
\mathbb{I} =
\begin{pmatrix}
  1 & 0\\[1mm]
  0 & 1
\end{pmatrix},
\quad
\sigma_x =
\begin{pmatrix}
  0 & 1\\[1mm]
  1 & 0
\end{pmatrix},
\quad
\sigma_y =
\begin{pmatrix}
  0 & -i\\[1mm]
  i & 0
\end{pmatrix},
\quad
\sigma_z =
\begin{pmatrix}
  1 & 0\\[1mm]
  0 & -1
\end{pmatrix}.
\]
Specifically, we conclude that
\[
\vert\psi\rangle\langle\psi\vert
= \frac{\mathbb{I} + \sin(\theta) \cos(\phi)\sigma_x + \sin(\theta)\sin(\phi)
  \sigma_y + \cos(\theta) \sigma_z}{2}.
\]

The coefficients of $\sigma_x,$ $\sigma_y,$ and $\sigma_z$ in the numerator of
this expression are all real numbers, so we can collect them together to form a
vector in an ordinary, three-dimensional Euclidean space.
\[
\bigl(\sin(\theta) \cos(\phi), \sin(\theta)\sin(\phi), \cos(\theta)\bigr)
\]
In fact, this is a unit vector.
Using \emph{spherical coordinates} it can be written as $(1,\theta,\phi).$
The first coordinate, $1,$ represents the \emph{radius} or
\emph{radial distance} (which is always $1$ in this case), $\theta$ represents
the \emph{polar angle}, and $\phi$ represents the \emph{azimuthal angle}.

\begin{figure}[!ht]
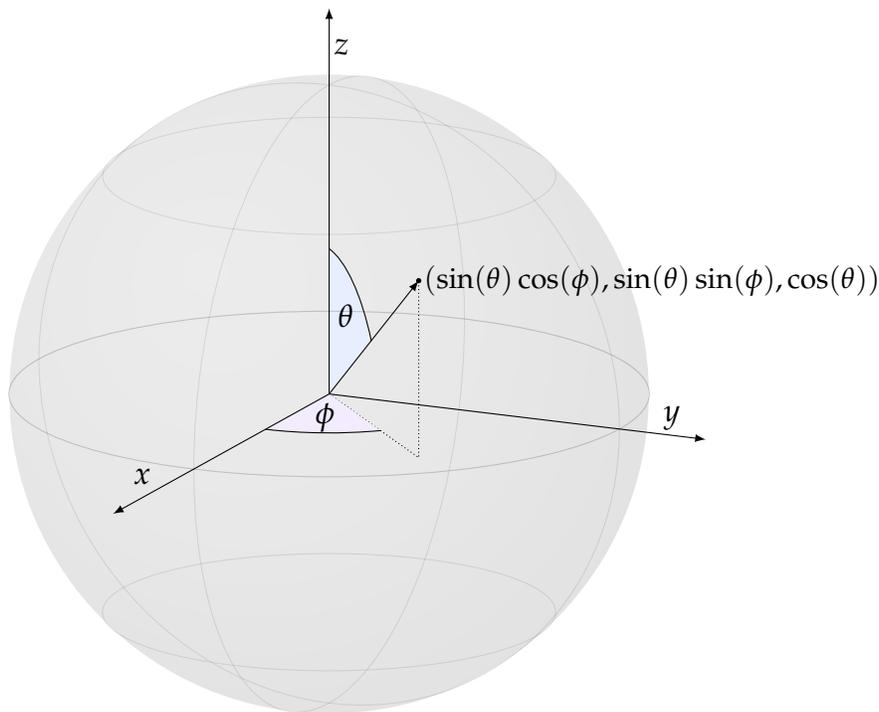

  \begin{center}
    \def\rotationSphere{-115}
    \def\radiusSphere{4.25cm}
    \def\psiLat{35}
    \def\psiLon{45}
    
    \begin{blochsphere}[
        color=Gray,
        tilt = 15,
        radius=\radiusSphere,
        opacity=0.075,
        rotation=\rotationSphere]
      
      \drawBallGrid[style={opacity=0.1}]{45}{90}
      
      \labelLatLon{ket0}{90}{0}
      \labelLatLon{ket1}{-90}{0}
      \labelLatLon{ketminus}{0}{180}
      \labelLatLon{ketplus}{00}{0}
      \labelLatLon{ketpluspi2}{0}{-90}
      
      
      \labelLatLon{ketplus3pi2}{0}{-270}
      \labelLatLon{psi}{\psiLat}{-\psiLon}
      
      \coordinate (origin) at (0,0);
      
      {
        \setLongitudinalDrawingPlane{\psiLon}
        \pic[
          current plane,
          angle radius = 2cm,
          draw,
          fill = DataColor0!10,
          fill opacity=1,
          text opacity=1,
          "$\theta$",
          angle eccentricity=0.7
        ]{angle=psi--origin--ket0};
        
      }

      {
        \setDrawingPlane{0}{0}
        
        \draw[current plane,densely dotted]
        (0,0) -- (-90+\psiLon:{cos(\psiLat)*\radiusSphere})
        coordinate (psiProjectedEquat) -- (psi);
        
        \pic[
          current plane,
          draw,
          text opacity=1,
          "$\phi$",
          angle radius = 2cm,
          fill = DataColor1!10,
          fill opacity=1,
          angle eccentricity=0.6
        ]{
          angle=ketplus--origin--psiProjectedEquat
        };            
        
        \draw[current plane,->] (origin) -- ++(1.3*\radiusSphere,0)
        node[pos=0.91, above] {$y$};            
      }
      
      {
        \setDrawingPlane{90}{90}
        
        \draw[current plane,->] (origin) -- ++(1.6*\radiusSphere,0)
        node[pos=0.8, above left] {$x$};
        
        \draw[current plane,->] (origin) -- ++(0,1.25*\radiusSphere)
        node[pos=0.9, right] {$z$};            
      }

      \draw[-latex] (0,0) -- (psi) node[right]{\small
        $(\sin(\theta)\cos(\phi),\sin(\theta)\sin(\phi),\cos(\theta))$};
      
      \node[circle, fill, inner sep=0.7] at (psi) {};

    \end{blochsphere}
  \end{center}
  \caption{Illustration of the Cartesian coordinates of a point on the
    unit $2$-sphere with polar angle $\theta$ and azimuthal angle $\phi$.
  }
  \label{fig:spherical-coordinates}
\end{figure}

In words, thinking about a sphere as the planet Earth, the polar angle $\theta$
is how far we rotate south from the north pole to reach the point being
described, from $0$ to $\pi = 180^{\circ},$ while the azimuthal angle $\phi$ is
how far we rotate east from the prime meridian, from $0$ to
$2\pi = 360^{\circ},$
as is illustrated in Figure~\ref{fig:spherical-coordinates}.
This assumes that we define the prime meridian to be the curve on the surface
of the sphere from one pole to the other that passes through the positive
$x$-axis.

Every point on the sphere can be described in this way --- which is to say that
the points we obtain when we range over all possible pure states of a qubit
correspond precisely to a sphere in $3$ real dimensions.
(This sphere is typically called the \emph{unit $2$-sphere} because the surface
of this sphere is two-dimensional.)

When we associate points on the unit $2$-sphere with pure states of qubits, we
obtain the \emph{Bloch sphere} representation these states.

\subsection{Six important examples}

\begin{figure}[t]
  \begin{center}
    \def\rotationSphere{-115}
    \def\radiusSphere{4cm}
    \def\psiLat{45}
    \def\psiLon{45}
    
    \begin{blochsphere}[
        color=black!30,
        tilt = 15,
        radius=\radiusSphere,
        opacity=0.075,
        rotation=\rotationSphere]
      
      \drawBallGrid[style={opacity=0.1}]{45}{90}
      
      \labelLatLon{0}{90}{0}
      \labelLatLon{1}{-90}{90}
      \labelLatLon{plus}{0}{0}
      \labelLatLon{minus}{0}{180}
      \labelLatLon{plusi}{0}{270}
      \labelLatLon{minusi}{0}{90}
      
      \node[yshift = 4mm] at
      (0) {$\ket{0}$};
      
      \node[yshift = -4mm] at
      (1) {$\ket{1}$};
      
      \node[anchor = north east, yshift = -0.5mm] at
      (plus) {$\ket{+}$};
      
      \node[anchor = south west, yshift = 0.5mm] at
      (minus) {$\ket{-}$};
      
      \node[anchor = west, xshift = 1mm, yshift = -0.5mm] at
      (plusi) {$\ket{+i}$};
      
      \node[anchor = east, xshift = -1mm, yshift = 0.5mm] at
      (minusi) {$\ket{-i}$};
      
      \draw[->] (0,0) -- (0);
      \draw[->] (0,0) -- (1);
      \draw[->] (0,0) -- (plus);
      \draw[->] (0,0) -- (minus);
      \draw[->] (0,0) -- (plusi);
      \draw[->] (0,0) -- (minusi);
      
      \filldraw (0) circle (1pt);
      \filldraw (1) circle (1pt);
      \filldraw (plus) circle (1pt);
      \filldraw (minus) circle (1pt);
      \filldraw (plusi) circle (1pt);
      \filldraw (minusi) circle (1pt);
      \filldraw (0,0) circle (1pt);
      
    \end{blochsphere}
  \end{center}
  \caption{The states $\ket{0}$, $\ket{1}$, $\ket{+}$, $\ket{-}$,
    $\ket{+i}$, and $\ket{-i}$ on the Bloch sphere.}
  \label{fig:six-Bloch-sphere-points}
\end{figure}

\begin{trivlist}

\item
  \emph{The standard basis:}
  $\{\vert 0\rangle,\vert 1\rangle\}.$
  Let's start with the state $\vert 0\rangle.$
  As a density matrix it can be written like this.
  \[
  \vert 0 \rangle \langle 0 \vert = \frac{\mathbb{I} + \sigma_z}{2}
  \]
  By collecting the coefficients of the Pauli matrices in the numerator, we see
  that the corresponding point on the unit $2$-sphere using Cartesian
  coordinates is $(0,0,1).$
  
  In spherical coordinates this point is $(1,0,\phi),$ where $\phi$ can be any
  angle.
  This is consistent with the expression
  \[
  \vert 0\rangle = \cos(0) \vert 0\rangle + e^{i \phi} \sin(0) \vert 1\rangle,
  \]
  which also works for any $\phi.$
  Intuitively speaking, the polar angle $\theta$ is zero, so we're at the
  north pole of the Bloch sphere, where the azimuthal angle is irrelevant.
  
  Along similar lines, the density matrix for the state $\vert 1\rangle$ can be
  written like so.
  \[
  \vert 1 \rangle \langle 1 \vert = \frac{\mathbb{I} - \sigma_z}{2}
  \]
  This time the Cartesian coordinates are $(0,0,-1).$ In spherical coordinates
  this point is $(1,\pi,\phi)$ where $\phi$ can be any angle. In this case the
  polar angle is all the way to $\pi,$ so we're at the south pole where the
  azimuthal angle is again irrelevant.

\item
  \emph{The basis $\{\vert + \rangle, \vert - \rangle\}$.}
  We have these expressions for the density matrices corresponding to these
  states.
  \[
  \begin{aligned}
    \vert {+} \rangle\langle {+} \vert & = \frac{\mathbb{I} +
      \sigma_x}{2}\\[2mm]
    \vert {-} \rangle\langle {-} \vert & = \frac{\mathbb{I} - \sigma_x}{2}
  \end{aligned}
  \]
  The corresponding points on the unit $2$-sphere have Cartesian coordinates
  $(1,0,0)$ and $(-1,0,0),$ and spherical coordinates $(1,\pi/2,0)$ and
  $(1,\pi/2,\pi),$ respectively.
  
  In words, $\vert +\rangle$ corresponds to the point where the positive
  $x$-axis intersects the unit $2$-sphere and $\vert -\rangle$ corresponds to
  the point where the negative $x$-axis intersects it. More intuitively,
  $\vert +\rangle$ is on the equator of the Bloch sphere where it meets the
  prime meridian, and $\vert - \rangle$ is on the equator on the opposite side
  of the sphere.
  
\item
  \emph{The basis $\{\vert {+i} \rangle, \vert {-i} \rangle\}$.}
  As we saw earlier in the lesson, these two states are defined like this:
  \[
  \begin{aligned}
    \vert {+i} \rangle & = \frac{1}{\sqrt{2}} \vert 0 \rangle +
    \frac{i}{\sqrt{2}} \vert 1 \rangle\\[2mm]
    \vert {-i} \rangle & = \frac{1}{\sqrt{2}} \vert 0 \rangle -
    \frac{i}{\sqrt{2}} \vert 1 \rangle.
  \end{aligned}
  \]
  This time we have these expressions.
  \[
  \begin{aligned}
    \vert {+i} \rangle\langle {+i} \vert & = \frac{\mathbb{I} +
      \sigma_y}{2}\\[2mm]
    \vert {-i} \rangle\langle {-i} \vert & = \frac{\mathbb{I} - \sigma_y}{2}
  \end{aligned}
  \]
  The corresponding points on the unit $2$-sphere have Cartesian coordinates
  $(0,1,0)$ and $(0,-1,0),$ while the spherical coordinates of these points
  are $(1,\pi/2,\pi/2)$ and $(1,\pi/2,3\pi/2),$ respectively.

  In words, $\vert {+i} \rangle$ corresponds to the point where the positive
  $y$-axis intersects the unit $2$-sphere and $\vert {-i} \rangle$ to the point
  where the negative $y$-axis intersects it.
\end{trivlist}

\begin{figure}[t]
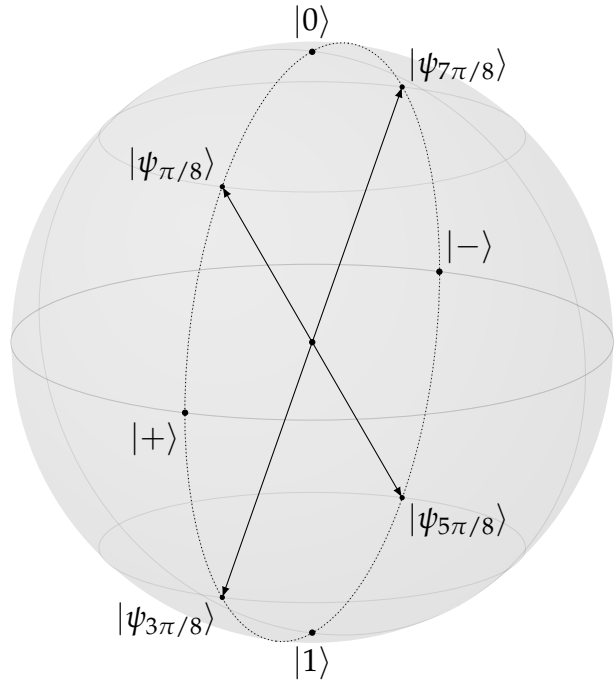

  \begin{center}
    \def\rotationSphere{-115}
    \def\radiusSphere{4cm}
    \def\psiLat{45}
    \def\psiLon{45}
    
    \begin{blochsphere}[
        color=Gray,
        tilt = 15,
        radius=\radiusSphere,
        opacity=0.075,
        rotation=\rotationSphere]

      \drawBallGrid[style={opacity=0.1}]{45}{90}
      
      \labelLatLon{0}{90}{0}
      \labelLatLon{1}{-90}{90}
      \labelLatLon{plus}{0}{0}
      \labelLatLon{minus}{0}{180}
      \labelLatLon{plusi}{0}{270}
      \labelLatLon{minusi}{0}{90}

      \node[yshift = 4mm] at
      (0) {$\ket{0}$};

      \node[yshift = -4mm] at
      (1) {$\ket{1}$};

      \node[anchor = north east, yshift = -0.5mm] at
      (plus) {$\ket{+}$};

      \node[anchor = south west, yshift = 0.5mm] at
      (minus) {$\ket{-}$};

      \labelLatLon{psi-pi-8}{45}{0}
      \labelLatLon{psi-3pi-8}{-45}{0}
      \labelLatLon{psi-5pi-8}{-45}{180}
      \labelLatLon{psi-7pi-8}{45}{180}

      \node[circle, fill, inner sep=0.7] at (psi-pi-8) {};
      \node[circle, fill, inner sep=0.7] at (psi-3pi-8) {};
      \node[circle, fill, inner sep=0.7] at (psi-5pi-8) {};
      \node[circle, fill, inner sep=0.7] at (psi-7pi-8) {};
      
      \draw[draw,->] (0,0) -- (psi-pi-8) node[above left]{
        $\ket{\psi_{\pi/8}}$};
      
      \draw[draw,->] (0,0) -- (psi-3pi-8) node[below left]{
        $\ket{\psi_{3\pi/8}}$};
      
      \draw[draw,->] (0,0) -- (psi-5pi-8) node[below right]{
        $\ket{\psi_{5\pi/8}}$};
      
      \draw[draw,->] (0,0) -- (psi-7pi-8) node[above right]{
        $\ket{\psi_{7\pi/8}}$};
      
      \drawLongitudeCircle[style={densely dotted}]{0}
      
      \filldraw (0) circle (1pt);
      \filldraw (1) circle (1pt);
      \filldraw (plus) circle (1pt);
      \filldraw (minus) circle (1pt);
      \filldraw (0,0) circle (1pt);
      
    \end{blochsphere}   
  \end{center}
  \caption{Qubit states of the form
    $\ket{\psi_{\alpha}} = \cos(\alpha) \ket{0} + \sin(\alpha) \ket{1}$
    on the Bloch sphere.}
    \label{fig:real-Bloch-sphere-points}
\end{figure}

Here's another class of quantum state vectors that has appeared from time to
time throughout this course, including previously in this lesson.
\[
\vert \psi_{\alpha} \rangle = \cos(\alpha) \vert 0\rangle + \sin(\alpha) \vert
1\rangle \qquad \text{(for $\alpha \in [0,\pi)$)}
\]
The density matrix representation of each of these states is as follows.
\[
\vert \psi_{\alpha} \rangle \langle \psi_{\alpha} \vert =
\begin{pmatrix}
\cos^2(\alpha) & \cos(\alpha)\sin(\alpha)\\[2mm]
\cos(\alpha)\sin(\alpha) & \sin^2(\alpha)
\end{pmatrix}
= \frac{\mathbb{I} + \sin(2\alpha) \sigma_x + \cos(2\alpha) \sigma_z}{2}
\]
Figure~\ref{fig:real-Bloch-sphere-points} illustrates the corresponding points
on the Bloch sphere for a few choices for $\alpha.$

\subsection{Convex combinations of points}

Similar to what we have already discussed for density matrices, we can take
convex combinations of points on the Bloch sphere to obtain representations of
qubit density matrices.
In general, this results in points \emph{inside} of the Bloch sphere, which
represent density matrices of states that are not pure.
Sometimes we refer to the \emph{Bloch ball} when we wish to be explicit about
the inclusion of points inside of the Bloch sphere as representations of qubit
density matrices.

For example, we've seen that the density matrix $\frac{1}{2}\mathbb{I},$ which
represents the completely mixed state of a qubit, can be written in these two
alternative ways:
\[
\frac{1}{2} \mathbb{I} = \frac{1}{2} \vert 0\rangle\langle 0\vert + \frac{1}{2}
\vert 1\rangle\langle 1\vert
\quad\text{and}\quad
\frac{1}{2} \mathbb{I} = \frac{1}{2} \vert +\rangle\langle +\vert + \frac{1}{2}
\vert -\rangle\langle -\vert.
\]
We also have
\[
\frac{1}{2} \mathbb{I} =
\frac{1}{2} \vert {+i}\rangle\langle {+i} \vert
+ \frac{1}{2} \vert {-i} \rangle\langle {-i}\vert,
\]
and more generally we can use any two orthogonal qubit state vectors (which
will always correspond to two antipodal points on the Bloch sphere).
If we average the corresponding points on the Bloch sphere in a similar way, we
obtain the same point, which is at the center of the sphere.
This is consistent with the observation that
\[
\frac{1}{2} \mathbb{I} = \frac{\mathbb{I} + 0 \cdot \sigma_x + 0 \cdot \sigma_y
  + 0 \cdot \sigma_z}{2},
\]
giving us the Cartesian coordinates $(0,0,0).$

A different example concerning convex combinations of Bloch sphere points is
the one discussed in the previous subsection.
\begin{multline*}
  \qquad\qquad
  \frac{1}{2} \vert 0\rangle\langle 0 \vert
  + \frac{1}{2} \vert +\rangle\langle + \vert
  = \begin{pmatrix} \frac{3}{4} & \frac{1}{4}\\[2mm]
    \frac{1}{4} & \frac{1}{4}
  \end{pmatrix}\\[1mm]
  = \cos^2(\pi/8) \vert \psi_{\pi/8} \rangle \langle \psi_{\pi/8}\vert +
  \sin^2(\pi/8) \vert \psi_{5\pi/8} \rangle \langle \psi_{5\pi/8}\vert
  \qquad\qquad
\end{multline*}
Figure~\ref{fig:Bloch-sphere-zero-plus-average} illustrates these two different
ways of obtaining this density matrix as a convex combination of pure states.

\begin{figure}[!ht]
  \begin{center}
    \def\rotationSphere{-115}
    \def\radiusSphere{4cm}
    \def\psiLat{45}
    \def\psiLon{45}
    
    \begin{blochsphere}[
        color=Gray,
        tilt = 15,
        radius=\radiusSphere,
        opacity=0.075,
        rotation=\rotationSphere]

      \drawBallGrid[style={opacity=0.1}]{45}{90}
      \drawLongitudeCircle[style={densely dotted}]{0}

      \labelLatLon{0}{90}{0}
      \labelLatLon{1}{-90}{0}
      \labelLatLon{plus}{0}{0}
      \labelLatLon{minus}{0}{180}
      \labelLatLon{plusi}{0}{270}
      \labelLatLon{minusi}{0}{90}

      \node[yshift = 4mm] at
      (0) {$\ket{0}$};

      \node[yshift = -4mm] at
      (1) {$\ket{1}$};

      \node[anchor = north east, yshift = -0.5mm] at
      (plus) {$\ket{+}$};

      \node[anchor = south west, yshift = 0.5mm] at
      (minus) {$\ket{-}$};
      
      \draw (0,0) -- (0);
      \draw (0,0) -- (1);
      \draw (0,0) -- (plus);
      \draw (0,0) -- (minus);

      \labelLatLon{psi-pi-8}{45}{0}
      \labelLatLon{psi-5pi-8}{-45}{180}
      
      \draw[draw=DataColor0, thick] (0,0) -- (psi-pi-8) node[above left]{
        $\ket{\psi_{\pi/8}}$};

      \filldraw (psi-pi-8) circle (1pt);
      
      \draw[draw=DataColor0, thick] (0,0) -- (psi-5pi-8) node[below right]{
        $\ket{\psi_{5\pi/8}}$};

      \filldraw (psi-5pi-8) circle (1pt);

      \draw[draw=DataColor0,thick] (0) -- (plus);

      \coordinate (intersect) at ($(0)!0.5!(plus)$);
      \filldraw (intersect) circle (1pt);

      \node (state_label) at ([xshift = -4cm,yshift = 5mm]intersect) {\small
        $\begin{pmatrix}
          \frac{3}{4} & \frac{1}{4}\\[1mm]
          \frac{1}{4} & \frac{1}{4}
        \end{pmatrix}$};

      \node[inner sep=1pt] (intersect_node) at (intersect) {};
      \draw[->] (state_label) -- (intersect_node);
      
      \filldraw (0) circle (1pt);
      \filldraw (1) circle (1pt);
      \filldraw (plus) circle (1pt);
      \filldraw (minus) circle (1pt);
      
    \end{blochsphere}
  \end{center}
  \caption{An illustration of the density matrix
    $\frac{1}{2}\ket{0}\bra{0} + \frac{1}{2}\ket{+}\bra{+}$ inside the
    Bloch sphere.}
  \label{fig:Bloch-sphere-zero-plus-average}
\end{figure}

\section{Multiple systems and reduced states}

Now we'll turn our attention to how density matrices work for multiple systems,
including examples of different types of correlations they can express and how
they can be used to describe the states of isolated parts of compound systems.

\subsection{Multiple systems}

Density matrices can represent states of multiple systems in an analogous way
to state vectors in the simplified formulation of quantum information,
following the same basic idea that multiple systems can be viewed as if they're
single, compound systems.
In mathematical terms, the rows and columns of density matrices representing
states of multiple systems are placed in correspondence with the Cartesian
product of the classical state sets of the individual systems.

For example, recall the state vector representations of the four Bell states.
\begin{alignat*}{2}
  \vert \phi^+ \rangle & = \frac{1}{\sqrt{2}}
  \vert 00 \rangle + \frac{1}{\sqrt{2}} \vert 11 \rangle & \qquad
  \vert \phi^- \rangle & = \frac{1}{\sqrt{2}}
  \vert 00 \rangle - \frac{1}{\sqrt{2}} \vert 11 \rangle \\[3mm]
  \vert \psi^+ \rangle & = \frac{1}{\sqrt{2}}
  \vert 01 \rangle + \frac{1}{\sqrt{2}} \vert 10 \rangle &
  \vert \psi^- \rangle & = \frac{1}{\sqrt{2}}
  \vert 01 \rangle - \frac{1}{\sqrt{2}} \vert 10 \rangle
\end{alignat*}
The density matrix representations of these states are as follows.
\begin{alignat*}{2}
  \vert \phi^+ \rangle \langle \phi^+ \vert & =
  \begin{pmatrix}
    \frac{1}{2} & 0 & 0 & \frac{1}{2}\\[2mm]
    0 & 0 & 0 & 0\\[2mm]
    0 & 0 & 0 & 0\\[2mm]
    \frac{1}{2} & 0 & 0 & \frac{1}{2}
  \end{pmatrix} \qquad
  &
  \vert \phi^- \rangle \langle \phi^- \vert & =
  \begin{pmatrix}
    \frac{1}{2} & 0 & 0 & -\frac{1}{2}\\[2mm]
    0 & 0 & 0 & 0\\[2mm]
    0 & 0 & 0 & 0\\[2mm]
    -\frac{1}{2} & 0 & 0 & \frac{1}{2}
  \end{pmatrix}
  \\[4mm]
  \vert \psi^+ \rangle \langle \psi^+ \vert & =
  \begin{pmatrix}
    0 & 0 & 0 & 0\\[2mm]
    0 & \frac{1}{2} & \frac{1}{2} & 0\\[2mm]
    0 & \frac{1}{2} & \frac{1}{2} & 0\\[2mm]
    0 & 0 & 0 & 0
  \end{pmatrix} \qquad
  &
  \vert \psi^- \rangle \langle \psi^- \vert & =
  \begin{pmatrix}
    0 & 0 & 0 & 0\\[2mm]
    0 & \frac{1}{2} & -\frac{1}{2} & 0\\[2mm]
    0 & -\frac{1}{2} & \frac{1}{2} & 0\\[2mm]
    0 & 0 & 0 & 0
  \end{pmatrix}.
\end{alignat*}

\subsubsection{Product states}

Similar to what we had for state vectors, tensor products of density matrices
represent \emph{independence} between the states of multiple systems.
For instance, if $\mathsf{X}$ is prepared in the state represented by the
density matrix $\rho$ and $\mathsf{Y}$ is independently prepared in the state
represented by $\sigma,$ then the density matrix describing the state of
$(\mathsf{X},\mathsf{Y})$ is the tensor product $\rho\otimes\sigma.$

The same terminology is used here as in the simplified formulation of quantum
information: states of this form are referred to as \emph{product states}.

\subsubsection{Correlated and entangled states}

States that cannot be expressed as product states represent \emph{correlations}
between systems.
There are, in fact, different types of correlations that can be represented by
density matrices.
Here are a few examples.

\begin{description}
\item[Correlated classical states.]
  For example, we can express the situation in which Alice and Bob share a
  random bit like this:
  \[
  \frac{1}{2} \vert 0 \rangle \langle 0 \vert \otimes \vert 0 \rangle \langle 0
  \vert +
  \frac{1}{2} \vert 1 \rangle \langle 1 \vert \otimes \vert 1 \rangle \langle 1
  \vert
  =
  \begin{pmatrix}
    \frac{1}{2} & 0 & 0 & 0\\[2mm]
    0 & 0 & 0 & 0\\[2mm]
    0 & 0 & 0 & 0\\[2mm]
    0 & 0 & 0 & \frac{1}{2}
  \end{pmatrix}
  \]

\item[Ensembles of quantum states.]
  Suppose we have $m$ density matrices $\rho_0,\ldots,\rho_{m-1},$ all
  representing states of a system $\mathsf{X},$ and we  randomly choose one of
  these states according to a probability vector $(p_0,\ldots,p_{m-1}).$ Such a
  process is represented by an \emph{ensemble} of states, which includes the
  specification of the density matrices $\rho_0,\ldots,\rho_{m-1},$ as well as
  the probabilities $(p_0,\ldots,p_{m-1}).$ We can associate an ensemble of
  states with a single density matrix, describing both the random choice of $k$
  and the corresponding density matrix $\rho_k,$ like this:
  \[
  \sum_{k = 0}^{m-1} p_k \vert k\rangle \langle k \vert \otimes \rho_k.
  \]
  To be clear, this is the state of a pair $(\mathsf{Y},\mathsf{X})$ where
  $\mathsf{Y}$ represents the classical selection of $k$ --- so we're assuming
  its classical state set is $\{0,\ldots,m-1\}.$ States of this form are
  sometimes called \emph{classical-quantum states}.

\item[Separable states.]
  We can imagine situations in which we have a classical correlation among the
  quantum states of two systems like this:
  \[
  \sum_{k = 0}^{m-1} p_k \rho_k \otimes \sigma_k.
  \]
  In words, for each $k$ from $0$ to $m-1,$ we have that with probability $p_k$
  the system on the left is in the state $\rho_k$ and the system on the right
  is in the state~$\sigma_k.$
  States like this are called \emph{separable states}.
  This concept can also be extended to more than two systems.

\item[Entangled states.]
  Not all states of pairs of systems are separable. In the general formulation
  of quantum information, this is how entanglement is defined: states that are
  not separable are said to be \emph{entangled}.

  Note that this terminology is consistent with the terminology we used in
  Lesson~\ref{lesson:entanglement-in-action} \emph{(Entanglement in Action)}.
  There we said that quantum state vectors that are not product states
  represent entangled states --- and indeed, for any quantum state vector
  $\vert\psi\rangle$ that is not a product state, we find that the state
  represented by the density matrix $\vert\psi\rangle\langle\psi\vert$ is not
  separable. Entanglement is much more complicated than this for states that
  are not pure.
\end{description}

\pagebreak

\subsection{Reduced states and the partial trace}

There's a simple but important thing we can do with density matrices in the
context of multiple systems, which is to describe the states we obtain by
ignoring some of the systems.
When multiple systems are in a quantum state and we discard or choose to ignore
one or more of the systems, the state of the remaining systems is called the
\emph{reduced state} of those systems.
Density matrix descriptions of reduced states are easily obtained through a
mapping, known as the \emph{partial trace}, from the density matrix describing
the state of the whole.

\subsubsection{Example: reduced states for an e-bit}

Suppose that we have a pair of qubits $(\mathsf{A},\mathsf{B})$ that are
together in the state
\[
\vert\phi^+\rangle = \frac{1}{\sqrt{2}} \vert 00 \rangle + \frac{1}{\sqrt{2}}
\vert 11 \rangle.
\]
We can imagine that Alice holds the qubit $\mathsf{A}$ and Bob holds
$\mathsf{B},$ which is to say that together they share an e-bit.
We'd like to have a density matrix description of Alice's qubit $\mathsf{A}$ in
isolation, as if Bob decided to take his qubit and visit the stars, never to be
seen again.

First let's think about what would happen if Bob decided somewhere on his
journey to measure his qubit with respect to a standard basis measurement.
If he did this, he would obtain the outcome $0$ with probability
\[
\bigl\| \bigl( \mathbb{I}_{\mathsf{A}} \otimes \langle 0\vert \bigr) \vert
\phi^+ \rangle \bigr\|^2
= \Bigl\| \frac{1}{\sqrt{2}} \vert 0 \rangle \Bigr\|^2 = \frac{1}{2},
\]
in which case the state of Alice's qubit becomes $\vert 0\rangle;$ and he would
obtain the outcome $1$ with probability
\[
\bigl\| \bigl( \mathbb{I}_{\mathsf{A}} \otimes \langle 1\vert \bigr) \vert
\phi^+ \rangle \bigr\|^2
= \Bigl\| \frac{1}{\sqrt{2}} \vert 1 \rangle \Bigr\|^2 = \frac{1}{2},
\]
in which case the state of Alice's qubit becomes $\vert 1\rangle.$

So, if we ignore Bob's measurement outcome and focus on Alice's qubit, we
conclude that she obtains the state $\vert 0\rangle$ with probability $1/2$ and
the state $\vert 1\rangle$ with probability $1/2.$
This leads us to describe the state of Alice's qubit in isolation by the
density matrix
\[
\frac{1}{2} \vert 0\rangle\langle 0\vert + \frac{1}{2} \vert 1\rangle\langle
1\vert = \frac{1}{2} \mathbb{I}_{\mathsf{A}}.
\]
That is, Alice's qubit is in the completely mixed state.
To be clear, this description of the state of Alice's qubit doesn't include
Bob's measurement outcome; we're ignoring Bob altogether.

Now, it might seem like the density matrix description of Alice's qubit in
isolation that we've just obtained relies on the assumption that Bob has
measured his qubit, but this is not actually so.
What we've done is to use the possibility that Bob measures his qubit to argue
that the completely mixed state arises as the state of Alice's qubit, based on
what we've already learned.
Of course, nothing says that Bob must measure his qubit --- but nothing says
that he doesn't.
And if he's light years away, then nothing he does or doesn't do can possibly
influence the state of Alice's qubit viewed it in isolation.
That is to say, the description we've obtained for the state of Alice's qubit
is the only description consistent with the impossibility of faster-than-light
communication.

We can also consider the state of Bob's qubit $\mathsf{B},$ which happens to be
the completely mixed state as well.
Indeed, for all four Bell states we find that the reduced state of both Alice's
qubit and Bob's qubit is the completely mixed state.

\subsubsection{Reduced states for a general quantum state vector}

Now let's generalize the example just discussed to two arbitrary systems
$\mathsf{A}$ and $\mathsf{B},$ not necessarily qubits in the state $\vert
\phi^+\rangle.$
We'll assume the classical state sets of $\mathsf{A}$ and $\mathsf{B}$ are
$\Sigma$ and $\Gamma,$ respectively.
A density matrix $\rho$ representing a state of the combined system
$(\mathsf{A},\mathsf{B})$ therefore has row and column indices corresponding to
the Cartesian product $\Sigma\times\Gamma.$

Suppose that the state of $(\mathsf{A},\mathsf{B})$ is described by the quantum
state vector $\vert\psi\rangle,$ so the density matrix describing this state is
$\rho = \vert\psi\rangle\langle\psi\vert.$
We'll obtain a density matrix description of the state of $\mathsf{A}$ in
isolation, which is conventionally denoted $\rho_{\mathsf{A}}.$
(A superscript is also sometimes used rather than a subscript.)

The state vector $\vert\psi\rangle$ can be expressed in the form
\[
\vert\psi\rangle = \sum_{b\in\Gamma} \vert\phi_b\rangle \otimes \vert b\rangle
\]
for a uniquely determined collection of vectors
$\{\vert\phi_b\rangle : b\in\Gamma\}.$
In particular, these vectors can be determined through a simple formula.
\[
\vert\phi_b\rangle = \bigl(\mathbb{I}_{\mathsf{A}} \otimes \langle
b\vert\bigr)\vert\psi\rangle
\]

Reasoning similarly to the previous example of an e-bit, if we were to measure
the system $\mathsf{B}$ with a standard basis measurement, we would obtain each
outcome $b\in\Gamma$ with probability $\|\vert\phi_b\rangle\|^2,$ in which case
the state of $\mathsf{A}$ becomes
\[
\frac{\vert \phi_b \rangle}{\|\vert\phi_b\rangle\|}.
\]
As a density matrix, this state can be written as follows.
\[
\biggl(\frac{\vert \phi_b \rangle}{\|\vert\phi_b\rangle\|}\biggr)
\biggl(\frac{\vert \phi_b \rangle}{\|\vert\phi_b\rangle\|}\biggr)^{\dagger}
= \frac{\vert \phi_b \rangle\langle\phi_b\vert}{\|\vert\phi_b\rangle\|^2}
\]
Averaging the different states according to the probabilities of the respective
outcomes, we arrive at the density matrix
\[
\rho_{\mathsf{A}} = \sum_{b\in\Gamma}
\|\vert\phi_b\rangle\|^2 \frac{\vert \phi_b
  \rangle\langle\phi_b\vert}{\|\vert\phi_b\rangle\|^2}
= \sum_{b\in\Gamma} \vert \phi_b \rangle\langle\phi_b\vert
= \sum_{b\in\Gamma}
\bigl(\mathbb{I}_{\mathsf{A}} \otimes \langle b\vert\bigr)
\vert\psi\rangle\langle\psi\vert
\bigl(\mathbb{I}_{\mathsf{A}} \otimes \vert b\rangle\bigr)
\]

\subsubsection{The partial trace}

The formula
\[
\rho_{\mathsf{A}}
= \sum_{b\in\Gamma}
\bigl(\mathbb{I}_{\mathsf{A}} \otimes \langle b\vert\bigr)
\vert\psi\rangle\langle\psi\vert
\bigl(\mathbb{I}_{\mathsf{A}} \otimes \vert b\rangle\bigr)
\]
leads us to the description of the reduced state of $\mathsf{A}$ for any
density matrix $\rho$ of the pair $(\mathsf{A},\mathsf{B}),$ not just a pure
state.
\[
\rho_{\mathsf{A}} = \sum_{b\in\Gamma}
\bigl( \mathbb{I}_{\mathsf{A}} \otimes \langle b \vert\bigr)
\rho
\bigl( \mathbb{I}_{\mathsf{A}} \otimes \vert b \rangle\bigr)
\]
This formula must work, simply by linearity together with the fact that every
density matrix can be written as a convex combination of pure states.

The operation being performed on $\rho$ to obtain $\rho_{\mathsf{A}}$ in this
equation is known as the \emph{partial trace}, and to be more precise we say
that the partial trace is performed on~$\mathsf{B},$ or that $\mathsf{B}$ is
\emph{traced out}.
This operation is denoted $\operatorname{Tr}_{\mathsf{B}},$ so we can write
\[
\operatorname{Tr}_{\mathsf{B}} (\rho) =
\sum_{b\in\Gamma}
\bigl( \mathbb{I}_{\mathsf{A}} \otimes \langle b \vert\bigr)
\rho
\bigl( \mathbb{I}_{\mathsf{A}} \otimes \vert b \rangle\bigr).
\]
We can also define the partial trace on $\mathsf{A},$ so it's the system
$\mathsf{A}$ that gets traced out rather than $\mathsf{B},$ like this.
\[
\operatorname{Tr}_{\mathsf{A}} (\rho) =
\sum_{a\in\Sigma}
\bigl(\langle a \vert\otimes\mathbb{I}_{\mathsf{B}}\bigr)
\rho
\bigl(\vert a \rangle\otimes\mathbb{I}_{\mathsf{B}}\bigr)
\]
This gives us the density matrix description $\rho_{\mathsf{B}}$ of the state
of $\mathsf{B}$ in isolation rather than $\mathsf{A}.$

To recapitulate, if $(\mathsf{A},\mathsf{B})$ is any pair of systems and we
have a density matrix $\rho$ describing a state of $(\mathsf{A},\mathsf{B}),$
the \emph{reduced states} of the systems $\mathsf{A}$ and $\mathsf{B}$ are as
follows.
\[
\begin{aligned}
  \rho_{\mathsf{A}} & = \operatorname{Tr}_{\mathsf{B}}(\rho)
  = \sum_{b\in\Gamma} \bigl( \mathbb{I}_{\mathsf{A}} \otimes \langle b
  \vert\bigr) \rho \bigl( \mathbb{I}_{\mathsf{A}} \otimes \vert b
  \rangle\bigr)\\[2mm]
  \rho_{\mathsf{B}} & = \operatorname{Tr}_{\mathsf{A}}(\rho) =
  \sum_{a\in\Sigma} \bigl( \langle a \vert \otimes
  \mathbb{I}_{\mathsf{B}}\bigr) \rho \bigl( \vert a \rangle\otimes
  \mathbb{I}_{\mathsf{B}} \bigr)
\end{aligned}
\]
If $\rho$ is a density matrix, then $\rho_{\mathsf{A}}$ and $\rho_{\mathsf{B}}$
will also necessarily be density matrices.

These notions can be generalized to any number of systems in place of two in a
natural way.
In general, we can put the names of whatever systems we choose in the subscript
of a density matrix $\rho$ to describe the reduced state of just those systems.
For example, if $\mathsf{A},$ $\mathsf{B},$ and $\mathsf{C}$ are systems and
$\rho$ is a density matrix describing a state of
$(\mathsf{A},\mathsf{B},\mathsf{C}),$ then we can define
\[
\begin{aligned}
  \rho_{\mathsf{AC}} & = \operatorname{Tr}_{\mathsf{B}}(\rho) =
  \sum_{b\in\Gamma} \bigl( \mathbb{I}_{\mathsf{A}} \otimes \langle b \vert
  \otimes \mathbb{I}_{\mathsf{C}} \bigr) \rho
  \bigl( \mathbb{I}_{\mathsf{A}} \otimes \vert b \rangle \otimes
  \mathbb{I}_{\mathsf{C}} \bigr) \\[2mm]
  \rho_{\mathsf{C}} & = \operatorname{Tr}_{\mathsf{AB}}(\rho) =
  \sum_{a\in\Sigma} \sum_{b\in\Gamma} \bigl( \langle a \vert \otimes \langle b
  \vert \otimes \mathbb{I}_{\mathsf{C}} \bigr) \rho \bigl( \vert a \rangle
  \otimes \vert b \rangle \otimes \mathbb{I}_{\mathsf{C}} \bigr)
\end{aligned}
\]
and similarly for other choices for the systems.

\subsubsection{Alternative description of the partial trace}

An alternative way to describe the partial trace mappings
$\operatorname{Tr}_{\mathsf{A}}$ and $\operatorname{Tr}_{\mathsf{B}}$ is that
they are the \emph{unique} linear mappings that satisfy the formulas
\[
\begin{aligned}
  \operatorname{Tr}_{\mathsf{A}}(M \otimes N) & = \operatorname{Tr}(M) N \\[2mm]
  \operatorname{Tr}_{\mathsf{B}}(M \otimes N) & = \operatorname{Tr}(N) M.
\end{aligned}
\]
In these formulas, $N$ and $M$ are square matrices of the appropriate sizes:
the rows and columns of $M$ correspond to the classical states of $\mathsf{A}$
and the rows and columns of $N$ correspond to the classical states of
$\mathsf{B}.$

This characterization of the partial trace is not only fundamental from a
mathematical viewpoint, but can also allow for quick calculations in some
situations.
For example, consider this state of a pair of qubits $(\mathsf{A},\mathsf{B}).$
\[
\rho =
\frac{1}{2} \vert 0\rangle\langle 0\vert \otimes \vert 0\rangle\langle 0\vert +
\frac{1}{2} \vert 1\rangle\langle 1\vert \otimes \vert +\rangle\langle +\vert
\]
To compute the reduced state $\rho_{\mathsf{A}}$ for instance, we can use
linearity together with the fact that
$\vert 0\rangle\langle 0\vert$ and $\vert +\rangle\langle +\vert$ have unit
trace.
\[
\rho_{\mathsf{A}} =
\operatorname{Tr}_{\mathsf{B}}(\rho) =
\frac{1}{2} \operatorname{Tr}\bigl(\vert 0\rangle\langle 0\vert\bigr)\, \vert
0\rangle\langle 0\vert + \frac{1}{2} \operatorname{Tr}\bigl(\vert
+\rangle\langle +\vert\bigr) \vert 1\rangle\langle 1\vert = \frac{1}{2} \vert
0\rangle\langle 0\vert + \frac{1}{2} \vert 1\rangle\langle 1\vert
\]
The reduced state $\rho_{\mathsf{B}}$ can be computed similarly.
\[
\rho_{\mathsf{B}} =
\operatorname{Tr}_{\mathsf{A}}(\rho) =
\frac{1}{2} \operatorname{Tr}\bigl(\vert 0\rangle\langle 0\vert\bigr)\, \vert
0\rangle\langle 0\vert + \frac{1}{2} \operatorname{Tr}\bigl(\vert
1\rangle\langle 1\vert\bigr) \vert +\rangle\langle +\vert = \frac{1}{2} \vert
0\rangle\langle 0\vert + \frac{1}{2} \vert +\rangle\langle +\vert
\]

\subsubsection{The partial trace for two qubits}

The partial trace can also be described explicitly in terms of matrices.
Here we'll do this just for two qubits, but this can also be generalized to
larger systems.
Assume that we have two qubits $(\mathsf{A},\mathsf{B}),$ so that any density
matrix describing a state of these two qubits can be written as
\[
\rho = 

\]
for some choice of complex numbers $\{\alpha_{jk} : 0\leq j,k\leq 3\}.$

The partial trace over the first system has the following formula.
\[
\operatorname{Tr}_{\mathsf{A}} %
\]
One way to think about this formula begins by viewing $4\times 4$ matrices as
$2\times 2$ block matrices, where each block is $2\times 2.$
That is,
\[
\rho = %
.
\end{alignat*}
We then have
\[
\operatorname{Tr}_{\mathsf{A}}%
= M_{0,0} + M_{1,1}.
\]

Here's the formula when the second system is traced out rather than the first.
\begin{align*}
\operatorname{Tr}_{\mathsf{B}}
%
\end{align*}
In terms of block matrices of a form similar to before, we have this formula.
\[
\operatorname{Tr}_{\mathsf{B}}
%
\]
The block matrix descriptions of these functions can be extended to systems
larger than qubits in a natural and direct way.

To finish the lesson, let's apply these formulas to the same state we
considered above.
\[
\rho = \frac{1}{2} \vert 0\rangle \langle 0 \vert \otimes \vert 0\rangle
\langle 0 \vert
+ \frac{1}{2} \vert 1\rangle \langle 1 \vert \otimes \vert +\rangle \langle +
\vert
=
%
.
\]
The reduced state of the first system $\mathsf{A}$ is
\[
\operatorname{Tr}_{\mathsf{B}}
%
\]
and the reduced state of the second system $\mathsf{B}$ is
\[
\operatorname{Tr}_{\mathsf{A}}
%
.
\]


\lesson{Quantum Channels}
\label{lesson:quantum-channels}

In the general formulation of quantum information, operations on quantum states
are represented by a special class of mappings called \emph{channels}.
This includes useful operations, such as ones corresponding to unitary gates
and circuits, as well as operations we deem as noise and would prefer to avoid.
We can also describe measurements as channels, which we'll do in the next
lesson.
In short, any discrete-time change in states that is physically realizable (in
an idealized sense) can be described by a channel.

The term \emph{channel} comes to us from information theory, which (among other
things) studies the information-carrying capacities of noisy
\emph{communication channels}.
In this context, a quantum channel could specify the quantum state that's
received when a given quantum state is sent, perhaps through a quantum network
of some sort.

It should be understood, however, that the terminology merely reflects this
historical motivation and is used in a more general way.
Indeed, we can describe a wide variety of things (such as complicated quantum
computations) as channels, even though they have nothing to do with
communication and would be unlikely to arise naturally in such a setting.

We'll begin the lesson with a discussion of some basic aspects of channels,
along with a small selection of examples.
Then we'll move on to three different ways to represent channels in
mathematical terms later in the lesson.
We'll see that, although these representations are different, they all offer
equivalent mathematical characterizations of channels.

\section{Quantum channel basics}

In mathematical terms, channels are linear mappings from density matrices to
density matrices that satisfy certain requirements.
Throughout this lesson we'll use uppercase Greek letters, including $\Phi$ and
$\Psi,$ as well as some other letters in specific cases, to refer to channels.

Every channel $\Phi$ has an input system and an output system, and we'll
typically use the name $\mathsf{X}$ to refer to the input system and
$\mathsf{Y}$ to refer to the output system.
It's common that the output system of a channel is the same as the input
system, and in this case we can use the same letter $\mathsf{X}$ to refer to
both.

\subsection{Channels are linear mappings}

Channels are described by \emph{linear} mappings, just like probabilistic
operations in the standard formulation of classical information and unitary
operations in the simplified formulation of quantum information.

If a channel $\Phi$ is performed on an input system $\mathsf{X}$ whose state is
described by a density matrix $\rho,$ then the output system of the channel is
described by the density matrix $\Phi(\rho).$
In the situation in which the output system of $\Phi$ is also $\mathsf{X},$ we
can simply view that the channel represents a change in the state of
$\mathsf{X},$ from $\rho$ to $\Phi(\rho).$
When the output system of $\Phi$ is a different system, $\mathsf{Y},$ rather
than $\mathsf{X},$ it should be understood that $\mathsf{Y}$ is a new system
that is created by the process of applying the channel, and that the input
system $\mathsf{X}$ is no longer available once the channel is applied --- as
if the channel itself transformed $\mathsf{X}$ into $\mathsf{Y},$ leaving it in
the state $\Phi(\rho).$

The assumption that channels are described by \emph{linear} mappings can be
viewed as being an axiom --- or, in other words, a basic postulate of the
theory rather than something that is proved.
We can, however, see the need for channels to act linearly on convex
combinations of density matrix inputs in order for them to be consistent with
probability theory and what we've already learned about density matrices.

To be more specific, suppose that we have a channel $\Phi$ and we apply it to a
system when it's in one of the two states represented by the density matrices
$\rho$ and~$\sigma.$
If we apply the channel to $\rho$ we obtain the density matrix $\Phi(\rho)$ and
if we apply it to $\sigma$ we obtain the density matrix $\Phi(\sigma).$
Thus, if we randomly choose the input state of $\mathsf{X}$ to be $\rho$ with
probability $p$ and $\sigma$ with probability $1-p,$ we'll obtain the output
state $\Phi(\rho)$ with probability $p,$ and $\Phi(\sigma)$ with probability
$1-p,$ which we represent by a weighted average of density matrices as
$p\Phi(\rho) + (1-p)\Phi(\sigma).$

On the other hand, we could think about the input state of the channel as being
represented by the weighted average $p\rho + (1-p)\sigma,$ in which case the
output is $\Phi(p\rho + (1-p)\sigma).$
It's the same state regardless of how we choose to think about it, so we must
have
\[
\Phi(p\rho + (1-p)\sigma) = p\Phi(\rho) + (1-p)\Phi(\sigma).
\]
Whenever we have a mapping that satisfies this condition for every choice of
density matrices $\rho$ and $\sigma$ and scalars $p\in [0,1],$ there's always a
unique way to extend that mapping to every matrix input (i.e., not just density
matrix inputs) so that it's linear.

\subsection{Channels transform density matrices into density matrices}

Naturally, in addition to being linear mappings, channels must also transform
density matrices into density matrices.
If a channel $\Phi$ is applied to an input system while this system is in a
state represented by a density matrix $\rho,$ then we obtain a system whose
state is represented by $\Phi(\rho),$ which must be a valid density matrix in
order for us to interpret it as a state.

It is critically important, though, that we consider a more general situation,
where a channel $\Phi$ transforms a system $\mathsf{X}$ into a system
$\mathsf{Y}$ in the presence of an additional system $\mathsf{Z}$ to which
nothing happens.
That is, if we start with the pair of systems $(\mathsf{Z},\mathsf{X})$ in a
state described by some density matrix, and then apply $\Phi$ just to
$\mathsf{X},$ transforming it into $\mathsf{Y},$ we must obtain a density
matrix describing a state of the pair $(\mathsf{Z},\mathsf{Y}).$

We can describe in mathematical terms how a channel $\Phi,$ having an input
system $\mathsf{X}$ and an output system $\mathsf{Y},$ transforms a state of
the pair $(\mathsf{Z},\mathsf{X})$ into a state of $(\mathsf{Z},\mathsf{Y})$
when nothing is done to $\mathsf{Z}.$
To keep things simple, we'll assume that the classical state set of
$\mathsf{Z}$ is $\{0,\ldots,m-1\}.$
This allows us to write an arbitrary density matrix $\rho,$ representing a
state of $(\mathsf{Z},\mathsf{X}),$ in the following form.
\[
\rho = \sum_{a,b = 0}^{m-1} \vert a\rangle\langle b\vert \otimes \rho_{a,b}
= \begin{pmatrix}
  \rho_{0,0} & \rho_{0,1} & \cdots & \rho_{0,m-1} \\[1mm]
  \rho_{1,0} & \rho_{1,1} & \cdots & \rho_{1,m-1} \\[1mm]
  \vdots & \vdots & \ddots & \vdots\\[1mm]
  \rho_{m-1,0} & \rho_{m-1,1} & \cdots & \rho_{m-1,m-1}
\end{pmatrix}
\]
On the right-hand side of this equation we have a block matrix, which can
alternatively be described using Dirac notation as we have in the middle
expression.
Each matrix $\rho_{a,b}$ has rows and columns corresponding to the classical
states of $\mathsf{X},$ and these matrices can be determined by a simple
formula.
\[
\rho_{a,b} = \bigl(\langle a \vert \otimes \mathbb{I}_{\mathsf{X}} \bigr) \rho
\bigl(\vert b \rangle \otimes \mathbb{I}_{\mathsf{X}} \bigr)
\]
Note that these are not density matrices in general --- it's only when they're
arranged together to form $\rho$ that we obtain a density matrix.

The following equation describes the state of $(\mathsf{Z},\mathsf{Y})$ that is
obtained when $\Phi$ is applied to $\mathsf{X}.$
\[
\sum_{a,b = 0}^{m-1} \vert a\rangle\langle b\vert \otimes \Phi(\rho_{a,b})
= \begin{pmatrix}
\Phi(\rho_{0,0}) & \Phi(\rho_{0,1}) & \cdots & \Phi(\rho_{0,m-1}) \\[1mm]
\Phi(\rho_{1,0}) & \Phi(\rho_{1,1}) & \cdots & \Phi(\rho_{1,m-1}) \\[1mm]
\vdots & \vdots & \ddots & \vdots\\[1mm]
\Phi(\rho_{m-1,0}) & \Phi(\rho_{m-1,1}) & \cdots & \Phi(\rho_{m-1,m-1})
\end{pmatrix}
\]
Notice that, in order to evaluate this expression for a given choice of $\Phi$
and $\rho,$ we must understand how $\Phi$ works as a linear mapping on
non-density matrix inputs, as each $\rho_{a,b}$ generally won't be a density
matrix on its own.

The previous equation is consistent with the expression
$(\operatorname{Id}_{\mathsf{Z}} \otimes \,\Phi)(\rho),$ in which
$\operatorname{Id}_{\mathsf{Z}}$ denotes the \emph{identity channel} on the
system $\mathsf{Z}.$
This presumes that we've extended the notion of a tensor product to linear
mappings from matrices to matrices, which is straightforward --- but it isn't
really essential to the lesson and won't be explained further.

Reiterating a statement made above, in order for a linear mapping $\Phi$ to be
a valid channel it must be the case that, for every choice for $\mathsf{Z}$ and
every density matrix $\rho$ of the pair $(\mathsf{Z},\mathsf{X}),$ we always
obtain a density matrix when $\Phi$ is applied to~$\mathsf{X}.$
In mathematical terms, the properties a mapping must possess to be a channel
are that it must be \emph{trace-preserving} --- so that the matrix we obtain by
applying the channel has trace equal to one --- as well as \emph{completely
positive} --- so that the resulting matrix is positive semidefinite.
These are both important properties that can be considered and studied
separately, but it isn't critical for the sake of this lesson to consider them
in isolation.

There are, in fact, linear mappings that always output a density matrix when
given a density matrix as input, but fail to map density matrices to density
matrices for compound systems, so we do eliminate some linear mappings from the
class of channels in this way.
(The linear mapping given by matrix transposition is the simplest example.)

We have an analogous formula to one above in the case that the two systems
$\mathsf{X}$ and $\mathsf{Z}$ are swapped, so that $\Phi$ is applied to the
system on the left rather than the right.
\[
\bigl(\Phi\otimes\operatorname{Id}_{\mathsf{Z}}\bigr)(\rho)
= \sum_{a,b = 0}^{m-1} \Phi(\rho_{a,b}) \otimes \vert a\rangle\langle b\vert
\]
This assumes that $\rho$ is a state of $(\mathsf{X},\mathsf{Z})$ rather than
$(\mathsf{Z},\mathsf{X}).$
This time the block matrix description doesn't work because the matrices
$\rho_{a,b}$ don't fall into consecutive rows and columns in $\rho,$ but it's
the same underlying mathematical structure.

Any linear mapping that satisfies the requirement that it always transforms
density matrices into density matrices, even when it's applied to just one part
of a compound systems, represents a valid channel.
So, in an abstract sense, the notion of a channel is determined by the notion
of a density matrix, together with the assumption that channels act linearly.
In this regard, channels are analogous to unitary operations in the simplified
formulation of quantum information, which are precisely the linear mappings
that always transform quantum state vectors to quantum state vectors for a
given system; as well as to probabilistic operations (represented by stochastic
matrices) in the standard formulation of classical information, which are
precisely the linear mappings that always transform probability vectors into
probability vectors.

\subsection{Unitary operations as channels}

Suppose $\mathsf{X}$ is a system and $U$ is a unitary matrix representing an
operation on $\mathsf{X}.$
The channel $\Phi$ that describes this operation on density matrices is defined
as follows for every
density matrix $\rho$ representing a quantum state of $\mathsf{X}.$
\begin{equation}
  \Phi(\rho) = U \rho U^{\dagger}
  \label{eq:unitary-conjugation}
\end{equation}
This action, where we multiply by $U$ on the left and $U^{\dagger}$ on the
right, is commonly referred to as \emph{conjugation} by the matrix $U.$

This description is consistent with the fact that the density matrix that
represents a given
quantum state vector $\vert\psi\rangle$ is $\vert\psi\rangle\langle\psi\vert.$
In particular, if the unitary operation $U$ is performed on $\vert\psi\rangle,$
then the output state is represented by the vector $U\vert\psi\rangle,$ and so
the density matrix describing this state is equal to
\[
(U \vert \psi \rangle )( U \vert \psi \rangle )^{\dagger} = U
\vert\psi\rangle\langle\psi\vert U^{\dagger}.
\]

Once we know that, as a channel, the operation $U$ has the action
\[
\vert\psi\rangle\langle \psi\vert \mapsto U
\vert\psi\rangle\langle\psi\vert U^{\dagger}
\]
on pure states, we can conclude by linearity that it must work as is specified
by the equation \eqref{eq:unitary-conjugation} for any
density matrix $\rho.$

The particular channel we obtain when we take $U = \mathbb{I}$ is the
\emph{identity channel} $\operatorname{Id},$ which we can also give a subscript
(such as $\operatorname{Id}_{\mathsf{Z}},$ as we've already encountered) when
we wish to indicate explicitly what system this channel acts on.
Its output is always equal to its input: $\operatorname{Id}(\rho) = \rho.$
This might not seem like an interesting channel, but it's actually a very
important one --- and it's fitting that this is our first example.
The identity channel is the \emph{perfect} channel in some contexts,
representing an ideal memory or a perfect, noiseless transmission of
information from a sender to a receiver.

Every channel defined by a unitary operation in this way is indeed a valid
channel:
conjugation by a matrix $U$ gives us a linear map; and if $\rho$ is a density
matrix of a system $(\mathsf{Z},\mathsf{X})$ and $U$ is unitary, then the
result, which we can express as
\[
(\mathbb{I}_{\mathsf{Z}} \otimes U) \rho (\mathbb{I}_{\mathsf{Z}} \otimes
U^{\dagger}),
\]
is also a density matrix.
Specifically, this matrix must be positive semidefinite, for if $\rho =
M^{\dagger} M$ then
\[
(\mathbb{I}_{\mathsf{Z}} \otimes U) \rho (\mathbb{I}_{\mathsf{Z}} \otimes
U^{\dagger}) = K^{\dagger} K
\]
for $K = M (\mathbb{I}_{\mathsf{Z}} \otimes U^{\dagger}),$ and it must have
unit trace by the cyclic property of the trace.
\[
\operatorname{Tr}\bigl((\mathbb{I}_{\mathsf{Z}} \otimes U) \rho
(\mathbb{I}_{\mathsf{Z}} \otimes U^{\dagger})\bigr)
= \operatorname{Tr}\bigl((\mathbb{I}_{\mathsf{Z}} \otimes
U^{\dagger})(\mathbb{I}_{\mathsf{Z}} \otimes U) \rho \bigr)
= \operatorname{Tr}\bigl((\mathbb{I}_{\mathsf{Z}} \otimes
\mathbb{I}_{\mathsf{X}}) \rho \bigr)
= \operatorname{Tr}(\rho) = 1
\]

\subsection{Convex combinations of channels}

Suppose we have two channels, $\Phi_0$ and $\Phi_1,$ that share the same input
system and the same output system.
For any real number $p\in[0,1],$ we could decide to apply $\Phi_0$ with
probability $p$ and $\Phi_1$ with probability $1-p,$ which gives us a new
channel that can be written as $p \Phi_0 + (1-p) \Phi_1.$
Explicitly, the way that this channel acts on a given density matrix is
specified by the following simple equation.
\[
(p \Phi_0 + (1-p) \Phi_1)(\rho) = p \Phi_0(\rho) + (1-p) \Phi_1(\rho)
\]

More generally, if we have channels $\Phi_{0},\ldots,\Phi_{m-1}$ and a
probability vector $(p_0,\ldots, p_{m-1}),$ then we can average these channels
together to obtain a new channel.
\[
\sum_{k = 0}^{m-1} p_k \Phi_k
\]
This is a \emph{convex combination} of channels, and we always obtain a valid
channel through this process.
A simple way to say this in mathematical terms is that, for a given choice of
an input and output system, the set of all channels is a \emph{convex set}.

As an example, we could choose to apply one of a collection of \emph{unitary}
operations to a certain system.
We obtain what's known as a \emph{mixed unitary} channel, which is a channel
that can be expressed in the following form.
\[
\Phi(\rho) = \sum_{k=0}^{m-1} p_k U_k \rho U_k^{\dagger}
\]
Mixed unitary channels for which all of the unitary operations are Pauli
matrices (or tensor products of Pauli matrices) are called \emph{Pauli
channels}, and are commonly encountered in quantum computing.

\subsection{Examples of qubit channels}

Now we'll take a look at a few specific examples of channels that aren't
unitary.
For all of these examples, the input and output systems are both single qubits,
which is to say that these are examples of \emph{qubit channels}.

\subsubsection{The qubit reset channel}

This channel does something very simple: it resets a qubit to the $\vert
0\rangle$ state.
As a linear mapping this channel can be expressed as follows for every qubit
density matrix $\rho.$
\[
\Lambda(\rho) = \operatorname{Tr}(\rho) \vert 0\rangle\langle 0\vert
\]

Although the trace of every density matrix $\rho$ is equal to $1,$ writing the
channel in this way makes it clear that it's a linear mapping that could be
applied to any $2\times 2$ matrix, not just a density matrix.
As we already observed, we need to understand how channels work as linear
mappings on non-density matrix inputs to describe what happens when they're
applied to just one part of a compound system.

For example, suppose that $\mathsf{A}$ and $\mathsf{B}$ are qubits and together
the pair $(\mathsf{A},\mathsf{B})$ is in the Bell state $\vert \phi^+\rangle.$
As a density matrix, this state is given by
\[
\vert \phi^+\rangle\langle \phi^+ \vert =
\begin{pmatrix}
\frac{1}{2} & 0 & 0 & \frac{1}{2} \\[1mm]
0 & 0 & 0 & 0 \\[1mm]
0 & 0 & 0 & 0 \\[1mm]
\frac{1}{2} & 0 & 0 & \frac{1}{2}
\end{pmatrix}.
\]
Using Dirac notation we can alternatively express this state as follows.
\[
\vert \phi^+\rangle\langle \phi^+ \vert =
\frac{1}{2} \vert 0 \rangle \langle 0 \vert \otimes \vert 0 \rangle \langle 0
\vert +
\frac{1}{2} \vert 0 \rangle \langle 1 \vert \otimes \vert 0 \rangle \langle 1
\vert +
\frac{1}{2} \vert 1 \rangle \langle 0 \vert \otimes \vert 1 \rangle \langle 0
\vert +
\frac{1}{2} \vert 1 \rangle \langle 1 \vert \otimes \vert 1 \rangle \langle 1
\vert
\]
By applying the qubit reset channel to $\mathsf{A}$ and doing nothing to
$\mathsf{B}$ we obtain the following state.
\[
\begin{aligned}
  & \frac{1}{2} \Lambda(\vert 0 \rangle \langle 0 \vert) \otimes \vert 0 \rangle
  \langle 0 \vert +
  \frac{1}{2} \Lambda(\vert 0 \rangle \langle 1 \vert) \otimes \vert 0 \rangle
  \langle 1 \vert \\[1mm]
  + & \frac{1}{2} \Lambda(\vert 1 \rangle \langle 0 \vert) \otimes \vert 1
  \rangle \langle 0 \vert + \frac{1}{2} \Lambda(\vert 1 \rangle \langle 1
  \vert) \otimes \vert 1 \rangle \langle 1 \vert
  \\[2mm]
  & \qquad\qquad
  = \frac{1}{2} \vert 0 \rangle \langle 0 \vert \otimes \vert 0 \rangle \langle
  0 \vert + \frac{1}{2} \vert 0 \rangle \langle 0 \vert \otimes \vert 1 \rangle
  \langle 1 \vert\\[1mm]
  & \qquad\qquad
  = \vert 0\rangle \langle 0\vert \otimes \frac{\mathbb{I}}{2} &
\end{aligned}
\]

It might be tempting to say that resetting $\mathsf{A}$ has had an effect on
$\mathsf{B},$ causing it to become completely mixed --- but in some sense it's
actually the opposite.
Before $\mathsf{A}$ was reset, the reduced state of $\mathsf{B}$ was the
completely mixed state, and that doesn't change as a result of resetting
$\mathsf{A}.$

\subsubsection{The completely dephasing channel}

Here's an example of a qubit channel called $\Delta,$ described by its action
on $2\times 2$ matrices:
\[
\Delta
\begin{pmatrix}
   \alpha_{00} & \alpha_{01}\\[1mm]
   \alpha_{10} & \alpha_{11}
\end{pmatrix}
= \begin{pmatrix}
  \alpha_{00} & 0\\[1mm]
  0 & \alpha_{11}
\end{pmatrix}.
\]
In words, $\Delta$ zeros out the off-diagonal entries of a $2\times 2$ matrix.
This example can be generalized to arbitrary systems, as opposed to qubits: for
whatever density matrix is input, the channel zeros out all of the off-diagonal
entries and leaves the diagonal alone.

This channel is called the \emph{completely dephasing channel}, and it can be
thought of as representing an extreme form of the process known as
\emph{decoherence} --- which essentially ruins quantum superpositions and
turns them into classical probabilistic states.

Another way to think about this channel is that it describes a standard basis
measurement on a qubit, where an input qubit is measured and then discarded,
and where the output is a density matrix describing the measurement outcome.
Alternatively, but equivalently, we can imagine that the measurement outcome is
discarded, leaving the qubit in its post-measurement state.

Let us again consider an e-bit, and see what happens when $\Delta$ is applied
to just one of the two qubits.
Specifically, we have qubits $\mathsf{A}$ and $\mathsf{B}$ for which
$(\mathsf{A},\mathsf{B})$ is in the state $\vert\phi^+\rangle,$ and this time
let's apply the channel to the second qubit.
Here's the state we obtain.
\[
\begin{aligned}
  & \frac{1}{2} \vert 0 \rangle \langle 0 \vert \otimes \Delta(\vert 0 \rangle
  \langle 0 \vert) +
  \frac{1}{2} \vert 0 \rangle \langle 1 \vert \otimes \Delta(\vert 0 \rangle
  \langle 1 \vert) \\[2mm]
  + & \frac{1}{2} \vert 1 \rangle \langle 0 \vert \otimes \Delta(\vert 1
  \rangle \langle 0 \vert) +
  \frac{1}{2} \vert 1 \rangle \langle 1 \vert \otimes \Delta(\vert 1 \rangle
  \langle 1 \vert) \\[2mm]
  & \qquad\qquad
  = \frac{1}{2} \vert 0 \rangle \langle 0 \vert \otimes \vert 0 \rangle \langle
  0 \vert
  + \frac{1}{2} \vert 1 \rangle \langle 1 \vert \otimes \vert 1 \rangle \langle
  1 \vert
\end{aligned}
\]
Alternatively we can express this equation using block matrices.
\[
\begin{pmatrix}
  \Delta\begin{pmatrix}
  \frac{1}{2} & 0\\[1mm]
  0 & 0
  \end{pmatrix}
  & \Delta\begin{pmatrix}
  0 & \frac{1}{2}\\[1mm]
  0 & 0
  \end{pmatrix} \\[5mm]
  \Delta\begin{pmatrix}
  0 & 0\\[1mm]
  \frac{1}{2} & 0
  \end{pmatrix}
  & \Delta\begin{pmatrix}
  0 & 0\\[1mm]
  0 & \frac{1}{2}
  \end{pmatrix}
\end{pmatrix}
= \begin{pmatrix}
  \frac{1}{2} & 0 & 0 & 0\\[1mm]
  0 & 0 & 0 & 0\\[1mm]
  0 & 0 & 0 & 0\\[1mm]
  0 & 0 & 0 & \frac{1}{2}
\end{pmatrix}
\]

We can also consider a qubit channel that only slightly dephases a qubit, as
opposed to completely dephasing it, which is a less extreme form of decoherence
than what is represented by the completely dephasing channel.
In particular, suppose that $\varepsilon \in (0,1)$ is a small but nonzero real
number.
We can define a channel
\[
\Delta_{\varepsilon} = (1 - \varepsilon) \operatorname{Id} + \varepsilon \Delta,
\]
which transforms a given qubit density matrix $\rho$ like this:
\[
\Delta_{\varepsilon}(\rho) = (1 - \varepsilon) \rho + \varepsilon \Delta(\rho).
\]
That is, nothing happens with probability $1-\varepsilon,$ and with probability
$\varepsilon,$ the qubit dephases.
In terms of matrices, this action can be expressed as follows, where the
diagonal entries are left alone and the off-diagonal entries are multiplied by
$1-\varepsilon.$
\[
\rho =
\begin{pmatrix}
  \langle 0\vert \rho \vert 0 \rangle & \langle 0\vert \rho \vert 1 \rangle
  \\[1mm]
  \langle 1\vert \rho \vert 0 \rangle & \langle 1\vert \rho \vert 1 \rangle
\end{pmatrix}
\mapsto
\begin{pmatrix}
  \langle 0\vert \rho \vert 0 \rangle & (1-\varepsilon) \langle 0\vert \rho
  \vert 1 \rangle \\[1mm]
  (1-\varepsilon) \langle 1\vert \rho \vert 0 \rangle & \langle 1\vert \rho
  \vert 1 \rangle
\end{pmatrix}
\]

\subsubsection{The completely depolarizing channel}

Here's another example of a qubit channel called $\Omega.$
\[
\Omega(\rho) = \operatorname{Tr}(\rho) \frac{\mathbb{I}}{2}
\]
Here, $\mathbb{I}$ denotes the $2\times 2$ identity matrix.
In words, for any density matrix input~$\rho,$ the channel $\Omega$ outputs the
completely mixed state.
It doesn't get any noisier than this!
This channel is called the \emph{completely depolarizing channel}, and like the
completely dephasing channel it can be generalized to arbitrary systems in
place of qubits.

We can also consider a less extreme variant of this channel where depolarizing
happens with probability $\varepsilon,$ similar to what we saw for the
dephasing channel.
\[
\Omega_{\varepsilon}(\rho) = (1 - \varepsilon) \rho + \varepsilon \Omega(\rho).
\]

\section{Channel representations}

Next, we'll discuss mathematical representations of channels.

Linear mappings from vectors to vectors can be represented by matrices in a
familiar way, where the action of the linear mapping is described by
matrix-vector multiplication.
But channels are linear mappings from matrices to matrices, not vectors to
vectors.
So, \emph{in general}, how can we express channels in mathematical terms?

For some channels, we may have a simple formula that describes them, like for
the three examples of non-unitary qubit channels described previously.
But an arbitrary channel may not have such a nice formula, so it isn't
practical in general to express a channel in this way.

As a point of comparison, in the simplified formulation of quantum information
we use \emph{unitary matrices} to represent operations on quantum state
vectors: every unitary matrix represents a valid operation and every valid
operation can be expressed as a unitary matrix.
In essence, the question being asked is: How can we do something analogous for
channels?

To answer this question, we'll require some additional mathematical machinery.
We'll see that channels can, in fact, be described mathematically in a few
different ways, including representations named in honor of three individuals
who played key roles in their development:
Stinespring, Kraus, and Choi.
Together, these different ways of describing channels offer different angles
from which they can be viewed and analyzed.

\subsection{Stinespring representations}

Stinespring representations are based on the idea that every channel can be
implemented in a standard way, where an input system is first combined with an
initialized workspace system, forming a compound system;
then a unitary operation is performed on the compound system;
and finally the workspace system is discarded (or traced out), leaving the
output of the channel.

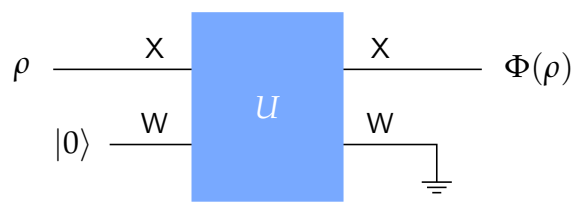
\begin{figure}[b]
  \begin{center}
    \begin{tikzpicture}[
        line width = 0.6pt,
        gate/.style={%
          inner sep = 0,
          fill = CircuitBlue,
          draw = CircuitBlue,
          text = white,
          minimum size = 6mm
      }]
      
      \node (Xin) at (-3,1) {};
      \node (Yin) at (-2.25,0) {};
      \node (Xout) at (3,1) {};
      \node (BendY) at (2.25,0) {};
      \node (GroundY) at (2.25,-0.5) {};
      
      \draw (Xin) -- (Xout);          
      \draw (Yin) -- (BendY.center) -- (GroundY.center);
      
      \node[gate, minimum height=25mm, minimum width=20mm]
      (U) at (0,0.5) {$U$};          
      
      \pic at (GroundY) {ground};
      
      \node[anchor = south] at (-1.5,1) {\small $\mathsf{X}$};
      \node[anchor = south] at (-1.5,0) {\small $\mathsf{W}$};
      \node[anchor = south] at (1.5,1) {\small $\mathsf{X}$};
      \node[anchor = south] at (1.5,0) {\small $\mathsf{W}$};
      
      \node[anchor = east] at (Xin) {$\rho$};
      \node[anchor = west] at (Xout) {$\Phi(\rho)$};
      \node[anchor = east, xshift = 0.75mm] at (Yin) {$\ket{0}$};
      
    \end{tikzpicture}
  \end{center}
  \caption{An implementation of a channel from $\mathsf{X}$ to $\mathsf{X}$
    using a unitary operation $U$ and a workspace system $\mathsf{W}$.}
  \label{fig:Stinespring-X-to-X}
\end{figure}

Figure~\ref{fig:Stinespring-X-to-X} depicts such an implementation, in the form
of a circuit diagram, for a channel whose input and output systems are the same
system, $\mathsf{X}.$
In this diagram, the wires represent arbitrary systems, as indicated by the
labels above the wires, and not necessarily single qubits.
Also, the \emph{ground} symbol commonly used in electrical engineering
indicates explicitly that $\mathsf{W}$ is discarded.

In words, the way the implementation works is as follows.
The input system $\mathsf{X}$ begins in some state $\rho,$ while a workspace
system $\mathsf{W}$ is initialized to the standard basis state $\vert
0\rangle.$
A unitary operation $U$ is performed on the pair $(\mathsf{W},\mathsf{X}),$ and
finally the workspace system $\mathsf{W}$ is \emph{traced out}, leaving
$\mathsf{X}$ as the output.

A mathematical expression of the resulting channel, $\Phi,$ is as follows.
\[
\Phi(\rho) = \operatorname{Tr}_{\mathsf{W}} \bigl( U (\vert 0\rangle \langle 0
\vert_{\mathsf{W}} \otimes \rho) U^{\dagger} \bigr)
\]
As usual, we're using Qiskit's ordering convention:
the system $\mathsf{X}$ is on top in the diagram, and therefore corresponds to
the right-hand tensor factor in the formula.

Note that we're presuming that $0$ is a classical state of $\mathsf{W},$ and we
choose it to be the initialized state of this system, which will help to
simplify the mathematics.
One could, however, choose any fixed pure state to represent the initialized
state of~$\mathsf{W}$ without changing the basic properties of the
representation.

\begin{figure}[!ht]
  \begin{center}
    \begin{tikzpicture}[
        line width = 0.6pt,       
        gate/.style={%
          inner sep = 0,
          fill = CircuitBlue,
          draw = CircuitBlue,
          text = white,
          minimum size = 6mm
      }]
      
      \node (Xin) at (-3,1) {};
      \node (Yin) at (-2.25,0) {};
      \node (Xout) at (3,1) {};
      \node (BendY) at (2.25,0) {};
      \node (GroundY) at (2.25,-0.5) {};
      
      \draw (Xin) -- (Xout);          
      \draw (Yin) -- (BendY.center) -- (GroundY.center);
      
      \node[gate, minimum height=25mm, minimum width=20mm]
      (U) at (0,0.5) {$U$};          
      
      \pic at (GroundY) {ground};
      
      \node[anchor = south] at (-1.5,1) {\small $\mathsf{X}$};
      \node[anchor = south] at (-1.5,0) {\small $\mathsf{W}$};
      \node[anchor = south] at (1.5,1) {\small $\mathsf{Y}$};
      \node[anchor = south] at (1.5,0) {\small $\mathsf{G}$};
      
      \node[anchor = east] at (Xin) {$\rho$};
      \node[anchor = west] at (Xout) {$\Phi(\rho)$};
      \node[anchor = east, xshift = 0.75mm] at (Yin) {$\ket{0}$};
      
    \end{tikzpicture}   
  \end{center}
  \caption{An implementation of a channel from $\mathsf{X}$ to $\mathsf{Y}$.
    The unitary operation $U$ transforms $(\mathsf{X},\mathsf{W})$ to
    $(\mathsf{Y},\mathsf{G})$, where $\mathsf{W}$ represents a workspace system
    and $\mathsf{G}$ represents a garbage system that is traced out.}
  \label{fig:Stinespring-X-to-Y}
\end{figure}
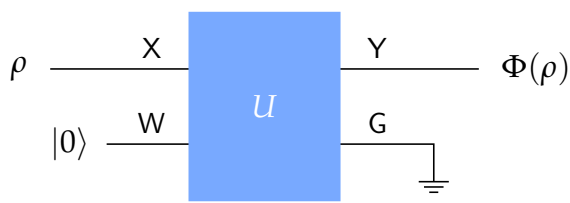

In general, the input and output systems of a channel need not be the same.
Figure~\ref{fig:Stinespring-X-to-Y} shows an implementation of a channel $\Phi$
whose input system is $\mathsf{X}$ and whose output system is $\mathsf{Y}.$
This time the unitary operation transforms $(\mathsf{W},\mathsf{X})$ into a
pair $(\mathsf{G},\mathsf{Y}),$ where $\mathsf{G}$ is a new ``garbage'' system
that gets traced out, leaving $\mathsf{Y}$ as the output system.

In order for $U$ to be unitary, it must be a square matrix.
This requires that the pair $(\mathsf{G},\mathsf{Y})$ has the same number of
classical states as the pair $(\mathsf{W},\mathsf{X}),$ and so the systems
$\mathsf{W}$ and $\mathsf{G}$ must be chosen in a way that allows this.
We obtain a mathematical expression of the resulting channel, $\Phi,$ that is
similar to what we had before.
\[
\Phi(\rho) = \operatorname{Tr}_{\mathsf{G}} \bigl( U (\vert 0\rangle \langle 0
\vert_{\mathsf{W}} \otimes \rho) U^{\dagger} \bigr)
\]

When a channel is described in this way, as a unitary operation along with a
specification of how the workspace system is initialized and how the output
system is selected, we say that it is expressed in \emph{Stinespring form} or
that it's a \emph{Stinespring representation} of the channel.

It's not at all obvious, but every channel does in fact have a Stinespring
representation, as we will see by the end of the lesson.
We'll also see that Stinespring representations aren't unique; there will
always be different ways to implement the same channel in the manner that's
been described.

\begin{trivlist}
\item
  \textbf{Remark.}
  In the context of quantum information, the term \emph{Stinespring
  representation} commonly refers to a slightly more general expression of a
  channel having the form
  \[
  \Phi(\rho) = \operatorname{Tr}_{\mathsf{G}} \bigl( A \rho A^{\dagger} \bigr)
  \]
  for an \emph{isometry} $A,$ which is a matrix whose columns are orthonormal
  but that might not be a square matrix.
  For Stinespring representations having the form that we've adopted as a
  definition, we can obtain an expression of this other
  form by taking
  \[
  A = U (\vert 0\rangle_{\mathsf{W}} \otimes \mathbb{I}_{\mathsf{X}}).
  \]
\end{trivlist}

\subsubsection{Completely dephasing channel}

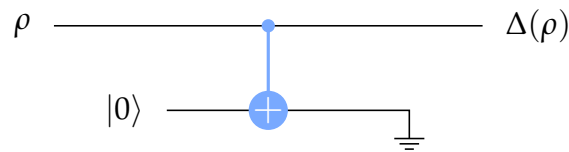
\begin{figure}[!ht]
  \begin{center}
    \begin{tikzpicture}[
        scale=1.5,
        line width = 0.6pt,
        control/.style={%
          circle,
          fill=CircuitBlue,
          minimum size = 5pt,
          inner sep=0mm},
        not/.style={%
          circle,
          fill = CircuitBlue,
          draw = CircuitBlue,
          text = white,
          minimum size = 5mm,
          inner sep=0mm,
          label = {center:\textcolor{white}{\large $+$}}
        }
      ]
      
      \node (In1) at (-2,0.75) {};
      \node (In2) at (-1,0) {};
      \node (Out1) at (2,0.75) {};
      \node (Bend) at (1.25,0) {};
      \node (Ground) at (1.25,-0.25) {};
      
      \draw (In1) -- (Out1);
      \draw (In2) -- (Bend.center) -- (Ground.center);            
      \pic at (Ground) {ground};
      
      \node[control] (Control1) at (0,0.75) {};
      \node[not] (CNot) at (0,0) {};
      
      \draw[very thick,draw=CircuitBlue] (Control1.center) --
      ([yshift=-0.1pt]CNot.north);
      
      \node[anchor = east] at (In1) {$\rho$};
      \node[anchor = east] at (In2) {$\ket{0}$};
      
      \node[anchor = west] at (Out1) {$\Delta(\rho)$};

    \end{tikzpicture}
  \end{center}
  \caption{A Stinespring representation of the completely dephasing
    channel.}
  \label{fig:dephasing-circuit}
\end{figure}

Figure~\ref{fig:dephasing-circuit} shows a Stinespring representation of the
qubit dephasing channel $\Delta.$
In this diagram, both wires represent single qubits --- so this is an ordinary
quantum circuit diagram.

To see that the effect that this circuit has on the input qubit is indeed
described by the completely dephasing channel, we can go through the circuit
one step at a time, using the explicit matrix representation of the partial
trace discussed in the previous lesson.
We'll refer to the top qubit as $\mathsf{X}$ --- this is the input and output
of the channel --- and we'll assume that $\mathsf{X}$ starts in some arbitrary
state $\rho.$

The first step is the introduction of a workspace qubit $\mathsf{W}.$
Prior to the controlled-NOT gate being performed, the state of the pair
$(\mathsf{W},\mathsf{X})$ is represented by the following density matrix.
\[
\vert 0\rangle \langle 0 \vert_{\mathsf{W}} \otimes \rho
= 

\]
As per Qiskit's ordering convention, the top qubit $\mathsf{X}$ is on the right
and the bottom qubit $\mathsf{W}$ is on the left.
We're using density matrices rather than quantum state vectors, but they're
tensored together in a similar way to what's done in the simplified formulation
of quantum information.

The next step is to perform the controlled-NOT operation, where $\mathsf{X}$ is
the control and $\mathsf{W}$ is the target.
Still keeping in mind the Qiskit ordering convention, the matrix representation
of this gate is as follows.
\[
%
\]
This is a unitary operation, and to apply it to a density matrix we conjugate
by the unitary matrix.
The conjugate transpose doesn't happen to change this particular matrix, so the
result is as follows.
\begin{multline*}
  \qquad
  \qquad
  %
  \qquad
  \qquad
\end{multline*}

Finally, the partial trace is performed on $\mathsf{W}.$
Recalling the action of this operation on $4\times 4$ matrices, which was
described in the previous lesson, we obtain the following density matrix
output.
\begin{multline*}
  \quad
\operatorname{Tr}_{\mathsf{W}} %
= \Delta(\rho)
\quad
\end{multline*}
We can alternatively compute the partial trace by first converting to Dirac
notation.
\[

=
\begin{array}{r}
  \langle 0\vert \rho \vert 0\rangle \, \vert 0\rangle\langle 0\vert \otimes
  \vert 0\rangle\langle 0\vert \\[1mm]
  +\, \langle 0\vert \rho \vert 1\rangle \, \vert 0\rangle\langle 1\vert
  \otimes \vert 0\rangle\langle 1\vert \\[1mm]
  +\, \langle 1\vert \rho \vert 0\rangle \, \vert 1\rangle\langle 0\vert
  \otimes \vert 1\rangle\langle 0\vert \\[1mm]
  +\, \langle 1\vert \rho \vert 1\rangle \, \vert 1\rangle\langle 1\vert
  \otimes \vert 1\rangle\langle 1\vert
\end{array}
\]
Tracing out the qubit on the left-hand side yields the same answer as before.
\[
\langle 0\vert \rho \vert 0\rangle \, \vert 0\rangle\langle 0\vert
+\, \langle 1\vert \rho \vert 1\rangle \, \vert 1\rangle\langle 1\vert
= \Delta(\rho)
\]

An intuitive way to think about this circuit is that the controlled-NOT
operation effectively copies the classical state of the input qubit, and when
the copy is thrown in the trash the input qubit ``collapses'' probabilistically
to one of the two possible classical states, which is equivalent to complete
dephasing.

\subsubsection{Completely dephasing channel (alternative)}

\begin{figure}[t]
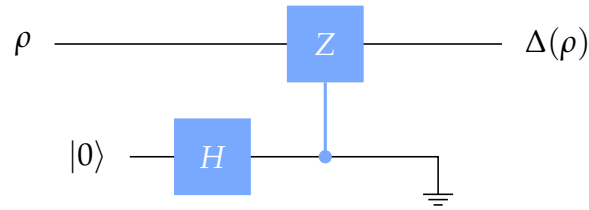

  \begin{center}

  \end{center}
  \caption{An alternative Stinespring representation of the completely
    dephasing channel.}
  \label{fig:dephasing-alternative}
\end{figure}

The circuit described above is not the only way to implement the completely
dephasing channel.
Figure~\ref{fig:dephasing-alternative} illustrates a different way to do it.

Here's a quick analysis showing that this implementation works.
After the Hadamard gate is performed we have this two-qubit state as a density
matrix:
\[
\begin{aligned}
  \vert + \rangle\langle + \vert \otimes \rho
  & = \frac{1}{2}%
.
\end{aligned}
\]
The controlled-$\sigma_z$ gate operates by conjugation as follows.
\begin{multline*}
\frac{1}{2}
%
\end{multline*}
Finally the workspace system $\mathsf{W}$ is traced out.
\begin{align*}
\frac{1}{2}
\operatorname{Tr}_{\mathsf{W}}
%
\end{align*}

This implementation is based on a simple idea:
dephasing is equivalent to either doing nothing (i.e., applying an identity
operation) or applying a $\sigma_z$ gate, each with probability $1/2.$
\[
\begin{aligned}
  \frac{1}{2} \rho + \frac{1}{2} \sigma_z \rho \sigma_z
  & = \frac{1}{2}
  %
\\[2mm]
  & = \Delta(\rho)
\end{aligned}
\]
That is, the completely dephasing channel is an example of a mixed-unitary
channel, and more specifically, a Pauli channel.

\subsubsection{Qubit reset channel}

\begin{figure}[!ht]
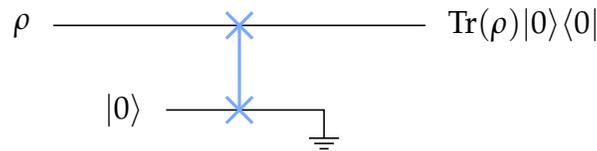

  \begin{center}
    %
  \end{center}
  \caption{A Stinespring representation of the qubit reset channel.}
  \label{fig:reset-stinespring-1}
\end{figure}

The qubit reset channel can be implemented as is illustrated in
Figure~\ref{fig:reset-stinespring-1}.
The swap gate simply shifts the $\vert 0\rangle$ initialized state of the
workspace qubit so that it gets output, while the input state $\rho$ gets moved
to the bottom qubit and then traced out.

\begin{figure}[b]
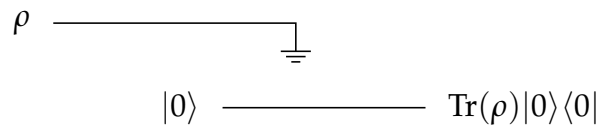

  \begin{center}
    %
  \end{center}
  \caption{An alternative representation of the qubit reset channel.}
  \label{fig:reset-stinespring-2}
\end{figure}

Alternatively, if we don't demand that the output of the channel is left on
top, we can take the very simple circuit shown in
Figure~\ref{fig:reset-stinespring-2} as our representation.
In words, resetting a qubit to the $\vert 0\rangle$ state is equivalent to
throwing the qubit in the trash and getting a new one.

\subsection{Kraus representations}

Now we'll discuss \emph{Kraus representations}, which offer a convenient
formulaic way to express the action of a channel through matrix multiplication
and addition.
In particular, a Kraus representation is a specification of a channel, $\Phi,$
in the following form.
\[
\Phi(\rho) = \sum_{k = 0}^{N-1} A_k \rho A_k^{\dagger}
\]
Here, $A_0,\ldots,A_{N-1}$ are matrices that all have the same dimensions:
their columns correspond to the classical states of the input system,
$\mathsf{X},$ and their rows correspond to the classical states of the output
system, whether it's $\mathsf{X}$ or some other system~$\mathsf{Y}.$
In order for $\Phi$ to be a valid channel, these matrices must satisfy the
following condition.
\[
\sum_{k = 0}^{N-1} A_k^{\dagger} A_k = \mathbb{I}_{\mathsf{X}}
\]
This condition is equivalent to the condition that $\Phi$ preserves trace.
The other property required of a channel --- which is complete positivity ---
follows from the general form of the equation for $\Phi,$ as a sum of
conjugations.

Sometimes it's convenient to name the matrices $A_0,\ldots,A_{N-1}$ in a
different way.
For instance, we could number them starting from $1,$ or we could use states in
some arbitrary classical state set $\Gamma$ instead of numbers as subscripts:
\[
\Phi(\rho) = \sum_{a\in\Gamma} A_a \rho A_a^{\dagger}
\quad
\text{where}
\quad
\sum_{a\in\Gamma} A_a^{\dagger} A_a  = \mathbb{I}.
\]
These different ways of naming these matrices, which are called \emph{Kraus
matrices}, are all common and can be convenient in different situations --- but
we'll stick with the names $A_0,\ldots,A_{N-1}$ in this lesson for the sake of
simplicity.

The number $N$ can be an arbitrary positive integer, but it never needs to be
too large:
if the input system $\mathsf{X}$ has $n$ classical states and the output system
$\mathsf{Y}$ has $m$ classical states, then any given channel from $\mathsf{X}$
to $\mathsf{Y}$ will always have a Kraus representation for which $N$ is at
most the product $nm.$

\subsubsection{Completely dephasing channel}

We obtain a Kraus representation of the completely dephasing channel by taking
$A_0 = \vert 0\rangle\langle 0\vert$ and $A_1 = \vert 1\rangle\langle 1\vert.$
\[
\begin{aligned}
  \sum_{k = 0}^1 A_k \rho A_k^{\dagger}
  & =
  \vert 0\rangle\langle 0 \vert \rho \vert 0\rangle\langle 0 \vert
  + \vert 1\rangle\langle 1 \vert \rho \vert 1\rangle\langle 1 \vert\\
  & = \langle 0 \vert \rho \vert 0\rangle \, \vert 0\rangle\langle 0 \vert
  + \langle 1 \vert \rho \vert 1\rangle \, \vert 1\rangle\langle 1 \vert \\[2mm]
  & =
  \begin{pmatrix}
    \langle 0 \vert \rho \vert 0 \rangle & 0 \\[1mm]
    0 & \langle 1 \vert \rho \vert 1 \rangle
  \end{pmatrix}
\end{aligned}
\]
These matrices satisfy the required condition.
\[
\sum_{k = 0}^1 A_k^{\dagger} A_k
= \vert 0\rangle\langle 0\vert 0\rangle\langle 0\vert + \vert 1\rangle\langle
1\vert 1\rangle\langle 1\vert
= \vert 0\rangle\langle 0\vert
+ \vert 1\rangle\langle 1\vert
= \mathbb{I}
\]

Alternatively we can take $A_0 = \frac{1}{\sqrt{2}}\mathbb{I}$ and $A_1 =
\frac{1}{\sqrt{2}}\sigma_z,$ so that
\[
\sum_{k = 0}^1 A_k \rho A_k^{\dagger} = \frac{1}{2} \rho + \frac{1}{2} \sigma_z
\rho \sigma_z = \Delta(\rho),
\]
as was computed previously.
This time the required condition can be verified as follows.
\[
\sum_{k = 0}^1 A_k^{\dagger} A_k = \frac{1}{2} \mathbb{I} + \frac{1}{2}
\sigma_z^2 = \frac{1}{2} \mathbb{I} + \frac{1}{2} \mathbb{I} = \mathbb{I}
\]

\subsubsection{Qubit reset channel}

We obtain a Kraus representation of the qubit reset channel by taking $A_0 =
\vert 0\rangle\langle 0\vert$ and $A_1 = \vert 0\rangle\langle 1\vert.$
\[
\begin{aligned}
  \sum_{k = 0}^1 A_k \rho A_k^{\dagger}
  & =
  \vert 0\rangle\langle 0 \vert \rho \vert 0\rangle\langle 0 \vert
  + \vert 0\rangle\langle 1 \vert \rho \vert 1\rangle\langle 0 \vert\\
  & = \langle 0 \vert \rho \vert 0\rangle \, \vert 0\rangle\langle 0 \vert
  + \langle 1 \vert \rho \vert 1\rangle \, \vert 0\rangle\langle 0 \vert\\[2mm]
  & = \operatorname{Tr}(\rho) \vert 0\rangle \langle 0 \vert
\end{aligned}
\]
These matrices satisfy the required condition.
\[
\sum_{k = 0}^1 A_k^{\dagger} A_k
= \vert 0\rangle\langle 0\vert 0\rangle\langle 0\vert + \vert 1\rangle\langle
0\vert 0\rangle\langle 1\vert
= \vert 0\rangle\langle 0\vert + \vert 1\rangle\langle 1\vert = \mathbb{I}
\]

\subsubsection{Completely depolarizing channel}

One way to obtain a Kraus representation for the completely depolarizing
channel is to choose Kraus matrices $A_0,\ldots,A_3$ as follows.
\[
A_0 = \frac{\vert 0\rangle\langle 0\vert}{\sqrt{2}} \quad
A_1 = \frac{\vert 0\rangle\langle 1\vert}{\sqrt{2}} \quad
A_2 = \frac{\vert 1\rangle\langle 0\vert}{\sqrt{2}} \quad
A_3 = \frac{\vert 1\rangle\langle 1\vert}{\sqrt{2}}
\]
For any qubit density matrix $\rho$ we then have
\[
\begin{aligned}
  \sum_{k = 0}^3 A_k \rho A_k^{\dagger}
  & = \frac{1}{2} \bigl(\vert 0\rangle\langle 0\vert \rho \vert 0\rangle\langle
  0\vert
  + \vert 0\rangle\langle 1\vert \rho \vert 1\rangle\langle 0\vert
  + \vert 1\rangle\langle 0\vert \rho \vert 0\rangle\langle 1\vert
  + \vert 1\rangle\langle 1\vert \rho \vert 1\rangle\langle 1\vert\bigr)\\
  & = \operatorname{Tr}(\rho) \frac{\mathbb{I}}{2}\\[1mm]
  & = \Omega(\rho).
\end{aligned}
\]
An alternative Kraus representation is obtained by choosing Kraus matrices like
so.
\[
A_0 = \frac{\mathbb{I}}{2} \quad
A_1 = \frac{\sigma_x}{2} \quad
A_2 = \frac{\sigma_y}{2} \quad
A_3 = \frac{\sigma_z}{2}
\]
To verify that these Kraus matrices do in fact represent the completely
depolarizing channel, let's first observe that conjugating an arbitrary
$2\times 2$ matrix by a Pauli matrix works as follows.
\[
\begin{aligned}
  \sigma_x
  \begin{pmatrix}
    \alpha_{0,0} & \alpha_{0,1}\\[1mm]
    \alpha_{1,0} & \alpha_{1,1}
  \end{pmatrix}
  \sigma_x
  & =
  \begin{pmatrix}
    \alpha_{1,1} & \alpha_{1,0}\\[1mm]
    \alpha_{0,1} & \alpha_{0,0}
  \end{pmatrix}\\[5mm]
  \sigma_y
  \begin{pmatrix}
    \alpha_{0,0} & \alpha_{0,1}\\[1mm]
    \alpha_{1,0} & \alpha_{1,1}
  \end{pmatrix}
  \sigma_y
  & =
  \begin{pmatrix}
    \alpha_{1,1} & -\alpha_{1,0}\\[1mm]
    -\alpha_{0,1} & \alpha_{0,0}
  \end{pmatrix}\\[5mm]
  \sigma_z
  \begin{pmatrix}
    \alpha_{0,0} & \alpha_{0,1}\\[1mm]
    \alpha_{1,0} & \alpha_{1,1}
  \end{pmatrix}
  \sigma_z
  & =
  \begin{pmatrix}
    \alpha_{0,0} & -\alpha_{0,1}\\[1mm]
    -\alpha_{1,0} & \alpha_{1,1}
  \end{pmatrix}
\end{aligned}
\]
This allows us to verify the correctness of our Kraus representation.
\[
\begin{aligned}
  \sum_{k = 0}^3 A_k \rho A_k^{\dagger}
  & = \frac{\rho + \sigma_x \rho \sigma_x + \sigma_y \rho \sigma_y + \sigma_z
    \rho \sigma_z}{4} \\
  & =
  \scalebox{0.84}{$\displaystyle
  \frac{1}{4}
  \begin{pmatrix}
    \langle 0\vert\rho\vert 0\rangle
    + \langle 1\vert\rho\vert 1\rangle
    + \langle 1\vert\rho\vert 1\rangle
    + \langle 0\vert\rho\vert 0\rangle
    &
    \langle 0\vert\rho\vert 1\rangle
    + \langle 1\vert\rho\vert 0\rangle
    - \langle 1\vert\rho\vert 0\rangle
    - \langle 0\vert\rho\vert 1\rangle
    \\[2mm]
    \langle 1\vert\rho\vert 0\rangle
    + \langle 0\vert\rho\vert 1\rangle
    - \langle 0\vert\rho\vert 1\rangle
    - \langle 1\vert\rho\vert 0\rangle
    &
    \langle 1\vert\rho\vert 1\rangle
    + \langle 0\vert\rho\vert 0\rangle
    + \langle 0\vert\rho\vert 0\rangle
    + \langle 1\vert\rho\vert 1\rangle
  \end{pmatrix}$}
  \\[4mm]
  & = \operatorname{Tr}(\rho) \frac{\mathbb{I}}{2}
\end{aligned}
\]
This Kraus representation expresses an important idea, which is that the state
of a qubit can be completely randomized by applying to it one of the four Pauli
matrices (including the identity matrix) chosen uniformly at random.
Thus, the completely depolarizing channel is another example of a Pauli
channel.

It is not possible to find a Kraus representation for the completely
depolarizing channel $\Omega$ having three or fewer Kraus matrices; at least
four are required for this channel.

\subsubsection{Unitary channels}

If we have a unitary matrix $U$ representing an operation on a system
$\mathsf{X},$ we can express the action of this unitary operation as a channel:
\[
\Phi(\rho) = U \rho U^{\dagger}.
\]
This expression is already a valid Kraus representation of the channel $\Phi$
where we happen to have just one Kraus matrix $A_0 = U.$ In this case, the
required condition
\[
\sum_{k = 0}^{N-1} A_k^{\dagger} A_k = \mathbb{I}_{\mathsf{X}}
\]
takes the much simpler form $U^{\dagger} U = \mathbb{I}_{\mathsf{X}},$ which we
know is true because $U$ is unitary.

\subsection{Choi representations}

Now we'll discuss a third way that channels can be described, through the
\emph{Choi representation}.
The way it works is that each channel is represented by a single matrix known
as its \emph{Choi matrix}.
If the input system has $n$ classical states and the output system has $m$
classical states, then the Choi matrix of the channel will have $nm$ rows and
$nm$ columns.

Choi matrices provide a \emph{faithful} representation of channels, meaning
that two channels are the same if and only if they have the same Choi matrix.
One reason why this is important is that it provides us with a way of
determining whether two different descriptions correspond to the same channel
or to different channels: we simply compute the Choi matrices and compare them
to see if they're equal.
In contrast, Stinespring and Kraus representations are not unique in this way,
as we have seen.

Choi matrices are also useful in other regards for uncovering various
mathematical properties of channels.

\subsubsection{Definition}

Let $\Phi$ be a channel from a system $\mathsf{X}$ to a system $\mathsf{Y},$
and assume that the classical state set of the input system $\mathsf{X}$ is
$\Sigma.$
The Choi representation of $\Phi,$ which is denoted $J(\Phi),$ is defined by
the following equation.
\[
J(\Phi) = \sum_{a,b\in\Sigma} \vert a\rangle\langle b \vert \otimes \Phi\bigl(
\vert a\rangle\langle b \vert\bigr)
\]
If we assume that $\Sigma = \{0,\ldots, n-1\}$ for some positive integer $n,$
then we can alternatively express $J(\Phi)$ as a block matrix:
\[
J(\Phi)
= \begin{pmatrix}
\Phi\bigl(\vert 0\rangle\langle 0\vert\bigr) & \Phi\bigl(\vert 0\rangle\langle
1\vert\bigr) & \cdots & \Phi\bigl(\vert 0\rangle\langle n-1\vert\bigr) \\[1mm]
\Phi\bigl(\vert 1\rangle\langle 0\vert\bigr) & \Phi\bigl(\vert 1\rangle\langle
1\vert\bigr) & \cdots & \Phi\bigl(\vert 1\rangle\langle n-1\vert\bigr) \\[1mm]
\vdots & \vdots & \ddots & \vdots\\[1mm]
\Phi\bigl(\vert n-1\rangle\langle 0\vert\bigr) & \Phi\bigl(\vert
n-1\rangle\langle 1\vert\bigr) & \cdots & \Phi\bigl(\vert n-1\rangle\langle
n-1\vert\bigr)
\end{pmatrix}
\]
That is, as a block matrix, the Choi matrix of a channel has one block
$\Phi(\vert a\rangle\langle b\vert)$ for each pair $(a,b)$ of classical states
of the input system, with the blocks arranged in a natural way.

Notice that the set $\{\vert a\rangle\langle b\vert \,:\, 0\leq a,b < n\}$
forms a basis for the space of all $n\times n$ matrices.
Because $\Phi$ is linear, it follows that its action can be recovered from its
Choi matrix by taking linear combinations of the blocks.

\subsubsection{The Choi state of a channel}

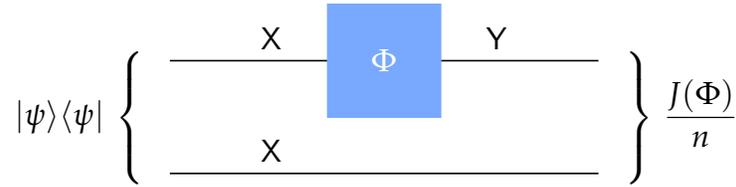
\begin{figure}[t]
  \begin{center}
    \begin{tikzpicture}[
        scale = 1.5,
        line width = 0.6pt,
        gate/.style={%
          inner sep = 0,
          fill = CircuitBlue,
          draw = CircuitBlue,
          text = white,
          minimum size = 10mm
        }
      ]
      
      \node (In0) at (-2,1) {};
      \node (In1) at (-2,0) {};
      \node (Out0) at (2,1) {};
      \node (Out1) at (2,0) {};
      
      \draw (In0) -- (Out0);
      \draw (In1) -- (Out1);
      
      \node[gate, minimum size = 15mm] (Phi) at (0,1) {$\Phi$};
      \node[anchor = south] at (-1,1) {$\mathsf{X}$};
      \node[anchor = south] at (-1,0) {$\mathsf{X}$};
      \node[anchor = south] at (1,1) {$\mathsf{Y}$};
      
      \node[anchor = east] at ($(In0)!0.5!(In1)$) {%
        $\ket{\psi}\bra{\psi}\,\left\{\rule{0mm}{10mm}\right.$};
      
      \node[anchor = west] at ($(Out0)!0.5!(Out1)$) {%
        $\left.\rule{0mm}{10mm}\right\}\,\displaystyle
        \frac{J(\Phi)}{n}$};
      
    \end{tikzpicture}
  \end{center}
  \caption{Evaluating a channel on one-half of the maximally entangled
    state $\vert\psi\rangle$ yields the normalized Choi matrix of the channel.}
  \label{fig:Choi-state}
\end{figure}

Another way to think about the Choi matrix of a channel is that it's a density
matrix if we divide by $n = \vert\Sigma\vert.$
Let's focus on the situation that $\Sigma = \{0,\ldots,n-1\}$ for simplicity,
and imagine that we have two identical copies of $\mathsf{X}$ that are together
in the entangled state
\[
\vert \psi \rangle = \frac{1}{\sqrt{n}} \sum_{a = 0}^{n-1} \vert a \rangle
\otimes \vert a \rangle.
\]
As a density matrix, this state is as follows.
\[
\vert \psi \rangle \langle \psi \vert = \frac{1}{n} \sum_{a,b = 0}^{n-1}
\vert a\rangle\langle b \vert \otimes \vert a\rangle\langle b \vert
\]
If we apply $\Phi$ to the copy of $\mathsf{X}$ on the right-hand side, we
obtain the Choi matrix divided by $n.$
\[
(\operatorname{Id}\otimes \,\Phi) \bigl(\vert \psi \rangle \langle \psi
\vert\bigr)
= \frac{1}{n} \sum_{a,b = 0}^{n-1} \vert a\rangle\langle b \vert \otimes
\Phi\bigl(\vert a\rangle\langle b \vert\bigr)
= \frac{J(\Phi)}{n}
\]
In words, up to a normalization factor $1/n,$ the Choi matrix of $\Phi$ is the
density matrix we obtain by evaluating $\Phi$ on one-half of a \emph{maximally
entangled} pair of input systems, as Figure~\ref{fig:Choi-state} depicts.
Notice in particular that this implies that the Choi matrix of a channel must
always be positive semidefinite.

We also see that, because the channel $\Phi$ is applied to the right/top system
alone, it cannot affect the reduced state of the left/bottom system.
In the case at hand, that state is the completely mixed state
$\mathbb{I}_{\mathsf{X}}/n,$ and therefore
\[
\operatorname{Tr}_{\mathsf{Y}} \biggl(\frac{J(\Phi)}{n}\biggr) =
\frac{\mathbb{I}_{\mathsf{X}}}{n}.
\]
Clearing the denominator $n$ from both sides yields
$\operatorname{Tr}_{\mathsf{Y}} (J(\Phi)) = \mathbb{I}_{\mathsf{X}}.$
We can alternatively draw this same conclusion by using the fact that channels
must always preserve trace, and therefore
\[
\begin{aligned}
  \operatorname{Tr}_{\mathsf{Y}} (J(\Phi))
  & = \sum_{a,b\in\Sigma} \operatorname{Tr}\bigl(\Phi( \vert a\rangle\langle b
  \vert)\bigr) \, \vert a\rangle\langle b \vert \\
  & = \sum_{a,b\in\Sigma} \operatorname{Tr}\bigl(\vert a\rangle\langle b
  \vert\bigr) \, \vert a\rangle\langle b \vert \\
  & = \sum_{a\in\Sigma} \vert a\rangle\langle a \vert \\
  & = \mathbb{I}_{\mathsf{X}}.
\end{aligned}
\]

In summary, the Choi representation $J(\Phi)$ for any channel $\Phi$ must be
positive semidefinite and must satisfy
\[
\operatorname{Tr}_{\mathsf{Y}} (J(\Phi)) = \mathbb{I}_{\mathsf{X}}.
\]
As we will see by the end of the lesson, these two conditions are not only
necessary but also sufficient, meaning that any linear mapping $\Phi$ from
matrices to matrices that satisfies these requirements must, in fact, be a
channel.

\subsubsection{Completely dephasing channel}

The Choi representation of the completely dephasing channel $\Delta$ is
\[
J(\Delta) = \sum_{a,b = 0}^{1} \vert a\rangle\langle b \vert \otimes
\Delta\bigl(\vert a\rangle\langle b \vert\bigr)
= \sum_{a = 0}^{1} \vert a\rangle\langle a \vert \otimes \vert
a\rangle\langle a \vert
= \begin{pmatrix}
  1 & 0 & 0 & 0\\[1mm]
  0 & 0 & 0 & 0\\[1mm]
  0 & 0 & 0 & 0\\[1mm]
  0 & 0 & 0 & 1
  \end{pmatrix}.
\]

\subsubsection{Completely depolarizing channel}

The Choi representation of the completely depolarizing channel is
\[
J(\Omega) = \sum_{a,b = 0}^{1} \vert a\rangle\langle b \vert \otimes
\Omega\bigl(\vert a\rangle\langle b \vert\bigr)
= \sum_{a = 0}^{1} \vert a\rangle\langle a \vert \otimes \frac{1}{2}
\mathbb{I}
= \frac{1}{2} \mathbb{I} \otimes \mathbb{I}
= \begin{pmatrix}
    \frac{1}{2} & 0 & 0 & 0\\[1mm]
    0 & \frac{1}{2} & 0 & 0\\[1mm]
    0 & 0 & \frac{1}{2} & 0\\[1mm]
    0 & 0 & 0 & \frac{1}{2}
  \end{pmatrix}.
\]

\subsubsection{Qubit reset channel}

The Choi representation of the qubit reset channel $\Phi$ is
\[
J(\Lambda) = \sum_{a,b = 0}^{1} \vert a\rangle\langle b \vert \otimes
\Lambda\bigl(\vert a\rangle\langle b \vert\bigr)
= \sum_{a = 0}^{1} \vert a\rangle\langle a \vert \otimes \vert 0\rangle
\langle 0\vert
= \mathbb{I} \otimes \vert 0\rangle \langle 0\vert
= \begin{pmatrix}
  1 & 0 & 0 & 0\\[1mm]
  0 & 0 & 0 & 0\\[1mm]
  0 & 0 & 1 & 0\\[1mm]
  0 & 0 & 0 & 0
\end{pmatrix}.
\]

\subsubsection{The identity channel}

The Choi representation of the qubit identity channel $\operatorname{Id}$ is
\[
J(\operatorname{Id})
= \sum_{a,b = 0}^{1} \vert a\rangle\langle b \vert \otimes
\operatorname{Id}\bigl(\vert a\rangle\langle b \vert\bigr)
= \sum_{a,b = 0}^{1} \vert a \rangle\langle b \vert \otimes \vert a\rangle
\langle b \vert 
= \begin{pmatrix}
  1 & 0 & 0 & 1\\[1mm]
  0 & 0 & 0 & 0\\[1mm]
  0 & 0 & 0 & 0\\[1mm]
  1 & 0 & 0 & 1
\end{pmatrix}.
\]
Notice in particular that $J(\operatorname{Id})$ is not the identity matrix.
The Choi representation does not directly describe a channel's action in the
usual way that a matrix represents a linear mapping.

\section{Equivalence of the representations}

We've now discussed three different ways to represent channels in mathematical
terms, namely Stinespring representations, Kraus representations, and Choi
representations.
We also have the definition of a channel, which states that a channel is a
linear mapping that always transforms density matrices into density matrices,
even when the channel is applied to just part of a compound system.
The remainder of the lesson is devoted to a mathematical proof that the three
representations are equivalent and precisely capture the definition.

\subsection{Overview of the proof}

Our goal is to establish the equivalence of a collection of four statements,
and we'll begin by writing them down precisely.
All four statements follow the same conventions that have been used throughout
the lesson, namely that $\Phi$ is a linear mapping from square matrices to
square matrices, the rows and columns of the input matrices have been placed in
correspondence with the classical states of a system $\mathsf{X}$ (the input
system), and the rows and columns of the output matrices have been placed in
correspondence with the classical states of a system $\mathsf{Y}$ (the output
system).

\begin{enumerate}
  \item
    $\Phi$ is a channel from $\mathsf{X}$ to $\mathsf{Y}.$ That is, $\Phi$
    always transforms density matrices to density matrices, even when it acts
    on one part of a larger compound system.

  \item
    The Choi matrix $J(\Phi)$ is positive semidefinite and satisfies the
    condition $\operatorname{Tr}_{\mathsf{Y}}(J(\Phi)) =
    \mathbb{I}_{\mathsf{X}}.$

  \item
    There is a Kraus representation for $\Phi.$ That is, there exist matrices
    $A_0,\ldots,A_{N-1}$ for which the equation $\Phi(\rho) = \sum_{k =
      0}^{N-1} A_k \rho A_k^{\dagger}$ is true for every input $\rho,$ and that
    satisfy the condition $\sum_{k = 0}^{N-1} A_k^{\dagger} A_k =
    \mathbb{I}_{\mathsf{X}}.$

  \item
    There is a Stinespring representation for $\Phi.$ That is, there exist
    systems $\mathsf{W}$ and $\mathsf{G}$ for which the pairs
    $(\mathsf{W},\mathsf{X})$ and $(\mathsf{G},\mathsf{Y})$ have the same
    number of classical states, along with a unitary matrix $U$ representing a
    unitary operation from $(\mathsf{W},\mathsf{X})$ to
    $(\mathsf{G},\mathsf{Y}),$ such that $\Phi(\rho) =
    \operatorname{Tr}_{\mathsf{G}}\bigl( U (\vert 0\rangle\langle 0 \vert
    \otimes \rho) U^{\dagger} \bigr).$
\end{enumerate}

The way the proof works is that a cycle of implications is proved:
the first statement in our list implies the second, the second implies the
third, the third implies the fourth, and the fourth statement implies the
first.
This establishes that all four statements are equivalent --- which is to say
that they're either all true or all false for a given choice of $\Phi$ ---
because the implications can be followed transitively from any one statement to
any other.

This is a common strategy when proving that a collection of statements are
equivalent, and a useful trick to use in such a context is to set up the
implications in a way that makes them as easy to prove as possible.
That is the case here --- and in fact we've already encountered two of the four
implications.

\subsection{Channels to Choi matrices}

Referring to the statements listed above by their numbers, the first
implication to be proved is 1 $\Rightarrow$ 2.
This implication was already discussed in the context of the Choi state of a
channel.
Here we'll summarize the mathematical details.

Assume that the classical state set of the input system $\mathsf{X}$ is
$\Sigma$ and let $n = \vert\Sigma\vert.$
Consider the situation in which $\Phi$ is applied to the second of two copies
of $\mathsf{X}$ together in the state
\[
\vert \psi \rangle = \frac{1}{\sqrt{n}} \sum_{a \in \Sigma} \vert a \rangle
\otimes \vert a \rangle,
\]
which, as a density matrix, is given by
\[
\vert \psi \rangle \langle \psi \vert = \frac{1}{n} \sum_{a,b \in \Sigma}
\vert a\rangle\langle b \vert \otimes \vert a\rangle\langle b \vert.
\]
The result can be written as
\[
(\operatorname{Id}\otimes \,\Phi) \bigl(\vert \psi \rangle \langle \psi
\vert\bigr)
= \frac{1}{n} \sum_{a,b = 0}^{n-1} \vert a\rangle\langle b \vert \otimes
\Phi\bigl(\vert a\rangle\langle b \vert\bigr)
= \frac{J(\Phi)}{n},
\]
and by the assumption that $\Phi$ is a channel this must be a density matrix.
Like all density matrices it must be positive semidefinite, and multiplying a
positive semidefinite matrix by a positive real number yields another positive
semidefinite matrix, and therefore $J(\Phi) \geq 0.$

Moreover, under the assumption that $\Phi$ is a channel, it must preserve
trace, and therefore
\[
\begin{aligned}
  \operatorname{Tr}_{\mathsf{Y}} (J(\Phi))
  & = \sum_{a,b\in\Sigma} \operatorname{Tr}\bigl(\Phi( \vert a\rangle\langle b
  \vert)\bigr) \, \vert a\rangle\langle b \vert\\
  & = \sum_{a,b\in\Sigma} \operatorname{Tr}\bigl(\vert a\rangle\langle b
  \vert\bigr) \, \vert a\rangle\langle b \vert\\
  & = \sum_{a\in\Sigma} \vert a\rangle\langle a \vert\\
  & = \mathbb{I}_{\mathsf{X}}.
\end{aligned}
\]

\subsection{Choi to Kraus representations}

The second implication, again referring to the statements in our list by their
numbers, is 2 $\Rightarrow$ 3.
To be clear, we're ignoring the other statements --- and in particular we
cannot make the assumption that $\Phi$ is a channel.
All we have to work with is that $\Phi$ is a linear mapping whose Choi
representation satisfies $J(\Phi) \geq 0$ and
$\operatorname{Tr}_{\mathsf{Y}} (J(\Phi)) = \mathbb{I}_{\mathsf{X}}.$
This, however, is all we need to conclude that $\Phi$ has a Kraus
representation
\[
\Phi(\rho) = \sum_{k = 0}^{N-1} A_k \rho A_k^{\dagger}
\]
for which the condition
\[
\sum_{k = 0}^{N-1} A_k^{\dagger} A_k = \mathbb{I}_{\mathsf{X}}
\]
is satisfied.

We begin with the critically important assumption that $J(\Phi)$ is positive
semidefinite, which means that it is possible to express it in the form
\begin{equation}
  J(\Phi) = \sum_{k = 0}^{N-1} \vert \psi_k \rangle \langle \psi_k \vert
  \label{eq:psd-form}
\end{equation}
for some way of choosing the vectors
$\vert\psi_0\rangle,\ldots,\vert\psi_{N-1}\rangle.$
In general there will be multiple ways to do this --- and in fact this directly
mirrors the freedom one has in choosing a Kraus representation for $\Phi.$

One way to obtain such an expression is to first use the spectral theorem to
write
\[
J(\Phi) = \sum_{k = 0}^{N-1} \lambda_k \vert \gamma_k \rangle \langle \gamma_k
\vert,
\]
in which $\lambda_0,\ldots,\lambda_{N-1}$ are the eigenvalues of $J(\Phi)$
(which are necessarily nonnegative real numbers because $J(\Phi)$ is positive
semidefinite) and $\vert\gamma_0\rangle,\ldots,\vert\gamma_{N-1}\rangle$ are
unit eigenvectors corresponding to the eigenvalues
$\lambda_0,\ldots,\lambda_{N-1}.$

Note that, while there's no freedom in choosing the eigenvalues (except for how
they're ordered), there is freedom in the choice of the eigenvectors,
particularly when there are eigenvalues with multiplicity larger than one.
So, this is not a unique expression of $J(\Phi)$ --- we're just assuming we
have one such expression.
Regardless, because the eigenvalues are nonnegative real numbers, they have
nonnegative square roots, and so we can select
\[
\vert\psi_k\rangle = \sqrt{\lambda_k} \vert \gamma_k\rangle
\]
for each $k = 0,\ldots,N-1$ to obtain an expression of the form
\eqref{eq:psd-form}.

It is, however, not essential that the expression \eqref{eq:psd-form} comes
from a spectral decomposition in this way, and in particular the vectors
$\vert\psi_0\rangle,\ldots,\vert\psi_{N-1}\rangle$ need not be orthogonal in
general.
It is noteworthy, though, that we can choose these vectors to be orthogonal if
we wish --- and moreover we never need $N$ to be larger than $nm$
(recalling that $n$ and $m$ denote the numbers of classical states of
$\mathsf{X}$ and $\mathsf{Y},$ respectively).

Next, each of the vectors $\vert\psi_0\rangle,\ldots,\vert\psi_{N-1}\rangle$
can be further decomposed as
\[
\vert\psi_k\rangle = \sum_{a\in\Sigma} \vert a\rangle \otimes \vert
\phi_{k,a}\rangle,
\]
where the vectors $\{ \vert \phi_{k,a}\rangle \}$ have entries corresponding to
the classical states of $\mathsf{Y}$ and can be explicitly determined by the
equation
\[
\vert \phi_{k,a}\rangle = \bigl( \langle a \vert \otimes
\mathbb{I}_{\mathsf{Y}}\bigr) \vert \psi_k\rangle
\]
for each $a\in\Sigma$ and $k=0,\ldots,N-1.$
Although $\vert\psi_0\rangle,\ldots,\vert\psi_{N-1}\rangle$ are not necessarily
unit vectors, this is the same process we would use to analyze what would
happen if a standard basis measurement was performed on the system $\mathsf{X}$
given a quantum state vector of the pair $(\mathsf{X},\mathsf{Y}).$

And now we come to the trick that makes this part of the proof work.
We define our Kraus matrices $A_0,\ldots,A_{N-1}$ according to the following
equation.
\[
A_k = \sum_{a\in\Sigma} \vert \phi_{k,a}\rangle\langle a \vert
\]
We can think about this formula purely symbolically: $\vert a\rangle$
effectively gets flipped around to form $\langle a\vert$ and moved to
right-hand side, forming a matrix.
For the purposes of verifying the proof, the formula is all we need.

There is, however, a simple and intuitive relationship between the vector
$\vert\psi_k\rangle$ and the matrix $A_k,$ which is that by \emph{vectorizing}
$A_k$ we get $\vert\psi_k\rangle.$
What it means to vectorize $A_k$ is that we stack the columns on top of one
another (with the leftmost column on top proceeding to the rightmost on the
bottom), in order to form a vector.
For instance, if $\mathsf{X}$ and $\mathsf{Y}$ are both qubits, and for some
choice of $k$ we have
\[
\vert\psi_k\rangle = \alpha_{00} \vert 0\rangle \otimes \vert 0\rangle +
\alpha_{01} \vert 0\rangle \otimes \vert 1\rangle +
\alpha_{10} \vert 1\rangle \otimes \vert 0\rangle +
\alpha_{11} \vert 1\rangle \otimes \vert 1\rangle
= \begin{pmatrix}
  \alpha_{00} \\[1mm]
  \alpha_{01} \\[1mm]
  \alpha_{10} \\[1mm]
  \alpha_{11}
\end{pmatrix},
\]
then
\[
A_k = \alpha_{00} \vert 0\rangle\langle 0\vert +
\alpha_{01} \vert 1\rangle\langle 0\vert +
\alpha_{10} \vert 0\rangle\langle 1\vert +
\alpha_{11} \vert 1\rangle\langle 1\vert
= \begin{pmatrix}
  \alpha_{00} & \alpha_{10}\\[1mm]
  \alpha_{01} & \alpha_{11}
\end{pmatrix}.
\]
(Beware: sometimes the vectorization of a matrix is defined in a slightly
different way, which is that the \emph{rows} of the matrix are transposed and
stacked on top of one another to form a column vector.)

First we'll verify that this choice of Kraus matrices correctly describes the
mapping $\Phi,$ after which we'll verify the other required condition.
To keep things straight, let's define a new mapping $\Psi$ as follows.
\[
\Psi(\rho) = \sum_{k = 0}^{N-1} A_k \rho A_k^{\dagger}
\]
Thus, our goal is to verify that $\Psi = \Phi.$

The way we can do this is to compare the Choi representations of these
mappings.
Choi representations are faithful, so we have $\Psi = \Phi$ if and only if
$J(\Phi) = J(\Psi).$
At this point we can simply compute $J(\Psi)$ using the expressions
\[
\vert\psi_k\rangle = \sum_{a\in\Sigma} \vert a\rangle \otimes \vert
\phi_{k,a}\rangle
\quad\text{and}\quad
A_k = \sum_{a\in\Sigma} \vert \phi_{k,a}\rangle\langle a \vert
\]
together with the bilinearity of tensor products to simplify.
\[
\begin{aligned}
  J(\Psi) & = \sum_{a,b\in\Sigma}  \vert a\rangle \langle b \vert \otimes
  \sum_{k = 0}^{N-1} A_k \vert a\rangle \langle b \vert A_k^{\dagger}\\[2mm]
  & = \sum_{a,b\in\Sigma}  \vert a\rangle \langle b \vert \otimes
  \sum_{k = 0}^{N-1}  \vert \phi_{k,a} \rangle \langle \phi_{k,b} \vert \\[2mm]
  & = \sum_{k = 0}^{N-1} \biggl(\sum_{a\in\Sigma} \vert a\rangle \otimes \vert
  \phi_{k,a} \rangle\biggr)
  \biggl(\sum_{b\in\Sigma} \langle b\vert \otimes \langle \phi_{k,b}
  \vert\biggr)\\[2mm]
  & = \sum_{k = 0}^{N-1} \vert \psi_k \rangle \langle \psi_k \vert \\[2mm]
  & = J(\Phi)
\end{aligned}
\]
So, our Kraus matrices correctly describe $\Phi.$

It remains to check the required condition on $A_0,\ldots,A_{N-1},$ which turns
out to be equivalent to the assumption $\operatorname{Tr}_{\mathsf{Y}}(J(\Phi))
= \mathbb{I}_{\mathsf{X}}$ (which we haven't used yet).
What we'll show is this relationship:
\begin{equation}
\Biggl( \sum_{k = 0}^{N-1} A_k^{\dagger} A_k \Biggr)^{T} = \operatorname{Tr}_{\mathsf{Y}}(J(\Phi))
\label{eq:Choi-relation}
\end{equation}
(in which we're referring the \emph{matrix transpose} on the left-hand side).

Starting on the left, we can first observe that
\[
\begin{aligned}
  \Biggl(\sum_{k = 0}^{N-1} A_k^{\dagger} A_k\Biggr)^T
  & = \Biggl(\sum_{k = 0}^{N-1} \sum_{a,b\in\Sigma} \vert b \rangle \langle
  \phi_{k,b} \vert \phi_{k,a} \rangle \langle a \vert\Biggr)^T\\
  & = \sum_{k = 0}^{N-1} \sum_{a,b\in\Sigma} \langle \phi_{k,b} \vert
  \phi_{k,a} \rangle \vert a \rangle  \langle b \vert.
\end{aligned}
\]
The last equality follows from the fact that the transpose is linear and maps
$\vert b\rangle\langle a \vert$ to $\vert a\rangle\langle b \vert.$

Moving to the right-hand side of our equation, we have
\[
J(\Phi) = \sum_{k = 0}^{N-1} \vert \psi_k\rangle\langle\psi_k \vert
= \sum_{k = 0}^{N-1} \sum_{a,b\in\Sigma} \vert a\rangle \langle b \vert \otimes
\vert\phi_{k,a}\rangle\langle \phi_{k,b} \vert
\]
and therefore
\[
\begin{aligned}
  \operatorname{Tr}_{\mathsf{Y}}(J(\Phi))
  & = \sum_{k = 0}^{N-1} \sum_{a,b\in\Sigma}
  \operatorname{Tr}\bigl(\vert\phi_{k,a}\rangle\langle \phi_{k,b} \vert
  \bigr)\,
  \vert a\rangle \langle b \vert\\
  & = \sum_{k = 0}^{N-1} \sum_{a,b\in\Sigma} \langle \phi_{k,b} \vert
  \phi_{k,a} \rangle \vert a \rangle  \langle b \vert.
\end{aligned}
\]

We've obtained the same result, and therefore the equation
\eqref{eq:Choi-relation} has been verified.
It follows, by the assumption $\operatorname{Tr}_{\mathsf{Y}} (J(\Phi)) =
\mathbb{I}_{\mathsf{X}},$ that
\[
\Biggl(\sum_{k = 0}^{N-1} A_k^{\dagger} A_k\Biggr)^T = \mathbb{I}_{\mathsf{X}}
\]
and therefore, because the identity matrix is its own transpose, the required
condition is true.
\[
\sum_{k = 0}^{N-1} A_k^{\dagger} A_k = \mathbb{I}_{\mathsf{X}}
\]

\subsection{Kraus to Stinespring representations}

Now suppose that we have a Kraus representation of a mapping
\[
\Phi(\rho) = \sum_{k = 0}^{N-1} A_k \rho A_k^{\dagger}
\]
for which
\[
\sum_{k = 0}^{N-1} A_k^{\dagger} A_k = \mathbb{I}_{\mathsf{X}}.
\]
Our goal is to find a Stinespring representation for $\Phi.$

What we'd like to do first is to choose the garbage system $\mathsf{G}$ so that
its classical state set is $\{0,\ldots,N-1\}.$
For $(\mathsf{W},\mathsf{X})$ and $(\mathsf{G},\mathsf{Y})$ to have the same
size, however, $n$ must divide $m N,$ allowing us to take
$\mathsf{W}$ to have classical states $\{0,\ldots,d-1\}$ for $d = mN/n.$
For an arbitrary choice of $n,$ $m,$ and $N,$ it may not be the case that
$mN/n$ is an integer, so we're not actually free to choose $\mathsf{G}$ so that
it's classical state set is $\{0,\ldots,N-1\}.$
But we can always increase $N$ arbitrarily in the Kraus representation by
choosing $A_k = 0$ for however many additional values of $k$ that we wish.

And so, if we tacitly assume that $mN/n$ is an integer, which is equivalent to
$N$ being a multiple of $m/\operatorname{gcd}(n,m),$ then we're free to take
$\mathsf{G}$ so that its classical state set is $\{0,\ldots,N-1\}.$
In particular, if it is the case that $N = nm,$ then we may take $\mathsf{W}$
to have $m^2$ classical states.

It remains to choose $U,$ and we'll do this by matching the following pattern.
\[
U =
\begin{pmatrix}
  A_{0} & \fbox{?} & \cdots & \fbox{?} \\[1mm]
  A_{1} & \fbox{?} & \cdots & \fbox{?} \\[1mm]
  \vdots & \vdots & \ddots & \vdots\\[1mm]
  A_{N-1} & \fbox{?} & \cdots & \fbox{?}
\end{pmatrix}
\]
To be clear, this pattern is meant to suggest a block matrix, where each block
(including $A_{0},\ldots,A_{N-1}$ as well as the blocks marked with a question
mark) has $m$ rows and $n$ columns.
There are $N$ rows of blocks, which means that there are $d = mN/n$ columns of
blocks.

Expressed in more formulaic terms, we will define $U$ as
\[
U = \sum_{k=0}^{N-1} \sum_{j=0}^{d-1} \vert k \rangle \langle j \vert
\otimes M_{k,j}
= \begin{pmatrix}
  M_{0,0} & M_{0,1} & \cdots & M_{0,d-1} \\[1mm]
  M_{1,0} & M_{1,1} & \cdots & M_{1,d-1} \\[1mm]
  \vdots & \vdots & \ddots & \vdots\\[1mm]
  M_{N-1,0} & M_{N-1,1} & \cdots & M_{N-1,d-1}
\end{pmatrix}
\]
where each matrix $M_{k,j}$ has $m$ rows and $n$ columns, and in particular we
shall take $M_{k,0} = A_k$ for $k = 0,\ldots,N-1.$

This must be a unitary matrix, and the blocks labeled with a question mark, or
equivalently $M_{k,j}$ for $j>0,$ must be selected with this in mind --- but
aside from allowing $U$ to be unitary, the blocks labeled with a question mark
won't have any relevance to the proof.

Let's momentarily disregard the concern that $U$ is unitary and focus on the
expression
\[
\operatorname{Tr}_{\mathsf{G}} \bigl( U (\vert 0\rangle \langle 0
\vert_{\mathsf{W}} \otimes \rho)U^{\dagger}\bigr)
\]
that describes the output state of $\mathsf{Y}$ given the input state $\rho$ of
$\mathsf{X}$ for our Stinespring representation.
We can alternatively write
\[
U(\vert 0\rangle\langle 0 \vert \otimes \rho)U^{\dagger}
= U(\vert 0\rangle\otimes\mathbb{I}_{\mathsf{W}}) \rho (\langle 0\vert \otimes
\mathbb{I}_{\mathsf{W}}) U^{\dagger},
\]
and we see from our choice of $U$ that
\[
U(\vert 0\rangle\otimes\mathbb{I}_{\mathsf{W}}) =
\sum_{k = 0}^{N-1} \vert k\rangle \otimes A_k.
\]
We therefore find that
\[
U(\vert 0\rangle\langle 0 \vert \otimes \rho)U^{\dagger}
= \sum_{j,k = 0}^{N-1} \vert k\rangle\langle j\vert \otimes A_k \rho
A_j^{\dagger},
\]
and so
\[
\begin{aligned}
  \operatorname{Tr}_{\mathsf{G}} \bigl( U (\vert 0\rangle \langle 0
  \vert_{\mathsf{W}} \otimes \rho) U^{\dagger}\bigr)
  & = \sum_{j,k = 0}^{N-1} \operatorname{Tr}\bigl(\vert k\rangle\langle
  j\vert\bigr) \, A_k \rho A_j^{\dagger} \\
  & = \sum_{k = 0}^{N-1} A_k \rho A_k^{\dagger} \\
  & = \Phi(\rho).
\end{aligned}
\]

We therefore have a correct representation for the mapping $\Phi,$ and it
remains to verify that we can choose $U$ to be unitary.
Consider the first $n$ columns of $U$ when it's selected according to the
pattern above.
Taking these columns alone, we have a block matrix
\[
\begin{pmatrix}
  A_0\\[1mm]
  A_1\\[1mm]
  \vdots\\[1mm]
  A_{N-1}
\end{pmatrix}.
\]
There are $n$ columns, one for each classical state of $\mathsf{X},$ and as
vectors let us name the columns as $\vert \gamma_a \rangle$ for each
$a\in\Sigma.$
Here's a formula for these vectors that can be matched to the block matrix
representation above.
\[
\vert \gamma_a\rangle = \sum_{k = 0}^{N-1} \vert k\rangle \otimes A_k \vert a
\rangle
\]

Now let's compute the inner product between any two of these vectors, meaning
the ones corresponding to any choice of $a,b\in\Sigma.$
\[
\langle \gamma_a \vert \gamma_b \rangle =
\sum_{j,k = 0}^{N-1} \langle k \vert j \rangle \, \langle a \vert A_k^{\dagger}
A_j \vert b\rangle =
\langle a \vert \Biggl( \sum_{k = 0}^{N-1} A_k^{\dagger} A_k \Biggr) \vert
b\rangle
\]
By the assumption
\[
\sum_{k = 0}^{m-1} A_k^{\dagger} A_k = \mathbb{I}_{\mathsf{X}}
\]
we conclude that the $n$ column vectors
$\{\vert\gamma_a\rangle\,:\,a\in\Sigma\}$ form an orthonormal set:
\[
\langle \gamma_a \vert \gamma_b \rangle = \begin{cases}
1 & a = b\\
0 & a\neq b
\end{cases}
\]
for all $a,b\in\Sigma.$

This implies that it is possible to fill out the remaining columns of $U$ so
that it becomes a unitary matrix.
In particular, the Gram--Schmidt orthogonalization process can be used to select
the remaining columns, as discussed in Lesson~\ref{lesson:quantum-circuits}
\emph{(Quantum Circuits)}.

\subsection{Stinespring representations back to the definition}

The final implication is 4 $\Rightarrow$ 1.
That is, we assume that we have a unitary operation transforming a pair of
systems $(\mathsf{W},\mathsf{X})$ into a pair
$(\mathsf{G},\mathsf{Y}),$ and our goal is to conclude that the mapping
\[
\Phi(\rho) = \operatorname{Tr}_{\mathsf{G}} \bigl( U (\vert 0\rangle \langle 0
\vert_{\mathsf{W}} \otimes \rho)U^{\dagger}\bigr)
\]
is a valid channel.
From its form, it is evident that $\Phi$ is linear, and it remains to verify
that it always transforms density matrices into density matrices.
This is pretty straightforward and we've already discussed the key points.

In particular, if we start with a density matrix $\sigma$ of a compound system
$(\mathsf{Z},\mathsf{X}),$ and then add on an additional workspace system
$\mathsf{W},$ we will certainly be left with a density matrix.
If we reorder the systems $(\mathsf{W},\mathsf{Z},\mathsf{X})$ for convenience,
we can write this state as
\[
\vert 0\rangle\langle 0\vert_{\mathsf{W}} \otimes \sigma.
\]
We then apply the unitary operation $U,$ and as we already discussed this is a
valid channel, and hence maps density matrices to density matrices.
Finally, the partial trace of a density matrix is another density matrix.

Another way to say this is to observe first that each of these things is a
valid channel:
\begin{enumerate}
\item Introducing an initialized workspace system.
\item Performing a unitary operation.
\item Tracing out a system.
\end{enumerate}
And finally, any composition of channels is another channel --- which is
immediate from the definition, but is also a fact worth observing in its own
right.

This completes the proof of the final implication, and therefore we've
established the equivalence of the four statements listed at the start of the
section.


\lesson{General Measurements}
\label{lesson:general-measurements}

Measurements provide an interface between quantum and classical information.
When a measurement is performed on a system in a quantum state, classical
information is extracted, revealing something about that quantum state --- and
generally changing or destroying it in the process.
In the simplified formulation of quantum information, as presented in
Unit~\ref{unit:basics-of-quantum-information}
\emph{(Basics of Quantum Information)}, we typically limit our attention
to \emph{projective measurements}, including the simplest type of measurement:
\emph{standard basis measurements}.
The concept of a measurement, however, can be generalized beyond projective
measurements.

In this lesson we'll consider measurements in greater generality.
We'll discuss a few different ways that general measurements can be described
in mathematical terms, and connect them to concepts discussed previously
in the course.

We'll also take a look at a couple of notions connected with measurements,
namely \emph{quantum state discrimination} and \emph{quantum state tomography}.
Quantum state discrimination refers to a situation that arises commonly in
quantum computing and cryptography, where a system is prepared in one of a
known collection of states, and the goal is to determine, by means of a
measurement, which state was prepared.
For quantum state tomography, on the other hand, many independent copies of a
single, unknown quantum state are made available, and the goal is to
reconstruct a density matrix description of that state by performing
measurements on the copies.

\section{Mathematical formulations of measurements}

The lesson begins with two equivalent mathematical descriptions of
measurements:
\begin{enumerate}
\item
  General measurements can be described by \emph{collections of matrices}, one
  for each measurement outcome, in a way that generalizes the description of
  projective measurements.
\item
  General measurements can be described as \emph{channels} whose outputs are
  always classical states (represented by diagonal density matrices).
\end{enumerate}

We'll restrict our attention to measurements having \emph{finitely many}
possible outcomes.
Although it is possible to define measurements with infinitely many possible
outcomes, they're much less typically encountered in the context of computation
and information processing, and they also require some additional mathematics
(namely measure theory) to be properly formalized.

Our initial focus will be on so-called \emph{destructive} measurements, where
the output of the measurement is a classical measurement outcome alone --- with
no specification of the post-measurement quantum state of whatever system was
measured.
Intuitively speaking, we can imagine that such a measurement destroys the
quantum system itself, or that the system is immediately discarded once the
measurement is made.
Later in the lesson we'll broaden our view and consider \emph{non-destructive}
measurements, where there's both a classical measurement outcome and a
post-measurement quantum state of the measured system.

\subsection{Measurements as collections of matrices}

Suppose $\mathsf{X}$ is a system that is to be measured, and assume for
simplicity that the classical state set of $\mathsf{X}$ is $\{0,\ldots, n-1\}$
for some positive integer $n,$ so that density matrices representing quantum
states of $\mathsf{X}$ are $n\times n$ matrices.
We won't actually have much need to refer to the classical states of
$\mathsf{X},$ but it will be convenient to refer to $n,$ the number of
classical states of $\mathsf{X}.$
We'll also assume that the possible outcomes of the measurement are the
integers $0,\ldots,m-1$ for some positive integer~$m.$

Note that we're just using these names to keep things simple;
it's straightforward to generalize everything that follows to other finite sets
of classical states and measurement outcomes, renaming them as desired.

\subsubsection{Projective measurements}

Recall that a \emph{projective measurement} is described by a collection of
\emph{projection matrices} that sum to the identity matrix.
In symbols,
\[
\{\Pi_0,\ldots,\Pi_{m-1}\}
\]
describes a projective measurement of $\mathsf{X}$ if each $\Pi_a$ is an
$n\times n$ projection matrix and the following condition is met.
\[
\Pi_0 + \cdots + \Pi_{m-1} = \mathbb{I}_{\mathsf{X}}
\]

When such a measurement is performed on a system $\mathsf{X}$ while it's in a
state described by some quantum state vector $\vert\psi\rangle,$ each outcome
$a$ is obtained with probability equal to $\|\Pi_a\vert\psi\rangle\|^2.$
We also have that the post-measurement state of $\mathsf{X}$ is obtained by
normalizing the vector $\Pi_a\vert\psi\rangle,$ but we're ignoring the
post-measurement state for now.

If the state of $\mathsf{X}$ is described by a density matrix $\rho$ rather
than a quantum state vector $\vert\psi\rangle,$ then we can alternatively
express the probability to obtain the outcome $a$ as $\operatorname{Tr}(\Pi_a
\rho).$
If $\rho = \vert \psi\rangle\langle\psi\vert$ is a pure state, then the two
expressions are equal:
\[
\operatorname{Tr}(\Pi_a \rho)
= \operatorname{Tr}(\Pi_a \vert \psi\rangle\langle\psi \vert)
= \langle \psi \vert \Pi_a \vert \psi \rangle
= \langle \psi \vert \Pi_a \Pi_a \vert \psi \rangle
= \|\Pi_a\vert\psi\rangle\|^2.
\]
Here we're using the cyclic property of the trace for the second equality, and
for the third equality we're using the fact that each $\Pi_a$ is a projection
matrix, and therefore satisfies $\Pi_a^2 = \Pi_a.$

In general, if $\rho$ is a convex combination
\[
\rho = \sum_{k = 0}^{N-1} p_k \vert \psi_k\rangle\langle \psi_k \vert
\]
of pure states, then the expression $\operatorname{Tr}(\Pi_a \rho)$ coincides
with the average probability for the outcome $a,$ owing to the fact that this
expression is linear in $\rho.$
\[
\operatorname{Tr}(\Pi_a \rho)
= \sum_{k = 0}^{N-1} p_k \operatorname{Tr}(\Pi_a \vert
\psi_k\rangle\langle\psi_k\vert)
= \sum_{k = 0}^{N-1} p_k \|\Pi_a\vert\psi_k\rangle\|^2
\]

\subsubsection{General measurements}

A mathematical description for general measurements is obtained by relaxing the
definition of projective measurements.
Specifically, we allow the matrices in the collection describing the
measurement to be arbitrary \emph{positive semidefinite} matrices rather than
projections.
(Projections are always positive semidefinite; they can alternatively be
defined as positive semidefinite matrices whose eigenvalues are all either 0 or
1.)

In particular, a general measurement of a system $\mathsf{X}$ having outcomes
$0,\ldots,m-1$ is specified by a collection of positive semidefinite matrices
$\{P_0,\ldots,P_{m-1}\}$ whose rows and columns correspond to the classical
states of $\mathsf{X}$ and that meet the condition
\[
P_0 + \cdots + P_{m-1} = \mathbb{I}_{\mathsf{X}}.
\]
If the system $\mathsf{X}$ is measured while it is in a state described by the
density matrix $\rho,$ then each outcome $a\in\{0,\ldots,m-1\}$ appears with
probability $\operatorname{Tr}(P_a \rho).$

As we must naturally demand, the vector of outcome probabilities
\[
\bigl(\operatorname{Tr}(P_0 \rho),\ldots,\operatorname{Tr}(P_{m-1} \rho)\bigr)
\]
of a general measurement always forms a probability vector, for any choice of a
density matrix $\rho.$
The following two observations establish that this is the case.
\begin{enumerate}
\item
  Each value $\operatorname{Tr}(P_a \rho)$ must be nonnegative, owing to the
  fact that the trace of the product of any two positive semidefinite matrices
  is always nonnegative:
  \[
  Q, R \geq 0 \; \Rightarrow \: \operatorname{Tr}(QR) \geq 0.
  \]
  One way to argue this fact is to use spectral decompositions of $Q$ and $R$
  together with the cyclic property of the trace to express the trace of the
  product $QR$ as a sum of nonnegative real numbers, which must therefore be
  nonnegative.
\item
  The condition $P_0 + \cdots + P_{m-1} = \mathbb{I}_{\mathsf{X}}$ together
  with the linearity of the trace ensures that the probabilities sum to $1.$
  \[
  \sum_{a = 0}^{m-1} \operatorname{Tr}(P_a \rho)
  = \operatorname{Tr}\Biggl(\sum_{a = 0}^{m-1} P_a \rho\Biggr)
  = \operatorname{Tr}(\mathbb{I}\rho) = \operatorname{Tr}(\rho) = 1
  \]
\end{enumerate}

\subsubsection{Example: any projective measurement}

Projections are always positive semidefinite, so every projective measurement
is an example of a general measurement.

For example, a standard basis measurement of a qubit can be represented by
$\{P_0,P_1\}$ where
\[
P_0 = \vert 0\rangle\langle 0\vert =
\begin{pmatrix}
  1 & 0 \\ 0 & 0
\end{pmatrix}
\quad\text{and}\quad
P_1 = \vert 1\rangle\langle 1\vert =
\begin{pmatrix}
  0 & 0 \\ 0 & 1
\end{pmatrix}.
\]
Measuring a qubit in the state $\rho$ results in outcome probabilities as
follows.
\[
\begin{aligned}
  \operatorname{Prob}(\text{outcome} = 0)
  & = \operatorname{Tr}(P_0 \rho) =
  \operatorname{Tr}\bigl(\vert 0\rangle\langle 0\vert \rho\bigr) =
  \langle 0\vert \rho \vert 0 \rangle \\[1mm]
  \operatorname{Prob}(\text{outcome} = 1)
  & = \operatorname{Tr}(P_1 \rho) =
  \operatorname{Tr}\bigl(\vert 1\rangle\langle 1\vert\rho\bigr) =
  \langle 1 \vert \rho \vert 1 \rangle
\end{aligned}
\]

\subsubsection{Example: a non-projective qubit measurement}

Suppose $\mathsf{X}$ is a qubit, and define two matrices as follows.
\[
P_0 =
\begin{pmatrix}
  \frac{2}{3} & \frac{1}{3}\\[2mm]
  \frac{1}{3} & \frac{1}{3}
\end{pmatrix}
\qquad
P_1 =
\begin{pmatrix}
  \frac{1}{3} & -\frac{1}{3}\\[2mm]
  -\frac{1}{3} & \frac{2}{3}
\end{pmatrix}
\]
These are both positive semidefinite matrices: they're Hermitian, and in both
cases the eigenvalues happen to be $1/2 \pm \sqrt{5}/6,$ which are both
positive.
We also have that $P_0 + P_1 = \mathbb{I},$ and therefore $\{P_0,P_1\}$
describes a measurement.

If the state of $\mathsf{X}$ is described by a density matrix $\rho$ and we
perform this measurement, then the probability of obtaining the outcome $0$ is
$\operatorname{Tr}(P_0 \rho)$ and the probability of obtaining the outcome $1$
is $\operatorname{Tr}(P_1 \rho).$
For instance, if $\rho = \vert + \rangle \langle + \vert$ then the
probabilities for the two outcomes $0$ and $1$ are as follows.
\[
\begin{aligned}
  \operatorname{Tr}(P_0 \rho)
  & = \operatorname{Tr}\left(
  \begin{pmatrix}
    \frac{2}{3} & \frac{1}{3}\\[2mm]
    \frac{1}{3} & \frac{1}{3}
  \end{pmatrix}
  \begin{pmatrix}
    \frac{1}{2} & \frac{1}{2}\\[2mm]
    \frac{1}{2} & \frac{1}{2}
  \end{pmatrix}
  \right)
  = \frac{5}{6}\\[4mm]
  \operatorname{Tr}(P_1 \rho)
  & = \operatorname{Tr}\left(
  \begin{pmatrix}
    \frac{1}{3} & -\frac{1}{3}\\[2mm]
    -\frac{1}{3} & \frac{2}{3}
  \end{pmatrix}
  \begin{pmatrix}
    \frac{1}{2} & \frac{1}{2}\\[2mm]
    \frac{1}{2} & \frac{1}{2}
  \end{pmatrix}
  \right)
  = \frac{1}{6}
\end{aligned}
\]

\subsubsection{Example: tetrahedral measurement}

\begin{figure}[!ht]
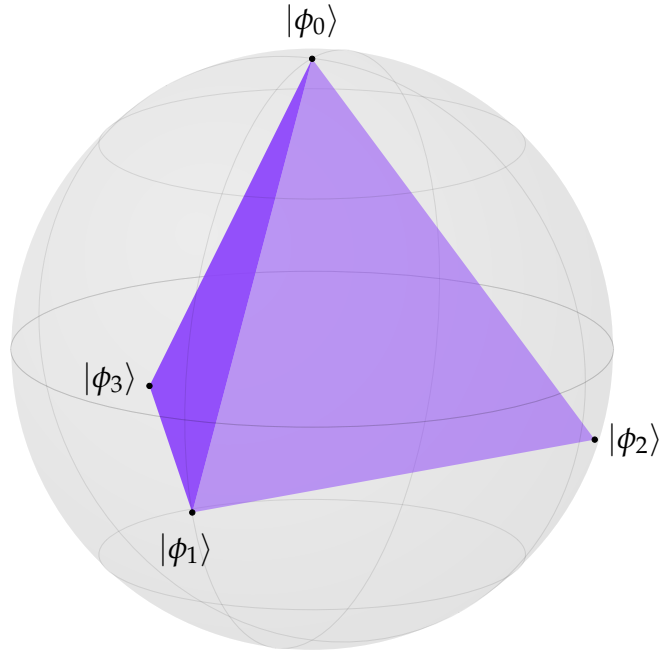

  \begin{center}
    \begin{blochsphere}[
        color=Gray,
        tilt = -15,
        radius=4cm,
        opacity=0.075,
        rotation=115]
      
      \drawBallGrid[style={opacity=0.1}]{45}{90}
      
      \coordinate (origin) at (0,0);
      
      \labelLatLon{0}{90}{0}
      \labelLatLon{1}{-19.47}{0}
      \labelLatLon{2}{-19.47}{120}
      \labelLatLon{3}{-19.47}{240}
      
      \node[anchor = -90, circle, inner sep=0mm] at (0) {$\ket{\phi_0}$};
      \node[anchor = 80, circle, inner sep=0mm] at (1) {$\ket{\phi_1}$};
      \node[anchor = 180, circle, inner sep=0.25mm] at (2) {$\ket{\phi_2}$};
      \node[anchor = 350, circle, inner sep=0.25mm] at (3) {$\ket{\phi_3}$};
         
      \fill[color = DataColor1, opacity = 0.5] (0) -- (1) -- (2) -- (0);
      \fill[color = DataColor1, opacity = 0.9] (0) -- (1) -- (3) -- (0);
      
      \filldraw (0) circle (1pt);
      \filldraw (1) circle (1pt);
      \filldraw (2) circle (1pt);
      \filldraw (3) circle (1pt);

    \end{blochsphere}
  \end{center}
  \caption{The tetrahedral states form the vertices of a regular tetrahedron
    inscribed within the Bloch sphere.}
  \label{fig:tetrahedral-states}
\end{figure}

Define four single-qubit quantum state vectors as follows.
\[
\begin{aligned}
  \vert\phi_0\rangle & = \vert 0 \rangle\\
  \vert\phi_1\rangle & = \frac{1}{\sqrt{3}}\vert 0 \rangle + \sqrt{\frac{2}{3}}
  \vert 1\rangle \\
  \vert\phi_2\rangle & = \frac{1}{\sqrt{3}}\vert 0 \rangle + \sqrt{\frac{2}{3}}
  e^{2\pi i/3} \vert 1\rangle \\
  \vert\phi_3\rangle & = \frac{1}{\sqrt{3}}\vert 0 \rangle + \sqrt{\frac{2}{3}}
  e^{-2\pi i/3} \vert 1\rangle
\end{aligned}
\]
These four states are sometimes known as \emph{tetrahedral} states because
they're vertices of a \emph{regular tetrahedron} inscribed within the Bloch
sphere, as illustrated in Figure~\ref{fig:tetrahedral-states}.

The Cartesian coordinates of these four states on the Bloch sphere are
\[
(0,0,1),\;
\left( \frac{2\sqrt{2}}{3} , 0 , -\frac{1}{3} \right),\;
\left( -\frac{\sqrt{2}}{3} , \sqrt{\frac{2}{3}} , -\frac{1}{3} \right),\;
\left( -\frac{\sqrt{2}}{3} , -\sqrt{\frac{2}{3}} , -\frac{1}{3} \right),
\]
which can be verified by expressing the density matrices representations of
these states as linear combinations of Pauli matrices.
\[
\begin{aligned}
  \vert \phi_0 \rangle\langle \phi_0 \vert & =
  \begin{pmatrix}
    1 & 0\\[1mm] 0 & 0
  \end{pmatrix}
  = \frac{\mathbb{I} + \sigma_z}{2} \\[2mm]
  \vert \phi_1 \rangle\langle \phi_1 \vert & =
  \begin{pmatrix}
    \frac{1}{3} & \frac{\sqrt{2}}{3} \\[2mm]
    \frac{\sqrt{2}}{3} & \frac{2}{3}
  \end{pmatrix}
  = \frac{\mathbb{I} + \frac{2\sqrt{2}}{3} \sigma_x - \frac{1}{3}\sigma_z}{2}
  \\[2mm]
  \vert \phi_2 \rangle\langle \phi_2 \vert & =
  \begin{pmatrix}
    \frac{1}{3} & -\frac{1}{3\sqrt{2}} - \frac{i}{\sqrt{6}} \\[2mm]
    -\frac{1}{3\sqrt{2}} + \frac{i}{\sqrt{6}} & \frac{2}{3}
  \end{pmatrix}
  = \frac{\mathbb{I} - \frac{\sqrt{2}}{3} \sigma_x + \sqrt{\frac{2}{3}}
    \sigma_y - \frac{1}{3}\sigma_z}{2}
  \\[2mm]
  \vert \phi_3 \rangle\langle \phi_3 \vert & =
  \begin{pmatrix}
    \frac{1}{3} & -\frac{1}{3\sqrt{2}} + \frac{i}{\sqrt{6}} \\[2mm]
    -\frac{1}{3\sqrt{2}} - \frac{i}{\sqrt{6}} & \frac{2}{3}
  \end{pmatrix}
  = \frac{\mathbb{I} - \frac{\sqrt{2}}{3} \sigma_x - \sqrt{\frac{2}{3}}
    \sigma_y - \frac{1}{3}\sigma_z}{2}
\end{aligned}
\]
These four states are perfectly spread out on the Bloch sphere, each one
equidistant from the other three and with the angles between any two of them
always being the same.

Now let us define a measurement $\{P_0,P_1,P_2,P_3\}$ of a qubit by setting
$P_a$ as follows for each $a=0,\ldots,3.$
\[
P_a = \frac{\vert\phi_a\rangle\langle\phi_a\vert}{2}
\]
We can verify that this is a valid measurement as follows.
\begin{enumerate}
\item
Each $P_a$ is evidently positive semidefinite, being a pure state divided by
one-half.
That is, each one is a Hermitian matrix and has one eigenvalue equal to $1/2$
and all other eigenvalues zero.
\item
  The sum of these matrices is the identity matrix: $P_0 + P_1 + P_2 + P_3 =
  \mathbb{I}.$
  The expressions of these matrices as linear combinations of Pauli matrices
  makes this straightforward to verify.
\end{enumerate}

\subsection{Measurements as channels}

A second way to describe measurements in mathematical terms is as channels.

Classical information can be viewed as a special case of quantum information,
insofar as we can identify probabilistic states with diagonal density matrices.
So, in operational terms, we can think about measurements as being channels
whose inputs are matrices describing states of whatever system is being
measured and whose outputs are \emph{diagonal} density matrices describing the
resulting distribution of measurement outcomes.

We'll see shortly that any channel having this property can always be written
in a simple, canonical form that ties directly to the description of
measurements as collections of positive semidefinite matrices.
Conversely, given an arbitrary measurement as a collection of matrices, there's
always a valid channel having the diagonal output property that describes the
given measurement as suggested in the previous paragraph.
Putting these observations together, we find that the two descriptions of
general measurements are equivalent.

Before proceeding further, let's be more precise about the measurement, how
we're viewing it as a channel, and what assumptions we're making about it.
As before, we'll suppose that $\mathsf{X}$ is the system to be measured, and
that the possible outcomes of the measurement are the integers $0,\ldots,m-1$
for some positive integer $m.$
We'll let $\mathsf{Y}$ be the system that stores measurement outcomes, so its
classical state set is $\{0,\ldots,m-1\},$ and we represent the measurement as
a channel named $\Phi$ from $\mathsf{X}$ to $\mathsf{Y}.$

Our assumption is that $\mathsf{Y}$ is \emph{classical} --- which is to say
that no matter what state we start with for $\mathsf{X},$ the state of
$\mathsf{Y}$ we obtain is represented by a diagonal density matrix.
We can express in mathematical terms that the output of $\Phi$ is always
diagonal in the following way.
First define the completely dephasing channel $\Delta_m$ on $\mathsf{Y}.$
\[
\Delta_m(\sigma) = \sum_{a = 0}^{m-1} \langle a \vert \sigma \vert a\rangle
\,\vert a\rangle\langle a\vert
\]
This channel is analogous to the completely dephasing qubit channel $\Delta$
from the previous lesson.
As a linear mapping, it zeros out all of the off-diagonal entries of an input
matrix and leaves the diagonal alone.

And now, a simple way to express that a given density matrix $\sigma$ is
diagonal is by the equation $\sigma = \Delta_m(\sigma).$
In words, zeroing out all of the off-diagonal entries of a density matrix has
no effect if and only if the off-diagonal entries were all zero to begin with.
The channel $\Phi$ therefore satisfies our assumption --- that $\mathsf{Y}$ is
classical --- if and only if
\[
\Phi(\rho) = \Delta_m(\Phi(\rho))
\]
for every density matrix $\rho$ representing a state of $\mathsf{X}.$

\subsection{Equivalence of the formulations}

\subsubsection{Channels to matrices}

Suppose that we have a channel from $\mathsf{X}$ to $\mathsf{Y}$ with the
property that
\[
\Phi(\rho) = \Delta_m(\Phi(\rho))
\]
for every density matrix $\rho.$
This may alternatively be expressed as follows.
\begin{equation}
  \Phi(\rho) =
  \sum_{a = 0}^{m-1} \langle a \vert \Phi(\rho) \vert a\rangle\, \vert
  a\rangle\langle a \vert
  \label{eq:dephased-channel}
\end{equation}

Like all channels, we can express $\Phi$ in Kraus form for some way of choosing
Kraus matrices $A_0,\ldots,A_{N-1}.$
\[
\Phi(\rho) = \sum_{k = 0}^{N-1} A_k \rho A_k^{\dagger}
\]
This provides us with an alternative expression for the diagonal entries of
$\Phi(\rho)$:
\[
\begin{aligned}
  \langle a \vert \Phi(\rho) \vert a\rangle
  & = \sum_{k = 0}^{N-1} \langle a \vert A_k \rho A_k^{\dagger} \vert a\rangle
  \\
  & = \sum_{k = 0}^{N-1} \operatorname{Tr}\bigl( A_k^{\dagger} \vert
  a\rangle\langle a \vert A_k \rho\bigr)\\
  & = \operatorname{Tr}\bigl(P_a\rho\bigr)
\end{aligned}
\]
for
\[
P_a = \sum_{k = 0}^{N-1} A_k^{\dagger} \vert a\rangle\langle a \vert A_k.
\]

Thus, for these same matrices $P_0,\ldots,P_{m-1}$ we can express the channel
$\Phi$ as follows.
\[
\Phi(\rho) = \sum_{a = 0}^{m-1} \operatorname{Tr}(P_a \rho) \vert
a\rangle\langle a\vert
\]
This expression is consistent with our description of general measurements in
terms of matrices, as we see each measurement outcome appearing with
probability $\operatorname{Tr}(P_a \rho).$

Now let's observe that the two properties required of the collection of
matrices $\{P_0,\ldots,P_{m-1}\}$ to describe a general measurement are indeed
satisfied.
The first property is that they're all positive semidefinite matrices.
One way to see this is to observe that, for every vector $\vert \psi\rangle$
having entries in correspondence with the classical state of $\mathsf{X}$ we
have
\[
\langle \psi \vert P_a \vert \psi\rangle
= \sum_{k = 0}^{N-1} \langle \psi \vert A_k^{\dagger} \vert a\rangle\langle a
\vert A_k\vert \psi\rangle
=  \sum_{k = 0}^{N-1} \bigl\vert\langle a \vert A_k\vert
\psi\rangle\bigr\vert^2 \geq 0.
\]
The second property is that if we sum these matrices we get the identity
matrix.
\[
\begin{aligned}
  \sum_{a = 0}^{m-1} P_a
  & = \sum_{a = 0}^{m-1} \sum_{k = 0}^{N-1} A_k^{\dagger} \vert a\rangle\langle
  a \vert A_k \\
  & = \sum_{k = 0}^{N-1} A_k^{\dagger} \Biggl(\sum_{a = 0}^{m-1} \vert
  a\rangle\langle a \vert\Biggr) A_k \\
  & = \sum_{k = 0}^{N-1} A_k^{\dagger} A_k \\
  & = \mathbb{I}_{\mathsf{X}}
\end{aligned}
\]
The last equality follows from the fact that $\Phi$ is a channel, so its Kraus
matrices must satisfy this condition.

\subsubsection{Matrices to channels}

Now let's verify that for any collection $\{P_0,\ldots,P_{m-1}\}$ of positive
semidefinite matrices satisfying $P_0 + \cdots + P_{m-1} =
\mathbb{I}_{\mathsf{X}},$ the mapping defined by
\[
\Phi(\rho) = \sum_{a = 0}^{m-1} \operatorname{Tr}(P_a \rho) \vert a
\rangle\langle a\vert
\]
is indeed a valid channel from $\mathsf{X}$ to $\mathsf{Y}.$

One way to do this is to compute the Choi representation of this mapping.
\[
\begin{aligned}
  J(\Phi) & = \sum_{b,c = 0}^{n-1} \vert b \rangle \langle c \vert \otimes
  \Phi(\vert b \rangle \langle c \vert)\\[1mm]
  & = \sum_{b,c = 0}^{n-1} \sum_{a = 0}^{m-1} \vert b \rangle \langle c \vert
  \otimes
  \operatorname{Tr}(P_a \vert b \rangle \langle c \vert) \vert a \rangle\langle
  a\vert\\[1mm]
  & = \sum_{b,c = 0}^{n-1} \sum_{a = 0}^{m-1} \vert b \rangle \langle b \vert
  P_a^T \vert c \rangle \langle c \vert \otimes
  \vert a \rangle\langle a\vert\\[1mm]
  & = \sum_{a = 0}^{m-1}  P_a^T \otimes \vert a \rangle\langle a\vert
\end{aligned}
\]
The transpose of each $P_a$ is introduced for the third equality because
\[
\langle c \vert P_a \vert b\rangle = \langle b \vert P_a^T \vert c\rangle.
\]
This allows for the expressions $\vert b \rangle \langle b \vert$ and $\vert c
\rangle \langle c \vert$ to appear, which simplify to the identity matrix upon
summing over $b$ and $c,$ respectively.

By the assumption that $P_0,\ldots,P_{m-1}$ are positive semidefinite, so too
are the matrices $P_0^{T},\ldots,P_{m-1}^{T}.$
In particular, transposing a Hermitian matrix results in another Hermitian
matrix, and the eigenvalues of any square matrix and its transpose always
agree.
It follows that $J(\Phi)$ is positive semidefinite.
Tracing out the output system $\mathsf{Y}$ (which is the system on the right)
yields
\[
\operatorname{Tr}_{\mathsf{Y}} (J(\Phi)) = \sum_{a = 0}^{m-1}  P_a^T =
\mathbb{I}_{\mathsf{X}}^T = \mathbb{I}_{\mathsf{X}},
\]
and so we conclude that $\Phi$ is a channel.

\subsection{Partial measurements}

Suppose that we have multiple systems that are collectively in a quantum state,
and a general measurement is performed on one of the systems.
This results in one of the measurement outcomes, selected at random according
to probabilities determined by the measurement and the state of the system
prior to the measurement.
The resulting state of the remaining systems will then, in general, depend on
which measurement outcome was obtained.

Let's examine how this works for a pair of systems $(\mathsf{X},\mathsf{Z})$
when the system $\mathsf{X}$ is measured.
(We're naming the system on the right $\mathsf{Z}$ because we'll take
$\mathsf{Y}$ to be a system representing the classical output of the
measurement when we view it as a channel.)
We can then easily generalize to the situation in which the systems are swapped
as well as to three or more systems.

Suppose the state of $(\mathsf{X},\mathsf{Z})$ prior to the measurement is
described by a density matrix $\rho,$ which we can write as follows.
\[
\rho = \sum_{b,c = 0}^{n-1} \vert b\rangle\langle c\vert \otimes \rho_{b,c}
\]
In this expression we're assuming the classical states of $\mathsf{X}$ are
$0,\ldots,n-1.$

We'll assume that the measurement itself is described by the collection of
matrices $\{P_0,\ldots,P_{m-1}\}.$
This measurement may alternatively be described as a channel $\Phi$ from
$\mathsf{X}$ to $\mathsf{Y},$ where $\mathsf{Y}$ is a new system having
classical state set $\{0,\ldots,m-1\}.$
Specifically, the action of this channel can be expressed as follows.
\[
\Phi(\xi) = \sum_{a = 0}^{m-1} \operatorname{Tr}(P_a \xi)\, \vert a \rangle
\langle a \vert
\]

\subsubsection{Outcome probabilities}

We're considering a measurement of the system $\mathsf{X},$ so the
probabilities with which different measurement outcomes are obtained can depend
only on $\rho_{\mathsf{X}},$ the reduced state of $\mathsf{X}.$
In particular, the probability for each outcome $a\in\{0,\ldots,m-1\}$ to
appear can be expressed in three equivalent ways.
\[
\operatorname{Tr}\bigl( P_a \rho_{\mathsf{X}}\bigr) =
\operatorname{Tr}\bigl( P_a \operatorname{Tr}_{\mathsf{Z}}(\rho)\bigr) =
\operatorname{Tr}\bigl( (P_a \otimes \mathbb{I}_{\mathsf{Z}}) \rho \bigr)
\]
The first expression naturally represents the probability to obtain the outcome
$a$ based on what we already know about measurements of a single system.
To get the second expression we're simply using the definition
$\rho_{\mathsf{X}} = \operatorname{Tr}_{\mathsf{Z}}(\rho).$

To get the third expression requires more thought --- and learners are
encouraged to convince themselves that it is true.
Here's a hint: The equivalence between the second and third expressions does
not depend on $\rho$ being a density matrix or on each $P_a$ being positive
semidefinite. Try showing it first for tensor products of the form
$\rho = M\otimes N$ and then conclude that it must be true in general by
linearity.

While the equivalence of the first and third expressions in the previous
equation may not be immediate, it does make sense.
Starting from a measurement on $\mathsf{X},$ we're effectively defining a
measurement of $(\mathsf{X},\mathsf{Z}),$ where we simply throw away
$\mathsf{Z}$ and measure $\mathsf{X}.$
Like all measurements, this new measurement can be described by a collection of
matrices, and it's not surprising that this measurement is described by the
collection
\[
\{P_0\otimes\mathbb{I}_{\mathsf{Z}}, \ldots,
P_{m-1}\otimes\mathbb{I}_{\mathsf{Z}}\}.
\]

\subsubsection{States conditioned on measurement outcomes}

If we want to determine not only the probabilities for the different outcomes
but also the resulting state of $\mathsf{Z}$ conditioned on each measurement
outcome, we can look to the channel description of the measurement.
In particular, let's examine the state we get when we apply $\Phi$ to
$\mathsf{X}$ and do nothing to $\mathsf{Z}.$
\[
\begin{aligned}
  (\Phi\otimes\operatorname{Id}_{\mathsf{Z}})(\rho)
  & = \sum_{b,c = 0}^{n-1} \Phi(\vert b\rangle\langle c\vert) \otimes
  \rho_{b,c}\\
  & = \sum_{a = 0}^{m-1} \sum_{b,c = 0}^{n-1} \operatorname{Tr}(P_a \vert
  b\rangle\langle c\vert) \,\vert a\rangle \langle a \vert \otimes \rho_{b,c}\\
  & = \sum_{a = 0}^{m-1}  \vert a\rangle \langle a \vert \otimes
  \sum_{b,c = 0}^{n-1} \operatorname{Tr}(P_a \vert b\rangle\langle c\vert)
  \rho_{b,c}\\
  & = \sum_{a = 0}^{m-1}  \vert a\rangle \langle a \vert \otimes
  \sum_{b,c = 0}^{n-1}
  \operatorname{Tr}_{\mathsf{X}}\bigl((P_a\otimes\mathbb{I}_{\mathsf{Z}})
  (\vert b\rangle\langle  c\vert\otimes\rho_{b,c})\bigr)\\
  & = \sum_{a = 0}^{m-1} \vert a\rangle \langle a \vert \otimes
  \operatorname{Tr}_{\mathsf{X}}\bigl((P_a \otimes \mathbb{I}_{\mathsf{Z}})
  \rho\bigr)
\end{aligned}
\]
Note that this is a density matrix by virtue of the fact that $\Phi$ is a
channel, so each matrix
$\operatorname{Tr}_{\mathsf{X}}\bigl((P_a \otimes \mathbb{I}_{\mathsf{Z}})
\rho)$ is necessarily positive semidefinite.

One final step transforms this expression into one that reveals what we're
looking for.
\[
\sum_{a = 0}^{m-1} \operatorname{Tr}\bigl((P_a \otimes \mathbb{I}_{\mathsf{Z}})
\rho)\, \vert a\rangle \langle a \vert \otimes
\frac{\operatorname{Tr}_{\mathsf{X}}\bigl((P_a \otimes \mathbb{I}_{\mathsf{Z}})
  \rho)}{\operatorname{Tr}\bigl((P_a \otimes \mathbb{I}_{\mathsf{Z}}) \rho)}
\]
This is an example of a \emph{classical-quantum state},
\[
\sum_{a = 0}^{m-1} p(a)\, \vert a\rangle\langle a\vert \otimes \sigma_a,
\]
like we saw in Lesson~\ref{lesson:density-matrices} \emph{(Density Matrices)}.
For each measurement outcome $a\in\{0,\ldots,m-1\},$ we have with probability
\[
p(a) = \operatorname{Tr}\bigl((P_a \otimes \mathbb{I}_{\mathsf{Z}}) \rho)
\]
that $\mathsf{Y}$ is in the classical state $\vert a \rangle \langle a \vert$
and $\mathsf{Z}$ is in the state
\begin{equation}
  \sigma_a = \frac{\operatorname{Tr}_{\mathsf{X}}\bigl((P_a \otimes
    \mathbb{I}_{\mathsf{Z}}) \rho)}{\operatorname{Tr}\bigl((P_a \otimes
    \mathbb{I}_{\mathsf{Z}}) \rho)}.
  \label{eq:post-measurement-state}
\end{equation}
That is, this is the density matrix we obtain by normalizing
\[
\operatorname{Tr}_{\mathsf{X}}\bigl((P_a \otimes \mathbb{I}_{\mathsf{Z}}) \rho)
\]
by dividing it by its trace.
(Formally speaking, the state $\sigma_a$ is only defined when the probability
$p(a)$ is nonzero; when $p(a) = 0$ this state is irrelevant, for it refers to a
discrete event that occurs with probability zero.)
Naturally, the outcome probabilities are consistent with our previous
observations.

In summary, this is what happens when the measurement $\{P_0,\ldots,P_{m-1}\}$
is performed on $\mathsf{X}$ when $(\mathsf{X},\mathsf{Z})$ is in the state
$\rho.$
\begin{enumerate}
\item
  Each outcome $a$ appears with probability
  $p(a) = \operatorname{Tr}\bigl((P_a \otimes \mathbb{I}_{\mathsf{Z}}) \rho).$
\item
  Conditioned on obtaining outcome $a,$ the state of $\mathsf{Z}$ is then
  represented by the density matrix $\sigma_a$ shown in the equation
  \eqref{eq:post-measurement-state}, which is obtained by normalizing
  $\operatorname{Tr}_{\mathsf{X}}\bigl((P_a \otimes \mathbb{I}_{\mathsf{Z}})
  \rho).$
\end{enumerate}

\subsubsection{Generalization}

We can adapt this description to other situations, such as when the ordering of
the systems is reversed or when there are three or more systems.
Conceptually it is straightforward, although it can become cumbersome to write
down the formulas.

In general, if we have $r$ systems $\mathsf{X}_1,\ldots,\mathsf{X}_r,$ the
state of the compound system $(\mathsf{X}_1,\ldots,\mathsf{X}_r)$ is $\rho,$
and the measurement $\{P_0,\ldots,P_{m-1}\}$ is performed on $\mathsf{X}_k$,
the following happens.
\begin{enumerate}
\item
  Each outcome $a$ appears with probability
  \[
  p(a) = \operatorname{Tr}\bigl((\mathbb{I}_{\mathsf{X}_1}\otimes \cdots
  \otimes\mathbb{I}_{\mathsf{X}_{k-1}} \otimes P_a \otimes
  \mathbb{I}_{\mathsf{X}_{k+1}} \otimes \cdots
  \otimes\mathbb{I}_{\mathsf{X}_r}) \rho\bigr).
  \]
\item
  Conditioned on obtaining outcome $a,$ the state of
  $(\mathsf{X}_1,\ldots,\mathsf{X}_{k-1},\mathsf{X}_{k+1},\ldots,\mathsf{X}_r)$
  is then represented by the following density matrix.
  \[
  \frac{\operatorname{Tr}_{\mathsf{X}_k}\bigl((\mathbb{I}_{\mathsf{X}_1}\otimes
    \cdots \otimes\mathbb{I}_{\mathsf{X}_{k-1}} \otimes P_a \otimes
    \mathbb{I}_{\mathsf{X}_{k+1}} \otimes \cdots
    \otimes\mathbb{I}_{\mathsf{X}_r})
    \rho\bigr)}{\operatorname{Tr}\bigl((\mathbb{I}_{\mathsf{X}_1}\otimes \cdots
    \otimes\mathbb{I}_{\mathsf{X}_{k-1}} \otimes P_a \otimes
    \mathbb{I}_{\mathsf{X}_{k+1}} \otimes \cdots
    \otimes\mathbb{I}_{\mathsf{X}_r}) \rho\bigr)}
  \]
\end{enumerate}

\section{Naimark's theorem}

Naimark's theorem is a fundamental fact concerning measurements.
It states that every general measurement can be implemented in a simple way
that's reminiscent of Stinespring representations of channels:
\begin{enumerate}
\item
  The system to be measured is first combined with an initialized workspace
  system, forming a compound system.
\item
  A unitary operation is then performed on the compound system.
\item
  Finally, the workspace system is \emph{measured} with respect to a
  \emph{standard basis measurement}, yielding the outcome of the original
  general measurement.
\end{enumerate}

\subsection{Theorem statement and proof}

Here's a statement of Naimark's theorem.

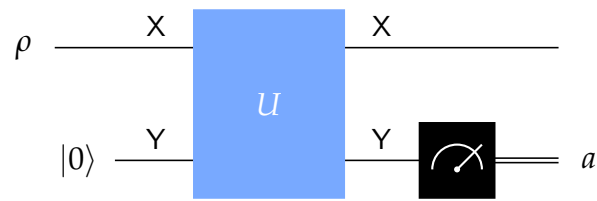
\begin{figure}[!ht]
  \begin{center}
    \begin{tikzpicture}[
        line width = 0.6pt,
        gate/.style={%
          inner sep = 0,
          fill = CircuitBlue,
          draw = CircuitBlue,
          text = white,
          minimum size = 6mm
        },
        blackgate/.style={%
          inner sep = 0,
          fill = black,
          draw = black,
          text = white,
          minimum size = 10mm}]
      
      \node (Xin) at (-3,1.5) {};
      \node (Yin) at (-2.2,0) {};
      \node (Xout) at (4,1.5) {};
      \node (M) at (2.5,0) {};
      \node (Mout) at (4,0) {};
      
      \draw (Xin) -- (Xout);          
      \draw (Yin) -- (M.west);

      \draw ([yshift=0.3mm]M.east) -- ([yshift=0.3mm]Mout.west);
      \draw ([yshift=-0.3mm]M.east) -- ([yshift=-0.3mm]Mout.west);
      
      \node[gate, minimum height=25mm, minimum width=20mm]
      (U) at (0,0.75) {$U$};          
      
      \node[anchor = south] at (-1.5,1.5) {\small $\mathsf{X}$};
      \node[anchor = south] at (-1.5,0) {\small $\mathsf{Y}$};
      \node[anchor = south] at (1.5,1.5) {\small $\mathsf{X}$};
      \node[anchor = south] at (1.5,0) {\small $\mathsf{Y}$};
      
      \node[anchor = east] at (Xin) {$\rho$};
      \node[anchor = west] at (Xout) {};
      \node[anchor = east, xshift = 0.75mm] at (Yin) {$\ket{0}$};

      \node[anchor = west] at (Mout) {$a$};
      
      \node[blackgate] at (M) {};
      \readout{M}
      
    \end{tikzpicture}
  \end{center}
  \caption{The implementation of a general measurement using a workspace
    system, a unitary operation, and a standard basis measurement, as in
    Naimark's theorem.}
  \label{fig:Naimark}
\end{figure}

\begin{callout}[title = {Naimark's theorem}]
  Let $\mathsf{X}$ be a system and let $\{P_0,\ldots,P_{m-1}\}$ be a collection
  of positive semidefinite matrices describing a general measurement on
  $\mathsf{X}.$
  Also let $\mathsf{Y}$ be a system whose classical state set is
  $\{0,\ldots,m-1\}$ (i.e., the set of possible outcomes of this measurement).
  \vspace{2mm}

  There exists a unitary operation $U$ on the compound system
  $(\mathsf{Y},\mathsf{X})$ so that the implementation suggested
  by Figure~\ref{fig:Naimark} has measurement outcome probabilities that
  agree precisely with the measurement $\{P_0,\ldots,P_{m-1}\}$,
  for all choices of an input state $\rho$.
\end{callout}

To be clear, the system $\mathsf{X}$ starts out in some arbitrary state $\rho$
while $\mathsf{Y}$ is initialized to the $\vert 0\rangle$ state.
The unitary operation $U$ is applied to $(\mathsf{Y},\mathsf{X})$ and then the
system $\mathsf{Y}$ is measured with a standard basis measurement, yielding
some outcome $a\in\{0,\ldots,m-1\}.$
The system $\mathsf{X}$ is pictured as part of the output of the circuit, but
for now we won't concern ourselves with the state of $\mathsf{X}$ after $U$ is
performed, and can imagine that it is traced out.
We'll be interested in the state of $\mathsf{X}$ after $U$ is performed later
in the lesson, though.

An implementation of a measurement in this way is clearly reminiscent of a
Stinespring representation of a channel, and the mathematical underpinnings are
similar as well.
The difference here is that the workspace system is measured rather than being
traced out like in the case of a Stinespring representation.

The fact that every measurement can be implemented in this way is pretty simple
to prove, but we're going to need a fact concerning positive semidefinite
matrices first.

\begin{callout}[title={Square root of a positive semidefinite matrix}]
  Suppose $P$ is a positive semidefinite matrix.
  There exists a unique positive semidefinite matrix $Q$ for which $Q^2 = P.$
  This unique positive semidefinite matrix is called the \emph{square root} of
  $P$ and is denoted $\sqrt{P}.$
\end{callout}
\noindent
One way to find the square root of a positive semidefinite matrix is to first
compute a spectral decomposition.
\[
P = \sum_{k=0}^{n-1} \lambda_k \vert \psi_k \rangle \langle \psi_k \vert
\]
Because $P$ is positive semidefinite, its eigenvalues must be nonnegative real
numbers, and by replacing them with their square roots we obtain an expression
for the square root of $P.$
\[
\sqrt{P} = \sum_{k=0}^{n-1} \sqrt{\lambda_k} \vert \psi_k \rangle \langle
\psi_k \vert
\]

With this concept in hand, we're ready to prove Naimark's theorem.
Under the assumption that $\mathsf{X}$ has $n$ classical states, a unitary
operation $U$ on the pair $(\mathsf{Y},\mathsf{X})$ can be represented by an
$nm\times nm$ matrix, which we can view as an $m\times m$ block matrix whose
blocks are $n\times n.$
The key to the proof is to take $U$ to be any unitary matrix that matches the
following pattern.
\[
U =
\begin{pmatrix}
\sqrt{P_0} & \fbox{?} & \cdots & \fbox{?} \\[1mm]
\sqrt{P_1} & \fbox{?} & \cdots & \fbox{?} \\[1mm]
\vdots & \vdots & \ddots & \vdots\\[1mm]
\sqrt{P_{m-1}} & \fbox{?} & \cdots & \fbox{?}
\end{pmatrix}
\]

For it to be possible to fill in the blocks marked with a question mark so that
$U$ is unitary, it's both necessary and sufficient that the first $n$ columns,
which are formed by the blocks $\sqrt{P_0},\ldots,\sqrt{P_{m-1}},$ are
orthonormal.
We can then use the Gram--Schmidt orthogonalization process to fill in the
remaining columns.

The first $n$ columns of $U$ can be expressed as vectors in the following way,
where $c = 0,\ldots,n-1$ refers to the column number starting from $0.$
\[
\vert\gamma_c\rangle = \sum_{a = 0}^{m-1} \vert a \rangle \otimes \sqrt{P_a}
\vert c\rangle
\]
We can compute the inner product between any two of them as follows.
\[
\langle \gamma_c \vert \gamma_d \rangle =
\sum_{a,b = 0}^{m-1} \langle a \vert b \rangle \cdot \langle c \vert
\sqrt{P_a}\sqrt{P_b}\, \vert d\rangle
= \langle c \vert \Biggl(\sum_{a = 0}^{m-1}  P_a \Biggr) \vert d\rangle
= \langle c \vert d\rangle
\]
This shows that these columns are in fact orthonormal, so we can fill in the
remaining columns of $U$ in a way that guarantees the entire matrix is unitary.

It remains to check that the measurement outcome probabilities for the
simulation are consistent with the original measurement.
For a given initial state $\rho$ of $\mathsf{X},$ the measurement described by
the collection $\{P_0,\ldots,P_{m-1}\}$ results in each outcome
$a\in\{0,\ldots,m-1\}$ with probability $\operatorname{Tr}(P_a \rho).$

To obtain the outcome probabilities for the simulation, let's first give the
name $\sigma$ to the state of $(\mathsf{Y},\mathsf{X})$ after $U$ has been
performed.
This state can be expressed as follows.
\[
\sigma =
U \bigl(\vert 0\rangle \langle 0 \vert \otimes \rho\bigr) U^{\dagger}
= \sum_{a,b=0}^{m-1} \vert a\rangle \langle b \vert \otimes \sqrt{P_a} \rho
\sqrt{P_b}
\]
Equivalently, in a block matrix form, we have the following equation.
\[
\begin{aligned}
  \sigma & =
  \begin{pmatrix}
    \sqrt{P_0} & \fbox{?} & \cdots & \fbox{?} \\[1mm]
    \sqrt{P_1} & \fbox{?} & \cdots & \fbox{?} \\[1mm]
    \vdots & \vdots & \ddots & \vdots\\[1mm]
    \sqrt{P_{m-1}} & \fbox{?} & \cdots & \fbox{?}
  \end{pmatrix}
  \begin{pmatrix}
    \rho & 0 & \cdots & 0 \\[1mm]
    0 & 0 & \cdots & 0 \\[1mm]
    \vdots & \vdots & \ddots & \vdots\\[1mm]
    0 & 0 & \cdots & 0
  \end{pmatrix}
  \begin{pmatrix}
    \sqrt{P_0} & \sqrt{P_1} & \cdots & \sqrt{P_{m-1}} \\[1mm]
    \fbox{?} & \fbox{?} & \cdots & \fbox{?} \\[1mm]
    \vdots & \vdots & \ddots & \vdots\\[1mm]
    \fbox{?} & \fbox{?} & \cdots & \fbox{?}
  \end{pmatrix}\\[1mm]
  & = \begin{pmatrix}
    \sqrt{P_0}\rho\sqrt{P_0} & \cdots & \sqrt{P_0}\rho\sqrt{P_{m-1}} \\[1mm]
    \vdots & \ddots & \vdots\\[1mm]
    \sqrt{P_{m-1}}\rho\sqrt{P_0} & \cdots & \sqrt{P_{m-1}}\rho\sqrt{P_{m-1}}
  \end{pmatrix}
\end{aligned}
\]
Notice that the entries of $U$ falling into the blocks marked with a question
mark have no influence on the outcome by virtue of the fact that we're
conjugating a matrix of the form $\vert 0 \rangle \langle 0 \vert \otimes \rho$
--- so the question mark entries are always multiplied by zero entries of
$\vert 0 \rangle \langle 0 \vert \otimes \rho$ when the matrix product is
computed.

Now we can analyze what happens when a standard basis measurement is performed
on $\mathsf{Y}.$
The probabilities of the possible outcomes are given by the diagonal entries of
the reduced state $\sigma_{\mathsf{Y}}$ of $\mathsf{Y}.$
\[
\sigma_{\mathsf{Y}} = \sum_{a,b=0}^{m-1} \operatorname{Tr}\Bigl(\sqrt{P_a} \rho
\sqrt{P_b}\Bigr) \vert a\rangle \langle b \vert
\]
In particular, using the cyclic property of the trace, we see that the
probability to obtain a given outcome $a\in\{0,\ldots,m-1\}$ is as follows.
\[
\langle a \vert \sigma_{\mathsf{Y}} \vert a \rangle
= \operatorname{Tr}\Bigl(\sqrt{P_a} \rho \sqrt{P_a}\Bigr)
= \operatorname{Tr}(P_a \rho)
\]
This matches with the original measurement, establishing the correctness of the
simulation.

\subsection{Non-destructive measurements}

So far in the lesson, we've concerned ourselves with \emph{destructive}
measurements, where the output consists of the classical measurement result
alone and there is no specification of the post-measurement quantum state of
the system that was measured.

\emph{Non-destructive} measurements, on the other hand, do precisely this.
Specifically, non-destructive measurements describe not only the classical
measurement outcome probabilities, but also the state of the system that was
measured conditioned on each possible measurement outcome.
Note that the term \emph{non-destructive} refers to the \emph{system} being
measured but not necessarily its state, which could change significantly as a
result of the measurement.

In general, for a given destructive measurement, there will be multiple (in
fact infinitely many) non-destructive measurements that are \emph{compatible}
with the given destructive measurement, meaning that the classical measurement
outcome probabilities match precisely with the destructive measurement.
So, there isn't a unique way to define the post-measurement quantum state of a
system for a given measurement.

It is, in fact, possible to generalize non-destructive measurements even
further, so that they produce a classical measurement outcome along with a
quantum state output of a system that isn't necessarily the same as the input
system.

The notion of a non-destructive measurement is an interesting and useful
abstraction.
It should, however, be recognized that non-destructive measurements can always
be described as compositions of channels and destructive measurements --- so
there is a sense in which the notion of a destructive measurement is the more
fundamental one.

\subsubsection{From Naimark's theorem}

Consider the simulation of a general measurement like we have in Naimark's
theorem.
A simple way to obtain a non-destructive measurement from this simulation is
revealed by Figure~\ref{fig:Naimark}, where the system $\mathsf{X}$ is not
traced out, but is part of the output.
This yields both a classical measurement outcome $a\in\{0,\ldots,m-1\}$ as well
as a post-measurement quantum state of $\mathsf{X}.$

Let's describe these states in mathematical terms.
We're assuming that the initial state of $\mathsf{X}$ is $\rho,$ so that after
the initialized system $\mathsf{Y}$ is introduced and $U$ is performed, we have
that $(\mathsf{Y},\mathsf{X})$ is in the state
\[
\sigma =
U \bigl(\vert 0\rangle \langle 0 \vert \otimes \rho\bigr) U^{\dagger}
= \sum_{a,b=0}^{m-1} \vert a\rangle \langle b \vert \otimes \sqrt{P_a} \rho
\sqrt{P_b}.
\]
The probabilities for the different classical outcomes to appear are the same
as before --- they can't change as a result of us deciding to ignore or not
ignore $\mathsf{X}.$
That is, we obtain each $a\in\{0,\ldots,m-1\}$ with probability
$\operatorname{Tr}(P_a \rho).$

Conditioned upon having obtained a particular measurement outcome $a,$ the
resulting state of $\mathsf{X}$ is given by this expression.
\[
\frac{\sqrt{P_a} \rho \sqrt{P_a}}{\operatorname{Tr}(P_a \rho)}
\]
One way to see this is to represent a standard basis measurement of
$\mathsf{Y}$ by the completely dephasing channel $\Delta_m,$ where the channel
output describes classical measurement outcomes as (diagonal) density matrices.
An expression of the state we obtain is as follows.
\[
\sum_{a,b=0}^{m-1} \Delta_m(\vert a\rangle \langle b \vert) \otimes \sqrt{P_a}
\rho \sqrt{P_b}
= \sum_{a=0}^{m-1} \vert a\rangle \langle a \vert \otimes \sqrt{P_a} \rho
\sqrt{P_a}.
\]
We can then write this state as a convex combination,
\[
\sum_{a=0}^{m-1} \operatorname{Tr}(P_a \rho)\, \vert a\rangle \langle a \vert
\otimes \frac{\sqrt{P_a} \rho \sqrt{P_a}}{\operatorname{Tr}(P_a \rho)},
\]
which is consistent with the expression we've obtained for the state of
$\mathsf{X}$ conditioned on each possible measurement outcome.

\subsubsection{From a Kraus representation}

There are alternative selections for $U$ in the context of Naimark's theorem
that produce the same measurement outcome probabilities but give entirely
different output states of $\mathsf{X}.$

For instance, one option is to substitute $(\mathbb{I}_{\mathsf{Y}} \otimes V)
U$ for $U,$ where $V$ is any unitary operation on $\mathsf{X}.$
The application of $V$ to $\mathsf{X}$ commutes with the measurement of
$\mathsf{Y}$ so the classical outcome probabilities do not change, but now the
state of $\mathsf{X}$ conditioned on the outcome $a$ becomes
\[
\frac{V \sqrt{P_a} \rho \sqrt{P_a}V^{\dagger}}{\operatorname{Tr}(P_a \rho)}.
\]

More generally, we could replace $U$ by the unitary matrix
\[
\Biggl(\sum_{a=0}^{m-1} \vert a\rangle\langle a \vert \otimes V_a\Biggr) U
\]
for any choice of unitary operations $V_0,\ldots,V_{m-1}$ on $\mathsf{X}.$
Again, the classical outcome probabilities are unchanged, but now the state of
$\mathsf{X}$ conditioned on the outcome $a$ becomes
\[
\frac{V_a \sqrt{P_a} \rho \sqrt{P_a}V_a^{\dagger}}{\operatorname{Tr}(P_a
  \rho)}.
\]

An equivalent way to express this freedom is connected with Kraus
representations.
That is, we can describe an $m$-outcome non-destructive measurement of a system
having $n$ classical states by a selection of $n\times n$ Kraus matrices
$A_0,\ldots,A_{m-1}$ satisfying the typical condition for Kraus matrices.
\begin{equation}
  \sum_{a = 0}^{m-1} A_a^{\dagger} A_a = \mathbb{I}_{\mathsf{X}}
  \label{eq:Kraus-equation}
\end{equation}
Assuming that the initial state of $\mathsf{X}$ is $\rho,$ the classical
measurement outcome is $a$ with probability
\[
\operatorname{Tr}\bigl(A_a \rho A_a^{\dagger}\bigr)
= \operatorname{Tr}\bigl(A_a^{\dagger} A_a \rho \bigr)
\]
and conditioned upon the outcome being $a$ the state of $\mathsf{X}$ becomes
\[
\frac{A_a \rho A_a^{\dagger}}{\operatorname{Tr}(A_a^{\dagger}A_a \rho)}.
\]
This is equivalent to choosing the unitary operation $U$ in Naimark's theorem
as follows.
\[
U =
\begin{pmatrix}
A_{0} & \fbox{?} & \cdots & \fbox{?} \\[1mm]
A_{1} & \fbox{?} & \cdots & \fbox{?} \\[1mm]
\vdots & \vdots & \ddots & \vdots\\[1mm]
A_{m-1} & \fbox{?} & \cdots & \fbox{?}
\end{pmatrix}
\]
We've already observed, in the previous lesson, that the columns formed by the
blocks $A_0,\ldots,A_{m-1}$ are necessarily orthogonal, by virtue of the
condition \eqref{eq:Kraus-equation}.

\subsubsection{Generalizations}

There are even more general ways to formulate non-destructive measurements than
the ways we've discussed.
The notion of a \emph{quantum instrument} (which won't be described here)
represents one way to do this.

\section{Quantum state discrimination and tomography}

In the last part of the lesson, we'll briefly consider two tasks associated
with measurements: \emph{quantum state discrimination} and \emph{quantum state
tomography}.

\begin{trivlist}
\item \textbf{Quantum state discrimination.}
   For quantum state discrimination, we have a known collection of quantum
   states $\rho_0,\ldots,\rho_{m-1}$ and probabilities
   $p_0,\ldots,p_{m-1}$ associated with these states.
   A succinct way of expressing this is to say that we have an \emph{ensemble}
   \[
   \{(p_0,\rho_0),\ldots,(p_{m-1},\rho_{m-1})\}
   \]
   of quantum states.

   A number $a\in\{0,\ldots,m-1\}$ is chosen randomly according to the
   probabilities $(p_0,\ldots,p_{m-1})$ and the system $\mathsf{X}$
   is prepared in the state $\rho_a.$
   The goal is to determine, by means of a measurement of $\mathsf{X}$ alone,
   which value of $a$ was chosen.

   Thus, we have a finite number of alternatives, along with a \emph{prior} ---
   which is our knowledge of the probability for each $a$ to
   be selected --- and the goal is to determine which alternative actually
   happened.
   This may be easy for some choices of states and probabilities, and for
   others it may not be possible without some chance of making an error.

 \item \textbf{Quantum state tomography.}
   For quantum state tomography, we have an \emph{unknown} quantum state of a
   system --- so unlike in quantum state discrimination there's typically no
   prior or any information about possible alternatives.

   This time, however, it's not a single copy of the state that's made
   available, but rather many \emph{independent} copies are made available.
   That is, $N$ identical systems $\mathsf{X}_1,\ldots,\mathsf{X}_N$ are each
   independently prepared in the state $\rho$ for some (possibly large) number
   $N.$
   The goal is to find an approximation of the unknown state, as a density
   matrix, by measuring the systems.
\end{trivlist}

\subsection{Discriminating between two states}

The simplest case for quantum state discrimination is that there are two
states, $\rho_0$ and $\rho_1,$ that are to be discriminated.

Imagine a situation in which a bit $a$ is chosen randomly: $a = 0$ with
probability $p$ and $a = 1$ with probability $1 - p.$
A system $\mathsf{X}$ is prepared in the state $\rho_a,$ meaning $\rho_0$ or
$\rho_1$ depending on the value of $a,$ and given to us.
Our goal is to correctly guess the value of $a$ by means of a measurement on
$\mathsf{X}.$
To be precise, we shall aim to maximize the probability that our guess is
correct.

\subsubsection{An optimal measurement}

An optimal way to solve this problem begins with a spectral decomposition of a
weighted difference between $\rho_0$ and $\rho_1,$ where the weights are the
corresponding probabilities.
\[
p \rho_0 - (1-p) \rho_1 = \sum_{k = 0}^{n-1} \lambda_k \vert \psi_k \rangle
\langle \psi_k \vert
\]
Notice that we have a minus sign rather than a plus sign in this expression:
this is a weighted \emph{difference} not a weighted sum.

We can maximize the probability of a correct guess by selecting a projective
measurement $\{\Pi_0,\Pi_1\}$ as follows.
First, partition the elements of $\{0,\ldots,n-1\}$ into two disjoint sets
$S_0$ and $S_1$ depending upon whether the corresponding eigenvalue of the
weighted difference is nonnegative or negative.
\[
\begin{gathered}
  S_0 = \{k\in\{0,\ldots,n-1\} : \lambda_k \geq 0 \}\\[2mm]
  S_1 = \{k\in\{0,\ldots,n-1\} : \lambda_k < 0 \}
\end{gathered}
\]
(It doesn't actually matter in which set $S_0$ or $S_1$ we include the values
of $k$ for which $\lambda_k = 0.$
Here we're choosing arbitrarily to include these values in $S_0.$)
We can then choose a \emph{projective} measurement as follows.
\[
\Pi_0 = \sum_{k \in S_0} \vert \psi_k \rangle \langle \psi_k \vert
\quad\text{and}\quad
\Pi_1 = \sum_{k \in S_1} \vert \psi_k \rangle \langle \psi_k \vert
\]
This is an optimal measurement in the situation at hand that minimizes the
probability of an incorrect determination of the selected state.

\subsubsection{Correctness probability}

Now we will determine the probability of correctness for the measurement
$\{\Pi_0,\Pi_1\}.$

As we begin, we won't really need to be concerned with the specific choice
we've made for $\Pi_0$ and $\Pi_1,$ though it may be helpful to keep it in
mind.
For \emph{any} measurement $\{P_0,P_1\}$ (not necessarily projective) we can
write the correctness probability as follows.
\[
p \operatorname{Tr}(P_0 \rho_0) + (1 - p) \operatorname{Tr}(P_1 \rho_1)
\]
Using the fact that $\{P_0,P_1\}$ is a measurement, so
$P_1 = \mathbb{I} - P_0,$ we can rewrite this expression as follows.
\begin{align*}
p \operatorname{Tr}(P_0 \rho_0) + (1 - p) \operatorname{Tr}((\mathbb{I} - P_0)
\rho_1)\hspace{-4cm}\\
& = p \operatorname{Tr}(P_0 \rho_0) - (1 - p) \operatorname{Tr}(P_0 \rho_1) +
(1-p) \operatorname{Tr}(\rho_1)\\[1mm]
& = \operatorname{Tr}\bigl( P_0 (p \rho_0 - (1-p)\rho_1) \bigr) + 1 - p
\end{align*}
On the other hand, we could have made the substitution
$P_0 = \mathbb{I} - P_1$ instead.
That wouldn't change the value but it does give us an alternative expression.
\begin{align*}
p \operatorname{Tr}((\mathbb{I} - P_1) \rho_0) + (1 - p) \operatorname{Tr}(P_1
\rho_1)\hspace{-4cm}\\[1mm]
& = p \operatorname{Tr}(\rho_0) - p \operatorname{Tr}(P_1 \rho_0) + (1 - p)
\operatorname{Tr}(P_1 \rho_1)\\[1mm]
& = p - \operatorname{Tr}\bigl( P_1 (p \rho_0 - (1-p)\rho_1) \bigr)
\end{align*}
The two expressions have the same value, so we can average them to give yet
another expression for this value.
Averaging the two expressions is just a trick to simplify the resulting
expression.
\begin{multline*}
  \quad
  \frac{1}{2} \bigl(\operatorname{Tr}\bigl( P_0 (p \rho_0 - (1-p)\rho_1) \bigr)
  + 1-p\bigr)
  + \frac{1}{2} \bigl(p - \operatorname{Tr}\bigl( P_1 (p \rho_0 - (1-p)\rho_1)
  \bigr)\bigr)\\
  = \frac{1}{2} \operatorname{Tr}\bigl( (P_0-P_1) (p \rho_0 - (1-p)\rho_1)\bigr)
  + \frac{1}{2}
  \quad
\end{multline*}

Now we can see why it makes sense to choose the projections $\Pi_0$ and $\Pi_1$
(as specified above) for $P_0$ and $P_1,$ respectively --- because that's how
we can make the trace in the final expression as large as possible.
In particular,
\[
(\Pi_0-\Pi_1) (p \rho_0 - (1-p)\rho_1) = \sum_{k = 0}^{n-1} \vert\lambda_k\vert
\cdot \vert \psi_k \rangle \langle \psi_k \vert.
\]
So, when we take the trace, we obtain the sum of the \emph{absolute values} of
the eigenvalues --- which is equal to what's known as the \emph{trace norm} of
the weighted difference.
\[
\operatorname{Tr}\bigl( (\Pi_0-\Pi_1) (p \rho_0 - (1-p)\rho_1)\bigr)
= \sum_{k = 0}^{n-1} \vert\lambda_k\vert = \bigl\| p \rho_0 - (1-p)\rho_1
\bigr\|_1
\]
Thus, the probability that the measurement $\{\Pi_0,\Pi_1\}$ leads to a correct
discrimination of $\rho_0$ and $\rho_1,$ given with probabilities $p$ and
$1-p,$ respectively, is as follows.
\[
\frac{1}{2} + \frac{1}{2} \bigl\| p \rho_0 - (1-p)\rho_1 \bigr\|_1
\]

The fact that this is the optimal probability for a correct discrimination of
$\rho_0$ and $\rho_1,$ given with probabilities $p$ and $1-p,$ is commonly
referred to as the \emph{Helstrom--Holevo theorem} (or sometimes just
\emph{Helstrom's theorem}).

\subsection{Discriminating three or more states}

For quantum state discrimination when there are three or more states, there is
no known closed-form solution for an optimal measurement, although it is
possible to formulate the problem as a \emph{semidefinite program} --- which
allows for efficient numerical approximations of optimal measurements with the
help of a computer.

It is also possible to \emph{verify} (or \emph{falsify}) optimality of a given
measurement in a state discrimination task through a condition known as the
\emph{Holevo--Yuen--Kennedy--Lax} condition.
In particular, for the state discrimination task defined by the ensemble
\[
\{(p_0,\rho_0),\ldots,(p_{m-1},\rho_{m-1})\},
\]
the measurement $\{P_0,\ldots,P_{m-1}\}$ is optimal if and only if the matrix
\[
Q_a = \sum_{b = 0}^{m-1} p_b \rho_b P_b - p_a \rho_a
\]
is positive semidefinite for every $a\in\{0,\ldots,m-1\}.$

For example, consider the quantum state discrimination task in which one of the
four tetrahedral states $\vert\phi_0\rangle,\ldots,\vert\phi_3\rangle$ is
selected uniformly at random.
The tetrahedral measurement $\{P_0,P_1,P_2,P_3\}$ succeeds with probability
\[
\frac{1}{4} \operatorname{Tr}(P_0 \vert\phi_0\rangle\langle \phi_0 \vert) +
\frac{1}{4} \operatorname{Tr}(P_1 \vert\phi_1\rangle\langle \phi_1 \vert) +
\frac{1}{4} \operatorname{Tr}(P_2 \vert\phi_2\rangle\langle \phi_2 \vert) +
\frac{1}{4} \operatorname{Tr}(P_3 \vert\phi_3\rangle\langle \phi_3 \vert)
= \frac{1}{2}.
\]
This is optimal by the Holevo--Yuen--Kennedy--Lax condition, as a calculation
reveals that
\[
Q_a = \frac{1}{4}(\mathbb{I} - \vert\phi_a\rangle\langle\phi_a\vert) \geq 0
\]
for $a = 0,1,2,3.$

\subsection{Quantum state tomography}

Finally, we'll briefly discuss the problem of \emph{quantum state tomography}.
For this problem, we're given a large number $N$ of independent copies of an
unknown quantum state $\rho,$ and the goal is to reconstruct an approximation
$\tilde{\rho}$ of $\rho.$
To be clear, this means that we wish to find a classical description of a
density matrix $\tilde{\rho}$ that is as close as possible to $\rho.$

We can alternatively describe the set-up in the following way.
An unknown density matrix $\rho$ is selected, and we're given access to $N$
quantum systems $\mathsf{X}_1,\ldots,\mathsf{X}_N,$ each of which has been
\emph{independently} prepared in the state $\rho.$
Thus, the state of the compound system $(\mathsf{X}_1,\ldots,\mathsf{X}_N)$ is
\[
\rho^{\otimes N} = \rho \otimes \rho \otimes \cdots \otimes \rho \quad
\text{($N$ times)}
\]
The goal is to perform measurements on the systems
$\mathsf{X}_1,\ldots,\mathsf{X}_N$ and, based on the outcomes of those
measurements, to compute a density matrix $\tilde{\rho}$ that closely
approximates $\rho.$
This turns out to be a fascinating problem and there is ongoing research on it.

Different types of strategies for approaching the problem may be considered.
For example, we can imagine a strategy where each of the systems
$\mathsf{X}_1,\ldots,\mathsf{X}_N$ is measured separately, in turn, producing a
sequence of measurement outcomes.
Different specific choices for which measurements are performed can be made,
including \emph{adaptive} and \emph{non-adaptive} selections.
In other words, the choice of what measurement is performed on a particular
system might or might not depend on the outcomes of prior measurements.
Based on the sequence of measurement outcomes, a guess $\tilde{\rho}$ for the
state $\rho$ is derived --- and again there are different methodologies for
doing this.

An alternative approach is to perform a single \emph{joint measurement} of the
entire collection, where we think about $(\mathsf{X}_1,\ldots,\mathsf{X}_N)$ as
a single system and select a single measurement whose output is a guess
$\tilde{\rho}$ for the state $\rho.$
This can lead to an improved estimate over what is possible for separate
measurements of the individual systems, although a joint measurement on all of
the systems together is likely to be much more difficult to implement.

\subsubsection{Qubit tomography using Pauli measurements}

We'll now consider quantum state tomography in the simple case where $\rho$ is
a qubit density matrix.
We assume that we're given qubits $\mathsf{X}_1,\ldots,\mathsf{X}_N$ that are
each independently in the state $\rho,$ and our goal is to compute an
approximation $\tilde{\rho}$ that is close to $\rho.$

Our strategy will be to divide the $N$ qubits
$\mathsf{X}_1,\ldots,\mathsf{X}_N$ into three roughly equal-size collections,
one for each of the three Pauli matrices $\sigma_x,$ $\sigma_y,$ and
$\sigma_z.$
Each qubit is then measured independently as follows.

\begin{enumerate}
\item
  For each of the qubits in the collection associated with $\sigma_x$ we
  perform a $\sigma_x$ measurement. This means that the qubit is measured with
  respect to the basis $\{\vert + \rangle, \vert -\rangle\},$ which is an
  orthonormal basis of eigenvectors of $\sigma_x,$ and the corresponding
  measurement outcomes are the eigenvalues associated with the two
  eigenvectors: $+1$ for the state $\vert + \rangle$ and $-1$ for the state
  $\vert -\rangle.$
  By averaging together the outcomes over all of the states
  in the collection associated with~$\sigma_x,$ we obtain an approximation of
  the expectation value
  \[
  \langle + \vert \rho \vert + \rangle - \langle - \vert \rho \vert - \rangle =
  \operatorname{Tr}(\sigma_x \rho).
  \]

\item
  For each of the qubits in the collection associated with $\sigma_y$ we
  perform a $\sigma_y$ measurement. Such a measurement is similar to a
  $\sigma_x$ measurement, except that the measurement basis is $\{\vert\! +\!i
  \rangle, \vert\! -\!i \rangle\},$ the eigenvectors of $\sigma_y.$ Averaging
  the outcomes over all of the states in the collection associated with
  $\sigma_y,$ we obtain an approximation of the expectation value
  \[
  \bra{+i} \rho \ket{+i} - \bra{-i} \rho \ket{-i}
  = \operatorname{Tr}(\sigma_y \rho).
  \]

\item
  For each of the qubits in the collection associated with $\sigma_z$ we
  perform a $\sigma_z$ measurement. This time the measurement basis is the
  standard basis $\{\vert 0\rangle, \vert 1 \rangle\},$ the eigenvectors of
  $\sigma_z.$ Averaging the outcomes over all of the states in the collection
  associated with $\sigma_z,$ we obtain an approximation of the expectation
  value
  \[
  \langle 0 \vert \rho \vert 0 \rangle - \langle 1 \vert \rho \vert 1 \rangle =
  \operatorname{Tr}(\sigma_z \rho).
  \]
\end{enumerate}

Once we have obtained approximations
\[
\alpha_x \approx \operatorname{Tr}(\sigma_x \rho),\;
\alpha_y \approx \operatorname{Tr}(\sigma_y \rho),\;
\alpha_z \approx \operatorname{Tr}(\sigma_z \rho)
\]
by averaging the measurement outcomes for each collection, we can approximate
$\rho$ as
\[
\tilde{\rho} = \frac{\mathbb{I} + \alpha_x \sigma_x + \alpha_y \sigma_y +
  \alpha_z \sigma_z}{2} \approx
\frac{\mathbb{I} + \operatorname{Tr}(\sigma_x \rho) \sigma_x +
  \operatorname{Tr}(\sigma_y \rho) \sigma_y + \operatorname{Tr}(\sigma_z \rho)
  \sigma_z}{2}
= \rho.
\]
In the limit as $N$ approaches infinity, this approximation converges in
probability to the true density matrix $\rho$ by the \emph{law of large
numbers,} and well-known statistical bounds (such as \emph{Hoeffding's
inequality}) can be used to bound the probability that the approximation
$\tilde{\rho}$ deviates from $\rho$ by varying amounts.

An important thing to recognize, however, is that the matrix $\tilde{\rho}$
obtained in this way may fail to be a density matrix.
In particular, although it will always have trace equal to $1,$ it may fail to
be positive semidefinite.
There are different known strategies for \emph{rounding} such an approximation
$\tilde{\rho}$ to a density matrix, one of them being to compute a spectral
decomposition, replace any negative eigenvalues with $0,$ and then renormalize
(by dividing the matrix we obtain by its trace).

\subsubsection{Qubit tomography using the tetrahedral measurement}

Another option for performing qubit tomography is to individually measure every
qubit $\mathsf{X}_1,\ldots,\mathsf{X}_N$ using the tetrahedral measurement
$\{P_0,P_1,P_2,P_3\}$ described earlier.
That is,
\[
P_0 = \frac{\vert \phi_0 \rangle \langle \phi_0 \vert}{2}, \quad
P_1 = \frac{\vert \phi_1 \rangle \langle \phi_1 \vert}{2}, \quad
P_2 = \frac{\vert \phi_2 \rangle \langle \phi_2 \vert}{2}, \quad
P_3 = \frac{\vert \phi_3 \rangle \langle \phi_3 \vert}{2}
\]
for
\[
\begin{aligned}
  \vert \phi_0 \rangle & = \vert 0 \rangle\\
  \vert \phi_1 \rangle & = \frac{1}{\sqrt{3}} \vert 0 \rangle +
  \sqrt{\frac{2}{3}} \vert 1 \rangle\\
  \vert \phi_2 \rangle & = \frac{1}{\sqrt{3}} \vert 0 \rangle +
  \sqrt{\frac{2}{3}} e^{2\pi i/3} \vert 1 \rangle\\
  \vert \phi_3 \rangle & = \frac{1}{\sqrt{3}} \vert 0 \rangle +
  \sqrt{\frac{2}{3}} e^{-2\pi i/3} \vert 1 \rangle.
\end{aligned}
\]

Each outcome is obtained some number of times, which we will denote as $n_a$
for each $a\in\{0,1,2,3\},$ so that $n_0 + n_1 + n_2 + n_3 = N.$
The ratio of these numbers with $N$ provides an estimate of the probability
associated with each possible outcome:
\[
\frac{n_a}{N} \approx \operatorname{Tr}(P_a \rho).
\]

Finally, we shall make use of the following remarkable formula:
\[
\rho = \sum_{a=0}^3 \Bigl( 3 \operatorname{Tr}(P_a \rho) - \frac{1}{2}\Bigr)
\vert \phi_a \rangle \langle \phi_a \vert.
\]
To establish this formula, we can use the following equation for the absolute
values squared of inner products of tetrahedral states, which can be checked
through direct calculations.
\[
\bigl\vert \langle \phi_a \vert \phi_b \rangle \bigr\vert^2 =
\begin{cases}
  1 & a=b\\
  \frac{1}{3} & a\neq b.
\end{cases}
\]
The four matrices
\begingroup\allowdisplaybreaks
\begin{align*}
  \vert\phi_0\rangle \langle \phi_0 \vert
  & = \begin{pmatrix}
    1 & 0\\[2mm]
    0 & 0
  \end{pmatrix}\\[3mm]
  \vert\phi_1\rangle \langle \phi_1 \vert
  & = \begin{pmatrix}
    \frac{1}{3} & \frac{\sqrt{2}}{3}\\[2mm]
    \frac{\sqrt{2}}{3} & \frac{2}{3}
  \end{pmatrix}\\[3mm]
  \vert\phi_2\rangle \langle \phi_2 \vert
  & = \begin{pmatrix}
    \frac{1}{3} & \frac{\sqrt{2}}{3}e^{-2\pi i/3}\\[2mm]
    \frac{\sqrt{2}}{3}e^{2\pi i/3} & \frac{2}{3}
  \end{pmatrix}\\[3mm]
  \vert\phi_3\rangle \langle \phi_3 \vert
  & = \begin{pmatrix}
    \frac{1}{3} & \frac{\sqrt{2}}{3}e^{2\pi i/3}\\[2mm]
    \frac{\sqrt{2}}{3}e^{-2\pi i/3} & \frac{2}{3}
  \end{pmatrix}
\end{align*}
\endgroup
are linearly independent, so it suffices to prove that the formula is true when
$\rho = \vert\phi_b\rangle\langle\phi_b\vert$ for $b = 0,1,2,3.$
In particular,
\[
3 \operatorname{Tr}(P_a \vert\phi_b\rangle\langle\phi_b\vert) - \frac{1}{2}
= \frac{3}{2} \vert \langle \phi_a \vert \phi_b \rangle \vert^2 - \frac{1}{2}
= \begin{cases}
  1 & a=b\\
  0 & a\neq b
\end{cases}
\]
and therefore
\[
\sum_{a=0}^3 \biggl( 3 \operatorname{Tr}(P_a
\vert\phi_b\rangle\langle\phi_b\vert) -
\frac{\operatorname{Tr}(\vert\phi_b\rangle\langle\phi_b\vert)}{2}\biggr) \vert
\phi_a \rangle \langle \phi_a \vert = \vert \phi_b\rangle\langle \phi_b \vert.
\]

We arrive at an approximation of $\rho$.
\[
\tilde{\rho} = \sum_{a=0}^3 \Bigl( \frac{3 n_a}{N} - \frac{1}{2}\Bigr) \vert
\phi_a \rangle \langle \phi_a \vert
\]
This approximation will always be a Hermitian matrix having trace equal to one,
but it may fail to be positive semidefinite.
In this case, the approximation must be rounded to a density matrix, similar to
the strategy involving Pauli measurements.
   

\lesson{Purifications and Fidelity}
\label{lesson:purifications-and-fidelity}

This lesson is centered around a fundamentally important concept in the theory
of quantum information, which is that of a \emph{purification} of a state.
A purification of a quantum state, represented by a density matrix $\rho,$ is a
pure state of a larger compound system that leaves us with $\rho$ when the rest
of the compound system is traced out.
As we'll see, \emph{every} state $\rho$ has a purification, provided that the
portion of the compound system that gets traced out is large enough.

It's both common and useful to consider purifications of states when reasoning
about them.
Intuitively speaking, quantum state vectors are simpler mathematical objects
than density matrices, and we can often conclude interesting things about
density matrices by thinking about them as representing parts of larger systems
whose states are pure --- and therefore simpler (at least in some regards).
This is an example of a \emph{dilation} in mathematics, where something
relatively complicated is obtained by restricting or reducing something larger
yet simpler.

The lesson also discusses the \emph{fidelity} between two quantum states, which
is a value that quantifies the similarity between the states.
We'll see how fidelity is defined by a mathematical formula and discuss how it
connects to the notion of a purification through \emph{Uhlmann's theorem}.

\section{Purifications}

\subsection{Definition of purifications}

Let us begin with a precise mathematical definition for purifications.

\begin{callout}[title={Purifications}]
  Suppose $\mathsf{X}$ is a system in a state represented by a density matrix
  $\rho,$ and $\vert\psi\rangle$ is a quantum state vector of a pair
  $(\mathsf{X},\mathsf{Y})$ that leaves $\rho$ when $\mathsf{Y}$ is traced out:
  \[
  \rho = \operatorname{Tr}_{\mathsf{Y}} \bigl( \vert
  \psi\rangle\langle\psi\vert\bigr).
  \]
  The state vector $\vert\psi\rangle$ is then said to be a \emph{purification}
  of $\rho.$
\end{callout}

\noindent
The pure state $\vert\psi\rangle\langle\psi\vert,$ expressed as a density
matrix rather than a quantum state vector, is also commonly referred to as a
purification of $\rho$ when the equation in the definition is true, but we'll
generally use the term to refer to a quantum state vector.

The term \emph{purification} is also used more generally when the ordering of
the systems is reversed, when the names of the systems and states are different
(of course), and when there are more than two systems.
For instance, if $\vert \psi \rangle$ is a quantum state vector representing a
pure state of a compound system $(\mathsf{A},\mathsf{B},\mathsf{C}),$ and the
equation
\[
\rho = \operatorname{Tr}_{\mathsf{B}}
\bigl(\vert\psi\rangle\langle\psi\vert\bigr)
\]
is true for a density matrix $\rho$ representing a state of the system
$(\mathsf{A},\mathsf{C}),$ then $\vert\psi\rangle$ is still referred to as a
purification of $\rho.$

For the purposes of this lesson, however, we'll focus on the specific form
described in the definition.
Properties and facts concerning purifications, according to this definition,
can typically be generalized to more than two systems by re-ordering and
partitioning the systems into two compound systems, one playing the role of
$\mathsf{X}$ and the other playing the role of $\mathsf{Y}.$

\subsection{Existence of purifications}

Suppose that $\mathsf{X}$ and $\mathsf{Y}$ are any two systems and $\rho$ is a
given state of $\mathsf{X}.$
We will prove that there exists a quantum state vector $\vert\psi\rangle$ of
$(\mathsf{X},\mathsf{Y})$ that \emph{purifies} $\rho$ --- which is another way
of saying that $\vert\psi\rangle$ is a purification of $\rho$ --- provided that
the system $\mathsf{Y}$ is large enough.
In particular, if $\mathsf{Y}$ has at least as many classical states as
$\mathsf{X},$ then a purification of this form necessarily exists for every
state $\rho.$
Fewer classical states of $\mathsf{Y}$ are required for some states $\rho;$
in general, $\operatorname{rank}(\rho)$ classical states of $\mathsf{Y}$ are
necessary and sufficient for the existence of a quantum state vector of
$(\mathsf{X},\mathsf{Y})$ that purifies $\rho.$

Consider first any expression of $\rho$ as a convex combination of $n$ pure
states, for any positive integer $n.$
\[
\rho = \sum_{a = 0}^{n-1} p_a \vert\phi_a\rangle\langle\phi_a\vert
\]
In this expression, $(p_0,\ldots,p_{n-1})$ is a probability vector and
$\vert\phi_0\rangle,\ldots,\vert\phi_{n-1}\rangle$ are quantum state vectors of
$\mathsf{X}.$

One way to obtain such an expression is through the spectral theorem, in which
case $n$ is the number of classical states of $\mathsf{X},$
$p_0,\ldots,p_{n-1}$ are the eigenvalues of $\rho,$ and
$\vert\phi_0\rangle,\ldots,\vert\phi_{n-1}\rangle$ are orthonormal eigenvectors
corresponding to these eigenvalues.
There's actually no need to include the terms corresponding to the zero
eigenvalues of $\rho$ in the sum, which allows us to alternatively choose $n =
\operatorname{rank}(\rho)$ and $p_0,\ldots,p_{n-1}$ to be the nonzero
eigenvalues of $\rho.$
This is the minimum value of $n$ for which an expression of $\rho$ taking the
form above exists.

To be clear, it is \emph{not necessary} that the chosen expression of $\rho,$
as a convex combination of pure states, comes from the spectral theorem ---
this is just one way to obtain such an expression.
In particular, $n$ could be any positive integer, the unit vectors
$\vert\phi_0\rangle,\ldots,\vert\phi_{n-1}\rangle$ need not be orthogonal, and
the probabilities $p_0,\ldots,p_{n-1}$ need not be eigenvalues of $\rho.$

We can now identify a purification of $\rho$ as follows.
\[
\vert\psi\rangle = \sum_{a = 0}^{n-1} \sqrt{p_a} \, \vert\phi_a\rangle \otimes
\vert a \rangle
\]
Here we're making the assumption that the classical states of $\mathsf{Y}$
include $0,\ldots,n-1.$
If they do not, an arbitrary choice for $n$ distinct classical states of
$\mathsf{Y}$ can be substituted for $0,\ldots,n-1.$
Verifying that this is indeed a purification of $\rho$ is a simple matter of
computing the partial trace, which can be done in the following two equivalent
ways.
\[
\begin{aligned}
  \operatorname{Tr}_{\mathsf{Y}} \bigl(\vert\psi\rangle\langle\psi\vert\bigr)
  & = \sum_{a = 0}^{n-1} (\mathbb{I}_{\mathsf{X}} \otimes \langle a\vert)
  \vert\psi\rangle\langle\psi\vert
  (\mathbb{I}_{\mathsf{X}} \otimes \vert a\rangle) = \sum_{a = 0}^{n-1} p_a
  \vert\phi_a\rangle\langle\phi_a\vert = \rho\\[2mm]
  \operatorname{Tr}_{\mathsf{Y}} \bigl(\vert\psi\rangle\langle\psi\vert\bigr)
  & = \sum_{a,b = 0}^{n-1} \sqrt{p_a} \sqrt{p_b} \, \vert\phi_a\rangle\langle
  \phi_b\vert \, \operatorname{Tr}(\vert a \rangle \langle b \vert)
  = \sum_{a = 0}^{n-1} p_a \, \vert\phi_a\rangle\langle \phi_a\vert = \rho
\end{aligned}
\]

More generally, for any orthonormal set of vectors
$\{\vert\gamma_0\rangle,\ldots,\vert\gamma_{n-1}\rangle\},$ the quantum state
vector
\[
\vert\psi\rangle = \sum_{a = 0}^{n-1} \sqrt{p_a} \, \vert\phi_a\rangle \otimes
\vert \gamma_a \rangle
\]
is a purification of $\rho.$

\subsubsection{Example: two purifications of a density matrix}

Suppose that $\mathsf{X}$ and $\mathsf{Y}$ are both qubits and
\[
\rho = \begin{pmatrix}
  \frac{3}{4} & \frac{1}{4}\\[2mm]
  \frac{1}{4} & \frac{1}{4}
\end{pmatrix}
\]
is a density matrix representing a state of $\mathsf{X}.$

We can use the spectral theorem to express $\rho$ as
\[
\rho =
\cos^2(\pi/8) \vert \psi_{\pi/8}\rangle\langle\psi_{\pi/8}\vert +
\sin^2(\pi/8) \vert \psi_{5\pi/8}\rangle\langle\psi_{5\pi/8}\vert,
\]
where $\vert \psi_{\theta} \rangle = \cos(\theta) \vert 0\rangle +
\sin(\theta)\vert 1\rangle.$
The quantum state vector
\[
\cos(\pi/8) \vert \psi_{\pi/8}\rangle \otimes \vert 0\rangle +
\sin(\pi/8) \vert \psi_{5\pi/8}\rangle \otimes \vert 1\rangle,
\]
which describes a pure state of the pair $(\mathsf{X},\mathsf{Y}),$ is
therefore a purification of $\rho.$

Alternatively, we can write
\[
\rho = \frac{1}{2} \vert 0\rangle\langle 0\vert + \frac{1}{2} \vert
+\rangle\langle +\vert.
\]
This is a convex combination of pure states but not a spectral decomposition
because $\vert 0\rangle$ and $\vert +\rangle$ are not orthogonal and $1/2$ is
not an eigenvalue of $\rho.$
Nevertheless, the quantum state vector
\[
\frac{1}{\sqrt{2}} \vert 0 \rangle \otimes \vert 0\rangle +
\frac{1}{\sqrt{2}} \vert + \rangle \otimes \vert 1\rangle
\]
is a purification of $\rho.$

\subsection{Schmidt decompositions}

Next, we will discuss \emph{Schmidt decompositions}, which are expressions of
quantum state vectors of \emph{pairs} of systems that take a certain form.
Schmidt decompositions are closely connected with purifications, and they're
very useful in their own right.
Indeed, when reasoning about a given quantum state vector $\vert\psi\rangle$ of
a pair of systems, the first step is often to identify or consider a Schmidt
decomposition of this state.

\begin{callout}[title={Schmidt decompositions}]
  Let $\vert \psi\rangle$ be a given quantum state vector of a pair of systems
  $(\mathsf{X},\mathsf{Y}).$ A \emph{Schmidt decomposition} of
  $\vert\psi\rangle$ is an expression of the form
  \[
  \vert \psi\rangle = \sum_{a = 0}^{r-1} \sqrt{p_a}\, \vert x_a\rangle \otimes
  \vert y_a \rangle,
  \]
  where $p_0,\ldots,p_{r-1}$ are positive real numbers summing to $1$ and
  \emph{both} of the sets $\{\vert x_0\rangle,\ldots,\vert x_{r-1}\rangle\}$
  and $\{\vert y_0\rangle,\ldots,\vert y_{r-1}\rangle\}$ are orthonormal.
\end{callout}

The values
\[
\sqrt{p_0},\ldots,\sqrt{p_{r-1}}
\]
in a Schmidt decomposition of $\vert\psi\rangle$ are known as its \emph{Schmidt
coefficients}, which are uniquely determined (up to their ordering) --- they're
the only positive real numbers that can appear in such an expression of
$\vert\psi\rangle.$
The sets
\[
\{\vert x_0\rangle,\ldots,\vert x_{r-1}\rangle\} \quad\text{and}\quad
\{\vert y_0\rangle,\ldots,\vert y_{r-1}\rangle\},
\]
on the other hand, are not uniquely determined, and the freedom one has in
choosing these sets of vectors will be clarified in the explanation that
follows.

We'll now verify that a given quantum state vector $\vert\psi\rangle$ does
indeed have a Schmidt decomposition, and in the process, we'll learn how to
find one.
Consider first an arbitrary (not necessarily orthogonal) basis
$\{\vert x_0\rangle, \ldots, \vert x_{n-1}\rangle\}$ of the vector space
corresponding to the system $\mathsf{X}.$
Because this is a basis, there will always exist a uniquely determined
selection of vectors $\vert z_0\rangle,\ldots,\vert z_{n-1}\rangle$ for which
the following equation is true.
\begin{equation}
  \vert \psi\rangle = \sum_{a = 0}^{n-1} \vert x_a\rangle \otimes \vert z_a
  \rangle
  \label{eq:Schmidt-form-1}
\end{equation}

For example, suppose $\{\vert x_0\rangle,\ldots,\vert x_{n-1}\rangle\}$ is the
standard basis associated with $\mathsf{X}.$
Assuming the classical state set of $\mathsf{X}$ is $\{0,\ldots,n-1\},$ this
means that $\vert x_a\rangle = \vert a\rangle$ for each $a\in\{0,\ldots,n-1\},$
and we find that
\[
\vert\psi\rangle = \sum_{a = 0}^{n-1} \vert a\rangle \otimes \vert z_a\rangle
\]
when
\[
\vert z_a \rangle = ( \langle a \vert \otimes \mathbb{I}_{\mathsf{Y}}) \vert
\psi\rangle
\]
for each $a\in\{0,\ldots,n-1\}.$
We frequently consider expressions like this when contemplating a standard
basis measurement of $\mathsf{X}.$

It's important to note that the formula
\[
\vert z_a \rangle = ( \langle a \vert \otimes \mathbb{I}_{\mathsf{Y}}) \vert
\psi\rangle
\]
for the vectors $\vert z_0\rangle,\ldots,\vert z_{n-1}\rangle$ in this example
only works because $\{\vert 0\rangle,\ldots,\vert n-1\rangle\}$ is an
\emph{orthonormal} basis.
In general, if $\{\vert x_0\rangle,\ldots,\vert x_{n-1}\rangle\}$ is a basis
that is not necessarily orthonormal, then the vectors $\vert
z_0\rangle,\ldots,\vert z_{n-1}\rangle$ are still uniquely determined by the
equation \eqref{eq:Schmidt-form-1}, but a different formula is needed.
One way to find them is first to identify vectors $\vert
w_0\rangle,\ldots,\vert w_{n-1}\rangle$ so that the equation
\[
\langle w_a \vert x_b \rangle =
\begin{cases}
  1 & a=b\\
  0 & a\neq b
\end{cases}
\]
is satisfied for all $a,b\in\{0,\ldots,n-1\},$ at which point we have
\[
\vert z_a \rangle = (\langle w_a \vert \otimes \mathbb{I}_{\mathsf{Y}}) \vert
\psi\rangle.
\]

For a given basis $\{\vert x_0\rangle,\ldots,\vert x_{n-1}\rangle\}$ of the
vector space corresponding to $\mathsf{X},$ the uniquely determined vectors
$\vert z_0\rangle,\ldots,\vert z_{n-1}\rangle$ for which the equation
\eqref{eq:Schmidt-form-1} is satisfied won't necessarily satisfy any special
properties, even if $\{\vert x_0\rangle,\ldots,\vert x_{n-1}\rangle\}$ happens
to be an orthonormal basis.
If, however, we choose $\{\vert x_0\rangle, \ldots, \vert x_{n-1}\rangle\}$ to
be an orthonormal basis of \emph{eigenvectors} of the reduced state
\[
\rho = \operatorname{Tr}_{\mathsf{Y}} \bigl( \vert \psi\rangle \langle \psi
\vert \bigr),
\]
then something interesting happens.
Specifically, for the uniquely determined collection
$\{\vert z_0\rangle,\ldots,\vert z_{n-1}\rangle\}$ for which the equation
\eqref{eq:Schmidt-form-1} is true, we find that this collection must be
\emph{orthogonal.}

In greater detail, consider a spectral decomposition of $\rho.$
\[
\rho = \sum_{a = 0}^{n-1} p_a \vert x_a \rangle \langle x_a \vert
\]
Here we're denoting the eigenvalues of $\rho$ by $p_0,\ldots,p_{n-1}$ in
recognition of the fact that $\rho$ is a density matrix --- so the vector of
eigenvalues $(p_0,\ldots,p_{n-1})$ forms a probability vector --- while
$\{\vert x_0\rangle,\ldots,\vert x_{n-1}\rangle\}$ is an orthonormal basis of
eigenvectors corresponding to these eigenvalues.
To see that the unique collection
$\{\vert z_0\rangle,\ldots,\vert z_{n-1}\rangle\}$ for which the equation
\eqref{eq:Schmidt-form-1} is true is necessarily orthogonal, we can begin by
computing the partial trace.
\[
\operatorname{Tr}_{\mathsf{Y}} (\vert\psi\rangle\langle\psi\vert)
= \sum_{a,b = 0}^{n-1} \vert x_a\rangle\langle x_b\vert
\operatorname{Tr}(\vert z_a\rangle\langle z_b\vert)
= \sum_{a,b = 0}^{n-1} \langle z_b\vert z_a\rangle \, \vert
x_a\rangle\langle x_b\vert.
\]
This expression must agree with the spectral decomposition of $\rho.$
We conclude from the fact that
$\{\vert x_0\rangle,\ldots,\vert x_{n-1}\rangle\}$ is a basis
that the set of matrices
\[
\bigl\{ \vert x_a\rangle\langle x_b\vert \,:\, a,b\in\{0,\ldots,n-1\} \bigr\}
\]
is linearly independent, and so it follows that
\[
\langle z_b \vert z_a\rangle =
\begin{cases}
  p_a & a=b\\[1mm]
  0 & a\neq b,
\end{cases}
\]
establishing that $\{\vert z_0\rangle,\ldots,\vert z_{n-1}\rangle\}$ is
orthogonal.

We've nearly obtained a Schmidt decomposition of $\vert\psi\rangle.$
It remains to discard those terms in \eqref{eq:Schmidt-form-1} for which
$p_a = 0$ and then write
\[
\vert z_a\rangle = \sqrt{p_a}\vert y_a\rangle
\]
for a unit vector $\vert y_a\rangle$ for each of the remaining terms.
A convenient way to do this begins with the observation that we're free to
number the eigenvalue/eigenvector pairs in a spectral decomposition of the
reduced state $\rho$ however we wish --- so we may assume that the eigenvalues
are sorted in decreasing order:
\[
p_0 \geq p_1 \geq \cdots \geq p_{n-1}.
\]
Letting $r = \operatorname{rank}(\rho),$ we find that $p_0,\ldots,p_{r-1} > 0$
and $p_r = \cdots = p_{n-1} = 0.$
So, we have
\[
\rho = \sum_{a = 0}^{r-1} p_a \vert x_a \rangle \langle x_a \vert,
\]
and we can write the quantum state vector $\vert \psi \rangle$ as
\[
\vert\psi\rangle = \sum_{a = 0}^{r-1} \vert x_a\rangle \otimes \vert
z_a\rangle.
\]

Finally, given that
\[
\| \vert z_a \rangle \|^2 = \langle z_a \vert z_a \rangle = p_a > 0
\]
for $a=0,\ldots,r-1,$ we can define unit vectors
$\vert y_0\rangle,\ldots,\vert y_{r-1}\rangle$ as 
\[
\vert y_a\rangle = \frac{\vert z_a\rangle}{\|\vert z_a\rangle\|}
= \frac{\vert z_a\rangle}{\sqrt{p_a}}, 
\]
so that $\vert z_a\rangle = \sqrt{p_a}\vert y_a\rangle$ for each
$a\in\{0,\ldots,r-1\}.$
The vectors $\{\vert z_0\rangle, \ldots, \vert z_{r-1}\rangle\}$ are
orthogonal and nonzero, so it follows that
$\{\vert y_0\rangle, \ldots, \vert y_{r-1}\rangle\}$ is an \emph{orthonormal}
set, and so we have obtained a Schmidt decomposition of $\vert\psi\rangle.$
\[
\vert \psi\rangle = \sum_{a = 0}^{r-1} \sqrt{p_a}\, \vert x_a\rangle \otimes
\vert y_a \rangle
\]

Concerning the choice of the vectors
$\{\vert x_0\rangle,\ldots,\vert x_{r-1}\rangle\}$ and
$\{\vert y_0\rangle,\ldots,\vert y_{r-1}\rangle\},$
we can select $\{\vert x_0\rangle,\ldots,\vert x_{r-1}\rangle\}$ to be any
orthonormal set of eigenvectors corresponding to the nonzero eigenvalues of the
reduced state
$\operatorname{Tr}_{\mathsf{Y}}(\vert\psi\rangle\langle\psi\vert)$ (as we have
done above), in which case the vectors $\{\vert y_0\rangle,\ldots,\vert
y_{r-1}\rangle\}$ are uniquely determined.
The situation is symmetric between the two systems, so we can alternatively
choose $\{\vert y_0\rangle,\ldots,\vert y_{r-1}\rangle\}$ to be any orthonormal
set of eigenvectors corresponding to the nonzero eigenvalues of the reduced
state $\operatorname{Tr}_{\mathsf{X}}(\vert\psi\rangle\langle\psi\vert),$ in
which case the vectors $\{\vert x_0\rangle,\ldots,\vert x_{r-1}\rangle\}$ will
be uniquely determined.

Notice, however, that once one of the sets is selected, as a set of
eigenvectors of the corresponding reduced state as just described, the other is
determined --- so they cannot be chosen independently.

Although it won't come up again in this course, it is noteworthy that the
nonzero eigenvalues $p_0,\ldots,p_{r-1}$ of the reduced state
$\operatorname{Tr}_{\mathsf{X}}(\vert\psi\rangle\langle\psi\vert)$ must always
agree with the nonzero eigenvalues of the reduced state
$\operatorname{Tr}_{\mathsf{Y}}(\vert\psi\rangle\langle\psi\vert)$ for any pure
state $\vert\psi\rangle$ of a pair of systems $(\mathsf{X},\mathsf{Y}).$
Intuitively speaking, the reduced states of $\mathsf{X}$ and $\mathsf{Y}$ have
exactly the same amount of randomness in them when the pair
$(\mathsf{X},\mathsf{Y})$ is in a pure state.
This fact is revealed by the Schmidt decomposition: in both cases the
eigenvalues of the reduced states must agree with the squares of the Schmidt
coefficients of the pure state.

\subsection{Unitary equivalence of purifications}

We can use Schmidt decompositions to establish a fundamentally important fact
concerning purifications known as the \emph{unitary equivalence of
purifications}.

\begin{callout}[title = {Unitary equivalence of purifications}]
  Suppose that $\mathsf{X}$ and $\mathsf{Y}$ are systems, and
  $\vert\psi\rangle$ and $\vert\phi\rangle$ are quantum state vectors of
  $(\mathsf{X},\mathsf{Y})$ that both purify the same state of $\mathsf{X}.$
  In symbols,
  \[
  \operatorname{Tr}_{\mathsf{Y}} (\vert\psi\rangle\langle\psi\vert)
  = \rho = \operatorname{Tr}_{\mathsf{Y}} (\vert\phi\rangle\langle\phi\vert)
  \]
  for some density matrix $\rho$ representing a state of $\mathsf{X}.$
  There must then exist a unitary operation $U$ on $\mathsf{Y}$ alone that
  transforms the first purification into the second:
  \[
  (\mathbb{I}_{\mathsf{X}} \otimes U) \vert\psi\rangle = \vert\phi\rangle.
  \]
\end{callout}

\noindent
We'll discuss a few implications of this theorem as the lesson continues, but
first let's see how it follows from our previous discussion of Schmidt
decompositions.

Our assumption is that $\vert\psi\rangle$ and $\vert\phi\rangle$ are quantum
state vectors of a pair of systems $(\mathsf{X},\mathsf{Y})$ that satisfy the
equation
\[
\operatorname{Tr}_{\mathsf{Y}} (\vert\psi\rangle\langle\psi\vert) = \rho =
\operatorname{Tr}_{\mathsf{Y}} (\vert\phi\rangle\langle\phi\vert)
\]
for some density matrix $\rho$ representing a state of $\mathsf{X}.$
We shall consider a spectral decomposition of $\rho.$
\[
\rho = \sum_{a = 0}^{n-1} p_a \vert x_a\rangle\langle x_a\vert
\]
Here $\{\vert x_0\rangle,\ldots,\vert x_{n-1}\rangle\}$ is an orthonormal basis
of eigenvectors of $\rho.$

By following the prescription described previously we can obtain Schmidt
decompositions for both $\vert\psi\rangle$ and $\vert\phi\rangle$ having the
following form.
\[
\begin{aligned}
  \vert\psi\rangle & = \sum_{a = 0}^{r-1} \sqrt{p_a} \,
  \vert x_a\rangle \otimes \vert u_a\rangle\\[1mm]
  \vert\phi\rangle & = \sum_{a = 0}^{r-1} \sqrt{p_a} \,
  \vert x_a\rangle \otimes \vert v_a\rangle
\end{aligned}
\]
In these expressions $r$ is the rank of $\rho$ and
$\{\vert u_0\rangle,\ldots,\vert u_{r-1}\rangle\}$ and
$\{\vert v_0\rangle,\ldots,\vert v_{r-1}\rangle\}$ are orthonormal sets of
vectors in the space corresponding to $\mathsf{Y}.$

For any two orthonormal sets in the same space that have the same number of
elements, there's always a unitary matrix that transforms the first set into
the second, so we can choose a unitary matrix $U$ so that
$U \vert u_a\rangle = \vert v_a\rangle$ for $a = 0,\ldots,r-1.$
In particular, to find such a matrix $U$ we can first use the Gram--Schmidt
orthogonalization process to extend our orthonormal sets to orthonormal bases
$\{\vert u_0\rangle,\ldots,\vert u_{m-1}\rangle\}$ and
$\{\vert v_0\rangle,\ldots,\vert v_{m-1}\rangle\},$ where $m$ is the dimension
of the space corresponding to $\mathsf{Y},$ and then take
\[
U = \sum_{a = 0}^{m-1} \vert v_a\rangle\langle u_a\vert.
\]
We now find that
\[
(\mathbb{I}_{\mathsf{X}} \otimes U) \vert\psi\rangle
= \sum_{a = 0}^{r-1} \sqrt{p_a} \, \vert x_a\rangle \otimes U \vert
u_a\rangle
= \sum_{a = 0}^{r-1} \sqrt{p_a} \, \vert x_a\rangle \otimes \vert v_a\rangle
= \vert\phi\rangle,
\]
which completes the proof.

Here are just a few of many interesting examples and implications connected
with the unitary equivalence of purifications.
We'll see another critically important one later in the lesson, in the context
of fidelity, known as \emph{Uhlmann's theorem}.

\subsubsection{Superdense coding}

In the superdense coding protocol, Alice and Bob share an e-bit, meaning that
Alice holds a qubit $\mathsf{A},$ Bob holds a qubit $\mathsf{B},$ and together
the pair $(\mathsf{A},\mathsf{B})$ is in the $\vert\phi^{+}\rangle$ Bell state.
The protocol describes how Alice can transform this shared state into any one
of the four Bell states, $\vert\phi^+\rangle,$ $\vert\phi^-\rangle,$
$\vert\psi^+\rangle,$ and $\vert\psi^-\rangle,$ by applying a unitary operation
to her qubit $\mathsf{A}.$
Once she has done this, she sends $\mathsf{A}$ to Bob, and then Bob performs a
measurement on the pair $(\mathsf{A},\mathsf{B})$ to see which Bell state he
holds.

For all four Bell states, the reduced state of Bob's qubit $\mathsf{B}$ is
completely mixed.
\[
\operatorname{Tr}_{\mathsf{A}}(\vert\phi^+\rangle\langle\phi^+\vert) =
\operatorname{Tr}_{\mathsf{A}}(\vert\phi^-\rangle\langle\phi^-\vert) =
\operatorname{Tr}_{\mathsf{A}}(\vert\psi^+\rangle\langle\psi^+\vert) =
\operatorname{Tr}_{\mathsf{A}}(\vert\psi^-\rangle\langle\psi^-\vert) =
\frac{\mathbb{I}}{2}
\]
By the unitary equivalence of purifications, we immediately conclude that for
each Bell state there must exist a unitary operation on Alice's qubit
$\mathsf{A}$ alone that transforms $\vert\phi^+\rangle$ into the chosen Bell
state.
Although this does not reveal the precise details of the protocol, the unitary
equivalence of purifications does immediately imply that superdense coding is
possible.

We can also conclude that generalizations of superdense coding to larger
systems are always possible, provided that we replace the Bell states with any
orthonormal basis of purifications of the completely mixed state.

\subsubsection{Cryptographic implications}

The unitary equivalence of purifications has implications concerning the
implementation of cryptographic primitives using quantum information.
For instance, the unitary equivalence of purifications reveals that it is
impossible to implement an ideal form of \emph{bit commitment} using quantum
information.

The bit commitment primitive involves two participants, Alice and Bob (who
don't trust one another), and has two phases.
\begin{itemize}
\item
  The first phase is the \emph{commit} phase, through which Alice commits to a
  binary value $b\in\{0,1\}.$
  This commitment must be \emph{binding}, which means that Alice cannot change
  her mind, as well as \emph{concealing}, which means that Bob can't tell which
  value Alice has committed to.
\item
  The second phase is the \emph{reveal} phase, in which the bit committed by
  Alice becomes known to Bob, who should then be convinced that it was truly
  the committed value that was revealed.
\end{itemize}

In intuitive, operational terms, the first phase of bit commitment should
function as if Alice writes a binary value on a piece of paper, locks the paper
inside of a safe, and gives the safe to Bob while keeping the key for herself.
Alice has committed to the binary value written on the paper because the safe
is in Bob's possession (so it's binding), but because Bob can't open the safe
he can't tell which value Alice committed to (so it's concealing).
The second phase should work as if Alice hands the key to the safe to Bob, so
that he can open the safe to reveal the value to which Alice committed.

As it turns out, it is impossible to implement a perfect bit commitment
protocol by means of quantum information alone, for this contradicts the
unitary equivalence of purifications.
Here is a high-level summary of an argument that establishes this.

To begin, we can assume Alice and Bob only perform unitary operations or
introduce new initialized systems as the protocol is executed.
The fact that every channel has a Stinespring representation allows us to make
this assumption.

At the end of the commit phase of the protocol, Bob holds in his possession
some compound system that must be in one of two quantum states: $\rho_0$ if
Alice committed to the value $0$ and $\rho_1$ if Alice committed to the value
$1.$
In order for the protocol to be perfectly concealing, Bob should not be able to
tell the difference between these two states --- so it must be that $\rho_0 =
\rho_1.$
(Otherwise there would be a measurement that discriminates these states
probabilistically.)

However, because Alice and Bob have only used unitary operations, the state of
all of the systems involved in the protocol together after the commit phase
must be in a pure state.
In particular, suppose that $\vert\psi_0\rangle$ is the pure state of all of
the systems involved in the protocol when Alice commits to $0,$ and
$\vert\psi_1\rangle$ is the pure state of all of the systems involved in the
protocol when Alice commits to $1.$
If we write $\mathsf{A}$ and $\mathsf{B}$ to denote Alice and Bob's (possibly
compound) systems, then
\[
\begin{aligned}
  \rho_0 & =
  \operatorname{Tr}_{\mathsf{A}}(\vert\psi_0\rangle\langle\psi_0\vert)\\[1mm]
  \rho_1 & =
  \operatorname{Tr}_{\mathsf{A}}(\vert\psi_1\rangle\langle\psi_1\vert).
\end{aligned}
\]

Given the requirement that $\rho_0 = \rho_1$ for a perfectly concealing
protocol, we find that $\vert\psi_0\rangle$ and $\vert\psi_1\rangle$ are
purifications of the same state --- and so, by the unitary equivalence of
purifications, there must exist a unitary operation $U$ on $\mathsf{A}$ alone
such that
\[
(U\otimes\mathbb{I}_{\mathsf{B}})\vert\psi_0\rangle = \vert\psi_1\rangle.
\]
Alice is therefore free to change her commitment from $0$ to $1$ by applying
$U$ to $\mathsf{A},$ or from $1$ to $0$ by applying $U^{\dagger},$ and so the
hypothetical protocol being considered completely fails to be binding.

\subsubsection{Hughston--Jozsa--Wootters theorem}

The last implication of the unitary equivalence of purifications that we'll
discuss in this lesson is a theorem known as the
Hughston--Jozsa--Wootters theorem.

\begin{callout}[title = {Hughston--Jozsa--Wootters theorem}]
  Let $\mathsf{X}$ and $\mathsf{Y}$ be systems and let $\vert\phi\rangle$ be a
  quantum state vector of the pair $(\mathsf{X},\mathsf{Y}).$
  Also let $N$ be an arbitrary positive integer, let $(p_0,\ldots,p_{N-1})$ be
  a probability vector, and let
  $\vert\psi_0\rangle,\ldots,\vert\psi_{N-1}\rangle$ be quantum state vectors
  representing states of $\mathsf{X}$ such that
  \[
  \operatorname{Tr}_{\mathsf{Y}}\bigl(\vert\phi\rangle\langle\phi\vert\bigr) =
  \sum_{a = 0}^{N-1} p_a \vert\psi_a\rangle\langle\psi_a\vert.
  \]
  There exists a measurement $\{P_0,\ldots,P_{N-1}\}$ on $\mathsf{Y}$ such that
  the following two statements are true when this measurement is performed on
  $\mathsf{Y}$ when $(\mathsf{X},\mathsf{Y})$ is in the state
  $\vert\phi\rangle:$
  \begin{enumerate}
  \item
    Each outcome $a\in\{0,\ldots,N-1\}$ appears with probability~$p_a$.
  \item
    Conditioned on obtaining the outcome $a,$ the state of
    $\mathsf{X}$ becomes $\vert\psi_a\rangle.$
  \end{enumerate}
\end{callout}

\noindent
(This is, in fact, a slightly simplified version of this theorem.)

Intuitively speaking, this theorem says that as long as we have a pure state of
two systems, then for \emph{any} way of thinking about the reduced state of the
first system as a convex combination of pure states, there is a measurement of
the second system that effectively makes this way of thinking about the first
system a reality.
Notice that the number $N$ is not necessarily bounded by the number of
classical states of $\mathsf{X}$ or $\mathsf{Y}.$
For instance, it could be that $N = 1,000,000$ while $\mathsf{X}$ and
$\mathsf{Y}$ are qubits.

We shall prove this theorem using the unitary equivalence of purifications,
beginning with the introduction of a new system $\mathsf{Z}$ whose classical
state set is $\{0,\ldots,N-1\}.$
Consider the following two quantum state vectors of the triple
$(\mathsf{X},\mathsf{Y},\mathsf{Z}).$
\[
\begin{aligned}
  \vert\gamma_0\rangle & = \vert\phi\rangle_{\mathsf{XY}}\otimes\vert
  0\rangle_{\mathsf{Z}}\\[1mm]
  \vert\gamma_1\rangle & = \sum_{a = 0}^{N-1} \sqrt{p_a}\,
  \vert\psi_a\rangle_{\mathsf{X}} \otimes \vert 0\rangle_{\mathsf{Y}} \otimes
  \vert a\rangle_{\mathsf{Z}}
\end{aligned}
\]
The first vector $\vert\gamma_0\rangle$ is simply the given quantum state
vector $\vert\phi\rangle$ tensored with $\vert 0\rangle$ for the new system
$\mathsf{Z}.$
For the second vector $\vert\gamma_1\rangle,$ we essentially have a quantum
state vector that would make the theorem trivial --- at least if $\mathsf{Y}$
were replaced by $\mathsf{Z}$ --- because a standard basis measurement
performed on $\mathsf{Z}$ clearly yields each outcome $a$ with probability
$p_a,$ and conditioned on obtaining this outcome the state of $\mathsf{X}$
becomes $\vert\psi_a\rangle.$

By thinking about the pair $(\mathsf{Y},\mathsf{Z})$ as a single, compound
system that can be traced out to leave $\mathsf{X},$ we find that we have
identified two different purifications of the state
\[
\rho = \sum_{a = 0}^{N-1} p_a \vert\psi_a\rangle\langle\psi_a\vert.
\]
Specifically, for the first one we have
\[
\operatorname{Tr}_{\mathsf{YZ}} (\vert\gamma_0\rangle\langle\gamma_0\vert)
= \operatorname{Tr}_{\mathsf{Y}} (\vert\phi\rangle\langle\phi\vert) = \rho
\]
and for the second one we have
\[
\begin{aligned}
  \operatorname{Tr}_{\mathsf{YZ}} (\vert\gamma_1\rangle\langle\gamma_1\vert)
  & = \sum_{a,b = 0}^{N-1} \sqrt{p_a}\sqrt{p_b} \,
  \vert\psi_a\rangle\langle\psi_a\vert
  \operatorname{Tr}(\vert 0\rangle\langle 0\vert \otimes \vert a\rangle\langle
  b\vert)\\
  & = \sum_{a = 0}^{N-1} p_a \, \vert\psi_a\rangle\langle\psi_a\vert\\
  & = \rho.
\end{aligned}
\]
There must therefore exist a unitary operation $U$ on $(\mathsf{Y},\mathsf{Z})$
satisfying
\[
(\mathbb{I}_{\mathsf{X}} \otimes U) \vert \gamma_0 \rangle =
\vert\gamma_1\rangle
\]
by the unitary equivalence of purifications.

\begin{figure}[!ht]
  \begin{center}
    \begin{tikzpicture}[
        line width = 0.6pt,        
        gate/.style={%
          inner sep = 0,
          fill = CircuitBlue,
          draw = CircuitBlue,
          text = white,
          minimum size = 10mm
        },
        blackgate/.style={%
          inner sep = 0,
          fill = black,
          draw = black,
          text = white,
          minimum size = 10mm}
      ]
      
      \node (Zin) at (-1.75,1) {};
      \node (Yin) at (-4,0) {};
      \node (Xin) at (-4,-1.25) {};

      \node (MeasureZ) at (2,1) {};
      \node (BendY) at (2,0) {};
      \node (GroundY) at (2,-0.5) {};
      \node (Xout) at (3.75,-1.25) {};
      \node (Zout) at (3.75,1) {};
      
      \draw (Xin) -- (Xout);
      \draw (Yin) -- (BendY.center) -- (GroundY.center);
      \draw (Zin) -- (MeasureZ);
      
      \node[gate, minimum height=20mm, minimum width=16mm]
      (U) at (0,0.5) {$U$};     
      
      \pic at (GroundY) {ground};

      \node[anchor = south east] at (-.875,1) {$\mathsf{Z}$};
      \node[anchor = south east] at (-3,0) {$\mathsf{Y}$};
      \node[anchor = south east] at (-3,-1.25) {$\mathsf{X}$};
      \node[anchor = east] at (Xin) {};
      \node[anchor = east] at (Zin) {$\ket{0}$};

      \draw ([yshift=0.3mm]MeasureZ.east) -- ([yshift=0.3mm]Zout.west);
      \draw ([yshift=-0.3mm]MeasureZ.east) -- ([yshift=-0.3mm]Zout.west);
      
      \node[blackgate] at (MeasureZ) {};
      \readout{MeasureZ}
      
      \node[anchor = east] at ($(Xin)!0.5!(Yin)$) {%
        $\ket{\phi}\,\left\{\rule{0mm}{9mm}\right.$};

      \node[anchor = west] at (Zout) {$a$ (with probability $p_a$)};
      \node[anchor = west] at (Xout) {$\ket{\psi_a}$};

      \node[
        draw,
        densely dotted,
        minimum height = 28mm,
        minimum width=55mm
      ] at (0.12,0.45) {};

    \end{tikzpicture}   
  \end{center}
  \caption{An implementation of a measurement for the
    Hughston--Jozsa--Wootters theorem.}
  \label{fig:HSW-measurement}
\end{figure}
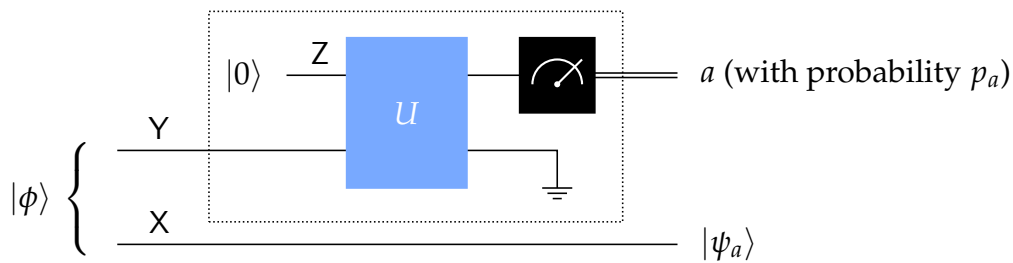

Using this unitary operation $U,$ we can implement a measurement that satisfies
the requirements of the theorem as Figure~\ref{fig:HSW-measurement}
illustrates.
In words, we introduce the new system $\mathsf{Z}$ initialized to the
$\vert 0\rangle$ state, apply $U$ to $(\mathsf{Y},\mathsf{Z}),$ which
transforms the state of $(\mathsf{X},\mathsf{Y},\mathsf{Z})$ from
$\vert\gamma_0\rangle$ into $\vert\gamma_1\rangle,$ and then measure
$\mathsf{Z}$ with a standard basis measurement, which we've already observed
gives the desired behavior.

The dotted rectangle in the figure represents an implementation of this
measurement, which can be described as a collection of positive semidefinite
matrices $\{P_0,\ldots,P_{N-1}\}$ as follows.
\[
P_a = (\mathbb{I}_{\mathsf{Y}} \otimes \langle 0\vert) U^{\dagger}
(\mathbb{I}_{\mathsf{Y}} \otimes \vert a\rangle\langle a \vert)U
(\mathbb{I}_{\mathsf{Y}} \otimes \vert 0\rangle)
\]

\section{Fidelity}

In this part of the lesson, we'll discuss the \emph{fidelity} between quantum
states, which is a measure of their similarity --- or how much they
``overlap.''

Given two quantum state vectors, the fidelity between the pure states
associated with these quantum state vectors equals the absolute value of the
inner product between the quantum state vectors.
This provides a basic way to measure their similarity: the result is a value
between $0$ and $1,$ with larger values indicating greater similarity.
In particular, the value is zero for orthogonal states (by definition), while
the value is $1$ for states equivalent up to a global phase.

Intuitively speaking, the fidelity can be seen as an extension of this basic
measure of similarity, from quantum state vectors to density matrices.

\subsection{Definition of fidelity}

It's fitting to begin with a definition of fidelity.
At first glance, the definition that follows might look unusual or mysterious,
and perhaps not easy to work with.
The function it defines, however, turns out to have many interesting properties
and multiple alternative formulations, making it much nicer to work with than
it may initially appear.

\begin{callout}[title={Fidelity}]
  Let $\rho$ and $\sigma$ be density matrices representing quantum states of
  the same system.
  The \emph{fidelity} between $\rho$ and $\sigma$ is defined as
  \[
  \operatorname{F}(\rho,\sigma) = \operatorname{Tr}\sqrt{\sqrt{\rho} \sigma
    \sqrt{\rho}}.
  \]
\end{callout}

\begin{trivlist}
\item \textbf{Remark.}
  Although this is a common definition, it is also common that the fidelity is
  defined as the \emph{square} of the quantity defined here, which is then
  referred to as the \emph{root-fidelity}.
  Neither definition is right or wrong --- it's essentially a matter of
  preference.
  Nevertheless, one must always be careful to understand or clarify which
  definition is being used.
\end{trivlist}

To make sense of the formula in the definition, notice first that
$\sqrt{\rho} \sigma \sqrt{\rho}$ is a positive semidefinite matrix:
\[
\sqrt{\rho} \sigma \sqrt{\rho} = M^{\dagger} M
\]
for $M = \sqrt{\sigma}\sqrt{\rho}.$
Like all positive semidefinite matrices, this positive semidefinite matrix has
a unique positive semidefinite square root, the trace of which is the fidelity.

For every square matrix $M,$ the eigenvalues of the two positive semidefinite
matrices $M^{\dagger} M$ and $M M^{\dagger}$ are always the same, and hence the
same is true for the square roots of these matrices.
Choosing $M = \sqrt{\sigma}\sqrt{\rho}$ and using the fact that the trace of a
square matrix is the sum of its eigenvalues, we find that
\[
\operatorname{F}(\rho,\sigma)
= \operatorname{Tr}\sqrt{\sqrt{\rho} \sigma \sqrt{\rho}}
= \operatorname{Tr}\sqrt{M^{\dagger} M} = \operatorname{Tr}\sqrt{M M^{\dagger}}
= \operatorname{Tr}\sqrt{\sqrt{\sigma} \rho \sqrt{\sigma}}
= \operatorname{F}(\sigma,\rho).
\]
So, although it is not immediate from the definition, the fidelity is symmetric
in its two arguments.

\subsubsection{Fidelity in terms of the trace norm}

An equivalent way to express the fidelity is by this formula:
\[
\operatorname{F}(\rho,\sigma) = \bigl\|\sqrt{\sigma}\sqrt{\rho}\bigr\|_1.
\]
Here we see the \emph{trace norm}, which we encountered in the previous lesson
in the context of state discrimination.
The trace norm of a (not necessarily square) matrix $M$ can be defined as
\[
\| M \|_1 = \operatorname{Tr}\sqrt{M^{\dagger} M},
\]
and by applying this definition to the matrix $\sqrt{\sigma}\sqrt{\rho}$ we
obtain the formula in the definition.

An alternative way to express the trace norm of a (square) matrix $M$ is
through this formula.
\[
\| M \|_1 = \max_{U\:\text{unitary}} \bigl\vert \operatorname{Tr}(M U)
\bigr\vert.
\]
Here the maximum is over all \emph{unitary} matrices $U$ having the same number
of rows and columns as $M.$
Applying this formula in the situation at hand reveals another expression of
the fidelity.
\[
\operatorname{F}(\rho,\sigma)
= \max_{U\:\text{unitary}} \bigl\vert\operatorname{Tr}\bigl(
\sqrt{\sigma}\sqrt{\rho}\, U\bigr) \bigr\vert
\]

\subsubsection{Fidelity for pure states}

One last point on the definition of fidelity is that every pure state is (as a
density matrix) equal to its own square root, which allows the formula for the
fidelity to be simplified considerably when one or both of the states is pure.
In particular, if one of the two states is pure we have the following formula.
\[
\operatorname{F}\bigl( \vert\phi\rangle\langle\phi\vert, \sigma \bigr)
= \sqrt{\langle \phi\vert \sigma \vert \phi \rangle}
\]

If both states are pure, the formula simplifies to the absolute value of the
inner product of the corresponding quantum state vectors, as was mentioned at
the start of the section.
\[
\operatorname{F}\bigl( \vert\phi\rangle\langle\phi\vert,
\vert\psi\rangle\langle\psi\vert \bigr)
= \bigl\vert \langle \phi\vert \psi \rangle \bigr\vert
\]

\subsection{Basic properties of fidelity}

The fidelity has many remarkable properties and several alternative
formulations.
Here are just a few basic properties listed without proofs.

\begin{enumerate}
\item
  For any two density matrices $\rho$ and $\sigma$ having the same size, the
  fidelity $\operatorname{F}(\rho,\sigma)$ lies between zero and one: $0\leq
  \operatorname{F}(\rho,\sigma) \leq 1.$ It is the case that
  $\operatorname{F}(\rho,\sigma)=0$ if and only if $\rho$ and $\sigma$ have
  orthogonal images (so they can be discriminated without error), and
  $\operatorname{F}(\rho,\sigma)=1$ if and only if $\rho = \sigma.$
\item
  The fidelity is \emph{multiplicative,} meaning that the fidelity between two
  product states is equal to the product of the individual fidelities:
  \[
  \operatorname{F}(\rho_1\otimes\cdots\otimes\rho_m,
  \sigma_1\otimes\cdots\otimes\sigma_m)
  = \operatorname{F}(\rho_1,\sigma_1)\cdots \operatorname{F}(\rho_m,\sigma_m).
  \]
\item
  The fidelity between states is nondecreasing under the action of any
  channel. That is, if $\rho$ and $\sigma$ are density matrices and $\Phi$ is a
  channel that can take these two states as input, then it is necessarily the
  case that
  \[
  \operatorname{F}(\rho,\sigma) \leq \operatorname{F}(\Phi(\rho),\Phi(\sigma)).
  \]
\item
  The \emph{Fuchs-van de Graaf inequalities} establish a close (though not
  exact) relationship between fidelity and trace distance: for any two states
  $\rho$ and $\sigma$ we have
  \[
  1 - \frac{1}{2}\|\rho - \sigma\|_1 \leq \operatorname{F}(\rho,\sigma)
  \leq \sqrt{1 - \frac{1}{4}\|\rho - \sigma\|_1^2}.
  \]
\end{enumerate}

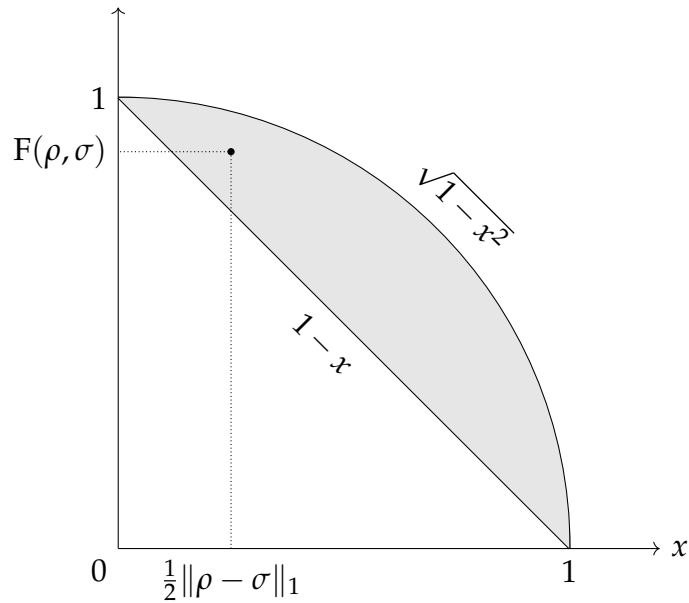
\begin{figure}[!ht]
  \begin{center}
    \begin{tikzpicture}[scale=6]
      \draw[->] (0,0) -- (1.2,0) node[right] {$x$};
      \draw[->] (0,0) -- (0,1.2);
      
      \begin{scope}
        \clip (0,0) rectangle (1.2,1.2);
        \draw[black, thick] (0,0) circle (1);
      \end{scope}
      
      \draw[black, thick] (0,1) -- (1,0);
      
      \begin{scope}
        \clip (0,1) -- (1,0) -- (1,1) -- cycle;
        \fill[LightGray] (0,0) circle (1);
      \end{scope}
      
      \node[rotate=-45, anchor=south] at (0.72,0.72) {$\sqrt{1 - x^2}$};
      \node[rotate=-45, anchor=north] at (0.49,0.49) {$1 - x$};
      
      \draw[densely dotted] (0.25,0) -- (0.25,0.88) -- (0,0.88);
      \fill[black] (0.25,0.88) circle (0.008);
      
      \node[anchor=north] at (0.25,0) {$\frac{1}{2}\|\rho-\sigma\|_1$};
      \node[anchor=east] at (0,0.88) {$\mathrm{F}(\rho,\sigma)$};
      \node[anchor=north] at (1,0) {$1$};
      \node[anchor=east] at (0,1) {$1$};
      \node[anchor=north east] at (0,0) {$0$};
    \end{tikzpicture}
    
  \end{center}
  \caption{The horizontal line corresponding to the fidelity and the vertical
    line corresponding to the trace distance between two states must intersect
    inside the shaded region.}
  \label{fig:FvdG-plot}
\end{figure}

\noindent
The final property can be expressed graphically as shown in
Figure~\ref{fig:FvdG-plot}.
Specifically, for any choice of states $\rho$ and $\sigma$ of the same system,
the horizontal line that crosses the $y$-axis at
$\operatorname{F}(\rho,\sigma)$ and the vertical line that crosses the $x$-axis
at $\frac{1}{2}\|\rho-\sigma\|_1$ (which is sometimes called the \emph{trace
distance} between $\rho$ and $\sigma$) must intersect within the gray region
bordered below by the line $y = 1-x$ and above by the unit circle.
The most interesting region of this figure from a practical viewpoint is the
upper left-hand corner of the gray region:
if the fidelity between two states is close to one, then their trace distance
is close to zero, and vice versa.

\subsection{Gentle measurement lemma}

Next we'll take a look at a simple but important fact, known as the
\emph{gentle measurement lemma}, which connects fidelity to non-destructive
measurements.
It's a very useful lemma that comes up from time to time, and it's also
noteworthy because the seemingly clunky definition for the fidelity actually
makes the lemma very easy to prove.

The set-up is as follows.
Let $\mathsf{X}$ be a system in a state $\rho$ and let $\{P_0,\ldots,P_{m-1}\}$
be a collection of positive semidefinite matrices representing a general
measurement of $\mathsf{X}.$
Suppose further that if this measurement is performed on the system
$\mathsf{X}$ while it's in the state $\rho,$ one of the outcomes is highly
likely.
To be concrete, let's assume that the likely measurement outcome is $0,$ and
specifically let's assume that
\[
\operatorname{Tr}(P_0 \rho) > 1 - \varepsilon
\]
for a small positive real number $\varepsilon > 0.$

What the gentle measurement lemma states is that, under these assumptions, the
non-destructive measurement obtained from $\{P_0,\ldots,P_{m-1}\}$ through
Naimark's theorem causes only a small disturbance to $\rho$ in case the likely
measurement outcome $0$ is observed.

More specifically, the lemma states that the fidelity-squared between $\rho$
and the state we obtain from the non-destructive measurement, conditioned on
the outcome being $0,$ is greater than $1-\varepsilon.$
\[
\operatorname{F}\Biggl(\rho,\frac{\sqrt{P_0}\rho
  \sqrt{P_0}}{\operatorname{Tr}(P_0\rho)}\Biggr)^2
> 1-\varepsilon.
\]

We'll need a basic fact about measurements to prove this.
The measurement matrices $P_0, \ldots, P_{m-1}$ are positive semidefinite and
sum to the identity, which allows us to conclude that all of the eigenvalues of
$P_0$ are real numbers between $0$ and~$1.$
This follows from the fact that, for any unit vector $\vert\psi\rangle,$ the
value $\langle \psi \vert P_a \vert \psi \rangle$ is a nonnegative real number
for each $a\in\{0,\ldots,m-1\}$ (because each $P_a$ is positive semidefinite),
together with these numbers summing to one.
\[
\sum_{a = 0}^{m-1} \langle \psi \vert P_a \vert \psi \rangle
= \langle \psi \vert \Biggl(\sum_{a = 0}^{m-1}  P_a \Biggr) \vert \psi \rangle
= \langle \psi \vert \mathbb{I} \vert \psi \rangle = 1.
\]
Hence $\langle \psi \vert P_0 \vert \psi \rangle$ is always a real number
between $0$ and $1,$ and this implies that every eigenvalue of $P_0$ is a real
number between $0$ and $1$ because we can choose $\vert\psi\rangle$
specifically to be a unit eigenvector corresponding to whichever eigenvalue is
of interest.

From this observation we can conclude the following inequality for every
density matrix $\rho.$
\[
\operatorname{Tr}\bigl( \sqrt{P_0} \rho\bigr) \geq \operatorname{Tr}\bigl( P_0
\rho\bigr)
\]
In greater detail, starting from a spectral decomposition
\[
P_0 = \sum_{k=0}^{n-1} \lambda_k \vert\psi_k\rangle\langle\psi_k\vert
\]
we conclude that
\[
\operatorname{Tr}\bigl( \sqrt{P_0} \rho\bigr)
= \sum_{k = 0}^{n-1} \sqrt{\lambda_k} \langle \psi_k \vert \rho \vert \psi_k
\rangle \geq \sum_{k = 0}^{n-1} \lambda_k \langle \psi_k \vert \rho \vert
\psi_k \rangle = \operatorname{Tr}\bigl( P_0 \rho\bigr)
\]
from the fact that $\langle \psi_k \vert \rho \vert \psi_k \rangle$ is a
nonnegative real number and $\sqrt{\lambda_k} \geq \lambda_k$ for each
$k = 0,\ldots,n-1.$ (Squaring numbers between $0$ and $1$ can never make them
larger.)

Now we can prove the gentle measurement lemma by evaluating the fidelity and
then using our inequality.
First, let's simplify the expression of interest.
\[
\begin{aligned}
  \operatorname{F}\Biggl(\rho,\frac{\sqrt{P_0}\rho
    \sqrt{P_0}}{\operatorname{Tr}(P_0\rho)}\Biggr)
  & = \operatorname{Tr}\sqrt{\frac{\sqrt{\rho}\sqrt{P_0}\rho
      \sqrt{P_0}\sqrt{\rho}}{\operatorname{Tr}(P_0\rho)}}\\
  & = \operatorname{Tr}\sqrt{\Biggl(\frac{\sqrt{\rho}\sqrt{P_0}\sqrt{\rho}}{
      \sqrt{\operatorname{Tr}(P_0\rho)}}\Biggr)^2}\\
  & = \operatorname{Tr}\Biggl(\frac{\sqrt{\rho}\sqrt{P_0}\sqrt{\rho}}{
    \sqrt{\operatorname{Tr}(P_0\rho)}}\Biggr)\\
  & = \frac{\operatorname{Tr}\bigl(\sqrt{P_0}\rho
    \bigr)}{\sqrt{\operatorname{Tr}(P_0\rho)}}
\end{aligned}
\]
Notice that these are all equalities --- we've not used our inequality (or any
other inequality) at this point, so we have an exact expression for the
fidelity.
We can now use our inequality to conclude
\[
\operatorname{F}\Biggl(\rho,\frac{\sqrt{P_0}\rho
  \sqrt{P_0}}{\operatorname{Tr}(P_0\rho)}\Biggr)
= \frac{\operatorname{Tr}\bigl(\sqrt{P_0}\rho
  \bigr)}{\sqrt{\operatorname{Tr}(P_0\rho)}}
\geq \frac{\operatorname{Tr}\bigl(P_0\rho
  \bigr)}{\sqrt{\operatorname{Tr}(P_0\rho)}}
= \sqrt{\operatorname{Tr}\bigl(P_0\rho\bigr)}
\]
and therefore, by squaring both sides,
\[
\operatorname{F}\Biggl(\rho,\frac{\sqrt{P_0}\rho
  \sqrt{P_0}}{\operatorname{Tr}(P_0\rho)}\Biggr)^2
\geq \operatorname{Tr}\bigl(P_0\rho\bigr) > 1-\varepsilon.
\]

\subsection{Uhlmann's theorem}

To conclude the lesson, we'll take a look at \emph{Uhlmann's theorem}, which is
a fundamental fact about the fidelity that connects it with the notion of a
purification.
What the theorem says, in simple terms, is that the fidelity between any two
quantum states is equal to the \emph{maximum} inner product (in absolute value)
between two purifications of those states.

\begin{callout}[title = {Uhlmann's theorem}]
  Let $\rho$ and $\sigma$ be density matrices representing states of a system
  $\mathsf{X},$ and let $\mathsf{Y}$ be a system having at least as many
  classical states as $\mathsf{X}.$ The fidelity between $\rho$ and $\sigma$ is
  given by
  \[
  \operatorname{F}(\rho,\sigma) = \max\bigl\{ \vert \langle \phi \vert \psi
  \rangle \vert \,:\,
  \operatorname{Tr}_{\mathsf{Y}}\bigl(\vert\phi\rangle\langle\phi\vert\bigr) =
  \rho,\;
  \operatorname{Tr}_{\mathsf{Y}}\bigl(\vert\psi\rangle\langle\psi\vert\bigr) =
  \sigma\bigr\},
  \]
  where the maximum is over all quantum state vectors $\vert\phi\rangle$
  and $\vert\psi\rangle$ of $(\mathsf{X},\mathsf{Y}).$
\end{callout}

\noindent
We can prove this theorem using the unitary equivalence of purifications ---
but it isn't completely straightforward and we'll make use of a trick along the
way.

To begin, consider spectral decompositions of the two density matrices $\rho$
and $\sigma.$
\[
\begin{aligned}
  \rho & = \sum_{a = 0}^{n-1} p_a \vert u_a\rangle\langle u_a\vert \\[2mm]
  \sigma & = \sum_{b = 0}^{n-1} q_b \vert v_b\rangle\langle v_b\vert
\end{aligned}
\]
The two collections $\{\vert u_0 \rangle,\ldots,\vert u_{n-1}\rangle\}$ and
$\{\vert v_0 \rangle,\ldots,\vert v_{n-1}\rangle\}$ are orthonormal bases of
eigenvectors of $\rho$ and $\sigma,$ respectively, and $p_0,\ldots,p_{n-1}$ and
$q_0,\ldots,q_{n-1}$ are the corresponding eigenvalues.

We'll also define
$\vert \overline{u_0} \rangle,\ldots,\vert\overline{u_{n-1}}\rangle$ and
$\vert \overline{v_0} \rangle,\ldots,\vert \overline{v_{n-1}}\rangle$ to be the
vectors obtained by taking the complex conjugate of each entry of
$\vert u_0 \rangle,\ldots,\vert u_{n-1}\rangle$ and
$\vert v_0 \rangle,\ldots,\vert v_{n-1}\rangle.$
That is, for an arbitrary vector $\vert w\rangle$ we can define
$\vert\overline{w}\rangle$ according to the following equation for each
$c\in\{0,\ldots,n-1\}.$
\[
\langle c \vert \overline{w}\rangle = \overline{\langle c \vert w\rangle}
\]
Notice that for any two vectors $\vert u\rangle$ and $\vert v\rangle$ we have
$\langle \overline{u} \vert \overline{v}\rangle = \langle v\vert u\rangle.$
More generally, for any square matrix $M$ we have the following formula.
\[
\langle \overline{u} \vert M \vert \overline{v}\rangle = \langle v\vert M^T
\vert u\rangle
\]
It follows that $\vert u\rangle$ and $\vert v\rangle$ are orthogonal if and
only if $\vert \overline{u}\rangle$ and $\vert \overline{v}\rangle$ are
orthogonal, and therefore
$\{\vert \overline{u_0} \rangle,\ldots,\vert \overline{u_{n-1}}\rangle\}$ and
$\{\vert \overline{v_0} \rangle,\ldots,\vert \overline{v_{n-1}}\rangle\}$ are
both orthonormal bases.

Now consider the following two vectors $\vert\phi\rangle$ and
$\vert\psi\rangle,$ which are purifications of $\rho$ and $\sigma,$
respectively.
\[
\begin{aligned}
  \vert\phi\rangle & = \sum_{a = 0}^{n-1} \sqrt{p_a}\, \vert u_a\rangle \otimes
  \vert \overline{u_a}\rangle \\[2mm]
  \vert\psi\rangle & = \sum_{b = 0}^{n-1} \sqrt{q_b}\, \vert v_b\rangle \otimes
  \vert \overline{v_b}\rangle
\end{aligned}
\]
This is the trick referred to previously.
Nothing indicates explicitly at this point that it's a good idea to make these
particular choices for purifications of $\rho$ and $\sigma,$ but they are valid
purifications, and the complex conjugations will allow the algebra to work out
the way we need.

By the unitary equivalence of purifications, we know that every purification of
$\rho$ for the pair of systems $(\mathsf{X},\mathsf{Y})$ must take the form
$(\mathbb{I}_{\mathsf{X}}\otimes U)\vert\phi\rangle$ for some unitary matrix
$U,$ and likewise every purification of $\sigma$ for the pair
$(\mathsf{X},\mathsf{Y})$ must take the form
$(\mathbb{I}_{\mathsf{X}}\otimes V)\vert\psi\rangle$ for some unitary matrix
$V.$
The inner product of two such purifications can be simplified as follows.
\[
\begin{aligned}
  \langle \phi \vert (\mathbb{I}\otimes U^{\dagger}) (\mathbb{I}\otimes V)
  \vert \psi \rangle
  \hspace{-2.5cm}\\
  & = \sum_{a,b = 0}^{n-1} \sqrt{p_a} \sqrt{q_b}\, \langle u_a \vert v_b \rangle
  \langle \overline{u_a} \vert U^{\dagger} V \vert \overline{v_b} \rangle \\
  & = \sum_{a,b = 0}^{n-1} \sqrt{p_a} \sqrt{q_b}\, \langle u_a \vert v_b \rangle
  \langle v_b \vert (U^{\dagger} V)^T \vert u_a \rangle \\
  & = \operatorname{Tr}\Biggl(
  \sum_{a,b = 0}^{n-1} \sqrt{p_a} \sqrt{q_b}\, \vert u_a \rangle\langle u_a
  \vert v_b \rangle
  \langle v_b \vert (U^{\dagger} V)^T\Biggr)\\
  & = \operatorname{Tr}\Bigl(
  \sqrt{\rho}\sqrt{\sigma}\, (U^{\dagger} V)^T\Bigr)
\end{aligned}
\]

As $U$ and $V$ range over all possible unitary matrices, the matrix
$(U^{\dagger} V)^T$ also ranges over all possible unitary matrices.
Thus, maximizing the absolute value of the inner product of two purifications
of $\rho$ and $\sigma$ yields the following equation.
\begin{multline*}
  \qquad\qquad
  \max_{U,V\:\text{unitary}} \biggl\vert \operatorname{Tr}\Bigl(
  \sqrt{\rho}\sqrt{\sigma}\, (U^{\dagger} V)^T\Bigr)\biggr\vert \\
  = \max_{W\:\text{unitary}} \biggl\vert
  \operatorname{Tr}\Bigl(\sqrt{\rho}\sqrt{\sigma}\, W\Bigr)\biggr\vert
  = \bigl\| \sqrt{\rho}\sqrt{\sigma} \bigr\|_1
  = \operatorname{F}(\rho,\sigma).\qquad\qquad
\end{multline*}

\stopcontents[part]


\unit[Foundations of Quantum Error Correction]{Foundations of\\[-1mm]
Quantum Error Correction}
\label{unit:foundations-of-quantum-error-correction}

This final unit of the course is on quantum error correction.
It begins with an explanation of what quantum error correcting codes are and
how they work.
It then moves on to the stabilizer formalism for describing quantum error
correcting codes, CSS codes, and several key examples
of quantum error correcting codes.
The unit concludes with fault-tolerant quantum computation, in which quantum
computations are performed on error-corrected quantum information.

\begin{trivlist}
  \setlength{\parindent}{0mm}
  \setlength{\parskip}{2mm}
  \setlength{\itemsep}{1mm}
\item
  \textbf{Lesson 13: Correcting Quantum Errors}

  This lesson takes a first look at quantum error correction, including the
  first quantum error correcting code discovered --- the 9-qubit Shor code
  --- and the foundational concept in quantum error correction known as the
  discretization of errors.
  
  Lesson video URL: \url{https://youtu.be/OoQSdcKAIZc}

\item
  \textbf{Lesson 14: The Stabilizer Formalism}

  This lesson introduces the stabilizer formalism, which is a mathematical tool
  through which a broad class of quantum error correcting codes, known as
  stabilizer codes, can be specified and analyzed.
  
  Lesson video URL: \url{https://youtu.be/3ib2JP_LeIU}
  
\item
  \textbf{Lesson 15: Quantum Code Constructions}

  This lesson focuses on more sophisticated quantum error correcting codes,
  including ones that can tolerate relatively high error rates.
  It begins with a general class of quantum error correcting codes known as CSS
  codes, then moves on to the toric code, and concludes with a brief discussion
  surface codes and color codes.
  
  Lesson video URL: \url{https://youtu.be/9TCIOm8gcVQ}
  
\item
  \textbf{Lesson 16: Fault-Tolerant Quantum Computation}

  This lesson describes a basic methodology for fault tolerant implementations
  of quantum circuits and how to control error propagation.
  It concludes with a high-level discussion of the threshold theorem, which
  states that arbitrarily large quantum circuits can be implemented reliably so
  long as the error rate falls below a certain finite threshold value.
  
  Lesson video URL: \url{https://youtu.be/aeaqXh2XXMk}

\end{trivlist}


\lesson{Correcting Quantum Errors}
\label{lesson:correcting-quantum-errors}

Quantum computing has the potential to enable efficient solutions to
computational tasks for which efficient classical algorithms are not known, and
possibly don't exist.
There are, however, very significant challenges that must be overcome before we
can reliably implement the sorts of large-scale quantum computations we hope
will one day be possible.

The heart of the matter is that quantum information is extremely fragile;
you can literally ruin it just by looking at it.
For this reason, to correctly operate, quantum computers need to isolate the
quantum information they store from the environment around them to an extreme
degree.
But, at the same time, quantum computers must provide very precise control over
this quantum information, including proper initialization, accurate and
reliable unitary operations, and the ability to perform measurements so that
the results of the computation can be obtained.
There's clearly some tension between these requirements, and in the early days
of quantum computing some viewed that the fragility of quantum information, and
its susceptibility to both inaccuracies and environmental noise, would
ultimately make quantum computing impossible.

Today, there's little doubt that building an accurate and reliable large-scale
quantum computer is a monumental challenge.
But, we have a key tool to help us in this endeavor --- quantum error
correction --- which leads most people who are knowledgeable about the field to
be optimistic about large-scale quantum computing one day becoming a reality.

We'll study quantum error correction in this unit, with a focus on the
fundamentals.
In this lesson, we'll take a first look at quantum error correction, including
the very first quantum error correcting code discovered --- the
\emph{9-qubit Shor code} --- and we'll also discuss a foundational concept in
quantum error correction known as the \emph{discretization of errors}.

\section{Repetition codes}

We'll begin the lesson with a discussion of \emph{repetition codes.}
Repetition codes don't protect quantum information against every type of error
that can occur on qubits, but they do form the basis for the 9-qubit Shor code,
which we'll see in the next lesson, and they're also useful for explaining the
basics of error correction.

\subsection{Classical encoding and decoding}

Repetition codes are extremely basic examples of error correcting codes.
The idea is that we can protect bits against errors by simply repeating each
bit some fixed number of times.

In particular, let's first consider the 3-bit repetition code, just in the
context of classical information to start.
This code encodes one bit into three by repeating the bit three times, so $0$
is encoded as $000$ and $1$ is encoded as $111.$
\[
\begin{aligned}
  0 & \mapsto 000\\
  1 & \mapsto 111
\end{aligned}
\]

If nothing goes wrong, we can obviously distinguish the two possibilities for
the original bit from their encodings.
The point is that if there was an error and one of the three bits flipped,
meaning that a 0 changes into a 1 or a 1 changes to a 0, then we can still
figure out what the original bit was by determining which of the two binary
values appears twice.
Equivalently, we can \emph{decode} by computing the majority value (i.e., the
binary value that appears most frequently).
\[
a b c \mapsto \operatorname{majority}(a,b,c)
\]

Of course, if 2 or 3 bits of the encoding flip, then the decoding won't work
properly and the wrong bit will be recovered, but if at most 1 of the 3 bits
flip, the decoding will be correct.
This is a typical property of error correcting codes in general: they may allow
for the correction of errors, but only if there aren't too many of them.

\subsubsection{Noise reduction for the binary symmetric channel}

For an example of a situation in which the chances of making an error can be
decreased using a repetition code, suppose that our goal is to communicate a
single bit to a hypothetical receiver, and we're able to transmit bits through
a so-called \emph{binary symmetric channel}, which flips each bit sent through
it independently with some probability $p.$
That is, with probability $1-p,$ the receiver gets whatever bit was sent
through the channel, but with probability $p,$ the bit-flips and the receiver
gets the opposite bit value.

So, if we choose not to use the 3-bit repetition code, and simply send whatever
bit we have in mind through the channel, the receiver therefore receives the
wrong bit with probability $p.$
On the other hand, if we first encode the bit we want to send using the 3-bit
repetition code, and then send each of the three bits of the encoding through
the channel, then each one of them flips independently with probability $p.$
The chances of a bit-flip are now greater because there are now three bits that
might flip rather than one, but if at most one of the bits flips, then the
receiver will decode correctly.
An error therefore persists after decoding only if two or three of the bits
flip during transmission.

The probability that two bits flip during transmission is $3p^2(1-p),$ which is
$p^2(1-p)$ for each of the three choices for the bit that doesn't flip, while
the probability that all three bits flip is $p^3.$
The total probability of two or three bit-flips is therefore
\[
3 p^2 (1 - p) + p^3 = 3 p^2 - 2 p^3.
\]
For values of $p$ smaller than one-half, this results in a decrease in the
probability that the receiver ends up with the wrong bit. 
There will still be a chance of an error in this case, but the code
\emph{decreases} the likelihood.
(For values of $p$ \emph{greater} than one-half, on the other hand, the code
actually \emph{increases} the likelihood that the receiver gets the wrong bit.)

\begin{figure}[!ht]
  \begin{center}
    \begin{tikzpicture}
      \begin{axis}[
          scale = 1.5,
          samples=\samples,
          domain = 0:1,
          y = 4cm,
          x = 4cm,
          xtick = {0.25, 0.5, 0.75, 1},
          ytick = {0.25, 0.5, 0.75, 1},
          axis lines=center,
          axis line style={-},
          x label style={at={(axis description cs:0.5,-0.15)},
            anchor=north},                    
          y label style={at={(axis description cs:-0.25,.5)},
            rotate=90,anchor=south},                    
          xlabel={$p$},
          ylabel={error probability},
          legend style={at={(0.85,0.4)},anchor=south west},
          legend style={draw=none},
          legend cell align=left,              
          clip=false
        ]
        
        \draw[densely dotted] (0.5,0) -- (0.5, 0.5);
        \addplot[black] {x};        
        \addplot[very thick, DataColor0] {3*x*x - 2*x*x*x};

        \legend{ \;$p$, \;$3p^2 - 2p^3$}
      \end{axis}
      
    \end{tikzpicture}
  \end{center}
  \caption{The probability that two or three bits flip during transmission
    for the binary symmetric channel, leading to a decoding error for the 3-bit
    repetition code, is drawn in blue.}
  \label{fig:binary-symmetric-error}
\end{figure}
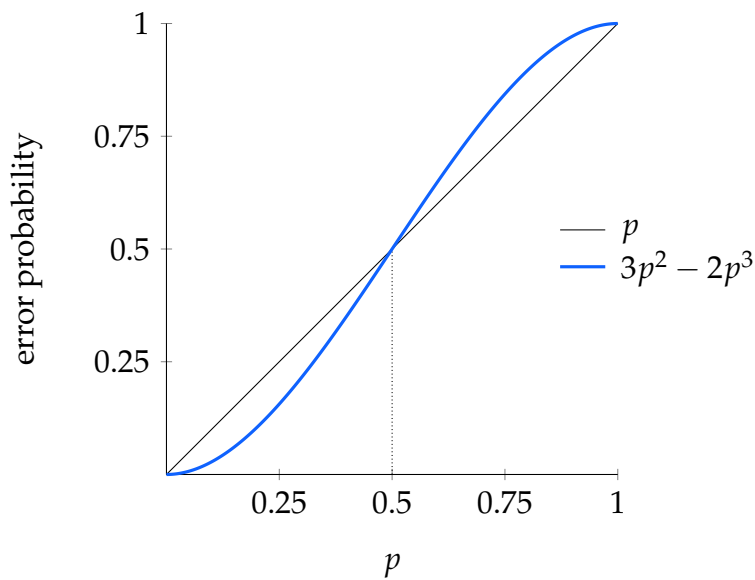

\subsection{Encoding qubits}

The 3-bit repetition code is a classical error correcting code, but we can
consider what happens if we try to use it to protect \emph{qubits} against
errors.
As we'll see, it's not a very impressive quantum error correcting code, because
it actually makes some errors more likely.
It is, however, the first step toward the Shor code, and will serve us well
from a pedagogical viewpoint.

\begin{figure}[!ht]
  \begin{center}
    \begin{tikzpicture}[
        line width = 0.6pt,
        scale = 1.4,
        control/.style={%
          circle,
          fill=CircuitBlue,
          minimum size = 5pt,
          inner sep=0mm},
        gate/.style={%
          inner sep = 0,
          fill = CircuitBlue,
          draw = CircuitBlue,
          text = white,
          minimum size = 10mm},
        not/.style={%
          circle,
          fill = CircuitBlue,
          draw = CircuitBlue,
          text = white,
          minimum size = 5mm,
          inner sep=0mm,
          label = {center:\textcolor{white}{\large $+$}}
        }
      ]
      
      \node (In0) at (-2.5,0.7) {};
      \node (In1) at (-1.5,0) {};
      \node (In2) at (-1.5,-0.7) {};
      \node (Out0) at (1.5,0.7) {};
      \node (Out1) at (1.5,0) {};
      \node (Out2) at (1.5,-0.7) {};
      
      \draw (In0) -- (Out0);
      \draw (In1) -- (Out1);
      \draw (In2) -- (Out2);
      
      \node[control] (Control1) at (-0.5,0.7) {};
      \node[control] (Control2) at (0.5,0.7) {};
      
      \node[not] (Not1) at (-0.5,0) {};
      \node[not] (Not2) at (0.5,-0.7) {};
      
      \draw[very thick,draw=CircuitBlue] (Control1.center) -- (Not1.north);
      \draw[very thick,draw=CircuitBlue] (Control2.center) -- (Not2.north);
      
      \node[anchor = east] at (In0) {$\alpha\ket{0} + \beta\ket{1}$};
      \node[anchor = east] at (In1) {$\ket{0}$};
      \node[anchor = east] at (In2) {$\ket{0}$};
      
      \node[anchor = west] at (Out1) {%
        $\left.\rule{0mm}{13mm}\right\}\;
        \alpha\ket{000} + \beta\ket{111}$};

    \end{tikzpicture}
  \end{center}
  \caption{An encoding circuit for the $3$-bit repetition code.}
  \label{fig:repetition-encoder}
\end{figure}
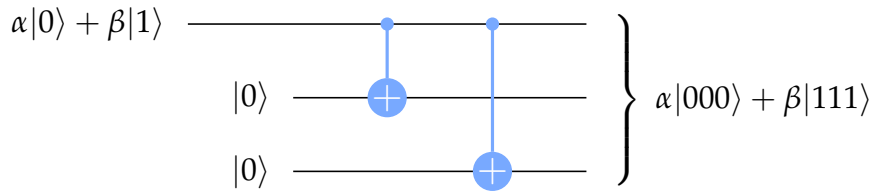

To be clear, when we refer to the 3-bit repetition code being used for qubits,
we have in mind an encoding of a qubit where \emph{standard basis states} are
repeated three times, so that a single-qubit state vector is encoded as
follows.
\[
\alpha \vert 0\rangle + \beta \vert 1\rangle \mapsto \alpha \vert 000\rangle +
\beta \vert 111\rangle
\]
This encoding is easily implemented by the quantum circuit in
Figure~\ref{fig:repetition-encoder}, which makes use of two initialized
workspace qubits and two controlled-NOT gates.

Notice, in particular, that this encoding is not the same as repeating the
quantum state three times, as in a given qubit state vector being encoded as
$\vert\psi\rangle \mapsto \vert\psi\rangle\vert\psi\rangle\vert\psi\rangle.$
Such an encoding cannot be implemented for an unknown quantum state
$\vert\psi\rangle$ by the no cloning theorem.

\subsubsection{Bit-flip errors}

Now suppose that an error takes place after the encoding has been performed.
Specifically, let's suppose that an $X$ gate, or in other words a bit-flip,
occurs on one of the qubits.
For instance, if the middle qubit experiences a bit-flip, the state of the
three qubits is transformed into this state:
\[
\alpha \vert 010\rangle + \beta \vert 101\rangle.
\]
This isn't the only sort of error that could occur --- and it's also reasonable
to question the assumption that an error takes the form of a perfect, unitary
operation.
We'll return to these issues in the last section of the lesson, and for now we
can view an error of this form as being just one possible type of error (albeit
a fundamentally important one).

We can see clearly from the mathematical expression for the state above that
the middle bit is the one that's different inside of each ket.
But suppose that we had the three qubits in our possession and didn't know
their state.
If we suspected that a bit-flip may have occurred, one option to verify that a
bit flipped would be to perform a standard basis measurement, which, in the
case at hand, would cause us to see $010$ or $101$ with probabilities
$\vert\alpha\vert^2$ and $\vert\beta\vert^2,$ respectively.
In either case, our conclusion would be that the middle bit flipped --- but,
unfortunately, we would lose the original quantum state $\alpha\vert 0\rangle +
\beta \vert 1\rangle.$
This is the state we're trying to protect, so measuring in the standard basis
is an unsatisfactory option.

\begin{figure}[!ht]
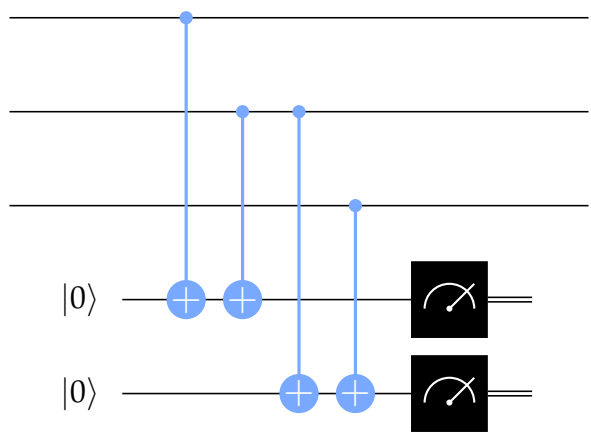

  \begin{center}

  \end{center}
  \caption{An error detection circuit for the $3$-bit repetition code.}
  \label{fig:repetition-error-detection}
\end{figure}

What we can do instead is to use the quantum circuit shown in
Figure~\ref{fig:repetition-error-detection}, feeding the encoded state into the
top three qubits.
This circuit nondestructively measures the \emph{parity} of the standard basis
states of the top two qubits as well as the bottom two qubits of the
three-qubit encoding.

\begin{figure}[!ht]
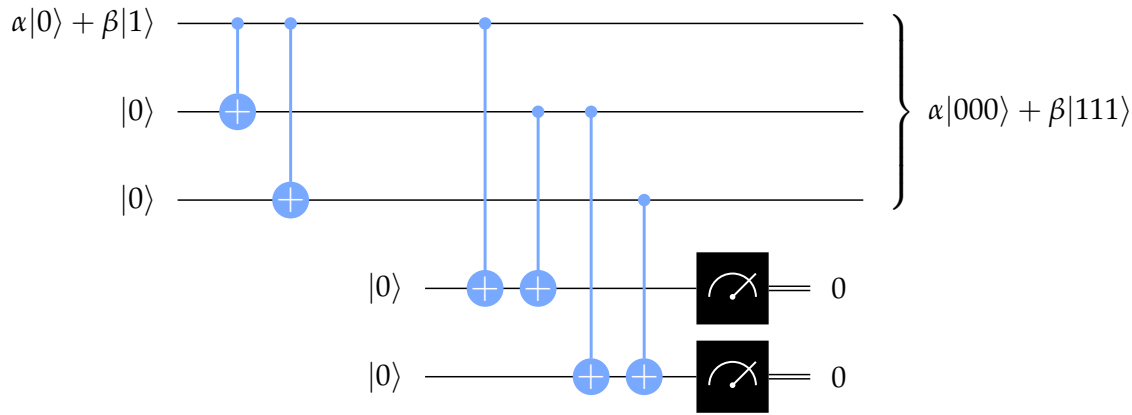

  \begin{center}
    \scalebox{0.94}{%
%
    }
  \end{center}
  \caption{If no errors occur, the error detection circuit results in the
    outcome $00$ and the encoded state is unchanged.}
  \label{fig:repetition-decode}
\end{figure}

\begin{figure}[p]
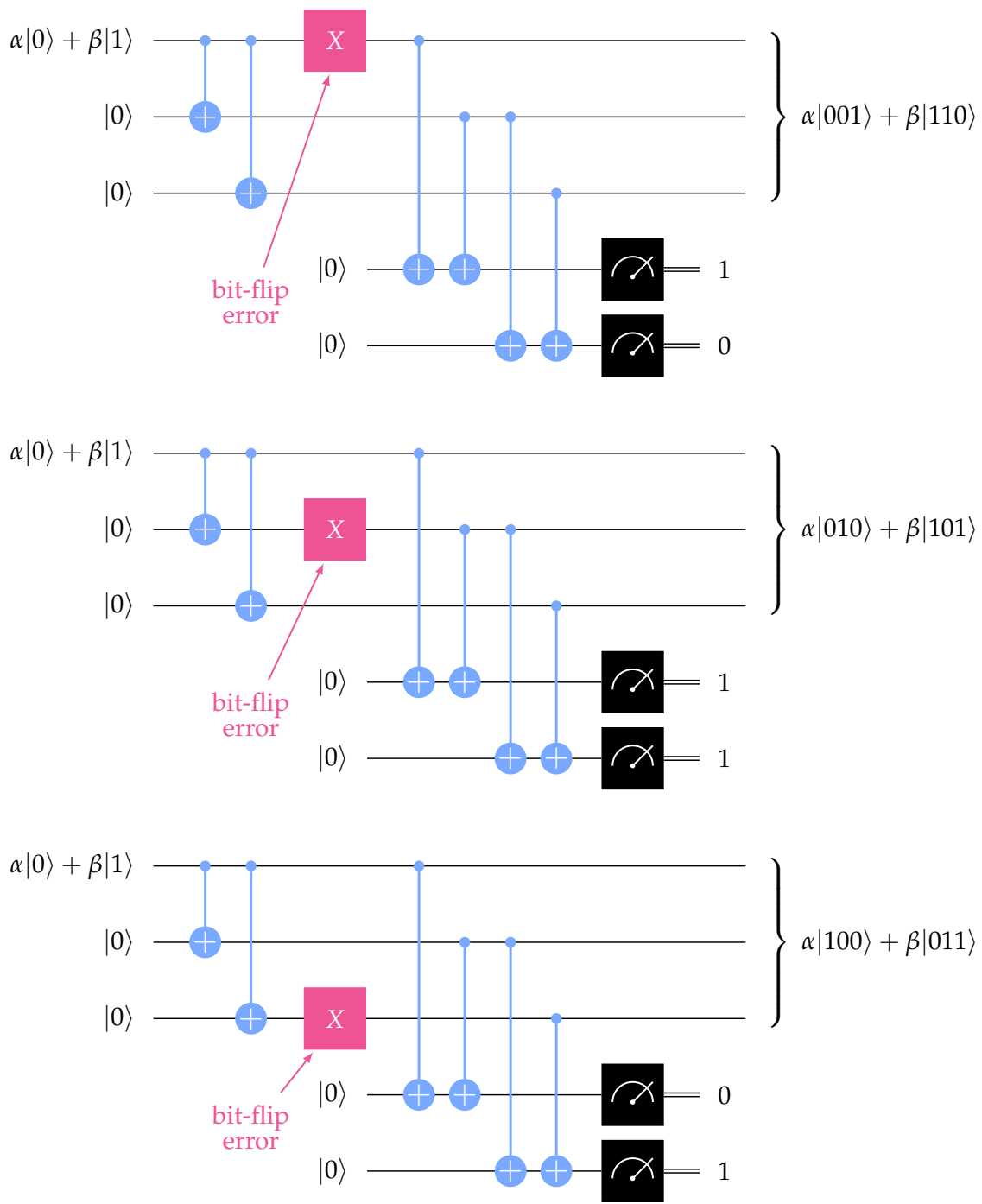

  \begin{center}
    \scalebox{0.94}{%
      %
%
    }
    
  \end{center}
  \caption{A single bit-flip error is detected by the $3$-bit repetition code,
    with the measurement outcomes revealing which qubit was affected.}
  \label{fig:repetition-decode-errors}
\end{figure}

Under the assumption that at most one bit flipped, one can easily deduce from
the measurement outcomes the location of the bit-flip (or the absence of one).
In particular, as Figure~\ref{fig:repetition-decode-errors} illustrates, the
three possible locations for a bit-flip error on the encoded state are
revealed by the measurement outcomes.
If no bit-flips occur, on the other hand, the measurement outcomes are $00$, as
shown in Figure~\ref{fig:repetition-decode}.

Crucially, the state of the top three qubits does not collapse in any of the
cases, which allows us to correct a bit-flip error if one has occurred --- by
simply applying the same bit-flip again with an $X$ gate.
The following table summarizes the states we obtain from at most one bit-flip,
the measurement outcomes (which are called the \emph{syndrome} in the context
of error correction), and the correction needed to get back to the original
encoding.

\begin{center}
  \begin{tabular}{c@{\hspace{4mm}}c@{\hspace{4mm}}c}
    State & Syndrome & Correction \\\hline
    $\alpha\vert 000\rangle + \beta \vert 111\rangle$ & $00$ &
    $\mathbb{I}\otimes\mathbb{I}\otimes\mathbb{I}$ \\
    $\alpha\vert 001\rangle + \beta \vert 110\rangle$ & $01$ &
    $\mathbb{I}\otimes\mathbb{I}\otimes X$ \\
    $\alpha\vert 010\rangle + \beta \vert 101\rangle$ & $11$ &
    $\mathbb{I}\otimes X\otimes\mathbb{I}$ \\
    $\alpha\vert 100\rangle + \beta \vert 011\rangle$ & $10$ &
    $X\otimes\mathbb{I}\otimes\mathbb{I}$
  \end{tabular}
\end{center}

\noindent
Once again, we're only considering the possibility that at most one bit-flip
occurred.
This wouldn't work correctly if two or three bit-flips occurred, and we also
haven't considered other possible errors besides bit-flips.

\subsubsection{Phase-flip errors}

In the quantum setting, bit-flip errors aren't the only errors we need to worry
about.
For instance, we also have to worry about \emph{phase-flip errors}, which are
described by $Z$ gates.
Along the same lines as bit-flip errors, we can think about phase-flip errors
as representing just another possibility for an error that can affect a qubit.

However, as we will see in the last section of the lesson, which is on the
so-called \emph{discretization of errors} for quantum error correcting codes, a
focus on bit-flip errors and phase-flip errors turns out to be well-justified.
Specifically, the ability to correct a bit-flip error, a phase-flip error, or
both of these errors simultaneously automatically implies the ability to
correct an arbitrary quantum error on a single qubit.

Unfortunately, the 3-bit repetition code doesn't protect against phase-flips at
all.
For instance, suppose that a qubit state $\alpha\vert 0\rangle + \beta\vert
1\rangle$ has been encoded using the 3-bit repetition code, and a phase-flip
error occurs on the middle qubit.
This results in the state
\[
(\mathbb{I} \otimes Z \otimes \mathbb{I}) ( \alpha \vert 000\rangle + \beta
\vert 111\rangle) = \alpha \vert 000\rangle - \beta \vert 111\rangle,
\]
which is exactly the state we would have obtained from encoding the qubit state
$\alpha\vert 0\rangle - \beta\vert 1\rangle.$
Indeed, a phase-flip error on any one of the three qubits of the encoding has
this same effect, which is equivalent to a phase-flip error occurring on the
original qubit prior to encoding.
Under the assumption that the original quantum state is an unknown state,
there's therefore no way to detect that an error has occurred, because the
resulting state is a perfectly valid encoding of a different qubit state.
In particular, running the error detection circuit from before on the state
$\alpha \vert 000\rangle - \beta \vert 111\rangle$ is certain to result in the
syndrome $00,$ which wrongly suggests that no errors have occurred.

Meanwhile, there are now three qubits rather than one that could potentially
experience phase-flip errors.
So, in a situation in which phase-flip errors are assumed to occur
independently on each qubit with some nonzero probability $p$ (similar to a
binary symmetric channel except for phase-flips rather than bit-flips), this
code actually increases the likelihood of a phase-flip error after decoding for
small values of $p.$
To be more precise, we'll get a phase-flip error on the original qubit after
decoding whenever there are an odd number of phase-flip errors on the three
qubits of the encoding, which happens with probability
\[
3 p (1 - p)^2 + p^3.
\]
This value is larger than $p$ when $0<p<1/2,$ so the code increases the
probability of a phase-flip error for values of $p$ in this range.

\subsubsection{Modified repetition code for phase-flip errors}

We've observed that the 3-bit repetition code is completely oblivious to
phase-flip errors, so it doesn't seem to be very helpful for dealing with this
sort of error.
We can, however, modify the 3-bit repetition code in a simple way so that it
does detect phase-flip errors.
This modification will render the code oblivious to bit-flip errors --- but, as
we'll see in the next section, we can combine together the 3-bit repetition
code with this modified version to obtain the Shor code, which can correct both
bit-flip and phase-flip errors.

\begin{figure}[!ht]
  \begin{center}
       \begin{tikzpicture}[
        line width = 0.6pt,
        scale = 1.4,
        control/.style={%
          circle,
          fill=CircuitBlue,
          minimum size = 5pt,
          inner sep=0mm},
        gate/.style={%
          inner sep = 0,
          fill = CircuitBlue,
          draw = CircuitBlue,
          text = white,
          minimum size = 8.25mm},
        not/.style={%
          circle,
          fill = CircuitBlue,
          draw = CircuitBlue,
          text = white,
          minimum size = 5mm,
          inner sep=0mm,
          label = {center:\textcolor{white}{\large $+$}}
        }
      ]
      
      \node (In0) at (-2.25,0.7) {};
      \node (In1) at (-1.25,0) {};
      \node (In2) at (-1.25,-0.7) {};
      \node (Out0) at (2.5,0.7) {};
      \node (Out1) at (2.5,0) {};
      \node (Out2) at (2.5,-0.7) {};
      
      \draw (In0) -- (Out0);
      \draw (In1) -- (Out1);
      \draw (In2) -- (Out2);
      
      \node[control] (Control1) at (-0.25,0.7) {};
      \node[control] (Control2) at (0.5,0.7) {};
      
      \node[not] (Not1) at (-0.25,0) {};
      \node[not] (Not2) at (0.5,-0.7) {};
      
      \draw[very thick,draw=CircuitBlue] (Control1.center) -- (Not1.north);
      \draw[very thick,draw=CircuitBlue] (Control2.center) -- (Not2.north);

      \node[gate] at (1.5,0.7) {$H$};
      \node[gate] at (1.5,0) {$H$};
      \node[gate] at (1.5,-0.7) {$H$};
      
      \node[anchor = east] at (In0) {$\alpha\ket{0} + \beta\ket{1}$};
      \node[anchor = east] at (In1) {$\ket{0}$};
      \node[anchor = east] at (In2) {$\ket{0}$};
      
      \node[anchor = west] at (Out1) {%
        $\left.\rule{0mm}{13mm}\right\}\;
        \alpha\ket{+++} + \beta\ket{---}$};

    \end{tikzpicture}

  \end{center}
  \caption{An encoding circuit for the modified $3$-bit repetition code
    for phase-flip errors.}
  \label{fig:repetition-encoder-modified}
\end{figure}
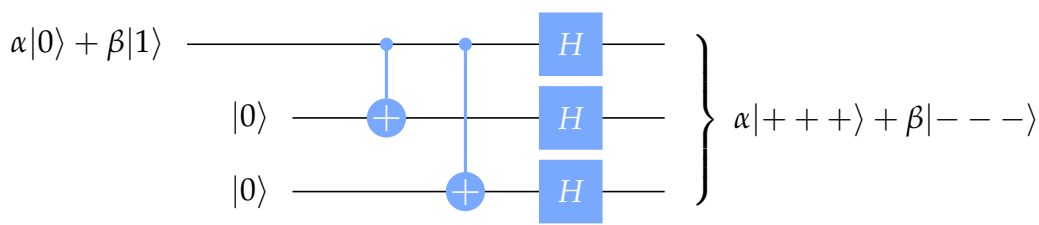

Figure~\ref{fig:repetition-encoder-modified} shows a modified version of the
encoding circuit from above, which will now be able to protect against
phase-flip errors.
The modification is very simple: we simply apply a Hadamard gate to each qubit
after performing the two controlled-NOT gates.

A Hadamard gate transforms a $\vert 0\rangle$ state into a $\vert + \rangle$
state, and a $\vert 1\rangle$ state into a $\vert - \rangle$ state, so the net
effect is that the single qubit state $\alpha\vert 0\rangle + \beta \vert
1\rangle$ is encoded as
\[
\alpha \vert {+}\,{+}\,{+} \rangle + \beta \vert {-}\,{-}\,{-} \rangle
\]
where $\vert {+}\,{+}\,{+} \rangle = \vert + \rangle \otimes \vert + \rangle
\otimes\vert + \rangle$ and
$\vert {-}\,{-}\,{-} \rangle = \vert - \rangle \otimes \vert - \rangle
\otimes\vert - \rangle.$

\begin{figure}[t]
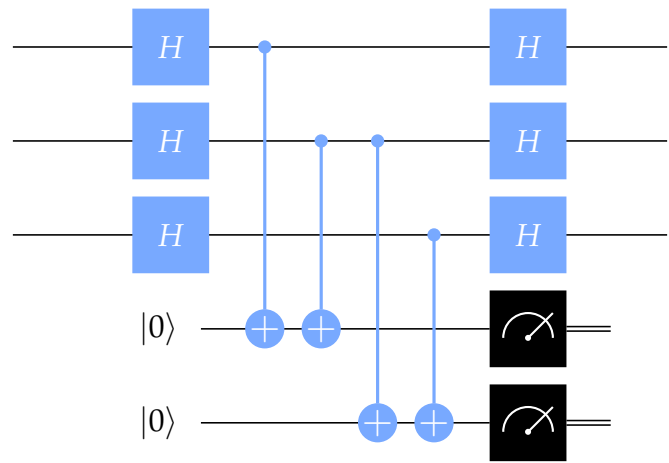

  \begin{center}

   
  \end{center}
  \caption{An error detection circuit for the modified $3$-bit repetition
    code for phase-flip errors.}
  \label{fig:three-bit-modified-error-detect}
\end{figure}

\begin{figure}[t]
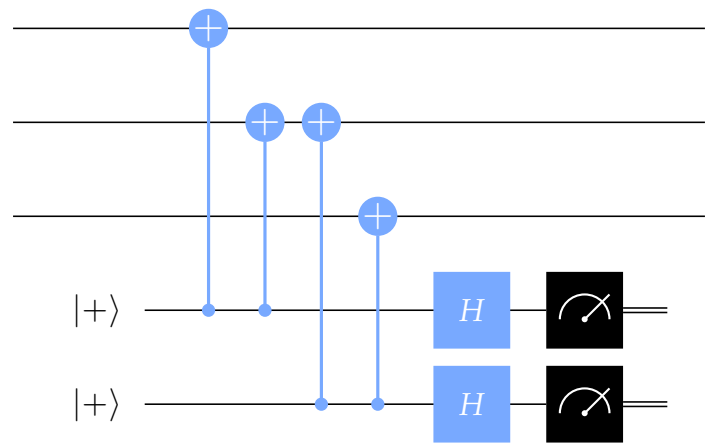

  \begin{center}
    %
  \end{center}
  \caption{A simplification of the error detection circuit for the modified
    $3$-bit repetition code for phase-flip errors shown in
    Figure~\ref{fig:three-bit-modified-error-detect}.}
  \label{fig:simplified-phase-error-detect}
\end{figure}

A phase-flip error, or equivalently a $Z$ gate, flips between the states $\vert
+ \rangle$ and $\vert - \rangle,$ so this encoding will be useful for detecting
(and correcting) phase-flip errors.
Specifically, the error-detection circuit from earlier can be modified as
in Figure~\ref{fig:three-bit-modified-error-detect}.
In words, we take the circuit from before and simply put Hadamard gates on the
top three qubits at both the beginning and the end.
The idea is that the first three Hadamard gates transform $\vert + \rangle$ and
$\vert - \rangle$ states back into $\vert 0\rangle$ and $\vert 1\rangle$
states, the same parity checks as before take place, and then the second layer
of Hadamard gates transforms the state back to $\vert + \rangle$ and $\vert -
\rangle$ states so that we recover our encoding.
For future reference, let's observe that this phase-flip detection circuit can
be simplified as is shown in Figure~\ref{fig:simplified-phase-error-detect}.

\begin{figure}[!ht]
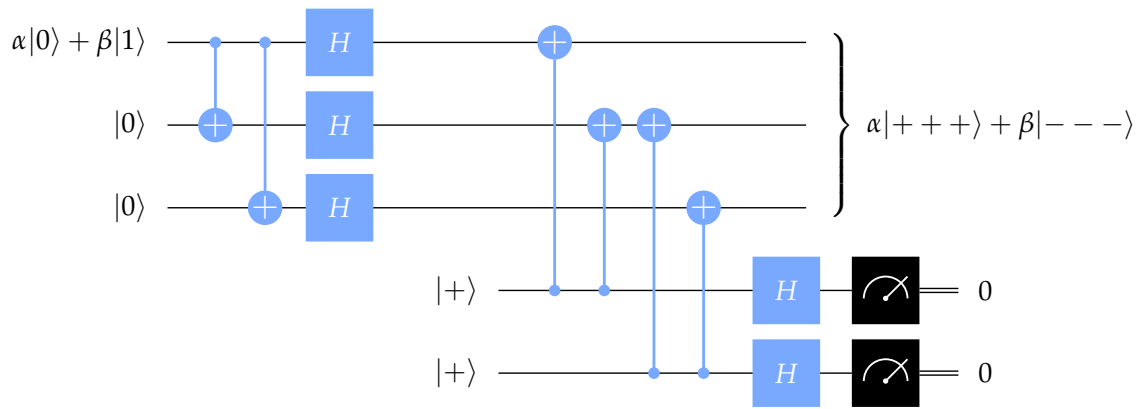

  \begin{center}
    \scalebox{0.88}{%
%
    }
  \end{center}
  \caption{If no errors occur, the error detection circuit results in the
    outcome $00$ and the encoded state is unchanged.}
  \label{fig:no-phase-error}
\end{figure}

Figures~\ref{fig:no-phase-error} and \ref{fig:phase-repetition-decode-errors}
describe how our modified version of the 3-bit repetition code, including the
encoding step and the error detection step, functions when at most one
phase-flip error occurs.
The behavior is similar to the ordinary 3-bit repetition code for bit-flips.

\begin{figure}[p]
  \begin{center}
    \scalebox{0.88}{%
      %
%
    }
  \end{center}
  \caption{A single phase-flip error is detected by the modified $3$-bit
    repetition code, with the measurement outcomes revealing which qubit was
    affected.}
  \label{fig:phase-repetition-decode-errors}
\end{figure}%
Here's an analogous table to the one from above, this time considering the
possibility of at most one phase-flip error.
\nopagebreak
\begin{center}
  \begin{tabular}{c@{\hspace{4mm}}c@{\hspace{4mm}}c}
    State & Syndrome & Correction \\\hline
    $\alpha\vert {+}\,{+}\,{+} \rangle + \beta \vert {-}\,{-}\,{-}\rangle$
    & $00$ & $\mathbb{I}\otimes\mathbb{I}\otimes\mathbb{I}$ \\
    $\alpha\vert {+}\,{+}\,{-}\rangle + \beta \vert {-}\,{-}\,{+}\rangle$
    & $01$ & $\mathbb{I}\otimes\mathbb{I}\otimes Z$ \\
    $\alpha\vert {+}\,{-}\,{+}\rangle + \beta \vert {-}\,{+}\,{-}\rangle$
    & $11$ & $\mathbb{I}\otimes Z\otimes\mathbb{I}$ \\
    $\alpha\vert {-}\,{+}\,{+} \rangle + \beta \vert {+}\,{-}\,{-}\rangle$
    & $10$ & $Z\otimes\mathbb{I}\otimes\mathbb{I}$
  \end{tabular}
\end{center}
 
Unfortunately, this modified version of the 3-bit repetition code can now no
longer correct bit-flip errors.
All is not lost, however.
As suggested previously, we'll be able to combine the two codes we've just seen
into one code --- the 9-qubit Shor code --- that can correct both bit-flip and
phase-flip errors, and indeed any error on a single qubit.

\section{The 9-qubit Shor code}

Now we turn to the 9-qubit Shor code, which is a quantum error correcting code
obtained by combining together the two codes considered in the previous
section: the 3-bit repetition code for qubits, which allows for the correction
of a single bit-flip error, and the modified version of that code, which allows
for the correction of a single phase-flip error.

\subsection{Code description}

The 9-qubit Shor code is the code we obtain by \emph{concatenating} the two
codes from the previous section.
This means that we first apply one encoding, which encodes one qubit into
three, and then we apply the other encoding to \emph{each} of the three qubits
used for the first encoding, resulting in nine qubits in total.

\begin{figure}[!ht]
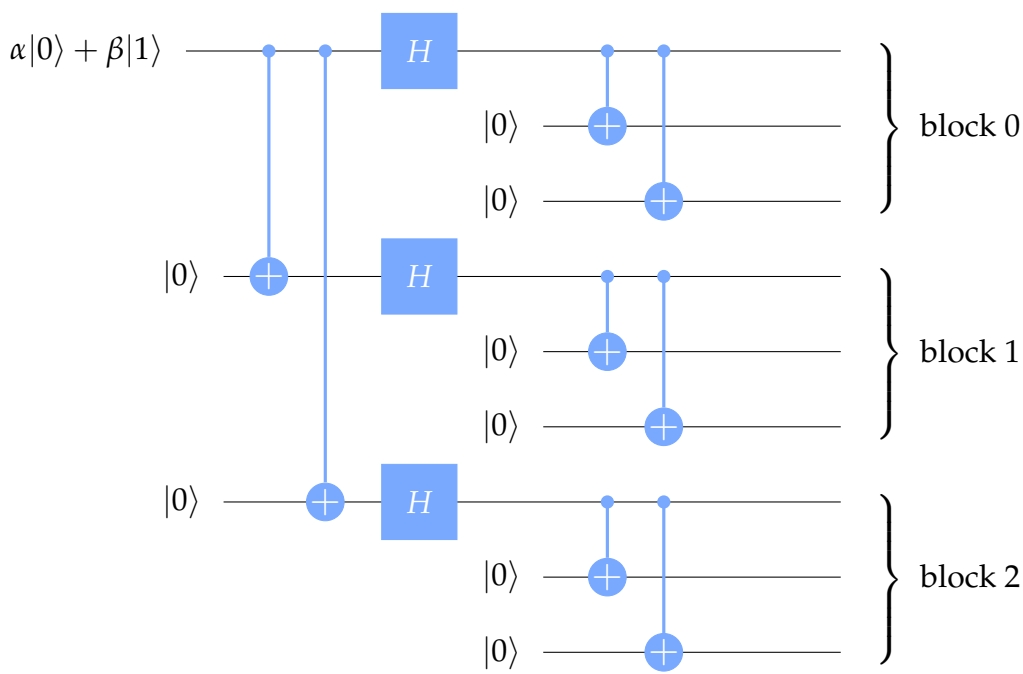

  \begin{center}

    
  \end{center}
  \caption{An encoding circuit for the $9$-qubit Shor code.}
  \label{fig:Shor-encoder}
\end{figure}

To be more precise, while we could apply the two codes in either order in this
particular case, we'll make the choice to first apply the modified version of
the 3-bit repetition code (which detects phase-flip errors), and then we'll
encode \emph{each} of the resulting three qubits independently using the
original 3-bit repetition code (which detects bit-flip errors).
Figure~\ref{fig:Shor-encoder} shows a circuit diagram representation of this
encoding.

As the figure suggests, we'll think about the nine qubits of the Shor code as
being grouped into three blocks of three qubits, where each block is obtained
from the \emph{second} encoding step (which is the ordinary 3-bit repetition
code).
The ordinary 3-bit repetition code, which here is applied three times
independently, is called the \emph{inner code} in this context, whereas the
\emph{outer code} is the code used for the first encoding step, which is the
modified version of the 3-bit repetition code that detects phase-flip errors.

We can alternatively specify the code by describing how the two standard basis
states for our original qubit get encoded.
\[
\begin{aligned}
  \vert 0\rangle &
  \:\mapsto\:
  \frac{1}{2\sqrt{2}}
  (\vert 000\rangle + \vert 111\rangle) \otimes
  (\vert 000\rangle + \vert 111\rangle) \otimes
  (\vert 000\rangle + \vert 111\rangle) \\[4mm]
  \vert 1\rangle &
  \:\mapsto\:
  \frac{1}{2\sqrt{2}}
  (\vert 000\rangle - \vert 111\rangle) \otimes
  (\vert 000\rangle - \vert 111\rangle) \otimes
  (\vert 000\rangle - \vert 111\rangle)
\end{aligned}
\]
Once we know this, we can determine by linearity how an arbitrary qubit state
vector is encoded.

\subsection{Correcting bit-flip and phase-flip errors}

\subsubsection{Errors and CNOT gates}

To analyze how $X$ and $Z$ errors affect encodings of qubits, both for the
9-qubit Shor code as well as other codes, it will be helpful to observe a few
simple relationships between these errors and CNOT gates.
As we begin to analyze the 9-qubit Shor code, this is a reasonable moment to
pause to do this.

\begin{figure}[p]
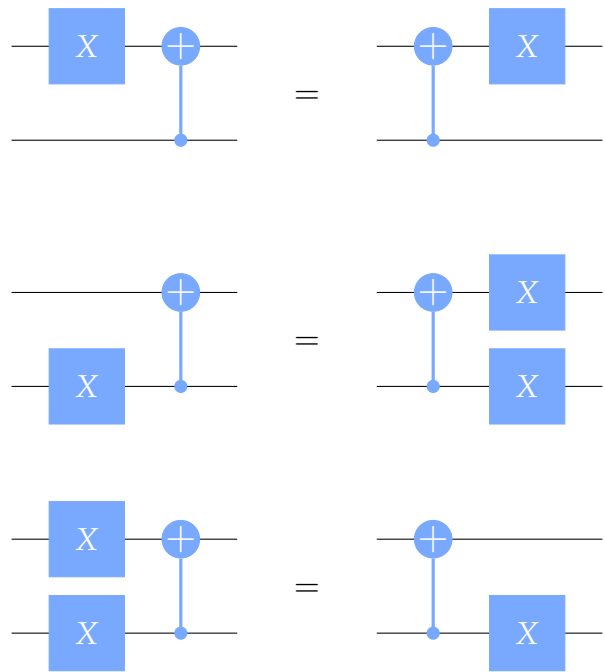

  \begin{center}

    
  \end{center}
  \caption{Relationships among $X$ and CNOT gates.}
  \label{fig:X-and-CNOT}
\end{figure}

Figure~\ref{fig:X-and-CNOT} illustrate three basic relationships among $X$
gates and CNOT gates.
Specifically, applying an $X$ gate to the \emph{target} qubit prior to a CNOT
is equivalent to swapping the order and performing the CNOT first, but applying
an $X$ gate to the \emph{control} qubit prior to a CNOT is equivalent to
applying $X$ gates to both qubits after the CNOT.
Finally, applying $X$ gates to both qubits prior to a CNOT is equivalent to
applying the CNOT first and then applying an $X$ gate to the control qubit.
These relationships can be verified by performing the required matrix
multiplications or computing the effect of the circuits on standard basis
states.

\begin{figure}[p]
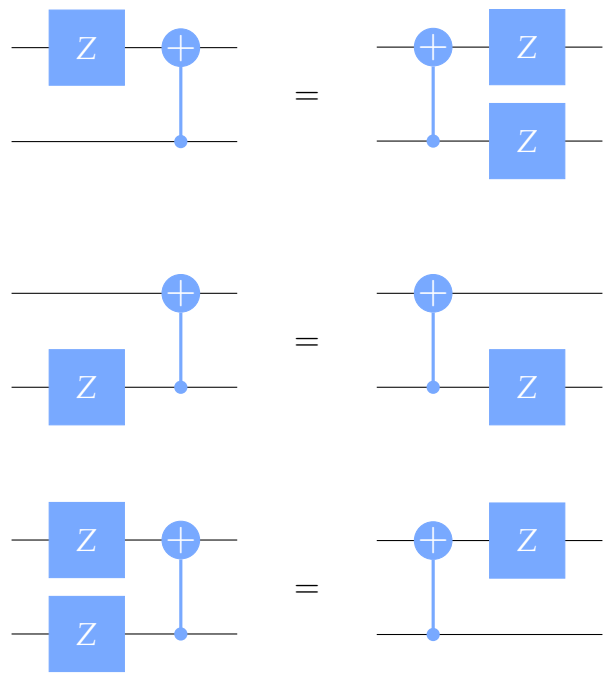

  \vspace{3mm} 
  
  \begin{center}
    %
       
  \end{center}
  \caption{Relationships among $Z$ and CNOT gates.}
  \label{fig:Z-and-CNOT}
\end{figure}

The situation is similar for $Z$ gates, except that the roles of the control
and target qubits switch.
In particular, we have the three relationships depicted by
Figure~\ref{fig:Z-and-CNOT}.

\subsubsection{Correcting bit-flip errors}

Now we'll consider how errors can be detected and corrected using the 9-qubit
Shor code, starting with bit-flip errors --- which we'll generally refer to as
$X$ errors hereafter for the sake of brevity.

To detect and correct $X$ errors, we can simply treat each of the three blocks
in the encoding separately.
Each block is an encoding of a qubit using the 3-bit repetition code, which
protects against $X$ errors --- so by performing the syndrome measurements and
$X$ error corrections described previously to each block, we can detect and
correct up to one $X$ error per block.
In particular, if there is at most one $X$ error on the nine qubits of the
encoding, this error will be detected and corrected by this procedure.
In short, correcting bit-flip errors is a simple matter for this code, due to
the fact that the inner code corrects bit-flip errors.

\subsubsection{Correcting phase-flip errors}

Next we'll consider phase-flip errors, or $Z$ errors for brevity.
This time it's not quite as clear what we should do because the outer code is
the one that detects $Z$ errors, but the inner code seems to be somehow ``in
the way,'' making the detection and correction of these errors slightly more
difficult.

\begin{figure}[p]
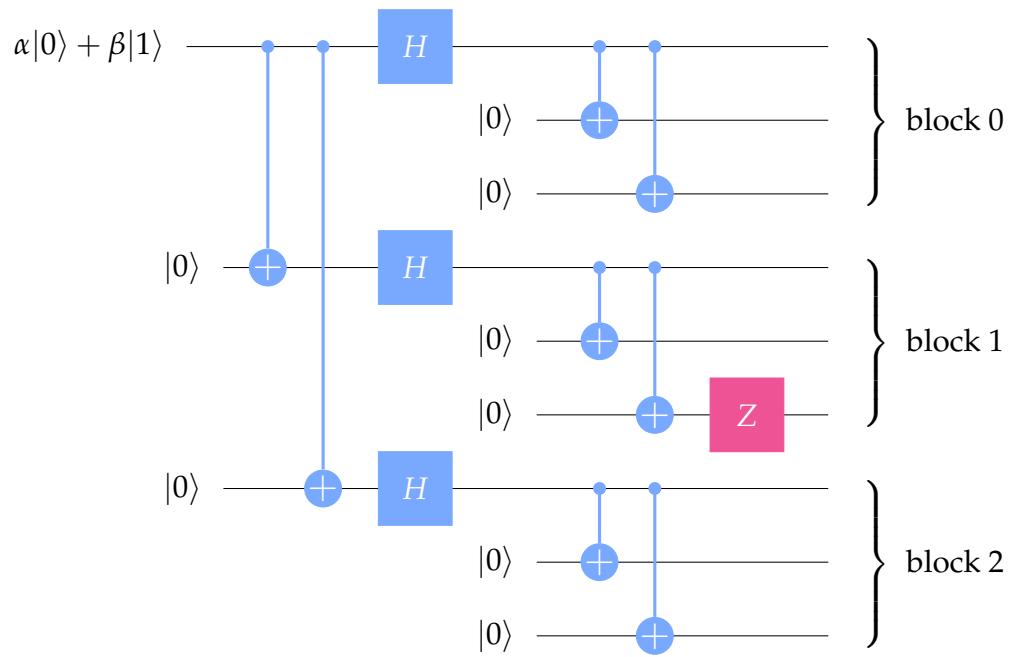

  \begin{center}
    \scalebox{0.98}{%
%
    }

  \end{center}
  \caption{A phase-flip error on one of the qubits of the $9$-qubit Shor code.}
  \label{fig:Shor-phase-error-1}
\end{figure}

\begin{figure}[p]
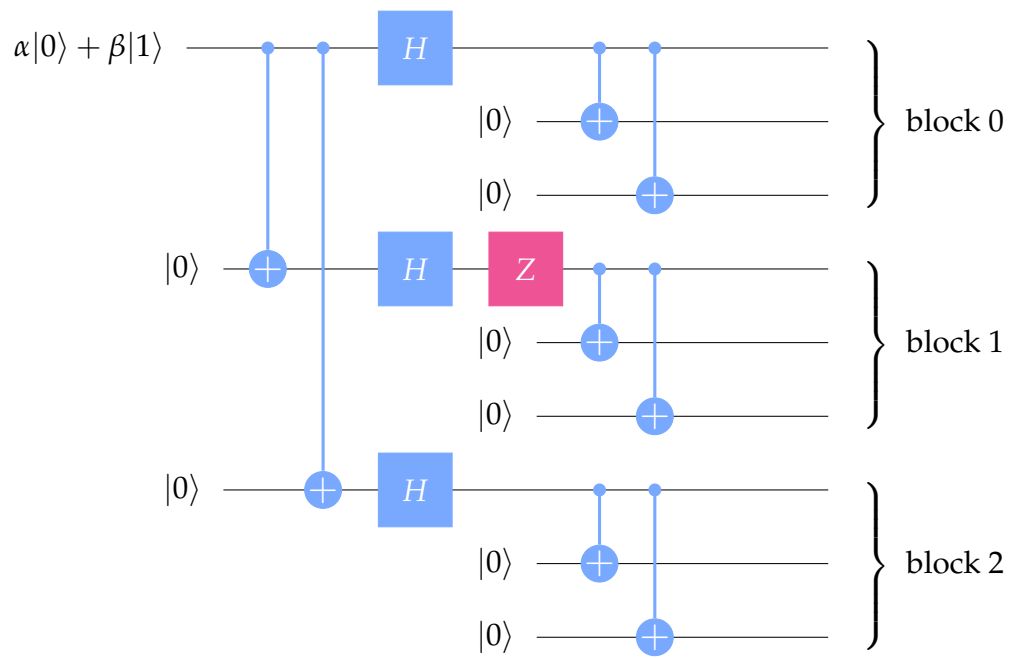

  \begin{center}
    \scalebox{0.98}{%
    %
%
    }
  \end{center}
  \caption{A phase-flip error within the middle block, such as the one
    indicated in Figure~\ref{fig:Shor-phase-error-1}, is equivalent to
    one on the middle qubit prior to the inner encoding.}
  \label{fig:Shor-phase-error-2}
\end{figure}

Suppose that a $Z$ error occurs on one of the 9 qubits of the Shor code, such
as the one indicated in Figure~\ref{fig:Shor-phase-error-1}.
We've already observed what happens when a $Z$ error occurs when we're using
the 3-bit repetition code --- it's equivalent to a $Z$ error occurring prior to
encoding.
In the context of the 9-qubit Shor code, this means that a $Z$ error on any one
of the three qubits within a block always has the same effect, which is
equivalent to a $Z$ error occurring on the corresponding qubit prior to the
inner code being applied.
For example, the error in Figure~\ref{fig:Shor-phase-error-1} is equivalent to
the one suggested in Figure~\ref{fig:Shor-phase-error-2}.
This can be reasoned using the relationships between $Z$ and CNOT gates
described above, or by simply evaluating the circuits on an arbitrary qubit
state $\alpha \vert 0\rangle + \beta \vert 1\rangle.$

This suggests one option for detecting and correcting $Z$ errors, which is to
\emph{decode} the inner code, leaving us with the three qubits used for the
outer encoding along with six initialized workspace qubits.
We can then check these three qubits of the outer code for $Z$ errors, and then
finally we can re-encode using the inner code, to bring us back to the 9-qubit
encoding we get from the Shor code.
If we do detect a $Z$ error, we can either correct it prior to re-encoding with
the inner code, or we can correct it after re-encoding, by applying a $Z$ gate
to any of the qubits in that block.

\begin{figure}[t]
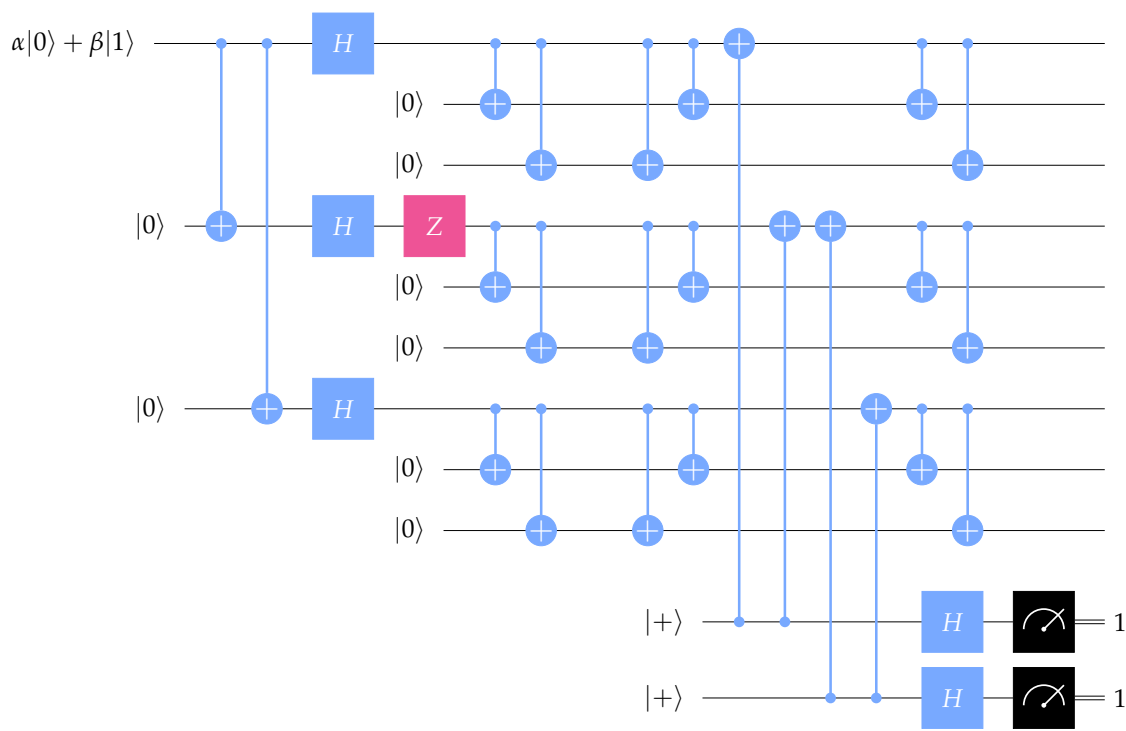

  \begin{center}
    \scalebox{0.81}{%
%
    }
  \end{center}
  \caption{To detect phase-flip errors, we can decode the inner code,
    run the error detection circuit on the three qubits of the outer code,
    and then re-encode the inner code.}
  \label{fig:Shor-phase-error-detect-1}
\end{figure}

\begin{figure}[t]
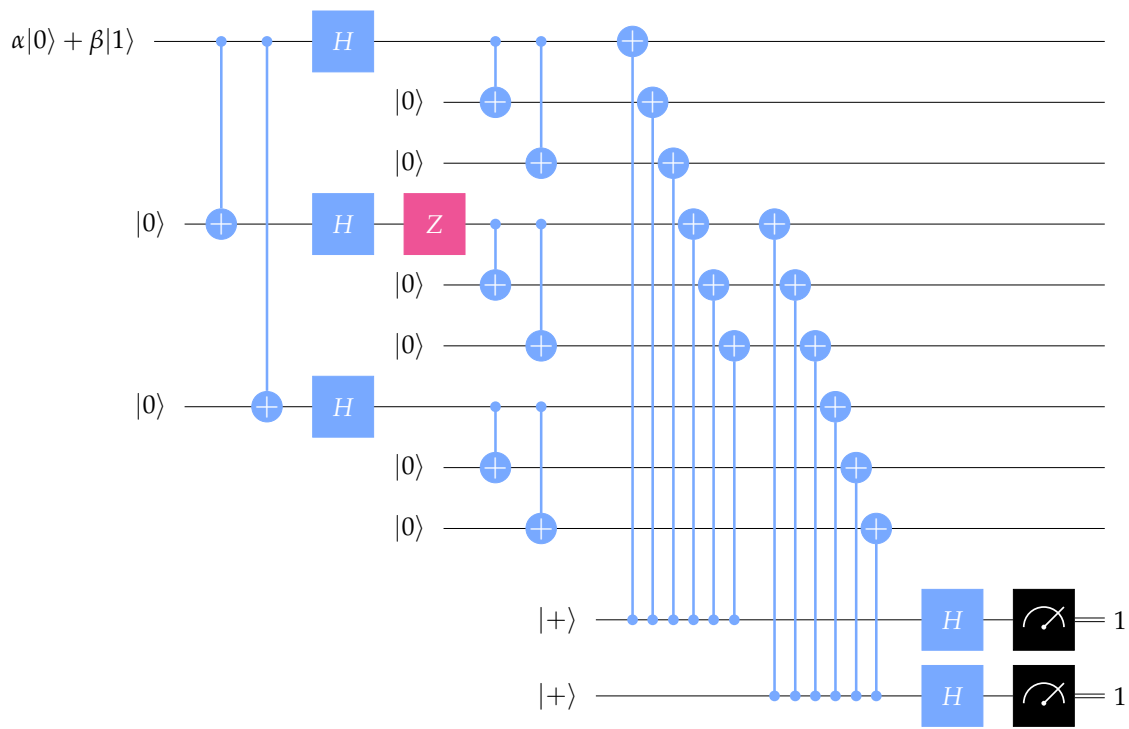

  \begin{center}
    \scalebox{0.81}{%
%
    }
   
  \end{center}
  \caption{A simplification of the circuit in
    Figure~\ref{fig:Shor-phase-error-detect-1} using fewer CNOT gates.}
  \label{fig:Shor-phase-error-detect-2}
\end{figure}

Figure~\ref{fig:Shor-phase-error-detect-1} is a circuit diagram that includes
the encoding circuit and the error suggested above together with the steps just
described (but not the actual correction step).
In this particular example, the syndrome measurement is $11,$ which locates the
$Z$ error as having occurred on one of the qubits in the middle block.
An advantage of correcting $Z$ errors after the re-encoding step rather than
before is that we can simplify the circuit above.
The circuit is Figure~\ref{fig:Shor-phase-error-detect-2} equivalent, but
requires four fewer CNOT gates.
Again, the syndrome doesn't indicate which qubit has been affected by a
$Z$ error, but rather which block has experienced a $Z$ error, with the effect
being the same regardless of which qubit within the block was affected.
We can then correct the error by applying a $Z$ gate to any of the three qubits
of the affected block.

As an aside, here we see an example of \emph{degeneracy} in a quantum
error-correcting code, where we're able to correct certain errors without being
able to identify them uniquely.

\subsubsection{Simultaneous bit- and phase-flip errors}

We've now seen how both $X$ and $Z$ errors can be detected and corrected using
the 9-qubit Shor code, and in particular how at most one $X$ error or at most
one $Z$ error can be detected and corrected.
Now let's suppose that both a bit-flip and a phase-flip error occur, possibly
on the same qubit.
As it turns out, nothing different needs to be done in this situation from what
has already been discussed ---
the code is able to detect and correct up to one $X$ error and one $Z$ error
simultaneously, without further modification.

To be more specific, $X$ errors are detected by applying the ordinary 3-bit
repetition code syndrome measurement, which is performed separately on each of
the three blocks of three qubits; and $Z$ errors are detected through the
procedure described just above, which is equivalent to decoding the inner code,
performing the syndrome measurement for the modified 3-bit repetition code for
phase-flips, and then re-encoding.
These two error detection steps --- as well as the corresponding corrections
--- can be performed completely independently of one another, and in fact it
doesn't matter in which order they're performed.

\begin{figure}[t]
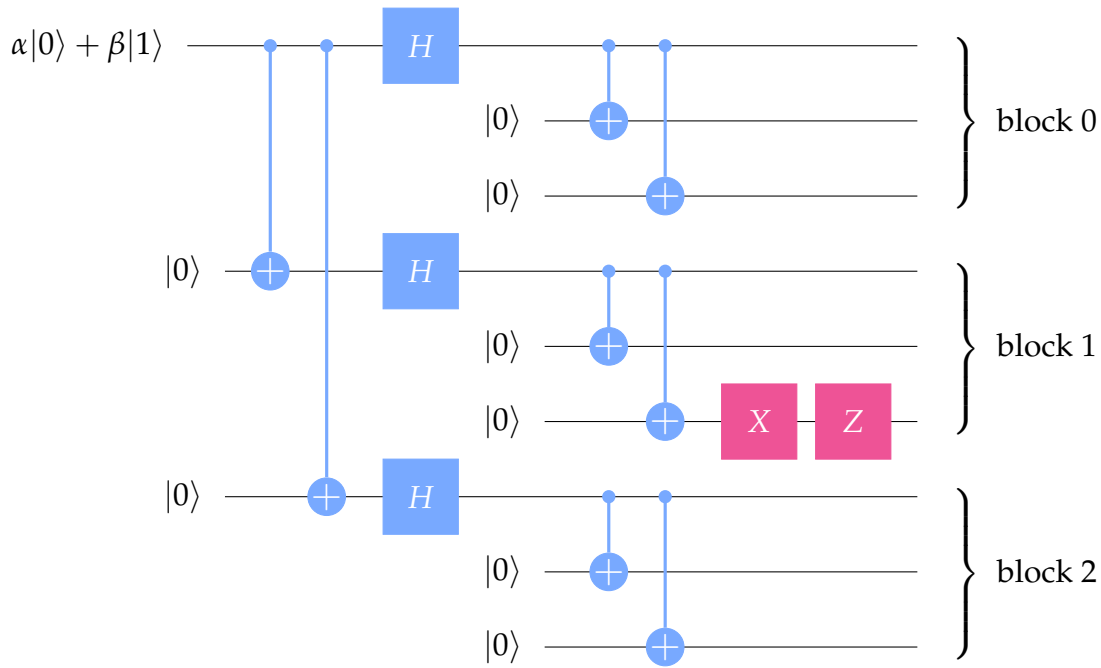

  \begin{center}

   
  \end{center}
  \caption{A bit-flip error and a phase-flip error on the same qubit
    in the $9$-qubit Shor code.}
  \label{fig:Shor-XZ-error}
\end{figure}

\begin{figure}[t]
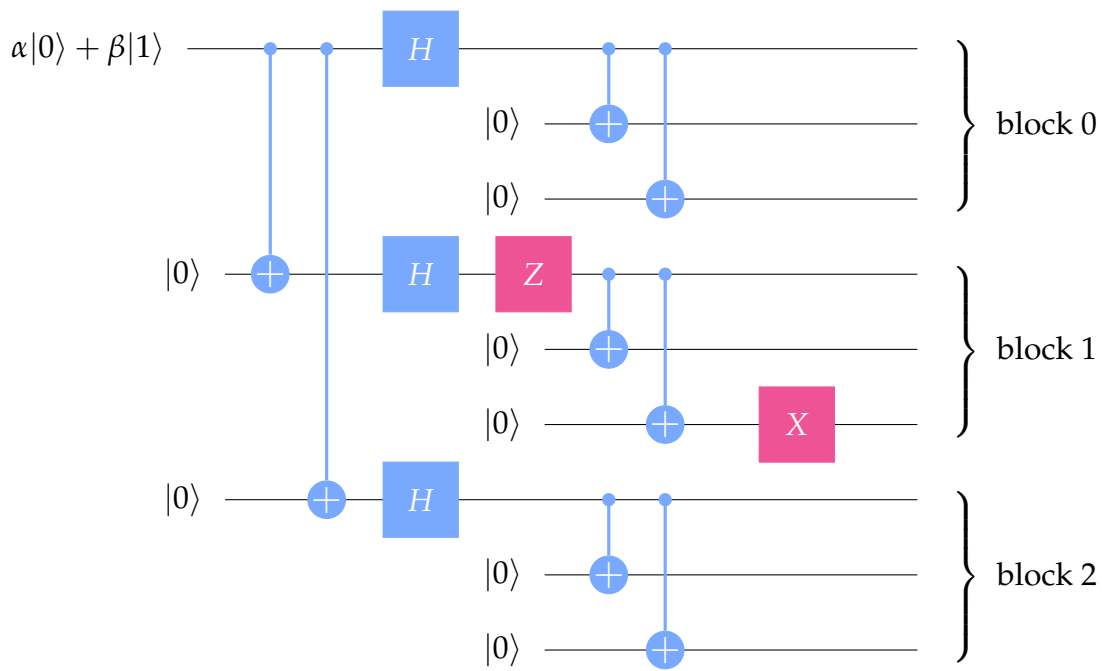

  \begin{center}
    %
   
  \end{center}
  \caption{An equivalent circuit to the one in Figure~\ref{fig:Shor-XZ-error}
    up to a $-1$ global phase factor.}
  \label{fig:Shor-ZX-error}
\end{figure}

To see why this is, consider the example depicted in the circuit diagram
in Figure~\ref{fig:Shor-XZ-error}, where both an $X$ and a $Z$ error have
affected the bottom qubit of the middle block.
Let's first observe that the ordering of the errors doesn't matter, in the
sense that reversing the position of the $X$ and $Z$ errors yields an
equivalent circuit.
To be clear, $X$ and $Z$ do not commute, they \emph{anti-commute}:
\[
XZ =
%
= -ZX.
\]
This implies that changing the ordering leads to an irrelevant $-1$ global
phase factor.
We can then move the $Z$ error just like before to obtain another equivalent
circuit, shown in Figure~\ref{fig:Shor-ZX-error}, which is equivalent to the
one above up to a global phase factor.

At this point it's evident that if the procedure to detect and correct $X$
errors is performed first, the $X$ error will be corrected, after which the
procedure for detecting and correcting $Z$ errors can be performed to eliminate
the $Z$ error as before.

\pagebreak

Alternatively, the procedure to detect and correct $Z$ errors can be performed
first.
The fact that this procedure works as expected, even in the presence of one or
more $X$ errors, follows from the fact that $X$ gates on any of the nine qubits
used for the encoding commute with all of the gates in our simplified circuit
for measuring the syndrome for $Z$ errors.
Thus, this syndrome measurement will still correctly identify which block has
been affected by a $Z$ error.
The fact that a $Z$ error on any block is corrected by applying a $Z$ gate to
any qubit of that block, even if an $X$ error has also occurred, follows from
the same argument as above concerning the ordering of $X$ and $Z$ gates giving
us equivalent circuits up to a global phase.

It follows that the 9-qubit Shor code can correct an $X$ error, a $Z$ error, or
both, on any one of the nine qubits used for this code.
In fact, we can correct more errors than that, including multiple $X$ errors
(as long as they fall into different blocks) or multiple $Z$ errors (as long as
at most one block experiences an odd number of them) --- but going forward,
what will be most relevant for the purposes of this lesson is that we can
correct an $X$ error, a $Z$ error, or both on any one qubit.

\subsection{Error reduction for random errors}

Before we move on to the last section of the lesson, which concerns arbitrary
quantum errors, let's briefly consider the performance of the 9-qubit Shor code
when errors represented by Pauli matrices occur \emph{randomly} on the qubits.

To be more concrete, let's consider a simple noise model where errors occur
\emph{independently} on the qubits, with each qubit experiencing an error with
probability $p$, and with no correlation between errors on different qubits
--- along similar lines to a binary symmetric channel for classical bits.
We could assign different probabilities for $X,$ $Y,$ and $Z$ errors to occur,
but to keep things simple, we'll consider the worst case scenario for the
9-qubit Shor code, which is that a $Y$ error occurs on each of the affected
qubits.
A $Y$ error, by the way, is equivalent (up to an irrelevant global phase
factor) to both an $X$ and a $Z$ error occurring on the same qubit, given that
$Y = iXZ.$
This explains our apparent disregard of $Y$ errors up to this point.

Now, supposing that $\mathsf{Q}$ is a qubit in some particular state that we'd
like to protect against errors, we can consider the option to use the 9-qubit
Shor code.
A natural question to ask is, ``Should we use it?''
The answer is not necessarily ``yes.''
If there's too much noise, meaning in this context that $p$ is too large, using
the Shor code could actually make things worse --- just like the 3-bit
repetition code is worse than no code when $p$ is larger than one-half.
But, if $p$ is small enough, then the answer is ``yes,'' we should use the
code, because it will decrease the likelihood that the encoded state becomes
corrupted.
Let's see why this is, and what it means for $p$ to be too large or small
enough for this code.

The Shor code corrects any Pauli error on a single qubit, including a $Y$ error
of course, but it doesn't properly correct two or more $Y$ errors.
To be clear, we're assuming that we're using the $X$ and $Z$ error corrections
described earlier in the section.
(Of course, if we knew in advance that we only had to worry about $Y$ errors,
we would naturally choose our corrections differently --- but that's cheating
the noise model, and we'd always be able to change the model by selecting
different Pauli errors to make this new choice of corrections fail whenever two
or more qubits are affected by errors.)

So, the code protects $\mathsf{Q}$ so long as at most one of the nine qubits is
affected by an error, which happens with probability
\[
(1-p)^9 + 9 p (1-p)^8.
\]
Otherwise, with probability
\[
1 - (1-p)^9 - 9 p (1-p)^8,
\]
the code fails to protect $\mathsf{Q}.$

Specifically, what that means in this context is that, up to a global phase, a
non-identity Pauli operation will be applied to our qubit $\mathsf{Q}$ (as a
so-called \emph{logical qubit}).
That is, if $X$ and $Z$ errors are detected and corrected for the Shor code as
described earlier in the lesson, we'll be left with the encoding of a state
that's equivalent, up to a global phase, to the encoding of a non-identity
Pauli operation applied to the original state of $\mathsf{Q}.$
A more succinct way to say this is that a \emph{logical} error will have
occurred.
That may or may not have an effect on the original state of $\mathsf{Q}$ --- or
in other words the \emph{logical qubit} we've encoded with nine \emph{physical
qubits} --- but, for the sake of this analysis, we're considering this event to
mean failure.

\begin{figure}[b]
  \begin{center}
    \begin{tikzpicture}
      \begin{axis}[
          scale = 1,
          samples=\samples,
          domain = 0:0.035,
          xmin = 0,
          xmax = 0.035,
          xtick = {0.01, 0.02, 0.03, 0.04, 0.05},
          xticklabels = {0.01, 0.02, 0.03, 0.04, 0.05},
          ytick = {0.01, 0.02, 0.03, 0.04, 0.05},
          yticklabels = {0.01, 0.02, 0.03, 0.04, 0.05},
          axis lines=center,
          axis line style={-},
          x label style={%
            at={(axis description cs:.5,-0.14)},anchor=north},
          xlabel={$p$},
          y label style={at={(axis description cs:-0.18,.5)},
            rotate=90,anchor=south},                    
          ylabel={error probability},
          clip=false,
          legend style={at={(0.98,0.4)},anchor=south west},
          legend style={draw=none},
          legend cell align=left,
          scaled ticks=false
        ]

        \addplot[black] {x};

        \addplot[thick, DataColor0, domain = -0.000034:0.035] {%
          1 - ((1-x)*(1-x)*(1-x)*(1-x)*(1-x)*(1-x)*(1-x)*(1-x)*(1-x)
          + 9*(1-x)*(1-x)*(1-x)*(1-x)*(1-x)*(1-x)*(1-x)*(1-x)*x)
        };
        
        \legend{\;$p$, \;$1 - (1-p)^9 - 9p(1-p)^8$}
        \draw[densely dotted] (axis cs:0.0323,0) -- (axis cs:0.0323,0.0323);
        
      \end{axis}
    \end{tikzpicture}
  \end{center}
  \caption{A plot illustrating the break-even point for the $9$-qubit Shor
    code.}
  \label{fig:breakeven-Shor}
\end{figure}
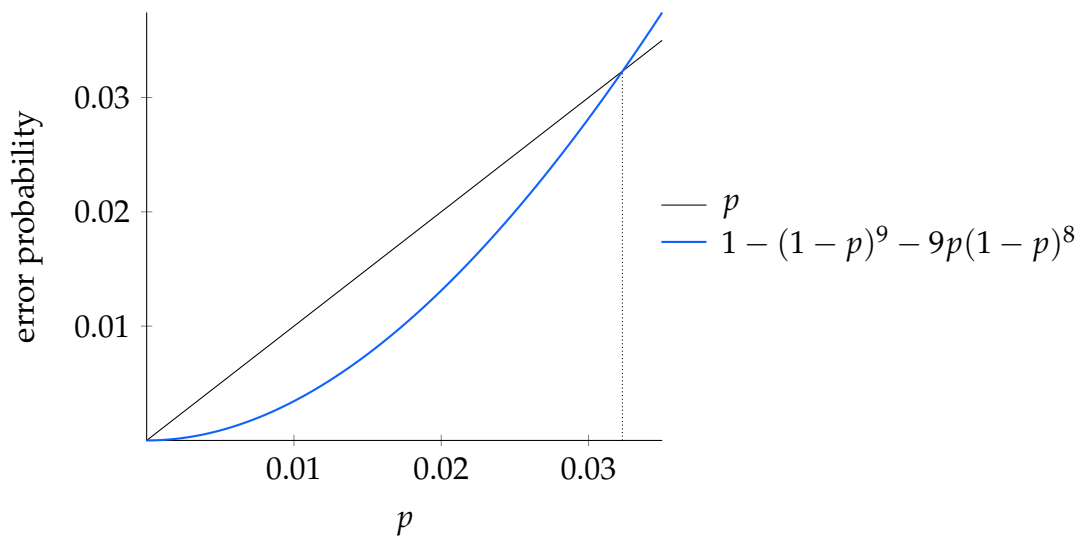

On the other hand, if we didn't bother to use the code, our one and only qubit
would suffer a similar fate (of being subject to a non-identity Pauli
operation) with probability $p.$
The code helps when the first probability is smaller than the second:
\[
1 - (1-p)^9 - 9 p (1-p)^8 < p.
\]
Figure~\ref{fig:breakeven-Shor} illustrates, for very small values of $p,$ that
the code provides an advantage, with the break-even point occurring at about
$0.0323.$

If $p$ is smaller than this break-even point, then the code helps;
at the break-even point the probabilities are equal, so we're just wasting our
time along with 8 qubits if we use the code;
and beyond the break-even point we should absolutely not be using this code
because it's \emph{increasing} the chance of a logical error on $\mathsf{Q}.$

Three and a quarter percent or so may not seem like a very good break-even
point, particularly when compared to $50\%,$ which is the analogous break-even
point for the 3-bit repetition code for classical information.
This difference is, in large part, due to the fact that quantum information is
more delicate and harder to protect than classical information.
But also --- while recognizing that the 9-qubit Shor code represents a
brilliant discovery, as the world's first quantum error correcting code --- it
should be acknowledged that it isn't actually a very good code in practical
terms.

\section{Discretization of errors}

So far we've considered $X$ errors and $Z$ errors in the context of the 9-qubit
Shor code, and in this section we'll consider arbitrary errors.
What we'll find is that, to handle such errors, we don't need to do anything
different from what we've already discussed;
the ability to correct $X$ errors, $Z$ errors, or both, implies the ability to
correct arbitrary errors.
This phenomenon is sometimes called the \emph{discretization of errors}.

\subsection{Unitary qubit errors}

Let's begin with single-qubit \emph{unitary} errors.
For example, such an error could correspond to a very small rotation of the
Bloch sphere, possibly representing an error incurred by a gate that isn't
perfect, for instance.
Or it could be any other unitary operation on a qubit and not necessarily one
that's close to the identity.

It might seem like correcting for such errors is difficult.
After all, there are infinitely many possible errors like this, and it's
inconceivable that we could somehow identify each error exactly and then undo
it.
However, as long as we can correct for a bit-flip, a phase-flip, or both, then
we will succeed in correcting an arbitrary single-qubit unitary error using the
procedures described earlier in the lesson.

To see why this is the case, let us recognize first that we can express an
arbitrary $2 \times 2$ unitary matrix $U,$ representing an error on a single
qubit, as a linear combination of the four Pauli matrices (including the
identity matrix).
\[
U = \alpha \mathbb{I} + \beta X + \gamma Y + \delta Z
\]
As we will see, when the error detection circuits are run, the measurements
that give us the syndrome bits effectively collapse the state of the encoding
probabilistically to one where an error (or lack of an error) represented by
one of the four Pauli matrices has taken place.
(It follows from the fact that $U$ is unitary that the numbers $\alpha,$
$\beta,$ $\gamma,$ and $\delta$ must satisfy $\vert\alpha\vert^2 +
\vert\beta\vert^2 + \vert\gamma\vert^2 + \vert\delta\vert^2 = 1,$
and indeed, the values $\vert\alpha\vert^2,$ $\vert\beta\vert^2,$
$\vert\gamma\vert^2,$ and $\vert\delta\vert^2$ are the probabilities with which
the encoded state collapses to one for which the corresponding Pauli error has
occurred.)

To explain how this works in greater detail, it will be convenient to use
subscripts to indicate which qubit a given qubit unitary operation acts upon.
For example, using Qiskit's qubit numbering convention
$(\mathsf{Q}_8,\mathsf{Q}_7,\ldots,\mathsf{Q}_0)$ to number the 9 qubits used
for the Shor code, we have these expressions for various unitary operations on
single qubits, where in each case we tensor the unitary matrix with the
identity matrix on every other qubit.
\[
\begin{aligned}
  X_0 & = \mathbb{I} \otimes \mathbb{I} \otimes \mathbb{I} \otimes \mathbb{I}
  \otimes \mathbb{I} \otimes \mathbb{I} \otimes \mathbb{I} \otimes \mathbb{I}
  \otimes X \\[1.5mm]
  Z_4 & = \mathbb{I} \otimes \mathbb{I} \otimes \mathbb{I} \otimes \mathbb{I}
  \otimes Z \otimes \mathbb{I} \otimes \mathbb{I} \otimes \mathbb{I} \otimes
  \mathbb{I} \\[1.5mm]
  U_7 & = \mathbb{I} \otimes U \otimes \mathbb{I} \otimes \mathbb{I} \otimes
  \mathbb{I} \otimes \mathbb{I} \otimes \mathbb{I} \otimes \mathbb{I} \otimes
  \mathbb{I}
\end{aligned}
\]
So, in particular, for a given qubit unitary operation $U,$ we can specify the
action of $U$ applied to qubit $k$ by the following formula, which is similar
to the one before except that each matrix represents an operation applied to
qubit $k.$
\[
U_k = \alpha \mathbb{I}_k + \beta X_k + \gamma Y_k + \delta Z_k
\]

Now suppose that $\vert\psi\rangle$ is the 9-qubit encoding of a qubit state.
If the error $U$ takes place on qubit $k,$ we obtain the state $U_k
\vert\psi\rangle,$ which can be expressed as a linear combination of Pauli
operations acting on $\vert\psi\rangle$ as follows.
\[
U_k \vert\psi\rangle = \alpha \vert\psi\rangle + \beta X_k\vert\psi\rangle +
\gamma Y_k\vert\psi\rangle + \delta Z_k\vert\psi\rangle
\]
At this point let's make the substitution $Y = iXZ.$
\[
U_k \vert\psi\rangle = \alpha \vert\psi\rangle + \beta X_k\vert\psi\rangle + i
\gamma X_kZ_k\vert\psi\rangle + \delta Z_k\vert\psi\rangle
\]

Now consider the error-detection and correction steps described previously.
We can think about the measurement outcomes for the three inner code parity
checks along with the one for the outer code collectively as a single syndrome
consisting of 8 bits.
Just prior to the actual standard basis measurements that produce these
syndrome bits, the state has the following form.
\[
\begin{gathered}
  \alpha\,\vert \mathbb{I} \text{ syndrome}\rangle \otimes \vert\psi\rangle \\
  + \beta\,\vert X_k \text{ syndrome}\rangle \otimes X_k\vert\psi\rangle \\
  + i \gamma\,\vert X_k Z_k \text{ syndrome}\rangle \otimes X_k
  Z_k\vert\psi\rangle \\
  + \delta\,\vert Z_k \text{ syndrome}\rangle \otimes Z_k\vert\psi\rangle
\end{gathered}
\]

To be clear, we have two systems at this point.
The system on the left is the 8 qubits we'll measure to get the syndrome, where
$\vert \mathbb{I} \text{ syndrome}\rangle,$
$\vert X_k \text{ syndrome}\rangle,$ and so on, refer to whatever 8-qubit
standard basis state is consistent with the corresponding error (or non-error).
The system on the right is the 9 qubits we're using for the encoding.

Notice that these two systems are now correlated (in general), and this is the
key to why this works.
By measuring the syndrome, the state of the 9 qubits on the right effectively
collapses to one in which a Pauli error consistent with the measured syndrome
has been applied to one of the qubits.
Moreover, the syndrome itself provides enough information so that we can undo
the error and recover the original encoding $\vert\psi\rangle.$

In particular, if the syndrome qubits are measured and the appropriate
corrections are made, we obtain a state that can be expressed as a density
matrix,
\[
\xi \otimes \vert\psi\rangle\langle\psi\vert,
\]
where
\[
\begin{aligned}
  \xi = & \vert\alpha\vert^2 \vert \mathbb{I} \text{ syndrome}\rangle\langle
  \mathbb{I} \text{ syndrome}\vert \\[1mm]
  & + \vert\beta\vert^2 \vert X_k \text{ syndrome}\rangle\langle X_k \text{
    syndrome}\vert\\[1mm]
  & + \vert\gamma\vert^2 \vert X_k Z_k \text{ syndrome}\rangle\langle X_k Z_k
  \text{ syndrome}\vert\\[1mm]
  & + \vert\delta\vert^2 \vert Z_k \text{ syndrome}\rangle\langle Z_k \text{
    syndrome}\vert.
\end{aligned}
\]
Critically, this is a product state:
we have our original, uncorrupted encoding as the right-hand tensor factor,
and on the left we have a density matrix $\xi$ that describes a random error
syndrome.
There is no longer any correlation with the system on the right, which is the
one we care about, because the errors have been corrected.

At this point we can throw the syndrome qubits away or reset them so we can use
them again.
This is how the randomness --- or \emph{entropy} --- created by errors is
removed from the system.

This is the discretization of errors for the special case of unitary errors.
In essence, by measuring the syndrome, we effectively \emph{project} the error
onto an error that's described by a Pauli matrix.

At first glance it may seem too good to be true that we can correct for
arbitrary unitary errors like this, even errors that are tiny and hardly
noticeable on their own.
But, what's important to realize here is that this is a unitary error on a
\emph{single} qubit, and by the design of the code, a single-qubit operation
can't change the state of the logical qubit that's been encoded.
All it can possibly do is to move the state out of the subspace of valid
encodings, but then the error detections collapse the state and the corrections
bring it back to where it started.

\subsection{Arbitrary qubit errors}

Finally, let's consider arbitrary errors that are not necessarily unitary.
To be precise, we'll consider an error described by an arbitrary qubit channel
$\Phi.$
For example, this could be a dephasing or depolarizing channel, a reset
channel, or a strange channel that we've never thought about before.

The first step is to consider any Kraus representation of $\Phi.$
\[
\Phi(\sigma) = \sum_j A_j \sigma A_j^{\dagger}
\]
This is a qubit channel, so each $A_j$ is a $2\times 2$ matrix, which we can
express as a linear combination of Pauli matrices.
\[
A_j = \alpha_j \mathbb{I} + \beta_j X + \gamma_j Y + \delta_j Z
\]
This allows us to express the action of the error $\Phi$ on a chosen qubit $k$
in terms of Pauli matrices as follows.
\[
\Phi_k \bigl( \vert\psi\rangle\langle\psi\vert\bigr) =
\sum_j (\alpha_j \mathbb{I}_k + \beta_j X_k + \gamma_j Y_k + \delta_j Z_k)
\vert\psi\rangle\langle\psi\vert
(\alpha_j \mathbb{I}_k + \beta_j X_k + \gamma_j Y_k + \delta_j Z_k)^{\dagger}
\]
In short, we've simply expanded out all of our Kraus matrices as linear
combinations of Pauli matrices.

If we now compute and measure the error syndrome, and correct for any errors
that are revealed, we'll obtain a similar sort of state to what we had in the
case of a unitary error:
\[
\xi \otimes \vert\psi\rangle\langle\psi\vert,
\]
where this time we have
\[
\begin{aligned}
  \xi = & \sum_j \Bigl(\vert\alpha_j\vert^2 \vert \mathbb{I} \text{
    syndrome}\rangle\langle \mathbb{I} \text{ syndrome}\vert \\[-3mm]
  & \qquad + \vert\beta_j\vert^2 \vert X_k \text{ syndrome}\rangle\langle X_k
  \text{ syndrome}\vert\\[2mm]
& \qquad + \vert\gamma_j\vert^2 \vert X_k Z_k \text{ syndrome}\rangle\langle
  X_k Z_k \text{ syndrome}\vert\\[2mm]
  & \qquad + \vert\delta_j\vert^2 \vert Z_k \text{ syndrome}\rangle\langle Z_k
  \text{ syndrome}\vert \Bigr).
\end{aligned}
\]
The details are a bit messier and are not shown here.
Conceptually speaking, the idea is identical to the unitary case.

\subsection{Generalization}

The discretization of errors generalizes to other quantum error-correcting
codes, including ones that can detect and correct errors on multiple qubits.
In such cases, errors on multiple qubits can be expressed as \emph{tensor
products} of Pauli matrices, and correspondingly different syndromes specify
Pauli operation corrections that might be performed on multiple qubits rather
than just one qubit.

Again, by measuring the syndrome, errors are effectively projected or collapsed
onto a discrete set of possibilities represented by tensor products of Pauli
matrices, and by correcting for those Pauli errors, we can recover the original
encoded state.
Meanwhile, whatever randomness is generated in the process is moved into the
syndrome qubits, which are discarded or reset, thereby removing the randomness
generated in this process from the system that stores the encoding.


\lesson{The Stabilizer Formalism}
\label{lesson:the-stabilizer-formalism}

In the previous lesson, we took a first look at quantum error correction,
focusing specifically on the 9-qubit Shor code.
In this lesson, we'll introduce the \emph{stabilizer formalism}, which is a
mathematical framework through which a broad class of quantum error correcting
codes, known as \emph{stabilizer codes}, can be specified and analyzed.
This includes the 9-qubit Shor code along with many other examples, including
codes that seem likely to be well-suited to real-world quantum devices.
Not every quantum error correcting code is a stabilizer code, but many are,
including every example that we'll see in this course.

The lesson begins with a short discussion of Pauli matrices, and tensor
products of Pauli matrices more generally, which can represent not only
operations on qubits, but also measurements of qubits --- in which case they're
typically referred to as \emph{observables}.
We'll then go back and take a second look at the repetition code and see how it
can be described in terms of Pauli matrix observables.
This will both inform and lead into a general discussion of stabilizer codes,
including several examples, basic properties of stabilizer codes, and how the
fundamental tasks of encoding, detecting errors, and correcting those errors
can be performed.

\section{Pauli operations and observables}

Pauli matrices play a central role in the stabilizer formalism.
We'll begin the lesson with a discussion of Pauli matrices, including some of
their basic algebraic properties, and we'll also discuss how Pauli matrices
(and tensor products of Pauli matrices) can describe measurements.

\subsection{Pauli operation basics}

Here are the Pauli matrices, including the $2\times 2$ identity matrix and the
three non-identity Pauli matrices.
\[
\mathbb{I} =
\begin{pmatrix}
  1 & 0\\
  0 & 1
\end{pmatrix}
\qquad
X =
\begin{pmatrix}
  0 & 1\\
  1 & 0
\end{pmatrix}
\qquad
Y =
\begin{pmatrix}
  0 & -i\\
  i & 0
\end{pmatrix}
\qquad
Z =
\begin{pmatrix}
  1 & 0\\
  0 & -1
\end{pmatrix}
\]

\subsubsection{Properties of Pauli matrices}

All four of the Pauli matrices are both unitary and Hermitian.
We used the names $\sigma_x,$ $\sigma_y,$ and $\sigma_z$ to refer to the
non-identity Pauli matrices earlier in the course, but it is conventional to
instead use the capital letters $X,$ $Y,$ and $Z$ in the context of error
correction.
This convention was followed in the previous lesson, and we'll continue to do
this for the remaining lessons.

Different non-identity Pauli matrices \emph{anti-commute} with one another.
\[
XY = -YX \qquad XZ = -ZX \qquad YZ = -ZY
\]
These anti-commutation relations are simple and easy to verify by performing
the multiplications, but they're critically important, in the stabilizer
formalism and elsewhere.
As we will see, the minus signs that emerge when the ordering between two
different non-identity Pauli matrices is reversed in a matrix product
correspond precisely to the detection of errors in the stabilizer formalism.

We also have the multiplication rules listed here.
\[
XX = YY = ZZ = \mathbb{I} \qquad
XY = iZ \qquad YZ = iX \qquad ZX = iY
\]
That is, each Pauli matrix is its own inverse (which is always true for any
matrix that is both unitary and Hermitian), and multiplying two different
non-identity Pauli matrices together is always $\pm i$ times the remaining
non-identity Pauli matrix.
In particular, up to a phase factor, $Y$ is equivalent to $X Z,$ which explains
our focus on $X$ and $Z$ errors and apparent lack of interest in $Y$ errors in
quantum error correction;
$X$ represents a bit-flip, $Z$ represents a phase-flip, and so (up to a global
phase factor) $Y$ represents both of those errors occurring simultaneously on
the same qubit.

\subsubsection{Pauli operations on multiple qubits}

The four Pauli matrices all represent operations (which could be errors) on a
single qubit --- and by tensoring them together we obtain operations on
multiple qubits.
As a point of terminology, when we refer to an \emph{n-qubit Pauli operation},
we mean a tensor product of any $n$ Pauli matrices, such as the examples shown
here, for which $n=9.$
\[
\begin{gathered}
  \mathbb{I} \otimes \mathbb{I} \otimes \mathbb{I} \otimes \mathbb{I} \otimes
  \mathbb{I} \otimes \mathbb{I} \otimes \mathbb{I} \otimes \mathbb{I} \otimes
  \mathbb{I} \\[1mm]
  X \otimes X \otimes \mathbb{I} \otimes \mathbb{I} \otimes \mathbb{I} \otimes
  \mathbb{I} \otimes \mathbb{I} \otimes \mathbb{I} \otimes \mathbb{I} \\[1mm]
  X \otimes Y \otimes Z \otimes \mathbb{I} \otimes \mathbb{I} \otimes
  \mathbb{I} \otimes X \otimes Y \otimes Z
\end{gathered}
\]
Often, the term \emph{Pauli operation} refers to a tensor product of Pauli
matrices \emph{along with a phase factor,} or sometimes just certain phase
factors such as $\pm 1$ and $\pm i.$
There are good reasons to allow for phase factors like this from a mathematical
viewpoint --- but, to keep things as simple as possible, we'll use the term
\emph{Pauli operation} in this course to refer to a tensor product of Pauli
matrices without the possibility of a phase factor different than $1$.

The \emph{weight} of an $n$-qubit Pauli operation is the number of non-identity
Pauli matrices in the tensor product.
For instance, the first example above has weight $0,$ the second has weight
$2,$ and the third has weight $6.$
Intuitively speaking, the weight of an $n$-qubit Pauli operation is the number
of qubits on which it acts non-trivially.
It's typical that quantum error correcting codes are designed so that they can
detect and correct errors represented by Pauli operations so long as their
weight isn't too high.

\subsubsection{Pauli operations as generators}

It's sometimes useful to consider collections of Pauli operations as
\emph{generators} of sets (more specifically, \emph{groups}) of operations, in
an algebraic sense that you may recognize if you're familiar with group theory.
If you're not familiar with group theory, that's OK --- it's not essential for
the lesson.
A familiarity with the basics of group theory is, however, strongly recommended
for those interested in exploring quantum error correction in greater depth.

Suppose that $P_1, \ldots,  P_r$ are $n$-qubit Pauli operations.
When we refer to the \emph{set generated} by $P_1, \ldots, P_r,$ we mean the
set of all matrices that can be obtained by multiplying these matrices
together, in any combination and in any order we choose, taking each one as
many times as we like.
The notation used to refer to this set is $\langle P_1, \ldots, P_r \rangle.$

For example, the set generated by the three non-identity Pauli matrices is as
follows.
\[
\langle X, Y, Z \rangle = \bigl\{\alpha P\,:\,\alpha\in\{1,i,-1,-i\},\;
P\in\{\mathbb{I},X,Y,Z\} \bigr\}
\]
This can be reasoned through the multiplication rules listed earlier.
There are 16 different matrices in this set, which is commonly called the
\emph{Pauli group}.
For a second example, if we remove $Y,$ we obtain half of the Pauli group.
\[
\langle X, Z\rangle = \{ \mathbb{I}, X, Z, -iY, -\mathbb{I}, -X, -Z, iY \}
\]
Here's one final example (for now), where this time we have $n=2.$
\[
\langle X \otimes X, Z \otimes Z\rangle = \{ \mathbb{I}\otimes\mathbb{I},
X\otimes X, Z\otimes Z, -Y\otimes Y \}
\]
In this case we obtain just four elements, owing to the fact that $X\otimes X$
and $Z\otimes Z$ commute:
\[
\begin{aligned}
  (X\otimes X)(Z\otimes Z) & = (XZ) \otimes (XZ)\\
  & = (-ZX)\otimes (-ZX)\\
  & = (ZX)\otimes (ZX)\\
  & = (Z\otimes Z)(X\otimes X).
\end{aligned}
\]

\subsection{Pauli observables}

Pauli matrices, and $n$-qubit Pauli operations more generally, are unitary, and
therefore they describe unitary operations on qubits.
But they're also \emph{Hermitian} matrices, and for this reason they describe
\emph{measurements,} as will now be explained.

\subsubsection{Hermitian matrix observables}

Consider first an arbitrary Hermitian matrix $A.$
When we refer to $A$ as an \emph{observable}, we're associating with $A$ a
certain uniquely defined projective measurement.
In words, the possible outcomes are the distinct eigenvalues of $A,$ and the
projections that define the measurement are the ones that project onto the
spaces spanned by the corresponding eigenvectors of $A.$
So, the outcomes for such a measurement happen to be real numbers --- but
because matrices have only finitely many eigenvalues, there will only be
finitely many different measurement outcomes for a given choice of $A.$

In greater detail, it follows by the spectral theorem that it is possible to
write
\[
A = \sum_{k = 0}^{m-1} \lambda_k \Pi_k
\]
for \emph{distinct} real number eigenvalues $\lambda_0,\ldots,\lambda_{m-1}$
and projections $\Pi_0,\ldots,\Pi_{m-1}$ satisfying
\[
\Pi_0 + \cdots + \Pi_{m-1} = \mathbb{I}.
\]
Such an expression of a matrix is unique up to the ordering of the eigenvalues.
Another way to say this is that, if we insist that the eigenvalues are ordered
in decreasing value $\lambda_0 > \lambda_1 > \cdots > \lambda_{m-1},$ then
there's only one way to write $A$ in the form above.

Based on this expression, the measurement we associate with the observable $A$
is the projective measurement described by the projections
$\Pi_0,\ldots,\Pi_{m-1},$ and the eigenvalues $\lambda_0,\ldots,\lambda_{m-1}$
are understood to be the measurement outcomes corresponding to these
projections.

\subsubsection{Measurements from Pauli operations}

Let's see what measurements of the sort just described look like for Pauli
operations, starting with the three non-identity Pauli matrices.
These matrices have spectral decompositions as follows.
\[
\begin{gathered}
  X = \vert {+} \rangle\langle {+} \vert - \vert {-} \rangle\langle {-} \vert\\
  Y = \vert {+i} \rangle\langle {+i} \vert - \vert {-i} \rangle\langle {-i}
  \vert\\
  Z = \vert {0} \rangle\langle {0} \vert - \vert {1} \rangle\langle {1} \vert
\end{gathered}
\]
The measurements defined by $X,$ $Y,$ and $Z,$ viewed as observables, are
therefore the projective measurements defined by the following sets of
projections, respectively.
\[
\begin{gathered}
  \bigl\{\vert {+} \rangle\langle {+} \vert, \vert {-} \rangle\langle {-} \vert
  \bigr\} \\
  \bigl\{\vert {+i} \rangle\langle {+i} \vert, \vert {-i} \rangle\langle {-i}
  \vert\bigr\} \\
  \bigl\{\vert {0} \rangle\langle {0} \vert, \vert {1} \rangle\langle {1}
  \vert\bigr\}
\end{gathered}
\]
In all three cases, the two possible measurement outcomes are the eigenvalues
$+1$ and $-1.$
Such measurements are called $X$ measurements, $Y$ measurements, and
$Z$ measurements.
We encountered these measurements in Lesson~\ref{lesson:general-measurements}
\emph{(General Measurements)}, where they arose in the context of quantum state
tomography.

Of course, a $Z$ measurement is essentially just a standard basis measurement
and an $X$ measurement is a measurement with respect to the plus/minus basis of
a qubit --- but, as these measurements are described here, we're taking the
eigenvalues $+1$ and $-1$ to be the actual measurement outcomes.

The same prescription can be followed for Pauli operations on $n\geq 2$ qubits,
though it must be stressed that there will still be just two possible outcomes
for the measurements described in this way: $+1$ and $-1,$ which are the only
possible eigenvalues of Pauli operations.
The two corresponding projections will therefore have rank higher than one in
this case.
More precisely, for every non-identity $n$-qubit Pauli operation, the $2^n$
dimensional state space always splits into two subspaces of eigenvectors having
equal dimension, so the two projections that define the associated measurement
will both have rank $2^{n-1}.$

The measurement described by an $n$-qubit Pauli operation, considered as an
observable, is therefore not the same thing as a measurement with respect to an
\emph{orthonormal basis} of eigenvectors of that operation, nor is it the same
thing as independently measuring each of the corresponding Pauli matrices
independently, as observables, on $n$ qubits.
Both of those alternatives would necessitate $2^n$ possible measurement
outcomes, but here we have just the two possible outcomes $+1$~and~$-1.$

For example, consider the 2-qubit Pauli operation $Z\otimes Z$ as an
observable.
We can effectively take the tensor product of the spectral decompositions to
obtain one for the tensor product.
\[
\begin{aligned}
  Z\otimes Z & = (\vert 0\rangle\langle 0\vert - \vert 1\rangle\langle 1\vert)
  \otimes (\vert 0\rangle\langle 0\vert - \vert 1\rangle\langle 1\vert)\\
  & = \bigl( \vert 00\rangle\langle 00\vert + \vert 11\rangle\langle 11\vert
  \bigr) - \bigl( \vert 01\rangle\langle 01\vert + \vert 10\rangle\langle
  10\vert \bigr)
\end{aligned}
\]
That is, we have $Z\otimes Z = \Pi_0 - \Pi_1$ for
\[
\Pi_0 = \vert 00\rangle\langle 00\vert + \vert 11\rangle\langle 11\vert
\quad\text{and}\quad
\Pi_1 = \vert 01\rangle\langle 01\vert + \vert 10\rangle\langle 10\vert,
\]
so these are the two projections that define the measurement.
If, for instance, we were to measure a $\vert\phi^+\rangle$ Bell state
nondestructively using this measurement, then we would be certain to obtain the
outcome $+1,$ and the state would be unchanged as a result of the measurement.
In particular, the state would not collapse to $\vert 00\rangle$
or~$\vert 11\rangle.$

\subsubsection{Nondestructive implementation through phase estimation}

For any $n$-qubit Pauli operation, we can perform the measurement associated
with that observable nondestructively using phase estimation.

\begin{figure}[t]
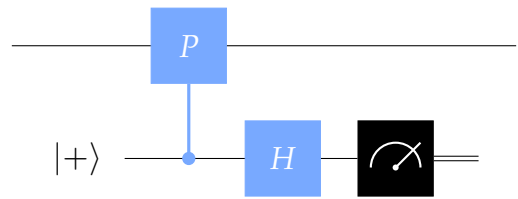

  \begin{center}

  \end{center}
  \caption{A quantum circuit based on phase estimation for non-destructively
    measuring a Pauli observable $P$ on the top qubit.}
  \label{fig:phase-estimation-Pauli}
\end{figure}

Figure~\ref{fig:phase-estimation-Pauli} shows a circuit based on phase
estimation that works for any Pauli matrix $P,$ where the measurement is being
performed on the top qubit.
The outcomes $0$ and $1$ of the standard basis measurement in the circuit
correspond to the eigenvalues $+1$ and $-1,$ just like we usually have for
phase estimation with one control qubit.
Note that the control qubit is on the bottom in this diagram, whereas in
Lesson~\ref{lesson:phase-estimation-and-factoring}
\emph{(Phase Estimation and Factoring)} the control qubits were drawn on
the top.

\begin{figure}[t]
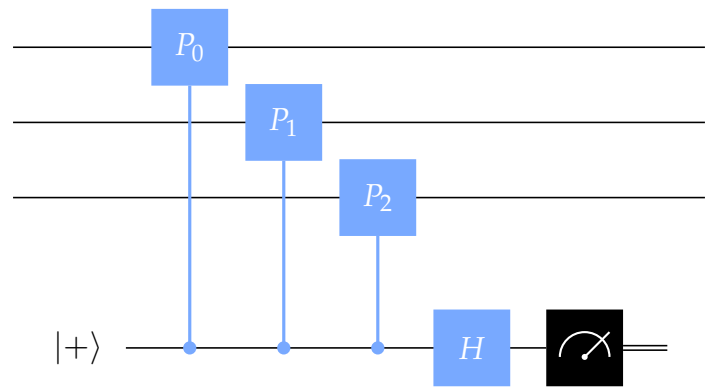

  \vspace*{3mm}
  \begin{center}
    %
   
  \end{center}
  \caption{A quantum circuit performing $3$-qubit Pauli observable
    $P_2\otimes P_1\otimes P_0$ non-destructively on the top three qubits.}
  \label{fig:phase-estimation-3-qubit-Pauli}
\end{figure}

A similar method works for Pauli operations on multiple qubits.
For example, the circuit illustrated in
Figure~\ref{fig:phase-estimation-3-qubit-Pauli} performs a nondestructive
measurement of the $3$-qubit Pauli observable $P_2\otimes P_1\otimes P_0,$ for
any choice of $P_0,P_1,P_2 \in \{X,Y,Z\}.$
This approach generalizes to $n$-qubit Pauli observables, for any $n,$ in the
natural way.
Of course, we only need to include controlled-unitary gates for
\emph{non-identity} tensor factors of Pauli observables when implementing such
measurements; controlled-identity gates are simply identity gates and can
therefore be omitted.
This means that lower weight Pauli observables require smaller circuits to be
implemented through this approach.

\pagebreak

Notice that, irrespective of $n,$ these phase estimation circuits have just a
single control qubit, which is consistent with the fact that there are just two
possible measurement outcomes for these measurements.
Using more control qubits wouldn't reveal additional information because these
measurements are already perfect using a single control qubit.
(One way to see this is directly from the general procedure for phase
estimation: the assumption $U^2 = \mathbb{I}$ renders any additional control
qubits beyond the first pointless.)

\begin{figure}[t]
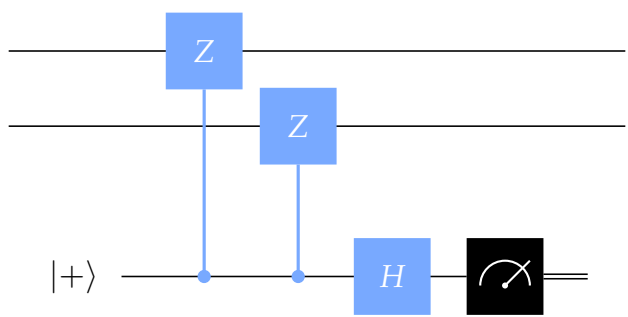

  \begin{center}
   
   
  \end{center}
  \caption{A quantum circuit implementing a $Z\otimes Z$ measurement on the top
    two qubits.}
  \label{fig:phase-estimation-ZZ}
\end{figure}

\begin{figure}[t]
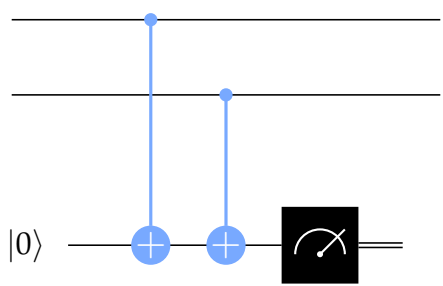

  \vspace*{3mm}
  \begin{center}
    %
  \end{center}
  \caption{A simplification of the circuit in
    Figure~\ref{fig:phase-estimation-ZZ}.}
  \label{fig:simplified-ZZ}
\end{figure}

Figure~\ref{fig:phase-estimation-ZZ} shows a specific example, of a
nondestructive implementation of a $Z\otimes Z$ measurement, which is relevant
to the description of the 3-bit repetition code as a stabilizer code that we'll
see shortly.
In this case, and for tensor products of more than two $Z$ observables more
generally, the circuit can be simplified, as is shown in
Figure~\ref{fig:simplified-ZZ}.
Thus, this measurement is equivalent to nondestructively measuring the parity
(or XOR) of the standard basis states of two qubits.

\section{Repetition code revisited}

Next, we'll take a second look at the 3-bit repetition code, this time phrasing
it in terms of Pauli operations.
This will be our first example of a \emph{stabilizer code}.

\subsection{Pauli observables for the repetition code}

Recall that, when we apply the 3-bit repetition code to qubits, a given qubit
state vector $\alpha\vert 0\rangle + \beta\vert 1\rangle$ is encoded as
\[
\vert\psi\rangle = \alpha\vert 000\rangle + \beta\vert 111\rangle.
\]
Any state $\vert\psi\rangle$ of this form is a valid 3-qubit encoding of a
qubit state --- but if we had a state that we weren't sure about, we could
verify that we have a valid encoding by checking the following two equations.
\[
\begin{aligned}
  (Z \otimes Z \otimes \mathbb{I}) \vert\psi\rangle & = \vert\psi\rangle\\[1mm]
  (\mathbb{I} \otimes Z \otimes Z) \vert\psi\rangle & = \vert\psi\rangle
\end{aligned}
\]
The first equation states that applying $Z$ operations to the leftmost two
qubits of $\vert\psi\rangle$ has no effect, which is to say that
$\vert\psi\rangle$ is an eigenvector of $Z\otimes Z\otimes \mathbb{I}$ with
eigenvalue $1.$
The second equation is similar except that $Z$ operations are applied to the
rightmost two qubits.
The idea is that, if we think about $\vert\psi\rangle$ as a linear combination
of standard basis states, then the first equation implies that we can only have
nonzero coefficients for standard basis states where the leftmost two bits have
even parity (or, equivalently, are equal), and the second equation implies that
we can only have nonzero coefficients for standard basis states for which the
rightmost two bits have even parity.

\begin{figure}[!ht]
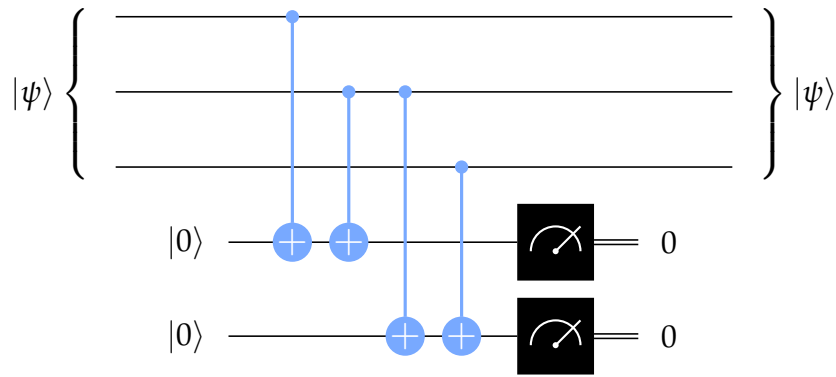

  \begin{center}
   
  \end{center}
  \caption{A circuit that simultaneously measures the Pauli observables
    $Z\otimes Z\otimes \mathbb{I}$ and $\mathbb{I}\otimes Z\otimes Z$
    is equivalent to the error detection circuit for the $3$-bit repetition
    code.
  }
  \label{fig:3-bit-repetition-parity-checks}
\end{figure}

Equivalently, if we view the two Pauli operations $Z\otimes Z\otimes
\mathbb{I}$ and $\mathbb{I}\otimes Z\otimes Z$ as observables, and measure both
using the circuits suggested at the end of the previous section, then we would
be certain to obtain measurement outcomes corresponding to $+1$ eigenvalues,
because $\vert\psi\rangle$ is an eigenvector of both observables with
eigenvalue $1.$
But, the simplified version of the (combined) circuit for independently
measuring both observables, shown in
Figure~\ref{fig:3-bit-repetition-parity-checks}, is none other than the parity
check circuit for the 3-bit repetition code.

The two equations above therefore imply that the parity check circuit outputs
$00,$ which is the syndrome that indicates that no errors have been detected.

The 3-qubit Pauli operations $Z\otimes Z\otimes \mathbb{I}$ and
$\mathbb{I}\otimes Z\otimes Z$ are called \emph{stabilizer generators} for this
code, and the \emph{stabilizer} of the code is the set generated by the
stabilizer generators.
\[
\langle Z\otimes Z\otimes \mathbb{I}, \mathbb{I}\otimes Z\otimes Z\rangle
= \{
  \mathbb{I}\otimes\mathbb{I}\otimes\mathbb{I},
  Z\otimes Z\otimes\mathbb{I},
  Z\otimes\mathbb{I}\otimes Z,
  \mathbb{I}\otimes Z\otimes Z
\}
\]
The stabilizer is a fundamentally important mathematical object associated with
this code, and the role that it plays will be discussed as the lesson
continues.
For now, let's observe that we could have made a different choice for the
generators and corresponding parity checks, specifically by taking
$Z\otimes\mathbb{I}\otimes Z$ in place of either of the generators we did
select, but the stabilizer and the code itself would be unchanged as a result.

\subsection{Error detection}

Next, we'll consider bit-flip detection for the 3-bit repetition code, with a
focus on the interactions and relationships among the Pauli operations that are
involved: the stabilizer generators and the errors themselves.

Suppose we've encoded a qubit using the 3-bit repetition code, and a bit-flip
error occurs on the leftmost qubit.
This causes the state $\vert\psi\rangle$ to be transformed according to the
action of an $X$ operation (or $X$ error).
\[
\vert\psi\rangle \mapsto (X \otimes \mathbb{I} \otimes \mathbb{I})
\vert\psi\rangle
\]
This error can be detected by performing the parity checks for the 3-bit
repetition code, as discussed in the previous lesson, which is equivalent to
nondestructively measuring the stabilizer generators $Z\otimes Z\otimes
\mathbb{I}$ and $\mathbb{I}\otimes Z\otimes Z$ as observables.

Let's begin with the first stabilizer generator.
The state $\vert\psi\rangle$ has been affected by an $X$ error on the leftmost
qubit, and our goal is to understand how the measurement of this stabilizer
generator, as an observable, is influenced by this error.
Because $X$ and $Z$ anti-commute, whereas every matrix commutes with the
identity matrix, it follows that
$Z\otimes Z\otimes \mathbb{I}$ anti-commutes with
$X\otimes\mathbb{I}\otimes\mathbb{I}.$
Meanwhile, because $\vert\psi\rangle$ is a valid encoding of a qubit, $Z\otimes
Z\otimes \mathbb{I}$ acts trivially on $\vert\psi\rangle.$
\[
\begin{aligned}
  (Z \otimes Z \otimes \mathbb{I})(X \otimes \mathbb{I} \otimes \mathbb{I})
  \vert\psi\rangle
  & = -(X \otimes \mathbb{I} \otimes \mathbb{I})(Z \otimes Z \otimes
  \mathbb{I})\vert\psi\rangle \\
  & = -(X \otimes \mathbb{I} \otimes \mathbb{I}) \vert\psi\rangle
\end{aligned}
\]

Therefore, $(X \otimes \mathbb{I} \otimes \mathbb{I}) \vert\psi\rangle$ is an
eigenvector of $Z \otimes Z \otimes \mathbb{I}$ with eigenvalue $-1.$
When the measurement associated with the observable $Z \otimes Z \otimes
\mathbb{I}$ is performed on the state $(X \otimes \mathbb{I} \otimes
\mathbb{I}) \vert\psi\rangle,$ the outcome is therefore certain to be the one
associated with the eigenvalue $-1.$

Similar reasoning can be applied to the second stabilizer generator, but this
time the error commutes with the stabilizer generator rather than
anti-commuting, and so the outcome for this measurement is the one associated
with the eigenvalue $+1.$
\[
\begin{aligned}
  (\mathbb{I} \otimes Z \otimes Z)(X \otimes \mathbb{I} \otimes \mathbb{I})
  \vert\psi\rangle
  & = (X \otimes \mathbb{I} \otimes \mathbb{I})(\mathbb{I} \otimes Z \otimes
  Z)\vert\psi\rangle\\
  & = (X \otimes \mathbb{I} \otimes \mathbb{I}) \vert\psi\rangle
\end{aligned}
\]

What we find when considering these equations is that, regardless of our
original state $\vert\psi\rangle,$ the corrupted state is an eigenvector of
both stabilizer generators, and whether the eigenvalue is $+1$ or $-1$ is
determined by whether the \emph{error} commutes or anti-commutes with each
stabilizer generator.
For errors represented by Pauli operations, it will always be one or the other,
because any two Pauli operations either commute or anti-commute.
Meanwhile, the actual state $\vert\psi\rangle$ doesn't play an important role,
except for the fact that the stabilizer generators act trivially on this state.

For this reason, we really don't need to concern ourselves in general with the
specific encoded state we're working with.
All that matters is whether the error commutes or anti-commutes with each
stabilizer generator.
In particular, these are the relevant equations with regard to this particular
error for this code.
\[
\begin{aligned}
  (Z \otimes Z \otimes \mathbb{I})(X \otimes \mathbb{I} \otimes \mathbb{I})
  & = -(X \otimes \mathbb{I} \otimes \mathbb{I})(Z \otimes Z \otimes
  \mathbb{I})\\[1mm]
  (\mathbb{I} \otimes Z \otimes Z)(X \otimes \mathbb{I} \otimes \mathbb{I})
  & = (X \otimes \mathbb{I} \otimes \mathbb{I})(\mathbb{I} \otimes Z \otimes Z)
\end{aligned}
\]

Here's a table with one row for each stabilizer generator and one column for
each error.
The entry in the table is either $+1$ or $-1$ depending on whether the error
and the stabilizer generator commute or anti-commute.
The table only includes columns for the errors corresponding to a single
bit-flip, as well as no error at all, which is described by the identity
tensored with itself three times.
We could add more columns for other errors, but for now our focus will be on
just these errors.
\[
\begin{array}{c|cccc}
  & \mathbb{I}\otimes\mathbb{I} \otimes\mathbb{I} & X \otimes \mathbb{I}
  \otimes \mathbb{I}
  & \mathbb{I}\otimes X\otimes\mathbb{I} & \mathbb{I} \otimes\mathbb{I} \otimes
  X \\ \hline
  Z\otimes Z\otimes\mathbb{I} & +1 & -1 & -1 & +1 \\
  \mathbb{I}\otimes Z\otimes Z & +1 & +1 & -1 & -1
\end{array}
\]

For each error in the table, the corresponding column therefore reveals how
that error transforms any given encoding into a $+1$ or $-1$ eigenvector of
each stabilizer generator.
Equivalently, the columns describe the \emph{syndrome} we would obtain from the
parity checks, which are equivalent to nondestructive measurements of the
stabilizer generators as observables.

Of course, the table has $+1$ and $-1$ entries rather than $0$ and $1$ entries
--- and it's common to think about a syndrome as being a binary string rather
than column of $+1$ and $-1$ entries --- but we can equally well think about
these vectors with $+1$ and $-1$ entries as syndromes to connect them directly
to the eigenvalues of the stabilizer generators.
In general, the syndromes tell us something about whatever error took place,
and if we know that one of the four possible errors listed in the table
occurred, the syndrome indicates which one it was.

\subsection{Syndromes}

Encodings for the 3-bit repetition code are 3-qubit states, so they're unit
vectors in an 8-dimensional complex vector space.
The four possible syndromes effectively split this 8 dimensional space into
four 2-dimensional subspaces, where quantum state vectors in each subspace
always result in the same syndrome.
The diagram in Figure~\ref{fig:stabilizer-generator-table} illustrates
specifically how the 8-dimensional space is divided up by the two stabilizer
generators.

\begin{figure}[!ht]
  \begin{center}
    \begin{tikzpicture}[scale = 2.5]
      \draw (-1,1) -- (1,1) -- (1,-1) -- (-1,-1) -- (-1,1);
      \draw (0,1) -- (0,-1);
      \draw (1,0) -- (-1,0);
      
      \node[anchor = east] at (-1,0.5) {$+1$};
      \node[anchor = east] at (-1,-0.5) {$-1$};
      \node[anchor = south] at (-0.5,1) {$+1$};
      \node[anchor = south] at (0.5,1) {$-1$};
      
      \node[anchor = east] at (-1.6,0) {$Z\otimes Z\otimes\mathbb{I}$};
      \node[anchor = south] at (0,1.45) {$\mathbb{I}\otimes Z\otimes Z$};
      
      \node at (-1.5,0) {$\left\{\rule{0mm}{27mm}\right.$};
      \node at (0,1.35) {$\overbrace{\rule{48mm}{0mm}}$};
      
      \node at (-0.5,0.5) {%
        $\begin{array}{cc}
          \ket{000}\\[1mm]
          \ket{111}
        \end{array}$};
      
      \node at (0.5,0.5) {%
        $\begin{array}{cc}
          \ket{001}\\[1mm]
          \ket{110}
        \end{array}$};
      
      \node at (0.5,-0.5) {%
        $\begin{array}{cc}
          \ket{010}\\[1mm]
          \ket{101}
        \end{array}$};
      
      \node at (-0.5,-0.5) {%
        $\begin{array}{cc}
          \ket{100}\\[1mm]
          \ket{011}
        \end{array}$};

    \end{tikzpicture}   
  \end{center}
  \caption{The stabilizer generators $Z\otimes Z\otimes \mathbb{I}$ and
    $\mathbb{I}\otimes Z\otimes Z$ split the $8$-dimensional space
    corresponding to three qubits into four $2$-dimensional subspaces
    spanned by the standard basis states indicated in the squares corresponding
    to the measurement outcomes.}
  \label{fig:stabilizer-generator-table}
\end{figure}

Each stabilizer generator splits the space into two subspaces of equal
dimension, namely the space of $+1$ eigenvectors and the space of $-1$
eigenvectors for that observable.
For example, the $+1$ eigenvectors of $Z\otimes Z\otimes\mathbb{I}$ are linear
combinations of standard basis states for which the leftmost two bits have even
parity, and the $-1$ eigenvectors are linear combinations of standard basis
states for which the leftmost two bits have odd parity.
The situation is similar for the other stabilizer generator, except that for
this one it's the rightmost two bits rather than the leftmost two bits.

The four 2-dimensional subspaces corresponding to the four possible syndromes
are easy to describe in this case, owing to the fact that this is a very simple
code.
In particular, the subspace corresponding to the syndrome $(+1,+1)$ is the
space spanned by $\vert 000\rangle$ and $\vert 111\rangle$, which is the space
of valid encodings (also known as the \emph{code space}), and in general the
spaces are spanned by the standard basis shown in the corresponding squares.

The syndromes also partition all of the 3-qubit \emph{Pauli operations} into 4
equal-size collections, depending upon which syndrome that operation (as an
error) would cause.
For example, any Pauli operation that commutes with both stabilizer generators
results in the syndrome $(+1,+1),$ and among the 64 possible 3-qubit Pauli
operations, there are exactly 16 of them in this category (including
$\mathbb{I}\otimes \mathbb{I}\otimes Z,$ $Z\otimes Z\otimes Z,$ and $X\otimes
X\otimes X$ for instance), and likewise for the other 3 syndromes.

Both of these properties --- that the syndromes partition both the state space
in which encodings live and all of the Pauli operations on this space into
equal-sized collections --- are true in general for stabilizer codes, which
we'll define precisely in the next section.

Although it's mainly an aside at this point, it's worth mentioning that Pauli
operations that commute with \emph{both} stabilizer generators, or equivalently
Pauli operations that result in the syndrome $(+1,+1),$ but are not themselves
proportional to elements of the stabilizer, turn out to behave just like
single-qubit Pauli operations on the encoded qubit (i.e., the logical qubit)
for this code.
For example, $X\otimes X \otimes X$ commutes with both stabilizer generators,
but is itself not proportional to any element in the stabilizer, and indeed the
effect of this operation on an encoding is equivalent to an $X$ gate on the
logical qubit being encoded.
\[
(X\otimes X \otimes X)(\alpha \vert 000\rangle + \beta \vert 111\rangle) =
\alpha \vert 111\rangle + \beta \vert 000\rangle
\]
Again, this is a phenomenon that generalizes to all stabilizer codes.

\section{Stabilizer codes}

Now we'll define stabilizer codes in general.
We'll also discuss some of their basic properties and how they work, including
how states can be encoded and how errors are detected and corrected using these
codes.

\subsection{Definition of stabilizer codes}

An $n$-qubit stabilizer code is specified by a list of $n$-qubit Pauli
operations, $P_1,\ldots,P_r.$
These operations are called \emph{stabilizer generators} in this context, and
they must satisfy the following three properties.
\begin{enumerate}
\item
  The stabilizer generators all \emph{commute} with one another.
  \[
  P_j P_k = P_k P_j \qquad \text{(for all $j,k\in\{1,\ldots,r\}$)}
  \]
\item
  The stabilizer generators form a \emph{minimal generating set}.
  \[
  P_k \notin \langle P_1,\ldots,P_{k-1},P_{k+1},\ldots,P_r\rangle \qquad
  \text{(for all $k\in\{1,\ldots,r\}$)}
  \]
\item
  At least one quantum state vector is fixed by all of the stabilizer
  generators.
  \[
  -\mathbb{I}^{\otimes n} \notin \langle P_1,\ldots, P_r\rangle
  \]
  (It's not obvious that the existence of a quantum state vector
  $\vert\psi\rangle$ fixed by all of the stabilizer generators, meaning $P_1
  \vert\psi\rangle = \cdots = P_r \vert\psi\rangle = \vert\psi\rangle,$ is
  equivalent to $-\mathbb{I}^{\otimes n} \notin \langle P_1,\ldots,
  P_r\rangle,$ but indeed this is the case, and we'll see why a bit later in
  the lesson.)
\end{enumerate}

\noindent
Assuming that we have such a list $P_1,\ldots,P_r,$ the \emph{code space}
defined by these stabilizer generators is the subspace $\mathcal{C}$ containing
every $n$-qubit quantum state vector fixed by all $r$ of these stabilizer
generators.
\[
\mathcal{C} = \bigl\{ \vert\psi\rangle \,:\, P_1 \vert\psi\rangle = \cdots =
P_r \vert\psi\rangle = \vert\psi\rangle \bigr\}
\]
Quantum state vectors in this subspace are precisely the ones that can be
viewed as \emph{valid encodings} of quantum states.
We'll discuss the actual process of encoding later.

Finally, the \emph{stabilizer} of the code defined by the stabilizer generators
$P_1, \ldots, P_r$ is the set generated by these operations:
\[
\langle P_1,\ldots,P_r\rangle.
\]

A natural way to think about a stabilizer code is to view the stabilizer
generators as observables, and to collectively interpret the outcomes of the
measurements associated with these observables as an error syndrome.
Valid encodings are $n$-qubit quantum state vectors for which the measurement
outcomes, as eigenvalues, are all guaranteed to be $+1.$
Any other syndrome, where at least one $-1$ measurement outcome occurs, signals
that an error has been detected.

We'll take a look at several examples shortly, but first just a few remarks
about the three conditions on stabilizer generators are in order.

The first condition is natural, in light of the interpretation of the
stabilizer generators as observables, for it implies that it doesn't matter in
what order the measurements are performed: the observables commute, so the
measurements commute.
This naturally imposes certain algebraic constraints on stabilizer codes that
are important to how they work.

The second condition requires that the stabilizer generators form a minimal
generating set, meaning that removing any one of them would result in a smaller
stabilizer.
Strictly speaking, this condition isn't really essential to the way stabilizer
codes work in an \emph{operational} sense --- and, as we'll see in the next
lesson, it does sometimes make sense to think about sets of stabilizer
generators for codes that actually don't satisfy this condition.
For the sake of \emph{analyzing} stabilizer codes and explaining their
properties, however, we will assume that this condition is in place.
In short, this condition guarantees that each observable that we measure to
obtain the error syndrome \emph{adds} information about possible errors, as
opposed to being redundant and producing results that could be inferred from
the other stabilizer generator measurements.

The third condition requires that at least one nonzero vector is fixed by all
of the stabilizer generators, which is equivalent to $-\mathbb{I}^{\otimes n}$
not being contained in the stabilizer.
The need for this condition comes from the fact that it actually is possible to
choose a minimal generating set of $n$-qubit Pauli operations that all commute
with one another, and yet no nonzero vectors are fixed by every one of the
operations.
We're not interested in ``codes'' for which there are no valid encodings, so we
rule out this possibility by requiring this condition as a part of the
definition.

\subsection{Examples}

Here are some examples of stabilizer codes for small values of $n.$
We'll see more examples, including ones for which $n$ can be much larger, in
the next lesson.

\subsubsection{3-bit repetition code}

The 3-bit repetition code is an example of a stabilizer code, where our
stabilizer generators are
$Z \otimes Z \otimes \mathbb{I}$ and $\mathbb{I} \otimes Z \otimes Z.$

We can easily check that these two stabilizer generators fulfill the required
conditions.
First, the two stabilizer generators $Z \otimes Z \otimes \mathbb{I}$ and
$\mathbb{I} \otimes Z \otimes Z$ commute with one another.
\[
(Z \otimes Z \otimes \mathbb{I})(\mathbb{I} \otimes Z \otimes Z) = Z \otimes
\mathbb{I} \otimes Z
= (\mathbb{I} \otimes Z \otimes Z)(Z \otimes Z \otimes \mathbb{I})
\]
Second, we have a minimal generating set (rather trivially in this case).
\[
\begin{aligned}
  Z \otimes Z \otimes \mathbb{I} \notin \langle\mathbb{I} \otimes Z \otimes
  Z\rangle
  & = \{\mathbb{I}\otimes\mathbb{I}\otimes\mathbb{I}, \mathbb{I} \otimes Z
  \otimes Z\}\\[1mm]
  \mathbb{I} \otimes Z \otimes Z \notin \langle Z \otimes Z \otimes
  \mathbb{I}\rangle
  & = \{\mathbb{I}\otimes\mathbb{I}\otimes\mathbb{I}, Z \otimes Z \otimes
  \mathbb{I}\}
\end{aligned}
\]
And third, we already know that $\vert 000\rangle$ and $\vert 111\rangle,$ as
well as any linear combination of these vectors, are fixed by both $Z \otimes Z
\otimes \mathbb{I}$ and $\mathbb{I} \otimes Z \otimes Z.$
Alternatively, we can conclude this using the equivalent condition from the
definition.
\[
-\mathbb{I}\otimes\mathbb{I}\otimes\mathbb{I}\notin
\langle Z \otimes Z \otimes \mathbb{I}, \mathbb{I} \otimes Z \otimes Z\rangle
= \{
\mathbb{I}\otimes\mathbb{I}\otimes\mathbb{I},
Z\otimes Z\otimes\mathbb{I},
Z\otimes\mathbb{I}\otimes Z,
\mathbb{I}\otimes Z\otimes Z
\}
\]
These conditions can be much more difficult to check for more complicated
stabilizer codes.

\subsubsection{Modified 3-bit repetition code}

In the previous lesson, we saw that it's possible to modify the 3-bit
repetition code so that it protects against phase-flip errors rather than
bit-flip errors.
As a stabilizer code, this new code is easy to describe:
its stabilizer generators are $X \otimes X \otimes \mathbb{I}$ and $\mathbb{I}
\otimes X \otimes X.$

This time the stabilizer generators represent $X\otimes X$ observables rather
than $Z\otimes Z$ observables, so they're essentially parity checks in the
plus/minus basis rather than the standard basis.
The three required conditions on the stabilizer generators are easily verified,
along similar lines to the ordinary 3-bit repetition code.

\subsubsection{9-qubit Shor code}

Here's the 9-qubit Shor code, which is also a stabilizer code, expressed by
stabilizer generators.
\[
\begin{gathered}
  Z \otimes Z \otimes \mathbb{I} \otimes
  \mathbb{I} \otimes \mathbb{I} \otimes \mathbb{I} \otimes
  \mathbb{I} \otimes \mathbb{I} \otimes \mathbb{I}\\[1mm]
  \mathbb{I} \otimes Z \otimes Z \otimes
  \mathbb{I} \otimes \mathbb{I} \otimes \mathbb{I} \otimes
  \mathbb{I} \otimes \mathbb{I} \otimes \mathbb{I}\\[1mm]
  \mathbb{I} \otimes \mathbb{I} \otimes \mathbb{I} \otimes
  Z \otimes Z \otimes \mathbb{I} \otimes
  \mathbb{I} \otimes \mathbb{I} \otimes \mathbb{I}\\[1mm]
  \mathbb{I} \otimes \mathbb{I} \otimes \mathbb{I} \otimes
  \mathbb{I} \otimes Z \otimes Z \otimes
  \mathbb{I} \otimes \mathbb{I} \otimes \mathbb{I}\\[1mm]
  \mathbb{I} \otimes \mathbb{I} \otimes \mathbb{I} \otimes
  \mathbb{I} \otimes \mathbb{I} \otimes \mathbb{I} \otimes
  Z \otimes Z \otimes \mathbb{I}\\[1mm]
  \mathbb{I} \otimes \mathbb{I} \otimes \mathbb{I} \otimes
  \mathbb{I} \otimes \mathbb{I} \otimes \mathbb{I} \otimes
  \mathbb{I} \otimes Z \otimes Z\\[1mm]
  X \otimes X \otimes X \otimes X \otimes X \otimes X \otimes
  \mathbb{I} \otimes \mathbb{I} \otimes \mathbb{I}\\[1mm]
  \mathbb{I} \otimes \mathbb{I} \otimes \mathbb{I}\otimes
  X \otimes X \otimes X \otimes X \otimes X \otimes X
\end{gathered}
\]
In this case, we basically have three copies of the 3-bit repetition code, one
for each of the three blocks of three qubits, as well as the last two
stabilizer generators, which take a form reminiscent of the circuit for
detecting phase-flips for this code.
An alternative way to think about the last two stabilizer generators is that
they take the same form as for the 3-bit repetition code for phase-flips,
except that $X\otimes X\otimes X$ is substituted for $X,$ which is consistent
with the fact that $X\otimes X\otimes X$ corresponds to an $X$ operation on
logical qubits encoded using the 3-bit repetition code.

Before we move on to other examples, it should be noted that tensor product
symbols are often omitted when describing stabilizer codes by lists of
stabilizer generators, because it tends to make them easier to read and to see
their patterns.
For example, the same stabilizer generators as above for the 9-qubit Shor code
look like this without the tensor product symbols being written explicitly.
\[
\begin{array}{ccccccccc}
  Z & Z & \mathbb{I} &
  \mathbb{I} & \mathbb{I} & \mathbb{I} &
  \mathbb{I} & \mathbb{I} & \mathbb{I}\\[1mm]
  \mathbb{I} & Z & Z &
  \mathbb{I} & \mathbb{I} & \mathbb{I} &
  \mathbb{I} & \mathbb{I} & \mathbb{I}\\[1mm]
  \mathbb{I} & \mathbb{I} & \mathbb{I} &
  Z & Z & \mathbb{I} &
  \mathbb{I} & \mathbb{I} & \mathbb{I}\\[1mm]
  \mathbb{I} & \mathbb{I} & \mathbb{I} &
  \mathbb{I} & Z & Z &
  \mathbb{I} & \mathbb{I} & \mathbb{I}\\[1mm]
  \mathbb{I} & \mathbb{I} & \mathbb{I} &
  \mathbb{I} & \mathbb{I} & \mathbb{I} &
  Z & Z & \mathbb{I}\\[1mm]
  \mathbb{I} & \mathbb{I} & \mathbb{I} &
  \mathbb{I} & \mathbb{I} & \mathbb{I} &
  \mathbb{I} & Z & Z\\[1mm]
  X & X & X & X & X & X &
  \mathbb{I} & \mathbb{I} & \mathbb{I}\\[1mm]
  \mathbb{I} & \mathbb{I} & \mathbb{I}&
  X & X & X & X & X & X
\end{array}
\]

\subsubsection{7-qubit Steane code}

Here's another example of a stabilizer code, known as the \emph{7-qubit Steane
code}.
It has some remarkable features, and we'll come back to this code from time to
time throughout the remaining lessons of the course.
\[
\begin{array}{ccccccc}
  Z & Z & Z & Z & \mathbb{I} & \mathbb{I} & \mathbb{I} \\[1mm]
  Z & Z & \mathbb{I} & \mathbb{I} & Z & Z & \mathbb{I} \\[1mm]
  Z & \mathbb{I} & Z & \mathbb{I} & Z & \mathbb{I} & Z \\[1mm]
  X & X & X & X & \mathbb{I} & \mathbb{I} & \mathbb{I} \\[1mm]
  X & X & \mathbb{I} & \mathbb{I} & X & X & \mathbb{I} \\[1mm]
  X & \mathbb{I} & X & \mathbb{I} & X & \mathbb{I} & X
\end{array}
\]
For now, let's simply observe that this is a valid stabilizer code.
The first three stabilizer generators clearly commute with one another, because
$Z$ commutes with itself and the identity commutes with everything, and the
situation is similar for the last three stabilizer generators.
It remains to check that if we take one of the $Z$ stabilizer generators (i.e.,
one of the first three) and one of the $X$ stabilizer generators (i.e., one of
the last three), then these two generators commute, and one can go through the
9 possible pairings to check that.
In all of these cases, an $X$ and a $Z$ Pauli matrix always line up in the same
position an even number of times, so the two generators will commute, just like
$X\otimes X$ and $Z\otimes Z$ commute.
This is also a minimal generating set, and it defines a nontrivial code space,
which are facts left to you to contemplate.

The 7-qubit Steane code is similar to the 9-qubit Shor code in that it encodes
a single qubit and allows for the correction of an arbitrary error on one
qubit, but it requires only 7 qubits rather than 9.

\subsubsection{5-qubit code}

Seven is not the fewest number of qubits required to encode one qubit and
protect it against an arbitrary error on one qubit --- here's a stabilizer code
that does this using just 5 qubits.
\[
\begin{array}{ccccc}
  X & Z & Z & X & \mathbb{I} \\[1mm]
  \mathbb{I} & X & Z & Z & X \\[1mm]
  X & \mathbb{I} & X & Z & Z \\[1mm]
  Z & X & \mathbb{I} & X & Z \\[1mm]
\end{array}
\]
This code is typically called \emph{the 5-qubit code.}
This is the smallest number of qubits in a quantum error correcting code that
can allow for the correction of an arbitrary single-qubit error.

\subsubsection{One-dimensional stabilizer codes}

Here's another example of a stabilizer code, though it doesn't actually encode
any qubits:
the code space is one-dimensional.
It is, however, still a valid stabilizer code by the definition.
\[
\begin{array}{cc}
  Z & Z \\[1mm]
  X & X
\end{array}
\]
Specifically, the code space is the one-dimensional space spanned by an e-bit
$\vert\phi^+\rangle.$

Here's a related example of a stabilizer code whose code space is the
one-dimensional space spanned by a GHZ state
$(\vert 000\rangle + \vert 111\rangle)/\sqrt{2}.$
\[
\begin{array}{ccc}
  Z & Z & \mathbb{I} \\[1mm]
  \mathbb{I} & Z & Z \\[1mm]
  X & X & X
\end{array}
\]

\subsection{Code space dimension}

Suppose that we have a stabilizer code, described by $n$-qubit stabilizer
generators $P_1,\ldots,P_r.$
Perhaps the very first question that comes to mind about this code is, ``How
many qubits does it encode?''

This question has a simple answer.
Assuming that the $n$-qubit stabilizer generators $P_1, \ldots, P_r$ satisfy
the three requirements of the definition (namely, that the stabilizer
generators all commute with one another, that this is a minimal generating set,
and that the code space is nonempty), it must then be that the code space for
this stabilizer code has dimension $2^{n-r},$ so $n-r$ qubits can be encoded
using this code.

Intuitively speaking, we have $n$ qubits to use for this encoding, and each
stabilizer generator effectively ``takes a qubit away'' in terms of how many
qubits we can encode.
Note that this is not about which or how many \emph{errors} can be detected or
corrected, it is only a statement about the dimension of the code space.

For example, for both the 3-bit repetition code and the modified version of
that code for phase-flip errors, we have $n=3$ qubits and $r=2$ stabilizer
generators, and therefore these codes can each encode 1 qubit.
For another example, consider the 5-qubit code: we have 5 qubits and 4
stabilizer generators, so once again the code space has dimension 2, meaning
that one qubit can be encoded using this code.
For one final example, the code whose stabilizer generators are $X\otimes X$
and $Z\otimes Z$ has a one-dimensional code space, spanned by the state
$\vert\phi^+\rangle,$ which is consistent with having $n=2$ qubits and $r=2$
stabilizer generators.

Now let's see how this fact can be proved.
The first step is to observe that, because the stabilizer generators commute,
and because every Pauli operation is its own inverse, every element in the
stabilizer can be expressed as a product
\[
P_1^{a_1} \cdots P_r^{a_r},
\]
where $a_1,\ldots,a_r\in\{0,1\}.$
Equivalently, each element of the stabilizer is obtained by multiplying
together some subset of the stabilizer generators.
Indeed, every stabilizer element can be expressed \emph{uniquely} in this way,
due to the condition that $\{P_1,\ldots,P_r\}$ is a minimal generating set.

Next, define $\Pi_k$ to be the projection onto the space of $+1$-eigenvectors
of $P_k,$ for each $k\in\{1,\ldots,r\}.$
These projections can be obtained by averaging the corresponding Pauli
operations with the identity operation as follows.
\[
\Pi_k = \frac{\mathbb{I}^{\otimes n} + P_k}{2}
\]
The code space $\mathcal{C}$ is the subspace of all vectors that are fixed by
all $r$ of the stabilizer generators $P_1,\ldots,P_r,$ or equivalently, all $r$
of the projections $\Pi_1,\ldots,\Pi_r.$

Given that the stabilizer generators all commute with one another, the
projections $\Pi_1,\ldots,\Pi_r$ must also commute.
This allows us to use a fact from linear algebra, which is that the
\emph{product} of these projections is the projection onto the
\emph{intersection} of the subspaces corresponding to the individual
projections.
That is to say, the product $\Pi_1\cdots\Pi_r$ is the projection onto the code
space $\mathcal{C}.$

We can now expand out the product $\Pi_1\cdots\Pi_r$ using the formulas for
these projections to obtain the following expression.
\[
\Pi_1\cdots\Pi_r
= \biggl(\frac{\mathbb{I}^{\otimes n} +
  P_1}{2}\biggr)\cdots\biggl(\frac{\mathbb{I}^{\otimes n} + P_r}{2}\biggr)
= \frac{1}{2^r}\sum_{a_1,\ldots,a_r \in \{0,1\}} P_1^{a_1}\cdots P_r^{a_r}
\]
In words, the projection onto the code space of a stabilizer code is equal, as
a matrix, to the \emph{average} over all of the elements in the stabilizer of
that code.

Finally, we can compute the dimension of the code space by using the fact that
the dimension of any subspace is equal to the trace of the projection onto that
subspace.
Thus, the dimension of the code space $\mathcal{C}$ is given by the following
formula.
\[
\operatorname{dim}(\mathcal{C}) =
\operatorname{Tr}(\Pi_1\cdots\Pi_r) =
\frac{1}{2^r} \sum_{a_1,\ldots,a_r \in \{0,1\}}
\operatorname{Tr}(P_1^{a_1}\cdots P_r^{a_r})
\]
We can evaluate this expression by making use of a couple of basic facts.
\begin{itemize}
  \item
    We have $P_1^0 \cdots P_r^0 = \mathbb{I}^{\otimes n}$ and therefore
    \[
    \operatorname{Tr}(P_1^{0}\cdots P_r^{0}) = 2^n.
    \]
  \item
    For $(a_1,\ldots,a_r) \neq (0,\ldots,0),$ the product
    $P_1^{a_1}\cdots P_r^{a_r}$ must be $\pm 1$ times a Pauli operation --- but
    we cannot obtain $\mathbb{I}^{\otimes n}$ because this would contradict the
    minimality of the set $\{P_1,\ldots,P_r\},$ and we cannot obtain
    $-\mathbb{I}^{\otimes n}$ because the third condition on the stabilizer
    generators forbids it.
    Therefore, because the trace of every non-identity Pauli operation is zero,
    we obtain
    \[
    \operatorname{Tr}(P_1^{a_1}\cdots P_r^{a_r}) = 0.
    \]
\end{itemize}

\noindent
The dimension of the code space is therefore $2^{n-r}$ as claimed:
\[
\operatorname{dim}(\mathcal{C}) =
\frac{1}{2^r} \sum_{a_1,\ldots,a_r \in \{0,1\}}
\operatorname{Tr}(P_1^{a_1}\cdots P_r^{a_r}) 
= \frac{1}{2^r} \operatorname{Tr}(P_1^{0}\cdots P_r^{0}) 
= 2^{n-r}.
\]

As an aside, we can now see that the assumption that $-\mathbb{I}^{\otimes n}$
is not contained in the stabilizer implies that the code space must contain at
least one quantum state vector.
This is because, as we've just verified, this assumption implies that the code
space has dimension $2^{n-r},$ which cannot be zero.
The converse implication happens to be trivial: if $-\mathbb{I}^{\otimes n}$ is
contained in the stabilizer, then the code space can't possibly contain any
quantum state vectors, because no nonzero vectors are fixed by this operation.

\subsection{Clifford operations and encodings}

Next, we'll briefly discuss how qubits can be encoded using stabilizer codes,
but to do that we first need to introduce \emph{Clifford operations}.

\begin{callout}[title = {Clifford operations}]
  Clifford operations are unitary operations, on any number of qubits, that can
  be implemented by quantum circuits with a restricted set of gates:
  \begin{itemize}
  \item Hadamard gates
  \item $S$ gates
  \item CNOT gates
  \end{itemize}
\end{callout}

Notice that $T$ gates are not included in the list, nor are Toffoli gates and
Fredkin gates.
Not only are those gates not included in the list, but in fact, it's not
possible to implement those gates using the ones listed here; they're not
Clifford operations.
Pauli operations, on the other hand, are Clifford operations because they can
be implemented with sequences of Hadamard and $S$ gates.

That's a simple way to define Clifford operations, but it doesn't explain why
they're defined like this or what's special about this particular collection of
gates.
The real reason Clifford operations are defined like this is that, up to
global phase factors, the Clifford operations are precisely the unitary
operations that always transform Pauli operations into Pauli operations by
conjugation.
To be more precise, an $n$-qubit unitary operation $U$ is equivalent to a
Clifford operation up to a phase factor if, and only if, for \emph{every}
$n$-qubit Pauli operation $P,$ we have
\[
U P U^{\dagger} = \pm Q
\]
for some $n$-qubit Pauli operation $Q.$
(Note that it is not possible to have $U P U^{\dagger} = \alpha Q$ for
$\alpha\notin\{+1,-1\}$ when $U$ is unitary and $P$ and $Q$ are Pauli
operations.
This follows from the fact that the matrix on the left-hand side of the
equation in question is both unitary and Hermitian, and $+1$ and $-1$ are the
only choices for $\alpha$ that allow the right-hand side to be unitary and
Hermitian as well.)

It is straightforward to verify the conjugation property just described when
$U$ is a Hadamard, $S,$ or CNOT gate.
In particular, this is easy for Hadamard gates,
\[
H X H^{\dagger} = Z, \qquad
H Y H^{\dagger} = -Y, \qquad
H Z H^{\dagger} = X,
\]
and $S$ gates,
\[
S X S^{\dagger} = Y, \qquad
S Y S^{\dagger} = -X, \qquad
S Z S^{\dagger} = Z.
\]
For CNOT gates, there are 15 non-identity Pauli operations on two qubits to
check.
Naturally, they can be checked individually --- but the relationships between
CNOT gates and $X$ and $Z$ gates listed (in circuit form) in the previous
lesson, together with the multiplication rules for Pauli matrices, offer a
shortcut to the same conclusion.

Once we know this conjugation property is true for Hadamard, $S,$ and CNOT
gates, we can immediately conclude that it is true for \emph{circuits} composed
of these gates --- which is to say, all Clifford operations.

It is more difficult to prove that the relationship works in the other
direction, which is that if a given unitary operation $U$ satisfies the
conjugation property for Pauli operations, then it must be possible to
implement it (up to a global phase) using just Hadamard, $S,$ and CNOT gates.
This won't be explained in this lesson, but it is true.

Clifford operations are not universal for quantum computation; unlike universal
sets of quantum gates, approximating arbitrary unitary operations to any
desired level of accuracy with Clifford operations is not possible.
Indeed, for a given value of $n,$ there are only \emph{finitely} many $n$-qubit
Clifford operations (up to phase factors).
Performing Clifford operations on standard basis states followed by standard
basis measurements also can't allow us to perform computations that are outside
of the reach of classical algorithms --- because we can \emph{efficiently
simulate} computations of this form classically.
This fact is known as the \emph{Gottesman--Knill theorem}.

\subsubsection{Encoders for stabilizer codes}

A stabilizer code defines a code space of a certain dimension, and we have the
freedom to use that code space however we choose --- nothing forces us to
encode qubits into this code space in a specific way.
It is always possible, however, to use a \emph{Clifford operation} as an
encoder, if we choose to do that.
To be more precise, for any stabilizer code that allows $m$ qubits to be
encoded into $n$ qubits, there's an $n$-qubit Clifford operation $U$ such that,
for any $m$-qubit quantum state vector $\vert\phi\rangle,$ we have that
\[
\vert\psi\rangle = U \bigl(\vert 0^{n-m} \rangle \otimes \vert \phi\rangle\bigr)
\]
is a quantum state vector in the code space of our code that we may interpret
as an encoding of $\vert\phi\rangle.$

This is good because Clifford operations are relatively simple, compared with
arbitrary unitary operations, and there are ways to optimize their
implementation using techniques similar to ones found in the proof of the
Gottesman--Knill theorem.
As a result, circuits for encoding states using stabilizer codes never need to
be too large.
In particular, it is always possible to perform an encoding for an $n$-qubit
stabilizer code using a Clifford operation that requires $O(n^2/\log(n))$
gates.
This is because \emph{every} Clifford operation on $n$ qubits can be
implemented by a circuit of this size.

\begin{figure}[!ht]
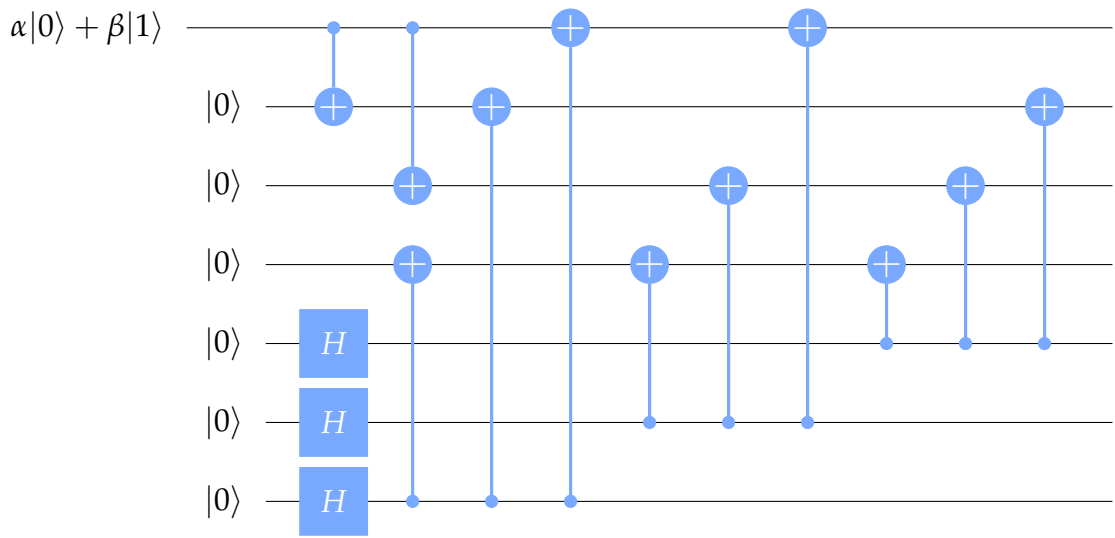

  \begin{center}
   
  \end{center}
  \caption{An encoding circuit for the $7$-qubit Steane code.}
  \label{fig:Steane-encoder}
\end{figure}

For example, Figure~\ref{fig:Steane-encoder} shows an encoder for the 7-qubit
Steane code.
It is indeed a Clifford operation, and as it turns out, this one doesn't even
need $S$ gates.

\subsection{Detecting errors}

For an $n$-qubit stabilizer code described by stabilizer generators
$P_1,\ldots, P_r,$ error detection works in the following way.

To detect errors, all of the stabilizer generators are measured as observables.
There are $r$ stabilizer generators, and therefore $r$ measurement outcomes,
each one being $+1$ or $-1$ (or a binary value if we choose to associate $0$
with $+1$ and $1$ with $-1,$ respectively).
We interpret the $r$ outcomes collectively, as a vector or string, as a
\emph{syndrome.}
The syndrome $(+1,\ldots,+1)$ indicates that no error has been detected, while
at least one $-1$ somewhere within the syndrome indicates an error has been
detected.

Suppose, in particular, that $E$ is an $n$-qubit Pauli operation, representing
a hypothetical error.
(We're only considering Pauli operations as errors, by the way, because the
discretization of errors works the same way for arbitrary stabilizer codes as
it does for the 9-qubit Shor code.)
There are three cases that determine whether or not $E$ is detected as an
error.

\subsubsection{Error detection cases}

\begin{enumerate}
\item
  The operation $E$ is proportional to an element in the stabilizer.
  \[
  E = \pm Q \; \text{for some}\; Q \in \langle P_1,\ldots,P_r\rangle
  \]
  In this case, $E$ must commute with every stabilizer generator, so we obtain
  the syndrome $(+1,\ldots,+1).$  This means that $E$ is not detected as an
  error.

\item
  The operation $E$ is not proportional to an element in the stabilizer, but it
  nevertheless commutes with every stabilizer generator.
  \[
  E\neq \pm Q\; \text{for} \; Q \in \langle P_1,\ldots,P_r\rangle,
  \;\text{but}\; E P_k = P_k E \;\text{for every}\; k\in\{1,\ldots,r\}
  \]
  This is an error that changes vectors in the code space in some nontrivial
  way. But, because $E$ commutes with every stabilizer generator, the syndrome
  is $(+1,\ldots,+1),$ so $E$ goes undetected by the code.

\item
  The operation $E$ anti-commutes with at least one of the stabilizer
  generators.
  \[
  P_k E = -E P_k \; \text{for at least one} \; k\in\{1,\ldots,r\}
  \]
  The syndrome is different than $(+1,\ldots,+1),$ so the error $E$ is detected
  by the code.
\end{enumerate}

In the first case, the error $E$ is not a concern because this operation does
nothing to vectors in the code space, except to possibly inject an irrelevant
global phase: $E \ket{\psi} = \pm\ket{\psi}$ for every encoded state
$\ket{\psi}.$ In essence, this is not actually an error --- whatever nontrivial
action $E$ may have happens outside of the code space --- so it's good that $E$
is not detected as an error, because nothing needs to be done about it.

The second case, intuitively speaking, is the bad case.
It's the \emph{anti-commutation} of an error with a stabilizer generator that
causes a $-1$ to appear somewhere in the syndrome, signaling an error, but that
doesn't happen in this case.
So, we have an error $E$ that does change vectors in the code space in some
nontrivial way, but it goes undetected by the code.
For example, for the 3-bit repetition code, the operation
$E = X\otimes X\otimes X$ falls into this category. 

The fact that such an error $E$ must change some vectors in the code space in a
non-trivial way can be argued as follows.
By the assumption that $E$ commutes with $P_1,\ldots,P_r$ but is not
proportional to a stabilizer element, we can conclude that we would obtain a
new, valid stabilizer code by including $E$ as a stabilizer generator along
with $P_1,\ldots,P_r.$
The code space for this new code, however, has only half the dimension of the
original code space, from which we can conclude that the action of $E$ on the
original code space cannot be proportional to the identity operation.

For the last of the three cases, which is that the error $E$ anti-commutes with
at least one stabilizer generator, the syndrome has at least one $-1$ somewhere
in it, which indicates that something is wrong.
As we have already discussed, the syndrome won't uniquely identify $E$ in
general, so it's still necessary to choose a correction operation for each
syndrome, which might or might not correct the error~$E.$
We'll discuss this step shortly, in the last part of the lesson.

\subsubsection{Distance of a stabilizer code}

As a point of terminology, when we refer to the \emph{distance} of a stabilizer
code, we mean the \emph{minimum weight} of a Pauli operation $E$ that falls
into the second category above --- meaning that it changes the code space in
some nontrivial way, but the code doesn't detect this.
When it is said that a stabilizer code is an $[[n,m,d]]$ stabilizer code, using
double square brackets, this means the following:
\begin{enumerate}
\item
  Encodings are $n$ qubits in length,
\item
  the code allows for the encoding $m$ qubits, and
\item
  the distance of the code is $d.$
\end{enumerate}

As an example, let's consider the 7-qubit Steane code.
Here are the stabilizer generators for this code:
\[
\begin{array}{ccccccc}
  Z & Z & Z & Z & \mathbb{I} & \mathbb{I} & \mathbb{I} \\[1mm]
  Z & Z & \mathbb{I} & \mathbb{I} & Z & Z & \mathbb{I} \\[1mm]
  Z & \mathbb{I} & Z & \mathbb{I} & Z & \mathbb{I} & Z \\[1mm]
  X & X & X & X & \mathbb{I} & \mathbb{I} & \mathbb{I} \\[1mm]
  X & X & \mathbb{I} & \mathbb{I} & X & X & \mathbb{I} \\[1mm]
  X & \mathbb{I} & X & \mathbb{I} & X & \mathbb{I} & X
\end{array}
\]
This code has distance 3, and we can argue this as follows.

First consider any Pauli operation $E$ having weight at most 2, and suppose
this operation commutes with all six stabilizer generators.
We will conclude that $E$ must be the identity operation, which (as always) is
an element of the stabilizer.
This will show that the distance of the code is strictly greater than 2.
Suppose, in particular, that $E$ takes the form
\[
E = P \otimes Q \otimes \mathbb{I} \otimes \mathbb{I} \otimes \mathbb{I}
\otimes \mathbb{I} \otimes \mathbb{I}
\]
for $P$ and $Q$ being possibly non-identity Pauli matrices.
This is just one case, and it is necessary to repeat the argument that follows
for all of the other possible locations for non-identity Pauli matrices among
the tensor factors of $E,$ but the argument is essentially the same for all of
the possible locations.

The operation $E$ commutes with all six stabilizer generators, so it commutes
with these two in particular:
\[
\begin{gathered}
  Z \otimes \mathbb{I} \otimes Z \otimes \mathbb{I} \otimes Z \otimes
  \mathbb{I} \otimes Z\\[1mm]
  X \otimes \mathbb{I} \otimes X \otimes \mathbb{I} \otimes X \otimes
  \mathbb{I} \otimes X
\end{gathered}
\]
The tensor factor $Q$ in our error $E$ lines up with the identity matrix in
both of these stabilizer generators (which is why they were selected).
Given that we have identity matrices in the rightmost 5 positions of $E,$ we
conclude that $P$ must commute with $X$ and $Z,$ for otherwise $E$ would
anti-commute with one of the two generators.
However, the only Pauli matrix that commutes with both $X$ and $Z$ is the
identity matrix, so $P = \mathbb{I}.$

Now that we know this, we can choose two more stabilizer generators that have
an $X$ and a $Z$ in the second position from left, and we draw a similar
conclusion: $Q = \mathbb{I}.$
It is therefore the case that $E$ is the identity operation.

So, there's no way for an error having weight at most 2 to go undetected by
this code, unless the error is the identity operation (which is in the
stabilizer and therefore not actually an error).
On the other hand, there are weight 3 Pauli operations that commute with all
six of these stabilizer generators, but aren't proportional to stabilizer
elements, such as
$\mathbb{I}\otimes\mathbb{I}\otimes\mathbb{I}\otimes\mathbb{I}\otimes X\otimes
X\otimes X$
and $\mathbb{I}\otimes\mathbb{I}\otimes\mathbb{I}\otimes\mathbb{I}\otimes
Z\otimes Z\otimes Z.$
This establishes that this code has distance 3, as claimed.

\subsection{Correcting errors}

The last topic of discussion for this lesson is the \emph{correction} of errors
for stabilizer codes.
As usual, assume that we have a stabilizer code specified by n-qubit stabilizer
generators $P_1, \ldots, P_r.$

The $n$-qubit Pauli operations, as errors that could affect states encoded
using this code, are partitioned into equal-sized collections according to
which syndrome they cause to appear.
There are $2^r$ distinct syndromes and $4^n$ Pauli operations, which means
there are $4^n/2^r$ Pauli operations causing each syndrome.
Any one of these errors could be responsible for the corresponding syndrome.

However, among the $4^n/2^r$ Pauli operations that cause each syndrome, there
are some that should be considered as being equivalent.
In particular, if the product of two Pauli operations is proportional to a
stabilizer element, then those two operations are effectively equivalent as
errors.

Another way to say this is that if we apply a correction operation $C$ to
attempt to correct an error $E,$ then this correction succeeds so long as the
composition $CE$ is proportional to a stabilizer element.
Given that there are $2^r$ elements in the stabilizer, it follows that each
correction operation $C$ corrects $2^r$ different Pauli errors.
This leaves $4^{n-r}$ inequivalent classes of Pauli operations, considered as
errors, that are consistent with each possible syndrome.

This means that, unless $n=r$ (in which case we have a trivial, one-dimensional
code space), we can't possibly correct every error detected by a stabilizer
code.
What we must do instead is to choose just \emph{one} correction operation for
each syndrome, in the hopes of correcting just one class of equivalent errors
that cause this syndrome.

One natural strategy for choosing which correction operation to perform for
each syndrome is to choose the \emph{lowest weight} Pauli operation that, as an
error, causes that syndrome.
There may in fact be multiple operations that tie for the lowest weight error
consistent with a given syndrome, in which case any one of them may be
selected.
The idea is that lower-weight Pauli operations represent more likely
explanations for whatever syndrome has been measured.
This might actually not be the case for some noise models, and one alternative
strategy is to compute the \emph{most likely} error that causes the given
syndrome, based on the chosen noise model.
For this lesson, however, we'll keep things simple and only consider
lowest-weight corrections.

For a distance $d$ stabilizer code, this strategy of choosing the correction
operation to be a lowest weight Pauli operation consistent with the measured
syndrome always allows for the correction of errors having weight strictly less
than half of $d,$ or in other words, weight at most $(d-1)/2.$
This shows, for instance, that the 7-qubit Steane code can correct for any
weight-one Pauli error, and by the discretization of errors, this means that
the Steane code can correct for an arbitrary error on one qubit.

\begin{figure}[b]
  \begin{center}
    \begin{tikzpicture}[>=latex, scale=1.45]

      \draw (-2.75,0) circle(2);
      
      \node[anchor = north] at (-2.75,-2.1) {syndrome $(+1,\ldots,+1)$};
      
      \begin{scope}
        \clip (-2.75,0) circle(2);
        \draw (-4.75,-0.5) -- (-0.75,-0.5);
      \end{scope}

      \node[anchor = north] at (-2.75,-0.8) {%
        \begin{tabular}{c}
          \textcolor{Highlight}{$\pm$ stabilizer}\\
          (weight $< d$)
        \end{tabular}%
      };
      
      \node[anchor = south] at (-2.75,0.3) {%
        \begin{tabular}{c}
          \textcolor{Highlight}{undetected}\\[-0.5mm]
          \textcolor{Highlight}{errors}\\
          (weight $\geq d$)
      \end{tabular}};
      
      \draw (2.75,0) circle(2);
      
      \node[anchor = north] at (2.75,-2.1) {%
        \begin{tabular}{c}
          syndrome $s \neq (+1,\ldots,+1)$\\[0.5mm]
          correction operation $= C$
        \end{tabular}%
      };

      \node[circle, fill, inner sep = 1pt, outer sep = 1.5pt] (E)
      at (1.8,0) {};
      
      \node[anchor = west] at (E) {$E$ (weight $<d/2$)};

      \node[circle, fill, inner sep = 1pt, outer sep = 1.5pt] (S)
      at (-1.3,-0.85) {};
      
      \node[anchor = east] at (S) {$CE$};

      \draw[->, thick, draw = Highlight] (E) -- (S)
      node[anchor = south, rotate = 10.3, pos=0.55] {$C$};

    \end{tikzpicture}   
  \end{center}
  \caption{Composing a lowest-weight correction operation $C$ with a Pauli
    error $E$ having weight less than half of the distance $d$ must give an
    operation $CE$ that is proportional to a stabilizer element.}
  \label{fig:lowest-weight-correction}
\end{figure}
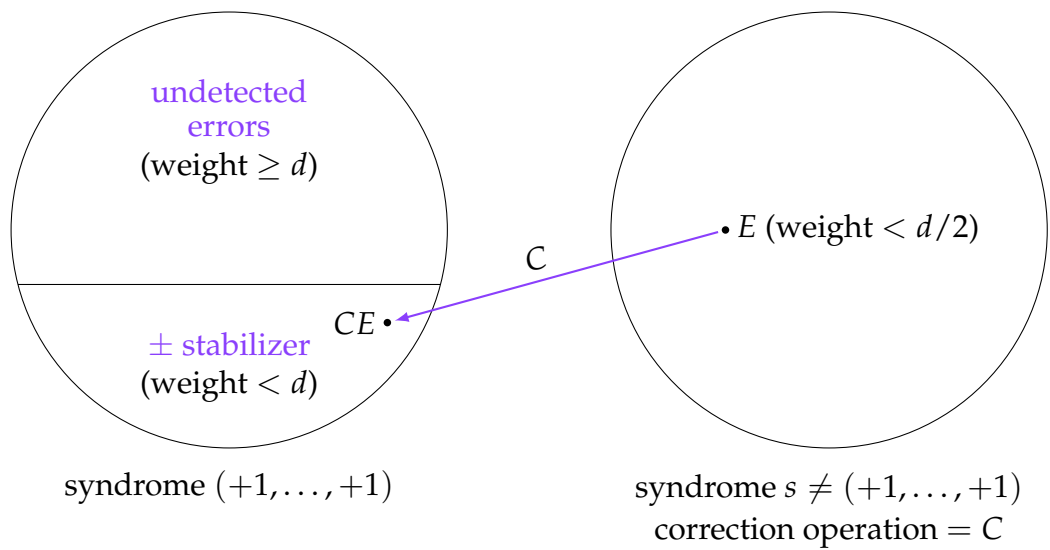

To see how this works, consider the diagram in
Figure~\ref{fig:lowest-weight-correction}.
The circle on the left represents all of the Pauli operations that result in
the syndrome $(+1,\ldots,+1),$ which is the syndrome that suggests that no
errors have occurred and nothing is wrong.
Among these operations we have elements that are proportional to elements of
the stabilizer, and we also have non-trivial errors that change the code space
in some way but aren't detected by the code.
By the definition of distance, every Pauli operation in this category must have
weight at least $d,$ because $d$ is defined as the minimum weight of these
operations.

The circle on the right represents the Pauli operations that result in a
different syndrome $s\neq(+1,\ldots,+1),$ including an error $E$ having weight
strictly less than $d/2$ that we will consider.
The correction operation $C$ chosen for the syndrome $s$ is the lowest weight
Pauli operation in the collection represented by the circle on the right in the
diagram (or any one of them in case there's a tie).
So, it could be that $C = E,$ but not necessarily.
What we can say for certain, however, is that $C$ cannot have weight larger
than the weight of $E,$ because $C$ has minimal weight among the operations in
this collection --- and therefore $C$ has weight strictly less than $d/2.$

Now consider what happens when the correction operation $C$ is applied to
whatever state we obtained after the error $E$ takes place.
Assuming that the original encoding was $\vert\psi\rangle,$ we're left with
$CE\vert\psi\rangle.$
Our goal will be to show that $CE$ is proportional to an element in the
stabilizer, implying that the correction is successful and (up to a global
phase) we're left with the original encoded state $\vert\psi\rangle.$

First, because $E$ and $C$ cause the same syndrome, the composition $CE$ must
commute with every stabilizer generator.
In particular, if $P_k$ is any one of the stabilizer generators, then we must
have
\[
P_k E = \alpha E P_k
\quad\text{and}\quad
P_k C = \alpha C P_k
\]
for the same value of $\alpha\in\{+1,-1\},$ because this is the $k$-th entry in
the syndrome $s$ that both $C$ and $E$ generate.
Hence, we have
\[
P_k (CE) = \alpha C P_k E = \alpha^2 (CE) P_k = (CE) P_k,
\]
so $P_k$ commutes with $CE.$
We've therefore shown that $CE$ belongs in the circle on the left in the
diagram, because it generates the syndrome $(+1,\ldots,+1).$

Second, the composition $CE$ must have weight at most the sum of the weights of
$C$ and $E$ --- which follows from a moment's thought about products of Pauli
operations --- and therefore the weight of $CE$ is strictly less than $d.$
This implies that $CE$ is proportional to an element in the stabilizer of our
code, which is what we wanted to show.
By choosing our correction operations to be lowest-weight representatives of
the set of errors that generate each syndrome, we're therefore guaranteed to
correct any Pauli errors having weight less than half of the distance of the
code.

There is one problem, however.
For stabilizer codes in general, it's a computationally difficult problem to
compute the lowest weight Pauli operation causing a given syndrome.
(Indeed, this is true even for classical codes, which in this context we can
think of as stabilizer codes where we only have $\mathbb{I}$ and $Z$ matrices
appearing as tensor factors within the stabilizer generators.)
So, unlike the encoding step, Clifford operations will not be coming to our
rescue this time.

The solution is to choose specific codes for which good corrections can be
computed efficiently, for which there's no simple recipe.
Simply put, devising stabilizer codes for which good correction operations can
be computed efficiently is part of the artistry of quantum code design.
We'll see this artistry on display in the next lesson.


\lesson{Quantum Code Constructions}
\label{lesson:quantum-code-constructions}

We've seen a few examples of quantum error correcting codes in previous lessons
of this unit, including the 9-qubit Shor code, the 7-qubit Steane code, and
the 5-qubit code.
These codes are undoubtedly interesting and represent a natural place to begin
an exploration of quantum error correction, but a problem with them is that
they can only tolerate a very low error rate.
Correcting an error on one qubit out of five, seven, or nine isn't bad, but in
all likelihood we're going to need to be able to tolerate a lot more errors
than that to make large-scale quantum computing a reality.

In this lesson, we'll take a first look at some more sophisticated quantum
error correcting code constructions, including codes that can tolerate a much
higher error rate than the ones we've seen so far, and that are viewed as
promising candidates for practical quantum error correction.

We'll begin with a class of quantum error correcting codes known as \emph{CSS
codes}, named for Robert Calderbank, Peter Shor, and Andrew Steane, who first
discovered them.
The CSS code construction allows one to take certain \emph{pairs} of classical
error correcting codes and combine them into a single quantum error correcting
code.

The second part of the lesson is on a code known as the \emph{toric code}.
This is a fundamental (and truly beautiful) example of a quantum error
correcting code that can tolerate relatively high error rates.
In fact, the toric code isn't a single example of a quantum error correcting
code, but rather it's an infinite family of codes, one for each positive
integer greater than one.

Finally, in the last part of the lesson, we'll briefly discuss a couple of
other families of quantum codes, including \emph{surface codes} (which are
closely connected to the toric code) and \emph{color codes}.

\section{CSS codes}

\subsection{Classical linear codes}

Classical error correcting codes were first studied in the 1940s, and many
different codes are now known, with the most commonly studied and used codes
falling into a category of codes known as \emph{linear codes.}
We'll see exactly what the word ``linear'' means in this context in just a
moment, but a very simple way to express what linear codes are at this point is
that they're stabilizer codes that happen to be classical.
CSS codes are essentially \emph{pairs} of classical linear codes that are
combined together to create a quantum error correcting code.
So, for the sake of the discussion that follows, we're going to need to
understand a few basic things about classical linear codes.

Let $\Sigma$ be the binary alphabet for this entire discussion.
When we refer to a \emph{classical linear code}, we mean a non-empty set
$\mathcal{C}\subseteq\Sigma^n$ of binary strings of length $n,$ for some
positive integer $n,$ which must satisfy just one basic property:
if $u$ and $v$ are binary strings in $\mathcal{C},$ then the string $u\oplus v$
is also in $\mathcal{C}.$
Here, $u\oplus v$ refers to the bitwise exclusive-OR of $u$ and $v,$ as we
encountered multiple times in
Unit~\ref{unit:fundamentals-of-quantum-algorithms}
\emph{(Fundamentals of Quantum Algorithms)}.

In essence, when we refer to a classical error correcting code as being
\emph{linear,} we're thinking about binary strings of length $n$ as being
$n$-dimensional vectors, where the entries are all either $0$ or $1,$ and
demanding that the code itself forms a linear subspace.
Instead of ordinary vector addition over the real or complex numbers, however,
we're using addition modulo $2,$ which is simply the exclusive-OR.
That is, if we have two \emph{codewords} $u$ and $v,$ meaning that $u$ and $v$
are binary strings in $\mathcal{C},$ then $u + v$ modulo 2, which is to say
$u\oplus v,$ must also be a codeword in $\mathcal{C}.$
Notice, in particular, that this implication must be true even if $u = v.$
This implies that $\mathcal{C}$ must include the all-zero string $0^n,$ because
the bitwise exclusive-OR of any string with itself is the all-zero string.

\subsubsection{Example: the 3-bit repetition code}

The 3-bit repetition code is an example of a classical linear code.
In particular, we have $\mathcal{C} = \{000,111\},$ so, with respect to the
linearity condition, there are just two possible choices for $u$ and two
possible choices for $v.$
It's a trivial matter to go through the four possible pairs to see that we
always get a codeword when we take the bitwise exclusive-OR:
\[
000 \oplus 000 = 000, \quad
000 \oplus 111 = 111, \quad
111 \oplus 000 = 111, \quad
111 \oplus 111 = 000.
\]

\subsubsection{Example: the $[7,4,3]$-Hamming code}

Here's another example of a classical linear code called the $[7,4,3]$-Hamming
code.
It was one of the very first classical error correcting codes ever discovered,
and it consists of these 16 binary strings of length 7.
(Sometimes the $[7,4,3]$-Hamming code is understood to mean the code with these
strings reversed, but we'll take it to be the code containing the strings shown
here.)
\[
\begin{array}{cccc}
  0000000 & 1100001 & 1010010 & 0110011\\[1mm]
  0110100 & 1010101 & 1100110 & 0000111\\[1mm]
  1111000 & 0011001 & 0101010 & 1001011\\[1mm]
  1001100 & 0101101 & 0011110 & 1111111
\end{array}
\]
There is very simple logic behind the selection of these strings, but it's
secondary to the lesson and won't be explained here.
For now, it's enough to observe that this is a classical linear code: XORing
any two of these strings together will always result in another string in the
code.

The notation $[7,4,3]$ (in single square brackets) means something analogous to
the double square bracket notation for stabilizer codes mentioned in the
previous lesson, but here it's for classical linear codes.
In particular, codewords have $7$ bits, we can encode $4$ bits using the code
(because there are $16 = 2^4$ codewords), and it happens to be a distance $3$
code, which means that any two distinct codewords must differ in at least $3$
positions --- so at least $3$ bits must be flipped to change one codeword into
another.
The fact that this is a distance $3$ code implies that it can correct for up to
one bit-flip error.

\subsubsection{Describing classical linear codes}

The examples just mentioned are very simple examples of classical linear codes,
but even the $[7,4,3]$-Hamming code looks somewhat mysterious when the
codewords are simply listed.
There are better, more efficient ways to describe classical linear codes,
including the following two ways.

\begin{trivlist}
  \item
    \textbf{Generators}.
    One way to describe a classical linear code is with a minimal list of
    codewords that \emph{generates} the code, meaning that by taking all of the
    possible subsets of these codewords and XORing them together, we get the
    entire code.

    That is, the strings $u_1,\ldots,u_m\in\Sigma^n$ generate the classical
    linear code $\mathcal{C}$ if
    \[
    \mathcal{C} = \bigl\{\alpha_1 u_1 \oplus \cdots \oplus \alpha_m
    u_m\,:\,\alpha_1,\ldots,\alpha_m\in\{0,1\}\bigr\},
    \]
    with the understanding that $\alpha u = u$ when $\alpha = 1$ and
    $\alpha u = 0^n$ when $\alpha = 0,$ and we say that this list is
    \emph{minimal} if removing one of the strings generates a smaller code.
    A natural way to think about such a description is that the collection
    $\{u_1,\ldots,u_m\}$ forms a \emph{basis} for $\mathcal{C}$ as a subspace,
    where we're thinking about strings as vectors with binary-valued entries,
    keeping in mind that we're working in a vector space where arithmetic is
    done modulo $2.$

  \item
    \textbf{Parity checks}.
    Another natural way to describe a classical linear code is by
    \emph{parity checks} --- meaning a minimal list of binary strings for which
    the strings in the code are precisely the ones whose \emph{binary dot
    product} with every one of these parity check strings is zero.
    Similar to the bitwise exclusive-OR, the binary dot product appeared
    several times in Unit~\ref{unit:fundamentals-of-quantum-algorithms}
    \emph{(Fundamentals of Quantum Algorithms)}.

    That is, the strings $v_1,\ldots,v_r\in\Sigma^n$ are parity check strings
    for the classical linear code $\mathcal{C}$ if
    \[
    \mathcal{C} = \bigl\{ u\in \Sigma^n\,:\, u\cdot v_1 = \cdots = u \cdot v_r
    = 0 \bigr\},
    \]
    and this set of strings is \emph{minimal} if removing one results in a
    larger code.
    These are called parity check strings because $u$ has binary dot product
    equal to zero with $v$ if and only if the bits of $u$ in positions where
    $v$ has 1s have even parity.
   So, to determine if a string $u$ is in the code $\mathcal{C},$ it suffices
   to check the parity of certain subsets of the bits of $u.$
\end{trivlist}

\noindent
An important thing to notice here is that the binary dot product is not an
inner product in a formal sense. In particular, when two strings have binary
dot product equal to zero, it doesn't mean that they're orthogonal in the usual
way we think about orthogonality. For example, the binary dot product of the
string $11$ with itself is zero --- so it is possible that a parity check
string for a classical linear code is itself in the code.

Classical linear codes over the binary alphabet always include a number of
strings that's a power of $2$ --- and for a single classical linear code
described in the two different ways just described, it will always be the case
that $n = m + r.$
In particular, if we have a minimal set of $m$ generators, then the code
encodes $m$ bits and we'll necessarily have $2^m$ codewords;
and if we have a minimal set of $r$ parity check strings, then we'll have
$2^{n-r}$ codewords.
So, each generator doubles the size of the code space while each parity check
string halves the size of the code space.

For example, the 3-bit repetition code is a linear code, so it can be described
in both of these ways.
In particular, there's only one choice for a generator that works: $111.$
We can alternatively describe the code with two parity check strings, such as
$110$ and $011$ --- which should look familiar from our previous discussions of
this code --- or we could instead take the parity check strings to be $110$ and
$101,$ or $101$ and $011.$ (Generators and parity check strings are generally
not unique for a given classical linear code.)

For a second example, consider the $[7,4,3]$-Hamming code.
Here's one choice for a list of generators that works.
\[
\begin{array}{c}
  1111000\\[1mm]
  0110100\\[1mm]
  1010010\\[1mm]
  1100001
\end{array}
\]
And here's a choice for a list of parity checks for this code.
\[
\begin{array}{c}
  1111000\\[1mm]
  1100110\\[1mm]
  1010101
\end{array}
\]
Here, by the way, we see that \emph{all} of our parity check strings are
themselves in the code.

One final remark about classical linear codes, which connects them to the
stabilizer formalism, is that parity check strings are equivalent to stabilizer
generators that only consist of $Z$ and identity Pauli matrices.
For instance, the parity check strings $110$ and $011$ for the 3-bit repetition
code correspond precisely to the stabilizer generators $Z\otimes Z\otimes
\mathbb{I}$ and $\mathbb{I}\otimes Z\otimes Z,$ which is consistent with the
discussions of Pauli observables from the previous lesson.

\subsection{Definition of CSS codes}

CSS codes are stabilizer codes obtained by combining together certain
\emph{pairs} of classical linear codes.
This doesn't work for two arbitrary classical linear codes --- the two codes
must have a certain relationship.
Nevertheless, this construction opens up many possibilities for quantum error
correcting codes, based in part on over 75 years of classical coding theory.

In the stabilizer formalism, stabilizer generators containing only $Z$ and
identity Pauli matrices are equivalent to parity checks, as we just observed
for the 3-bit repetition code.
For another example, consider the following parity check strings for the
$[7,4,3]$-Hamming code.
\[
\begin{array}{c}
  1111000\\[1mm]
  1100110\\[1mm]
  1010101
\end{array}
\]
These parity check strings correspond to the following stabilizer generators
(written without tensor product symbols), which we obtain by replacing each $1$
by a $Z$ and each $0$ by an $\mathbb{I}.$
These are three of the six stabilizer generators for the 7-qubit Steane code.
\[
\begin{array}{ccccccc}
  Z & Z & Z & Z & \mathbb{I} & \mathbb{I} & \mathbb{I} \\[1mm]
  Z & Z & \mathbb{I} & \mathbb{I} & Z & Z & \mathbb{I} \\[1mm]
  Z & \mathbb{I} & Z & \mathbb{I} & Z & \mathbb{I} & Z
\end{array}
\]
Let us give the name \emph{$Z$ stabilizer generators} to stabilizer generators
like this, meaning that they only have Pauli $Z$ and identity tensor factors
--- so $X$ and $Y$ Pauli matrices never occur in $Z$ stabilizer generators.

We can also consider stabilizer generators where only $X$ and identity Pauli
matrices appear as tensor factors.
Stabilizer generators like this can be viewed as being analogous to
$Z$ stabilizer generators, except that they describe parity checks in the
$\{\vert+\rangle,\vert-\rangle\}$ basis rather than the standard basis.
Stabilizer generators of this form are called \emph{$X$ stabilizer generators}
--- so no $Y$ or $Z$ Pauli matrices are allowed this time.

For example, consider the remaining three stabilizer generators from the
7-qubit Steane code.
\[
\begin{array}{ccccccc}
  X & X & X & X & \mathbb{I} & \mathbb{I} & \mathbb{I} \\[1mm]
  X & X & \mathbb{I} & \mathbb{I} & X & X & \mathbb{I} \\[1mm]
  X & \mathbb{I} & X & \mathbb{I} & X & \mathbb{I} & X
\end{array}
\]
They follow exactly the same pattern from the $[7,4,3]$-Hamming code as the
$Z$ stabilizer generators, except this time we substitute $X$ for $1$ rather
than $Z.$
What we obtain from just these three stabilizer generators is a code that
includes the 16 states shown here, which we get by applying Hadamard operations
to the standard basis states that correspond to the strings in the
$[7,4,3]$-Hamming code.
(Of course, the code space for this code also includes linear combinations of
these states.)
\[
\begin{array}{cccc}
  \vert {+++++++} \rangle \quad & \vert {--++++-} \rangle \quad
  & \vert {-+-++-+} \rangle \quad & \vert {+--++--} \rangle \\
  \vert {+--+-++} \rangle \quad & \vert {-+-+-+-} \rangle \quad
  & \vert {--++--+} \rangle \quad & \vert {++++---} \rangle \\
  \vert {----+++} \rangle \quad & \vert {++--++-} \rangle \quad
  & \vert {+-+-+-+} \rangle \quad & \vert {-++-+--} \rangle \\
  \vert {-++--++} \rangle \quad & \vert {+-+--+-} \rangle \quad
  & \vert {++----+} \rangle \quad & \vert {-------} \rangle
\end{array}
\]

We can now define CSS codes in very simple terms.

\begin{callout}[title = {CSS codes}]
  A CSS code is a stabilizer code that can be expressed using only $X$ and $Z$
  stabilizer generators.
\end{callout}

\noindent
That is, CSS codes are stabilizer codes for which we have stabilizer generators
in which no Pauli $Y$ matrices appear, and for which $X$ and $Z$ never appear
in the \emph{same} stabilizer generator.

To be clear, by this definition, a CSS code is one for which it is
\emph{possible} to choose just $X$ and $Z$ stabilizer generators --- but we
must keep in mind that there is freedom in how we choose stabilizer generators
for stabilizer codes.
Thus, there will generally be different choices for the stabilizer generators
of a CSS code that don't happen to be $X$ or $Z$ stabilizer generators (in
addition to at least one choice for which they are).

Here's a very simple example of a CSS code that includes both a $Z$ stabilizer
generator and an $X$ stabilizer generator:
\[
\begin{array}{cc}
  Z & Z \\[1mm]
  X & X
\end{array}
\]
It's clear that this is a CSS code, because the first stabilizer generator is a
$Z$ stabilizer generator and the second is an $X$ stabilizer generator.
Of course, a CSS code must also be a valid stabilizer code --- meaning that the
stabilizer generators must commute, form a minimal generating set, and fix at
least one quantum state vector.
These requirements happen to be simple to observe for this code.
As we noted in the previous lesson, the code space for this code is the
one-dimensional space spanned by the $\vert\phi^+\rangle$ Bell state.
The fact that both stabilizer generators fix this state is apparent by
considering the following two expressions of an e-bit, together with an
interpretation of these stabilizer generators as parity checks in the
$\{\vert 0\rangle, \vert 1\rangle\}$ and $\{\vert{+}\rangle, \vert{-}\rangle\}$
bases.
\[
\vert\phi^+\rangle
= \frac{\vert 0\rangle\vert 0\rangle + \vert 1\rangle\vert 1\rangle}{\sqrt{2}}
= \frac{\vert +\rangle\vert +\rangle + \vert -\rangle\vert -\rangle}{\sqrt{2}}
\]

The 7-qubit Steane code is another example of a CSS code.
\[
\begin{array}{ccccccc}
  Z & Z & Z & Z & \mathbb{I} & \mathbb{I} & \mathbb{I} \\[1mm]
  Z & Z & \mathbb{I} & \mathbb{I} & Z & Z & \mathbb{I} \\[1mm]
  Z & \mathbb{I} & Z & \mathbb{I} & Z & \mathbb{I} & Z \\[1mm]
  X & X & X & X & \mathbb{I} & \mathbb{I} & \mathbb{I} \\[1mm]
  X & X & \mathbb{I} & \mathbb{I} & X & X & \mathbb{I} \\[1mm]
  X & \mathbb{I} & X & \mathbb{I} & X & \mathbb{I} & X
\end{array}
\]
Here we have three $Z$ stabilizer generators and three $X$ stabilizer
generators, and we've already verified that this is a valid stabilizer code.

And the 9-qubit Shor code is another example.
\[
\begin{array}{ccccccccc}
  Z & Z & \mathbb{I} &
  \mathbb{I} & \mathbb{I} & \mathbb{I} &
  \mathbb{I} & \mathbb{I} & \mathbb{I}\\[1mm]
  \mathbb{I} & Z & Z &
  \mathbb{I} & \mathbb{I} & \mathbb{I} &
  \mathbb{I} & \mathbb{I} & \mathbb{I}\\[1mm]
  \mathbb{I} & \mathbb{I} & \mathbb{I} &
  Z & Z & \mathbb{I} &
  \mathbb{I} & \mathbb{I} & \mathbb{I}\\[1mm]
  \mathbb{I} & \mathbb{I} & \mathbb{I} &
  \mathbb{I} & Z & Z &
  \mathbb{I} & \mathbb{I} & \mathbb{I}\\[1mm]
  \mathbb{I} & \mathbb{I} & \mathbb{I} &
  \mathbb{I} & \mathbb{I} & \mathbb{I} &
  Z & Z & \mathbb{I}\\[1mm]
  \mathbb{I} & \mathbb{I} & \mathbb{I} &
  \mathbb{I} & \mathbb{I} & \mathbb{I} &
  \mathbb{I} & Z & Z\\[1mm]
  X & X & X & X & X & X &
  \mathbb{I} & \mathbb{I} & \mathbb{I}\\[1mm]
  \mathbb{I} & \mathbb{I} & \mathbb{I}&
  X & X & X & X & X & X
\end{array}
\]
This time we have six $Z$ stabilizer generators and just two $X$ stabilizer
generators.
This is fine, there doesn't need to be a balance or a symmetry between the two
types of generators (though there often is).

Once again, it is critical that CSS codes are valid stabilizer codes, and in
particular each $Z$ stabilizer generator must commute with each $X$ stabilizer
generator.
So, not every collection of $X$ and $Z$ stabilizer generators defines a valid
CSS code.

\subsection{Error detection and correction}

With regard to error detection and correction, CSS codes in general have a
similar characteristic to the 9-qubit Shor code, which is that $X$ and $Z$
errors can be detected and corrected completely independently; the $Z$
stabilizer generators describe a code that protects against bit-flips, and the
$X$ stabilizer generators describe a code that independently protects against
phase-flips.
This works because $Z$ stabilizer generators necessarily commute with $Z$
errors, as well as $Z$ operations that are applied as corrections, so they're
completely oblivious to both, and likewise for $X$ stabilizer generators,
errors, and corrections.

As an example, let's consider the 7-qubit Steane code.
\[
\begin{array}{ccccccc}
  Z & Z & Z & Z & \mathbb{I} & \mathbb{I} & \mathbb{I} \\[1mm]
  Z & Z & \mathbb{I} & \mathbb{I} & Z & Z & \mathbb{I} \\[1mm]
  Z & \mathbb{I} & Z & \mathbb{I} & Z & \mathbb{I} & Z \\[1mm]
  X & X & X & X & \mathbb{I} & \mathbb{I} & \mathbb{I} \\[1mm]
  X & X & \mathbb{I} & \mathbb{I} & X & X & \mathbb{I} \\[1mm]
  X & \mathbb{I} & X & \mathbb{I} & X & \mathbb{I} & X
\end{array}
\]
The basic idea for this code is now apparent: it's a $[7,4,3]$-Hamming code for
bit-flip errors and a $[7,4,3]$-Hamming code for phase-flip errors.
The fact that the $X$ and $Z$ stabilizer generators commute is perhaps good
fortune, for this wouldn't be a valid stabilizer code if they didn't.
But there are, in fact, many known examples of classical linear codes that
yield a valid stabilizer code when used in a similar way.

In general, suppose we have a CSS code for which the $Z$ stabilizer generators
allow for the correction of up to $j$ bit-flip errors, and the $X$ stabilizer
generators allow for the correction of up to $k$ phase-flip errors.
For example, $j = 1$ and $k = 1$ for the Steane code, given that the
$[7,4,3]$-Hamming code can correct one bit-flip.
It then follows, by the discretization of errors, that this CSS code can
correct for \emph{any error} on a number of qubits up to the \emph{minimum} of
$j$ and $k.$
This is because, when the syndrome is measured, an arbitrary error on this
number of qubits effectively collapses probabilistically into some combination
of $X$ errors, $Z$ errors, or both --- and then the $X$ errors and $Z$ errors
are detected and corrected independently.

In summary, provided that we have two classical linear codes (or two copies of
a single classical linear code) that are compatible, in that they define $X$
and $Z$ stabilizer generators that commute, the CSS code we obtain by combining
them inherits the error correction properties of those two codes, in the sense
just described.

Notice that there is a price to be paid though, which is that we can't
\emph{encode} as many qubits as we could bits with the two classical codes.
This is because the total number of stabilizer generators for the CSS code is
the \emph{sum} of the number of parity checks for the two classical linear
codes, and each stabilizer generator cuts the dimension of the code space in
half.
For example, the $[7,4,3]$-Hamming code allows for the encoding of four
classical bits, because we have just three parity check strings for this code,
whereas the 7-qubit Steane code only encodes one qubit, because it has six
stabilizer generators.

\subsection{Code spaces of CSS codes}

The last thing we'll do in this discussion of CSS codes is to consider the
\emph{code spaces} of these codes.
This will give us an opportunity to examine in greater detail the relationship
that must hold between two classical linear codes in order for them to be
compatible, in the sense that they can be combined together to form a CSS code.

Consider any CSS code on $n$ qubits, and let $z_1, \ldots, z_s \in \Sigma^n$ be
the $n$-bit parity check strings that correspond to the $Z$ stabilizer
generators of this code.
This means that the classical linear code described by just the $Z$ stabilizer
generators, which we'll name $\mathcal{C}_Z,$ takes the following form.
\[
\mathcal{C}_Z = \bigl\{ u \in \Sigma^n \,:\, u \cdot z_1 = \cdots = u \cdot z_s
= 0 \bigr\}
\]
In words, the classical linear code $\mathcal{C}_Z$ contains every string whose
binary dot product with every one of the parity check strings $z_1, \ldots,
z_s$ is zero.

Along similar lines, let us take $x_1,\ldots,x_t\in\Sigma^n$ to be the $n$-bit
parity check strings corresponding to the $X$ stabilizer generators of our
code.
Thus, the classical linear code corresponding to the $X$ stabilizer generators
takes this form.
\[
\mathcal{C}_X = \bigl\{ u \in \Sigma^n \,:\, u \cdot x_1 = \cdots = u \cdot x_t
= 0 \bigr\}
\]
The $X$ stabilizer generators alone therefore describe a code that's similar to
this code, but in the $\{\vert {+}\rangle,\vert {-}\rangle\}$ basis rather than
the standard basis.

Now we'll introduce two new classical linear codes that are derived from the
same choices of strings, $z_1,\ldots,z_s$ and $x_1,\ldots,x_t,$ but where we
take these strings as \emph{generators} rather than parity check strings.
In particular, we obtain these two codes.
\[
\begin{aligned}
  \mathcal{D}_Z & =
  \bigl\{ \alpha_1 z_1 \oplus \cdots \oplus
  \alpha_s z_s \,:\, \alpha_1,\ldots,\alpha_s \in \{0,1\}
  \bigr\}\\[1mm]
  \mathcal{D}_X & =
  \bigl\{ \alpha_1 x_1 \oplus \cdots \oplus
  \alpha_t x_t \,:\, \alpha_1,\ldots,\alpha_t \in \{0,1\}
  \bigr\}
\end{aligned}
\]
These are known as the \emph{dual codes} of the codes defined previously:
$\mathcal{D}_Z$ is the dual code of $\mathcal{C}_Z$ and $\mathcal{D}_X$ is the
dual code of $\mathcal{C}_X.$
It may not be clear at this point why these dual codes are relevant, but they
turn out to be quite relevant for multiple reasons, including the two reasons
explained in the following paragraphs.

First, the conditions that must hold for two classical linear codes
$\mathcal{C}_Z$ and $\mathcal{C}_X$ to be compatible, in the sense that they
can be paired together to form a CSS code, can be described in simple terms by
referring to the dual codes.
Specifically, it must be that $\mathcal{D}_Z\subseteq\mathcal{C}_X,$ or
equivalently, that $\mathcal{D}_X\subseteq\mathcal{C}_Z.$
In words, the dual code $\mathcal{D}_Z$ includes the strings corresponding to
$Z$ stabilizer generators, and their containment in $\mathcal{C}_X$ is
equivalent to the binary dot product of each of these strings with the ones
corresponding to the $X$ stabilizer generators being zero.
That, in turn, is equivalent to each $Z$ stabilizer generator commuting with
each $X$ stabilizer generator.
Alternatively, by reversing the roles of the $X$ and $Z$ stabilizer generators
and starting from the containment $\mathcal{D}_X\subseteq\mathcal{C}_Z,$ we can
reach the same conclusion.

Second, by referring to the dual codes, we can easily describe the code spaces
of a given CSS code.
In particular, the code space is spanned by vectors of the following form.
\[
\vert u \oplus \mathcal{D}_X\rangle
= \frac{1}{\sqrt{2^t}} \sum_{v\in\mathcal{D}_X} \vert u \oplus v\rangle \qquad
\text{(for all $u\in\mathcal{C}_Z$)}
\]
In words, these vectors are uniform superpositions over the strings in the dual
code $\mathcal{D}_X$ of the code corresponding to the $X$ stabilizer
generators, shifted by (in other words, bitwise XORed with) strings in the code
$\mathcal{C}_Z$ corresponding to the $Z$ stabilizer generators.
To be clear, different choices for the shift --- represented by the string $u$
in this expression --- can result in the same vector.
So, these states aren't all distinct, but collectively they span the entire
code space.

Here's an intuitive explanation for why such vectors are both in the code space
and span it.
Consider the $n$-qubit standard basis state $\vert u\rangle,$ for some
arbitrary $n$-bit string $u,$ and suppose that we \emph{project} this state
onto the code space.
That is to say, letting $\Pi$ denote the projection onto the code space of our
CSS code, consider the vector $\Pi\vert u\rangle.$
There are two cases:

\begin{trivlist}
\item \textbf{Case 1:} $u\in\mathcal{C}_Z$.
  This implies that each $Z$ stabilizer generator of our CSS code acts
  trivially on $\vert u\rangle.$
  The $X$ stabilizer generators, on the other hand, each simply flip some of
  the bits of $\vert u\rangle.$
  In particular, for each generator $v$ of $\mathcal{D}_X,$ the $X$ stabilizer
  generator corresponding to $v$ transforms $\vert u\rangle$ into $\vert
  u\oplus v\rangle.$
  By characterizing the projection $\Pi$ as the \emph{average} over the
  elements of the stabilizer (as we saw in the previous lesson), we obtain this
  formula:
  \[
  \Pi \vert u \rangle = \frac{1}{2^t} \sum_{v\in\mathcal{D}_{X}} \vert u \oplus
  v\rangle = \frac{1}{\sqrt{2^t}} \vert u \oplus \mathcal{D}_X\rangle.
  \]
\item \textbf{Case 2:} $u\notin\mathcal{C}_Z.$
  This implies that at least one of the parity checks corresponding to the $Z$
  stabilizer generators fails, which is to say that $\vert u\rangle$ must be a
  $-1$ eigenvector of at least one of the $Z$ stabilizer generators. The code
  space of the CSS code is the intersection of the $+1$ eigenspaces of the
  stabilizer generators. So, as a $-1$ eigenvector of at least one of the $Z$
  stabilizer generators, $\vert u\rangle$ is therefore orthogonal to the code
  space:
  \[
  \Pi\vert u\rangle = 0.
  \]
\end{trivlist}

And now, as we range over all $n$-bit strings $u,$ discard the ones for which
$\Pi\vert u\rangle = 0,$ and normalize the remaining ones, we obtain the
vectors described previously, which demonstrates that they span the code space.

We can also use the symmetry between $X$ and $Z$ stabilizer generators to
describe the code space in a similar but different way.
In particular, it is the space spanned by vectors having the following form.
\[
H^{\otimes n} \vert u \oplus \mathcal{D}_Z\rangle
= \frac{1}{\sqrt{2^s}} \sum_{v\in\mathcal{D}_Z}
H^{\otimes n}\vert u \oplus v\rangle \qquad \text{(for $u\in\mathcal{C}_X$)}
\]
In essence, $X$ and $Z$ have been swapped in each instance in which they appear
--- but we must also swap the standard basis for the
$\{\vert+\rangle,\vert-\rangle\}$ basis, which is why the Hadamard operations
are included.

As an example, let us consider the 7-qubit Steane code.
The parity check strings for both the $X$ and $Z$ stabilizer generators are the
same: $1111000,$ $1100110,$ and $1010101.$
The codes $\mathcal{C}_X$ and $\mathcal{C}_Z$ are therefore the same; both are
equal to the $[7,4,3]$-Hamming code.
\[
\mathcal{C}_X = \mathcal{C}_Z =
\left\{
\begin{array}{llll}
  0000000, & 0000111, & 0011001, & 0011110,\\
  0101010, & 0101101, & 0110011, & 0110100,\\
  1001011, & 1001100, & 1010010, & 1010101,\\
  1100001, & 1100110, & 1111000, & 1111111
\end{array}
\right\}
\]
The dual codes $\mathcal{D}_X$ and $\mathcal{D}_Z$ are therefore also the same.
We have three generators, so we obtain eight strings.
\[
\mathcal{D}_X = \mathcal{D}_Z =
\left\{
\begin{array}{llll}
  0000000, & 0011110, & 0101101, & 0110011,\\
  1001011, & 1010101, & 1100110, & 1111000
\end{array}
\right\}
\]
These strings are all contained in the $[7,4,3]$-Hamming code, and so the CSS
condition is satisfied: $\mathcal{D}_Z \subseteq \mathcal{C}_X,$ or
equivalently, $\mathcal{D}_X \subseteq \mathcal{C}_Z.$

Given that $\mathcal{D}_X$ contains half of all of the strings in
$\mathcal{C}_Z,$ there are only two different vectors $\vert u\oplus
\mathcal{D}_X\rangle$ that can be obtained by choosing $u\in\mathcal{C}_Z.$
This is expected, because the 7-qubit Steane code has a two-dimensional code
space.
We can use the two states we obtain in this way to encode the logical state
$\vert 0\rangle$ and $\vert 1\rangle$ as follows.
\[
\scalebox{0.78}{%
  $\displaystyle
  \begin{aligned}
    \vert 0\rangle & \mapsto
    \frac{
      \vert 0000000\rangle + \vert 0011110\rangle + \vert 0101101\rangle + \vert
      0110011\rangle + \vert 1001011\rangle + \vert 1010101\rangle + \vert
      1100110\rangle + \vert 1111000\rangle}{\sqrt{8}}\\[4mm]
    \vert 1\rangle & \mapsto
    \frac{
      \vert 0000111\rangle + \vert 0011001\rangle + \vert 0101010\rangle + \vert
      0110100\rangle + \vert 1001100\rangle + \vert 1010010\rangle + \vert
      1100001\rangle + \vert 1111111\rangle}{\sqrt{8}}
  \end{aligned}$}
\]

As usual, this choice isn't forced on us --- we're free to use the code space
to encode qubits however we choose.
This encoding is, however, consistent with the example of an encoding circuit
for the 7-qubit Steane code in the previous lesson.

\section{The toric code}

Next we'll discuss a specific CSS code known as the \emph{toric code}, which
was discovered by Alexei Kitaev in 1997.
In fact, the toric code isn't a single code, but rather it's a family of codes,
one for each positive integer starting from 2.
These codes possess a few key properties:

\begin{itemize}
\item
  The stabilizer generators have \emph{low weight}, and in particular they all
  have weight four. In coding theory parlance, the toric code is an example of a
  quantum low-density parity check code, or \emph{quantum LDPC code} (where
  \emph{low} means 4 in this case). This is nice because each stabilizer
  generator measurement doesn't need to involve too many qubits.

\item
  The toric code has \emph{geometric locality}. This means that not only do the
  stabilizer generators have low weight, but it's also possible to arrange the
  qubits spatially so that each of the stabilizer generator measurements only
  involves qubits that are close together. In principle, this makes these
  measurements easier to implement than if they involved spatially distant
  qubits.

\item
  Members of the toric code family have increasingly \emph{large distance} and
  can tolerate a relatively \emph{high error rate}.
\end{itemize}

\subsection{Toric code description}

\begin{figure}[!ht]
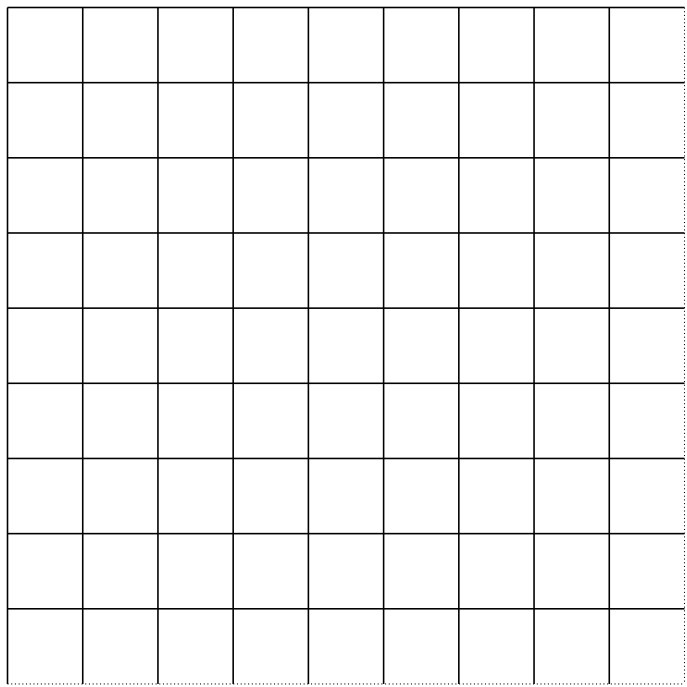

  \begin{center}
    
    
  \end{center}
  \caption{A $9\times 9$ lattice.}
  \label{fig:9-by-9-lattice}
\end{figure}

Let $L\geq 2$ be a positive integer, and consider an $L\times L$ lattice with
so-called \emph{periodic boundaries}.
For example, Figure~\ref{fig:9-by-9-lattice} depicts an $L\times L$ lattice for
$L=9.$
Notice that the lines on the right and on the bottom are dotted lines.
This is meant to indicate that dotted line on the right is the \emph{same line}
as the line all the way on the left, and similarly, the dotted line on the
bottom is the same line as the one on the very top.

To realize this sort of configuration physically requires three dimensions.
In particular, we could form the lattice into a cylinder by first matching up
the left and right sides, and then bend the cylinder around so that the circles
at the ends, which used to be the top and bottom edges of the lattice, meet.
Or we could match up the top and bottom first and then the sides; it works both
ways and doesn't matter which we choose for the purposes of this discussion.

\begin{figure}[t]
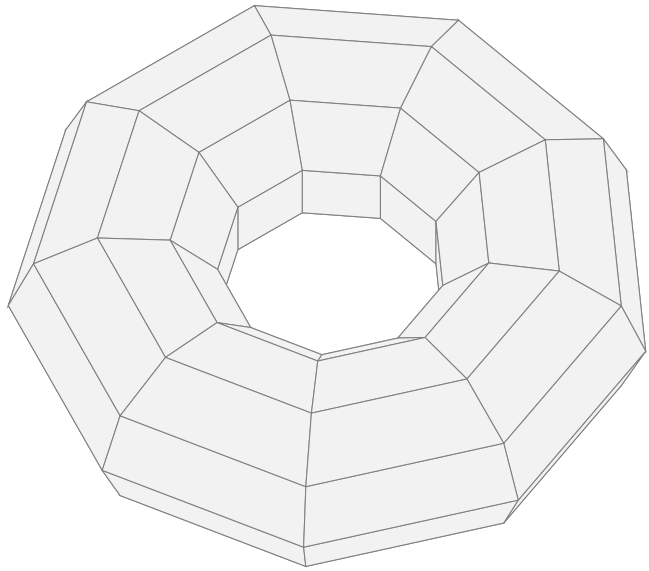

  \begin{center}
    %

  \end{center}
  \caption{A $9\times 9$ lattice with periodic boundaries embedded on the
    surface of a torus.}
  \label{fig:torus}
\end{figure}

What we obtain is a \emph{torus} --- or, in other words, a doughnut (although
thinking about it as an inner tube of a tire is perhaps a better image to have
in mind because this isn't a solid: the lattice has become just the
\emph{surface} of a torus). This is where the name toric code comes from.

The way one can ``move around'' on a torus like this, between adjacent points
on the lattice, will likely be familiar to those that have played old-school
video games, where moving off the top of the screen causes you emerge on the
bottom, and likewise for the left and right edges of the screen.
This is how we will view this lattice with periodic boundaries, as opposed to
speaking specifically about a torus in 3-dimensional space.

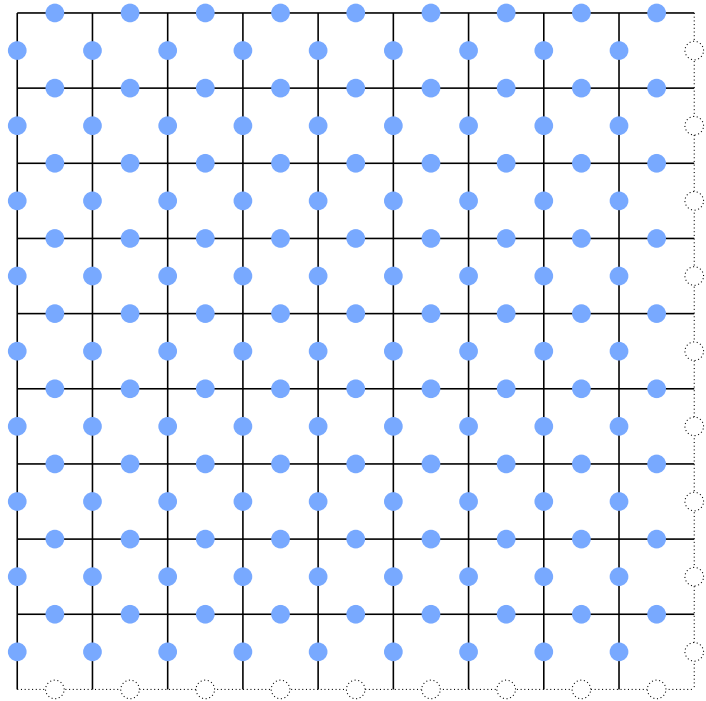
\begin{figure}[!ht]
  \begin{center}
    \begin{tikzpicture}[
        scale=1,
        qubit/.style={%
          circle,
          inner sep=2.5pt,
          fill=CircuitBlue
        },
        dotqubit/.style={%
          circle,
          inner sep=2.5pt,
          draw,
          densely dotted,
          fill=white}
      ]
      
      \def\L{9}
      \pgfmathtruncatemacro{\n}{\L - 1}

      \foreach \x in {0,1,...,\L} {
        \foreach \y in {0,1,...,\L} {
          \coordinate(L\x_\y) at (\x,\L-\y);
        }
      }

      \foreach \x in {0,...,\L} {
        \foreach \y in {0,...,\L} {
          \coordinate(Q\x_\y) at (\x+0.5,\L-\y);
          \coordinate(R\x_\y) at (\x,\L-\y-0.5);
        }
      }
      
      \foreach \z in {0,...,\n} {
        \draw[semithick] (L\z_0) -- (L\z_\L);
        \draw[semithick] (L0_\z) -- (L\L_\z);
      }
      
      \draw[densely dotted] (L0_\L) -- (L\L_\L);
      \draw[densely dotted] (L\L_0) -- (L\L_\L);
      
      \foreach \x in {0,...,\n} {
        \foreach \y in {0,...,\n} {                      
          \node[qubit] at (R\x_\y) {};
          \node[qubit] at (Q\x_\y) {};
        }
      }

      \foreach \z in {0,...,\n} {
        \node[dotqubit] at (Q\z_\L) {};
        \node[dotqubit] at (R\L_\z) {};
      }      
      
    \end{tikzpicture}
  \end{center}
  \caption{Qubits, indicated by blue circles, are placed on the edges of the
    lattice.}
  \label{fig:qubits-on-lattice-edges}
\end{figure}

Next, qubits are arranged on the \emph{edges} of this lattice, as illustrated
in Figure~\ref{fig:qubits-on-lattice-edges}, where qubits are indicated by
solid blue circles.
Note that the qubits placed on the dotted lines aren't solid because they're
already represented on the topmost and leftmost lines in the lattice.
In total there are $2L^2$ qubits: $L^2$ qubits on horizontal lines and $L^2$
qubits on vertical lines.

To describe the toric code itself, it remains to describe the stabilizer
generators:
\begin{itemize}
\item
  For each \emph{tile} formed by the lines in the lattice there is one $Z$
  stabilizer generator, obtained by tensoring $Z$ matrices on the four qubits
  touching that tile along with identity matrices on all other qubits.

\item
  For each \emph{vertex} formed by the lines in the lattice there is one $X$
  stabilizer generator, obtained by tensoring $X$ matrices on the four qubits
  adjacent to that vertex along with identity matrices on all other qubits.
\end{itemize}

\begin{figure}[t]
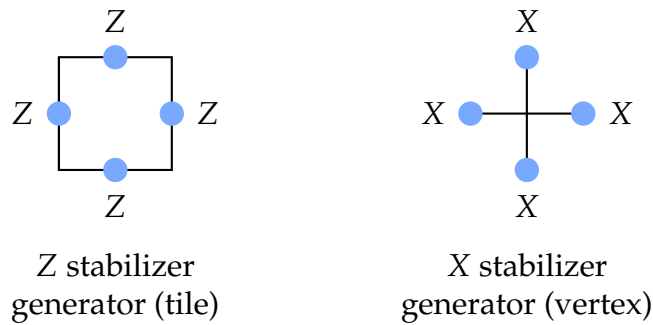

  \begin{center}
   
  \end{center}
  \caption{The two types of stabilizer generators for the toric code.}
  \label{fig:toric-code-stabilizer-generators}
\end{figure}

\noindent
In both cases we obtain a weight-4 Pauli operation. Individually, these
stabilizer generators may be illustrated as in
Figure~\ref{fig:toric-code-stabilizer-generators}.

\begin{figure}[!ht]
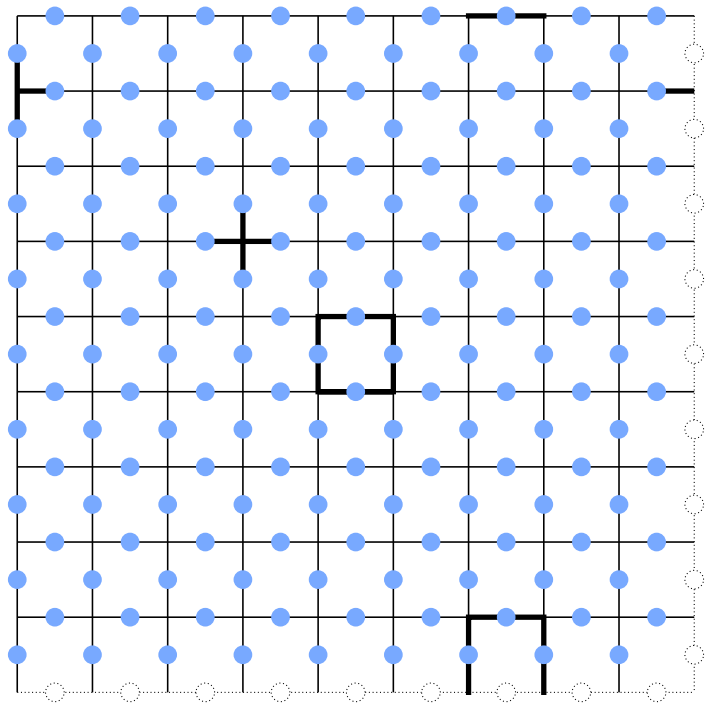

  \begin{center}
    %
    
  \end{center}
  \caption{Examples of stabilizer generators of the two types are indicated by
    thick lines. In total, there are $L^2$ stabilizer generators of each type.} 
  \label{fig:stabilizer-generator-examples}
\end{figure}

Figure~\ref{fig:stabilizer-generator-examples} shows some examples of
stabilizer generators in the context of the lattice itself.
Notice that the stabilizer generators that wrap around the periodic boundaries
are included.
These generators that wrap around the periodic boundaries are not special or in
any way distinguished from the ones that don't.

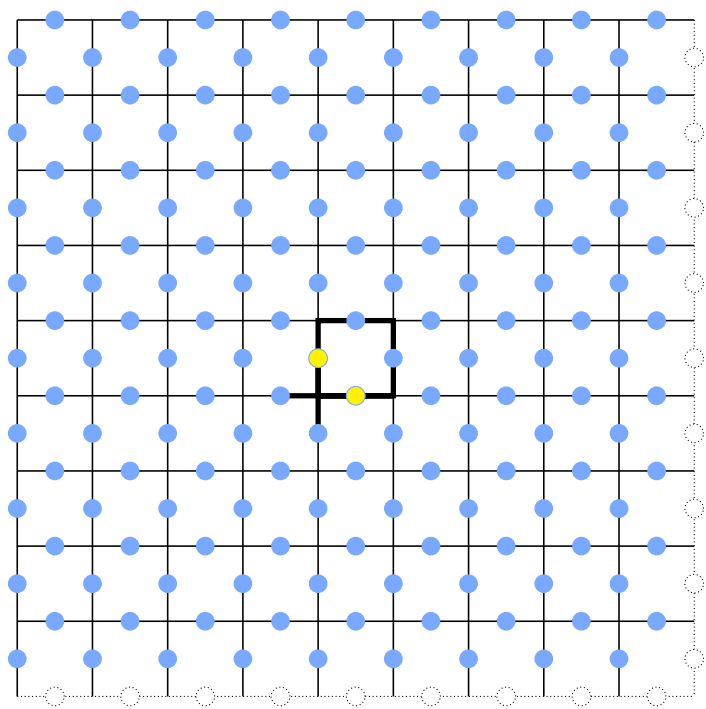
\begin{figure}[!ht]
  \begin{center}
    \begin{tikzpicture}[
        scale=1,
        qubit/.style={%
          circle,
          inner sep=2.5pt,
          fill=CircuitBlue
        },
        fixqubit/.style={%
          circle,
          draw=CircuitBlue,
          inner sep=2.5pt,
          fill=yellow
        },
        dotqubit/.style={%
          circle,
          inner sep=2.5pt,
          draw,
          densely dotted,
          fill=white}                    
      ]
      
      \def\L{9}
      \pgfmathtruncatemacro{\n}{\L - 1}

      \foreach \x in {0,1,...,\L} {
        \foreach \y in {0,1,...,\L} {
          \coordinate(L\x_\y) at (\x,\L-\y);
        }
      }

      \foreach \x in {0,...,\L} {
        \foreach \y in {0,...,\L} {
          \coordinate(Q\x_\y) at (\x+0.5,\L-\y);
          \coordinate(R\x_\y) at (\x,\L-\y-0.5);
        }
      }
      
      \foreach \z in {0,...,\n} {
        \draw[semithick] (L\z_0) -- (L\z_\L);
        \draw[semithick] (L0_\z) -- (L\L_\z);
      }
      
      \draw[densely dotted] (L0_\L) -- (L\L_\L);
      \draw[densely dotted] (L\L_0) -- (L\L_\L);
      

      \draw[line width=2pt]
      (L4_5) -- (L5_5) -- (L5_4) -- (L4_4) -- (L4_5) -- (L5_5);

      \draw[line width=2pt] (R4_4) -- (R4_5);
      \draw[line width=2pt] (Q4_5) -- (Q3_5);
      
      
      \foreach \x in {0,...,\n} {
        \foreach \y in {0,...,\n} {                      
          \node[qubit] at (R\x_\y) {};
          \node[qubit] at (Q\x_\y) {};
        }
      }

      \foreach \z in {0,...,\n} {
        \node[dotqubit] at (Q\z_\L) {};
        \node[dotqubit] at (R\L_\z) {};
      }      

      \node[fixqubit] at (Q4_5) {};
      \node[fixqubit] at (R4_4) {};
      
    \end{tikzpicture}
  \end{center}
  \caption{When an $X$ and a $Z$ stabilizer generator overlap, it is always on
    exactly two qubits --- implying that the stabilizer generators commute.}
  \label{fig:stabilizer-generators-overlap}
\end{figure}

The stabilizer generators must commute for this to be a valid stabilizer code.
As usual, the $Z$ stabilizer generators all commute with one another, because
$Z$ commutes with itself and the identity commutes with everything, and
likewise for the $X$ stabilizer generators.
The $Z$ and $X$ stabilizer generators clearly commute when they act
nontrivially on disjoint sets of qubits, like for the examples shown in
Figure~\ref{fig:stabilizer-generator-examples}.
The remaining possibility is that a $Z$ stabilizer generator and an $X$
stabilizer generator overlap on the qubits upon which they act nontrivially,
and whenever this happens the generators must always overlap on two qubits,
as shown in Figure~\ref{fig:stabilizer-generators-overlap}.
Consequently, two stabilizer generators like this commute, just like
$Z\otimes Z$ and $X\otimes X$ commute.
The stabilizer generators therefore all commute with one another.

The second required condition on the stabilizer generators for a stabilizer
code is that they form a minimal generating set.
This condition is actually \emph{not} satisfied by this collection: if we
multiply all of the $Z$ stabilizer generators together, we obtain the identity
operation, and likewise for the $X$ stabilizer generators.
Thus, any one of the $Z$ stabilizer generators can be expressed as the product
of all of the remaining ones, and similarly, any one of the $X$ stabilizer
generators can be expressed as the product of the remaining $X$ stabilizer
generators.
If we remove any one of the $Z$ stabilizer generators and any one of the $X$
stabilizer generators, however, we do obtain a minimal generating set.

To be clear about this, we do in fact care equally about all of the stabilizer
generators, and in a strictly operational sense there isn't any need to select
one stabilizer generator of each type to remove.
But, for the sake of \emph{analyzing} the code --- and counting the generators
in particular --- we can imagine that one stabilizer generator of each type has
been removed, so that we get a minimal generating set, keeping in mind that we
could always infer the results of these removed generators (thinking of them as
observables) from the results of all of the other stabilizer generator
observables of the same type.

This leaves $L^2 - 1$ stabilizer generators of each type, or $2L^2 - 2$ in
total, in a (hypothetical) minimal generating set.
Given that there are $2L^2$ qubits in total, this means that the toric code
encodes $2L^2 - 2 (L^2 - 1) = 2$ qubits.

The final condition required of stabilizer generators is that at least one
quantum state vector is fixed by all of the stabilizer generators.
We will see that this is the case as we proceed with the analysis of the code,
but it's also possible to reason that there's no way to generate $-1$ times the
identity on all $2L^2$ qubits from the stabilizer generators.

\subsection{Detecting errors}

The toric code has a simple and elegant description, but its quantum
error-correcting properties may not be at all clear from a first glance.
As it turns out, it's an amazing code!
To understand why and how it works, let's begin by considering different errors
and the syndromes they generate.

The toric code is a CSS code, because all of our stabilizer generators are
either $Z$ or $X$ stabilizer generators.
This means that $X$ errors and $Z$ errors can be detected (and possibly
corrected) separately.
In fact, there's a simple symmetry between the $Z$ and $X$ stabilizer
generators that allows us to analyze $X$ errors and $Z$ errors in essentially
the same way.
So, we shall focus on $X$ errors, which are possibly detected by the $Z$
stabilizer generators --- but the entire discussion that follows can be
translated from $X$ errors to $Z$ errors, which are analogously detected by the
$X$ stabilizer generators.

\begin{figure}[!ht]
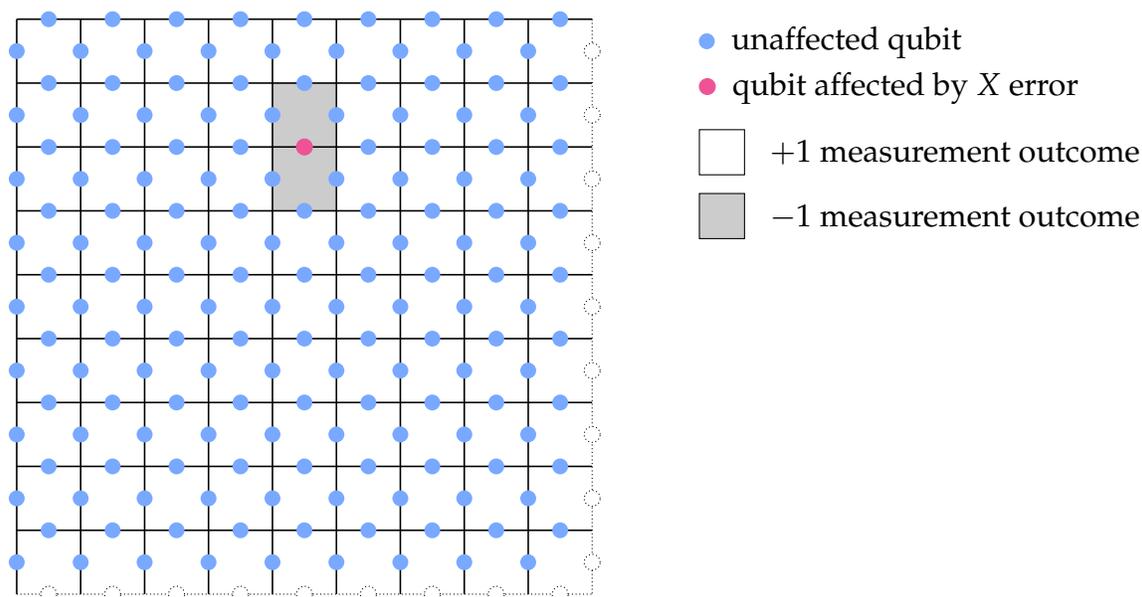

  \begin{center}
    \begin{invbox}{8cm}{8cm}
      \begin{center}
%
      \; qubit affected by $X$ error
      
      \vspace{3mm}
      
      %
      \; \raisebox{5pt}{$-1$ measurement outcome}

    \end{invbox}%
  \end{center}
  \caption{The effect of a single $X$ error on the $Z$ stabilizer
    generator measurement outcomes.}
  \label{fig:toric-code-bit-flip}
\end{figure}

Figure~\ref{fig:toric-code-bit-flip} depicts the effect of an $X$ error on a
single qubit.
Here, the assumption is that our $2L^2$ qubits were previously in a state
contained in the code space of the toric code, causing all of the stabilizer
generator measurements to output $+1.$
The $Z$ stabilizer generators detect $X$ errors, and there is one such
stabilizer generator for each tile in the figure, so we can indicate the
measurement outcome of the corresponding stabilizer generator with the color of
that tile:
$+1$ outcomes are indicated by white tiles and $-1$ outcomes are indicated by
gray tiles.
If a bit-flip error occurs on one of the qubits, the effect is that the
stabilizer generator measurements corresponding to the two tiles touching the
affected qubit now output $-1.$

This is intuitive when we consider $Z$ stabilizer generators and how they
behave.
In essence, each $Z$ stabilizer generator measures the \emph{parity} of the
four qubits that touch the corresponding tile (with respect to the standard
basis).
So, a $+1$ outcome doesn't indicate that no $X$ errors have occurred on these
four qubits, but rather it indicates that an \emph{even} number of $X$ errors
have occurred on these qubits, whereas a $-1$ outcome indicates that an
\emph{odd} number of $X$ errors have occurred.
A single $X$ error therefore flips the parity of the four bits on both of the
tiles it touches, causing the stabilizer generator measurements to output $-1.$

Next let's introduce multiple $X$ errors to see what happens.
In particular, we'll consider a chain of adjacent $X$ errors, where two $X$
errors are adjacent if they affect qubits touching the same tile.
\begin{figure}[!ht]
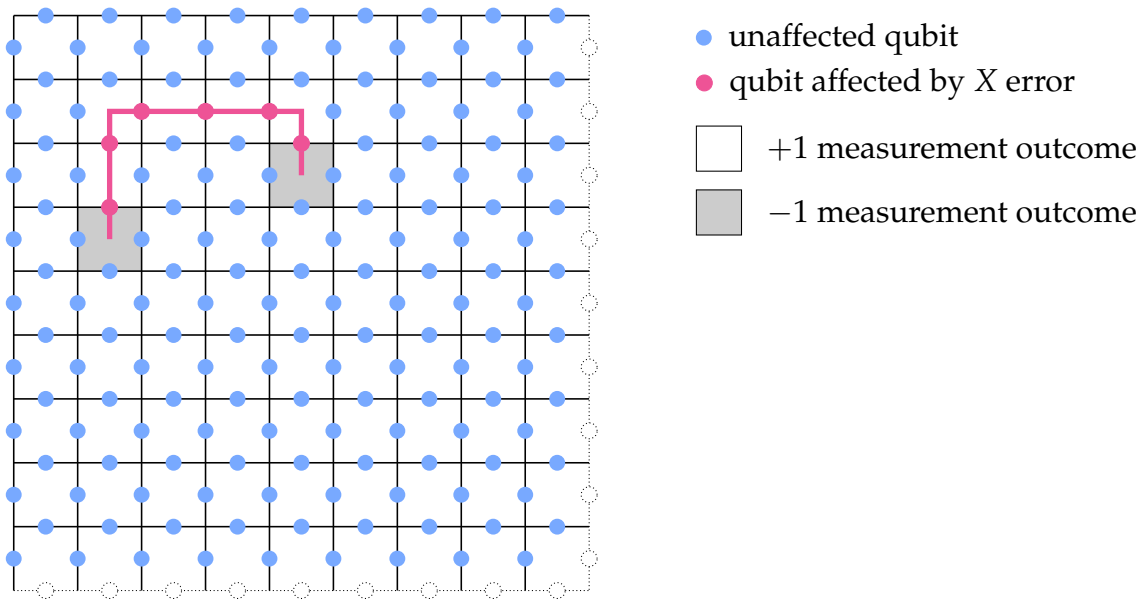

  \begin{center}
    \begin{invbox}{8cm}{8cm}
      \begin{center}
%
      \; qubit affected by $X$ error
      
      \vspace{3mm}
      
      %
      \; \raisebox{5pt}{$-1$ measurement outcome}

    \end{invbox}%
  \end{center}
  \caption{The effect of a chain of adjacent $X$ errors on the $Z$ stabilizer
    generator measurement outcomes.}
  \label{fig:X-error-chain}
\end{figure}%
As shown in Figure~\ref{fig:X-error-chain},
the two $Z$ stabilizer generators at the endpoints of the chain both give the
outcome $-1$ in this case, because an odd number of $X$ errors have occurred on
those two corresponding tiles.
All of the other $Z$ stabilizer generators, on the other hand, give the outcome
$+1,$ including the ones touching the chain but not at the endpoints, because
an even number of $X$ errors have occurred on the qubits touching the
corresponding tiles.

Thus, as long as we have a chain of $X$ errors that has endpoints, the toric
code will detect that errors have occurred, resulting in $-1$ measurement
outcomes for the $Z$ stabilizer generators corresponding to the endpoints of
the chain.
Note that the actual chain of errors is not revealed, only the endpoints!
This is OK --- in the next subsection we'll see that we don't need to know
exactly which qubits were affected by $X$ errors to correct them.
(The toric code is an example of a highly \emph{degenerate} code, in the sense
that it generally does not uniquely identify the errors it corrects.)

\begin{figure}[p]
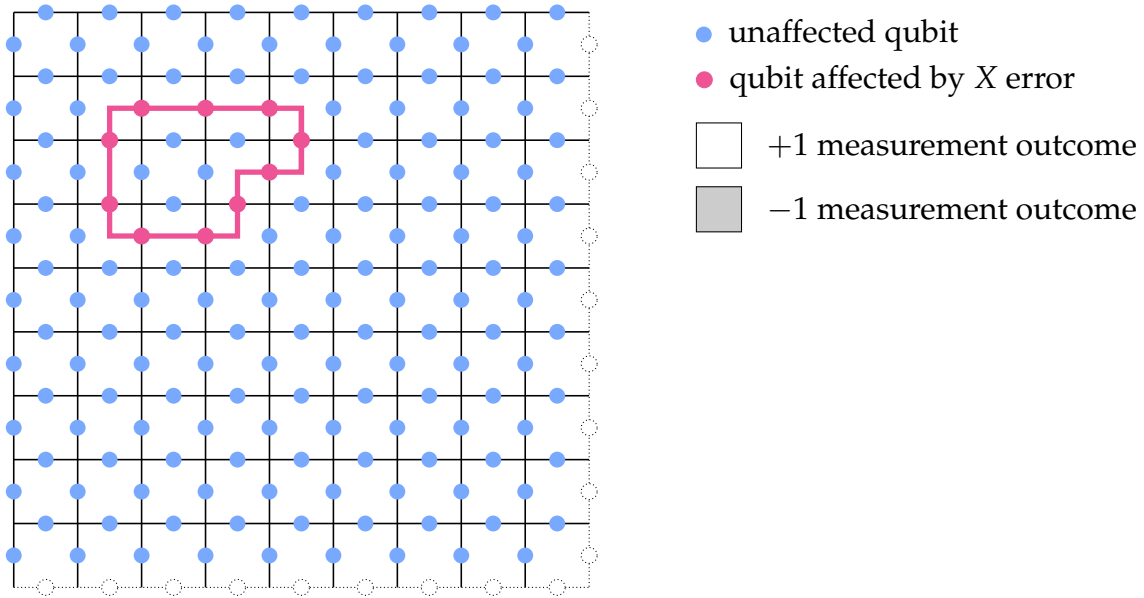

  \begin{center}
    \begin{invbox}{8cm}{8cm}
      \begin{center}
%
      \; qubit affected by $X$ error
      
      \vspace{3mm}
      
      %
      \; \raisebox{5pt}{$-1$ measurement outcome}

    \end{invbox}%
   
  \end{center}
  \caption{A closed loop of adjacent $X$ errors goes undetected by the
  toric code.}
  \label{fig:X-error-loop}
\end{figure}

It is, however, possible for a chain of adjacent $X$ errors not to have
endpoints, which is to say that a chain of errors could form a \emph{closed
loop}, like in Figure~\ref{fig:X-error-loop}.
\begin{figure}[p]
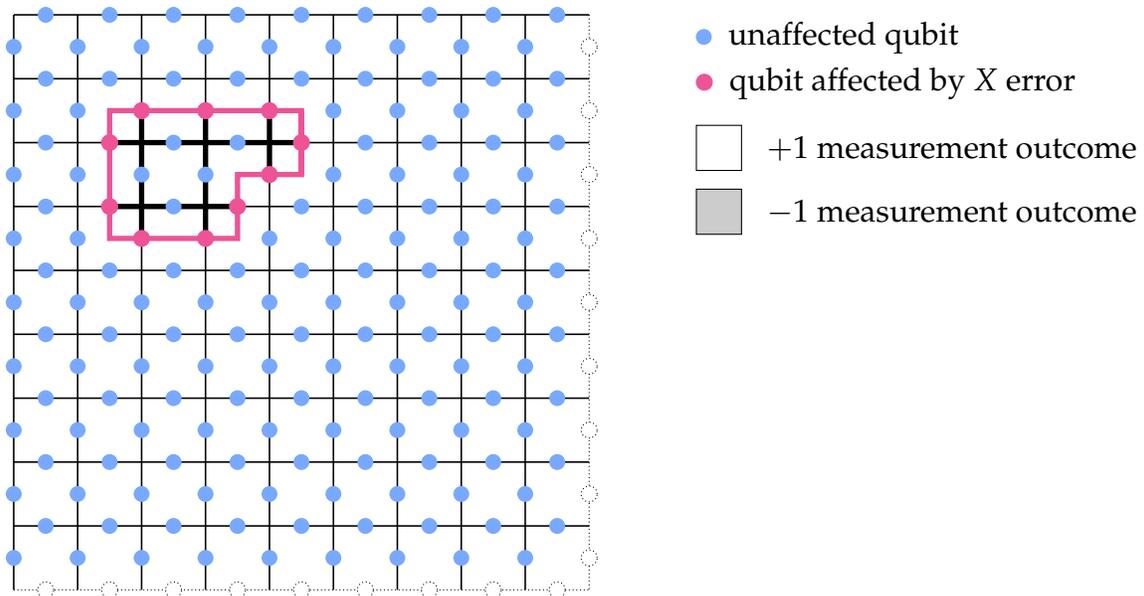

  \begin{center}
    \begin{invbox}{8cm}{8cm}
      \begin{center}
        %
%
      \; qubit affected by $X$ error
      
      \vspace{3mm}
      
      %
      \; \raisebox{5pt}{$-1$ measurement outcome}

    \end{invbox}%
   
  \end{center}
  \caption{The closed loop of adjacent $X$ errors illustrated in
    Figure~\ref{fig:X-error-loop} is generated by the $X$ stabilizer generators
    within the loop.}
  \label{fig:X-error-loop-generated}
\end{figure}%
In such a case, an even number of $X$ errors have occurred on every tile, so
every stabilizer generator measurement results in a $+1$ outcome.
Closed loops of adjacent $X$ errors are therefore not detected by the code.

This might seem disappointing, because we only need four $X$ errors to form a
closed loop (and we're hoping for better than a distance 4 code).
However, a closed loop of $X$ errors of the form depicted in
Figure~\ref{fig:X-error-loop} is not actually an error --- because it's in the
stabilizer!
Recall that, in addition to the $Z$ stabilizer generators, we also have an $X$
stabilizer generator for each vertex in the lattice.
And if we multiply adjacent $X$ stabilizer generators together, the result is
that we obtain closed loops of $X$ operations.
For instance, the closed loop in Figure~\ref{fig:X-error-loop} can be obtained
by multiplying together the $X$ stabilizer generators indicated in
Figure~\ref{fig:X-error-loop-generated}.

This is, however, not the only type of closed loop of $X$ errors that we can
have --- and it is not the case that every closed loop of $X$ errors is
included in the stabilizer.
In particular, the different types of loops can be characterized as follows.

\begin{enumerate}
\item
  Closed loops of $X$ errors with an \emph{even} number of $X$ errors on every
  horizontal line and every vertical line of qubits. (The example shown above
  falls into this category.) Loops of this form are always contained in the
  stabilizer, as they can effectively be shrunk down to nothing by multiplying
  them by $X$ stabilizer generators.
\item
  Closed loops of $X$ errors with an \emph{odd} number of $X$ errors on at
  least one horizontal line or at least one vertical line of qubits. Loops of
  this form are never contained in the stabilizer and therefore represent
  nontrivial errors that go undetected by the code.
\end{enumerate}

\noindent
An example of a closed loop of $X$ errors in the second category is shown in
the Figure~\ref{fig:X-error-logical}.
\begin{figure}[b]
  \begin{center}
    \begin{invbox}{8cm}{8cm}
      \begin{center}
%
      \; qubit affected by $X$ error
      
      \vspace{3mm}
      
      %
      \; \raisebox{5pt}{$-1$ measurement outcome}

    \end{invbox}%
    
  \end{center}
  \caption{An example of a closed loop of $X$ errors in the second category
    described above.}
  \label{fig:X-error-logical}
\end{figure}%
Such a chain of errors is not contained in the stabilizer because every $X$
stabilizer generator places an even number of $X$ operations on every
horizontal line and every vertical line of qubits.
This is therefore an actual example of a nontrivial error that the code fails
to detect.

The key is that the only way to form a loop of the second sort is to go around
the torus, meaning either around the hole in the middle of the torus, through
it, or both.
Intuitively speaking, a chain of $X$ errors like this cannot be shrunk down to
nothing by multiplying it by $X$ stabilizer generators because the topology of
a torus prohibits it.
The toric code is sometimes categorized as a \emph{topological} quantum error
correcting code for this reason.

The shortest that such a loop can be is $L,$ and therefore this is the distance
of the toric code: any closed loop of $X$ errors with length less than $L$ must
fall into the first category, and is therefore contained in the stabilizer; and
any chain of $X$ errors with endpoints is detected by the code.
Given that the toric code uses $2L^2$ qubits to encode $2$ qubits and has
distance $L,$ it follows that it's a $[[2L^2,2,L]]$ stabilizer code.

\subsection{Correcting errors}

We've discussed error \emph{detection} for the toric code, and now we'll
briefly discuss how to \emph{correct} errors.
The toric code is a CSS code, so $X$ errors and $Z$ errors can be detected and
corrected independently.
Keeping our focus on $Z$ stabilizer generators, which detect $X$ errors, let us
consider how a chain of $X$ errors can be corrected.
($Z$ errors are corrected in a symmetric way.)

If a syndrome different from the $(+1,\ldots,+1)$ syndrome appears when the $Z$
stabilizer generators are measured, the $-1$ outcomes reveal the endpoints of
one or more chains of $X$ errors.
We can attempt to correct these errors by pairing together the $-1$ outcomes
and forming a chain of $X$ corrections between them.
When doing this, it makes sense to choose \emph{shortest paths} along which the
corrections take place.

\begin{figure}[!ht]
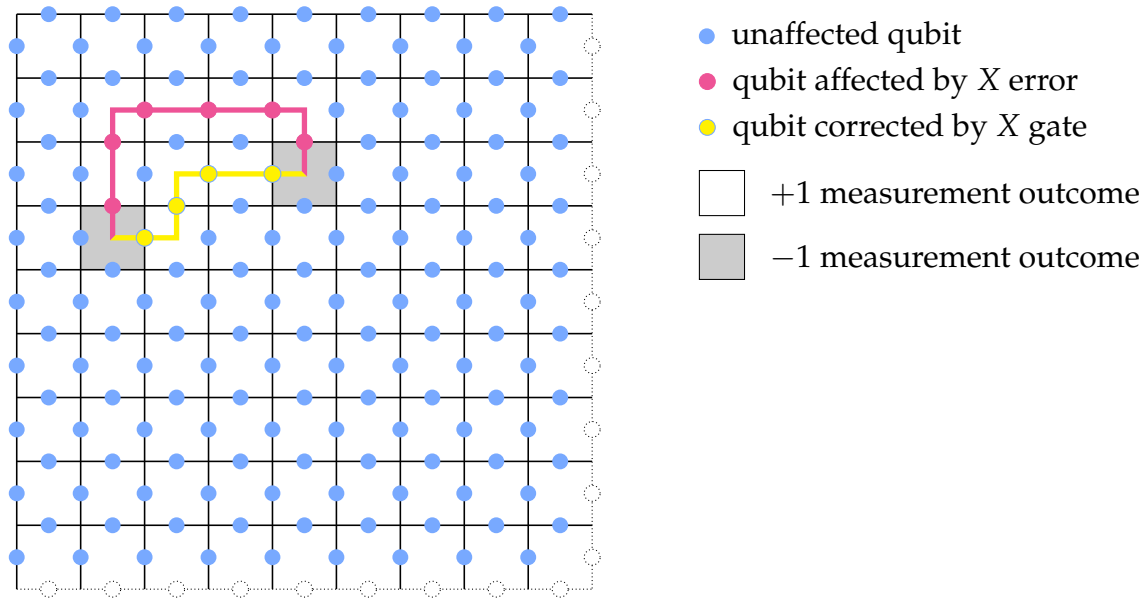

  \begin{center}
    \begin{invbox}{8cm}{8cm}
      \begin{center}
%
      \; qubit affected by $X$ error     
      \vspace{1mm}

      %
%
    \; qubit corrected by $X$ gate
    
    \vspace{3mm}
      
      %
      \; \raisebox{5pt}{$-1$ measurement outcome}

    \end{invbox}%
   
  \end{center}
  \caption{A chain of adjacent $X$ errors being corrected by an adjacent chain
    of $X$ corrections.}
  \label{fig:X-error-correction}
\end{figure}

For instance, consider the diagram in Figure~\ref{fig:X-error-correction},
which depicts a syndrome with two $-1$ outcomes, indicated by gray tiles,
caused by a chain of $X$ errors illustrated by the magenta line and circles.
As we have already remarked, the chain itself is not revealed by the syndrome;
only the endpoints are visible.

To attempt to correct this chain of errors, a shortest path between the $-1$
measurement outcomes is selected and $X$ gates are applied as corrections to
the qubits along this path (indicated in yellow in the figure).
While the corrections may not match up with the actual chain of errors, the
errors and corrections together form a closed loop of $X$ operations that is
contained in the stabilizer of the code.
The correction is therefore successful in this situation, as the combined
effect of the errors and corrections is to do nothing to an encoded state.

\begin{figure}[p]
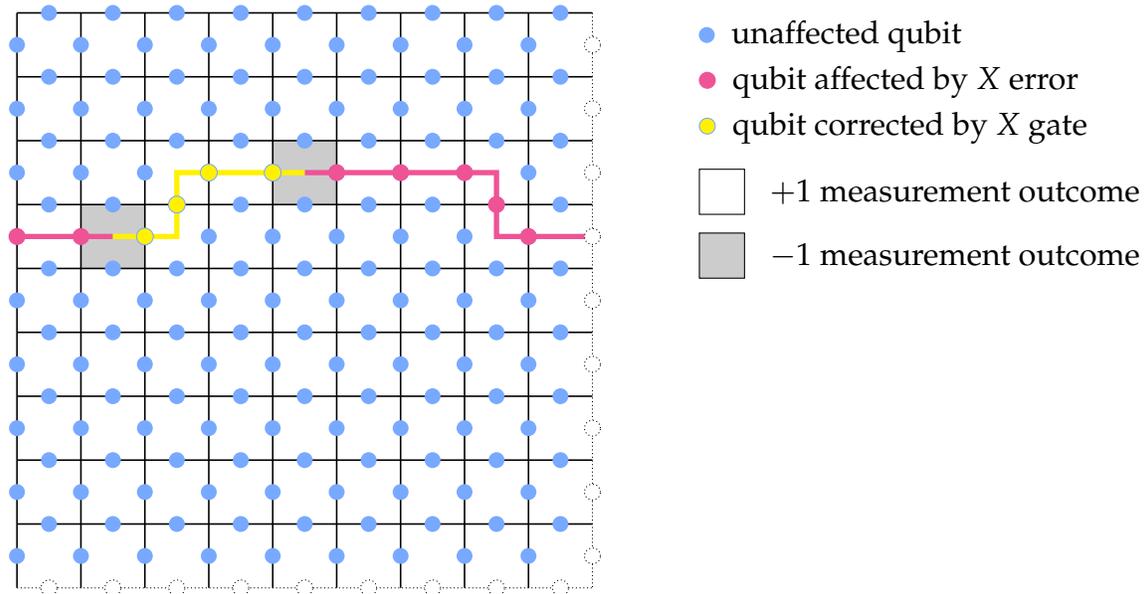

  \begin{center}
    \begin{invbox}{8cm}{8cm}
      \begin{center}
%
      \; qubit affected by $X$ error     
      \vspace{1mm}

      %
%
    \; qubit corrected by $X$ gate
    
    \vspace{3mm}
      
      %
      \; \raisebox{5pt}{$-1$ measurement outcome}

    \end{invbox}%
   
  \end{center}
  \caption{A chain of adjacent $X$ corrections failing to correct
  a chain of adjacent $X$ errors.}
  \label{fig:correction-complete-error}
\end{figure}

This strategy won't always be successful.
For example, a different explanation for the same syndrome as in the previous
figure is shown in Figure~\ref{fig:correction-complete-error}.
This time, the same chain of corrections as before fails to correct for this
chain of errors, because the combined effect of the errors and corrections is
that we obtain a closed loop of $X$ operations that wraps around the torus, and
therefore has a nontrivial effect on the code space.
So, there's no guarantee that the strategy just described, of choosing a
shortest path of $X$ corrections between two $-1$ syndrome measurement
outcomes, will properly correct the error that caused this syndrome.

\begin{figure}[p]
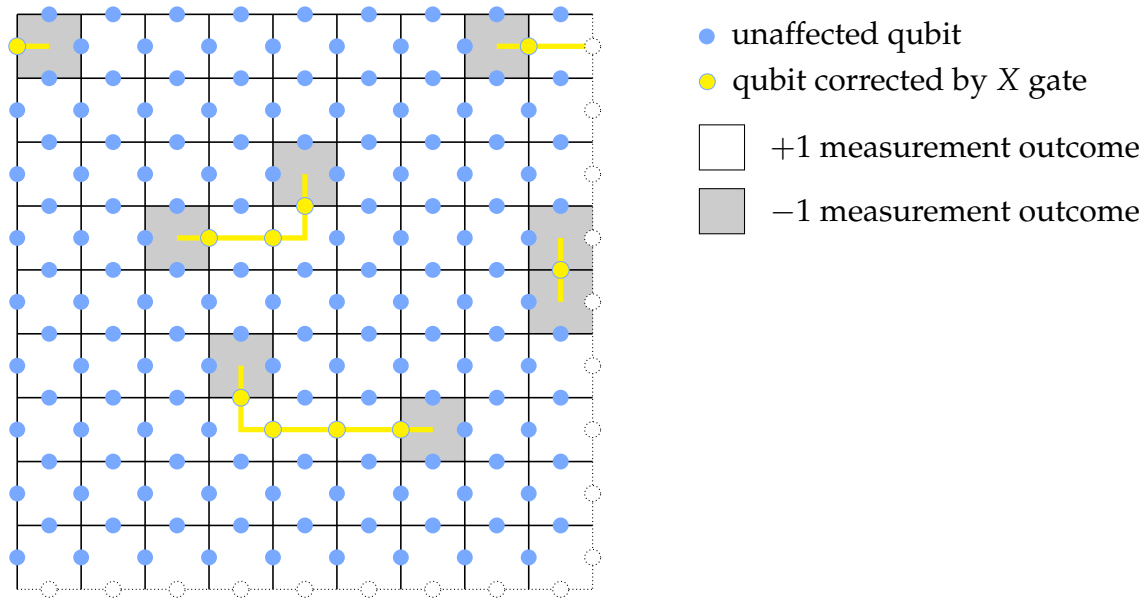

  \begin{center}
    \begin{invbox}{8cm}{8cm}
      \begin{center}
%
    \; qubit corrected by $X$ gate
    
    \vspace{3mm}
      
      %
      \; \raisebox{5pt}{$-1$ measurement outcome}

    \end{invbox}%
   
  \end{center}
  \caption{Multiple $X$ correction chains forming a minimum-weight perfect
    matching between $-1$ measurement outcomes.}
  \label{fig:multiple-correction-chains}
\end{figure}

Perhaps more likely, depending on the noise model, is that a syndrome with more
than two $-1$ entries is measured, like
Figure~\ref{fig:multiple-correction-chains} suggests.
In such a case, there are different correction strategies known.
One natural strategy is to attempt to pair up the $-1$ measurement outcomes and
perform corrections along shortest paths that connect the pairs, as is
indicated in the figure in yellow.
In particular, a \emph{minimum-weight perfect matching} between the $-1$
measurement outcomes can be computed, and then the pairs are connected by
shortest paths of $X$ corrections.
The computation of a minimum-weight perfect matching can be done efficiently
with a classical algorithm known as the \emph{blossom algorithm}, which was
discovered by Edmonds in the 1960s.

This approach is generally not optimal for the most typically studied noise
models, but based on numerical simulations it works very well in practice below
a noise rate of approximately 10\%, assuming independent Pauli errors where
$X,$ $Y,$ and $Z,$ are equally likely. 
Increasing $L$ doesn't have a significant effect on the break-even point at
which the code starts to help, but does lead to a faster decrease in the
probability for a logical error as the error rate decreases past the break-even
point.

\section{Other code families}

It's been over 25 years since the toric code was discovered, and there's been a
great deal of research into quantum error correcting codes since then,
including the discovery of other topological quantum codes inspired by the
toric code, as well as codes based on different ideas.
A comprehensive list of known quantum error correcting code constructions would
be impossible to include here --- but we will scratch the surface just a bit to
briefly examine a couple of prominent examples.

\subsection{Surface codes}

As it turns out, it isn't actually necessary that the toric code has periodic
boundaries.
That is to say, it's possible to cut out just a portion of the toric code and
lay it flat on a two-dimensional surface, rather than a torus, to obtain a
quantum error correcting code --- provided that the stabilizer generators on
the edges are properly defined.
What we obtain is called a \emph{surface code}.

\begin{figure}[!ht]
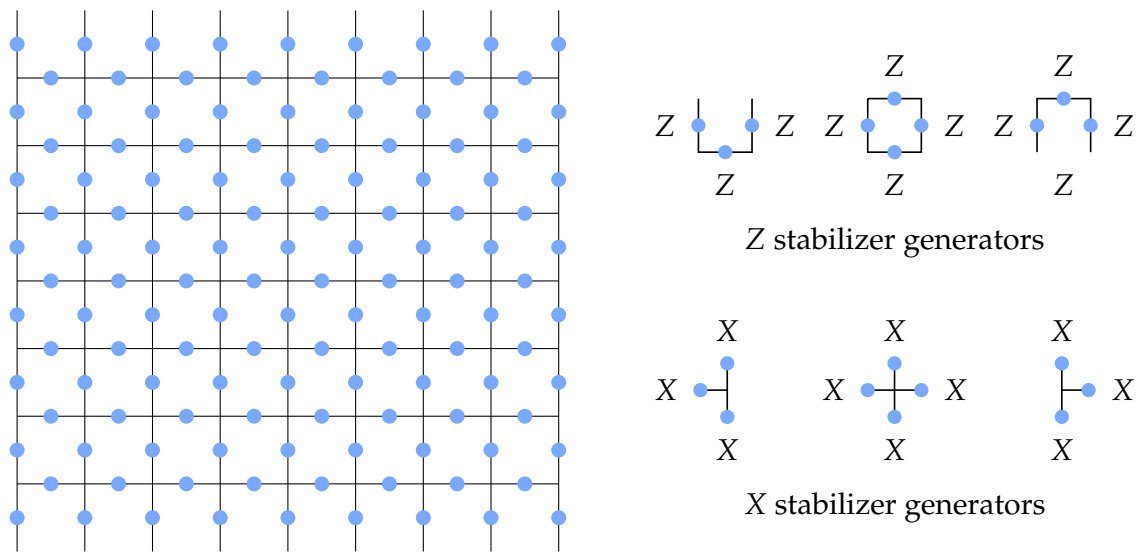

  \begin{center}
    \begin{invbox}{7.5cm}{7.25cm}
      \begin{center}
\\[1mm]
        $X$ stabilizer generators
      \end{center}
    \end{invbox}%
    }%
  \end{center}
  \caption{A surface code with smooth edges on the sides and
    rough edges on the top and bottom.}
  \label{fig:surface-code}
\end{figure}

For example, Figure~\ref{fig:surface-code} shows a diagram of a surface code,
where the lattice is cut with so-called rough edges at the top and bottom and
smooth edges at the sides.
The edge cases for the stabilizer generators are defined in the natural way,
which is that Pauli operations on ``missing'' qubits are simply omitted.
Surface codes of this form encode a single qubit, rather than two like the
toric code.
The stabilizer generators happen to form a minimal generating set in this case,
without the need to remove one of each type as with the toric code.
But, despite these differences, the important characteristics of the toric code
are inherited.
In particular, nontrivial undetected errors for this code correspond to chains
of errors that either stretch from the left edge to the right edge
(for chains of $X$ errors) or from top to bottom (for chains of $Z$ errors).

\begin{figure}[!hb]
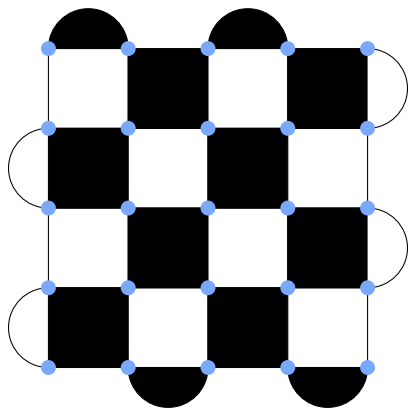

  \begin{center}

  \end{center}
  \caption{A diagram of a rotated surface code. Black faces denote $X$
    stabilizer generators and white faces denote $Z$ stabilizer generators.}
  \label{fig:rotated-surface-code}
\end{figure}

It's also possible to cut the edges for a surface code diagonally to obtain
what are sometimes called \emph{rotated} surface codes, which are so-named not
because the codes are rotated in a meaningful sense, but because the
\emph{diagrams} are rotated (by 45 degrees).
For example, Figure~\ref{fig:rotated-surface-code} shows a diagram of a rotated
surface code having distance~5.

For this type of diagram, black tiles (including the rounded ones on the edges)
indicate $X$ stabilizer generators, where $X$ operations are applied to the
(two or four) vertices of each tile, while white tiles represent $Z$ stabilizer
generators.
Rotated surface codes have similar properties to (non-rotated) surface codes,
but are more economical in terms of how many qubits are used.

\subsection{Color codes}

Color codes are another interesting class of codes, which also fall into the
general category of topological quantum codes.
They will only briefly be described here.

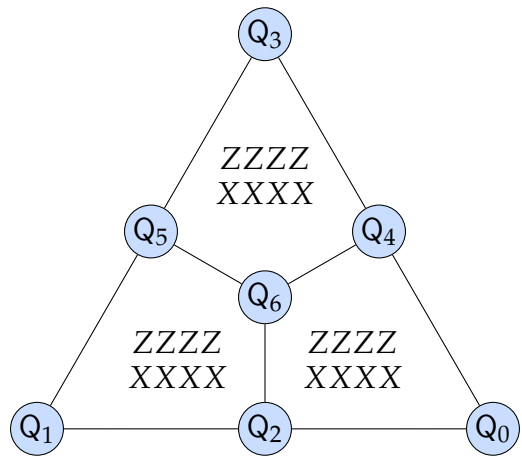
\begin{figure}[!ht]
  \begin{center}
    \begin{tikzpicture}[scale = 3.5,
        qubit/.style={%
          circle,
          draw,
          fill=CircuitBlue!40,
          minimum size=20pt,
          inner sep = 0mm}
      ]

      \node[qubit] (1) at (0,1) {$\mathsf{Q}_3$};
      \node[qubit] (5) at (-0.866,-0.5) {$\mathsf{Q}_1$};
      \node[qubit] (7) at (0.866,-0.5) {$\mathsf{Q}_0$};
      \node[qubit] (2) at ($(1)!0.5!(5)$) {$\mathsf{Q}_5$};
      \node[qubit] (4) at ($(1)!0.5!(7)$) {$\mathsf{Q}_4$};
      \node[qubit] (6) at ($(5)!0.5!(7)$) {$\mathsf{Q}_2$};
      \node[qubit] (3) at (0,0) {$\mathsf{Q}_6$};
      
      \draw (1) -- (2);
      \draw (1) -- (4);
      \draw (2) -- (3);
      \draw (2) -- (5);
      \draw (3) -- (4);
      \draw (3) -- (6);
      \draw (4) -- (7);
      \draw (5) -- (6);
      \draw (6) -- (7);

      \node at (0,0.45) {%
        $\begin{array}{c} ZZZZ\\[-0.5mm]XXXX \end{array}$
      };
      
      \node at (-0.333,-0.25) {%
        $\begin{array}{c} ZZZZ\\[-0.5mm]XXXX \end{array}$
      };
      
      \node at (0.333,-0.25) {%
        $\begin{array}{c} ZZZZ\\[-0.5mm]XXXX \end{array}$
      };
      
    \end{tikzpicture}%
  \end{center}
  \caption{A graphical representation of the $7$-qubit Steane code.}
  \label{fig:Steane-color-code}
\end{figure}

One way to think about color codes is to view them as geometric generalizations
of the 7-qubit Steane code.
With this in mind, let's consider the 7-qubit Steane code again, and suppose
that the seven qubits are named and ordered using Qiskit's numbering convention
as
$(\mathsf{Q}_6,\mathsf{Q}_5,\mathsf{Q}_4,\mathsf{Q}_3,
\mathsf{Q}_2,\mathsf{Q}_1,\mathsf{Q}_0).$
Recall that the stabilizer generators for this code are as follows.
\[
\begin{array}{ccccccc}
  Z & Z & Z & Z & \mathbb{I} & \mathbb{I} & \mathbb{I} \\[1mm]
  Z & Z & \mathbb{I} & \mathbb{I} & Z & Z & \mathbb{I} \\[1mm]
  Z & \mathbb{I} & Z & \mathbb{I} & Z & \mathbb{I} & Z \\[1mm]
  X & X & X & X & \mathbb{I} & \mathbb{I} & \mathbb{I} \\[1mm]
  X & X & \mathbb{I} & \mathbb{I} & X & X & \mathbb{I} \\[1mm]
  X & \mathbb{I} & X & \mathbb{I} & X & \mathbb{I} & X
\end{array}
\]
If we associate these seven qubits with the vertices of the graph
shown in Figure~\ref{fig:Steane-color-code}, we find that the stabilizer
generators match up precisely with the \emph{faces} formed by the edges of the
graph.

That is, for each face, there's both a $Z$ stabilizer generator and an $X$
stabilizer generator that act nontrivially on those qubits found at the
vertices of that face.
The 7-qubit Steane code therefore possesses geometric locality, so in principle
it's not necessary to move qubits over large distances to measure the
stabilizer generators.
The fact that the $Z$ and $X$ stabilizer generators always act nontrivially on
\emph{exactly} the same sets of qubits is also nice for reasons connected with
fault-tolerant quantum computation, which is the topic for the next lesson.

\begin{figure}[!ht]
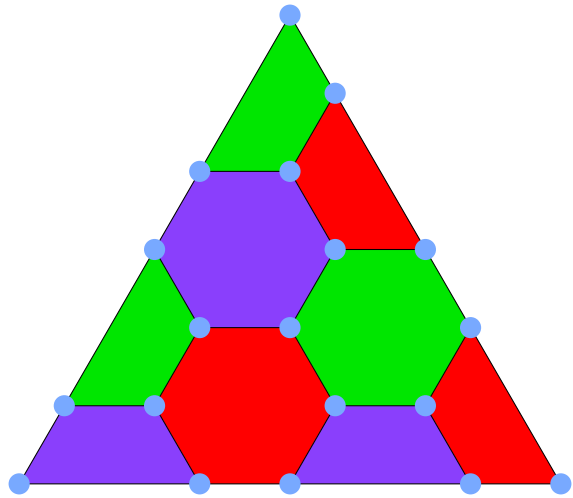

  \begin{center}

  \end{center}
  \caption{A graphical representation of a $[[19,1,5]]$ color code.}
  \label{fig:color-code}
\end{figure}

Color codes are quantum error correcting codes (CSS codes to be more precise)
that generalize this basic pattern, except that the underlying graphs may be
different.
For example, Figure~\ref{fig:color-code} shows a graph with 19 vertices that
works.
It defines a code that encodes one qubit into 19 qubits and has distance 5
(so it's a $[[19,1,5]]$ stabilizer code).
This can be done with many other graphs, including families of graphs that grow
in size and have interesting structures.

Color codes are so-named because one of the required conditions on the graphs
that define them is that the faces can be three-colored, meaning that the faces
can each be assigned one of three colors in such a way that no two faces of the
same color share an edge (as we have in the diagram above).
The colors don't actually matter for the definition of the code itself ---
there are always $Z$ and $X$ stabilizer generators for each face, regardless of
its color --- but the colors are important for analyzing how the codes work.

\subsection{Other codes}

Quantum error correction is an active and rapidly advancing area of research.
Those interested in exploring deeper may wish to consult the
\emph{Error Correction Zoo} (\url{https://errorcorrectionzoo.org/}), which
lists numerous examples and categorizations of quantum error correcting codes.

\begin{callout}[title = {Example: the gross code}]
  The gross code is a recently discovered $[[144,12,12]]$ stabilizer code.
  It is similar to the toric code, except each stabilizer generator acts
  nontrivially on two additional qubits, slightly further away from the tile or
  vertex for that generator (so each stabilizer generator has weight 6). The
  advantage of this code is that it can encode 12 qubits, compared with just
  two for the toric code.
\end{callout}


\lesson{Fault-Tolerant Quantum Computation}
\label{lesson:fault-tolerant-quantum-computation}

In the previous lessons of this unit, we've seen several examples of quantum
error correcting codes, which can detect and allow for the correction of errors
--- so long as not too many qubits are affected.
If we want to use error correction for quantum \emph{computing}, however, there
are still many issues to be reckoned with.
This includes the reality that, not only is quantum information fragile and
susceptible to noise, but the quantum gates, measurements, and state
initializations used to implement quantum computations will themselves be
imperfect.

For instance, if we wish to perform error correction on one or more qubits that
have been encoded using a quantum error correcting code, then this must be done
using gates and measurements that might not work correctly --- which means not
only failing to detect or correct errors, but possibly introducing new errors.

In addition, the actual computations we're interested in performing must be
implemented, again with gates that aren't perfect.
But, we certainly can't risk decoding qubits for the sake of performing these
computations, and then re-encoding once we're done, because errors might strike
when the protection of a quantum error correcting code is absent.
This means that quantum gates must somehow be performed on \emph{logical}
qubits that never go without the protection of a quantum error correcting code.

This all presents a major challenge.
But it is known that, as long as the level of noise falls below a certain
\emph{threshold value}, it is possible in theory to perform arbitrarily large
quantum computations reliably using noisy hardware.
We'll discuss this critically important fact, which is known as the
\emph{threshold theorem}, toward the end of the lesson.

The lesson starts with a basic framework for fault-tolerant quantum computing,
including a short discussion of noise models and a general methodology for
fault-tolerant implementations of quantum circuits.
We'll then move on to the issue of \emph{error propagation} in fault-tolerant
quantum circuits and how to control it.
In particular, we'll discuss \emph{transversal} implementations of gates, which
offer a very simple way to control error propagation --- though there is a
fundamental limitation that prevents us from using this method exclusively ---
and we'll also take a look at a different methodology involving so-called
\emph{magic states}, which offers a different path to controlling error
propagation in fault-tolerant quantum circuits.

And finally, the lesson concludes with a high-level discussion of the threshold
theorem, which states that arbitrarily large quantum circuits can be
implemented reliably, so long as the error rate for all of the components
involved falls below a certain finite threshold value.
This threshold value depends on the error correcting code that is used, as well
as the specific choices that are made for fault-tolerant implementations of
gates and measurements, but critically it does \emph{not} depend on the size of
the quantum circuit being implemented.

\section{An approach to fault tolerance}

We'll begin by outlining a basic approach to fault-tolerant quantum computing
based on quantum circuits and error correcting codes.

\begin{figure}[!ht]
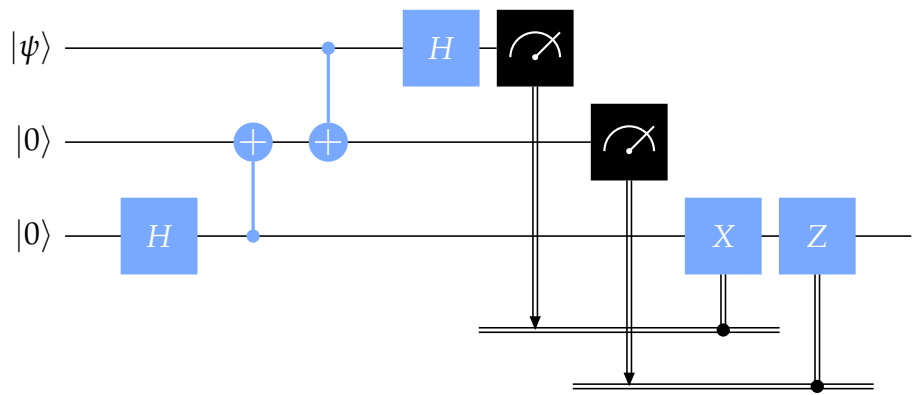

  \begin{center}

  \end{center}
  \caption{A teleportation circuit.}
  \label{fig:teleportation-circuit}
\end{figure}

For the sake of this discussion, let us consider the example of a
quantum circuit shown in Figure~\ref{fig:teleportation-circuit}.
This happens to be a teleportation circuit, including the preparation of the
e-bit, but the specific functionality of the circuit is immaterial --- it's
just an example, and in actuality we're likely to be interested in
significantly larger circuits.
A circuit like this one represents an ideal, and an actual implementation of it
won't be perfect.
So what could go wrong?

The fact of the matter is that quite a lot could go wrong!
In particular, the state initializations, unitary operations, and measurements
will all be imperfect;
and the qubits themselves will be susceptible to noise, including decoherence,
at every point in the computation, even when no operations are being performed
on them and they're simply storing quantum information.
That is to say, just about everything could go wrong.

There is one exception, though: Any \emph{classical} computations that are
involved are assumed to be perfect --- because, practically speaking, classical
computations are perfect.
For example, if we decide to use a surface code for error correction, and a
classical perfect matching algorithm is run to compute corrections, we really
don't need to concern ourselves with the possibility that errors in this
classical computation will lead to a faulty solution.
As another example, quantum computations often necessitate classical pre- and
post-processing, and these classical computations can safely be assumed to be
perfect as well.

\subsection{Noise models}

To analyze fault-tolerant implementations of quantum circuits, we require a
precise mathematical model --- a \emph{noise model} --- through which
\emph{probabilities} for various things to go wrong can be associated.
Hypothetically speaking, one could attempt to come up with a highly detailed,
complicated noise model that aims to reflect the reality of what happens in a
particular device.
But, if the noise model is too complicated or difficult to reason about, it
will likely be of limited use.
For this reason, simpler noise models are much more typically considered.

One example of a simple noise model is the \emph{independent stochastic noise
model}, where errors or faults affecting different components at different
moments in time --- or, in other words, different \emph{locations} in a quantum
circuit --- are assumed to be independent.
For instance, each gate might fail with a certain probability, an error might
strike each stored qubit per unit time with a different probability, and so on,
with \emph{no correlations} among the different possible errors.

Now, it is certainly reasonable to object to such a model, because there
probably will be correlations among errors in real physical devices.
For instance, there might be a small chance of a catastrophic error that wipes
out all the qubits at once.
Perhaps more likely, there could be errors that are localized but that
nevertheless affect multiple components in a quantum computer.
Nobody suggests otherwise!
Nevertheless, the independent stochastic noise model does provide a simple
baseline that captures the idea that nature is unpredictable but not malicious,
and it isn't intentionally trying to ruin quantum computations.

Other, less forgiving noise models are also commonly studied.
For example, a common relaxation of the assumption of independence among errors
affecting different locations in a quantum circuit is that \emph{just the
locations} of the errors are independent, but the actual errors affecting these
locations could be correlated.

Regardless of what noise model is chosen, it should be recognized that
\emph{learning} about the errors that affect specific devices, and formulating
new error models if the old ones lead us astray, could potentially be an
important part of the development of fault-tolerant quantum computation.

\subsection{Fault-tolerant circuit implementations}

Next we'll consider a basic strategy for fault-tolerant implementations of
quantum circuits.
We'll use the teleportation circuit above as a running example to illustrate
the strategy, though it could be applied to any quantum circuit.

\begin{figure}[t]
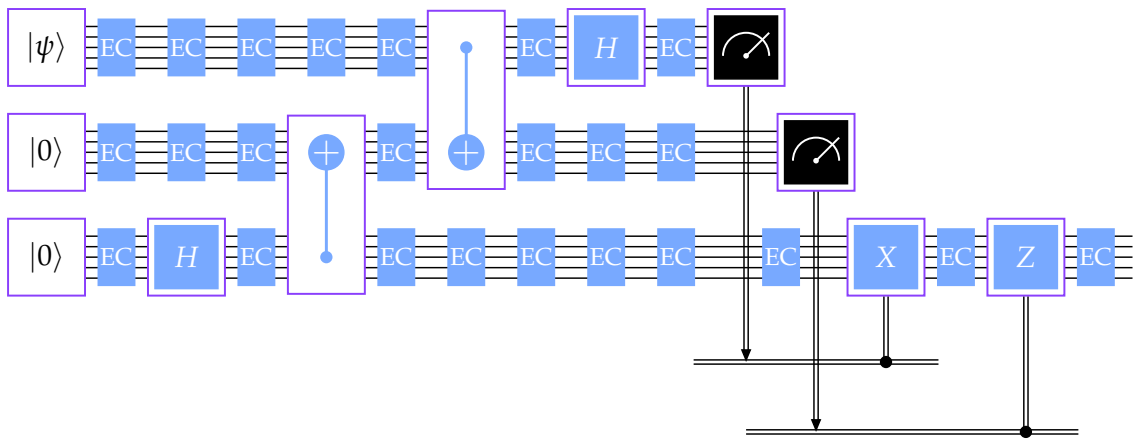

  \begin{center}
    \scalebox{0.93}{%
%
    }
  \end{center}
  \caption{A fault-tolerant implementation of the circuit in
    Figure~\ref{fig:teleportation-circuit}.}
  \label{fig:FT-teleportation-circuit}
\end{figure}

Figure~\ref{fig:FT-teleportation-circuit} shows a diagram of a fault-tolerant
implementation of our teleportation circuit.
The individual components in this diagram and their connection to the original
circuit are as follows.
\begin{enumerate}
\item
  State preparations, unitary gates, and measurements are not performed
  directly, as single operations, but rather are performed by so-called
  \emph{gadgets}, which could each involve multiple qubits and multiple
  operations. In the diagram, gadgets are indicated by purple boxes labelled by
  whatever state preparation, gate, or measurement is to be implemented.

\item
  The \emph{logical} qubits on which the original, ideal circuit is run are
  protected using a quantum error correcting code. Rather than acting directly
  on these logical qubits, the gadgets act on the \emph{physical} qubits that
  encode them. The diagram suggests that five physical qubits are used for each
  logical qubit, as if the $5$-qubit code were being used, but the number could
  naturally be different. It is worth stressing that these logical qubits are
  \emph{never} exposed; they spend their entire existence being protected by
  whatever quantum error correcting code we've chosen.

\item
  Error correction is performed repeatedly, as suggested by the blue boxes
  labeled ``EC'' in the diagram, throughout the computation. It is critically
  important that this is done both frequently and in parallel. As errors take
  place, entropy builds up, and constant work is required to remove it from the
  system at a high enough rate to allow the computation to function correctly.
\end{enumerate}

There are therefore specific choices that must be made, including the selection
of the gadgets as well as the quantum error correcting code itself.
Once these choices have been made, and assuming a particular noise model has
been adopted, there is a fundamental question that we may ask ourselves:
Is this actually helping?
That is, are we making things better, or might we actually be making things
worse?

If the rate of noise is too high, the entire process just suggested could very
well make things worse, just like the 9-qubit Shor code makes things worse for
independent errors if the error probability on each qubit is above the
break-even point.
If, however, the rate of noise is below a certain threshold, then all of this
extra work will get us somewhere --- and as we'll discuss toward the end of the
lesson, paths open up for further error reduction.

\section{Controlling error propagation}

Fault-tolerant quantum computation is akin to a race between errors and error
correction.
If the number of errors is small enough, error correction will successfully
correct them; but if there are too many errors, error correction will fail.

For this reason, sufficient care must be given to the way quantum computations
are performed in fault-tolerant implementations of circuits, to control
\emph{error propagation.}
That is, an error on one qubit can potentially be propagated to multiple qubits
through the action of gates in a quantum circuit, which can cause the number of
errors to increase dramatically.
This is a paramount concern, for if we don't manage to control error
propagation, our error-correction efforts will quickly be overwhelmed by
errors.
If, on the other hand, we're able to keep the propagation of errors under
control, then error correction stands a fighting chance of keeping up, allowing
errors to be corrected at a high enough rate to allow the quantum computation
to function as intended.

The starting point for a technical discussion of this issue is the recognition
that two-qubit gates (or multiple-qubit gates more generally) can propagate
errors, even when they function perfectly.
For instance, consider a controlled-NOT gate, and suppose that an $X$ error
occurs on the control qubit just prior to the controlled-NOT gate being
performed.
As we already observed in Lesson~\ref{lesson:correcting-quantum-errors}
\emph{(Correcting Quantum Errors)}, this is equivalent to an $X$ error
occurring on \emph{both} qubits after the controlled-NOT is performed.
And the situation is similar for a $Z$ error acting on the target rather than
the control prior to the controlled-NOT gate being performed.

\begin{figure}[b]
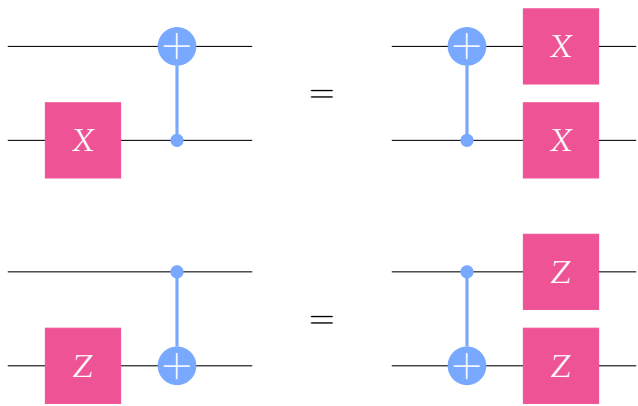

  \begin{center}
   
  \end{center}
  \caption{CNOT gates propagate $X$ and $Z$ errors.}
  \label{fig:CNOT-error-propagation}
\end{figure}

This is a \emph{propagation of errors}, because the unfortunate location of an
$X$ or $Z$ error prior to the controlled-NOT gate effectively turns it into two
errors after the controlled-NOT gate.
This happens even when the controlled-NOT gate is perfect, and we must not
forget that a given controlled-NOT gate may itself be noisy, which can
\emph{create} correlated errors on two qubits.

\begin{figure}[t]
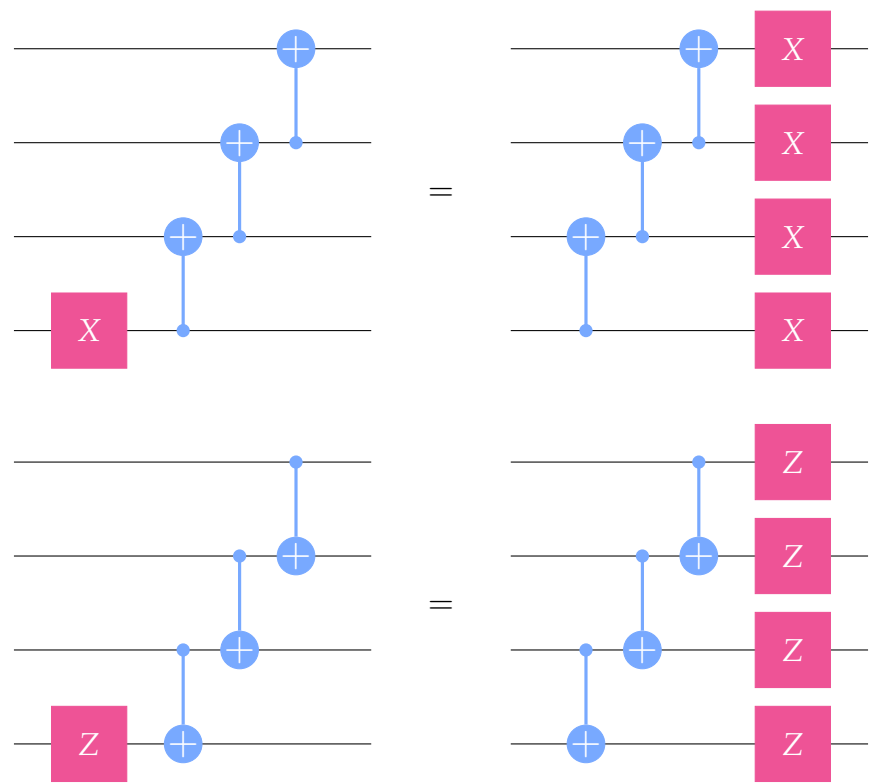

  \begin{center}

    
  \end{center}
  \caption{Multiple CNOT gates can further propagate $X$ and $Z$ errors.}
  \label{fig:multiple-CNOT-error-propagation}
\end{figure}

Adding to our concern is the fact that subsequent two-qubit gates might
propagate these errors even further, as
Figure~\ref{fig:multiple-CNOT-error-propagation} suggests.
In some sense, we can never get around this; so long as we use multiple-qubit
gates, there will be a potential for error propagation.
However, as we'll discuss in the subsections that follow, steps can be taken to
limit the damage this causes, allowing for propagated errors to be managed.

\subsection{Transversal gate implementations}

The simplest known way to mitigate error propagation in fault-tolerant quantum
circuits is to implement gates \emph{transversally}, which means building
gadgets for them that have a certain simple form.
Specifically, the gadgets must be a \emph{tensor product} of operations (or, in
other words, a depth-one quantum circuit), where each operation can only act on
a single qubit \emph{position} within each code block that it touches.
This is perhaps easiest to explain through some examples.

\subsubsection{Examples of transversal gate implementations}

\begin{figure}[t]
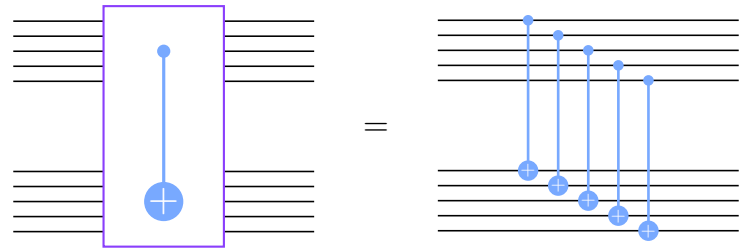

  \begin{center}
    \vspace{-2mm}
    

  \end{center}
  \caption{A transversal implementation of a CNOT gate for CSS codes.}
  \label{fig:transversal-CNOT}
\end{figure}

Consider Figure~\ref{fig:transversal-CNOT}, which suggests a transversal
implementation of a CNOT gate.
This particular implementation, where CNOTs are performed qubit by qubit, only
works for CSS codes --- but it does, in fact, work for \emph{all} CSS codes.

There are two code blocks in this figure, each depicted as consisting of five
qubits (although it could be more, as has already been suggested).
The circuit on the right has depth one, and each of the CNOT gates acts on a
single qubit position within each block:
both the control and target for the first CNOT is the topmost qubit (i.e.,
qubit 0 using the Qiskit numbering convention), both the control and target for
the second CNOT is the qubit second from top (i.e., qubit 1), and so on.
Hence, this is a transversal gadget.

For a second example --- really a class of examples --- consider any Pauli
gate.
Pauli gates can always be implemented transversally, for any stabilizer code,
by building gadgets that are composed of Pauli operations.
In particular, every Pauli operation on a logical qubit encoded by a stabilizer
code can be implemented transversally by choosing an appropriate Pauli
operation on the physical qubits used for the encoding.
This is consistent with a fact that was mentioned in passing in
Lesson~\ref{lesson:the-stabilizer-formalism}
\emph{(The Stabilizer Formalism)}: up to a global phase, Pauli operations that
commute with every stabilizer generator of a stabilizer code act like Pauli
operations on the qubit or qubits encoded by that code.

As a specific example, consider the $9$-qubit Shor code, for which standard
basis states could be encoded as follows.
\[
\begin{aligned}
  \vert 0\rangle &
  \:\mapsto\:
  \frac{1}{2\sqrt{2}}
  (\vert 000\rangle + \vert 111\rangle) \otimes
  (\vert 000\rangle + \vert 111\rangle) \otimes
  (\vert 000\rangle + \vert 111\rangle) \\[3mm]
  \vert 1\rangle &
  \:\mapsto\:
  \frac{1}{2\sqrt{2}}
  (\vert 000\rangle - \vert 111\rangle) \otimes
(\vert 000\rangle - \vert 111\rangle) \otimes
  (\vert 000\rangle - \vert 111\rangle)
\end{aligned}
\]
An $X$ gate on the logical qubit encoded by this code can be implemented
transversally by the $9$-qubit Pauli operation
\[
Z \otimes \mathbb{I} \otimes \mathbb{I} \otimes
Z \otimes \mathbb{I} \otimes \mathbb{I} \otimes
Z \otimes \mathbb{I} \otimes \mathbb{I}
\]
while a $Z$ gate on the logical qubit can be implemented transversally by the
$9$-qubit Pauli operation
\[
X \otimes X \otimes X \otimes
\mathbb{I} \otimes \mathbb{I} \otimes \mathbb{I} \otimes
\mathbb{I} \otimes \mathbb{I} \otimes \mathbb{I}.
\]
Both of these Pauli operations have weight $3,$ which is the minimum weight
required.
(The $9$-qubit Shor code has distance $3,$ so any non-identity Pauli operation
of weight $2$ or less is detected as an error.)

And, for a third example, the $7$-qubit Steane code (and indeed every color
code) allows for a transversal implementation of \emph{all} Clifford gates.
We've already seen how CNOT gates are implemented transversally for any CSS
code, so it remains to consider $H$ and $S$ gates.
A Hadamard gate applied to all $7$ qubits of the Steane code is equivalent to
$H$ being applied to the logical qubit it encodes, while an $S^{\dagger}$ gate
(as opposed to an $S$ gate) applied to all $7$ qubits is equivalent to a
logical $S$ gate.

\subsubsection{Error propagation for transversal gadgets}

Now that we know what transversal implementations of gates are, let us discuss
their connection to error propagation.

For a transversal implementation of a single-qubit gate, we simply have a
tensor product of single-qubit gates in our gadget, which acts on a code block
of physical qubits for the chosen quantum error correcting code.
Although any of these gates could fail and introduce an error, there will be no
\emph{propagation} of errors because no multiple-qubit gates are involved.
Immediately after the gadget is applied, error correction is performed; and if
the number of errors introduced by the gadget (or while the gadget is being
performed) is sufficiently small, the errors will be corrected.
So, if the rate of errors introduced by faulty gates is sufficiently small,
error correction has a good chance to succeed.

For a transversal implementation of a two-qubit gate, on the other hand, there
is the potential for a propagation of errors --- there is simply no way to
avoid this, as we have already observed.
The essential point, however, is that a transversal gadget can never cause a
propagation of errors \emph{within a single code block}.

For example, considering the transversal implementation of a CNOT gate for a
CSS code described above, an $X$ error could occur on the top qubit of the top
code block right before the gadget is performed, and the first CNOT within the
gadget will propagate that error to the top qubit in the lower block.
However, the two resulting errors are now in \emph{separate} code blocks.
So, assuming our code can correct an $X$ error, the error correction steps that
take place after the gadget will correct the two $X$ errors individually ---
because only a single error occurs within each code block.
In contrast, if error propagation were to happen inside of the \emph{same} code
block, it could turn a low-weight error into a high-weight error that the code
cannot handle.

\subsubsection{Non-universality of transversal gates}

For two different stabilizer codes, it may be that a particular gate can be
implemented transversally with one code but not the other.
For example, while it is not possible to implement a $T$ gate transversally
using the $7$-qubit Steane code, there are other codes for which this is
possible.

Unfortunately, it is never possible, for any non-trivial quantum error
correcting code, to implement a \emph{universal} set of gates transversally.
This fact is known as the \emph{Eastin--Knill theorem}.

\begin{callout}[title = {Eastin--Knill theorem}]
  For any quantum error correcting code with distance at least 2, the set of
  logical gates that can be implemented transversally generates a set of
  operations that (up to a global phase) is discrete, and is therefore not
  universal.
\end{callout}

The proof of this theorem will not be explained here.
It is not a complicated proof, but it does require a basic knowledge of Lie
groups and Lie algebras, which are not among the course prerequisites.
The basic idea, however, can be conveyed in intuitive terms:
Infinite families of transversal operations can't possibly stay within the code
space of a non-trivial code because minuscule differences in transversal
operations are well-approximated by low-weight Pauli operations, which the code
detects as errors.

In summary, transversal gadgets offer a simple and inherently fault-tolerant
implementation of gates --- but for any reasonable choice of a quantum error
correcting code, there will never be a universal gate set that can be
implemented in this way, which necessitates the use of alternative gadgets.

\subsection{Magic states}

Given that it is not possible, for any non-trivial choice for a quantum error
correcting code, to implement a universal set of quantum gates transversally,
we must consider other methods to implement gates fault-tolerantly.
One well-known method is based on the notion of \emph{magic states}, which are
quantum states of qubits that enable fault-tolerant implementations of certain
gates.

\subsubsection{Implementing gates with magic states}

Let us begin by considering $S$ and $T$ gates, which have matrix descriptions
as follows.
\[
S = \begin{pmatrix}
  1 & 0\\
  0 & i
\end{pmatrix}
= \begin{pmatrix}
  1 & 0\\
  0 & e^{i\pi/2}
\end{pmatrix}
\qquad\text{and}\qquad
T = \begin{pmatrix}
  1 & 0\\
  0 & \frac{1+i}{\sqrt{2}}
\end{pmatrix}
= \begin{pmatrix}
  1 & 0\\
  0 & e^{i\pi/4}
\end{pmatrix}
\]
By definition, $S$ is a Clifford operation, while $T$ is not;
it is not possible to implement a $T$ gate with a circuit composed of Clifford
gates ($H$ gates, $S$ gates, and CNOT gates).

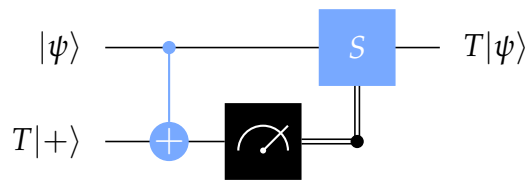
\begin{figure}[!ht]
  \begin{center}
    \begin{tikzpicture}[
        line width = 0.6pt,
        control/.style={%
          circle,
          fill=CircuitBlue,
          minimum size = 5pt,
          inner sep=0mm},
        gate/.style={%
          inner sep = 0,
          fill = CircuitBlue,
          draw = CircuitBlue,
          text = white,
          minimum size = 10mm},
        not/.style={%
          circle,
          fill = CircuitBlue,
          draw = CircuitBlue,
          text = white,
          minimum size = 5mm,
          inner sep=0mm,
          label = {center:\textcolor{white}{\large $+$}}
        },
        blackgate/.style={%
          inner sep = 0,
          fill = black,
          draw = black,
          text = white,
          minimum size = 10mm},
        gadget/.style={%
          fill = white,
          draw = Highlight,
          thick}
      ]

      \node (In0) at (-2.25,1.25) {};
      \node (In1) at (-2.25,0) {};
      \node (Out0) at (2.5,1.25) {};

      \node[blackgate] (M) at (0,0) {};
      \readout{M}
      
      \draw (In0.east) -- (Out0);
      \draw (In1) -- (M.west);
      
      \node[control] (C) at (-1.25,1.25) {};
      \node[control,fill=black] (CC) at (1.25,0) {};
      
      \node[not] (Not) at (-1.25,0) {};
      \draw[thick,draw=CircuitBlue] (C.center) -- (Not.north);
      
      \node[gate] (S) at (1.25,1.25) {$S$};

      \foreach \x in {-0.3mm,0.3mm} {
        \draw ([xshift=\x]CC.center) -- ([xshift=\x]S.south);
          \draw ([yshift=\x]M.east) -- ([yshift=\x]CC.center);
        }

        \node[anchor = east] at (In0) {$\ket{\psi}$};
        \node[anchor = east] at (In1) {$T\ket{+}$};
        \node[anchor = west] at (Out0) {$T\ket{\psi}$};
        
      \end{tikzpicture}
  \end{center}
  \caption{An implementation of a $T$ gate using a magic state.}
  \label{fig:magic-state-injection}
\end{figure}

However, it is possible to implement a $T$ gate (up to a global phase) with a
circuit composed of Clifford gates if, in addition, we have a copy of the state
\[
T\vert {+} \rangle = \frac{1}{\sqrt{2}} \vert 0 \rangle +
\frac{e^{i\pi/4}}{\sqrt{2}} \vert 1\rangle,
\]
and we allow for standard basis measurements and for gates to be classically
controlled.
In particular, the circuit in Figure~\ref{fig:magic-state-injection} represents
one way to do this.
The phenomenon on display here is a somewhat simplified example of
\emph{quantum gate teleportation}.

To check that this circuit works correctly, we can first compute the action of
the CNOT gate on the input.
\[
T \vert {+} \rangle \otimes \vert\psi\rangle
\stackrel{\text{CNOT}}{\longmapsto}
\frac{1}{\sqrt{2}} \vert 0\rangle \otimes T \vert \psi\rangle + \frac{1+i}{2}
\vert 1\rangle \otimes T^{\dagger} \vert \psi\rangle
\]

The measurement therefore gives the outcomes $0$ and $1$ with equal
probability.
If the outcome is $0,$ the $S$ gate is not performed, and the output state is
$T\vert\psi\rangle;$ and if the outcome is $1,$ the $S$ gate is performed, and
the output state is $ST^{\dagger}\vert\psi\rangle = T\vert \psi\rangle.$

The state $T\vert {+}\rangle$ is called a \emph{magic state} in this context,
although it's not unique in this regard: other states are also called magic
states when they can be used in a similar way (for possibly different gates and
using different circuits).
For example, exchanging the state $T\vert{+}\rangle$ for the state
$S\vert{+}\rangle$ and replacing the $S$ gate in the circuit above with a $Z$
gate implements an $S$ gate --- which is potentially useful for fault-tolerant
quantum computation using a code for which $S$ gates cannot be implemented
transversally.

\subsubsection{Fault-tolerant gadgets from magic states}

It may not be clear that using magic states to implement gates is helpful for
fault-tolerance.
For the $T$ gate implementation described above, for instance, it appears that
we still need to apply a $T$ gate to a $\vert{+}\rangle$ state to obtain a
magic state, which we then use to implement a $T$ gate.
So what is the advantage of using this approach for fault-tolerance?
Here are three key points that provide an answer to this question.

\begin{enumerate}
\item
  The creation of magic states does not necessitate applying the gate we're
  attempting to implement to a particular state. For example, applying a $T$
  gate to a $\vert {+} \rangle$ state is not the only way to obtain a
  $T\vert{+}\rangle$ state.

\item
  The creation of magic states can be done separately from the computation in
  which they're used. This means that errors that arise in the magic state
  creation process will not propagate to the actual computation being
  performed.

\item
  If the individual gates in the circuit implementing a chosen gate using a
  magic state can be implemented fault-tolerantly, and we assume the
  availability of magic states, we obtain a fault-tolerant implementation of
  the chosen gate.
\end{enumerate}

\begin{figure}[!ht]
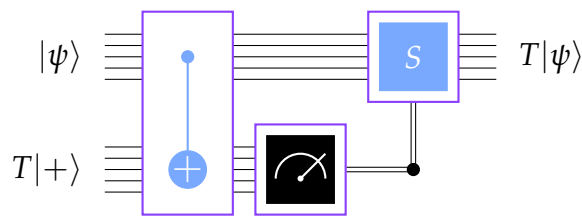

  \begin{center}

  \end{center}
  \caption{An implementation of a $T$ gate on an encoded qubit using an
  encoded magic state.}
  \label{fig:encoded-magic-state-injection}
\end{figure}

To simplify the discussion that follows, let's focus in on $T$ gates
specifically --- keeping in mind that the methodology can be extended to other
gates.
A fault-tolerant implementation of a $T$ gate using magic states takes the form
suggested by Figure~\ref{fig:encoded-magic-state-injection}.

Qubits in the original $T$-gate circuit correspond to logical qubits in this
diagram, which are encoded by whatever code we're using for fault-tolerance.
The inputs and outputs in the diagram should therefore be understood as
\emph{encodings} of these states.
This means, in particular, that we actually don't need just magic states --- we
need \emph{encoded} magic states.
The gates in the original $T$-gate circuit are here replaced by gadgets, which
we assume are fault-tolerant.

This particular figure therefore suggests that we already have fault-tolerant
gadgets for CNOT gates and $S$ gates.
For a color code, these gadgets could be transversal;
for a surface code (or any other CSS code), the CNOT can be performed
transversally, while the $S$ gate gadget might itself be implemented using
magic states, as was earlier suggested is possible.
(The figure also suggests that we have a fault-tolerant gadget for performing a
standard basis measurement, which we've ignored thus far.
This could actually be challenging for some codes selected to make it so, but
for a CSS code it's a matter of measuring each physical qubit followed by
classical post-processing.)

\begin{figure}[!ht]
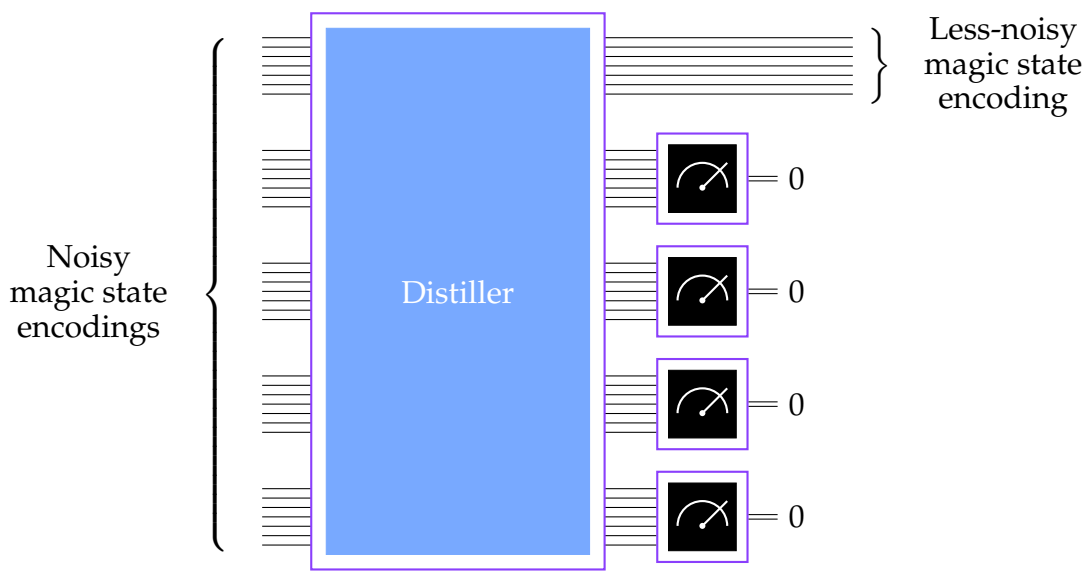

  \begin{center}

  \end{center}
  \caption{Magic state distillation on encoded states.}
  \label{fig:encoded-magic-state-distillation}
\end{figure}

The implementation is therefore fault-tolerant, assuming we have an encoding of
a magic state $T\vert{+}\rangle.$
But, we still haven't addressed the issue of how we obtain an encoding of this
state.
One way to obtain encoded magic states (or, perhaps more accurately, to make
them better) is through a process known as \emph{magic state distillation.}
The diagram in Figure~\ref{fig:encoded-magic-state-distillation} illustrates
what this process looks like at the highest level.

In words, a collection of noisy encoded magic states is fed into a special type
of circuit known as a \emph{distiller}.
All but one of the output blocks is measured --- meaning that \emph{logical}
qubits are measured with standard basis measurements.
If any of the measurement outcomes is $1,$ the process has failed and must be
restarted.
If, however, every measurement outcome is $0,$ the resulting state of the top
code block will be a less noisy encoded magic state.
This state could then join four more as inputs into another distiller, or used
to implement a $T$ gate if it is deemed to be sufficiently close to a true
encoded magic state.
Of course, the process must begin somewhere, with one possibility being to
prepare them non-fault-tolerantly.

There are different known ways to build the distiller itself, but they will not
be explained or analyzed here.
At a logical level, the typical approach --- remarkably and somewhat
coincidentally --- is to run an encoding circuit for a stabilizer code in
reverse!
This could, in fact, be a different stabilizer code from the one used for error
correction.
For example, one could potentially use a surface or color code for error
correction, but run an encoder for the $5$-qubit code in reverse for the sake
of magic state distillation.
Encoding circuits for stabilizer codes only require Clifford gates, which
simplifies the fault-tolerant implementation of a distiller.
In actuality, the specifics are dependent on the codes that are used.

In summary, this section has aimed to provide only a very high-level discussion
of magic states, with the intention being to provide just a basic idea of how
it works.

It is sometimes claimed that the overhead for using magic states to implement
gates fault-tolerantly along these lines would be extremely high, with the vast
majority of the work going into the distillation process.
However, this is actually not so clear --- there are many potential ways to
optimize these processes.
There are, in addition, alternative approaches to building fault-tolerant
gadgets for gates that cannot be implemented transversally.
For example, \emph{code deformation} and \emph{code switching} are keywords
associated with some of these schemes --- and new ways continue to be developed
and refined.

\subsection{Fault-tolerant error correction}

In addition to the implementation of the various gadgets required for a
fault-tolerant implementation of a given quantum circuit, there is another
important issue that must be recognized: the implementation of the error
correction steps themselves.
This goes back to the idea that anything involving quantum information is
susceptible to errors --- including the circuits that are themselves meant to
correct errors.

Consider, for instance, the type of circuit described in
Lesson~\ref{lesson:the-stabilizer-formalism} \emph{(The Stabilizer Formalism)}
for measuring stabilizer generators non-destructively using phase estimation.
These circuits are clearly not fault-tolerant because they can cause errors to
propagate within the code block on which they operate.
This might seem rather problematic, but there are multiple known ways to
perform error correction fault-tolerantly in a way that does not cause errors
to propagate within the code blocks being corrected.

\begin{figure}[!ht]
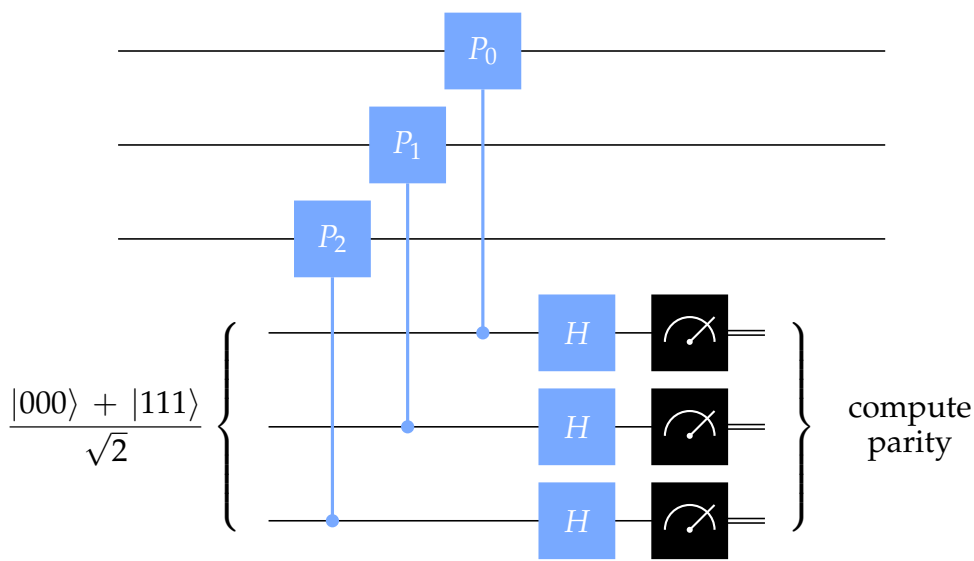

  \begin{center}
%
  \end{center}
  \caption{A circuit for measuring a stabilizer generator of the form
    $P_2\otimes P_1\otimes P_0$ using Shor error correction.}
  \label{fig:Shor-error-detection}
\end{figure}

One method is known as \emph{Shor error correction,} because it was first
discovered by Peter Shor.
The idea is to perform syndrome measurements using a so-called \emph{cat
state,} which is an $n$-qubit state of the form
\[
\frac{1}{\sqrt{2}} \vert 0^n \rangle + \frac{1}{\sqrt{2}} \vert 1^n \rangle,
\]
where $0^n$ and $1^n$ refer to the all-zero and all-one strings of length $n.$
For instance, this is a $\vert\phi^+\rangle$ state when $n=2$ and a GHZ state
when $n=3,$ but in general, Shor error correction requires a state like this
for $n$ being the weight of the stabilizer generator being measured.
As an example, the circuit shown in Figure~\ref{fig:Shor-error-detection}
measures a stabilizer generator of the form $P_2\otimes P_1 \otimes P_0.$

This necessitates the construction of the cat state itself, and to make it work
reliably in the presence of errors and potentially faulty gates, the method
actually requires repeatedly running circuits like this to make inferences
about where different errors may have occurred during the process.

\begin{figure}[!ht]
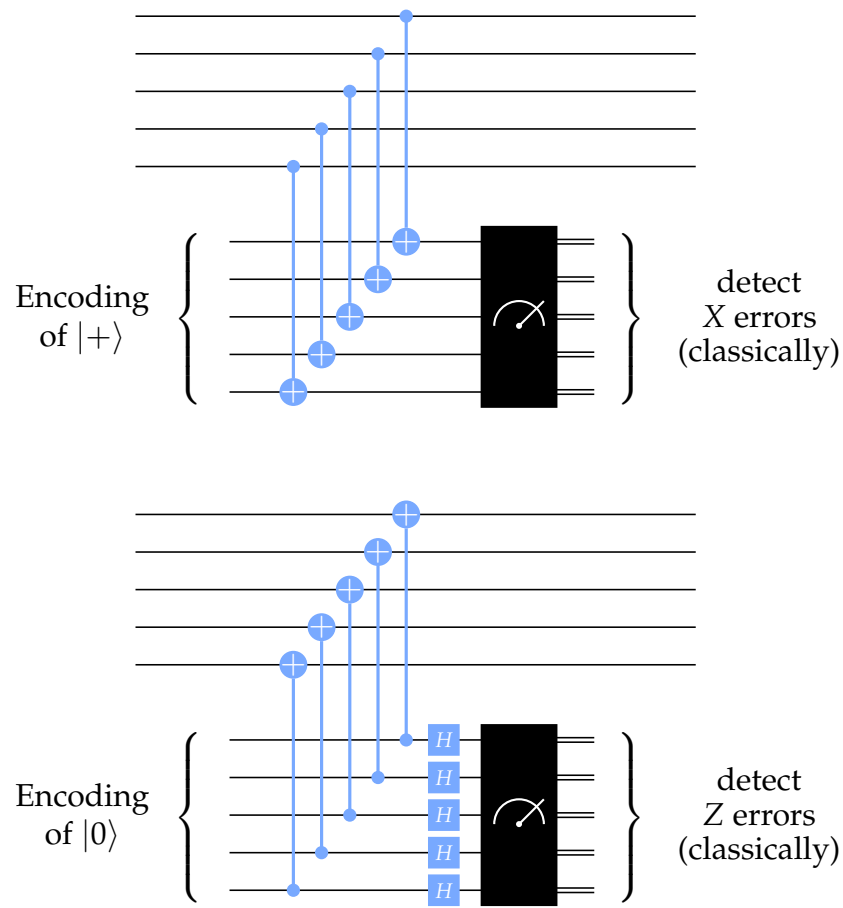

  \begin{center}
%
  \end{center}
  \caption{Circuits for detecting $X$ and $Z$ errors using Steane error
    correction.}
  \label{fig:Steane-error-detection}
\end{figure}

An alternative method is known as \emph{Steane error correction.}
This method works differently, and it only works for CSS codes.
The idea is that we don't actually perform the syndrome measurements on the
encoded quantum states in the circuit we're trying to run, but instead we
\emph{intentionally} propagate errors to a workspace system, and then measure
that system and \emph{classically} detect errors.
The circuit diagrams in Figure~\ref{fig:Steane-error-detection} illustrate how
this can be done for detecting $X$ and $Z$ errors, respectively.
A related method known as \emph{Knill error correction} extends this method to
arbitrary stabilizer codes using teleportation.

\section{Threshold theorem}

The final topic of discussion for the lesson is a very important theorem known
as the \emph{threshold theorem}.
Here is a somewhat informal statement of this theorem.

\begin{callout}[title = {Threshold theorem}]
  A quantum circuit having size $N$ can be implemented with high accuracy by a
  noisy quantum circuit, provided that the probability of error at each
  location in the noisy circuit is below a fixed, nonzero threshold value
  $p_{\text{th}} > 0.$ The size of the noisy circuit scales as $O(N \log^c(N))$
  for a positive constant $c.$
\end{callout}

\noindent
In simple terms, it says that if we have any quantum circuit having $N$ gates,
where $N$ can be as large as we like, then it's possible to implement that
circuit with high accuracy using a noisy quantum circuit, provided that the
level of noise is below a certain threshold value that is independent of N.
Moreover, it isn't too expensive to do this, in the sense that the size of the
noisy circuit required is on the order of $N$ times some constant power of the
logarithm of $N.$

To state the theorem more formally requires being specific about the noise
model, which will not be done in this lesson.
It can, for instance, be proved for the independent stochastic noise model that
was mentioned earlier, where errors occur independently at each possible
location in the circuit with some probability strictly smaller than the
threshold value, but it can also be proved for more general noise models where
there can be correlations among errors.

This is a theoretical result, and the most typical way it is proved doesn't
necessarily translate to a practical approach, but it does nevertheless have
great practical importance.
In particular, it establishes that there is no fundamental barrier to
performing quantum computations using noisy components; as long as the error
rate for these components is below the threshold value, they can be used to
build reliable quantum circuits of arbitrary size.
An alternative way to state its importance is to observe that, if the theorem
wasn't true, it would be hard to imagine large-scale quantum computing ever
becoming a reality.

There are many technical details involved in formal proofs of (formal
statements of) this theorem, and those details will not be communicated here
--- but the essential ideas can nevertheless be explained at an intuitive
level.
To make this explanation as simple as possible, let's imagine that we use the
$7$-qubit Steane code for error correction.
This would be an impractical choice for an actual physical implementation ---
as would be reflected by a minuscule threshold value $p_{\text{th}}$ --- but it
works well to convey the main ideas.
This explanation will also be rather cavalier about the noise model, with the
assumption being that an error strikes each location in a fault-tolerant
implementation independently with probability $p.$

Now, if the probability $p$ is larger than the reciprocal of $N,$ the size of
the circuit we aim to implement, chances are very good that an error will
strike somewhere.
So, we can attempt to run a fault-tolerant implementation of this circuit,
following the prescription outlined in the lesson.
We may then ask ourselves the question suggested earlier: Is this making things
better or worse?

If the probability $p$ of an error at each location is too large, then our
efforts will not help and may even make things worse, just like the $9$-qubit
Shor code doesn't help if the error probability is above 3.23\% or so.
In particular, the fault-tolerant implementation is considerably larger than
our original circuit, so there are a lot more locations where errors could
strike.

However, if $p$ is small enough, then we will succeed in reducing the error
probability for the \emph{logical} computation we're performing.
(In a formal proof, we would need to be very careful at this point: errors in
the logical computation will not necessarily be accurately described by the
original noise model.
This, in fact, motivates less forgiving noise models where errors might not be
independent --- but we will sweep this detail under the rug for the sake of
this explanation.)

In greater detail, in order for a logical error to occur in the original
circuit, at least two errors must fall into the same code block in the
fault-tolerant implementation, given that the Steane code can correct any
single error in a code block.
Keeping in mind there are many different ways to have two or more errors in the
same code block, it is possible to argue that the probability of a logical
error at each location in the original circuit is at most $C p^2$ for some
fixed positive real number $C$ that depends on the code and the gadgets we use,
but critically not on $N,$ the size of the original circuit.
If $p$ is smaller than $1/C,$ which is the number we can take as our threshold
value $p_{\text{th}},$ this translates to a reduction in error.

However, this new error rate might still be too high to allow the entire
circuit to work correctly.
A natural thing to do at this point is to choose a better code and better
gadgets to drive the error rate down to a point where the implementation is
likely to work.
Theoretically speaking, a simple way to argue that this is possible is to
\emph{concatenate}.
That is to say, we can think of the \emph{fault-tolerant} implementation of the
original circuit as if it were any other quantum circuit, and then implement
this new circuit fault-tolerantly, using the same scheme.
We can then do this again and again, as many times as we need to reduce the
error rate to a level that allows the original computation to work.

To get a rough idea for how the error rate decreases through this method, let's
consider how it works for a few iterations.
Note that a rigorous analysis would need to account for various technical
details we're omitting here.

We start with the error probability $p$ for locations in the original circuit.
Presuming that $p < p_{\text{th}} = 1/C,$ the logical error rate can be bounded
by $Cp^2 = (Cp) p$ after the first iteration.
By treating the fault-tolerant implementation as any other circuit, and
implementing it fault-tolerantly, we obtain a bound on the logical error rate
of
\[
C \bigl((Cp) p \bigr)^2 = (Cp)^3 p.
\]
Another iteration reduces the error bound further, to
\[
C \bigl((Cp)^3 p \bigr)^2 = (Cp)^7 p.
\]
Continuing in this manner for a total of $k$ iterations leads to a logical
error rate (for the original circuit) bounded by
\[
(Cp)^{2^k - 1} p,
\]
which is \emph{doubly exponential} in $k.$

This means that we don't need too many iterations to make the error rate
extremely small.
Meanwhile, the circuits are growing in size with each level of concatenation,
but the size only increases \emph{singly exponentially} in the number of
levels~$k.$
This is because, with each level of fault-tolerance, the size grows by at most
a factor determined by the maximum size of the gadgets being used.
When it is all put together, and an appropriate choice for the number of levels
of concatenation is made, we obtain the threshold theorem.

So, what is this threshold value in reality?
The answer depends on the code and the gadgets used.
For the Steane code together with magic state distillation, it is minuscule and
probably unlikely to be achievable in practice.
But, using surface codes and state of the art gadgets, the threshold has been
estimated to be on the order of 0.1\% to 1\%.

As new codes and methods are discovered, it is reasonable to expect the
threshold value to increase, while simultaneously the level of noise in actual
physical components will decrease.
Reaching the point at which large-scale quantum computations can be implemented
fault-tolerantly will not be easy, and will not happen overnight.
But, this theorem, together with advances in quantum codes and quantum
hardware, provide us with optimism as we continue to push forward to reach the
ultimate goal of building a large-scale, fault-tolerant quantum computer.


\backmatter

\appendix
\chapter*{Bibliography}

This bibliography includes numerous references that are relevant to this
course, including books, surveys, and research papers, divided into separate
lists: background and prerequisite material (such as linear algebra,
probability theory, and basic theoretical computer science); general references
that cover topics spanning or relevant to multiple units; and unit-specific
references.

Some of these references represent original research discoveries while others
are pedagogical in nature or are secondary sources that refine and/or simplify
the subject matter.
Some are connected directly to facts or discoveries mentioned in the text while
others are merely relevant or offer further explorations of various topics.
In some cases, only small portions of these sources may be relevant to this
course.
I have made no attempt to categorize them along these lines.

This bibliography should not be seen as a comprehensive list or a historical
record that aims to give proper attribution to discoveries and developments in
the field.
Rather, it's a list of suggestions for background or further reading.
After over 30 years studying, researching, and teaching quantum information and
computation, it would be extremely difficult for me to produce a comprehensive
list --- and the truth of the matter is that this course was informed by many
sources that are not listed here, as well as talks, presentations, and
personal conversations over the years.
I do regret any omissions, but this should be a good start for those wishing to
learn more.

\raggedright

\subsection*{Background and prerequisite material references}
\begin{trivlist}
  \setlength{\itemsep}{3mm}
\item Sheldon Axler. {\em Linear Algebra Done Right}. Springer, 3rd edition, 2015.
\item Rajendra Bhatia. {\em Matrix Analysis}. Springer, 1997.
\item Stephen Friedberg, Arnold Insel, and Lawrence Spence. {\em Linear Algebra}. Prentice Hall, 4th edition, 2003.
\item David Griffiths and Darrell Schroeter. {\em Introduction to Quantum Mechanics}. Cambridge University Press, 3rd edition, 2018.
\item Kenneth Hoffman and Ray Kunze. {\em Linear Algebra}. Prentice Hall, 2nd edition, 1971.
\item Roger Horn and Charles Johnson. {\em Matrix Analysis}. Cambridge University Press, 1985.
\item John Hunter. An introduction to real analysis, 2025. Available at \url{https://www.math.ucdavis.edu/~hunter/intro_analysis_pdf/intro_analysis.pdf}.
\item Sal Khan. Linear algebra. Khan Academy, 2025. Video series available at \url{https://www.khanacademy.org/math/linear-algebra}.
\item Tristan Needham. {\em Visual Complex Analysis}. Oxford University Press, 1997.
\item Sheldon Ross. {\em A First Course in Probability}. Pearson, 9th edition, 2014.
\item Michael Sipser. {\em Introduction to the Theory of Computation}. Cengage Learning, 3rd edition, 2013.
\end{trivlist}

\subsection*{General references}
\begin{trivlist}
  \setlength{\itemsep}{3mm}
\item Richard Feynman. Simulating physics with computers. {\em International Journal of Theoretical Physics}, 21(6/7):467--488, 1982.
\item Phillip Kaye, Raymond Laflamme, and Michele Mosca. {\em An Introduction to Quantum Computing}. Oxford University Press, 2007.
\item Alexei Kitaev. Quantum computations: algorithms and error correction. {\em Russian Mathematical Surveys}, 52(6):1191--1249, 1997.
\item Alexei Kitaev, Alexander Shen, and Mikhail Vyalyi. {\em Classical and Quantum Computation}, volume~47 of {\em Graduate Studies in Mathematics}. American Mathematical Society, 2002.
\item N.~David Mermin. {\em Quantum Computer Science: An Introduction}. Cambridge University Press, 2007.
\item Michael Nielsen and Isaac Chuang. {\em Quantum Computation and Quantum Information}. Cambridge University Press, 10th anniversary edition, 2010.
\item John Preskill. {\em Lecture Notes for Physics 229: Quantum Information and Computation}. California Institute of Technology, 2020. Available at \url{https://www.preskill.caltech.edu/ph229/}.
\end{trivlist}

\subsection*{Unit I references}
\begin{trivlist}
  \setlength{\itemsep}{3mm}
\item John Bell. On the {Einstein Podolsky Rosen} paradox. {\em Physics Physique Fizika}, 1(3):195--200, 1964.
\item Charles Bennett and Stephen Wiesner. Communication via one-and two-particle operators on {Einstein--Podolsky--Rosen} states. {\em Physical Review Letters}, 69(20):2881--2884, 1992.
\item Charles Bennett, Gilles Brassard, Claude Cr{\'e}peau, Richard Jozsa, Asher Peres, and William Wootters. Teleporting an unknown quantum state via dual classical and {Einstein--Podolsky--Rosen} channels. {\em Physical Review Letters}, 70(13):1895--1899, 1993.
\item Charles Bennett, Gilles Brassard, Sandu Popescu, Benjamin Schumacher, John Smolin, and William Wootters. Purification of noisy entanglement and faithful teleportation via noisy channels. {\em Physical Review Letters}, 76(5):722--725, 1996.
\item John Clauser, Michael Horne, Abner Shimony, and Richard Holt. Proposed experiment to test local hidden-variable theories. {\em Physical Review Letters}, 23(15):880--884, 1969.
\item Richard Cleve, Peter H{\o}yer, Benjamin Toner, and John Watrous. Consequences and limits of nonlocal strategies. In {\em Proceedings of the 19th Annual IEEE Conference on Computational Complexity}, pages 236--249, 2004.
\item Dennis Dieks. Communication by {EPR} devices. {\em Physics Letters A}, 92(6):271--272, 1982.
\item Paul Dirac. {\em The Principles of Quantum Mechanics}. Clarendon Press, fourth edition, 1958.
\item Ryszard Horodecki, Pawe{\l} Horodecki, Micha{\l} Horodecki, and Karol Horodecki. Quantum entanglement. {\em Reviews of Modern Physics}, 81(2):865--942, 2009.
\item Boris Tsirelson. Quantum generalizations of {Bell's} inequality. {\em Letters in Mathematical Physics}, 4(2):93--100, 1980.
\item William Wootters and Wojciech Zurek. A single quantum cannot be cloned. {\em Nature}, 299(5886):802--803, 1982.
\end{trivlist}

\subsection*{Unit II references}
\begin{trivlist}
  \setlength{\itemsep}{3mm}
\item Sanjeev Arora and Boaz Barak. {\em Computational Complexity: A Modern Approach}. Cambridge University Press, 2009.
\item Eric Bach and Jeffrey Shallit. {\em Algorithmic Number Theory, Volume I: Efficient Algorithms}. MIT Press, 1996.
\item Charles Bennett. Logical reversibility of computation. {\em IBM Journal of Research and Development}, 17:525--532, 1973.
\item Charles Bennett, Ethan Bernstein, Gilles Brassard, and Umesh Vazirani. Strengths and weaknesses of quantum computing. {\em SIAM Journal on Computing}, 26(5):1510--1523, 1997.
\item Ethan Bernstein and Umesh Vazirani. Quantum complexity theory. {\em SIAM Journal on Computing}, 26(5):1411--1473, 1997.
\item Michel Boyer, Gilles Brassard, Peter H{\o}yer, and Alain Tapp. Tight bounds on quantum searching. {\em Fortschritte der Physik}, 46(4-5):493--505, 1998.
\item Oscar Boykin, Tal Mor, Matthew Pulver, Vwani Roychowdhury, and Farrokh Vatan. A new universal and fault-tolerant quantum basis. {\em Information Processing Letters}, 75(3):101--107, 2000.
\item Gilles Brassard, Peter H{\o}yer, Michele Mosca, and Alain Tapp. Quantum amplitude amplification and estimation. {\em Contemporary Mathematics}, 305:53--74, 2002.
\item Richard Cleve, Artur Ekert, Chiara Macchiavello, and Michele Mosca. Quantum algorithms revisited. {\em Proceedings of the Royal Society of London A}, 454(1969):339--354, 1998.
\item James Cooley and John Tukey. An algorithm for the machine calculation of complex {Fourier} series. {\em Mathematics of Computation}, 19:297--301, 1965.
\item Don Coppersmith. An approximate {Fourier} transform useful in quantum factoring. arXiv:quant-ph/0201067, 1994.
\item David Deutsch. Quantum theory, the {Church-Turing} principle and the universal quantum computer. {\em Proceedings of the Royal Society of London A}, 400(1818):97--117, 1985.
\item David Deutsch and Richard Jozsa. Rapid solution of problems by quantum computation. {\em Proceedings of the Royal Society of London A}, 439(1907):553--558, 1992.
\item Edward Fredkin and Tommaso Toffoli. Conservative logic. {\em International Journal of Theoretical Physics}, 21(3/4):219--253, 1982.
\item Lov Grover. A fast quantum mechanical algorithm for database search. In {\em Proceedings of the 28th Annual ACM Symposium on Theory of Computing}, pages 212--219, 1996.
\item Alexei Kitaev. Quantum measurements and the {Abelian} stabilizer problem, 1996. arXiv:quant-ph/9511026.
\item Ronald Rivest, Adi Shamir, and Leonard Adleman. A method for obtaining digital signatures and public-key cryptosystems. {\em Communications of the ACM}, 21(2):120--126, 1978.
\item Peter Shor. Algorithms for quantum computation: discrete logarithms and factoring. In {\em Proceedings of the 35th Annual IEEE Symposium on Foundations of Computer Science}, pages 124--134, 1994. Conference version.
\item Peter Shor. Polynomial-time algorithms for prime factorization and discrete logarithms on a quantum computer. {\em SIAM Journal on Computing}, 26(5):1484--1509, 1997.
\item Daniel Simon. On the power of quantum computation. {\em SIAM Journal on Computing}, 26(5):1474--1483, 1997.
\item Andrew Yao. Quantum circuit complexity. In {\em Proceedings of the 34th Annual IEEE Symposium on Foundations of Computer Science}, pages 352--361, 1993.
\item Christof Zalka. Grover's quantum searching algorithm is optimal. {\em Physical Review A}, 60(4):2746--2751, 1999.
\end{trivlist}

\subsection*{Unit III references}
\begin{trivlist}
  \setlength{\itemsep}{3mm}
\item Man-Duen Choi. Completely positive linear maps on complex matrices. {\em Linear Algebra and its Applications}, 10(3):285--290, 1975.
\item Christopher Fuchs and Jeroen van~de Graaf. Cryptographic distinguishability measures for quantum-mechanical states. {\em IEEE Transactions on Information Theory}, 45(4):1216--1227, 1999.
\item Carl Helstrom. Quantum detection and estimation theory. {\em Mathematics in Science and Engineering}, 123, 1976.
\item Alexander Holevo. {\em Quantum Systems, Channels, Information: A Mathematical Introduction}. De Gruyter, 2012.
\item Lane Hughston, Richard Jozsa, and William Wootters. A complete classification of quantum ensembles having a given density matrix. {\em Physics Letters A}, 183:14--18, 1993.
\item Richard Jozsa. Fidelity for mixed quantum states. {\em Journal of Modern Optics}, 41(12):2315--2323, 1994.
\item Karl Kraus. States, effects, and operations: fundamental notions of quantum theory. {\em Lecture Notes in Physics}, 190, 1983.
\item Benjamin Schumacher. Sending quantum entanglement through noisy channels. {\em Physical Review A}, 54(4):2614--2628, 1996.
\item W.~Forrest Stinespring. Positive functions on {$C^{*}$}-algebras. {\em Proceedings of the American Mathematical Society}, 6(2):211--216, 1955.
\item Armin Uhlmann. The "transition probability" in the state space of a *-algebra. {\em Reports on Mathematical Physics}, 9(2):273--279, 1976.
\item John Watrous. {\em The Theory of Quantum Information}. Cambridge University Press, 2018.
\item Mark Wilde. {\em Quantum Information Theory}. Cambridge University Press, 2nd edition, 2017.
\item Andreas Winter. Coding theorem and strong converse for quantum channels. {\em IEEE Transactions on Information Theory}, 45(7):2481--2485, 1999.
\end{trivlist}

\subsection*{Unit IV references}
\begin{trivlist}
  \setlength{\itemsep}{3mm}
\item Dorit Aharonov and Michael Ben-Or. Fault-tolerant quantum computation with constant error. {\em Proceedings of the 29th Annual ACM Symposium on Theory of Computing}, pages 176--188, 1997.
\item Dave Bacon. Operator quantum error-correcting subsystems for self-correcting quantum memories. {\em Physical Review A}, 73(1):012340, 2006.
\item Hector Bombin and Miguel~Angel Martin-Delgado. Topological quantum distillation. {\em Physical Review Letters}, 97(18):180501, 2006.
\item Sergey Bravyi and Alexei Kitaev. Quantum codes on a lattice with boundary. {\em arXiv: quant-ph/9811052}, 1998.
\item Sergey Bravyi, Andrew Cross, Jay Gambetta, Dmitri Maslov, Patrick Rall, and Theodore Yoder. High-threshold and low-overhead fault-tolerant quantum memory. {\em Nature}, 627:778--782, 2024.
\item Robert Calderbank and Peter Shor. Good quantum error-correcting codes exist. {\em Physical Review A}, 54(2):1098--1105, 1996.
\item Eric Dennis, Alexei Kitaev, Andrew Landahl, and John Preskill. Topological quantum memory. {\em Journal of Mathematical Physics}, 43(9):4452--4505, 2002.
\item Bryan Eastin and Emanuel Knill. Restrictions on transversal encoded quantum gate sets. {\em Physical Review Letters}, 102(11):110502, 2009.
\item Austin Fowler, Matteo Mariantoni, John Martinis, and Andrew Cleland. Surface codes: Towards practical large-scale quantum computation. {\em Physical Review A}, 86(3):032324, 2012.
\item Daniel Gottesman. {\em Stabilizer codes and quantum error correction}. PhD thesis, California Institute of Technology, 1997. arXiv: quant-ph/9705052.
\item Alexei Kitaev. Fault-tolerant quantum computation by anyons. {\em Annals of Physics}, 303(1):2--30, 2003.
\item Emanuel Knill and Raymond Laflamme. Theory of quantum error-correcting codes. {\em Physical Review A}, 55(2):900--911, 1997.
\item Emanuel Knill, Raymond Laflamme, and Wojciech Zurek. Resilient quantum computation: error models and thresholds. {\em Proceedings of the Royal Society of London A}, 454(1969):365--384, 1998.
\item Emanuel Knill. Quantum computing with realistically noisy devices. {\em Nature}, 434(7029):39--44, 2005.
\item Daniel Lidar and Todd Brun. {\em Quantum Error Correction}. Cambridge University Press, 2013.
\item John Preskill. Reliable quantum computers. {\em Proceedings of the Royal Society of London A}, 454(1969):385--410, 1998.
\item Peter Shor. Scheme for reducing decoherence in quantum computer memory. {\em Physical Review A}, 52(4):R2493--R2496, 1995.
\item Peter Shor. Fault-tolerant quantum computation. In {\em Proceedings of the 37th Annual IEEE Symposium on Foundations of Computer Science}, pages 56--65, 1996.
\item Andrew Steane. Error correcting codes in quantum theory. {\em Physical Review Letters}, 77(5):793--797, 1996.
\item Andrew Steane. Active stabilization, quantum computation, and quantum state synthesis. {\em Physical Review Letters}, 78(11):2252--2255, 1997.
\end{trivlist}

\end{document}